\documentclass[12pt]{article}

\usepackage[margin=1in]{geometry}
\usepackage{amsthm,amsmath,amsfonts,amssymb}

\usepackage[colorlinks,citecolor=blue,urlcolor=blue]{hyperref}
\usepackage{graphicx}

\usepackage[numbers,sort&compress]{natbib}

\usepackage{caption}
\usepackage{algorithm}
\usepackage{algpseudocode}
\algrenewcommand\algorithmicrequire{\textbf{Input:}}
\algrenewcommand\algorithmicensure{\textbf{Output:}}

\theoremstyle{plain}

\newtheorem{theorem}{Theorem}[section]
\newtheorem{lemma}[theorem]{Lemma}

\theoremstyle{remark}
\newtheorem{definition}[theorem]{Definition}

\theoremstyle{plain}
\newtheorem{proposition}[theorem]{Proposition}

\theoremstyle{remark}
\newtheorem{remark}{Remark}

\newcommand{\n}{n}
\newcommand{\np}{n'}
\newcommand{\npln}[1]{\np_{#1}}
\newcommand{\m}{m}
\newcommand{\mln}[1]{\m_{#1}}
\newcommand{\kk}{k}
\newcommand{\kln}[1]{\kk_{#1}}

\newcommand{\bs}{S}
\newcommand{\bsln}[1]{\bs_{#1}}
\newcommand{\dd}{d}
\newcommand{\dln}[1]{\dd_{#1}}
\newcommand{\rr}{r}
\newcommand{\br}{R}
\newcommand{\bcnn}[2]{C_{#1}^{(#2)}}
\newcommand{\tconstant}{t}
\newcommand{\tln}[1]{\tconstant _{#1}}

\newcommand{\uu}{u}
\newcommand{\bu}{U}
\newcommand{\bt}{T}
\newcommand{\bv}{V}
\newcommand{\bw}{W}
\newcommand{\oo}{o}
\newcommand{\bo}{O}
\newcommand{\ooln}[1]{\oo_{#1}}
\newcommand{\bmo}{\mathcal{O}}
\newcommand{\lconstant}{l}

\newcommand{\uun}[1]{\uu_{#1}}
\newcommand{\iconstant}{i}
\newcommand{\jconstant}{j}
\newcommand{\bx}{X}
\newcommand{\bxp}{\bx '}
\newcommand{\bxpn}[1]{\bx '(#1)}
\newcommand{\bypn}[1]{\by '(#1)}
\newcommand{\by}{Y}
\newcommand{\byp}{\by '}
\newcommand{\ba}{A}

\newcommand{\bz}{Z}
\newcommand{\bzn}[1]{Z(#1)}
\newcommand{\bzln}[1]{Z_{#1}}
\newcommand{\bzlnn}[2]{Z_{#1}(#2)}
\newcommand{\bxn}[1]{\bx (#1)}
\newcommand{\byn}[1]{\by (#1)}
\newcommand{\bxln}[1]{\bx _{#1}}
\newcommand{\bxpln}[1]{X' _{#1}}
\newcommand{\bypln}[1]{Y' _{#1}}
\newcommand{\bxplnn}[2]{X' _{#1}(#2)}
\newcommand{\byplnn}[2]{Y' _{#1}(#2)}
\newcommand{\bxls}{\bx _{*}}
\newcommand{\bxlsn}[1]{\bx _{*}(#1)}
\newcommand{\byln}[1]{\by _{#1}}
\newcommand{\bxlnn}[2]{\bxln{#1}(#2)}
\newcommand{\bylnn}[2]{\byln{#1}(#2)}
\newcommand{\uuln}[1]{\uu_{#1}}
\newcommand{\bxs}{\bx_{*}}
\newcommand{\bxsp}{\bx_{*}'}
\newcommand{\bys}{\by_{*}}
\newcommand{\bysp}{\by_{*}'}
\newcommand{\bxsn}[1]{\bxs (#1)}
\newcommand{\bxspn}[1]{\bxsp (#1)}
\newcommand{\bxrs}[1]{\bxs^{(#1)}}
\newcommand{\bxrsn}[2]{\bxs^{(#1)}(#2)}
\newcommand{\byrs}[1]{\bys^{(#1)}}
\newcommand{\byrsn}[2]{\bys^{(#1)}(#2)}
\newcommand{\bysn}[1]{\bys (#1)}
\newcommand{\byspn}[1]{\bysp (#1)}
\newcommand{\bpn}[1]{P{(#1)}}
\newcommand{\xaa}{\bx '}
\newcommand{\vndistinct}{\mathcal{V}^{n}}

\newcommand{\bwp}{\bw'}
\newcommand{\bwpln}[1]{\bwp_{#1}}

\newcommand{\ssln}[1]{s_{#1}}
\newcommand{\grn}[1]{g^{(#1)}}
\newcommand{\qnn}[2]{q^{(#1)}_{#2}}
\newcommand{\pnn}[2]{p^{(#1)}_{#2}}

\newcommand{\bvln}[1]{V_{#1}}
\newcommand{\bvlnn}[2]{V_{#1}(#2)}
\newcommand{\corrcoef}{\gamma}
\newcommand{\rrln}[1]{r_{#1}}
\newcommand{\boconstant}{O}
\newcommand{\boln}[1]{O_{#1}}
\newcommand{\boconstantp}{O'}
\newcommand{\aconstant}{a}
\newcommand{\bconstant}{b}
\newcommand{\mathbs}{\mathbb{S}}
\newcommand{\mathban}[1]{\mathbb{A}_{#1}}
\newcommand{\natnum}{\mathbb{N}}    
\newcommand{\tonumber}[1]{[#1]}     
\newcommand{\svector}[3]{(#1(#2))_{#2 \in #3}}  
\newcommand{\mysetx}{\mathcal{Z}}
\newcommand{\myfirstset}[3]{\mysetx(#1, #2, #3)}

\newcommand{\mysecondset}[4]{\mathcal{X}(#1; #2, #3, #4)}
\newcommand{\mythirdset}[2]{\mathcal{Y}(#1, #2)}

\newcommand{\sfd}[2]{D\left(#1,#2\right)}      
\newcommand{\sfdsb}[2]{D(#1,#2)}      
\newcommand{\bd}{D}
\newcommand{\bdln}[1]{\bd_{#1}}
\newcommand{\indicator}[1]{I\left(#1\right)}   
\newcommand{\rank}[2]{R\left(#1, #2\right)}    

\newcommand{\rn}[1]{\mathrm{#1}}

\newcommand{\statement}{\mathbb{S}}

\newcommand{\mcase}[1]{\textit{Missing Case #1}}
\newcommand{\realR}{\mathbb{R}^{n}}
\newcommand{\bmcase}[1]{\textit{Missing Case #1}}
\newcommand{\sm}{Supplementary Material}

\newcommand{\bphi}{V}
\newcommand{\bpsi}{W}

\begin{document}

\title{Exact Bounds of Spearman's footrule in the Presence of Missing Data with Applications to Independence Testing}

\author{Yijin Zeng \and Niall M. Adams \and Dean A. Bodenham}

\date{{\normalsize Department of Mathematics, Imperial College London,} \\
{\normalsize South Kensington Campus, London SW7 2AZ, U.K.} \\
{\normalsize yijin.zeng20@imperial.ac.uk, 
\quad n.adams@imperial.ac.uk, 
\quad dean.bodenham@imperial.ac.uk}
}

\maketitle

\begin{abstract}
    This work studies exact bounds of Spearman's footrule between two partially 
    observed $n$-dimensional distinct real-valued vectors $X$ and $Y$. 
    The lower bound is obtained by sequentially constructing imputations of the 
    partially observed vectors, each with a non-increasing value of 
    Spearman's footrule. 
    The upper bound is found by first considering the set of all
    possible values of Spearman's footrule for imputations of $X$ and $Y$, 
    and then the size of this set is gradually reduced using several constraints.
    Algorithms with computational complexities 
    $\bmo(n^2)$ and $\bmo(n^3)$ are provided for 
    computing the lower and upper bound of 
    Spearman's footrule for $X$ and $Y$, respectively.
    As an application of the bounds, we propose a novel two-sample 
    independence testing method for data with missing values. 
    Improving on all existing approaches, our method controls the Type I 
    error under arbitrary missingness. 
    Simulation results demonstrate our method has good power, typically 
    when the proportion of pairs containing missing data is below 15\%.
\end{abstract}

\section{Introduction} \label{section:introduction}

We study exact bounds of Spearman's footrule given two partially observed 
vectors $\bx, \by \in \mathbb{R}^{\n}$ with distinct values.
Spearman's footrule is a measure of statistical association based on ranks, 
introduced by Spearman in \cite{spearman1906footrule} as a robust alternative to Spearman's rank correlation. 
This statistic is defined between two real-valued vectors $\bx$ and $\by$ 
and calculates the sum of absolute differences between the ranks of 
corresponding components in $\bx$ and $\by$. 
In this work we assume that the vectors contain distinct values and that 
there are no ties.

Among rank-based measures of association, Spearman's rank correlation 
and Kendall's $\tau$ coefficient \cite{kendall1948rank} are more
well-known. 
However, Spearman's footrule has experienced a resurgence of interest, 
perhaps due to \cite{diaconis1977spearman}, where the 
asymptotic normality of Spearman's footrule is established under the assumption 
of independence between $\bx$ and $\by$. Because of its simplicity 
and robustness to outliers, Spearman's footrule has found 
applications in various fields, including information 
retrieval \cite{fagin2003comparing,2010Generalized}, 
web search \cite{bar2005comparing, bar2006methods}, 
rank aggregation \cite{dwork2001rank, lin2010rank}, 
genomics \cite{kim2004spearman}, 
welfare studies \cite{P2016Measuring} 
and management science \cite{powell2010rank, bennett2020changes}.

In practice, it frequently occurs that a subset of the data are missing, 
posing a challenge to the direct computation of Spearman's footrule. 
One often attempts to define the Spearman's footrule using its lower
or upper bound, as \cite{fagin2003comparing}
sought to formalize. A common setting with missing data is called top-$k$ 
lists \cite{fagin2003comparing, lin2010rank,dwork2001rank,bar2006methods}, 
also known as right-censored rankings \cite{quade2006concordance}, 
censored rankings \cite{salama2004agreement} 
and progressive censoring \cite{sen2003spearman}. 
This setting assumes that all missing data are implicitly larger than 
any observed data, resulting in only the data with smaller ranks being 
observed. Under this setting, and additionally assume
that for any pair of $\bx$ and $\by$, at least one 
component in $\bx$ or $\by$ is observed,
Spearman's footrule is demonstrated to 
remain constant, regardless of the values of missing data \cite{fagin2003comparing}.

In this paper, we study the exact bounds of Spearman's footrule without the 
previously described assumption. To the best of our knowledge, the 
issue of defining Spearman's footrule under conditions of general missingness 
has received limited attention, with the exception of the work 
in \cite{alvo1997use}, where the authors define Spearman's footrule with 
missing data using the conditional expectation given all observed values. 

While one might either ignore all pairs with at least one missing 
observation or impute any missing data using the observed values, we 
show that these practices will result in a biased estimate of 
Spearman's footrule except in special cases such as when the missing 
data are missing completely at random. It appears intractable to give 
an unbiased estimate of Spearman's footrule based on observed values only 
under general cases of missingness. Our contribution fills this 
gap by establishing exact bounds for Spearman's footrule without relying 
on restrictive assumptions about the nature of the missing data.

A related issue to measuring 
the statistical association of $X$ and $Y$ is testing for independence
between the two quantities. 
In the presence of missing data,  
there is no default approach to testing for independence. When the 
missingness mechanisms are known, for example if the data are missing 
completely at random, the practice of ignoring the missing data may be 
justified and more sophisticated approaches may be viable, 
e.g. \cite{ParzenCorrelation}. However, in most real-world data analysis
situations, such information is unavailable, 
and such practices may not necessarily be valid. 
Perhaps the most common practice is to simply ignore all the pairs with 
missing data and conduct the independence testing only for completely 
observed pairs, as noted in \cite{2014Examining, Alvo1995RankCM}. 
The ``risk'' of ignoring pairs with missing values is considered by some 
authors to be inconsequential when the proportion of missing data is 
less than $5$\% \cite{Schafer1999Multiple, heymans2022handling} or 
$10\%$ \cite{Derrick2009How,Derrick2009How}, but others argue against 
such heuristics \cite{Madley2019The}. 
We show that even in certain cases where $5\%$ to $10\%$ of the data is 
missing, ignoring or imputing these missing values is perilous.

As an application, we explore the use of our new bounds for Spearman's footrule 
for independence testing in the presence of missing data.
Spearman's footrule 
is a natural statistic for testing independence 
\cite{P2016Measuring, genest2010spearman, luigi1999asymptotic, chen2023asymptotic}, 
and has greater statistical power than Spearman's rank correlation and 
Kendall's $\tau$ under certain alternatives 
\cite{genest2010spearman, luigi1999asymptotic, chen2023asymptotic}. 

The key idea of our testing method is to provide tight bounds of all possible 
$p$-values given the observed values. For a pre-specified significance 
level $\alpha$, the null hypothesis is rejected only when the upper bound of the
$p$-values, and hence all possible $p$-values, 
is less than or equal to $\alpha$. 
When at least one $p$-value is not signficant, then the null hypothesis fails 
to be rejected.

One motivation for this approach is that it allows us to have confidence in any 
significant result, because the significant result was obtained without making 
any assumption about the missing data. In other words, if a significant result
is obtained, then even if the missing data could have been observed or an imputation method
had been employed, the result would still have been significant. 
While this is a conservative method of testing, the derived bounds for Spearman's footrule 
are tight, and so the results cannot be improved while taking into account all possible values
for the missing data.
The idea of employing bounds to consider all possible outcomes  was previously used 
in \citep{horowitzmanski2000} to take into account the effect 
of missing covariates and outcomes in clinical trials and observational studies.

This approach of taking all possible $p$-values into accounts could also be interpreted
as a sensitivity analysis \citep{thabaneetal2013, goldberg2021data, cookzea2020, smuk2015} that considers all possible assumptions on the missing data.
As we mentioned before, if a significant result is obtained by our method, then this
result is consistent among all possible missing data assumptions, since every possible imputation
would lead to a significant result. If, however,  our method
fails to reject the null hypothesis, there are two possible outcomes: either all possible missing
data assumptions lead to insignificant results when the lower bound of the $p$-values
is greater than $\alpha$, or a subset of assumptions leads to
significant results when the lower bound of the $p$-value is less than or equal to $\alpha$, 
but the upper bound is larger than $\alpha$.
Hence our method answers whether different approaches to handling missing 
data lead to different conclusions \cite{thabaneetal2013}.

\subsection{Contributions}

We provide exact bounds of Spearman's footrule in the presence of missing data 
for two univariate vectors of distinct values. 
Let $\bx$ and $\by$ be $\n$-dimensional distinct real-valued vectors, 
which may only be partially observed.
If the missing values of $\bx$ and $\by$ are replaced with real values, 
we obtain \emph{imputations} $\bxs$ and $\bys$ which will be more formally
defined in Section~\ref{section:notation}.
Theorem~\ref{theorem:minimumCase4} provides conditions 
under which $\bxs$ and $\bys$ achieve  
the lower bound of 
Spearman's footrule,  
while Theorem~\ref{theorem:maximumCase4} deals with 
the upper bound.

While it is possible to find $\bxs$ and $\bys$ that give the lower and upper
bounds simply by enumerating all possible ranks of missing data, this strategy 
is computationally infeasible for even moderate sample sizes. 
Suppose $\bx$ and $\by$ are of length $n$ and have 
$\mln{\bx}$ and $\mln{\by}$ missing values, respectively. 
There are $(n!)^2/((n-\mln{\bx})!(n-\mln{\by})!)$ possible permutations of 
the ranks. Consequently, when $\n = 100, \mln{\bx} = \mln{\by} = 10$, the 
number of possible permutations of ranks is approximately $3.95 \times 10^{39}$. 
The second contribution is to provide efficient algorithms for 
computing our exact lower and upper bounds of Spearman's footrule.

To find $\bxs$ and $\bys$ for the lower bound, 
we propose Algorithm~\ref{alg:2}, which requires computation complexity 
of $\bmo(\n^2)$.
However, finding the upper bound is more challenging.
An algorithm with computational complexity $\bmo(\n^3)$ is provided in 
the Supplementary Material for finding the upper bound.

Our work offers insight to anyone using 
Spearman's footrule with missing data. 
Furthermore, 
given the close link between Spearman's footrule and Kendall's $\tau$, our 
results lead to non-trivial bounds for Kendall's $\tau$ with missing data.

We explore the use of the bounds for Spearman's footrule for independence 
testing in the presence of missing data. The core idea is to give the bounds 
of all possible $p$-values given the observed values. Unlike all other 
approaches for testing independence with missing data, our test is capable 
of controlling the Type I error regardless of the values of missing data and 
the missingness mechanisms. Extensive numerical simulations are conducted and 
confirm our method is particularly useful when data are missing not at random.

\subsection{Missing data mechanisms}

We briefly describe the three missing data scenarios outlined in \cite{little2019statistical}: missing completely at random (MCAR), missing at random (MAR) and missing not at random (MNAR) . Consider a vector of univariate real-value samples $z = (z_1,z_2,\cdots,z_N)$. Let $\iota = (\iota_1,\cdots,\iota_N)$ be an indicator function of $z$ such that $\iota_i$ taking value 1 if $z_i$ is missing and 0 if $z_i$ is observed. The core idea of \cite{little2019statistical} is to admit $\iota$ as a probabilistic phenomenon. Let $z$ be a realized value of a random variable $Z$. Let $f_{\theta}(\iota | z)$ denotes the probability of $I = \iota$ given $Z = z$, where $\theta$ denotes any unknown parameters of the distribution. Then, the missingness mechanism is MCAR if $f_{\theta}({\iota} | z) = f_{\theta}(\iota | \widetilde{z}), \forall \iota, z, \widetilde{z}.$ In such cases, the missingness mechanism $I$ is independent of the value of samples. Denote $z'$ as a sub-vector of $z$ including all observed samples in $z$, i.e. including all $z_i$ such that $\iota_i = 0$. Then, the missingness mechanism is MAR if $f_{\theta}({\iota} | z) = f_{\theta}(\iota | \widetilde{z}), \forall \iota, z, \widetilde{z} \text{ such that } z' = \widetilde{z}'.$ In such cases, the missingness mechanism $I$ is independent of the values of missing samples. If the missingness mechanisms is neither MCAR nor MAR, it is MNAR.

\subsection{Related Work} The contributions of this paper are providing exact 
bounds of Spearman's footrule in the presence of missing data and proposing 
a new independence testing method with missing data based on these bounds. A 
special missing data scenario often discussed in relation to Spearman's 
footrule is the so-called "top-$k$ lists" situation, as explored in 
\cite{fagin2003comparing, lin2010rank,dwork2001rank, quade2006concordance, 
salama2004agreement, sen2003spearman}, where all missing data are assumed 
larger than any observed data and only smaller ranks can be observed. 
Our study of Spearman's footrule with missing data makes no assumption about 
the missing data except that they are distinct values. 

To the best of our knowledge, only \cite{alvo1997use} considers the same general missing pattern as we do, where Spearman's footrule is defined as the conditional mean given all observed data. Another closely-related work \cite{papaioannou1984inequalities} assumes all data in $\bx$, $\by$ are distinct and establishes the exact lower bound of Spearman's footrule under the missing case where all pairs must either both be observed or missing, denoted \mcase{III} below. The bounds found in this special case are then extended to the case where observations in $X$ and $Y$ are potentially tied \cite{loukas1991rank, speevak2017inequalities}. Additional missing data scenarios are studied in \cite{Brandenburg2013The}, where the missing data are categorized into partial, interval, and bucket cases, with potential for ties or ranking within a defined range. Their findings indicate that computing the bounds of Spearman's footrule in ``partial missing'' cases is NP-hard, leaving the computational feasibility of other scenarios open. In this work, we assume all data are distinct, and values are not tied.

Another line of research closely related to our work is independence testing 
in the presence of missing data. Spearman's rank correlation and Kendall's 
$\tau$ coefficient in the presence of missing data are defined using 
conditional expectation given all observed values in \cite{Alvo1995RankCM}. 
Under the assumption that the incomplete rank vectors are uniformly 
distributed over all possible permutations, the distributions of these two 
statistics are then derived and used for hypothesis testing. The test results 
are analyzed empirically in \cite{cabilio1999power}. Using the same approach 
as in \cite{Alvo1995RankCM}, Spearman's footrule has been applied for 
trend testing in the presence of missing data \cite{alvo1997use}. 
The independence testing problem in the case where data are missing at 
random has been considered in  \cite{2014Examining, ParzenCorrelation}. 

Beyond the above test statistics adapted for independence testing with missing 
data, broader methodologies for handling missing data are available. These 
include case deletion (i.e. only using the fully observed data) 
\cite{scheffer2002dealing}, single imputation \cite{2002Missing}, multiple 
imputation \cite{Schafer1999Multiple}, and the expectation 
maximization algorithm \cite{1977Maximum}. All methods rely on assumptions 
about the nature of the missing data such as missing completely at random 
or missing at random. When data are missing not at random, 
knowledge of the specific 
missingness mechanisms are required \cite{2002Missing}. 
For comprehensive overviews of these strategies,  
\cite{little2019statistical, 2002Missing, Dong2013PrincipledMD, 
baraldi2010introduction} are recommended. Our unique contribution 
makes no assumption about the missing data except assuming that all values 
with a vector are distinct. 
Our testing method offers a quantifiable measure of impact of 
missing data on testing outcomes, through identifying bounds on $p$-values.

\section{Exact Bounds of Spearman's Footrule In the Presence of Missing Data} \label{section: Exact Bounds of Spearman's Footrule In the Presence of Missing Data }

\subsection{Notation and definitions} \label{section:notation}

For any $\n \in \natnum$, let $\tonumber{n}$ denote the set $\{1,\ldots,\n\}$. For any $\n$-dimensional vector $\bx$, let $\bxn{\iconstant}$ denote the \emph{component} of $\bx$ at \emph{index} $i \in \tonumber{n}$. 
Thus, an $\n$-dimensional vector $\bx$ can be written as $\bx = (\bxn{1}, \ldots, \bxn{\n})$.

We will only consider vectors with distinct real values, i.e. there are no ties.
Define $\vndistinct$ to be the set of $n$-dimensional vectors
with distinct, real values, i.e. if $\bx \in \vndistinct$ then $\bx \in \realR$
and all components of $\bx$ are distinct.

If $\bx \in \vndistinct$, given a set of indices $U \subset \tonumber{n}$, 
we shall consider $\bxs \in \vndistinct$ to be an \emph{imputation} of $\bx$ when $\bxs(i) = \bx(i)$ 
for all $i \in \tonumber{n} \setminus U$, and the \emph{imputed values} are $\bxs(u)$ for all $u \in U$.

For any subset $\bs \subset \tonumber{\n}$ of indices, we use the notation $\svector{\bx}{\lconstant}{\bs}$ to denote a vector including the components of $\bx$ corresponding to the indices in set $\bs$, with the order of components in $\svector{\bx}{\lconstant}{\bs}$ following the order of components in $\bx$. For example, if $\n = 5$ and $\bs = \{1,3,4\}$, we then have $\svector{\bx}{\lconstant}{\bs} = (\bxn{1}, \bxn{3}, \bxn{4})$.

We define the rank of a component using the indicator function. For a statement $\ba$, we denote $\indicator{\ba}$ as a indicator function such that if the statement $\ba$ is correct, $\indicator{\ba} = 1$, otherwise $\indicator{\ba} = 0$. Subsequently, for any $\n$-dimensional vector $\bx$ of distinct real numbers, the rank of any component $\bxn{i}$ in $\bx$ is defined as
\begin{align*}
    \rank{\bxn{\iconstant}}{\bx} = \sum_{\lconstant=1}^{\n} \indicator{\bxn{\lconstant} \le \bxn{\iconstant}}.
\end{align*}
Using this notation, Spearman's footrule for any two $\n$-dimensional real-valued, distinct vectors $\bx, \by$ is defined as
\begin{align*}
    \sfd{\bx}{\by} = \sum_{\iconstant=1}^{\n} |\rank{\bxn{\iconstant}}{\bx} - \rank{\byn{\iconstant}}{\by}|. 
\end{align*}

If a vector contains missing values, we will refer to it as \emph{partially observed}, 
while a vector with no missing values is \emph{fully observed}. 
Our objective is to establish the exact bounds of Spearman's footrule with partially observed vectors $\bx$ and $\by$, where the components of  $\bx$ and $\by$ can be missing in any configuration. We start by considering three specific cases which, when considered together, cover any possible case of a pair of partially observed vectors.

\begin{itemize}
    \item  \bmcase{I}. All missing values are either all in $\bx$ or all in $\by$. In other words, either $\bx$ or $\by$ is fully observed, while the other vector is partially observed.
    \item \bmcase{II}. For any pair $(\bxn{\iconstant}, \byn{\iconstant})$, $i \in \tonumber{n}$, at most one value is missing. Therefore, both $\bx$ and $\by$ may contain missing data, and if $\bxn{i}$ is missing, the paired $\byn{i}$ must be observed. On the other hand,  if $\byn{i}$ is missing, the paired $\bxn{i}$ must be observed.
    \item  \bmcase{III}. For any pair $(\bxn{\iconstant}, \byn{\iconstant})$, $i \in \tonumber{n}$, the two values are either both observed, or both missing. Therefore, both $\bx$ and $\by$ may contain missing values, and if $\bxn{\iconstant}$ is missing, the paired $\byn{\iconstant}$ is also missing. Similarly, if $\byn{\iconstant}$ is missing, the paired $\bxn{\iconstant}$ is also missing.
\end{itemize}

Note that these three missing cases are not mutually exclusive. In particular, the \mcase{I} can be considered as a special case of \mcase{II}, given that the latter includes the case where only one vector of $\bx$ and $\by$ can  have missing values.

Let $\bx, \by \in \vndistinct$, which may be only partially observed. We use the notation $\bu$ to denote the set of all indices where $\bxn{\iconstant}$ is missing but $\byn{\iconstant}$ is observed. Furthermore, we use the notation $\bv$ to denote the set of all indices where $\bxn{\iconstant}$ is observed but $\byn{\iconstant}$ is missing. Additionally, we use  the notation $\bw$ to denote the set of all indices where both $\bxn{\iconstant}$ and $\byn{\iconstant}$ are missing.

Finally, $\rank{\bx}{\bx}$ is used to denote the vector of ranks of all components in $\bx$, i.e.
\begin{align*}
    \rank{\bx}{\bx} = (\rank{\bxn{1}}{\bx}, \ldots, \rank{\bxn{n}}{\bx}).
\end{align*}
For example, if $X = (7, 2, 5)$, then $\rank{\bx}{\bx} = (3, 1, 2)$.
Note that, in practice, we often first convert the vectors $\bx$ and $\by$ into the rank vectors
$\rank{\bx}{\bx}$ and $\rank{\by}{\by}$, respectively, and work with these rank vectors directly.

\subsection{Lower Bound of Spearman's Footrule} \label{section:lowerbound}
In what follows, we first provide the exact lower bounds of Spearman's footrule. In Section \ref{section:lowerbound:caseI} and  \ref{section:lowerbound:caseII}, the problem for establishing exact lower bounds under \mcase{I} and \emph{II} are discussed, respectively. Algorithms~\ref{alg:1} and \ref{alg:2} provide computationally efficient methods for computing these bounds. The lower bound under \mcase{III} is discussed in \cite{papaioannou1984inequalities}, although it not shown there that the bound is tight. In Proposition~\ref{Proposition: A Special Case For Missing Case III}, we show this bound is tight. Finally, by synthesizing the results developed for \mcase{I}, \emph{II} and \emph{III}, we provide the lower bound in the general case. The discussion of the upper bound in Section~\ref{section:upperbound} follows the same structure. The proof of all results can be found in the \sm.

\subsubsection{\bmcase{I}} \label{section:lowerbound:caseI}

We begin by carefully studying \mcase{I}, where the missing values are assumed to be either all in $\bx$ or all in $\by$. Without loss of generality, we proceed by assuming only $\bx$ contains missing values. 

The following proposition lays the groundwork for future results. It investigates the conditions for taking the minimum Spearman's footrule when one component $\bxn{\uu}$ in $\bx$ is replaced by a new value $\bxsn{\uu}$.

\begin{proposition} \label{Prop:multiplesteps}
    Suppose $\bx, \by \in \vndistinct$ and  
    let $\bxs \in \vndistinct$ be 
    an imputation of $\bx$ for an index $\uu \in \tonumber{\n}$.
    If we choose 
    $\bxsn{\uu}$ such that		
    $\rank{\bxsn{\uuln{1}}}{\bxs} = \rank{\byn{\uuln{1}}}{\by}$,
    then
    \begin{align*}
        \sfd{\bxs}{\by} \le \sfd{\bx}{\by}.
    \end{align*}
    Moreover, for any other imputation $\xaa \in \vndistinct$ of $\bx$ for index $\uu$, 
    $\sfd{\bxs}{\by} \le \sfd{\xaa}{\by}.$
\end{proposition}

This proposition demonstrates that if one component of $\bx$ is changed, Spearman's footrule will be minimized
when the rank of that component equals to the rank of its paired component, i.e. $\rank{\bxsn{\uuln{1}}}{\bxs} = \rank{\byn{\uuln{1}}}{\by}$. 

When considering Proposition~\ref{Prop:multiplesteps}, a natural concern would be whether 
the required imputation $\bxs$ actually exists.
To this end, we verify the existence of such imputations $\bxs$ of $\bx$, 
in the more general case where multiple components are imputed 
so that the ranks of all imputed components match the ranks of their paired 
components in $\by$.

\begin{lemma} \label{Lemma:1}
    Suppose $\bx, \by \in \vndistinct$ and $
    \bu \subset \tonumber{\n}$ is a subset of indices. Then there exists $\bxs \in \vndistinct$ such that 
    \begin{align} \label{Lemma:1:eqn:1}
        \bxsn{\iconstant} = \bxn{\iconstant} 
        \,\,\text{for all }~\iconstant \in \tonumber{\n} \setminus \bu,
        \,\,\,\, \text{and} \,\,\,\,~\rank{\bxsn{\iconstant}}{\bxs} = \rank{\byn{\iconstant}}{\by} 
        \,\,\text{for all }~\iconstant \in \bu.
    \end{align}  
\end{lemma}

The proof in the \sm{} constructs $\bxs$ sequentially by considering the order of the values in the paired vector $\svector{\by}{u}{U}$. The first step is to identify $\uun{1}$, the index in $\bu$ with the minimum rank in $\by$, defined as $\uun{1} = \text{argmin} \{\rank{\byn{\uu}}{\by} | \uu \in \bu \}$.
Then the component $\bxn{\uun{1}}$ is imputed so that its rank satisfies
\begin{align}
    \rank{\bxn{\uun{1}}}{\svector{\bx}{\iconstant}{\left\{\tonumber{\n} \setminus \bu \right\} \cup \{\uun{1}\} }} = \rank{\byn{\uun{1}}}{\by}.
    \nonumber
\end{align}

Following this imputation, the ranks of all observed values in $\bx$ are updated. $\bu$ is then updated to $\bu \setminus \{\uu_1\}$, and this process is repeated until $\bu$ is empty.
Algorithm~\ref{alg:1} below formally describes the procedure of the proof of Lemma \ref{Lemma:1}. 
Note that in Algorithm~\ref{alg:1} we work with the rank vectors 
$\rank{\by}{\by}$ and $\rank{\bxs}{\bxs}$, 
rather than the actual vectors $\by$ and $\bxs$, 
which suffices for our purpose of computing the value of Spearman's footrule $D(\bxs,\by)$.

\begin{algorithm} 
    \caption{An Algorithm for Computing the Exact Lower Bound of Spearman's Footrule Under \bmcase{I}} \label{alg:1}
    \begin{algorithmic}[1]
        \Require{$\bx,\by \in \vndistinct$, where $\bx$ may be partially observed and $\by$ is fully observed.} 
        \Ensure{Minimum possible Spearman's footrule distance between $X$ and $Y$.}
        \State Rank all observed components in $\bx$ and $\by$. \label{alg:1:0}
        \While{$\bu \neq \emptyset$}
        \State Identify $\uun{1} = \text{argmin} \left\{\rank{\byn{\iconstant}}{\by}| \iconstant \in \bu \right\}$. \label{alg:1:1}
        \State Let $\rank{\bxn{\uun{1}}}{\svector{\bx}{\iconstant}{\left\{\tonumber{\n} \setminus \bu \right\} \cup \{\uun{1}\} }} = \rank{\byn{\uun{1}}}{\by}$.
        \For{$\iconstant \in \tonumber{\n} \setminus \bu$} \label{alg:1:2}
        \State Denote $r_{\iconstant} = \rank{\bxn{\iconstant }}{\svector{\bx}{\iconstant}{\tonumber{\n} \setminus \bu}}$
        \State Update
        $\rank{\bxn{\iconstant}}{\svector{\bx}{\jconstant}{\left\{\tonumber{\n} \setminus \bu \right\} \cup \{\uun{1}\} }}  \leftarrow r_{\iconstant} + \indicator{r_{\iconstant} \ge \rank{\byn{\uun{1}}}{\by}}.$
        \EndFor \label{alg:1:3}
        \State Update $\bu \leftarrow \bu \setminus \{\uun{1}\}$.
        \EndWhile
        \State Output $\sfd{\bx}{\by}$.
    \end{algorithmic}
\end{algorithm}

\begin{remark} \label{remark:1}
    The computational complexity of Algorithm~\ref{alg:1} is analyzed as follows. Ranking all observed components in $\bx$ and $\by$
    in line~\ref{alg:1:0} requires $\bmo(\n\log\n)$ steps.	
    Inside the while loop, identifying $\uu_{1}$ in line~\ref{alg:1:1} 
    requires the computational complexity $\bmo({\n})$, and updating ranks of observed components between line~\ref{alg:1:2} and line~\ref{alg:1:3} requires the computational complexity $\bmo({\n})$. Denote $\mln{1} = |\bu|$. The while loop run $\mln{1}$ times. Since each iteration is $\bmo(n)$, the computational complexity of  loop is $\bmo(\mln{1}{\n})$.
    Overall, the computational complexity for Algorithm~\ref{alg:1} is
    $\bmo(\n\log\n + \mln{1}{\n})$. 
\end{remark}

To further emphasise that, when computing Spearman's footrule, 
it is the \emph{ranks} $\rank{\bxs}{\bxs}$ of the imputed vector $\bxs$ that
are important, rather than the imputed values $\svector{\bxs}{\uu}{\bu}$ themselves, 
we have the following lemma which will be useful in proving later results.

\begin{lemma} \label{Lemma:2}
    Suppose $\bx, \by \in \vndistinct$ and $
    \bu \subset \tonumber{\n}$ is a subset of indices. Consider any $\bxln{1}, \bxln{2} \in \vndistinct$ 
    satisfying the conditions in \eqref{Lemma:1:eqn:1}, 
    Then $\sfd{\bx_1}{\by} = \sfd{\bx_2}{\by}$.
\end{lemma}

We are now ready to state the main result for \mcase{I}, which is an extension 
of Propostion~\ref{Prop:multiplesteps}.

\begin{theorem} \label{theorem:minimumCase1}
    Suppose $\bx, \by \in \vndistinct$ and $
    \bu \subset \tonumber{\n}$ is a subset of indices. Consider any $\bxs \in \vndistinct$  satisfying 
    the conditions in \eqref{Lemma:1:eqn:1}. Then
    \begin{align*}
        \sfd{\bxs}{\by} \le \sfd{\bx}{\by}.
    \end{align*}
    Moreover, for any $\xaa \in \vndistinct$ such that $\xaa(\iconstant) = \bxn{\iconstant}$ 
    for $\iconstant \in \tonumber{\n} \setminus \bu$, we have
    \begin{align*}
        \sfd{\bxs}{\by} \le \sfd{\xaa}{\by}.
    \end{align*}
\end{theorem}
The strategy for proving Theorem \ref{theorem:minimumCase1} is to sequentially construct imputations $\bxln{1}, \bxln{2}, \bxln{3} \in \vndistinct$ of $\bx$ for the set of indices $\bu$. 
As before, define $\uun{1} = \text{argmin} \{\rank{\byn{\uu}}{\by} | \uu \in \bu \}$.

The first vector $\bxln{1}$ is an imputation of $\bx$ for indices $\bu \setminus \{\uun{1}\}$ such 
that the ranks of $\bxln{1}$ and $\by$ are equal for the set $\bu \setminus \{\uun{1}\}$, 
i.e. $\rank{\svector{\bxln{1}}{\uu}{\bu \setminus \{\uun{1}\}}}{\bxln{1}} = \rank{\svector{\by}{\uu}{\bu \setminus \{\uun{1}\}}}{\by}$.
The second vector $\bxln{2}$ imputes the $\uun{1}$ component, so that its rank is equal to that of its 
paired component, i.e.
$\rank{\bxlnn{2}{\uun{1}}}{\bxln{2}} = \rank{\by(\uun{1})}{\by}$. 
However, it may now be the case that the ranks of the components of $\bxln{2}$ 
for the indices in $\bu \setminus \{\uun{1}\}$ may not equal the ranks of their paired components in $\by$, i.e.
it may be that
$\rank{\svector{\bxln{2}}{\uu}{\bu \setminus \{\uun{1}\}}}{\bxln{1}} \neq \rank{\svector{\by}{\uu}{\bu \setminus \{\uun{1}\}}}{\by}$.
Therefore, the third vector $\bxln{3}$ is constructed as an imputation of $\bxln{2}$ for the indices $\bu \setminus \{\uun{1}\}$, 
so that the ranks of its imputed components are equal to the ranks of its paired components in $\by$, 
and
$\rank{\svector{\bxln{3}}{\uu}{\bu}}{\bxln{3}} = \rank{\svector{\by}{\uu}{\bu}}{\by}$.

It is then demonstrated in the proof of Theorem \ref{theorem:minimumCase1} that Spearman's footrule decreases sequentially as follows:
\begin{align*}
    \sfd{\bx}{\by} \ge \sfd{\bxln{1}}{\by} \ge \sfd{\bxln{2}}{\by} \ge \sfd{\bxln{3}}{\by}.
\end{align*}
Since $\bxln{3}$ satisfies the conditions in \eqref{Lemma:1:eqn:1}, if we use Algorithm~\ref{alg:1} 
to construct $\bxs$ which also satisfies the conditions in \eqref{Lemma:1:eqn:1}, 
by Lemma \ref{Lemma:2} we must have $\sfd{\bxln{3}}{\by} = \sfd{\bxs}{\by}$.

Theorem \ref{theorem:minimumCase1} reveals an interesting property of Spearman's footrule. In order to minimize Spearman's footrule in the presence of missing data, or in other words, minimize Spearman's footrule for the pair 
$\bx, \by \in \vndistinct$ by imputing values for $\svector{\bx}{\uu}{\bu}$ 
for a subset of indices $\bu \subset \tonumber{n}$, one simply needs to
construct an imputation $\bxs$ that satisfies the conditions in \eqref{Lemma:1:eqn:1}.

This property, however, does not hold for minimizing Spearman's rank correlation $\rho$ 
and Kendall's $\tau$ coefficient defined for $\bx, \by \in \vndistinct$ as
\begin{align} 
    \rho(\bx, \by) &= \sum_{i=1}^{n} |\rank{\bxn{\iconstant}}{\bx} - \rank{\byn{\iconstant}}{\by}|^2,  \label{definition: Spearman Rank Correlation} \\
    \tau(\bx, \by) &= \sum_{i > j} \{ \indicator{\bxln{i} > \bxln{j}} \indicator{\byln{i} < \byln{j}} + \indicator{\bxln{i} < \bxln{j}} \indicator{\byln{i} > \byln{j}} \} \label{defintion: Kendall Tau},
\end{align} 
respectively. Note that $\rho(\bx, \by)$ and $\tau(\bx, \by)$ in Equations~\eqref{definition: Spearman Rank Correlation} 
and \eqref{defintion: Kendall Tau} are unscaled versions of the coefficients that measure the amount of 
discordance. Equation~\eqref{eqn:scaledtaurho} provides the more familiar versions of the coefficients
scaled to $[-1, 1]$.

As an example, consider the data shown in left-hand part of Table \ref{Tab:1}, 
where objects $a, b, c, d, e, f, g, h$
have been given two different sets of ranks, denoted by $\bx$ and $\by$, but the rank of object $d$ for $\bx$ is missing.

A natural question to ask is what rank of $d$ will maximize the
``agreement'' between the two referees $\bx$ and $\by$. 
For any of the three rank correlation statistics, 
maximizing the agreement is equivalent to minimizing the statistic,
since a perfect agreement would mean equal ranks and a zero-valued statistic.
Theorem \ref{theorem:minimumCase1} concludes that minimizing
Spearman's footrule requires the rank of $d$ to be $4$, to match the rank given by $\by$.
However, as shown in Table 3.1 in \sm, 
Spearman's rank correlation $\rho$ and Kendall's $\tau$ coefficient are both minimized
when the rank of $d$ is 1; the right-hand part of Table~\ref{Tab:1} provides partial results.

This special property of Spearman's footrule is further explored and extended 
in the context of \mcase{II}, as discussed in Theorem \ref{theorem:minimumCase2}.

\begin{table*}
    \caption{Left: example ranked data to compare Spearman's footrule $D$, Spearman's rank correlation $\rho$
    and Kendall's $\tau$ coefficient, where the $*$ indicates the rank of object $d$ for vector $\bx$ is missing.
    Right: values of these statistics for selected imputations  $\bxs$ of $\bx$, when the rank of $d$ is 
    imputed as either $3$ or $4$}
    \label{Tab:1}
    \begin{tabular}{lrrrrrrrr}
        \hline
        & $a$ & $b$ & $c$ & $d$ & $e$ & $f$ & $g$ & $h$ \\ 		
        \hline
        $\bx$    & 7 & 3  & 6  & *  &2  &5 & 4 & 1 \\
        $\by$    & 1 & 2  & 3  & 4  &5  &6 & 7 & 8 \\
        \hline
    \end{tabular}
    \qquad
    \qquad
    \begin{tabular}{cccc}
        \hline
        $\bxs$ & $\sfd{\bxs}{\by}$ & $\rho(\bxs, \by)$ & $\tau(\bxs, \by)$ \\ 		
        \hline
        $d=1$    & 26 & 122  & 19\\
        $d=4$    & 24 & 128  & 20 \\
        \hline
    \end{tabular}
\end{table*}

\subsubsection{\bmcase{II}} \label{section:lowerbound:caseII}

This section focuses on the lower bound of Spearman's footrule under \mcase{II}, where $\bx, \by \in \vndistinct$ may contain unobserved samples, but for each pair $(\bxn{\iconstant}, \byn{\iconstant})$, at least one value is observed.
We shall use $\bu \subset \tonumber{\n}$ to denote the set of indices where $\bxn{\iconstant}$ is missing but $\byn{\iconstant}$ is observed, and $\bv \subset \tonumber{\n}$ will denote the set of indices where $\bxn{\iconstant}$ is observed but $\byn{\iconstant}$ is missing. Given this notation, consider the following two conditions:

\begin{itemize}
    \item Condition A. $\bxs, \bys \in \vndistinct$ are imputations of $\bx$ and $\by$ for 
        indices $\bu, \bphi \subset \tonumber{\n}$ respectively, such that $\bu \cap \bphi = \emptyset$ and 
        \begin{align*}
            \bxsn{\iconstant} = \bxn{\iconstant}~\text{for all}~\iconstant \in \tonumber{\n} \setminus \bu,
            ~\text{and}~\bysn{\jconstant} = \byn{\jconstant}~\text{for all}~\jconstant \in \tonumber{\n} \setminus \bphi.
        \end{align*}
    \item  Condition B. $\bxs, \bys \in \vndistinct$ are such, for indices $\bu, \bphi \subset \tonumber{\n}$,
        \begin{align*}
            \rank{\bxsn{\iconstant}}{\bxs} &= \rank{\bysn{\iconstant}}{\bys}, \iconstant \in \bu \cup \bphi.
        \end{align*}
\end{itemize}
Condition A ensures that $\bxs$ and $\bys$ are imputations of $\bx$ and $\by$ satisfying \mcase{II}. 
Condition B ensures the ranks of the imputed components of $\bxs$ and $\bys$ 
are equal to the ranks of their paired components.

It is proved in Theorem \ref{theorem:minimumCase2} that the minimum possible value of
Spearman's footrule between $\bx$ and $\by$ equals $\sfd{\bxs}{\bys}$ 
when  $\bxs$ and $\bys$ satisfy both Conditions A and B. While finding $\bxs$ and $\bys$ 
that satisfy Condition A is straightforward, it is less obvious that we can always find 
imputations that satisfy both Conditions A and B. 
Hence, we first prove the existence of such imputations.

\begin{proposition} \label{Prop:existence}
    Suppose $\bx, \by \in \vndistinct$ and disjoint $\bu, \bv \subset \tonumber{n}$ are such that $\bu \cup \bphi \neq \emptyset$. Then, there exist imputations $\bxs, \bys \in \vndistinct$ satisfying both Conditions A and B.
\end{proposition}

The proof of Proposition \ref{Prop:existence} in the \sm{} follows a similar idea 
to the proof of Lemma \ref{Lemma:1}. The construction of $\bxs$ and $\bys$ is summarised
in Algorithm~\ref{alg:2} below.

\begin{algorithm} 
    \caption{An algorithm for computing the exact lower bound of Spearman's footrule under \mcase{II}} \label{alg:2}
    \begin{algorithmic}[1]
        \Require{$X, Y \in \vndistinct$, where both vectors may be partially observed under \mcase{II}.} 
        \Ensure{Minimum possible Spearman's footrule distance between $X$ and $Y$.}
        \State Rank all observed components in $\bx$, $\by$. \label{alg:2:0}
        \While{$\bu \neq \emptyset \text{ and } \bphi \neq \emptyset$}
        \State Let $\uun{1} = \text{argmin} \left\{\rank{\bxn{\iconstant}}{\svector{\bx}{\jconstant}{\tonumber{\n} \setminus \bu}}| \iconstant \in \bphi \right\}$. \label{alg:2:1}
        \State Let $\uun{2} = \text{argmin} \left\{\rank{\byn{\iconstant}}{\svector{\by}{\jconstant}{\tonumber{\n} \setminus \bphi}}| \iconstant \in \bu \right\}$. \label{alg:2:2}
        \If{$\rank{\bxn{\uun{1}}}{\svector{\bx}{\jconstant}{\tonumber{\n} \setminus \bu}} \le \rank{\byn{\uun{2}}}{\svector{\by}{\jconstant}{\tonumber{\n} \setminus \bphi}} $} 
        \State Let $\rank{\byn{\uun{1}}}{\svector{\by}{\jconstant}{ \left\{\tonumber{\n} \setminus \bphi \right\} \cup \{\uun{1}\} }} = \rank{\bxn{\uun{1}}}{\svector{\bx}{\jconstant}{\tonumber{\n} \setminus \bu}}$
        \For{$\iconstant \in \tonumber{\n} \setminus \bphi$} \label{alg:2:3}
        \State  Denote $\rank{\byn{\iconstant}}{\svector{\by}{\jconstant}{\tonumber{\n} \setminus \bphi}} = s_{\iconstant}$.
        \State  Update $\rank{\byn{\iconstant}}{\svector{\by}{\jconstant}{ \left\{\tonumber{\n} \setminus \bphi \right\} \cup \{\uun{1}\} }}  \leftarrow s_{\iconstant} + \indicator{s_{\iconstant} \ge \rank{\bxn{\uun{1}}}{\svector{\bx}{\jconstant}{\tonumber{\n} \setminus \bu}}}$. 
        \EndFor \label{alg:2:4}
        \State Update $\bphi \leftarrow \bphi \setminus \{\uun{1}\}$.
        \Else
        \State Let $\rank{\bxn{\uun{2}}}{\svector{\bx}{\jconstant}{\left\{\tonumber{\n} \setminus \bu\right\} \cup \{\uun{2}\} }} = \rank{\byn{\uun{2}}}{\svector{\by}{\jconstant}{\tonumber{\n} \setminus \bphi}}$
        \For{$\iconstant \in \tonumber{\n} \setminus \bu$} \label{alg:2:5}
        \State Denote $\rank{\bxn{\iconstant}}{\svector{\bx}{\jconstant}{\tonumber{\n} \setminus \bu}} = r_{\iconstant}$.
        \State Update $\rank{\bxn{\iconstant}}{\svector{\bx}{\jconstant}{\left\{\tonumber{\n} \setminus \bu\right\} \cup \{\uun{2}\} }} \leftarrow r_{\iconstant} + \indicator{r_{\iconstant} \ge \rank{\byn{\uun{2}}}{\svector{\by}{\jconstant}{\tonumber{\n} \setminus \bphi}}}.$ 
        \EndFor \label{alg:2:6}
        \State Update $\bu \leftarrow \bu \setminus \{\uun{2}\}$.
        \EndIf
        \EndWhile
        \If{$U \neq \emptyset$ or $\bphi \neq \emptyset$}
        \State Run Algorithm \ref{alg:1} when $U \neq \emptyset$. Reverse $\bx$ and $\by$ and run Algorithm \ref{alg:1} when $\bv \neq \emptyset$. \label{lineOfAlg2:alg1}
        \EndIf
    \end{algorithmic}
\end{algorithm}

\begin{remark}
    The computational complexity of Algorithm \ref{alg:2} is analyzed as follows. Ranking all observed components in $\bx$ and $\by$ in line~\ref{alg:2:1}takes $\bmo(\n\log\n)$ steps.
    Inside the while loop, the computational complexity for finding the minimum ranks in line \ref{alg:2:1} and \ref{alg:2:2} is $\bmo(\n)$, and for updating ranks of observed components from line~\ref{alg:2:3} to line~\ref{alg:2:4}, and line~\ref{alg:2:5} to line~\ref{alg:2:6} is $\bmo(n)$. Let $|\bu| = \mln{1}$, and $|\bv| = \mln{2}$. The loop can run up to $\mln{1} + \mln{2} - 1$ times. 
    Since each iteration is $\bmo(n)$,
    the computational complexity for the while loop is 
    $\bmo(\n(\mln{1} + \mln{2}))$.
    The computational complexity for Algorithm \ref{alg:1}, called in line~\ref{lineOfAlg2:alg1}, after ranking all observed components is $\bmo(\n\mln{1})$ when $\bu \neq \emptyset$ and 
    $\bmo(\n\mln{2})$ when $\bv \neq \emptyset$, according to Remark \ref{remark:1}. Therefore, the total computation complexity for Algorithm \ref{alg:2} is $\bmo(\n\log\n + \n(\mln{1} + \mln{2}))$.
\end{remark}

The following result shows that 
all imputations $\bxs$ and $\bys$ satisfying both Conditions A and B yield the same value of Spearman's footrule.

\begin{proposition} \label{Prop:equivalence}
    Suppose $\bx, \by \in \vndistinct$ and disjoint $\bu, \bv \subset \tonumber{\n}$ are such that $\bu \cup \bphi \neq \emptyset$. If $\bxln{1}, \byln{1}$ and $ \bxln{2}, \byln{2}$ 
    are pairs satisfing Conditions A and B, then $\sfd{\bxln{1}}{\byln{1}} = \sfd{\bxln{2}}{\byln{2}}$.
\end{proposition}

While the equivalence of \mcase{I} found in Lemma \ref{Lemma:2} can be proved 
in a relatively straightforward approach, the proof of 
Proposition~\ref{Prop:equivalence} is much more challenging. For \mcase{I}, the 
condition in \eqref{Lemma:1:eqn:1} explicitly provides the ranks of all 
components in $\bu$ for both $\bxln{1}$ and $\bxln{2}$. On the other hand, 
Condition~B for \mcase{II} only requires the ranks of all components of 
$\bxln{\jconstant}, \byln{\jconstant}$ for indices  $\bu \cup \bphi$ to be 
matched with their paired components, for $\jconstant \in \{1, 2 \}$. 
However, the rank of component $\bxlnn{1}{\iconstant}$ in $\bxln{1}$
is not necessary equal to the rank of $\bxlnn{2}{i}$ in $\bxln{2}$, for
$\iconstant \in \bu \cup \bphi$. 
Similarly, it is not necessarily true 
for all $\iconstant \in \bu \cup \bphi$ that
$\rank{\bylnn{1}{\iconstant}}{\byln{1}} = \rank{\bylnn{2}{\iconstant}}{\byln{2}}$.

Furthermore, the following result shows that
minimum possible Spearman's footrule is achieved when imputations 
$\bxs$ and $\bys$ satisfy both Conditions~A and B.
\begin{theorem} \label{theorem:minimumCase2}
    Suppose $\bx, \by \in \vndistinct$ and disjoint $\bu, \bv \subset \tonumber{\n}$ are such that $\bu \cup \bphi \neq \emptyset$. If $\bxs, \bys \in \vndistinct$ are imputations satisfying both Conditions~A and B, then
    \begin{align*}
        \sfd{\bxs}{\bys} \le \sfd{\bx}{\by}.
    \end{align*}
    For any other imputations $\bxp, \byp \in \vndistinct$ of $\bx, \by$ for indices $\bu, \bphi$, respectively,
    \begin{align*}
        \sfd{\bxs}{\bys} \le \sfd{\bxp}{\byp}.
    \end{align*}
\end{theorem}

\subsubsection{\bmcase{III}} \label{section:lowerbound:caseIII}

Given two vectors $\bx, \by \in \vndistinct$, 
recall that in \mcase{III}, for any pair of components $(\bxn{\iconstant}, \byn{\iconstant})$
at index $\iconstant \in \tonumber{\n}$, 
the two values are either both observed or both missing.  
This case for Spearman's footrule has been previously considered in the literature, and 
the following result is proved in Theorem~1 in \cite{papaioannou1984inequalities}.

\begin{theorem}[\cite{papaioannou1984inequalities}] \label{theeorem:minimumCase3} 
    Suppose $\bx, \by \in \vndistinct$ and $\bw \subset \tonumber{\n}$ . 
    For the subvectors, $\bxp = \svector{\bx}{\iconstant}{\tonumber{\n} \setminus \bpsi}$ and 
    $\byp = \svector{\by}{\iconstant}{\tonumber{\n} \setminus \bpsi}$, it can be shown that
    $\sfd{\bxp}{\byp} \le \sfd{\bx}{\by}$.
\end{theorem}

While this result provides a lower bound of Spearman's footrule, it does 
not directly prove that this bound is the minimum possible value of the 
Spearman's footrule. 
The following result fills this gap by identifying the 
conditions under which this lower bound is tight.
\begin{proposition} \label{Proposition: A Special Case For Missing Case III} 
    Suppose $\bx, \by \in \vndistinct$  and $\bw \subset \tonumber{\n}$. 
    Define the subvectors $\bxp = \svector{\bx}{\iconstant}{\tonumber{\n} \setminus \bpsi}$ and 
    $\byp = \svector{\by}{\iconstant}{\tonumber{\n} \setminus \bpsi}$. 
    If the following three conditions all hold,
    \begin{align}
        (i)&~\rank{\bxn{\iconstant}}{\bx} = \rank{\byn{\iconstant}}{\by}, \,\,\textrm{for all}~\iconstant \in \bpsi,
        \nonumber \\
        (ii)&~\min_{\jconstant \in \bpsi} \bxn{\jconstant} > 
        \max_{\jconstant \in \tonumber{\n} \setminus \bpsi} \bxn{\jconstant}, 
        \nonumber \\
        (iii)&~\min_{\jconstant \in \bpsi} \byn{\jconstant} > 
        \max_{\jconstant \in \tonumber{\n} \setminus \bpsi} \byn{\jconstant},
        \nonumber
    \end{align}
    then $\sfd{\bx}{\by} = \sfd{\bxp}{\byp}$.
\end{proposition}

\subsubsection{General missing case} We are now ready to provide the lower bound of 
Spearman's footrule in the general case, where the missing pattern of data is a combination
of \textit{Missing Cases I}, \textit{II}, and \textit{III}.
The previous results can be combined to give the following result.

\begin{theorem} \label{theorem:minimumCase4}
    Suppose $\bx, \by \in \vndistinct$ and $\bu$, $\bphi$, $\bpsi \subset \tonumber{\n}$ are 
    pairwise disjoint subsets. 
    Suppose $\bxs, \bys \in \vndistinct$ are imputations of $\bx, \by$ for indices $\bu \cup \bpsi$, and $\bphi \cup \bpsi$, respectively. If the following three conditions all hold
    \begin{align}
        (i)&~\rank{\bxsn{\iconstant}}{\bx} = \rank{\bysn{\iconstant}}{\by}, 
        \,\,\textrm{for all}~\iconstant \in \bu \cup \bv \cup \bw,
        \nonumber \\
        (ii)&~\min_{\jconstant \in \bpsi} \bxsn{\jconstant} > 
        \max_{\jconstant \in \tonumber{\n} \setminus \bpsi} \bxsn{\jconstant},
        \nonumber\\
        (iii)&~ \min_{\jconstant \in \bpsi} \bysn{\jconstant} > 
        \max_{\jconstant \in \tonumber{\n} \setminus \bpsi} \bysn{\jconstant}, 
        \nonumber
    \end{align}
    then 
    \begin{align*}
        \sfd{\bxs}{\bys} \le \sfd{\bx}{\by},
    \end{align*}
    and for any other imputations $\bxp, \byp \in \vndistinct$ of $\bx, \by$ for indices $\bu\cup \bpsi$, $\bphi\cup \bpsi$, respectively,
    \begin{align*}
        \sfd{\bxs}{\bys} \le \sfd{\bxp}{\byp}.
    \end{align*}
\end{theorem}
For the imputations $\bxs$ and $\bys$ in Theorem~\ref{theorem:minimumCase4}, 
Proposition~\ref{Proposition: A Special Case For Missing Case III} gives
$\sfd{\bxs}{\bys}  = \sfd{\svector{\bxs}{\iconstant}{\tonumber{\n} \setminus \bpsi}}{\svector{\bys}{\iconstant}{\tonumber{\n} \setminus \bpsi} }$, 
which suggests that, for indices in $\bpsi$, the components of both 
$\bx$ and $\by$ can be ignored for the purposes of calculating the lower 
bound. Hence, the minimum possible Spearman's footrule in the general 
missing case can be determined using Algorithm~\ref{alg:2} after ignoring 
all components with indices in $\bpsi$.

\subsection{Upper Bound of Spearman's Footrule} \label{section:upperbound}

This section provides the exact upper bound of Spearman's footrule in the 
presence of missing data. We consider \textit{Missing Cases I}, \textit{II}, 
and \textit{III} separately
and then combine our results to obtain the upper bound in the general 
missing case. 

\subsubsection{Missing Case I}

We start by considering \mcase{I} for $\bx, \by \in \vndistinct$,
where the missing components are assumed 
to be either all in $\bx$ or all in $\by$. Without loss of generality, 
let us assume only $\bx$ contains missing components 
while all values in $\by$ are observed. 

Suppose only a single component $\bxn{\uu}$ is missing. The 
following proposition is complementary to Proposition~\ref{Prop:multiplesteps}
which considers the lower bound for \mcase{I}.

\begin{proposition} \label{prop:multiplestepupperbound}
    Suppose $\bx, \by \in \vndistinct$ and  
    let $\bxln{1}, \bxln{2} \in \vndistinct$ be imputations of $\bx$ 
    for an index $\uu \in \tonumber{\n}$
    such that 
    $\rank{\bxlnn{1}{\uuln{1}}}{\bxln{1}} = 1$ and $\rank{\bxlnn{2}{\uuln{1}}}{\bxln{2}} = n$.
    Then
    \begin{align}
        \sfd{\bx}{\by} \le \max \{ \sfd{\bxln{1}}{\by}, \sfd{\bxln{2}}{\by}\},
        \nonumber
    \end{align}	
    and for any imputation $\xaa \in \vndistinct$ of $\bx$ for index $\uu$, 
    $\sfd{\bxp}{\by} \le \max \{ \sfd{\bxln{1}}{\by}, \sfd{\bxln{2}}{\by}\}$.
\end{proposition}

When a single component of $\bx$ is missing, Proposition \ref{prop:multiplestepupperbound} 
shows that the maximum possible value of Spearman's footrule is achieved when the 
missing component is imputed to have rank either $1$ or $\n$.

It is shown in the proof that in the special case where 
$\rank{\byn{\uuln{1}}}{\by} = 1$,
$\bxln{2}$ maximizes Spearman's footrule, and in the
special case where $\rank{\byn{\uuln{1}}}{\by} = \n$,
$\bxln{1}$ maximizes Spearman's footrule.

To extend Proposition~\ref{prop:multiplestepupperbound} to 
cases when multiple components of $\bx$ may be missing, we first make the following definition.

\begin{definition} \label{def:1}
    Let $\bu \subset \bt \subset \tonumber{\n}$ be a subset of indices. Define
    \begin{align*}
        \myfirstset{\bu}{\bt}{\n} =  \{ \bz \in \vndistinct|
        ~\text{for any}~\iconstant \in \bu,~ 
        \bzn{\iconstant} > \max_{\jconstant \in \bt \setminus \bu} \bzn{\jconstant} \text{ or } \bzn{\iconstant} < \min_{\jconstant \in \bt \setminus \bu} \bzn{\jconstant} \}.
    \end{align*}
\end{definition}
It will often happen that $\bt=\tonumber{\n}$ in $\myfirstset{\bu}{\bt}{\n}$, 
but this will not always be the case. This definition allows us to state the following result.

\begin{proposition} \label{prop:maximumcase1}
    Suppose $\bx, \by \in \vndistinct$, and let $\bu \subset \tonumber{\n}$.
    Then, there exists an imputation $\bxs \in \myfirstset{\bu}{\tonumber{\n}}{\n}$ of $\bx$ for indices $\bu$ such that	$\sfd{\bx}{\by} \le \sfd{\bxs}{\by}$, and for any other imputation $\xaa \in \vndistinct$ of $\bx$ for indices $\bu$, 
    $\sfd{\bxp}{\by} \le \sfd{\bxs}{\by}$.
\end{proposition}

When the components of $\bx$ with indices in $\bu \subset \tonumber{\n}$ are missing, Proposition~\ref{prop:maximumcase1} allows us to obtain the maximum possible Spearman's footrule by considering rank vectors of all possible imputations $\bxs$ of $\bx$ in the set $\myfirstset{\bu}{\tonumber{\n}}{\n}$.

We briefly consider the number of possible $\bxs \in \myfirstset{\bu}{\tonumber{\n}}{\n}$, 
up to equivalent rank vectors. An imputation $\bxs \in \myfirstset{\bu}{\tonumber{\n}}{\n}$ 
is such that all components in $\bu$ are either larger or smaller than all components at indices $\tonumber{\n}\setminus\bu$. Let $\mln{1} = |\bu|$ and let $\rr$ be the number of components smaller than all components in $\tonumber{\n}\setminus\bu$. Then, there are $\mln{1} - \rr$ components of $\bxs$ 
that are larger than its all components with indices in $\tonumber{\n}\setminus\bu$.

For the $\rr$ smaller components with indices in $\bu$, each component can take a distinct rank between 1 and $\rr$. Hence, there are $\rr !$ possibilities for the ranks of these components. Similarly, for the $\mln{1} - \rr$ larger components, each component can take a distinct rank between $\n - (\mln{1} - \rr) + 1$ and $\n$. Hence, there are $(\mln{1} - \rr)!$ possible ranks for these larger components. Since $0 \leq \rr \leq \mln{1}$, the number of possible rank vectors of imputations $\bxs \in \myfirstset{\bu}{\tonumber{\n}}{\n}$ is given by $\sum_{\rr = 0}^{\mln{1}} \rr !(\mln{1} - \rr)!$.

Even for a moderately small $\mln{1}$, such as $\mln{1} = 20$, the number of permutations can exceed $5.14 \times 10^{18}$. Therefore, further refinement of Proposition~\ref{prop:maximumcase1} is necessary to determine the maximum possible Spearman's footrule efficiently.
We therefore consider ordering the components of imputations $\bxs$ with indices in $\bu$ according to the values of their paired components. We start by introducing the following lemma.

\begin{lemma} \label{Lemma:5}
    Suppose $\bx,\by \in \vndistinct$ and for $\mln{1} \leq \n$ assume $\bu = \{1,\ldots,\mln{1}\}$ and 
    $\byn{1} < \cdots < \byn{\mln{1}}$. 
    Let $\bxp$ be any imputation of $\bx$ for indices $\bu$. Then, suppose 
    $\bxs$ is also an imputation of $\bx$ for indices $\bu$,
    and $\svector{\bxs}{\iconstant}{ \bu}$ is a permutation of 
    $\svector{\bxp}{\iconstant}{\setminus \bu}$ such that 
    $\bxsn{1} > \cdots > \bxsn{\mln{1}}$. Then, $\sfd{\bxp}{\by} \le \sfd{\bxs}{\by}$.
\end{lemma}

Consider any imputation $\bxp$ of $\bx$ for indices $\bu$, Lemma~\ref{Lemma:5} suggests the values of Spearman's footrule will always be larger or equal if the components of $\bxp$ with indices in $\bu$ are further arranged inversely according to their paired components. More formally, let us define
\begin{definition} \label{def:2}
    Suppose $\by \in \vndistinct$ and let $\bu \subset \bt \subset \tonumber{\n}$. Define
    \begin{align*}
        \mysecondset{\by}{\bu}{\bt}{\n} 
        = \left\{\bz \in \myfirstset{\bu}{\bt}{\n} |  ~\text{for any}~\iconstant, \jconstant \in \bu,~ \bzn{\iconstant} > \bzn{\jconstant}~\text{if}~\byn{\iconstant} < \byn{\jconstant} \right\}.
    \end{align*}
\end{definition}
Then, combining Proposition~\ref{prop:maximumcase1} and Lemma~\ref{Lemma:5}, the following result 
will allow us to efficiently determine an imputation $\bxs$ that maximises Spearman's footrule.
\begin{theorem} \label{theorem:maximumcase1}
    Suppose $\bx,\by \in \vndistinct$ and let $\bu \subset \tonumber{\n}$. 
    Then, there exists an imputation $\bxs \in \mysecondset{\by}{\bu}{\tonumber{\n}}{\n}$ of $\bx$ for indices $\bu$ such that	$\sfd{\bx}{\by} \le \sfd{\bxs}{\by}$, and for any other imputation $\xaa \in \vndistinct$ of $\bx$ for indices $\bu$, 
    $\sfd{\bxp}{\by} \le \sfd{\bxs}{\by}$.	
\end{theorem}

Compared with Proposition \ref{prop:maximumcase1}, Theorem \ref{theorem:maximumcase1} further narrows down the scope of possible imputations $\bxs$ to the set $\mysecondset{\by}{\bu}{\tonumber{\n}}{\n}$ for achieving the maximum possible Spearman's footrule.

\begin{remark} \label{remark:3}
    Suppose an imputation $\bxs \in \mysecondset{\by}{\bu}{\tonumber{\n}}{\n}$. Then $\bxs \in \myfirstset{\bu}{\tonumber{\n}}{\n}$, which means all components of $\bxs$ in $\bu$ either larger or smaller than all components in $\tonumber{\n}\setminus\bu$. Let $\mln{1} = |\bu|$ and let $\rr$ be the number of components smaller than all components in $\tonumber{\n}\setminus\bu$. Then, there are $\mln{1} - \rr$ components of $\bxs$ that are larger than its all components with indices in $\tonumber{\n}\setminus\bu$. The ranks of the $\rr$ smaller components in $\bu$ take all values between 1 and $\rr$, and the ranks of the $\mln{1} - \rr$ larger components in $\bu$ take all values between $\n - (\mln{1} - \rr) + 1$ and $\n$. Crucially, the rank of each component in $\bu$ is decided by the rank of its paired component. 
    Since $0 \le \rr \le \mln{1}$, the number of possible rank vectors for imputations $\bxs \in \mysecondset{\by}{\bu}{\tonumber{\n}}{\n}$ is $\mln{1} + 1$.
\end{remark}

Computing the value of Spearman's footrule between $\bxs$ and $\by$, for one possible $\bxs$,
requires computational complexity $\bmo(\n\log\n)$. Thus, computing Spearman's footrule for  
$\bxs \in \mysecondset{\by}{\bu}{\tonumber{\n}}{\n}$ with the distinct $(\mln{1} + 1)$ rank vectors has computational complexity of $\bmo(\n\mln{1}\log\n)$. However, it is not necessary to re-rank all components and re-calculate the value of Spearman's footrule every time the ranks of components of $\bxs$ in $\bu$ are altered. Section~\ref{section:efficientalgrithm} provides an efficient algorithm for computing the maximum Spearman's footrule under \mcase{I} with computational cost $\bmo(m_1 + \n \log \n)$.

\subsubsection{Missing Case II} \label{section:upperbound:caseII}

This section considers the upper bound of Spearman's footrule under \mcase{II}, where both $\bx$ and $\by$ may contain missing components, but each pair $(\bxn{\iconstant}, \byn{\iconstant})$ has \emph{at most one} value missing.

The following result is obtained by applying Theorem \ref{theorem:maximumcase1} under \mcase{I} twice.

\begin{theorem} \label{theorem:maximumcase2}
    Suppose $\bx,\by \in \vndistinct$ and $\bu, \bv \subset \tonumber{\n}$.
    Then, there exist imputations $(\bxs,\bys) \in (\mysecondset{\by}{\bu}{\tonumber{\n}}{\n}, \mysecondset{\bx}{\bv}{\tonumber{\n}}{\n})$ of $\bx, \by$ for indices $\bu$ and $\bv$, respectively, such that	$\sfd{\bx}{\by} \le \sfd{\bxs}{\bys}$. Furthermore, consider any imputation $\bxp,\byp \in \vndistinct$ of $\bx, \by$ for indices $\bu$ and $\bv$, respectively, we have 
    $\sfd{\bxp}{\byp} \le \sfd{\bxs}{\bys}$.
\end{theorem}

If we denote $\mln{1} = |\bu|$ and $\mln{2} = |\bv|$, 
then the number of possible rank vectors 
of imputations $\bxs \in \mysecondset{\by}{\bu}{\tonumber{\n}}{\n}$ is $\mln{1} + 1$ 
and the number of possible rank vectors of imputations 
$\bys \in \mysecondset{\bx}{\bv}{\tonumber{\n}}{\n}$ is $\mln{2} + 1$, 
following reasoning similar to that in Remark~\ref{remark:3} for \mcase{I}.
Hence, the number of possible combinations of rank vectors of imputations 
$(\bxs,\bys) \in (\mysecondset{\by}{\bu}{\tonumber{\n}}{\n}, \mysecondset{\bx}{\bv}{\tonumber{\n}}{\n})$ 
is $(\mln{1}+1)(\mln{2}+1)$.

\subsubsection{Missing Case III} \label{section:upperbound:caseIII}

This section provides the upper bound of Speaman's footrule under \mcase{III}, 
where for each pair $(\bxn{\iconstant}, \byn{\iconstant})$, 
the two values are either \emph{both} observed or \emph{both} missing. 

To start, we consider the case where there is only one pair 
$(\bxn{\uu}, \byn{\uu})$ where both values are missing. 
The following result is proved by applying 
Proposition~\ref{prop:multiplestepupperbound} repeatedly.

\begin{proposition} \label{prop:maximumcase3}
    Suppose $\bx,\by \in \vndistinct$, and let $\uu \in \tonumber{\n}$. Suppose $\bxln{1}, \byln{1} \in \vndistinct$ and $\bxln{2}, \byln{2} \in \vndistinct$ are imputations of $\bx$ and $\by$ for the index $\uu$ such that
    \begin{align*}
        &\rank{\bxlnn{1}{\uuln{1}}}{\bxln{1}} = 1,~\text{and}~ \rank{\bylnn{1}{\uuln{1}}}{\byln{1}} = \n,\\
        &\rank{\bxlnn{2}{\uuln{1}}}{\bxln{2}} = \n,~\text{and}~ \rank{\bylnn{2}{\uuln{1}}}{\byln{2}} = 1.
    \end{align*}
    Then, $\sfd{\bx}{\by} \le \max \{ \sfd{\bxln{1}}{\byln{1}}, \sfd{\bxln{2}}{\byln{2}}\}$.
\end{proposition}

When one pair $(\bxn{\uu}, \byn{\uu})$ is missing, 
Proposition~\ref{prop:maximumcase3} gives similar results as in 
Proposition~\ref{prop:multiplestepupperbound}, offering two possibilities 
for determining the maximum possible value of Spearman's footrule. 

Furthermore, to achieve the maximum possible Spearman's footrule, 
Proposition~\ref{prop:maximumcase3} requires the paired ranks of the 
imputations for the index $\uu$ to be at opposite ends of the sequence
$1,\ldots,\n$. 
For example, if the imputed rank of $\bxn{\uuln{1}}$ is 1, then the 
imputed rank of its paired component must be $\n$; conversely, if the 
imputed rank of $\bxn{\uuln{1}}$ is $\n$, then the rank of its paired 
component must be 1. Let $\bxs$ and $\bys$ be imputations of 
$\bx$ and $\by$, respectively. 
This condition can be expressed as
\begin{align*}
    \rank{\bxsn{\uu}}{\bxs} + \rank{\bysn{\uu}}{\bys} = \n + 1.
\end{align*}
It will be useful to define a set of indices for which this condition is
satisfied.
\begin{definition} \label{def:3}
    Let $\bw \subset \tonumber{\n}$ be a subset of indices.
    Define
    $\mythirdset{\bw}{\n}$ as 
    \begin{align*}
        \mythirdset{\bw}{\n}  =
        \left\{
            (\bzln{1},\bzln{2}) \in (\myfirstset{\bw}{\tonumber{\n}}{\n}, \vndistinct) |\rank{\bzlnn{1}{\iconstant}}{\bzln{1}} + \rank{\bzlnn{2}{\iconstant}}{\bzln{2}} = \n + 1,~\iconstant \in \bw
            \right\}.
    \end{align*}
\end{definition}
Using Definition~\ref{def:3}, we can state the following result.
\begin{theorem}  \label{theorem:maximumcase3}
    Let $\bx,\by \in \vndistinct$ and $\bw \subset \tonumber{\n}$.
    Then there exist imputations $(\bxs,\bys) \in \mythirdset{\bw}{\n}$ of $\bx$ and $\by$ for indices $\bw$ such that $\sfd{\bx}{\by} \le \sfd{\bxs}{\bys}$. Furthermore, consider any other imputations $\bxp,\byp \in \vndistinct$ of $\bx, \by$ for indices $\bw$. 
    Then $\sfd{\bxp}{\byp} \le \sfd{\bxs}{\bys}$.
\end{theorem}

Theorem~\ref{theorem:maximumcase3} establishes the existence of imputations
$(\bxs,\bys) \in \mythirdset{\bw}{\n}$ of $\bx$ and $\by$  that will maximize 
Spearman's footrule $\sfd{\bx}{\by}$.
In order to find this pair $(\bxs,\bys)$, 
we could enumerate all possible imputations. 
If $\bxs,\bys \in \realR$, there would be uncountably many imputations.
But, as already mentioned, we only need to consider $\bxs$ and $\bys$
to be \emph{rank} vectors, since 
the computation of Spearman's footrule only uses ranks.
Therefore, the problem is reduced to only considering rank vector imputations, 
and these can be enumerated.
However, different rank vector imputations may lead to the same value 
of Spearman's footrule statistic. 
Therefore, we will only count the number of
possible \emph{distinct values} of Spearman's footrule statistic 
using (ranked) imputations 
$(\bxs,\bys) \in \mythirdset{\bw}{\n}$.

For any imputations 
$(\bxs,\bys) \in \mythirdset{\bw}{\n}$ of $\bx$ and $\by$ for indices 
$\bw$, we have $\bxs \in \myfirstset{\bw}{\tonumber{\n}}{\n}$ using
Definition~\ref{def:3}.
Therefore, according to Definition~\ref{def:1} 
of $\myfirstset{\bw}{\tonumber{\n}}{\n}$, either
$\bxsn{\iconstant} > \max_{\jconstant \in \tonumber{\n} \setminus \bw} \bxsn{\jconstant}$
or
$\bxsn{\iconstant} < \min_{\jconstant \in \tonumber{\n} \setminus \bw} \bxsn{\jconstant}$, 
for any $\iconstant \in \bw$.

Denote $\mln{3} = |\bw|$ and $\rr = \sum_{\iconstant \in \bw} \indicator{\bxsn{\iconstant} < \svector{\bxs}{\lconstant}{\tonumber{\n} \setminus \bw} }$, i.e. the number of components of $\bxs$ with indices in $\bw$ smaller than all components of $\bxs$ with indices in $\tonumber{\n}\setminus\bw$. Then, there are $\mln{3} - \rr$ 
number of components of $\bxs$ with indices in $\bw$ that are 
larger than all components of $\bxs$ with indices in $\tonumber{\n}\setminus\bw$.

In fact, any imputations $(\bxs,\bys) \in \mythirdset{\bw}{\n}$ of $\bx$ and $\by$ for indices $\bw$ with the same value of $\rr$ give the same value of Spearman's footrule. 
For the moment, let $\rr \in \{1, 2, \dots, \mln{3}-1\}$.
The $\rr$ number of components of $\bxs$ with indices in $\bw$ smaller than all components of $\bxs$ with indices in $\tonumber{\n}\setminus\bw$ take the ranks between between 1 and 
$\rr$, and the $\mln{3} - \rr$ components of $\bxs$ with indices in $\bw$ larger than all components of $\bxs$ with indices in $\tonumber{\n}\setminus\bw$ take the 
ranks between $\n - (\mln{3} - \rr) + 1$ and $\n$. According to the definition of $\mythirdset{\bw}{\n}$, the ranks of components of $\bys$ in $\bw$ are decided by the equation $\rank{\bxsn{\iconstant}}{\bxs} + \rank{\bysn{\iconstant}}{\bys} = \n + 1$. Hence, without changing the values of Spearman's footrule, we can reorder the pairs of components of $\bxs$ and $\bys$ in $\bw$, 
so that
the components of $\bxs$ in $\bw$ are arranged from small to large as 
$1, \ldots, \rr, \n - (\mln{3} - \rr) +1, \ldots, \n,$
and then the components of $\bys$ in $\bw$ are arranged as
$\n, \ldots, \n+1-\rr, \mln{3} - \rr, 1$.

In the preceding paragraph, we considered $\rr \in \{1, 2, \dots, \mln{3}-1\}$, but 
it is also possible that $\rr=0$ or $\rr=\mln{3}$, 
in which case either all components of $\bxs$ with indices in $\bw$
are smaller or larger than all other components, which can be considered
similarly.

Since $\rr$ ranges between $0$ to $\mln{3}$, 
there are at most $\mln{3} + 1$ possible values
of Spearman's footrule statistic, given by all possible
imputations $(\bxs,\bys) \in \mythirdset{\bw}{\n}$.

The above procedure shows how to construct imputations $(\bxs,\bys)$ for each 
of these possible $\mln{3}+1$ values of Spearman's footrule. 
This is detailed in Algorithm~4 in the Supplementary Material.

\subsubsection{General Missing Case}

This section considers the upper bound of Spearman's footrule under the general missing case, where the missing pattern of data is a combination of \textit{Missing Cases I}, \textit{II}, and \textit{III}. The previous results can be combined to give the following result.

\begin{theorem} \label{theorem:maximumCase4}
    Suppose $\bx, \by \in \vndistinct$, and let $\bu$, $\bphi$, $\bpsi \subset \tonumber{\n}$ be pairwise disjoint subsets. 
    Then, there exist $(\bxs,\bys) \in (\mysecondset{\by}{\bu}{\tonumber{\n} \setminus \bw}{\n},  
    \mysecondset{\bx}{\bv}{\tonumber{\n} \setminus \bw}{\n}) \cap \mythirdset{\bw}{\n}$ of imputations $\bx$ and $\by$ for indices $\bu\cup\bw$ and $\bv \cup \bw$, respectively, such that $\sfd{\bx}{\by} \le \sfd{\bxs}{\bys}$. Furthermore, 
    {consider any imputation $\bxp,\byp \in \vndistinct$ of $\bx, \by$ for indices $\bu\cup\bw$ and $\bv \cup \bw$, respectively. Then we have} 
    $\sfd{\bxp}{\byp} \le \sfd{\bxs}{\bys}$.
\end{theorem}

The number of possible rank vectors of imputations $\bxs$ and $\bys$ for achieving the maximum possible Spearman's footrule under the general missing case can be analyzed by combining the analysis for \textit{Missing Cases I}, \textit{II}, and \textit{III}. 

Let us denote $|\bu| = \mln{1}, |\bphi| = \mln{2}, |\bpsi| = \mln{3}$. Consider any pair of imputations 
$(\bxs,\bys) \in \mythirdset{\bw}{\n} \cap 
(\mysecondset{\by}{\bu}{\tonumber{\n} \setminus \bw}{\n},  
\mysecondset{\bx}{\bv}{\tonumber{\n} \setminus \bw}{\n})$ of $\bx$ and $\by$ for indices $\bu\cup\bw$ and $\bv \cup \bw$, respectively, where $\bu$ and $\bv$ are disjoint from $\bw$.
In other words, consider a pair of imputations $(\bxs,\bys) \in \mythirdset{\bw}{\n}$, which means the components
in $\bw$ satisfy the condition in Definition~\ref{def:3}, and 
that $\bxs$ and $\bys$ satisfy the conditions of Definition~\ref{def:2} for the relevant components with indices
in $\bu$ and $\bv$, respectively.

Now suppose the components of $\bxs$ with indices in $\bu$ and the 
components of $\bys$ with indices in $\bv$ are fixed.
Then, following the discussion of Theorem~\ref{theorem:maximumcase3}, there are $\mln{3} + 1$ 
permutations of ranks of the imputations $\bxs$ and $\bys$ for indices in $\bw$ that would result in 
different values of Spearman's footrule. 

Next, suppose the components $\bxs$ and the components of $\bys$
with indices in $\bw$ are fixed.
Then there are $\mln{1} + 1$ and 
$\mln{2} + 1$ permutations of ranks of the imputations $\bxs$ and $\bys$ 
with indices in $\bu$ and $\bv$, respectively, that result in different values 
of Spearman's footrule, 
following a similar argument to Remark~\ref{remark:3}.

Hence, considering all imputations for indices $\bu\cup\bw$ and $\bv \cup \bw$,
the total number of imputations to consider in order to find the maximum
value of Spearman's footrule is $(\mln{1} + 1)(\mln{2} + 1)(\mln{3} + 1)$.

For calculating the values of Spearman's footrule of the $(\mln{1} + 1)(\mln{2} + 1)(\mln{3} + 1)$ 
number of imputations, an efficient algorithm is provided in Section 2 of the \sm, which 
has computational complexity of $\bmo(\n\log \n + \mln{1}\mln{2}\mln{3})$.

\subsection{Additional Results} \label{section: additional results} Obtaining the bounds of Spearman's footrule with partially observed data $\bx, \by$ allows us to derive the bounds of Kendall's $\tau$ for the same data.

Let $\bx, \by \in \vndistinct$, and define Kendall's $\tau$ coefficient according to \eqref{defintion: Kendall Tau}. The following result is proved in Theorem 2 in \cite{diaconis1977spearman}, which shows that the Kendall's $\tau$ coefficient and Spearman's footrule $D$ differ by at most a constant factor. 
\begin{theorem} [\cite{diaconis1977spearman}] \label{Theorem:KendallTau}
    Suppose $\bx, \by \in \vndistinct$. Then, the values of Spearman's footrule and Kendall's $\tau$ between $\bx$ and $\by$ are such that
    \begin{align*}
        \tau({\bx},{\by}) \le \sfd{\bx}{\by} \le 2\tau({\bx},{\by}).
    \end{align*} 
\end{theorem}

Suppose $\bx, \by \in \vndistinct$ are partially observed. Let us denote the exact lower and upper bounds of Spearman's footrule between $\bx, \by$ as $D_{\min}(\bx,\by)$ and $D_{\max}(\bx,\by)$, respectively, which can be computed according to Theorem~\ref{theorem:minimumCase4} and Theorem~\ref{theorem:maximumCase4}, respectively. Since $D_{\min}(\bx,\by) \le D(\bx,\by) \le D_{\max}(\bx,\by)$, Theorem \ref{Theorem:KendallTau} gives
\begin{align} \label{section:additionalresults:eqn:2}
    D_{\min}(\bx,\by)/2 \le \tau({\bx},{\by}) \le D_{\max}(\bx,\by).
\end{align}

In Section \ref{section: numerical simulations for bounds}, these bounds of the Kendall's $\tau$ coefficient given in \eqref{section:additionalresults:eqn:2} are investigated empirically using numerical simulations.

\subsection{An efficient algorithm for computing the upper bounds} \label{section:efficientalgrithm} 

This section provides an efficient algorithm for calculating exact upper bounds of Spearman's footrule in the presence of missing data. We consider \mcase{I} in this section, i.e. missing components are either all in $\bx$ or all in $\by$. The algorithms for \mcase{II}, \textit{III} and the general missing cases can be constructed similarly, and are provided in Section~2 of the \sm. 

Without loss of generality, let us assume $\bx$ is partially observed while 
$\by$ is fully observed. Denote the indices of all missing components in 
$\bx$ as $\bu$ and denote $\mln{1} = |\bu|$. Further, assume 
$\bu = \{1,\ldots, \mln{1}\}$ and the components of $\by$ with indices in $\bu$ 
are such that $\byn{1} < \ldots < \byn{\mln{1}}$. If this is not the case, 
simply relabel the relevant components in $\bx$ and $\by$.

Under \mcase{I}, Theorem~\ref{theorem:maximumcase1} allows us to obtain the maximum possible Spearman's footrule by considering imputations $\bxs$  
of $\bx$ for indices $\bu$ in $\mysecondset{\by}{\bu}{\tonumber{\n}}{\n}$. 
Remark~\ref{remark:3} further demonstrates 
that it is adequate to only consider $\mln{1} + 1$ imputations $\bxs \in \mysecondset{\by}{\bu}{\tonumber{\n}}{\n}$, each with a distinct value of $\rr =  \sum_{\iconstant \in \bu} \indicator{\bxsn{\iconstant} < \min \svector{\bxs}{\lconstant}{\tonumber{\n} \setminus \bu} }$, i.e. the number of components of $\bxs$ in $\bu$ that are smaller than all components of $\bxs$ with indices in $\tonumber{\n} \setminus \bu$, where $\rr \in \{0,\ldots, \mln{1}\}$. Let us denote the $\mln{1} + 1$ imputations as $\bxs^{(0)}, \ldots, \bxs^{(\mln{1})}$, corresponding to the imputations with $\rr = 0, \ldots, \mln{1}$. Then the maximum possible value of Spearman's footrule between partially observed $\bx$ and $\by$ is $\max \left\{\sfdsb{\bxrs{0} }{\by}, \ldots, \sfdsb{\bxrs{\mln{1}} }{\by}   \right\}$. 

The computational complexity of naively computing
all $\mln{1} + 1$ values in the above set 
$\left\{\sfdsb{\bxrs{0} }{\by}, \ldots, \sfdsb{\bxrs{\mln{1}} }{\by}   \right\}$ 
is $\bmo( (\mln{1}+1) \n\log \n)$, since computing the value of Spearman's footrule once is $\bmo (\n \log \n)$.  

However, we notice that it is not necessary to re-rank all components and re-calculate the value of Spearman's footrule every time the ranks of components of $\bxs$ in $\bu$ are altered. 
To construct an algorithm for efficiently calculating exact upper bounds of 
Spearman's footrule under \mcase{I} in $\bmo( \n\log \n)$, 
we provide the following result
\begin{proposition} \label{prop:alg}
    Suppose $\bx, \by \in \vndistinct$ and for $\mln{1} \in \tonumber{\n}$, 
    let $\bu = \{{1},\ldots,{\mln{1}} \} \subset \tonumber{\n}$ 
    be a subset of indices. Suppose
    $\byn{{1}} < \ldots < \byn{{\mln{1}}}$, and 
    let $\bxs^{(\rr)} \in \mysecondset{\by}{\bu}{\tonumber{\n}}{\n}$ 
    be an imputation of $\bx$ for indices in $\bu$ such that 
    $\sum_{\iconstant \in \bu} \indicator{ \bxs^{(\rr)}(\iconstant)   < \min \svector{\bxs^{(\rr)}}{\lconstant}{\tonumber{\n} \setminus \bu} } = \rr$.
    For any $\iconstant \in \tonumber{\n} \setminus \bu$, denote $d_i = \rank{\byn{\iconstant}}{\by} - \rank{\bxn{\iconstant}}{\svector{\bx}{\jconstant}{\tonumber{\n}\setminus \bu}}$,
    and for any $\rr \in \{0, \ldots, \mln{1} - 1\}$, denote $\ssln{\rr} = \sum_{\iconstant \in \tonumber{\n} 
    \setminus \bu} \indicator{d_i \le \rr}$. Then, for any $\rr \in \{0,\ldots, \mln{1}-1\}$, we have
    \begin{align*}
        \sfdsb{\bxrs{\rr+1} }{\by} &= \sfdsb{\bxrs{\rr} }{\by} +  2\ssln{\rr} - \n + \mln{1} + C^{(\rr)},
    \end{align*}
    where $C^{(\rr)} = |\rr + 1 - \rank{\byn{\mln{1} - \rr}}{\by}| - |\n - \mln{1} + \rr + 1 - \rank{\byn{\mln{1} - \rr}}{\by}|$.
\end{proposition}

Proposition~\ref{prop:alg} shows that the maximum possible values of Spearman's footrule can be found in linear time after ranking all observed components in $\bx$ and $\by$, as shown in Algorithm~\ref{alg:3}.

\begin{algorithm} 
    \caption{An efficient algorithm for computing exact upper bounds of Spearman's footrule under Missing Case I.}  \label{alg:3}
    \begin{algorithmic}[1]
        \Require{$\bx, \by \in \vndistinct$, where either $\bx$ or $\by$ is fully observed,
        while the other is partially observed. } 
        \Ensure{Maximum possible Spearman's footrule distance between $X$ and $Y$.}
        \State If $\bx$ is fully observed while $\by$ is partially observed, then change the label of $\bx$ and $\by$.
        \State Denote $\mln{1}$ as the number of missing components in $\bx$.
        If $\mln{1} = \n$, return $\sum_{\iconstant=1}^{\n} |\iconstant - (\n -\iconstant + 1)|$.
        \State Rank all observed components in $X$ and $Y$.  \label{alg:3:line:1}
        \State	Relabel $\bx$ and $\by$ such that $(\bxn{1}, \ldots, \bxn{\mln{1}})$ are unobserved and $\byn{1} < \ldots < \byn{\mln{1}}$.
        \State Let $\bu = \{1,2, \dots,\mln{1}\}.$ For $\iconstant \in \tonumber{\n} \setminus \bu$, let $\dln{\iconstant} = \rank{\byn{\iconstant}}{\by} - \rank{\bxn{\iconstant}}{\svector{\bx}{\lconstant}{\tonumber{\n} \setminus \bu}}.$ \label{alg:3:line:2}
        \State Computing $\ssln{\rr} =  \sum_{\iconstant \in \tonumber{\n} \setminus \bu} \indicator{\dln{\iconstant} \le \rr}$ for any $\rr = \{0, \ldots, \mln{1} - 1\}$. \label{alg:3:line:3}
        \State For any $\rr \in \{0, \ldots, \mln{1} - 1\}$, compute \label{alg:3:line:4}
        \begin{align*}
            C^{(\rr)} = |\rr + 1 - \rank{\byn{\mln{1} - \rr}}{\by}| - |\n - \mln{1} + \rr + 1 - \rank{\byn{\mln{1} - \rr}}{\by}|.
        \end{align*}
        \State Initialize $\bdln{0}$ = $\sum_{\iconstant = 1}^{\mln{1}} \left| (n - \iconstant + 1) - \rank{\byn{\iconstant}}{\svector{\by}{\lconstant}{\tonumber{\n}}}\right| + \sum_{i \in \tonumber{\n} \setminus \bu} |d_i|$. \label{alg:3:line:5}
        \State For $\rr \in \{0,\ldots,\mln{1}-1\}$, compute $D_{\rr+1} =  D_{\rr}  + 2\ssln{\rr} - \n + \mln{1} + C^{(\rr)}$. \label{alg:3:line:6}
        \State Return $\max \{\bdln{0}, \ldots, \bdln{\mln{1}}\}$.
    \end{algorithmic}
\end{algorithm}

\begin{remark}		
    The computational complexity of Algorithm~\ref{alg:3} is analyzed as follows. Ranking and relabeling 
    all observed components in $\bx$ and $\by$ in line~\ref{alg:3} requires $\bmo (\n \log \n)$ steps. 
    Using these rankings, in line~\ref{alg:3:line:2} computing each $\dln{\iconstant}$ is $\bmo(1)$, and so overall
    line~\ref{alg:3:line:2} is $\bmo(\n - \mln{1})$. 
    In line~\ref{alg:3:line:3}, $\ssln{0}, \dots, \ssln{\mln{1}-1}$ can be computed 
    collectively in $\bmo(\n - \mln{1})$ steps. See Algorithm~1 in the Supplementary Material.
    is $\bmo(\n + \mln{1}) = \bmo(\n)$. In line~\ref{alg:3:line:4}, each iteration of the for loop is $\bmo(1)$, and since the loop runs $\mln{1}$ times, the computational complexity for the loop is $\bmo(\mln{1})$. Line~\ref{alg:3:line:5} and \ref{alg:3:line:6}
    takes $\bmo(1)$ and $\bmo(\mln{1})$ steps, respectively.
    Therefore, the overall computational complexity for Algorithm~\ref{alg:3} is $\bmo( \n\log \n)$.
\end{remark}

\subsection{Simulation results for the bounds} \label{section: numerical simulations for bounds}

This section provides numerical simulations that evaluate the performance of the exact bounds of Spearman's footrule, and the bounds of Kendall's $\tau$ coefficient provided in \eqref{section:additionalresults:eqn:2}, between partially observed $\bx, \by \in \vndistinct$, with different proportions of the missing pairs.

\begin{figure}[h]
    \includegraphics[width=13.8cm]{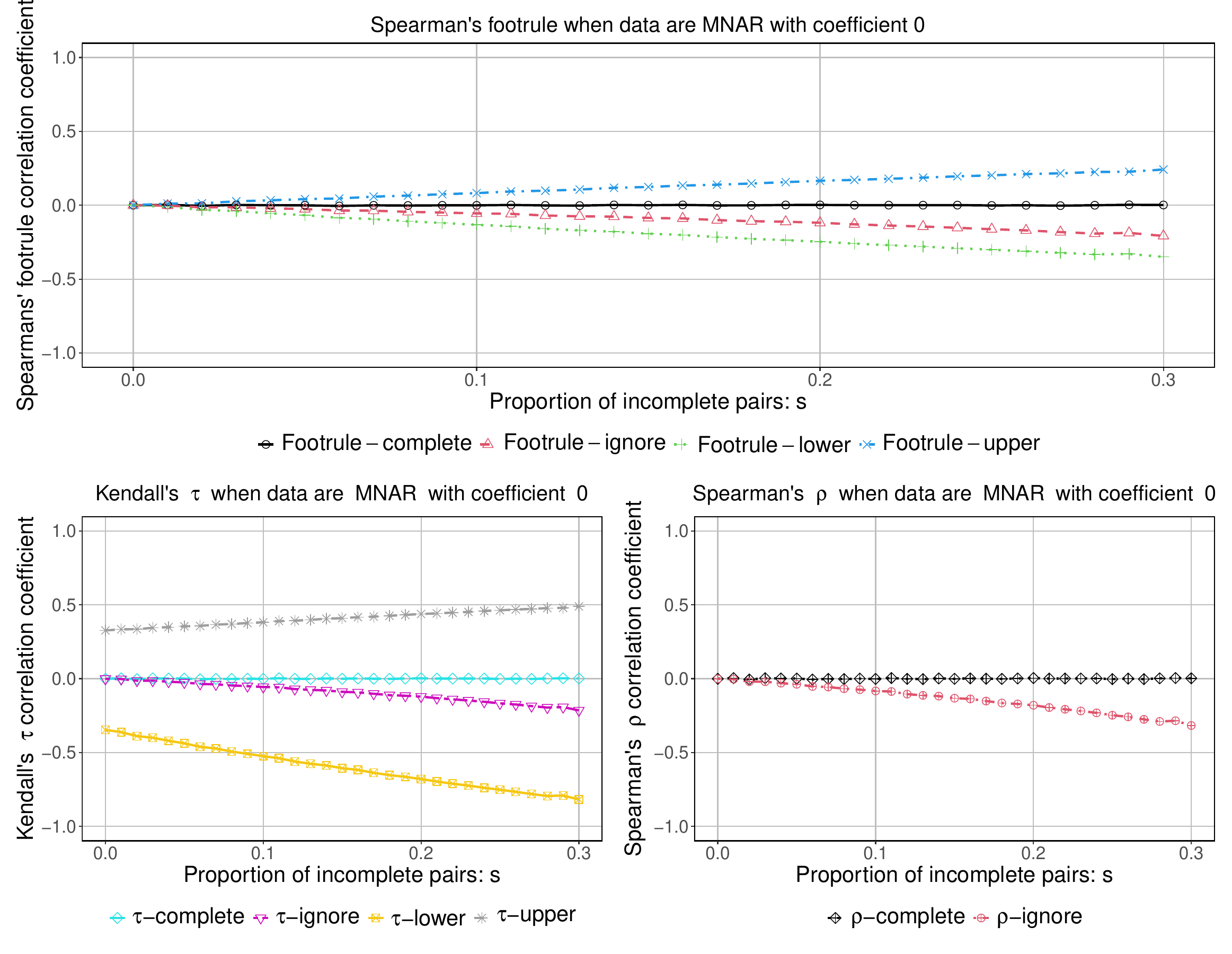}
    \caption{Rank correlation coefficients when data are missing completely at random (MNAR). 
    $\bx,\by$ are generated such that 
    $(\bx,\by) \sim^{iid} N(0, I_2)$. The methods are described in Table~\ref{tab:2}.
    The sample size for $\bx$ and $\by$ is $\n = 100$.
    Results represent the average of 1000 Monte Carlo simulations.}
    \label{fig:2}
\end{figure}

Additionally, we also investigate the behaviour of the methods when (i) the complete data are used and (ii) the missing values are ignored. When the missing values are ignored, the rank correlation coefficients are computed using only the pairs where both components from $\bx$ and $\by$ are observed.
When the data are missing not at random (MNAR), 
ignoring the missing data can lead to biased estimates of the correlation coefficients, 
as shown in Figure~\ref{fig:2}.

We first scale all the rank correlation coefficients between $[-1,1]$. For scaling Spearman's footrule, we define
\begin{align}
    D_{\text{Scale}}(\bx,\by) = 1 - {3\sfd{\bx}{\by}}/{(n^2-1)}.
    \label{eqn:scaledfootrule}
\end{align}
When $\n$ is odd, $D_{\text{Scale}}(\bx,\by) \in [-0.5, 1]$, but when $\n$ is even,
$D_{\text{Scale}}(\bx,\by)$ is in the range $[-0.5  \{1+3/(n^2-1)\}, 1]$ \cite{kendall1948rank}. For scaling Spearman's $\rho$ and Kendall's $\tau$, define
\begin{align}
    \rho_{\text{Scale}}(\bx,\by) = 1 - \rho(\bx,\by)/\{n(n^2-1)\}, 
    \,\,\,\,
    \tau_{\text{Scale}}(\bx,\by) =  1 - 4\tau(\bx,\by)/\{n(n-1)\},  
    \label{eqn:scaledtaurho}
\end{align}
respectively, where $\tau(\bx,\by)$ and $\rho(\bx, \by)$ are given in 
\eqref{definition: Spearman Rank Correlation} and \eqref{defintion: Kendall Tau}, 
respectively. It can be shown that both $\rho_{\text{Scale}}(\bx,\by)$ 
and $\tau_{\text{Scale}}(\bx,\by)$ are in the range $[-1,1]$ \cite{kendall1948rank}.

\subsubsection{The bounds when data are missing completely at random}
\label{sec:boundsmcar}

In the first simulation, data are assumed to be missing completely at random (MCAR). 
First, $\bx,\by$ are generated such that each pair $(\bx(\iconstant), \by(\iconstant))$ 
is an independent random sample from a two-dimensional standard normal distribution with covariance coefficient 0;
in other words. $\bx$ and $\by$ are uncorrelated. The sample size for $\bx$ and $\by$ is $\n = 100$.
Then, for each given proportion $s \in \{0, 0.05, \ldots, 0.30\}$ a subset of indices $\bt$ in  
$\{1, 2, \dots, n\}$ of size 
$\lfloor s \cdot \n  \rfloor $ is selected to be the set of components that are either partially observed or missing.
For each $\iconstant \in \bt$, 
there is probability $1/3$ such that $\bx(\iconstant)$ is missing but $\by(\iconstant)$ is observed, 
probability $1/3$  such that $\by(\iconstant)$ is missing but $\bx(\iconstant)$ is observed, 
and probability $1/3$ such that both $\bx(\iconstant)$ and $\by(\iconstant)$ are missing. 
In any of these three cases, where at least one of $\bx(\iconstant)$ and $\by(\iconstant)$
is missing, we say that $( \bx(\iconstant), \by(\iconstant) )$ is an \emph{incomplete pair}.

\begin{figure}
    \includegraphics[width=13.8cm]{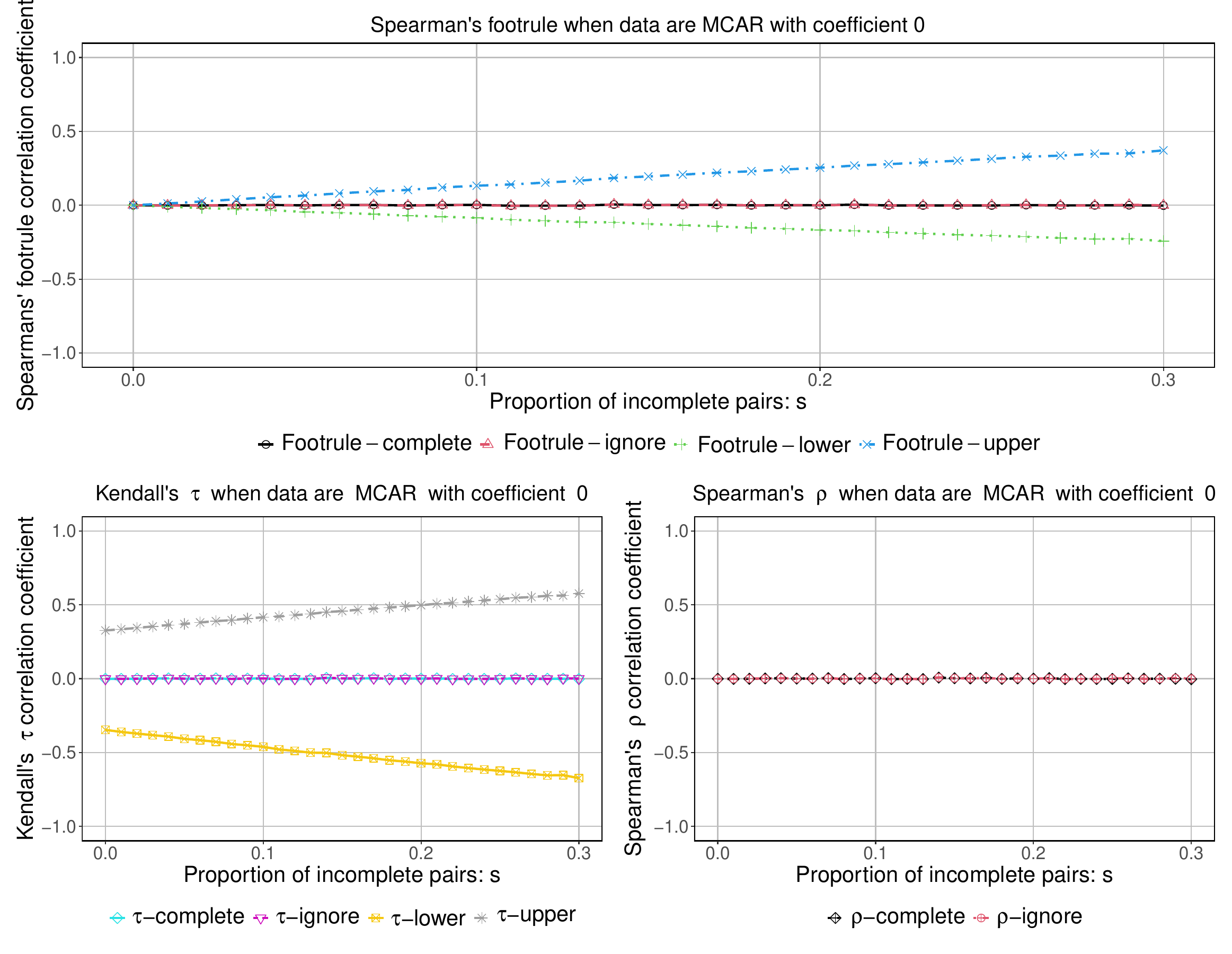}
    \caption{Rank correlation coefficients when data are missing completely at random (MCAR). 
    $\bx,\by$ are generated such that 
    $(\bx,\by) \sim^{iid} N(0, I_2)$. The methods are described in Table~\ref{tab:2}.
    The sample size for $\bx$ and $\by$ is $\n = 100$.
    Results represent the average of 1000 Monte Carlo simulations.}
    \label{fig:1}
\end{figure}

Subsequently, the bounds of Spearman's footrule 
for partially observed $\bx$ and $\by$ will be computed using
Theorems~\ref{theorem:minimumCase4} and \ref{theorem:maximumCase4}, while
the bounds for Kendall's $\tau$ for partially observed data will be computed using
\eqref{section:additionalresults:eqn:2}.
We do not have bounds for Spearman's $\rho$ for partially observed data, but we 
compute values for all three rank correlation coefficients in the cases where the
complete data is used or missing values are ignored.
Figure~\ref{fig:1} presents the results, which are an average over $1000$ Monte Carlo simulations.

Figure~\ref{fig:1} shows that 
when the (uncorrelated) data are MCAR, ignoring the missing values will result in the methods
producing unbiased estimates of the rank correlation coefficients, i.e. \emph{Footrule-ignore}, \emph{$\tau$-ignore}
and \emph{$\rho$-ignore} are all approximately zero.
Similarly, when the complete data are used, 
the three rank correlation coefficients will be unbiased, i.e.
\emph{Footrule-complete}, \emph{$\tau$-complete}
and \emph{$\rho$-complete} are all approximately zero.

The range of bounds of Spearman's footrule,
labelled \emph{Footrule-lower} and \emph{Footrule-upper},
increases gradually with the increasing proportion of incomplete pairs $s$. When $s = 0.3$, the lower and upper bounds of Spearman's footrule are around -0.2 and 0.4, respectively. 

For Kendall's $\tau$ coefficient, when the proportion of incomplete pairs is 0, 
the lower and upper bounds of Kendall's $\tau$,
labelled by \emph{$\tau$-lower} and \emph{$\tau$-upper},
range from approximately $-0.3$ to $0.3$. 
The bounds become wider as the proportion of incomplete pairs increases. 
When $s = 0.3$, the lower and upper bounds of Kendall's $\tau$ are approximately -0.6 and 0.5, respectively.
It appears that the bounds of Spearman's footrule and Kendall's $\tau$ coefficient
increase or decrease linearly, depending on the proportion $s$ of missing components.

\begin{table*}
    \caption{Description of the rank correlation coefficients or their bounds, often in the presence of 
    missing data, shown in Figure~\ref{fig:1} -- \ref{fig:2}.}
    \footnotesize
    \begin{tabular}{ll}
        \hline
        Coefficient/bound & Description \\
        \hline
        Footrule-upper & Upper bound for Spearman's footrule statistic $D$, when data is partially observed. \\
        Footrule-lower & Lower bound for Spearman's footrule statistic $D$, when data is partially observed. \\
        Footrule-ignore & Spearman's footrule statistic $D$, when any missing or partially observed data is ignored. \\
        Footrule-complete & Spearman's footrule statistic $D$, when data is fully observed. \\
        \hline
        $\tau$-upper & Upper bound for Kendall's $\tau$ coefficient, when data is partially observed. \\
        $\tau$-lower & Lower bound for Kendall's $\tau$ coefficient, when data is partially observed. \\
        $\tau$-ignore & Kendall's $\tau$ coefficient, when any missing or partially observed data is ignored. \\
        $\tau$-complete & Kendall's $\tau$ coefficient, when the data is fully observed. \\
        \hline
        $\rho$-ignore & Spearman's rank correlation $\rho$, when any missing or partially observed data is ignored. \\
        $\rho$-complete & Spearman's rank correlation $\rho$, when the data is fully observed. \\
        \hline
    \end{tabular}
    \label{tab:2}
\end{table*}

\subsubsection{The bounds when data are missing not at random}

\label{sec:boundsmnar}

The second simulation is designed in the same way as the first, except now the data
are \emph{missing not at random} (MNAR), rather than being MCAR.
There are many ways to defined a MNAR mechanism; we propose one approach here, 
and consider a second approach in the Supplementary Material.
Here, the index set $\bt \subset \{1, \dots, \n\}$ of the missing components 
is chosen now depends on the values of $\bx$ and $\by$.
Let $q = \sum_{\iconstant=1}^{\n} \indicator{\bxn{i}\byn{i} > 0}$
be the number of pairs of components in $\bx$ and $\by$ such that their product $\bxn{i}\byn{i} > 0$.
Each index $\iconstant \in \tonumber{\n}$ 
is selected to be in the set $\bt$ of indices of missing components according to the 
following probability conditional on $C = |T|$, the size of the set $T$:
\begin{align} \label{eqn:1}
    p\left( i \in T \, | \, C = \lfloor s \cdot \n  \rfloor  \right) = \left\{ \begin{array}{ll}
        \min\left\{1, sn/q   \right\},~~& \mbox{if}  \bxn{i}\byn{i} > 0,  \\  \max\{0, (sn - q) / (n - q ) \},~~& \mbox{otherwise,} 
    \end{array}\right.
\end{align}
for any given $s$.
If $sn < q$, then $\min\left\{1, sn/q   \right\} < 1$ and $\max\{0, (sn - q) / (n - q ) \} = 0$, 
and so some components with $\bxn{i}\byn{i} > 0$ 
will be fully observed, and no components with $\bxn{i}\byn{i} \leq 0$ will be partially observed.
However, if $sn  \geq q$, then all components with $\bxn{i}\byn{i} > 0$ will be partially observed
and, moreover, if $sn  > q$ then some components with $\bxn{i}\byn{i} \leq 0$ will be partially observed.
The probability is conditional in order to ensure we have subsets of indices of the desired cardinality
$C = \lfloor s \cdot \n  \rfloor$.

Figure~\ref{fig:2} shows that when the data are MNAR with the missingness mechanism 
specified in \eqref{eqn:1}, then 
\emph{Footrule-ignore}, \emph{$\tau$-ignore}
and \emph{$\rho$-ignore}
produce biased estimates of their rank correlation coefficients.
Furthermore, as the proportion of incomplete pairs, denoted by $s$, increases, 
the bias of 
\emph{Footrule-ignore}, \emph{$\tau$-ignore}
and \emph{$\rho$-ignore}
also increases.
On the other hand, 
\emph{Footrule-lower}, \emph{Footrule-upper}, \emph{$\tau$-lower} and \emph{$\tau$-upper}
have similar performance compared to when data are MCAR in Figure~\ref{fig:1}.
In Section G.2.4 of the Supplementary Material we repeat this experiment using a 
different MNAR missing mechanism, with similar results.

\section{Independence Testing with Missing Data} \label{section:test independence}

We now explore the use of the bounds of Spearman's footrule derived above for 
independence testing in the presence of missing data.
The core idea is to use the bounds for the statistic to obtain bounds for the 
$p$-value, and then to reject the null hypothesis when 
\emph{all possible} $p$-values will be significant.

More specifically, our method uses the bounds of the Spearman's footrule 
statistic to compute bounds for the possible $p$-values, 
given the missing data. 
In other words, if the missing data could have been fully observed, the 
resultant $p$-value would be within the computed bounds.
Then, for a given significance level $\alpha$, if the bounds show that all 
possible $p$-values 
are smaller than $\alpha$, the null hypothesis is rejected. On the other hand, 
if at least one $p$-value is not signficant, then the null hypothesis fails 
to be rejected. Our motivation for this approach is discussed in the 
introduction.

Below we will first show how our method bounds the $p$-values of 
Spearman's footrule in the presence of missing data using the bounds 
of the Spearman's footrule statistic. Then we will perform numerical 
simulations for investigating the Type I error and statistical power of our 
method. The unique contribution of our method is that, unlike all other 
existing methods, it provides an independence testing approach that 
controls the Type I error regardless of the missing data mechanism.
The only assumption we make about the data is that the values are
all distinct in order to avoid dealing with the issue of ties.

\subsection{Bounds of $p$-values in the presence of missing data}

For deriving the bounds of the $p$-value of Spearman's footrule with missing 
data, we begin by studying the distribution of Spearman's footrule statistic 
under the null hypothesis that $\bx$ and $\by$ are generated by independent 
continuous random variables \textit{without} missing data.

\begin{table*}
    \caption{Description of the independence testing methods, often in the presence of 
    missing data, shown in Figure~\ref{fig:3} -- \ref{fig:7}.}
    \footnotesize
    \begin{tabular}{ll}
        \hline
        Testing method & Description \\
        \hline
        Proposed & Based on $p$-values computed from bounds Footrule-upper and Footrule-lower.  \\
        Footrule-ignore & Based on $p$-value of Spearman's footrule, when partially-observed data is ignored. \\
        Footrule-complete & Based on $p$-value of Spearman's footrule, when data is fully observed. \\
        \hline
        Footrule-mean &  Based on $p$-value of Spearman's footrule, using mean imputation for missing values. \\
        Footrule-median & Based on $p$-value of Spearman's footrule, using median imputation for missing values.\\
        Footrule-hot deck &Based on $p$-value of Spearman's footrule, using hot deck imputation for missing values.\\
        \hline
        $\tau$-ignore & Based on $p$-value of Kendall's $\tau$ coefficient, when partially observed data is ignored. \\
        $\tau$-complete & Based on $p$-value of Kendall's $\tau$ coefficient, when the data is fully observed. \\
        \hline
        $\tau$-mean &  Based on $p$-value of Kendall's $\tau$, using mean imputation for missing values.\\
        $\tau$-median & Based on $p$-value of Kendall's $\tau$, using median imputation for missing values.\\
        $\tau$-hot deck & Based on $p$-value of Kendall's $\tau$, using hot deck imputation for missing values.\\
        \hline
        $\rho$-ignore & Based on $p$-value of Spearman's $\rho$, when partially observed data is ignored. \\
        $\rho$-complete & Based on $p$-value of Spearman's $\rho$, when the data is fully observed. \\
        \hline
        $\rho$-mean &  Based on $p$-value of Spearman's $\rho$, using mean imputation for missing values.\\
        $\rho$-median & Based on $p$-value of Spearman's $\rho$, using median imputation for missing values.\\
        $\rho$-hot deck & Based on $p$-value of Spearman's $\rho$, using hot deck imputation for missing values.\\
        \hline
        Alvo and Cabilio's $\rho$ & Based on $p$-value of estimate of Spearman's rank correlation $\rho$, from \cite{Alvo1995RankCM}.\\
        Alvo and Cabilio's $\tau$ & Based on $p$-value of estimate of Kendall's $\tau$ coefficent, from \cite{Alvo1995RankCM}.\\
        \hline
    \end{tabular}
    \label{tab:3}
\end{table*}

A well-known result \cite[Theorem 1]{diaconis1977spearman} shows that under the null hypothesis, the distribution of Spearman's footrule statistic for $n$ pairs of observations approximately follows a normal distribution with mean and variance equal to $n^2/3$ and $2n^3/45$, respectively. The accuracy of this approximation is investigated in \cite{salama1990note}, showing that when the sample size is at least $n = 40$, this normal approximation performs reasonably well: the difference in skewness between the exact and normal approximation distribution of Spearman's footrule  is $-0.075$ and the difference in kurtosis is $-0.074$.

Note that this result regarding the distribution of the test statistic is not 
making any assumption about the distribution of the data itself.
Furthermore, if one uses this result to compute a $p$-value for 
independence testing, the Type I error will be controlled.

Suppose the sample size $\n$ is large enough for employing normal approximation, e.g. $\n \ge 40$.  Denote $F_{n}$ as the cumulative distribution function of a normal distribution with mean equal to $n^2/3$ and variance equal to $2n^3/45$. 
The $p$-value of Spearman's footrule is then calculated as 
\begin{align} \label{pvalue}
    p(D(X,Y)) = 2 \cdot {\min}\{F_{n}(D(X,Y)), 1 - F_{n}(D(X,Y))\} .
\end{align}
Now suppose $\bx, \by \in \vndistinct$ are partially observed, and denote the minimum and maximum values of Spearman's footrule between $\bx$ and $\by$ as $D_{\min}$ and $D_{\max}$, respectively. The bounds of the $p$-value of Spearman's footrule are then determined using the following result.

\begin{proposition} \label{Proposition:boundsofpvalue}
    Suppose $\bx,\by \in \vndistinct$ are partially observed. Assume $\n$ is 
    sufficiently large. Let $D_{\min}$ and $D_{\max}$ be the minimum and 
    maximum possible values of Spearman's footrule between $\bx$ and $\by$, 
    respectively. Denote $p_1 = p(D_{\min})$ and $p_2 = p(D_{\max})$, where $p(\cdot)$ is defined in \eqref{pvalue}. Furthermore, define $p_{\min} = \min \{p_1,p_2\}$ and define 
    \begin{align}
        p_{\max}
        = \left\{ \begin{array}{ll}
            \max \{p_1,p_2\},~~& \mbox{if }  (D_{\min} - n^2/3)(D_{\max} - n^2/3) \geq 0,  \\  
            1,~~& \mbox{otherwise}. 
        \end{array}\right.
        \nonumber
    \end{align}
    Then, the $p$-value of $\sfd{\bx}{\by}$ is bounded such that $p(D(X,Y)) \in [p_{\min}, p_{\max}]$.
\end{proposition} 

In Proposition~\ref{Proposition:boundsofpvalue}, ``$n$ is sufficiently large'' is used to allow the use of normal approximation for computing the $p$-value of Spearman's footrule. 

We now describe our method for independence testing 
in the presence of missing data using 
Proposition~\ref{Proposition:boundsofpvalue}. 
Given a pre-specified significance level $\alpha$, 
our method will reject the null hypothesis if $p_{\max} < \alpha$, 
since this implies that all possible $p$-values are less than 
$\alpha$ and will be significant. However, our method
will fail to reject the null hypothesis if $p_{\min} > \alpha$, 
since this implies that all possible $p$-values are greater than 
$\alpha$ and so none will be significant.

Crucially, since our method rejects the null hypothesis only when $p(D(X,Y))$ 
is significant, the probability of our method making a Type I error is no larger 
than the probability of making a Type I error when data are fully observed, 
which is approximately equal to $\alpha$. 
In other words, our method is guaranteed to control 
the Type I error.

\subsection{Simulation results for independence testing} \label{section:simulation}

This section performs numerical simulations to investigate the Type I error and 
the statistical power of the proposed method for independence testing when 
data are partially observed.

For comparison with the proposed method, we also consider the three methods:
Spearman's footrule, Kendall's $\tau$ and Spearman's $\rho$, in the two cases
where the missing data is ignored and the complete data is observed.
When the missing data are ignored, the correlation coefficients are computed only
using pairs where both components from $\bx$ and $\by$ are observed.
We also consider these three methods when imputation methods are used;
we consider mean imputation, median imputation and hot deck imputation
\cite{2002Missing}.

Two independence testing methods that take missing data into account are also considered. 
These methods \cite{Alvo1995RankCM} are defined as 
the conditional expectations of Spearman's $\rho$ and Kendall's $\tau$, respectively, 
given the observed ranks, under the assumption that the ranks of the missing values
are uniformly distributed.
In the following, we will refer to the two methods 
as Alvo and Cabilio's $\rho$ and Alvo and Cabilio's $\tau$ methods, respectively.

\subsubsection{As the proportion of missing data increases}

In the first experiment, we explore how the methods perform as
the proportion $s$ of missing data increases. More precisely, 
we consider $s$ to be the proportion of incomplete pairs, where 
we recall that $(\bxn{\iconstant}, \byn{\iconstant} )$ is an incomplete pair if
at least one of $\bxn{\iconstant}$ or $\byn{\iconstant}$ is missing.

We start by considering the case when the data are missing completely at 
random (MCAR), as the proportion of incomplete pairs increases. 
For evaluating 
the Type I error of all methods, the data are generated following the same 
distribution and missingness mechanism for Figure~\ref{fig:1}, as described
in Section~\ref{sec:boundsmcar}, with sample size $\n = 500$. 
The significance level $\alpha = 0.05$ is used for all test 
methods. For evaluating the power, the data are generated in the same way,
except that the covariance matrix is now 
$\Sigma = \begin{pmatrix} 1 &  \corrcoef  \\  \corrcoef & 1 \end{pmatrix}$,
    with covariance coefficient $\corrcoef=0.5$.
    The Supplementary Material contains results for the same experiment, but
    with different values of $\alpha$, different values of 
    the covariance coefficient $\corrcoef$, and different sample sizes $\n$.

    \begin{figure}
        \includegraphics[width=14.5cm]{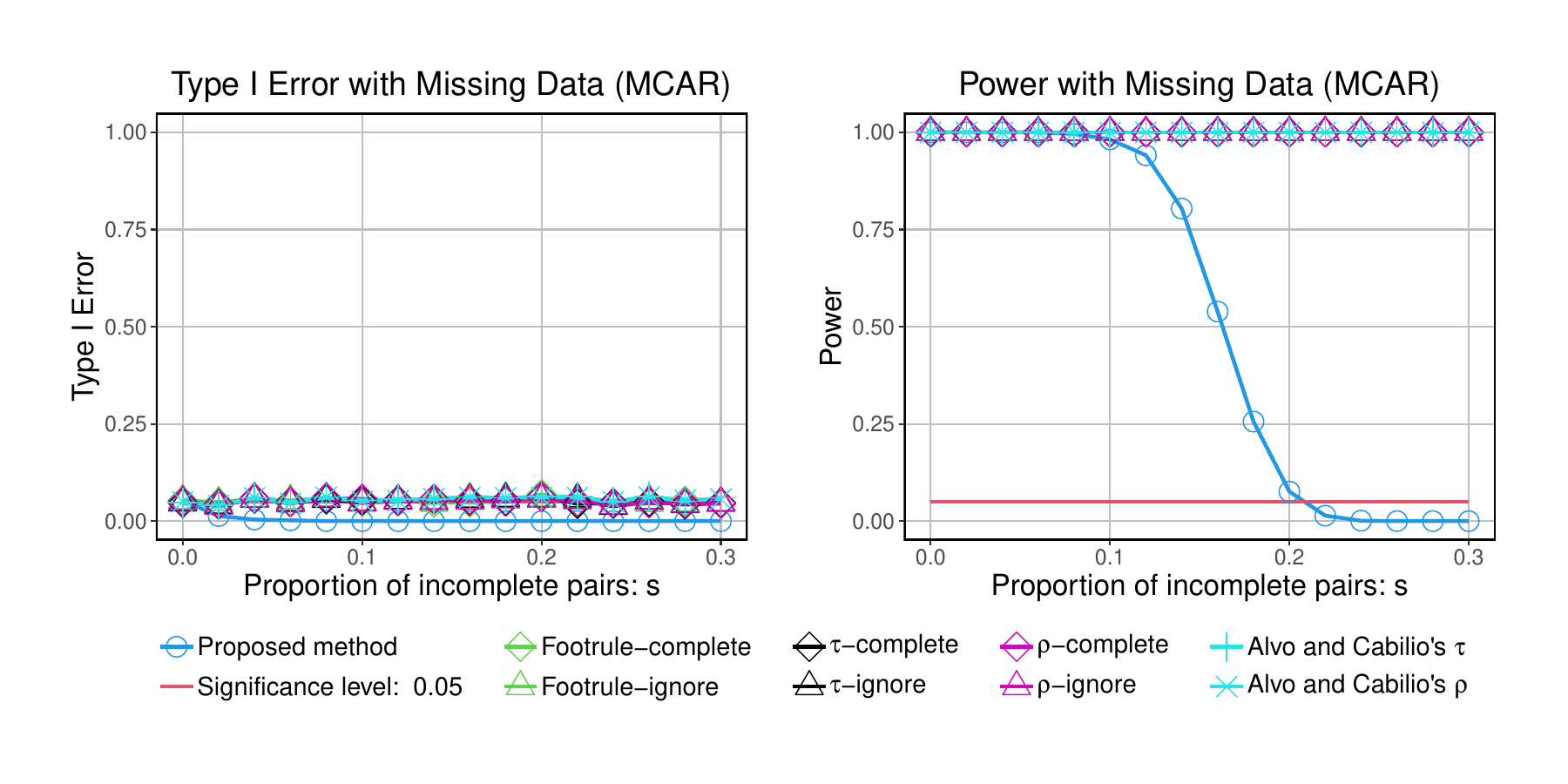}
        \vspace{-0.5cm}
        \caption{Statistical Type I error and power of the proposed method,
        as the proportion of missing data $s$ increases,
        where $s \in \{0.00, 0.02, \ldots, 0.30\}$.
        Other methods that ignore missing data or use complete data,
        and Alvo and Cabilio's $\rho$ and Alvo and Cabilio's $\tau$ methods
        are also considered.
        These methods are described in Table~\ref{tab:3}.
        The data is missing completely at random (MCAR).
        (Left) Type I error: $(\bx,\by) \overset{iid}{\sim} N(0, I_2)$;
        (Right) Power: 	$(X, Y) \overset{iid}{\sim} N(0, \Sigma)$, where
        $\Sigma = \begin{pmatrix} 1 &  \corrcoef  \\  \corrcoef & 1 \end{pmatrix}$,
            with covariance coefficient $\corrcoef=0.5$.
            For both figures, a significance level $\alpha = 0.05$ is used and the sample
            size $\n = 200$. The results in the figures are average of 1000 trials.}
            \label{fig:3}
    \end{figure}

    \begin{figure}[h]
        \includegraphics[width=14.5cm]{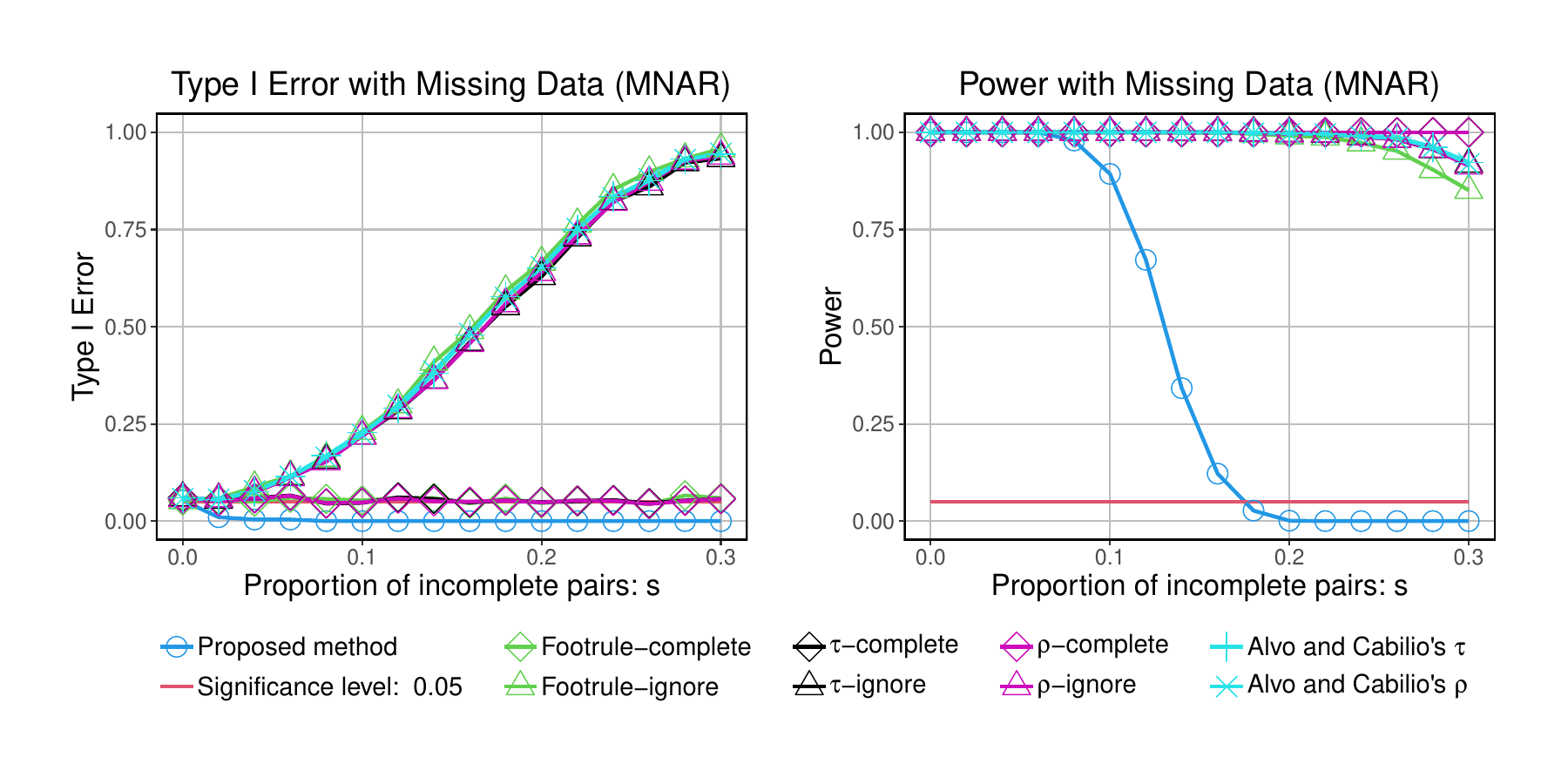}
        \vspace{-0.5cm}
        \caption{Statistical Type I error and power of the proposed method
        as the proportion of missing data $s$ increases,
        where $s \in \{0.00, 0.02, \ldots, 0.30\}$.
        Other methods that ignore missing data or use complete data,
        and Alvo and Cabilio's $\rho$ and Alvo and Cabilio's $\tau$ methods
        are also considered.
        These methods are described in Table~\ref{tab:3}.
        The data is missing completely at random (MNAR).
        (Left) Type I error: $(\bx,\by) \overset{iid}{\sim} N(0, I_2)$;
        (Right) Power: 	$(X, Y) \overset{iid}{\sim} N(0, \Sigma)$, where
        $\Sigma = \begin{pmatrix} 1 &  \corrcoef  \\  \corrcoef & 1 \end{pmatrix}$,
            with covariance coefficient $\corrcoef=0.5$.
            For both figures, a significance level $\alpha = 0.05$ is used and the sample
            size $\n = 200$. The results in the figures are average of 1000 trials.}
            \label{fig:4}
        \vspace{-0.5cm}
    \end{figure}

    Figure~\ref{fig:3} shows that all methods appear to control the Type I error 
    when the data are MCAR. 
    All methods appear to have good statistical power, but the power of
    the proposed method starts to drop
    when the proportion of incomplete pairs is larger than 10\%, and when
    $22\%$ of pairs are incomplete, the power is $0$.

    We next consider the case when the data are missing not at random (MNAR). 
    The parameters of this experiment are the same as for the MCAR case, but 
    now the missingness mechanism is 
    as described in in Section~\ref{sec:boundsmnar}.

    Figure~\ref{fig:4} shows that when data are MNAR, only the proposed 
    Spearman's footrule method and the complete data methods control
    the Type I error, while the other methods fail to control the Type I
    error when at least $4\%$ of the data is incomplete. 
    When the proportion of incomplete pairs is 
    $10\%$, the methods that ignore the missing data and Alvo and Cabilio's 
    $\rho$ and $\tau$ methods have Type I error close to $25\%$. 
    Meanwhile, all methods have good power with small proportion of incomplete pairs. 
    The power of the proposed method starts to drop when the proportion $s$ 
    at least $8\%$, and drops to $0$ when the proportion 
    $s$ is larger than $18\%$.

\begin{figure}[h]
        \includegraphics[width=14.5cm]{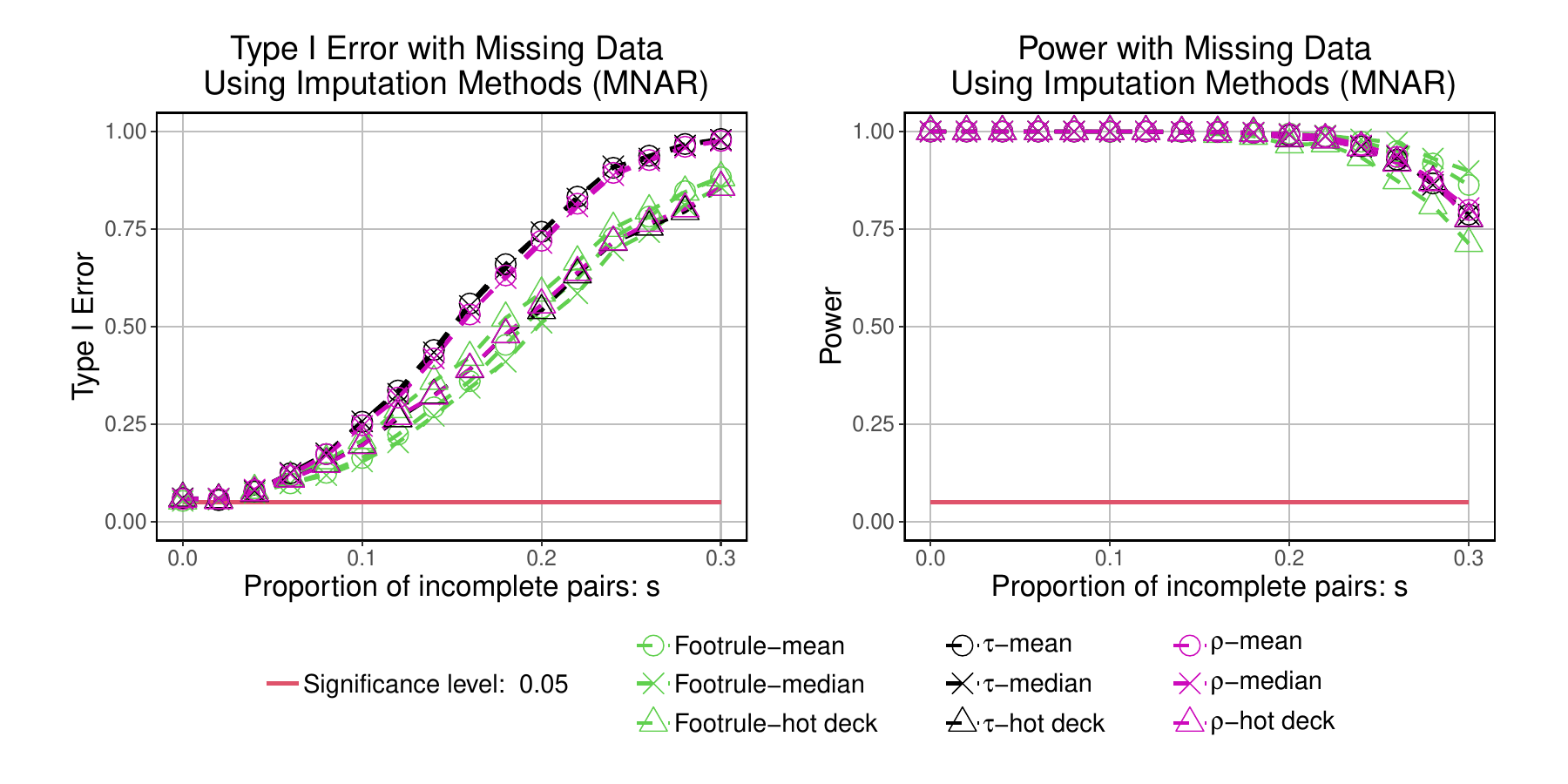}
        \vspace{-0.5cm}
        \caption{Statistical Type I error and power of imputation methods
        as the proportion of missing data $s$ increases,
        where $s \in \{0.00, 0.02, \ldots, 0.30\}$.
        The imputation methods impute missing data using either mean, median or 
        randomly selected (hot deck) values of observed data.
        These methods are described in Table~\ref{tab:3}.
        The data is missing completely at random (MNAR).
        (Left) Type I error: $(\bx,\by) \overset{iid}{\sim} N(0, I_2)$;
        (Right) Power: 	$(X, Y) \overset{iid}{\sim} N(0, \Sigma)$, where
        $\Sigma = \begin{pmatrix} 1 &  \corrcoef  \\  \corrcoef & 1 \end{pmatrix}$,
            with covariance coefficient $\corrcoef=0.5$.
            For both figures, a significance level $\alpha = 0.05$ is used and the sample
            size $\n = 200$. The results in the figures are average of 1000 trials.}
            \label{fig:5}
    \end{figure}

    We also investigate how imputation methods perform 
    in the experiment above, when data are missing not at random.
    We consider using mean imputation, median imputation and hot deck 
    imputation \cite{2002Missing} for Spearman's footrule, Kendall's $\tau$ and 
    Spearman's $\rho$. The results are shown in Figure~\ref{fig:5}, which
    demonstrates that these imputation methods also fail to control 
    the Type I error rate when at least $4\%$ of the data is incomplete.

    \subsubsection{As the sample size increases}

\begin{figure}
        \includegraphics[width=14cm]{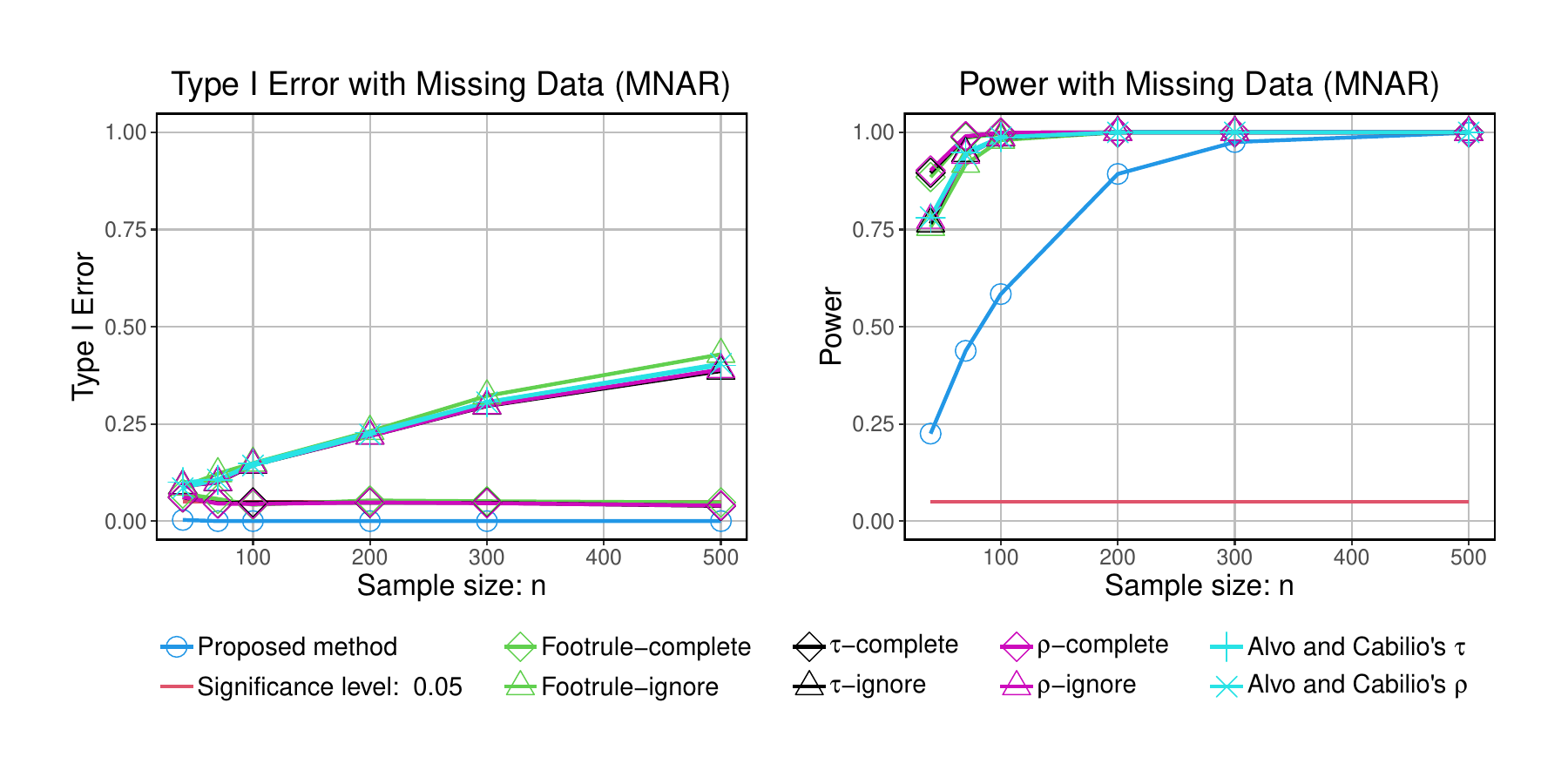}
        \vspace{-0.5cm}
        \caption{Statistical Type I error and power of the proposed method
        as the sample size $\n$ increases,
        where $\n \in \{40, 70, 100, 200, 300, 500\}$.
        Other methods that ignore missing data or use complete data,
        and Alvo and Cabilio's $\rho$ and Alvo and Cabilio's $\tau$ methods
        are also considered.
        These methods are described in Table~\ref{tab:3}.
        The data is missing completely at random (MNAR).
        (Left) Type I error: $(\bx,\by) \overset{iid}{\sim} N(0, I_2)$;
        (Right) Power: 	$(X, Y) \overset{iid}{\sim} N(0, \Sigma)$, where
        $\Sigma = \begin{pmatrix} 1 &  \corrcoef  \\  \corrcoef & 1 \end{pmatrix}$,
            with covariance coefficient $\corrcoef=0.5$.
            For both figures, a significance level $\alpha = 0.05$ is used and the proportion
            of missing pairs $s = 0.1$. The results in the figures are average of 1000 trials.}
            \label{fig:7}
    \end{figure}

    We again consider the above case when the data are missing not at random, 
    but now as the sample size $\n$ increases, with fixed proportion of 
    missing pairs $s=0.1$, and fixed correlation coefficient $\corrcoef=0.5$
    for the alternative hypothesis. 
    Figure~\ref{fig:7} shows that the proposed method still controls
    the Type I error as the sample size increases. Furthermore, 
    while the proposed method is not very powerful for smaller sample sizes, 
    as the sample size $\n$ increases, its power goes to $1$. 
    On the other hand, the other approaches have good power, but,
    excluding the complete data methods, all other approaches fail
    to control the Type I error, and their Type I error increases as
    $\n$ increases.

        \vspace{-0.4cm}

    \paragraph{Funding} Yijin Zeng is funded by a Roth Studentship from the 
    Department of Mathematics, Imperial College London and the EPSRC 
    CDT in Statistics and Machine Learning.

    \clearpage

\section*{Supplementary Material}

    \begin{appendix}

        \section{Some important lemmas} \label{appA}

        In this section, we provide several lemmas that will be important for 
        deriving exact bounds of the Spearman's footrule in the presence of missing data.

        \begin{lemma} \label{supp:lemma:2}
            Suppose $\bx \in \vndistinct$ and $\boconstant, \boconstantp \subset \tonumber{\n}$ are non-empty subsets of indices such that $\boconstant \subset \boconstantp$. Denote $|\boconstantp \setminus \boconstant| = \kk$. Then, for any $\iconstant \in \boconstant$, we have
            \begin{align*}
                \rank{\bxn{\iconstant}}{\svector{\bx}{\lconstant}{\boconstant}} + \kk \ge \rank{\bxn{\iconstant}}{\svector{\bx}{\lconstant}{\boconstantp}} \ge \rank{\bxn{\iconstant}}{\svector{\bx}{\lconstant}{\boconstant}}.
            \end{align*}
        \end{lemma}

        \begin{proof}
            Let $\iconstant \in \boconstant$ be a fixed index. Then, according to the definition of rank, we have
            \begin{align*}
                \rank{\bxn{\iconstant}}{\svector{\bx}{\lconstant}{\boconstantp}} &= \sum_{\lconstant \in \boconstantp} \indicator{\bxn{\lconstant} \le \bxn{\iconstant}}\\ 
                &=\sum_{\lconstant \in \boconstant} \indicator{\bxn{\lconstant} \le \bxn{\iconstant}} +  \sum_{\lconstant \in \boconstantp \setminus \boconstant} \indicator{\bxn{\lconstant} \le \bxn{\iconstant}}\\
                & =  \rank{\bxn{\iconstant}}{\svector{\bx}{\lconstant}{\boconstant}} +  \sum_{\lconstant \in \boconstantp \setminus \boconstant} \indicator{\bxn{\lconstant} \le \bxn{\iconstant}}.
            \end{align*}
            Since $|\boconstantp \setminus \boconstant| = \kk$, we have $\kk \ge \sum_{\lconstant \in \boconstantp \setminus \boconstant} \indicator{\bxn{\lconstant} \le \bxn{\iconstant}} \ge 0$. Thus, it follows that
            \begin{align*}
                \rank{\bxn{\iconstant}}{\svector{\bx}{\lconstant}{\boconstant}} + \kk \ge \rank{\bxn{\iconstant}}{\svector{\bx}{\lconstant}{\boconstantp}} \ge \rank{\bxn{\iconstant}}{\svector{\bx}{\lconstant}{\boconstant}},
            \end{align*}
            which completes our proof.
        \end{proof}

        \begin{lemma} \label{supp:lemma:3}
            Suppose $\bx \in \vndistinct$, and $\bu, \boconstant \subset \tonumber{\n}$ are non-empty disjoint subsets of indices such that $\bu \cup \boconstant = \tonumber{\n}$. Then, for any $\iconstant \in \boconstant$, if either of the following two conditions hold:
            \begin{align*}
                &(\rn{i}): \bxn{\jconstant} > \bxn{\iconstant}, \text{ for any } \jconstant \in \bu, \\
                \text{or } & (\rn{ii}): \rank{\bxn{\jconstant}}{\bx} > \rank{\bxn{\iconstant}}{\svector{\bx}{\lconstant}{\boconstant}}, \text{ for any } \jconstant \in \bu,
            \end{align*}
            we have $\rank{\bxn{\iconstant}}{\bx} = \rank{\bxn{\iconstant}}{\svector{\bx}{\lconstant}{\boconstant}}$.
        \end{lemma}

        \begin{proof}
            First, we show that if the condition $(\rn{i})$
            is true, then we have $\rank{\bxn{\iconstant}}{\bx} 
            = \rank{\bxn{\iconstant}}{\svector{\bx}{\lconstant}{\boconstant}}$.

            To start, let $\iconstant \in \boconstant$ be a fixed index such that 
            the condition $(\rn{i})$ is true, i.e.,
            for any $ \jconstant \in \bu$,
            $\bxn{\jconstant} > \bxn{\iconstant}$.
            According to the definition of rank, we have
            \begin{align*}
                \rank{\bxn{\iconstant}}{\bx} = \sum_{\lconstant \in \tonumber{\n}} \indicator{\bxn{\lconstant} \le \bxn{\iconstant}}. 
            \end{align*}
            Since $\bu$ and $\boconstant$ are non-empty disjoint subsets of $\tonumber{\n}$ such that $\bu \cup \boconstant = \tonumber{\n}$, then we have
            \begin{align*}
                \rank{\bxn{\iconstant}}{\bx} = \sum_{\lconstant \in \boconstant} \indicator{\bxn{\lconstant} \le \bxn{\iconstant}} + \sum_{\lconstant \in \bu} \indicator{\bxn{\lconstant} \le \bxn{\iconstant}}.
            \end{align*}	
            Now, according to the condition $(\rn{i})$, we have
            $\bxn{\jconstant} > \bxn{\iconstant}$, for any $\jconstant \in \bu$.
            Hence, we have
            \begin{align*}
                \sum_{\lconstant \in \bu} \indicator{\bxn{\lconstant} \le \bxn{\iconstant}} = 0,
            \end{align*}
            which then follows that
            \begin{align*}
                \rank{\bxn{\iconstant}}{\bx} = \sum_{\lconstant \in \boconstant} \indicator{\bxn{\lconstant} \le \bxn{\iconstant}} = \rank{\bxn{\iconstant}}{\svector{\bx}{\lconstant}{\boconstant}}. 
            \end{align*}
            Thus, we complete our prove when the condition $(\rn{i})$ is true.

            Next, we show that if the condition $(\rn{ii})$ is true, 
            we have $\rank{\bxn{\iconstant}}{\bx} = \rank{\bxn{\iconstant}}{\svector{\bx}{\lconstant}{\boconstant}}$.
            Let $\iconstant \in \boconstant$ be any fixed index such that the
            condition $(\rn{ii})$ is true, i.e., for any $ \jconstant \in \bu$,
            $\rank{\bxn{\jconstant}}{\bx} > \rank{\bxn{\iconstant}}{\svector{\bx}{\lconstant}{\boconstant}}$. 
            We prove the result for the condition $(\rn{ii})$ by showing that the condition $(\rn{ii})$ implies 
            the condition $(\rn{i})$, i.e.,
            \begin{align*}
                \rank{\bxn{\jconstant}}{\bx} > \rank{\bxn{\iconstant}}{\svector{\bx}{\lconstant}{\boconstant}}, 
                \text{ for any } \jconstant \in \bu \Rightarrow \bxn{\jconstant} > \bxn{\iconstant}, \text{ for any } \jconstant \in \bu.
            \end{align*}

            Without loss of generality, denote $\uu$ as the index of the minimum value in $ \svector{\bx}{\lconstant}{\bu}$, i.e.,
            \begin{align*}
                \bxn{\uu} = \min \svector{\bx}{\lconstant}{\bu}.
            \end{align*}
            Then, all components with indices in $\bu \setminus \{\uu\}$ are larger than $\bxn{\uu}$. Hence, we have 
            \begin{align*}
                \sum_{\lconstant \in \bu} \indicator{\bxn{\lconstant} \le \bxn{\uu}} = \indicator{\bxn{\uu} \le \bxn{\uu}} = 1.
            \end{align*}
            Subsequently, since $\bu$ and $\boconstant$ are non-empty disjoint subsets of $\tonumber{\n}$ such that $\bu \cup \boconstant = \tonumber{\n}$,		
            then according to the definition of rank, we have
            \begin{align} \label{supp:lemma:3:eqn:1}
                \begin{split}
                    \rank{\bxn{\uu}}{\bx} &= \sum_{\lconstant \in \boconstant} \indicator{\bxn{\lconstant} \le \bxn{\uu}} + \sum_{\lconstant \in \bu} \indicator{\bxn{\lconstant} \le \bxn{\uu}}\\
                    & = \sum_{\lconstant \in \boconstant} \indicator{\bxn{\lconstant} \le \bxn{\uu}} + \indicator{\bxn{\uu} \le \bxn{\uu}}\\
                    & = \sum_{\lconstant \in \boconstant \cup \{\uu\}} \indicator{\bxn{\lconstant} \le \bxn{\uu}} \\
                    & = \rank{\bxn{\uu}}{\svector{\bx}{\lconstant}{\boconstant \cup \{\uu\}}}.
                \end{split}
            \end{align}

            Next, according to Lemma~\ref{supp:lemma:2}, we have
            \begin{align*}
                \rank{\bxn{\iconstant}}{\svector{\bx}{\lconstant}{\boconstant  }} + 1\ge\rank{\bxn{\iconstant}}{\svector{\bx}{\lconstant}{\boconstant \cup \{\uu\}}}.
            \end{align*}
            According to the condition $(\rn{ii})$, we have $\rank{\bxn{\jconstant}}{\bx} > \rank{\bxn{\iconstant}}{\svector{\bx}{\lconstant}{\boconstant}}, 
            \text{ for any } \jconstant \in \bu$. Notice that $\uu \in \bu$,	
            then we have
            \begin{align*}
                &\rank{\bxn{\uu}}{\bx} > \rank{\bxn{\iconstant}}{\svector{\bx}{\lconstant}{\boconstant}} \\
                \Rightarrow&\rank{\bxn{\uu}}{\bx} \ge \rank{\bxn{\iconstant}}{\svector{\bx}{\lconstant}{\boconstant}} + 1,
            \end{align*}
            where the last $``\Rightarrow"$ holds because $\rank{\bxn{\iconstant}}{\svector{\bx}{\lconstant}{\boconstant}} $
            and $\rank{\bxn{\uu}}{\bx}$ are both integers.
            Hence, we have		
            \begin{align*}
                \rank{\bxn{\uu}}{\bx} \ge \rank{\bxn{\iconstant}}{\svector{\bx}{\lconstant}{\boconstant  }} + 1 \ge \rank{\bxn{\iconstant}}{\svector{\bx}{\lconstant}{\boconstant \cup \{\uu\}}}.
            \end{align*}
            According to \eqref{supp:lemma:3:eqn:1}, we further have
            \begin{align*}
                &\rank{\bxn{\uu}}{\svector{\bx}{\lconstant}{\boconstant \cup \{\uu\}}}
                = \rank{\bxn{\uu}}{\bx} \ge 
                \rank{\bxn{\iconstant}}{\svector{\bx}{\lconstant}{\boconstant \cup \{\uu\}}}.
            \end{align*}
            Since $\bx \in \vndistinct$ is a vector of distinct real values, and $\uu \notin \boconstant$, 
            then we have
            \begin{align*}
                &\rank{\bxn{\iconstant}}{\svector{\bx}{\lconstant}{\boconstant \cup \{\uu\}}} \neq \rank{\bxn{\uu}}{\svector{\bx}{\lconstant}{\boconstant \cup \{\uu\}}} \\
                \Rightarrow&\rank{\bxn{\uu}}{\svector{\bx}{\lconstant}{\boconstant \cup \{\uu\}}} > \rank{\bxn{\iconstant}}{\svector{\bx}{\lconstant}{\boconstant \cup \{\uu\}}}  \\
                \Rightarrow&\bxn{\uu} > \bxn{\iconstant}.
            \end{align*}
            Recall that $\bxn{\uu} = \min \svector{\bx}{\lconstant}{\bu}$. Therefore, we have $\bxn{\jconstant} > \bxn{\iconstant}$ for any $\jconstant \in \bu$. 

            Then, by applying 
            the result when the condition $(\rn{i})$ holds,
            we have $\rank{\bxn{\iconstant}}{\bx} = \rank{\bxn{\iconstant}}{\svector{\bx}{\lconstant}{\boconstant}}$,
            which completes our proof.
        \end{proof}

        \begin{lemma} \label{supp:lemma:4}
            Suppose $\bxln{1}, \bxln{2} \in \vndistinct$, and $ \bu, \bo$ are non-empty disjoint subsets of $ \tonumber{\n}$ such that $ \bu \cup \bo = \tonumber{\n}$. Then, if $\bxln{2}$ is an imputation of $\bxln{1}$
            for indices $\bu$ and 
            $\rank{\bxln{1}(i)}{\bxln{1}} = \rank{\bxln{2}(i)}{\bxln{2}}, \text{ for any } i \in \bu,$
            we have $\rank{\bxln{1}}{\bxln{1}} = \rank{\bxln{2}}{\bxln{2}}$.
        \end{lemma}

        \begin{proof}
            To start, let us denote
            \begin{align*}
                &\bu = \{\uuln{1}, \cdots, \uuln{\m}\}, \text{ and } \bo = \{\ooln{1}, \cdots, \ooln{\n - \m}\}.
            \end{align*}
            For both $ \jconstant = 1,2$, since $ \bxln{j} \in \vndistinct$ is a vector of $ \n$ distinct real numbers, the ranks of all components in $ \bxln{j}$ form a permutation of $ \{1,\cdots,\n\}$, i.e.,
            \begin{align*}
                \{\rank{\bxln{j}(1)}{\bxln{j}}, \cdots, \rank{\bxln{j}(\n)}{\bxln{j}}\} = \{1,\cdots,\n\}, \text{ for both } j = \{1,2\}.
            \end{align*}
            Let us denote
            \begin{align*}
                \bsln{j}= \{1,\cdots,\n\}\setminus \{\rank{\bxln{j}(\uuln{1})}{\bxln{j}}, \cdots, \rank{\bxln{j}(\uuln{\m})}{\bxln{j}}\}, \text{ for both } j = \{1,2\}.
            \end{align*}
            Then, since $ \bu \cup \bo = \tonumber{\n} = \{1,2,\cdots,\n\}$, we have
            \begin{align*}
                \{\rank{\bxln{j}(\ooln{1})}{\bxln{j}}, \cdots, \rank{\bxln{j}(\ooln{\n - \m})}{\bxln{j}} \} = \bsln{j}, \text{ for both } j = 1,2.
            \end{align*}
            Since $ \rank{\bxln{1}(i)}{\bxln{1}} = \rank{\bxln{2}(i)}{\bxln{2}}$, for any $i \in \bu$, we  have $ \bsln{2} = \bsln{1}$. 

            Notice that for any fixed $ i \in \{1,\cdots, \n - \m\}$, and $ \jconstant = 1,2$, the order of $ \rank{\bxln{j}(\ooln{i})}{\bxln{j}}$ in $ \bsln{j}$ equals to the order of $\bxln{j}(\ooln{i})$ in $ \{\bxln{j}(\ooln{1}), \ldots, \bxln{j}(\ooln{\n - \m})\}$. 
            Since $\bxln{2}$ is an imputation of $\bxln{1}$
            for indices $\bu$, then according to the definition of imputations,
            for any $ i \in \{1,2,\cdots, \n - \m\}$, we have $ \bxln{1}(\ooln{i}) = \bxln{2}(\ooln{i})$. 
            Hence, the order of $\bxln{1}(\ooln{i})$ in $ \{\bxln{1}(\ooln{1}), \ldots, \bxln{1}(\ooln{\n - \m})\}$
            equals to the order of $\bxln{2}(\ooln{i})$ in $ \{\bxln{2}(\ooln{1}), \ldots, \bxln{1}(\ooln{\n - \m})\}$.
            Thus, the order of $ \rank{\bxln{1}(\ooln{i})}{\bxln{1}}$ in $ \bsln{1}$ equals to the order of $ \rank{\bxln{2}(\ooln{i})}{\bxln{2}}$ in $ \bsln{2}$. Furthermore, since $ \bsln{1} = \bsln{2}$, 
            we have
            \begin{align*}
                \rank{\bxln{1}(\ooln{\iconstant})}{\bxln{1}} = \rank{\bxln{2}(\ooln{\iconstant})}{\bxln{2}},
            \end{align*}
            where $\iconstant \in \{1,\cdots, \n - \m\}$.	
            Since we also have $ \rank{\bxln{1}(i)}{\bxln{1}} = \rank{\bxln{2}(i)}{\bxln{2}}, \text{ for any } i \in \bu$, and $\boconstant \cup \bu = \tonumber{\n}$, then we have
            \begin{align*}
                \rank{\bxln{1}(i)}{\bxln{1}} = \rank{\bxln{2}(i)}{\bxln{2}}, \text{ for any } i \in \tonumber{\n}.
            \end{align*}
            In order words, we have $\rank{\bxln{1}}{\bxln{1}} = \rank{\bxln{2}}{\bxln{2}}$.
            This completes our proof.
        \end{proof}

        \section{Proof of lower bounds} \label{appB}

        This section provides results for deriving exact lower bounds of Spearman's footrule
        in the presence of missing data.

        \subsection{Proof of Proposition 2.1}
        This subsection proves Proposition 2.1. 
        First, we prove four lemmas that will be useful for proving
        Proposition 2.1. 

        \begin{lemma}  \label{supp:proposition:1:lemma:1}
            Suppose $\bx, \bxls \in \vndistinct$ and let $\bxs$ be an imputation of $\bx$ for an index $\uu \in \tonumber{\n}$. Then, if $\rank{\bxlsn{\uu}}{\bxls} = \rank{\bxn{\uu}}{\bx} + 1$, we have $\bxlsn{\uu} > \bxn{\uu}$ and
            \begin{align*}
                \sum_{\iconstant \in \tonumber{\n} \setminus \{\uu\}} \indicator{\bxn{\iconstant} > \bxn{\uu} } \indicator{\bxn{\iconstant} < \bxlsn{\uu}} = 1.
            \end{align*}
            However, if $\rank{\bxlsn{\uu}}{\bxls} = \rank{\bxn{\uu}}{\bx} - 1$, we have $\bxlsn{\uu} < \bxn{\uu}$ and
            \begin{align*}
                \sum_{\iconstant \in \tonumber{\n} \setminus \{\uu\}} \indicator{\bxn{\iconstant} > \bxlsn{\uu} } \indicator{\bxn{\iconstant} < \bxn{\uu}} = 1.
            \end{align*}
        \end{lemma}
        \begin{proof}
            To start, according to the definition of rank, we have
            \begin{align*}
                \rank{\bxn{\uu}}{\bx} = \sum_{\iconstant \in \tonumber{\n} } \indicator{\bxn{\iconstant} \le \bxn{\uu}} = \indicator{\bxn{\uu} \le \bxn{\uu}} + \sum_{\iconstant \in \tonumber{\n} \setminus \{\uu\}} \indicator{\bxn{\iconstant} \le \bxn{\uu}}.
            \end{align*}
            Since $\bx \in \vndistinct$ is a vector of distinct real values, we have
            \begin{align} \label{supp:proposition:1:lemma:1:eqn:1}
                \rank{\bxn{\uu}}{\bx} = 1 + \sum_{\iconstant \in \tonumber{\n} \setminus \{\uu\}} \indicator{\bxn{\iconstant} < \bxn{\uu}}.
            \end{align}
            Similarly, we can obtain
            \begin{align*}
                \rank{\bxlsn{\uu}}{\bxls} = 1 + \sum_{\iconstant \in \tonumber{\n} \setminus \{\uu\}} \indicator{\bxlsn{\iconstant} < \bxlsn{\uu}}.
            \end{align*} 
            Since $\bxs$ is an imputation of $\bx$ for the index $\uu$,
            then according to the definition of imputations,
            we have $\bxlsn{\iconstant} = \bxn{\iconstant}$ for any $\iconstant \in \tonumber{\n} \setminus \{\uu\}$. Hence, we further have
            \begin{align} \label{supp:proposition:1:lemma:1:eqn:2}
                \rank{\bxlsn{\uu}}{\bxls}  = 1 +  \sum_{\iconstant \in \tonumber{\n} \setminus \{\uu\}} \indicator{\bxn{\iconstant} < \bxlsn{\uu}}.
            \end{align}

            Subsequently, if $\rank{\bxlsn{\uu}}{\bxls} = \rank{\bxn{\uu}}{\bx} + 1$, by combining \eqref{supp:proposition:1:lemma:1:eqn:1} and \eqref{supp:proposition:1:lemma:1:eqn:2}, we have
            \begin{align*}
                \sum_{\iconstant \in \tonumber{\n} \setminus \{\uu\}} \indicator{\bxn{\iconstant} < \bxlsn{\uu}} = \sum_{\iconstant \in \tonumber{\n} \setminus \{\uu\}} \indicator{\bxn{\iconstant} < \bxn{\uu}} + 1.
            \end{align*}
            Hence, it can be seen that $\bxlsn{\uu} > \bxn{\uu}$.

            Now, since $\bx \in \vndistinct$ is a vector of distinct real values, we further have
            \begin{align*}
                &\sum_{\iconstant \in \tonumber{\n} \setminus \{\uu\}} \indicator{\bxn{\iconstant} < \bxlsn{\uu}} - \sum_{\iconstant \in \tonumber{\n} \setminus \{\uu\}} \indicator{\bxn{\iconstant} < \bxn{\uu}} = 1\\
                &\Leftrightarrow \sum_{\iconstant \in \tonumber{\n} \setminus \{\uu\}} \indicator{\bxn{\iconstant} < \bxn{\uu}} + \sum_{\iconstant \in \tonumber{\n} \setminus \{\uu\}} \indicator{\bxn{\uu} < \bxn{\iconstant} < \bxlsn{\uu}} - \sum_{\iconstant \in \tonumber{\n} \setminus \{\uu\}} \indicator{\bxn{\iconstant} < \bxn{\uu}} = 1 \\
                &\Leftrightarrow \sum_{\iconstant \in \tonumber{\n} \setminus \{\uu\}} \indicator{\bxn{\uu} < \bxn{\iconstant} < \bxlsn{\uu}} = 1\\
                &\Leftrightarrow \sum_{\iconstant \in \tonumber{\n} \setminus \{\uu\}} \indicator{\bxn{\iconstant} > \bxn{\uu}}\indicator{\bxn{\iconstant} < \bxlsn{\uu}} = 1,
            \end{align*}
            which proves our results when $\rank{\bxlsn{\uu}}{\bxls} = \rank{\bxn{\uu}}{\bx} + 1$.

            The case when $\rank{\bxlsn{\uu}}{\bxls} = \rank{\bxn{\uu}}{\bx} - 1$ can be proved similarly. Combining \eqref{supp:proposition:1:lemma:1:eqn:1} and \eqref{supp:proposition:1:lemma:1:eqn:2},
            we have 
            \begin{align*}
                \sum_{\iconstant \in \tonumber{\n} \setminus \{\uu\}} \indicator{\bxn{\iconstant} < \bxlsn{\uu}} = \sum_{\iconstant \in \tonumber{\n} \setminus \{\uu\}} \indicator{\bxn{\iconstant} < \bxn{\uu}} - 1,
            \end{align*}
            which follows $\bxlsn{\uu} < \bxn{\uu}$.

            Since $\bxls \in \vndistinct$ is a vector of distinct real values and $\bxlsn{\uu} \neq \bxlsn{\iconstant} = \bxn{\iconstant}$ for any $\iconstant \in \tonumber{\n} \setminus \{\uu\}$, we have
            \begin{align*}
                &\sum_{\iconstant \in \tonumber{\n} \setminus \{\uu\}} \indicator{\bxn{\iconstant} < \bxn{\uu}} - \sum_{\iconstant \in \tonumber{\n} \setminus \{\uu\}} \indicator{\bxn{\iconstant} < \bxlsn{\uu}} = 1\\
                &\Leftrightarrow \sum_{\iconstant \in \tonumber{\n} \setminus \{\uu\}} \indicator{\bxn{\iconstant} < \bxlsn{\uu}} + \sum_{\iconstant \in \tonumber{\n} \setminus \{\uu\}} \indicator{\bxlsn{\uu}< \bxn{\iconstant} < \bxn{\uu}} - \sum_{\iconstant \in \tonumber{\n} \setminus \{\uu\}} \indicator{\bxn{\iconstant} < \bxlsn{\uu}} = 1 \\
                &\Leftrightarrow \sum_{\iconstant \in \tonumber{\n} \setminus \{\uu\}} \indicator{\bxlsn{\uu}< \bxn{\iconstant} < \bxn{\uu}} = 1\\
                &\Leftrightarrow \sum_{\iconstant \in \tonumber{\n} \setminus \{\uu\}} \indicator{\bxn{\iconstant} > \bxlsn{\uu}}\indicator{\bxn{\iconstant} < \bxn{\uu}} = 1,
            \end{align*}
            which proves our results when $\rank{\bxlsn{\uu}}{\bxls} = \rank{\bxn{\uu}}{\bx} - 1$, 
            and completes out proof.
        \end{proof}

        \begin{lemma}  \label{supp:proposition:1:lemma:2}
            Suppose $\bx, \by \in \vndistinct$ and 
            let $\bxs \in \vndistinct$ be an imputation of $\bx$ for an index $\uu \in \tonumber{\n}$. 
            For any $\iconstant \in \tonumber{\n} \setminus \{\uu\}$, let us denote
            \begin{align*} 
                \begin{split}
                    &\xi_i = \left|\rank{\bxn{i}}{\svector{\bx}{\lconstant}{\tonumber{\n} \setminus \{\uu\}}} + 1 - \rank{\byn{i}}{\by}\right|, \\
                    \text{ and }&\psi_i = \left|\rank{\bxn{i}}{\svector{\bx}{\lconstant}{\tonumber{\n} \setminus \{\uu\}}} - \rank{\byn{i}}{\by}\right|.
                \end{split}
            \end{align*}
            Then, we have
            \begin{align*}
                \sfd{\bxs}{\by} - \sfd{\bx}{\by} &= \left|\rank{\bxsn{\uu}}{\bx} - \rank{\byn{\uu}}{\by}\right| -  \left|\rank{\bxn{\uu}}{\bx} - \rank{\byn{\uu}}{\by}\right| \\
                &+ \sum_{\iconstant \in \tonumber{\n} \setminus \{\uu\}} \indicator{\bxn{i} > \bxsn{\uu}} \indicator{\bxn{i} < \bxn{\uu}} (\xi_i - \psi_i)\\
                & + \sum_{\iconstant \in \tonumber{\n} \setminus \{\uu\}} \indicator{\bxn{i} < \bxsn{\uu}} \indicator{\bxn{i} > \bxn{\uu}} (\psi_i - \xi_i).
            \end{align*}
        \end{lemma}

        \begin{proof}
            For convenience, let us assume (after relabeling) $ \uu = 1$, and denote
            \begin{align*}
                &\alpha = \left|\rank{\bxn{1}}{\bx} - \rank{\byn{1}}{\by}\right|, \text{ and } \beta = \left|\rank{\bxsn{1}}{\bx} - \rank{\byn{1}}{\by}\right|.
            \end{align*}
            According to the definition of Spearman's footrule, we have
            \begin{align*}
                \sfd{\bx}{\by} &=  \sum_{i = 1}^{\n} \left|\rank{\bxn{i}}{\bx} - \rank{\byn{i}}{\by}\right| \\
                &= \left|\rank{\bxn{1}}{\bx} - \rank{\byn{1}}{\by}\right| + \sum_{i = 2}^{\n} \left|\rank{\bxn{i}}{\bx} - \rank{\byn{i}}{\by}\right| \\
                & = \alpha +   \sum_{i = 2}^{\n} \left|\rank{\bxn{i}}{\bx} - \rank{\byn{i}}{\by}\right|.
            \end{align*}
            According to the definition of rank, for any $\iconstant \in \{2,\ldots, \n\}$, we have 
            \begin{align*}
                \rank{\bxn{i}}{\bx} &= \sum_{\lconstant = 1}^{\n} \indicator{\bxn{1} \le \bxn{i}} \\
                & = \sum_{\lconstant = 2}^{\n} \indicator{\bxn{1} \le \bxn{i}}  + \indicator{\bxn{1} \le \bxn{i}}\\
                &=\rank{\bxn{i}}{\svector{\bx}{\lconstant}{\tonumber{\n} \setminus \{1\}}} + \indicator{\bxn{1} \le \bxn{i}}.
            \end{align*} 
            Since $\bx \in \vndistinct$ is a vector of distinct values, we have $\indicator{\bxn{1} \le \bxn{i}} = \indicator{\bxn{1} < \bxn{i}}$. Hence, for any $\iconstant \in \{2,\ldots, \n\}$, we have
            \begin{align*}
                \rank{\bxn{i}}{\bx} = \rank{\bxn{i}}{\svector{\bx}{\lconstant}{\tonumber{\n} \setminus \{1\}}} + \indicator{\bxn{1} < \bxn{i}}.
            \end{align*}
            Therefore, it follows that
            \begin{align*}
                \sfd{\bx}{\by} = \alpha + \sum_{i=2}^\n \left|\rank{\bxn{i}}{\svector{\bx}{\lconstant}{\tonumber{\n} \setminus \{1\}}} + \indicator{\bxn{1} < \bxn{i}} - \rank{\byn{i}}{\by}\right|.
            \end{align*}
            Subsequently,
            \begin{align*} 
                \begin{split}
                    &\sfd{\bx}{\by} \\
                    &= \alpha + \sum_{i=2}^\n \indicator{\bxn{1} < \bxn{i}} \left|\rank{\bxn{i}}{\svector{\bx}{\lconstant}{\tonumber{\n} \setminus \{1\}}} + 1 - \rank{\byn{i}}{\by}\right|\\ 
                    & + \sum_{i=2}^\n \indicator{\bxn{1} > \bxn{i}} \left|\rank{\bxn{i}}{\svector{\bx}{\lconstant}{\tonumber{\n} \setminus \{1\}}} - \rank{\byn{i}}{\by}\right|\\
                    & =  \alpha + 
                    \sum_{i=2}^\n \indicator{\bxn{i} > \bxn{1}} \xi_i
                    + \sum_{i=2}^\n \indicator{\bxn{i} < \bxn{1}} \psi_i.
                \end{split}
            \end{align*}
            Similarly, we can show that
            \begin{align*}
                \begin{split}
                    & \sfd{\bxs}{\by} \\
                    & = \beta + \sum_{i=2}^{\n} \indicator{\bxsn{1} < \bxsn{i}} \left|\rank{\bxsn{i}}{\svector{\bxs}{\lconstant}{\tonumber{\n} \setminus \{1\}}} + 1 - \rank{\byn{i}}{\by}\right| \\
                    & +  \sum_{i=2}^{\n} \indicator{\bxsn{1} > \bxsn{i}} \left|\rank{\bxsn{i}}{\svector{\bxs}{\lconstant}{\tonumber{\n} \setminus \{1\}}} - \rank{\byn{i}}{\by}\right|.
                \end{split}
            \end{align*}
            Since $\bxs$ is an imputation of $\bx$ for the index $1$,
            then according to the definition of imputations,
            we have $\bxlsn{\iconstant} = \bxn{\iconstant}$ 
            for any $\iconstant \in \tonumber{\n} \setminus \{1\}$. Thus,  we have
            \begin{align*}
                \begin{split}
                    \sfd{\bxs}{\by} &= \beta +  \sum_{i=2}^{\n} \indicator{\bxn{i} > \bxsn{1}} \left|\rank{\bxn{i}}{\svector{\bx}{\lconstant}{\tonumber{\n} \setminus \{1\}}} + 1 - \rank{\byn{i}}{\by}\right| \\
                    &+  \sum_{i=2}^{\n} \indicator{\bxn{i} < \bxsn{1}} \left|\rank{\bxn{i}}{\svector{\bx}{\lconstant}{\tonumber{\n} \setminus \{1\}}} - \rank{\byn{i}}{\by}\right| \\
                    & = \beta +  \sum_{i=2}^{\n} \indicator{\bxn{i} > \bxsn{1}} \xi_i +  \sum_{i=2}^{\n} \indicator{\bxn{i} < \bxsn{1}} \psi_i.
                \end{split}
            \end{align*}
            Now, we have
            \begin{align} \label{supp:proposition:1:lemma:2:eqn:1}
                \begin{split}
                    \sfd{\bxs}{\by} - \sfd{\bx}{\by} &= \beta - \alpha + \sum_{i=2}^{\n} \left\{\indicator{\bxn{i} > \bxsn{1}} - \indicator{\bxn{i} > \bxn{1}} \right\} \xi_i\\
                    &\quad +  \sum_{i=2}^{\n} \left\{\indicator{\bxn{i} < \bxsn{1}} - \indicator{\bxn{i} < \bxn{1}} \right\} \psi_i.
                \end{split}
            \end{align}

            Next, since $\bx, \bxs \in \vndistinct$ are both vectors of distinct real values, 
            then for any $i \in \{2,\cdots,\n\}$, we have $ \bxn{i} \neq \bxn{1}$, and $ \bxsn{i} \neq \bxsn{1}$. Further, since $\bxs$ is an imputation of $\bx$ for the index $1$, we have
            $\bxsn{i} = \bxn{i}$ for any $i \in \{2,\ldots,\n\}$. Hence, we also have $ \bxn{i} \neq \bxsn{1}$ for any $i \in \{2,\cdots,\n\}$. 
            Subsequently,  for any $i \in \{2,\cdots,\n\}$, we have
            \begin{align*}
                &\indicator{\bxn{i} > \bxsn{1}} - \indicator{\bxn{i} > \bxn{1}}\\
                &= \indicator{\bxn{i} > \bxsn{1}} \{ \indicator{\bxn{i} < \bxn{1}}  + \indicator{\bxn{i} \ge \bxn{1}} \} \\
                & - \indicator{\bxn{i} > \bxn{1}} \{   \indicator{\bxn{i} < \bxsn{1}} +  \indicator{\bxn{i} \ge \bxsn{1}}  \} \\
                &= \indicator{\bxn{i} > \bxsn{1}} \indicator{\bxn{i} < \bxn{1}} + \indicator{\bxn{i} > \bxsn{1}}\indicator{\bxn{i} \ge \bxn{1}}  \\
                & - \indicator{\bxn{i} > \bxn{1}}  \indicator{\bxn{i} < \bxsn{1}} - \indicator{\bxn{i} > \bxn{1}}\indicator{\bxn{i} \ge \bxsn{1}}
            \end{align*}
            By rearranging the above terms, we further have
            \begin{align*}
                & \indicator{\bxn{i} > \bxsn{1}} - \indicator{\bxn{i} > \bxn{1}}\\		
                &= \indicator{\bxn{i} > \bxsn{1}} \indicator{\bxn{i} < \bxn{1}} -  \indicator{\bxn{i} > \bxn{1}}\indicator{\bxn{i} < \bxsn{1}} \\
                & + \indicator{\bxn{i} > \bxsn{1}}\indicator{\bxn{i} \ge \bxn{1}}  - \indicator{\bxn{i} > \bxn{1}}\indicator{\bxn{i} \ge \bxsn{1}}
            \end{align*}
            Since for any $i \in \{2,\ldots,\n\}$, we have $ \bxn{i} \neq \bxn{1}, \bxsn{1}$. Hence,
            \begin{align*}
                &\indicator{\bxn{i} > \bxsn{1}} =  \indicator{\bxn{i} \ge \bxsn{1}}, \\
                &\indicator{\bxn{i} \ge \bxn{1}} = \indicator{\bxn{i} > \bxn{1}},
            \end{align*}
            which means
            \begin{align*}
                \indicator{\bxn{i} > \bxsn{1}}\indicator{\bxn{i} \ge \bxn{1}}  - \indicator{\bxn{i} > \bxn{1}}\indicator{\bxn{i} \ge \bxsn{1}} = 0
            \end{align*}
            Therefore, 	we have
            \begin{align} \label{supp:proposition:1:lemma:2:eqn:2}
                \begin{split}
                    & \indicator{\bxn{i} > \bxsn{1}} - \indicator{\bxn{i} > \bxn{1}}\\		
                    & = \indicator{\bxn{i} > \bxsn{1}} \indicator{\bxn{i} < \bxn{1}} - \indicator{\bxn{i} < \bxsn{1}} \indicator{\bxn{i} > \bxn{1}}.
                \end{split}
            \end{align}

            Similarly, for any $i \in \{2,\cdots,\n\}$, we have
            \begin{align*} 
                &\indicator{\bxn{i} < \bxsn{1}} - \indicator{\bxn{i} < \bxn{1}}\\
                &=\indicator{\bxn{i} < \bxsn{1}} \{\indicator{\bxn{i} > \bxn{1}}  + \indicator{\bxn{i} \le \bxn{1}} \}   \\
                & - \indicator{\bxn{i} < \bxn{1}} \{\indicator{\bxn{i} > \bxsn{1}} + \indicator{\bxn{i} \le \bxsn{1}}  \}   \\
                &= \indicator{\bxn{i} < \bxsn{1}} \indicator{\bxn{i} > \bxn{1}} + \indicator{\bxn{i} < \bxsn{1}}  \indicator{\bxn{i} \le \bxn{1}} \\
                & -\indicator{\bxn{i} < \bxn{1}}  \indicator{\bxn{i} > \bxsn{1}} -\indicator{\bxn{i} < \bxn{1}}  \indicator{\bxn{i} \le \bxsn{1}}. 
            \end{align*}
            By rearranging the above terms, we further have
            \begin{align*}
                &\indicator{\bxn{i} < \bxsn{1}} - \indicator{\bxn{i} < \bxn{1}}\\
                &= \indicator{\bxn{i} < \bxsn{1}} \indicator{\bxn{i} > \bxn{1}} 
                -\indicator{\bxn{i} < \bxn{1}}  \indicator{\bxn{i} > \bxsn{1}} \\
                & + \indicator{\bxn{i} < \bxsn{1}}  \indicator{\bxn{i} \le \bxn{1}} -\indicator{\bxn{i} < \bxn{1}}  \indicator{\bxn{i} \le \bxsn{1}}. 
            \end{align*}
            Since for any $i \in \{2,\ldots,\n\}$, we have $ \bxn{i} \neq \bxn{1}, \bxsn{1}$. Hence,
            \begin{align*}
                &\indicator{\bxn{i} < \bxsn{1}} =  \indicator{\bxn{i} \le \bxsn{1}}, \\
                &\indicator{\bxn{i} \le \bxn{1}} = \indicator{\bxn{i} < \bxn{1}},
            \end{align*}
            which means
            \begin{align*}
                \indicator{\bxn{i} < \bxsn{1}}  \indicator{\bxn{i} \le \bxn{1}} -\indicator{\bxn{i} < \bxn{1}}  \indicator{\bxn{i} \le \bxsn{1}} = 0.
            \end{align*}
            Therefore, we have
            \begin{align} \label{supp:proposition:1:lemma:2:eqn:3}
                \begin{split}
                    &\indicator{\bxn{i} < \bxsn{1}} - \indicator{\bxn{i} < \bxn{1}}\\
                    &= \indicator{\bxn{i} < \bxsn{1}} \indicator{\bxn{i} > \bxn{1}} -  \indicator{\bxn{i} < \bxn{1}}\indicator{\bxn{i} > \bxsn{1}}.
                \end{split}
            \end{align}

            Putting \eqref{supp:proposition:1:lemma:2:eqn:2} and \eqref{supp:proposition:1:lemma:2:eqn:3} back into \eqref{supp:proposition:1:lemma:2:eqn:1}, we obtain
            \begin{align*}
                \sfd{\bxs}{\by} - \sfd{\bx}{\by} &= \beta - \alpha + \sum_{i=2}^{\n} \indicator{\bxn{i} > \bxsn{1}} \indicator{\bxn{i} < \bxn{1}} \xi_i\\
                & - \sum_{i=2}^{\n} \indicator{\bxn{i} < \bxsn{1}} \indicator{\bxn{i} > \bxn{1}} \xi_i\\
                & + \sum_{i=2}^{\n} \indicator{\bxn{i} < \bxsn{1}} \indicator{\bxn{i} > \bxn{1}} \psi_i \\
                & - \sum_{i=2}^{\n} \indicator{\bxn{i} > \bxsn{1}} \indicator{\bxn{i} < \bxn{1}} \psi_i.
            \end{align*} 
            That is,
            \begin{align*}
                \sfd{\bxs}{\by} - \sfd{\bx}{\by} &= \beta - \alpha + \sum_{i=2}^{\n} \indicator{\bxn{i} > \bxsn{1}} \indicator{\bxn{i} < \bxn{1}} (\xi_i - \psi_i)\\
                & + \sum_{i=2}^{\n} \indicator{\bxn{i} < \bxsn{1}} \indicator{\bxn{i} > \bxn{1}} (\psi_i - \xi_i).
            \end{align*}
            This completes our proof.
        \end{proof}

        \begin{lemma}  \label{supp:proposition:1:lemma:3}
            Suppose $\bx, \by \in \vndistinct$, and let $\bxs \in \vndistinct$ be an imputation of $\bx$ for an index $\uu \in \tonumber{\n}$. Then, if either of the two following conditions hold		
            \begin{align*}
                &(\rn{i}) : \rank{\bxn{\uu}}{\bx} > \rank{\byn{\uu}}{\by} \text{ and } \rank{\bxsn{\uu}}{\bxs} = \rank{\bxn{\uu}}{\bx} - 1,\\
                \text{or }& (\rn{ii}) : \rank{\bxn{\uu}}{\bx} < \rank{\byn{\uu}}{\by} \text{ and } \rank{\bxsn{\uu}}{\bxs} = \rank{\bxn{\uu}}{\bx} + 1,
            \end{align*}
            we have $\sfd{\bxs}{\by} \le \sfd{\bx}{\by}$.
            However, if either of the two following conditions hold
            \begin{align*}
                &(\rn{iii}) : \rank{\bxn{\uu}}{\bx} \ge \rank{\byn{\uu}}{\by} \text{ and } \rank{\bxsn{\uu}}{\bxs} = \rank{\bxn{\uu}}{\bx} + 1,\\
                \text{or } &(\rn{iv}) : \rank{\bxn{\uu}}{\bx} \le \rank{\byn{\uu}}{\by} \text{ and } \rank{\bxsn{\uu}}{\bxs} = \rank{\bxn{\uu}}{\bx} - 1,
            \end{align*}
            we have $\sfd{\bxs}{\by} \ge \sfd{\bx}{\by}$.
        \end{lemma}

        \begin{proof}
            For convenience, let us assume (after relabeling) $ \uu = 1$, and denote
            \begin{align*}
                &\alpha = \left|\rank{\bxn{1}}{\bx} - \rank{\byn{1}}{\by}\right|, \text{ and } \beta = \left|\rank{\bxsn{1}}{\bx} - \rank{\byn{1}}{\by}\right|.
            \end{align*}
            For any $i \in \{2,\cdots,\n\}$, let us denote
            \begin{align*} 
                \begin{split}
                    &\xi_i = \left|\rank{\bxn{i}}{\svector{\bx}{\lconstant}{\tonumber{\n} \setminus \{1\}}} + 1 - \rank{\byn{i}}{\by}\right|, \\
                    \text{ and }&\psi_i = \left|\rank{\bxn{i}}{\svector{\bx}{\lconstant}{\tonumber{\n} \setminus \{1\}}} - \rank{\byn{i}}{\by}\right|.
                \end{split}
            \end{align*}	
            Then, for any $i \in \{2,\cdots,\n\}$, if
            \begin{align*}
                \rank{\bxn{i}}{\svector{\bx}{\lconstant}{\tonumber{\n} \setminus \{1\}}} - \rank{\byn{i}}{\by} \ge 0,
            \end{align*}
            we have 
            \begin{align*}
                \psi_i  &= 	\rank{\bxn{i}}{\svector{\bx}{\lconstant}{\tonumber{\n} \setminus \{1\}}} - \rank{\byn{i}}{\by}, \\
                \text{ and }\xi_i & = \rank{\bxn{i}}{\svector{\bx}{\lconstant}{\tonumber{\n} \setminus \{1\}}} + 1 - \rank{\byn{i}}{\by}.
            \end{align*}
            Hence we have $\xi_i = \psi_i + 1$. However, if 
            \begin{align*}
                \rank{\bxn{i}}{\svector{\bx}{\lconstant}{\tonumber{\n} \setminus \{1\}}} - \rank{\byn{i}}{\by} < 0,
            \end{align*}
            we have 
            \begin{align*}
                \psi_i  &=  \rank{\byn{i}}{\by} - \rank{\bxn{i}}{\svector{\bx}{\lconstant}{\tonumber{\n} \setminus \{1\}}}, \\
                \text{ and }\xi_i & = \rank{\byn{i}}{\by} -  \rank{\bxn{i}}{\svector{\bx}{\lconstant}{\tonumber{\n} \setminus \{1\}}} - 1.
            \end{align*}
            Hence we have $\xi_i  = \psi_i - 1$. Therefore, we have
            \begin{align} \label{supp:proposition:1:lemma:3:eqn:1}
                \xi_i - \psi_i \in \{-1, 1\}, \text{ for any } i \in \{2,\cdots,\n\}.
            \end{align}

            Next, according to Lemma~\ref{supp:proposition:1:lemma:2}, we have
            \begin{align*}
                \sfd{\bxs}{\by} - \sfd{\bx}{\by} &= \beta - \alpha + \sum_{i=2}^{\n} \indicator{\bxn{i} > \bxsn{1}} \indicator{\bxn{i} < \bxn{1}} (\xi_i - \psi_i)\\
                &+ \sum_{i=2}^{\n} \indicator{\bxn{i} < \bxsn{1}} \indicator{\bxn{i} > \bxn{1}} (\psi_i - \xi_i).
            \end{align*}
            For notation ease, let us denote
            \begin{align*}
                &\Sigma_1 = \sum_{i=2}^{\n} \indicator{\bxn{i} > \bxsn{1}} \indicator{\bxn{i} < \bxn{1}} (\xi_i - \psi_i),\\
                \text{and }&\Sigma_2 = \sum_{i=2}^{\n} \indicator{\bxn{i} < \bxsn{1}} \indicator{\bxn{i} > \bxn{1}} (\psi_i - \xi_i).
            \end{align*}
            Then we have
            \begin{align} \label{supp:proposition:1:lemma:3:eqn:2}
                \sfd{\bxs}{\by} - \sfd{\bx}{\by} = \beta - \alpha + \Sigma_1 + \Sigma_2.
            \end{align}

            Suppose the condition
            \begin{align*}
                (\rn{i}): \rank{\bxn{1}}{\bx} > \rank{\byn{1}}{\by} \text{ and } \rank{\bxsn{1}}{\bxs} = \rank{\bxn{1}}{\bx} - 1
            \end{align*}
            holds. Then we have
            \begin{align*}
                &\rank{\bxn{1}}{\bx} - \rank{\byn{1}}{\by} > 0\\
                \Rightarrow	&\alpha = |\rank{\bxn{1}}{\bx} - \rank{\byn{1}}{\by}| = \rank{\bxn{1}}{\bx} - \rank{\byn{1}}{\by},
            \end{align*}
            and
            \begin{align*}
                &\rank{\bxsn{1}}{\bxs} - \rank{\byn{1}}{\by} = \rank{\bxn{1}}{\bx} -1 -  \rank{\byn{1}}{\by} \ge 0\\
                \Rightarrow& 	\beta = |\rank{\bxsn{1}}{\bxs} - \rank{\byn{1}}{\by}| = \rank{\bxn{1}}{\bx} -1 -  \rank{\byn{1}}{\by}.
            \end{align*}
            Hence, we have 
            \begin{align*}
                \beta = \alpha - 1.
            \end{align*}
            Meanwhile, since $ \rank{\bxsn{1}}{\bxs} = \rank{\bxn{1}}{\bx} - 1$ and $ \bxs$ is an imputation
            of $\bx$ for the index $1$, then according to Lemma~\ref{supp:proposition:1:lemma:1}, we have
            \begin{align*}
                \bxsn{1} < \bxn{1} \Rightarrow  \indicator{\bxn{i} < \bxsn{1}} \indicator{\bxn{i} > \bxn{1}} = 0, \text{ for any } i \in \{2,\ldots,\n\}.
            \end{align*}
            Therefore,
            \begin{align*}
                \Sigma_2 = \sum_{i=2}^{\n} \indicator{\bxn{i} < \bxsn{1}} \indicator{\bxn{i} > \bxn{1}} (\psi_i - \xi_i) = 0.
            \end{align*}
            According to Lemma~\ref{supp:proposition:1:lemma:1}, we also have
            \begin{align*}
                \sum_{i=2}^{\n} \indicator{\bxn{i} > \bxsn{1}} \indicator{\bxn{i} < \bxn{1}} = 1.
            \end{align*}
            According to \eqref{supp:proposition:1:lemma:3:eqn:1}, we have $ \xi_i - \psi_i \in \{-1,1\}$, which then follows that
            \begin{align*}
                \Sigma_1 &= \sum_{i=2}^{\n} \indicator{\bxn{i} > \bxsn{1}} \indicator{\bxn{i} < \bxn{1}} (\xi_i - \psi_i) \\ &\le \sum_{i=2}^{\n} \indicator{\bxn{i} > \bxsn{1}} \indicator{\bxn{i} < \bxn{1}} = 1. 
            \end{align*}
            Recall that $ \beta = \alpha - 1$. Subsequently, following \eqref{supp:proposition:1:lemma:3:eqn:2}, we have 
            \begin{align*}
                \sfd{\bxs}{\by} - \sfd{\bx}{\by} = -1 + \Sigma_1 + \Sigma_2 \le -1 + 1 + 0 = 0,
            \end{align*}
            which proves our results when the condition $(\rn{i})$ holds. The other three cases can be proved similarly.

            Suppose the condition
            \begin{align*}
                (\rn{ii}) : \rank{\bxn{1}}{\bx} < \rank{\byn{1}}{\by} \text{ and } \rank{\bxsn{1}}{\bxs} = \rank{\bxn{1}}{\bx} + 1
            \end{align*}
            holds. Then we have
            \begin{align*}
                &\rank{\bxn{1}}{\bx} - \rank{\byn{1}}{\by} < 0\\
                \Rightarrow	&\alpha = |\rank{\bxn{1}}{\bx} - \rank{\byn{1}}{\by}| = \rank{\byn{1}}{\by} - \rank{\bxn{1}}{\bx},
            \end{align*}
            and
            \begin{align*}
                &\rank{\bxsn{1}}{\bxs} - \rank{\byn{1}}{\by} = \rank{\bxn{1}}{\bx} +1 -  \rank{\byn{1}}{\by} \le 0\\
                \Rightarrow& 	\beta = |\rank{\bxsn{1}}{\bxs} - \rank{\byn{1}}{\by}| = \rank{\byn{1}}{\by} - \rank{\bxn{1}}{\bx} - 1.
            \end{align*}
            Hence, we have 
            \begin{align*}
                \beta = \alpha - 1.
            \end{align*}
            Since $ \rank{\bxsn{1}}{\bxs} = \rank{\bxn{1}}{\bx} + 1$, and $\bxs$ is an imputation
            of $\bx$ for the index $1$, then according to Lemma~\ref{supp:proposition:1:lemma:1}, we have
            \begin{align*}
                \bxsn{1} > \bxn{1} \Rightarrow  \indicator{\bxn{i} > \bxsn{1}} \indicator{\bxn{i} < \bxn{1}} = 0 \text{ for any } i \in \{2,\ldots,\n\}.
            \end{align*}
            Therefore,
            \begin{align*}
                \Sigma_1 = \sum_{i=2}^{\n} \indicator{\bxn{i} > \bxsn{1}} \indicator{\bxn{i} < \bxn{1}} (\xi_i - \psi_i) = 0.
            \end{align*}
            According to Lemma~\ref{supp:proposition:1:lemma:1}, we also have
            \begin{align*}
                \sum_{i=2}^{\n} \indicator{\bxn{i} < \bxsn{1}} \indicator{\bxn{i} > \bxn{1}} = 1.
            \end{align*}
            According to \eqref{supp:proposition:1:lemma:3:eqn:1}, we have $ \xi_i - \psi_i \in \{-1,1\}$, which follows that
            \begin{align*}
                \Sigma_2 &=  \sum_{i=2}^{\n} \indicator{\bxn{i} < \bxsn{1}} \indicator{\bxn{i} > \bxn{1}} (\psi_i - \xi_i)\\ &\le \sum_{i=2}^{\n} \indicator{\bxn{i} < \bxsn{1}} \indicator{\bxn{i} > \bxn{1}} = 1. 
            \end{align*}
            Recall that $ \beta = \alpha - 1$. Subsequently, following \eqref{supp:proposition:1:lemma:3:eqn:2}, we have 
            \begin{align*}
                \sfd{\bxs}{\by} - \sfd{\bx}{\by} = -1 + \Sigma_1 + \Sigma_2 \le -1 + 0 + 1 = 0,
            \end{align*}
            which proves our results when the condition $(\rn{ii})$ holds.

            Suppose the condition
            \begin{align*}
                (\rn{iii}) : \rank{\bxn{1}}{\bx} \ge \rank{\byn{1}}{\by} \text{ and } \rank{\bxsn{1}}{\bxs} = \rank{\bxn{1}}{\bx} + 1
            \end{align*}
            holds. Then we have
            \begin{align*}
                &\rank{\bxn{1}}{\bx} - \rank{\byn{1}}{\by} \ge 0\\
                \Rightarrow	&\alpha = |\rank{\bxn{1}}{\bx} - \rank{\byn{1}}{\by}| = \rank{\bxn{1}}{\by} - \rank{\byn{1}}{\bx},
            \end{align*}
            and
            \begin{align*}
                &\rank{\bxsn{1}}{\bxs} - \rank{\byn{1}}{\by} = \rank{\bxn{1}}{\bx} +1 -  \rank{\byn{1}}{\by} > 0\\
                \Rightarrow& \beta = |\rank{\bxsn{1}}{\bxs} - \rank{\byn{1}}{\by}| = \rank{\bxn{1}}{\by} - \rank{\byn{1}}{\bx} + 1.
            \end{align*}
            Hence, we have 
            \begin{align*}
                \beta = \alpha + 1.
            \end{align*}
            Since $ \rank{\bxsn{1}}{\bxs} = \rank{\bxn{1}}{\bx} + 1$ and $ \bxs$ is an imputation
            of $\bx$ for the index $1$, then according to Lemma~\ref{supp:proposition:1:lemma:1}, we have
            \begin{align*}
                \bxsn{1} > \bxn{1} \Rightarrow \indicator{\bxn{i} > \bxsn{1}} \indicator{\bxn{i} < \bxn{1}} = 0  \text{ for any } i \in \{2,\ldots,\n\}.
            \end{align*}
            Therefore, we have
            \begin{align*}
                \Sigma_1 = \sum_{i=2}^{\n} \indicator{\bxn{i} > \bxsn{1}} \indicator{\bxn{i} < \bxn{1}} (\xi_i - \psi_i) = 0. 
            \end{align*}
            According to Lemma~\ref{supp:proposition:1:lemma:1}, we also have
            \begin{align*}
                \sum_{i=2}^{\n} \indicator{\bxn{i} < \bxsn{1}} \indicator{\bxn{i} > \bxn{1}} = 1.
            \end{align*}
            According to \eqref{supp:proposition:1:lemma:3:eqn:1}, we have $ \xi_i - \psi_i \in \{-1,1\}$, which follows that
            \begin{align*}
                \Sigma_2 &=  \sum_{i=2}^{\n} \indicator{\bxn{i} < \bxsn{1}} \indicator{\bxn{i} > \bxn{1}} (\psi_i - \xi_i) \\
                &\ge  \sum_{i=2}^{\n} \indicator{\bxn{i} < \bxsn{1}} \indicator{\bxn{i} > \bxn{1}}\cdot (-1) = -1. 
            \end{align*}
            Recall that $ \beta = \alpha + 1$. Following \eqref{supp:proposition:1:lemma:3:eqn:2}, we have
            \begin{align*}
                \sfd{\bxs}{\by} - \sfd{\bx}{\by} = 1 + \Sigma_1 + \Sigma_2 \ge 1 + 0 - 1 = 0,
            \end{align*}
            which proves our results when the condition $(\rn{iii})$ holds.

            Suppose the condition
            \begin{align*}
                (\rn{iv}) : \rank{\bxn{1}}{\bx} \le \rank{\byn{1}}{\by} \text{ and } \rank{\bxsn{1}}{\bxs} = \rank{\bxn{1}}{\bx} - 1
            \end{align*}
            holds. Then we have
            \begin{align*}
                &\rank{\bxn{1}}{\bx} - \rank{\byn{1}}{\by} \le 0,\\
                \Rightarrow	&\alpha = |\rank{\bxn{1}}{\bx} - \rank{\byn{1}}{\by}| =  \rank{\byn{1}}{\bx} - \rank{\bxn{1}}{\by},
            \end{align*}
            and
            \begin{align*}
                &\rank{\bxsn{1}}{\bxs} - \rank{\byn{1}}{\by} = \rank{\bxn{1}}{\bx} - 1 -  \rank{\byn{1}}{\by} < 0,\\
                \Rightarrow& 	\beta = |\rank{\bxsn{1}}{\bxs} - \rank{\byn{1}}{\by}| = \rank{\byn{1}}{\bx} - \rank{\bxn{1}}{\by} + 1.
            \end{align*}
            Hence, we have 
            \begin{align*}
                \beta = \alpha + 1.
            \end{align*}
            Since $ \rank{\bxsn{1}}{\bxs} = \rank{\bxn{1}}{\bx} - 1$ and $ \bxs$ is an imputation
            of $\bx$ for index $1$, then according to Lemma~\ref{supp:proposition:1:lemma:1}, we have
            \begin{align*}
                \bxsn{1} < \bxn{1} \Rightarrow  \indicator{\bxn{i} < \bxsn{1}} \indicator{\bxn{i} > \bxn{1}} = 0, \text{ for any } i \in \{2,\ldots,\n\}.
            \end{align*}
            Therefore,
            \begin{align*}
                \Sigma_2 = \sum_{i=2}^{\n} \indicator{\bxn{i} < \bxsn{1}} \indicator{\bxn{i} > \bxn{1}} (\psi_i - \xi_i) = 0.
            \end{align*}
            According to Lemma~\ref{supp:proposition:1:lemma:1}, we also have
            \begin{align*}
                \sum_{i=2}^{\n} \indicator{\bxn{i} > \bxsn{1}} \indicator{\bxn{i} < \bxn{1}} = 1.
            \end{align*}
            According to \eqref{supp:proposition:1:lemma:3:eqn:1}, we have $ \xi_i - \psi_i \in \{-1,1\}$, which follows that
            \begin{align*}
                \Sigma_1 &= \sum_{i=2}^{\n} \indicator{\bxn{i} > \bxsn{1}} \indicator{\bxn{i} < \bxn{1}} (\xi_i - \psi_i)\\
                & \ge \sum_{i=2}^{\n} \indicator{\bxn{i} > \bxsn{1}} \indicator{\bxn{i} < \bxn{1}} \cdot (-1) = -1. 
            \end{align*}
            Recall that $ \beta = \alpha + 1$. Following \eqref{supp:proposition:1:lemma:3:eqn:2}, we have
            \begin{align*}
                \sfd{\bxs}{\by} - \sfd{\bx}{\by} = 1 + \Sigma_1 + \Sigma_2 \ge 1 - 1 + 0 = 0,
            \end{align*}
            which proves our results when the condition $(\rn{iv})$ holds, and completes our proof.
        \end{proof}

        \begin{lemma} \label{supp:proposition:1:lemma:4}
            Suppose $\bx \in \vndistinct$ and $\uu \in \tonumber{\n}$ is an index. Let $\rr$ 
            be an integer between $1$ to $\n$. Then, there exists an imputation $\bxs \in \vndistinct$ 
            of $\bx$ for the index $\uu$ such that $\rank{\bxsn{\uu}}{\bxs} =~\rr$.
        \end{lemma}

        \begin{proof}
            For notation ease, let us assume (after relabeling) $\uu = 1$ and $\bxn{2}, \cdots, \bxn{\n}$ are ordered such that $\bxn{2} < \cdots < \bxn{\n}$.

            Suppose $\rr = 1$. Let $\bxsn{i} = \bxn{i}$ for any $i \in \{2, \cdots, \n\}$ and let $\bxsn{1}$ be a real number such that $\bxsn{1} < \bxn{2}$. Then $\bxs$ is an imputation of $\bx$ for the index $\uu$. Meanwhile, since $\bxsn{1} < \bxn{2}$, we have
            \begin{align*}
                &\bxsn{1}  < \bxn{2} < \cdots < \bxn{\n}\\
                \Rightarrow &\bxsn{1} < \bxsn{2} < \cdots < \bxsn{\n}.
            \end{align*}
            Thus, according to the definition of rank, we have 
            \begin{align*}
                \rank{\bxsn{1}}{\bxs}  = \sum_{i = 1}^{n} \indicator{\bxsn{i} \le \bxsn{1}} = 1.
            \end{align*}
            Therefore, $\rank{\bxsn{\uu}}{\bxs} = 1 = \rr$, which proves our result when $\rr = 1$. 

            Suppose $1 < \rr < \n$. Let $\bxsn{i} = \bxn{i}$ for any $i \in \{2, \cdots, \n\}$, and let $\bxsn{1}$ be a real number such that $\bxn{\rr} < \bxsn{1} < \bxn{\rr+1}$. Then $\bxs$ is an imputation of $\bx$ for the index~$\uu$. Meanwhile, since $\bxn{\rr} < \bxsn{1} < \bxn{\rr+1}$, we have
            \begin{align*}
                &\bxn{2} < \cdots < \bxn{\rr} < \bxn{1} < \bxn{\rr+1} < \cdots < \bxn{\n}\\
                \Rightarrow & \bxsn{2} < \cdots < \bxsn{\rr} < \bxsn{1} < \bxsn{\rr+1} < \cdots < \bxsn{\n}.
            \end{align*}
            Thus, according to the definition of rank, we have 
            \begin{align*}
                \rank{\bxsn{1}}{\bxs}  = \sum_{i = 1}^{\n} \indicator{\bxsn{i} \le \bxsn{1}} = \rr.
            \end{align*}
            Therefore, $\rank{\bxsn{\uu}}{\bxs} = \rr$, which proves our result when  $1 < \rr < \n$.

            Suppose $\rr = \n$. Let $\bxsn{i} = \bxn{i}$ for any $i \in \{2, \cdots, \n\}$ and let $\bxsn{1}$ be a real number such that $\bxsn{1} > \bxn{\n}$. Then, $\bxs$ is an imputation of $\bx$ for the index~$\uu$. Meanwhile, since $\bxsn{1} > \bxn{n}$, we have
            \begin{align*}
                &\bxsn{1} > \bxn{\n} > \cdots > \bxn{2}\\
                \Rightarrow & \bxsn{1} > \bxsn{\n} > \cdots > \bxsn{2}.
            \end{align*}
            Thus, according to the definition of rank, we have 
            \begin{align*}
                \rank{\bxsn{1}}{\bxs}  = \sum_{i = 1}^{n} \indicator{\bxsn{i} \le \bxsn{1}} = \n.
            \end{align*}
            Therefore, $\rank{\bxsn{\uu}}{\bxs} = \n$, which proves our result when $\rr = \n$,
            and completes our proof. 

        \end{proof}

        \begin{proposition} \label{supp:proposition:1}
            Suppose $\bx, \by \in \vndistinct$ and  
            let $\bxs \in \vndistinct$ be 
            an imputation of $\bx$ for an index $\uu \in \tonumber{\n}$.
            If we choose 
            $\bxsn{\uu}$ such that		
            $\rank{\bxsn{\uu}}{\bxs} = \rank{\byn{\uu}}{\by}$,
            then we have $\sfd{\bxs}{\by} \le \sfd{\bx}{\by}$.
            Moreover, for any other imputation $\xaa \in \vndistinct$ of $\bx$ for the index $\uu$, 
            we have $\sfd{\bxs}{\by} \le \sfd{\xaa}{\by}$.
        \end{proposition}

        \begin{proof}
            Notice that there are two statements in Proposition~\ref{supp:proposition:1}:
            \begin{itemize}
                \item[$(\rn{i})$] Suppose $\bx, \by \in \vndistinct$ and  
                    let $\bxs \in \vndistinct$ be 
                    an imputation of $\bx$ for an index $\uu \in \tonumber{\n}$.
                    If $\rank{\bxsn{\uu}}{\bxs} = \rank{\byn{\uu}}{\by}$,
                    then we have $\sfd{\bxs}{\by} \le \sfd{\bx}{\by}$. 
                \item[$(\rn{ii})$] Suppose $\bx, \by \in \vndistinct$ and  
                    let $\bxs \in \vndistinct$ be 
                    an imputation of $\bx$ for an index $\uu \in \tonumber{\n}$.
                    If $\rank{\bxsn{\uu}}{\bxs} = \rank{\byn{\uu}}{\by}$, then
                    for any other imputation $\xaa \in \vndistinct$ of $\bx$ for index $\uu$, 
                    we have $\sfd{\bxs}{\by} \le \sfd{\xaa}{\by}.$
            \end{itemize}
            Below, we first show that the statement $(\rn{i})$ is true.
            Then we prove the statement $(\rn{ii})$ is true using the statement~$(\rn{i})$.

            First, we prove the statement $(\rn{i})$ is true.
            Suppose $ \rank{\bxn{\uu}}{\bx} = \rank{\byn{\uu}}{\by}$. Then, we have
            \begin{align*}
                \rank{\bxn{\uu}}{\bx} = \rank{\byn{\uu}}{\by} = \rank{\bxsn{\uu}}{\bx}.
            \end{align*}
            Since $\bxs \in \vndistinct$ is 
            an imputation of $\bx$ for the index $\uu \in \tonumber{\n}$, 
            then according to Lemma~\ref{supp:lemma:4}, 
            we have $ \rank{\bxs}{\bxs} = \rank{\bx}{\bx}$, which implies that
            \begin{align*}
                \sfd{\bxs}{\by} = \sfd{\bx}{\by}.
            \end{align*}
            Hence, we have proved the statement $(\rn{i})$ 
            when $ \rank{\bxn{\uu}}{\bx} = \rank{\byn{\uu}}{\by}$.

            Suppose $ \rank{\bxn{\uu}}{\bx} > \rank{\byn{\uu}}{\by}$, let us denote $ \rank{\bxn{\uu}}{\bx} = \rank{\byn{\uu}}{\by} + a$, where $ a \in \mathbb{N}$ and $ a > 0$. According to Lemma~\ref{supp:proposition:1:lemma:2}, we can find vectors	
            $ \bxln{1}, \bxln{2}, \cdots, \bxln{a} \in \vndistinct$ such that each 
            vector is an imputation of $\bx$ for the index $\uu \in \tonumber{\n}$ and
            \begin{align*}
                \rank{\bxln{k}(\uu)}{\bxln{k}} = \rank{\bxn{\uu}}{\bx} - k, \text{ for any } k \in \{1, \cdots, a\}.
            \end{align*}
            Denote $\bxln{0} = \bx$. Then, for any $ k \in \{1, \cdots, a\}$, we have
            \begin{align*}
                \rank{\bxln{k-1}(\uu)}{\bxln{k-1}} = \rank{\bxn{\uu}}{\bx} - (k - 1) > \rank{\bxn{\uu}}{\bx} - a = \rank{\byn{\uu}}{\by},
            \end{align*}
            and
            \begin{align*}
                \rank{\bxln{k}(\uu)}{\bxln{k}} = \rank{\bxln{k-1}(\uu)}{\bxln{k-1}} - 1.
            \end{align*}
            Subsequently, by applying Lemma~\ref{supp:proposition:1:lemma:3} between each $\bxln{k}, \by$ and $ \bxln{k-1}, \by$, where $ k \in \{1, \cdots, a\}$, we have
            \begin{align*}
                \sfd{\bx}{\by} = \sfd{\bxln{0}}{\by} \ge \sfd{\bxln{1}}{\by} \ge \cdots \ge \sfd{\bxln{a}}{\by}.
            \end{align*}
            Notice that 
            \begin{align*}
                \rank{\bxln{a}(\uu)}{\bxln{a}} = \rank{\bxn{\uu}}{\bx} - a = \rank{\byn{\uu}}{\by} = \rank{\bxsn{\uu}}{\bxs}.
            \end{align*}
            Since $\bxln{a}$ is an imputation of $\bxs$ for the index $\uu$,
            then according to Lemma~\ref{supp:lemma:4}, we have $\rank{\bxs}{\bxs} = \rank{\bxln{a}}{\bxln{a}}$, which gives
            \begin{align*}
                \sfd{\bxs}{\by} = \sfd{\bxln{a}}{\by} \le \sfd{\bx}{\by}.
            \end{align*}
            This proves the statement $(\rn{i})$ when $ \rank{\bxn{\uu}}{\bx} > \rank{\byn{\uu}}{\by}$.

            The case when $ \rank{\bxn{\uu}}{\bx} < \rank{\byn{\uu}}{\by}$ can be proved similarly.
            Hence, we have shown the statement $(\rn{i})$ is true.

            Next, we show that the statement $(\rn{ii})$ is true using the statement $(\rn{i})$.
            Let $\xaa \in \vndistinct$ be an imputation of $\bx$ for the index $\uu$.
            According to Lemma~\ref{supp:proposition:1:lemma:2}, we can find an imputation	
            $\bxsp \in \vndistinct$ of $\bxp$ for the index $\uu \in \tonumber{\n}$,
            and $\rank{\bxspn{\uu}}{\bxs} = \rank{\byn{\uu}}{\by}$. Then, according to
            the statement $(\rn{i})$, we have 
            \begin{align*}
                \sfd{\bxsp}{\by} \le \sfd{\xaa}{\by}.
            \end{align*}
            Notice that $\bxs \in \vndistinct$ is an imputation of $\bx$ for the index $\uu \in \tonumber{\n}$, and $\rank{\bxsn{\uu}}{\bxs} = \rank{\byn{\uu}}{\by}$. Thus, according to Lemma~\ref{supp:lemma:4}, 
            we have $\rank{\bxsp}{\bxsp} = \rank{\bxs}{\bxs}$, which implies $\sfd{\bxsp}{\by} = \sfd{\bxs}{\by}$. 
            Thus, we have
            \begin{align*}
                \sfd{\bxs}{\by} = \sfd{\bxsp}{\by} \le \sfd{\xaa}{\by}.
            \end{align*}
            This proves the statement $(\rn{ii})$, and completes our proof.
        \end{proof}

        \subsection{Proof of Lemma 2.2} 

        This subsection proves Lemma 2.2. Specifically, we show the following result
        is true.

        \begin{lemma} \label{supp:theorem:2.2:lemma:1}
            Suppose $\bx, \by \in \vndistinct$ and $\bu \subset \tonumber{\n}$ 
            is a subset of indices. Then there exists an imputation $\bxs \in \vndistinct$ 
            of $\bx$ for indices $\bu$ such that
            \begin{align*}
                \bxsn{i} = \bxn{i}, i \in \tonumber{\n} \setminus \bu, \text{ and }  \rank{\bxsn{\iconstant}}{\bxs} = \rank{\byn{\iconstant}}{\by} 
                \,\,\text{for all }~\iconstant \in \bu.
            \end{align*}    
        \end{lemma}

        \begin{proof}
            We first consider the case when $\bu = \tonumber{\n}$. 
            Let $\bxs = \by$, then 
            according to the definition of imputations,   	
            we have that $\bxs \in \vndistinct$ is an 
            imputation of $\bx$ for $\bu = \tonumber{\n}$.
            Since $\bxs = \by$, we have $\rank{\bxs}{\bxs} = \rank{\by}{\by}$. 
            Hence, Lemma~\ref{supp:theorem:2.2:lemma:1} is true
            when $\bu = \tonumber{\n}$.

            However, if $\bu = \emptyset$, we can let $\bxs = \bx$.
            Then according to the definition of imputations,
            $\bxs \in \vndistinct$ is an 
            imputation of $\bx$ for $\bu = \emptyset$.
            Hence, Lemma~\ref{supp:theorem:2.2:lemma:1} is true
            when $\bu = \emptyset$.

            Now, we only need to prove 
            Lemma~\ref{supp:theorem:2.2:lemma:1}
            when $\bu \neq \tonumber{\n}$ and $\bu \neq \emptyset$.

            For any fixed $\n \in \mathbb{N}$, let $\bpn{\kk}$ be the statement of
            lemma~\ref{supp:theorem:2.2:lemma:1} when $|\bu| = \kk$. 
            We prove $\bpn{\kk}$ holds for any $\kk \in \{1, \ldots, \n-1\}$
            by induction on $\kk$.

            \emph{Base Case:} We show $\bpn{1}$ is true.
            Suppose $|\bu| = 1$. Then $\bpn{1}$
            is true according to Lemma~\ref{supp:proposition:1:lemma:2}.

            \emph{Induction Step: } We show the implication 
            $\bpn{1}, \bpn{\kk} \Rightarrow \bpn{\kk+1}$ for any
            $\kk \in \{1,\ldots, \n~-~2\}$. 

            Suppose $|\bu| = \kk + 1$, 
            and denote $\m = \kk + 1$.   	
            Without loss of generality, let us assume (after relabeling) $\bu = \{1, \cdots, \m\}$ and $\byn{1}, \cdots, \byn{\m}$ such that $\byn{1} < \cdots < \byn{\m}$.

            Notice that $|\bu \setminus \{1\}| = |\bu| - 1 = \m - 1 = \kk$. 
            Then since $\bpn{\kk}$ is true, there exists $\bxln{1} \in \vndistinct$ such that
            \begin{align}
                &\bxln{1}(i) = \bxn{i}, \quad i \in \tonumber{\n} \setminus (\bu \setminus \{1\}), \label{supp:theorem:2.2:lemma:1:eqn:1}\\
                &\rank{\bxln{1}(i)}{\bxln{1}} = \rank{\byn{i}}{\by}, \quad i \in \bu \setminus \{1\}. \label{supp:theorem:2.2:lemma:1:eqn:2}
            \end{align}
            According to Lemma~\ref{supp:lemma:2}, for any $i \in \bu \setminus \{1\}$, we have
            \begin{align*}
                \rank{\bxln{1}(i)}{\svector{\bxln{1}}{\lconstant}{\tonumber{\n} \setminus \{1\}}} \ge \rank{\bxln{1}(i)}{\bxln{1}} - 1 = \rank{\byn{i}}{\by} - 1,
            \end{align*} 
            where the last equation holds according to \eqref{supp:theorem:2.2:lemma:1:eqn:2}.
            Then, since $\byn{1} < \cdots < \byn{\m}$, we have
            $\rank{\byn{1}}{\by} < \cdots < \rank{\byn{\m}}{\by}$. 
            Hence, for any $i \in \bu \setminus \{1\}$, we have
            \begin{align}
                &\rank{\bxln{1}(i)}{\svector{\bxln{1}}{\lconstant}{\tonumber{\n} \setminus \{1\}}} \ge \rank{\byn{i}}{\by} - 1 > \rank{\byn{1}}{\by} - 1 \nonumber \\ 
                \Rightarrow &\rank{\bxln{1}(i)}{\svector{\bxln{1}}{\lconstant}{\tonumber{\n} \setminus \{1\}}} \ge \rank{\byn{1}}{\by}. \label{supp:theorem:2.2:lemma:1:eqn:3}
            \end{align}

            Since $\bpn{1}$ is true, then there exists $\bxln{2} \in \vndistinct$ such that
            \begin{align}
                &\bxln{2}(i) = \bxln{1}(i), \quad i \in \tonumber{\n} \setminus \{1\}, \label{supp:theorem:2.2:lemma:1:eqn:4}\\
                &\rank{\bxln{2}(1)}{\bxln{2}} = \rank{\byn{1}}{\by}. \label{supp:theorem:2.2:lemma:1:eqn:5}
            \end{align}
            According to Lemma~\ref{supp:lemma:2}, for any $i \in \bu \setminus \{1\}$, we have
            \begin{align*}
                \rank{\bxln{2}(i)}{\bxln{2}} &\ge \rank{\bxln{2}(i)}{\svector{\bxln{2}}{\lconstant}{\tonumber{\n} \setminus \{1\}}}.
            \end{align*}
            Notice that according to \eqref{supp:theorem:2.2:lemma:1:eqn:4}, for any $i \in \bu \setminus \{1\}$, we have 
            \begin{align*}
                \rank{\bxln{2}(i)}{\svector{\bxln{2}}{\lconstant}{\tonumber{\n} \setminus \{1\}}} = \rank{\bxln{1}(i)}{\svector{\bxln{1}}{\lconstant}{\tonumber{\n} \setminus \{1\}}}.
            \end{align*}
            Hence, for any $i \in \bu \setminus \{1\}$, we have
            \begin{align*}
                \rank{\bxln{2}(i)}{\bxln{2}} \ge \rank{\bxln{1}(i)}{\svector{\bxln{1}}{\lconstant}{\tonumber{\n} \setminus \{1\}}} \ge^{\eqref{supp:theorem:2.2:lemma:1:eqn:3}} \rank{\byn{1}}{\by} =^{\eqref{supp:theorem:2.2:lemma:1:eqn:5}} \rank{\bxln{2}(1)}{\bxln{2}}.
            \end{align*}
            Since $\bxln{2} \in \vndistinct$ is a vector of distinct real values, for any $i \in \bu \setminus \{1\}$ we have $\rank{\bxln{2}(i)}{\bxln{2}} \neq \rank{\bxln{2}(1)}{\bxln{2}}$. Hence, for any $i \in \bu \setminus \{1\}$,
            \begin{align}
                \rank{\bxln{2}(i)}{\bxln{2}} \ge \rank{\bxln{2}(1)}{\bxln{2}} \Rightarrow \rank{\bxln{2}(i)}{\bxln{2}} > \rank{\bxln{2}(1)}{\bxln{2}} \Rightarrow \bxln{2}(i) > \bxln{2}(1) \nonumber.
            \end{align}
            According to Lemma~\ref{supp:lemma:3}, we then have
            \begin{align} \label{supp:theorem:2.2:lemma:1:eqn:6}
                \rank{\bxln{2}(1)}{\svector{\bxln{2}}{\iconstant}{\tonumber{\n} \setminus (\bu \setminus \{1\})}} = \rank{\bxln{2}(1)}{\bxln{2}} =^{\eqref{supp:theorem:2.2:lemma:1:eqn:5}} \rank{\byn{1}}{\by}.
            \end{align}

            Now, since $\bpn{\kk}$ is true, there exists $\bxln{3} \in \vndistinct$ such that
            \begin{align}
                &\bxln{3}(i) = \bxln{2}(i), \quad i \in \tonumber{\n} \setminus (\bu \setminus \{1\}), \label{supp:theorem:2.2:lemma:1:eqn:7}\\
                &\rank{\bxln{3}(i)}{\bxln{3}} = \rank{\byn{i}}{\by}, \quad i \in \bu \setminus \{1\}. \label{supp:theorem:2.2:lemma:1:eqn:8}
            \end{align}
            Since $\byn{1} < \cdots < \byn{\m}$, we have $\rank{\byn{1}}{\by} < \cdots < \rank{\byn{\m}}{\by}$. Then, for any $i \in \bu \setminus \{1\}$, it follows that
            \begin{align*}
                \rank{\bxln{3}(i)}{\bxln{3}} =^{\eqref{supp:theorem:2.2:lemma:1:eqn:8}} \rank{\byn{i}}{\by} > \rank{\byn{1}}{\by} &=^{\eqref{supp:theorem:2.2:lemma:1:eqn:6}} \rank{\bxln{2}(1)}{\svector{\bxln{2}}{\lconstant}{\tonumber{\n} \setminus (\bu \setminus \{1\})}}\\
                &=^{\eqref{supp:theorem:2.2:lemma:1:eqn:7}} \rank{\bxln{3}(1)}{\svector{\bxln{3}}{\lconstant}{\tonumber{\n} \setminus (\bu \setminus \{1\})}}.
            \end{align*}
            Subsequently, according to Lemma \ref{supp:lemma:3}, we have
            \begin{align*}
                \rank{\bxln{3}(1)}{\bxln{3}} = \rank{\bxln{3}(1)}{\svector{\bxln{3}}{\lconstant}{\tonumber{\n} \setminus (\bu \setminus \{1\})}} =  \rank{\byn{1}}{\by}.
            \end{align*}
            Combining this result with \eqref{supp:theorem:2.2:lemma:1:eqn:8}, it follows that
            \begin{align*}
                \rank{\bxln{3}(i)}{\bxln{3}} = \rank{\byn{i}}{\by}, \text{ for any } i \in \bu.
            \end{align*}
            Notice that for any $\iconstant \in \tonumber{\n} \setminus \bu$, we have
            \begin{align*}
                \bxln{3}(i) =^{\eqref{supp:theorem:2.2:lemma:1:eqn:7}} \bxln{2}(i) =^{\eqref{supp:theorem:2.2:lemma:1:eqn:4}} \bxln{1}(i) =^{\eqref{supp:theorem:2.2:lemma:1:eqn:1}} \bxn{i}.
            \end{align*}   		
            Thus, we have found an imputation $\bxln{3} \in \vndistinct$ 
            of $\bx$ for the indices $\bu$ such that
            \begin{align*}
                \bxlnn{3}{i} = \bxn{i}, i \in \tonumber{\n} \setminus \bu, \text{ and }  \rank{\bxlnn{3}{\iconstant}}{\bxln{3}} = \rank{\byn{\iconstant}}{\by} 
                \,\,\text{for all }~\iconstant \in \bu.
            \end{align*}    	
            This proves $\bpn{\kk + 1}$ is true, and completes our proof.
        \end{proof}

        \subsection{Proof of Lemma 2.3}

        This subsection proves Lemma 2.3, which
        is a direct result obtained by applying Lemma~\ref{supp:lemma:4}.

        \begin{lemma} \label{supp:theorem:2.2:lemma:2}
            Suppose $\bx, \by \in \vndistinct$ and $\bu \subset \tonumber{\n}$ is a subset of indices. Let $\bxln{1}, \bxln{2} \in \vndistinct$ both be imputations of $\bx$ for indices $\bu$. Then, if for any $\iconstant \in \bu$,
            \begin{align*}
                \rank{\bxlnn{1}{i}}{\bxln{1}} = \rank{\bxlnn{2}{i}}{\bxln{2}} = \rank{\byn{i}}{\by},
            \end{align*}
            we have $\sfd{\bx_1}{\by} = \sfd{\bx_2}{\by}$.
        \end{lemma}   

        \begin{proof}
            Since $\bxln{1}, \bxln{2} \in \vndistinct$ are both imputations of $\bx$ for $\bu$,
            for any $i \in \tonumber{\n} \setminus \bu$, we have
            \begin{align*}
                \bxlnn{1}{i} = \bxlnn{2}{i} = \bxn{i}.
            \end{align*}
            Therefore, according to the definition of imputations,
            $\bxln{2}$ is an imputation of $\bxln{1}$ for indices $\bu$.	
            Since for any $\iconstant \in \bu$, we have
            \begin{align*}
                \rank{\bxlnn{1}{i}}{\bxln{1}} = \rank{\bxlnn{2}{i}}{\bxln{2}}. 
            \end{align*}
            Then, according to Lemma~\ref{supp:lemma:4}, we have
            \begin{align*}
                \rank{\bxln{1}}{\bxln{1}} = \rank{\bxln{2}}{\bxln{2}}.
            \end{align*}
            Hence, 
            \begin{align*}
                \sfd{\bxln{1}}{\by} = \sfd{\bxln{2}}{\by}.
            \end{align*}
            This completes our proof.
        \end{proof}

        \subsection{Proof of Theorem 2.4} 

        This subsection proves Theorem 2.4:

        \begin{theorem} \label{supp:theorem:2.4}
            Suppose $\bx, \by \in \vndistinct$, and $\bu \subset \tonumber{\n}$ is a subset of indices. Let $\bxs \in \vndistinct$  be
            an imputation of $\bx$ for indices $\bu$ such that $\rank{\bxsn{i}}{\bxs} = \rank{\byn{i}}{\by}$ for any $i \in \bu$. Then, we have $\sfd{\bxs}{\by} \le \sfd{\bx}{\by}$. Moreover, for any other imputation $\bxp \in \vndistinct$ of $\bx$ for $\bu$, we have
            $\sfd{\bxs}{\by} \le \sfd{\xaa}{\by}$.
        \end{theorem}

        \begin{proof}
            Notice that there are two statements in Theorem~\ref{supp:theorem:2.4}:
            \begin{itemize}
                \item[$(\rn{i})$] Suppose $\bx, \by \in \vndistinct$ and  
                    let $\bxs \in \vndistinct$ be 
                    an imputation of $\bx$ for indices $\bu \subset \tonumber{\n}$.
                    If $\rank{\bxsn{i}}{\bxs} = \rank{\byn{i}}{\by}$ for any $\iconstant \in \bu$,
                    then we have $\sfd{\bxs}{\by} \le \sfd{\bx}{\by}$. 
                \item[$(\rn{ii})$] Suppose $\bx, \by \in \vndistinct$ and  
                    let $\bxs \in \vndistinct$ be 
                    an imputation of $\bx$ for indices $\bu \in \tonumber{\n}$.
                    If $\rank{\bxsn{i}}{\bxs} = \rank{\byn{i}}{\by}$,  for any $\iconstant \in \bu$, then
                    for any other imputation $\xaa \in \vndistinct$ of $\bx$ for indices $\bu$, 
                    we have $\sfd{\bxs}{\by} \le \sfd{\xaa}{\by}.$
            \end{itemize}
            Below, we first show that the statement $(\rn{i})$ is true. Then, 
            we prove the statement $(\rn{ii})$ is true
            using the statement~$(\rn{i})$.

            First, we show that the statement $(\rn{i})$ is true.
            Denote $\m = |\bu|$. When $\m = \n$, we have 
            $\rank{\bxs}{\bxs} = \rank{\by}{\by}
            \Rightarrow \sfd{\bxs}{\by} = 0$.
            According to the definition of Spearman's footrule, we have $\sfd{\bx}{\by} \ge 0$.
            Hence, we have
            $\sfd{\bx}{\by} \ge \sfd{\bxs}{\by} = 0$.
            Therefore, we have shown the statement $(\rn{i})$ is true when $\m = \n$. 

            However, when $\m = 0$, then according to the definition of imputations,
            we have $\bx = \bxs$. Then, we have $\sfd{\bxs}{\by} = \sfd{\bx}{\by}$,
            which proves the statement $(\rn{i})$ when $\m = 0$.

            Now, in order to prove the statement $(\rn{i})$, we only need to
            consider cases when $1 \le \m \le \n - 1$. 
            For any fixed $\n \in \mathbb{N}$, let $\bpn{\kk}$ be the statement of
            statement $(\rn{i})$ when $|\bu| = \kk$. 
            We prove $\bpn{\kk}$ holds for any $\kk \in \{1, \ldots, \n-1\}$
            by induction on $\kk$.

            \emph{Base Case:} Notice that $\bpn{1}$ is true according to Proposition~\ref{supp:proposition:1}. 

            \emph{Induction Step: } We show the implication 
            $\bpn{1}, \bpn{\kk} \Rightarrow \bpn{\kk+1}$ for any
            $\kk \in \{1,\ldots, \n -2\}$

            For convenience, let us assume (after relabeling) $\bu = \{1, \cdots, \m\}$, and $\byn{1} < \cdots < \byn{\m}$. According to Lemma~\ref{supp:theorem:2.2:lemma:1}, we can find a vector $\bxln{1} \in \vndistinct$ such that
            \begin{align}
                &\bxln{1}(i) = \bxn{i}, \quad \text{for any } i \in \tonumber{\n} \setminus (\bu \setminus \{1\}), \label{supp:theorem:2.2:eqn:1}\\
                &\rank{\bxln{1}(i)}{\bxln{1}} = \rank{\byn{i}}{\by}, \quad \text{for any } i \in \bu \setminus \{1\}. \label{supp:theorem:2.2:eqn:2}
            \end{align} 
            That is, $\bxln{1}$ is an imputation of $\bx$ for indices $\bu \setminus \{1\}$ such that \eqref{supp:theorem:2.2:eqn:2} holds.
            Notice that $|\bu \setminus \{1\}| = |\bu| - 1 = \kk + 1 - 1 = \kk$. 
            Then since $\bpn{\kk}$ is true, we have 
            \begin{align*}
                \sfd{\bxln{1}}{\by} \le \sfd{\bx}{\by}.
            \end{align*}

            According to Lemma~\ref{supp:theorem:2.2:lemma:1}, we can find a vector $\bxln{2} \in \vndistinct$ such that
            \begin{align}
                &\bxln{2}(i) = \bxln{1}(i), \quad \text{for any } i \in \tonumber{\n} \setminus \{1\}, \label{supp:theorem:2.2:eqn:3}\\
                &\rank{\bxln{2}(1)}{\bxln{2}} = \rank{\byn{1}}{\by}. \label{supp:theorem:2.2:eqn:4}
            \end{align}
            That is, $\bxln{2}$ is an imputation of $\bxln{1}$ for index $1$ such that \eqref{supp:theorem:2.2:eqn:4} holds. Subsequently, since $\bpn{1}$ is true, we have
            \begin{align*}
                \sfd{\bxln{2}}{\by} \le \sfd{\bxln{1}}{\by}.
            \end{align*}
            According to~Lemma \ref{supp:lemma:2}, for any $i \in \tonumber{\n} \setminus \{1\}$, we have
            \begin{align*}
                \begin{split}
                    \rank{\bxln{2}(i)}{\bxln{2}}  \ge \rank{\bxln{2}(i)}{\svector{\bxln{2}}{\lconstant}{\tonumber{\n} \setminus \{1\}}} =^{\eqref{supp:theorem:2.2:eqn:3}} \rank{\bxln{1}(i)}{\svector{\bxln{1}}{\lconstant}{\tonumber{\n} \setminus \{1\}}}.
                \end{split}
            \end{align*}
            Subsequently, applying Lemma~\ref{supp:lemma:2} again, for any $i \in \bu \setminus \{1\}$, we have
            \begin{align*}
                \rank{\bxln{1}(i)}{\svector{\bxln{1}}{\lconstant}{\tonumber{\n} \setminus \{1\}}} \ge \rank{\bxln{1}(i)}{\bxln{1}} - 1 =^{\eqref{supp:theorem:2.2:eqn:2}} \rank{\byn{i}}{\by} - 1.
            \end{align*}
            Since $\byn{1} < \cdots < \byn{\m}$ and $\bu = \{1, \cdots, \m\}$, then for any $i \in \bu \setminus \{1\}$, we have
            \begin{align*}
                &\rank{\bxln{1}(i)}{\svector{\bxln{1}}{\lconstant}{\tonumber{\n} \setminus \{1\}}} \ge \rank{\byn{i}}{\by} - 1 > \rank{\byn{1}}{\by} - 1 \\
                \Rightarrow &\rank{\bxln{1}(i)}{\svector{\bxln{1}}{\lconstant}{\tonumber{\n} \setminus \{1\}}}  \ge \rank{\byn{1}}{\by}.
            \end{align*}
            Hence, for any $i \in \bu \setminus \{1\}$, we have 
            \begin{align*}
                \rank{\bxln{2}(i)}{\bxln{2}} \ge \rank{\bxln{1}(i)}{\svector{\bxln{1}}{\lconstant}{\tonumber{\n} \setminus \{1\}}} \ge \rank{\byn{1}}{\by} =^{\eqref{supp:theorem:2.2:eqn:4}} \rank{\bxln{2}(1)}{\bxln{2}}. 
            \end{align*}
            Since $ \bxln{2} \in \vndistinct$ is a vector of distinct real values, for any $i \in \bu \setminus \{1\}$, we further have
            \begin{align*}
                \rank{\bxln{2}(i)}{\bxln{2}} \neq \rank{\bxln{2}(1)}{\bxln{2}} \Rightarrow \rank{\bxln{2}(i)}{\bxln{2}} > \rank{\bxln{2}(1)}{\bxln{2}} \Rightarrow \bxln{2}(i) > \bxln{2}(1).
            \end{align*}
            Subsequently, according to Lemma \ref{supp:lemma:3}, 
            \begin{align} \label{supp:theorem:2.2:eqn:5}
                \rank{\bxln{2}(1)}{\svector{\bxln{2}}{\iconstant}{\tonumber{\n} \setminus (\bu \setminus \{1\})}} = \rank{\bxln{2}(1)}{\bxln{2}} =^{\eqref{supp:theorem:2.2:eqn:4}} \rank{\byn{1}}{\by}.
            \end{align}

            According to Lemma~\ref{supp:theorem:2.2:lemma:1}, we can find a vector  $\bxln{3} \in \vndistinct$ such that 
            \begin{align}
                &\bxln{3}(i) = \bxln{2}(i), \quad \text{for any } i \in \tonumber{\n} \setminus (\bu \setminus \{1\}), \label{supp:theorem:2.2:eqn:6} \\
                &\rank{\bxln{3}(i)}{\bxln{3}} = \rank{\byn{i}}{\by}, \quad \text{for any } i \in \bu \setminus \{1\}. \label{supp:theorem:2.2:eqn:7}  
            \end{align}
            That is, $\bxln{3}$ is an imputation of $\bxln{2}$ for indices
            $\bu \setminus \{1\}$ such that \eqref{supp:theorem:2.2:eqn:7} holds.
            Notice that $|\bu \setminus \{1\}| = |\bu| - 1 = \kk +1 - 1 = \kk$. 
            Then since $\bpn{\kk}$ is true, we have
            \begin{align*}
                \sfd{\bxln{3}}{\by} \le \sfd{\bxln{2}}{\by}.
            \end{align*}
            Since $\byn{1} < \cdots < \byn{\m}$, we have $\rank{\byn{i}}{\by} > \rank{\byn{1}}{\by}$ for any $i \in \bu \setminus \{1\}$. Hence, for any $i \in \bu \setminus \{1\}$, we have
            \begin{align*}
                \rank{\bxln{3}(i)}{\bxln{3}} =^{\eqref{supp:theorem:2.2:eqn:7}}  \rank{\byn{i}}{\by} > \rank{\byn{1}}{\by} =^{\eqref{supp:theorem:2.2:eqn:5}} \rank{\bxln{2}(1)}{\svector{\bxln{2}}{\iconstant}{\tonumber{\n} \setminus (\bu \setminus \{1\})}}.
            \end{align*}
            Further, according to \eqref{supp:theorem:2.2:eqn:6},
            \begin{align*}
                \rank{\bxln{2}(1)}{\svector{\bxln{2}}{\lconstant}{\tonumber{\n} \setminus (\bu \setminus \{1\})}} = \rank{\bxln{3}(1)}{\svector{\bxln{3}}{\lconstant}{\tonumber{\n} \setminus (\bu \setminus \{1\})}}.
            \end{align*}
            For any $i \in \bu \setminus \{1\}$, we then have
            \begin{align*}
                \rank{\bxln{3}(i)}{\bxln{3}} > \rank{\bxln{2}(1)}{\svector{\bxln{2}}{\lconstant}{\tonumber{\n} \setminus (\bu \setminus \{1\})}} = \rank{\bxln{3}(1)}{\svector{\bxln{3}}{\lconstant}{\tonumber{\n} \setminus (\bu \setminus \{1\})}}.
            \end{align*}
            By applying Lemma~\ref{supp:lemma:3}, we have
            \begin{align*}
                \rank{\bxln{3}(1)}{\bxln{3}} &= \rank{\bxln{3}(1)}{\svector{\bxln{3}}{\lconstant}{\tonumber{\n} \setminus (\bu \setminus \{1\})}} \\
                &=  \rank{\bxln{2}(1)}{\svector{\bxln{2}}{\lconstant}{\tonumber{\n} \setminus (\bu \setminus \{1\})}} =^{\eqref{supp:theorem:2.2:eqn:5}} \rank{\byn{1}}{\by}.
            \end{align*} 
            Combining this result with \eqref{supp:theorem:2.2:eqn:7}, we have
            \begin{align*}
                \rank{\bxln{3}(i)}{\bxln{3}} = \rank{\byn{i}}{\by}, \text{ for any } i \in \bu.
            \end{align*}
            Notice that for any $i \in \tonumber{\n} \setminus \bu$, we have
            \begin{align*}
                \bxln{3}(i) =^{\eqref{supp:theorem:2.2:eqn:6}} \bxln{2}(i) =^{ \eqref{supp:theorem:2.2:eqn:3}} \bxln{1}(i) =^{\eqref{supp:theorem:2.2:eqn:1}} \bxn{i}.
            \end{align*}
            Since $\bxs \in \vndistinct$ is an $\n$-dimensional vector of distinct real values such that
            \begin{align*} 
                &\bxsn{i} = \bxn{i}, \quad \text{for any } i \in \tonumber{\n} \setminus \bu,\\
                &\rank{\bxsn{i}}{\bxs} = \rank{\byn{i}}{\by}, \quad \text{for any } i \in \bu.
            \end{align*}
            Then according to Lemma~\ref{supp:lemma:4}, we have
            \begin{align*}
                \rank{\bxln{3}}{\bxln{3}} = \rank{\bxs}{\bxs} \Rightarrow \sfd{\bxln{3}}{\by} = \sfd{\bxs}{\by}.
            \end{align*}
            Notice that we have
            \begin{align*}
                \sfd{\bxs}{\by} = \sfd{\bxln{3}}{\by} \le \sfd{\bxln{2}}{\by} \le \sfd{\bxln{1}}{\by} \le \sfd{\bx}{\by}.
            \end{align*}
            Hence, we have shown $\bpn{\kk + 1}$ is true, this completes our proof for the statement $(\rn{i})$.

            Next, we show that the statement $(\rn{ii})$ is true using the statement $(\rn{i})$. According to Lemma~\ref{supp:theorem:2.2:lemma:1}, we can find a vector 
            $\bxsp \in \vndistinct$ such that it is an imputation of $\bxp$ for indices 
            $\bu \subset \tonumber{\n}$, and $\rank{\bxspn{i}}{\bxsp} = \rank{\byn{i}}{\by}$ for any $i \in \bu$. 
            Then, according to the statement~$(\rn{i})$, we have 
            \begin{align*}
                \sfd{\bxsp}{\by} \le \sfd{\xaa}{\by}.
            \end{align*}
            Since $\bxs \in \vndistinct$ is an imputation of $\bx$ for the  index $\bu \subset \tonumber{\n}$, and $\rank{\bxsn{i}}{\bxs} = \rank{\byn{i}}{\by}$ for any $i \in \bu$. 
            Then, according to Lemma~\ref{supp:lemma:4}, we have $\rank{\bxsp}{\bxsp} = \rank{\bxs}{\bxs}$, which means $\sfd{\bxsp}{\by} = \sfd{\bxs}{\by}$. Therefore, we have
            \begin{align*}
                \sfd{\bxs}{\by} = \sfd{\bxsp}{\by} \le \sfd{\xaa}{\by}.
            \end{align*}
            This proves the statement $(\rn{ii})$ and completes our proof.
        \end{proof}

        \subsection{Proof of Proposition 2.5}	

        This subsection proves Proposition 2.5. To start, we give the 
        following lemma, which is a direct result following Lemma~\ref{supp:theorem:2.2:lemma:1}.

        \begin{lemma} \label{supp:proposition:existence:lemma:1}
            Suppose $\bx \in \vndistinct$ and $\bu \subset \tonumber{\n}$ is a subset of
            indices. Denote $\m = |\bu|$, and let $(\sigma(\lconstant))_{\lconstant \in \bu}$
            be a vector of any $\m$ distinct values in $\tonumber{\n}$. 
            Then, there exist an imputation $\bxs \in \vndistinct$ of $\bx$ for $\bu$ such that
            $\rank{\bxsn{i}}{\bxs} = \sigma(i), \text{ for any } i \in \bu.$
        \end{lemma}

        \begin{proof}
            Let $\by \in \vndistinct$ be a vector such that 
            $\rank{\byn{i}}{\by} = \sigma(i), \text{ for any } i \in \bu$. 
            Subsequently, according to Lemma~\ref{supp:theorem:2.2:lemma:1}, 
            there exist an imputation $\bxs \in \vndistinct$ of $\bx$ for
            indices $\bu$ such that
            \begin{align*}
                \rank{\bxsn{i}}{\bxs} = \rank{\byn{i}}{\by} = \sigma(i), \text{ for any } i \in \bu.
            \end{align*}
            This completes our proof.
        \end{proof}

        We are now ready to prove Proposition 2.5.
        \begin{proposition} \label{supp:proposition:2.5}
            Suppose $\bx, \by \in \vndistinct$, and $\bu, \bv \subset \tonumber{\n}$ are disjoint subsets of $\tonumber{\n}$ such that $\bu \cap \bv = \emptyset$ and $\bu \cup \bv \neq \emptyset$. Then, there exist imputations $\bxs, \bys \in \vndistinct$ of $\bx$ and $\by$
            for $\bu$ and $\bv$, respectively, such that $\rank{\bxs(i)}{\bxs} = \rank{\bys(i)}{\bys}$ for any $i \in \bu \cup \bv$.
        \end{proposition}

        \begin{proof}
            For any given $\n \in \mathbb{N}$, let $\bpn{\kk}$ be the statement of Proposition~\ref{supp:proposition:2.5} when $|\bu| + |\bv| = \kk$.
            We prove $\bpn{\kk}$ holds for any $\kk \in \{1,\ldots, \n\}$ by
            induction on $\kk$.

            \textit{Base Case:} We show $\bpn{1}$ holds.    
            If $|\bu| + |\bv| = 1$, it is either 
            $\bu = \emptyset, |\bv| = 1$ 
            or $|\bu| = 1, \bv = \emptyset$. 
            Suppose $\bu = \emptyset, |\bv| = 1$. Then, let $\bxs = \bx$. 
            According to the definition of imputations,
            $\bxs \in \vndistinct$ is an imputation of $\bx$ for $\bu = \emptyset$. 
            Further, according to Lemma~\ref{supp:theorem:2.2:lemma:1}, 
            we can find an imputation 
            $\bys \in \vndistinct$ of $\by$ for indices $\bv$ such that 
            $\rank{\bysn{i}}{\bys} = \rank{\bxn{i}}{\bx}$
            for any $i \in \bv$. Notice that $\bxs = \bx$. Hence, we have
            $\rank{\bysn{i}}{\bys} = \rank{\bxsn{i}}{\bxs}$ for any $i \in \bv$.
            Recall that $\bu = \emptyset$. Therefore, we have
            $\rank{\bysn{i}}{\bys} = \rank{\bxsn{i}}{\bxs}$ for any $i \in \bu \cup \bv$.
            Hence, we have shown $\bpn{1}$ is true
            when $\bu = \emptyset, |\bv| = 1$.
            The case when $|\bu| = 1, \bv = \emptyset$ can be proved similarly.

            \textit{Induction Step:} We show the implication
            $\bpn{\kk} \Rightarrow \bpn{\kk + 1}$ for any 
            $\kk \in \{1,\ldots, \n-1\}$. 

            Suppose $|\bu| + |\bv| = \kk + 1$. Then, we have either
            \begin{align*}
                (\rn{i}): |\bu| = 0, |\bv| = \kk + 1 \text{ or } |\bu| = \kk + 1, |\bv| = 0,
            \end{align*}
            or
            \begin{align*}
                (\rn{ii}): |\bu| > 0, |\bv| > 0, |\bu| + |\bv| = \kk + 1,
            \end{align*}
            is true. Below, we consider the case $(\rn{i})$ and the case $(\rn{ii})$ separately.

            First, suppose the case 
            $(\rn{i}): |\bu| = 0, |\bv| = \kk + 1 \text{ or } |\bu| = \kk + 1, |\bv| = 0$ is true. 
            We consider $|\bu| = 0, |\bv| = \kk + 1$. Since $|\bu| = 0$, we have $\bu = \emptyset$.
            Let $\bxs = \bx$. According to the definition of imputations,
            $\bxs \in \vndistinct$ is an imputation of $\bx$ for $\bu = \emptyset$. 
            Further, according to Lemma~\ref{supp:theorem:2.2:lemma:1}, 
            we can find an imputation 
            $\bys \in \vndistinct$ of $\by$ for indices $\bv$ such that 
            $\rank{\bysn{i}}{\bys} = \rank{\bxn{i}}{\bx}$
            for any $i \in \bv$. Notice that $\bxs = \bx$. Hence, we have
            $\rank{\bysn{i}}{\bys} = \rank{\bxsn{i}}{\bxs}$ for any $i \in \bv$.
            Recall that $\bu = \emptyset$. Therefore, we have
            $\rank{\bysn{i}}{\bys} = \rank{\bxsn{i}}{\bxs}$ for any $i \in \bu \cup \bv$.
            Hence, we prove the result
            when $ |\bu| = 0, |\bv| = \kk + 1$. The case when $|\bu| = \kk + 1, |\bv| = 0$ can 
            be proved similarly. Hence, we have shown $\bpn{\kk + 1}$ is true under the case $(\rn{i})$.

            Now, we consider the case 
            $(\rn{ii}): |\bu| > 0, |\bv| > 0, |\bu| + |\bv| = \kk+1$. 	
            Without loss of generality, let us assume (after relabeling) 
            $\bv = \{1,\cdots,\tconstant\}$, 
            $\bu = \{\tconstant+1,\cdots,\tconstant+\m\}$,
            $\bx(1) < \cdots < \bx(\tconstant)$, and
            $\by(\tconstant+1) < \cdots < \by(\tconstant+\m)$.
            Further, let us assume	
            \begin{align*}
                \rank{\bx(1)}{\svector{\bx}{\lconstant}{\tonumber{\n} \setminus \bu}} \le \rank{\by(\tconstant+1)}{\svector{\by}{\lconstant}{\tonumber{\n} \setminus \bv}}.
            \end{align*}
            However, if
            $\rank{\bx(1)}{\svector{\bx}{\lconstant}{\tonumber{\n} \setminus \bu}} > \rank{\by(\tconstant+1)}{\svector{\by}{\lconstant}{\tonumber{\n} \setminus \bv}}$,
            we can then switch the label for $\bx$ and $\by$ and relabel the relevant components.

            Denote
            \begin{align} \label{supp:proposition:2.5:eqn:0}
                \alpha = \rank{\bx(1)}{\svector{\bx}{\lconstant}{\tonumber{\n} \setminus \bu}}.
            \end{align}
            Since $\bx(1) < \cdots < \bx(\tconstant)$ and $\by(\tconstant+1) \cdots < \by(\tconstant+\m)$, we have
            \begin{align}
                &\rank{\bx(1)}{\svector{\bx}{\lconstant}{\tonumber{\n} \setminus \bu}} <  \cdots < \rank{\bx(\tconstant)}{\svector{\bx}{\lconstant}{\tonumber{\n} \setminus \bu}},\label{supp:proposition:2.5:eqn:0.0} \\
                \text{ and }&\rank{\by(\tconstant+1)}{\svector{\by}{\lconstant}{\tonumber{\n} \setminus \bv}} < \cdots < \rank{\by(\tconstant+\m)}{\svector{\by}{\lconstant}{\tonumber{\n} \setminus \bv}}, \label{supp:proposition:2.5:eqn:0.1}
            \end{align}
            respectively. 
            Further, since 
            $\rank{\bx(1)}{\svector{\bx}{\lconstant}{\tonumber{\n} \setminus \bu}} \le \rank{\by(\tconstant+1)}{\svector{\by}{\lconstant}{\tonumber{\n} \setminus \bv}}$, 
            we have
            \begin{align} \label{supp:proposition:2.5:eqn:1}
                \alpha = \rank{\bx(1)}{\svector{\bx}{\lconstant}{\tonumber{\n} \setminus \bu}} \le \rank{\by(i)}{\svector{\by}{\lconstant}{\tonumber{\n} \setminus \bv}}, \text{ for any } i \in \bu.
            \end{align}

            Since $\bx \in \vndistinct$ is a vector of distinct real values,
            then $\{\rank{\bx(i)}{\svector{\bx}{\lconstant}{\tonumber{\n} \setminus \bu}} : i \in \bv\}$ is a set of $\tconstant$ distinct values in $\tonumber{\n}$.
            Subsequently, according to Lemma~\ref{supp:proposition:existence:lemma:1}, 
            there exist $\byln{1} \in \vndistinct$ such that
            \begin{align}
                &\byln{1}(i) = \by(i), \quad i \in \tonumber{\n} \setminus \bv, \label{supp:proposition:2.5:eqn:2}\\
                &\rank{\byln{1}(i)}{\byln{1}} = \rank{\bx(i)}{\svector{\bx}{\lconstant}{\tonumber{\n} \setminus \bu}}, \quad i \in \bv. \label{supp:proposition:2.5:eqn:3}
            \end{align}
            That is, $\byln{1}$ is an imputation of $\by$ for indices $\bv \in \tonumber{\n}$ 
            such that \eqref{supp:proposition:2.5:eqn:3} holds.

            Recall that $\bv = \{1,\cdots,\tconstant\}$.
            Let us assume $\bv \setminus \{1\} \neq \emptyset$, i.e., $\tconstant > 1$.
            Then, for any $i \in \bv \setminus \{1\}$, we have
            \begin{align*}
                &\rank{\byln{1}(i)}{\byln{1}} =^{\eqref{supp:proposition:2.5:eqn:3}} \rank{\bx(i)}{\svector{\bx}{\lconstant}{\tonumber{\n} \setminus \bu}} \\ &>^{\eqref{supp:proposition:2.5:eqn:0.0}} \rank{\bx(1)}{\svector{\bx}{\lconstant}{\tonumber{\n} \setminus \bu}} =^{\eqref{supp:proposition:2.5:eqn:3}} \rank{\byln{1}(1)}{\byln{1}} \\
                \Rightarrow &\byln{1}(i) > \byln{1}(1), \text{ for any } i \in \bv \setminus \{1\}.
            \end{align*}
            Subsequently, by applying Lemma \ref{supp:lemma:3}, we have
            \begin{align*}
                \rank{\byln{1}(1)}{\svector{\byln{1}}{\lconstant}{\tonumber{\n} \setminus (\bv \setminus \{1\})}} = \rank{\byln{1}(1)}{\byln{1}}.
            \end{align*}
            Notice the above equation still holds when $\bv \setminus \{1\} = \emptyset$.
            Recall that $\alpha = \rank{\bx(1)}{\svector{\bx}{\lconstant}{\tonumber{\n} \setminus \bu}}$. Hence, we further have 
            \begin{align} \label{supp:proposition:2.5:eqn:4}
                \begin{split}
                    \rank{\byln{1}(1)}{\svector{\byln{1}}{\lconstant}{\tonumber{\n} \setminus (\bv \setminus \{1\})}} &= \rank{\byln{1}(1)}{\byln{1}} \\
                    &=^{\eqref{supp:proposition:2.5:eqn:3}} \rank{\bx(1)}{\svector{\bx}{\lconstant}{\tonumber{\n} \setminus \bu}} = \alpha.
                \end{split}
            \end{align}

            Since $\bu$ and $\bv$ are non-empty disjoint subsets of $\tonumber{\n}$,
            we have $\bu \subset \tonumber{\n} \setminus \bv$. Hence, \eqref{supp:proposition:2.5:eqn:2} implies 
            $\byln{1}(\iconstant) = \byn{\iconstant}$ for any $\iconstant \in \bu$. 
            Thus, for any $\iconstant \in \bu$, we have
            \begin{align*}
                \rank{\byln{1}(\iconstant)}{\svector{\byln{1}}{\lconstant}{\tonumber{\n} \setminus \bv}} = \rank{\byn{\iconstant}}{\svector{\by}{\lconstant}{\tonumber{\n} \setminus \bv}}.
            \end{align*}
            Since $\tonumber{\n} \setminus \bv \subset \tonumber{\n} \setminus (\bv \setminus \{1\})$, we can apply Lemma \ref{supp:lemma:2} and get that, for any $\iconstant \in \bu$,
            \begin{align*}
                \begin{split}
                    \rank{\byln{1}(\iconstant)}{\svector{\byln{1}}{\lconstant}{\tonumber{\n} \setminus (\bv \setminus \{1\})}} 
                    &\ge \rank{\byln{1}(\iconstant)}{\svector{\byln{1}}{\lconstant}{\tonumber{\n} \setminus \bv}} \\
                    &= \rank{\byn{\iconstant}}{\svector{\by}{\lconstant}{\tonumber{\n} \setminus \bv}}\\
                    &\ge^{\eqref{supp:proposition:2.5:eqn:1}} \rank{\bxln{1}(1)}{\svector{\bxln{1}}{\lconstant}{\tonumber{\n} \setminus \bu}} = \alpha.
                \end{split}
            \end{align*} 
            Recall that $\byln{1} \in \vndistinct$ is a vector of distinct real numbers and $1 \notin \bu = \{\tconstant + 1, \ldots, \tconstant + \m\}$, where $1 \leq \tconstant, \m$. Hence, for any $\iconstant \in \bu$, we have
            \begin{align*}
                \rank{\byln{1}(\iconstant)}{\svector{\byln{1}}{\lconstant}{\tonumber{\n} \setminus (\bv \setminus \{1\})}} 
                \neq \rank{\byln{1}(1)}{\svector{\byln{1}}{\lconstant}{\tonumber{\n} \setminus (\bv \setminus \{1\})}} =^{\eqref{supp:proposition:2.5:eqn:4}} \alpha.
            \end{align*}
            Therefore, for any $\iconstant \in \bu$, we have
            \begin{align} \label{supp:proposition:2.5:eqn:5}
                \rank{\byln{1}(\iconstant)}{\svector{\byln{1}}{\lconstant}{\tonumber{\n} \setminus (\bv \setminus \{1\})}} > \alpha.
            \end{align}

            Since $|\bv \setminus \{1\}| = |\bv| - 1 = \m - 1$ and $|\bu| = \tconstant$,
            we have $|\bv \setminus \{1\}|  + |\bu| = \m - 1 + \tconstant = \kk$.
            Since $\bpn{\kk}$ is true, then there 
            exist $\bxln{2}, \byln{2} \in \vndistinct$ such that
            \begin{align}
                &\bxln{2}(i) = \bx(i), \text{ for any } i \in \tonumber{\n} \setminus \bu, \label{supp:proposition:2.5:eqn:6}\\
                &\byln{2}(i) = \byln{1}(i), \text{ for any } i \in \tonumber{\n} \setminus (\bv \setminus \{1\}), \label{supp:proposition:2.5:eqn:7}\\
                \text{ and } &\rank{\bxln{2}(i)}{\bxln{2}} = \rank{\byln{2}(i)}{\byln{2}}, \text{ for any } i \in \bu \cup (\bv \setminus \{1\}). \label{supp:proposition:2.5:eqn:8}
            \end{align}
            That is, $\bxln{2}, \byln{2}$ are imputations of $\bx$ and $\byln{1}$ for $\bu$ and $\bv \setminus \{1\}$, respectively, such that \eqref{supp:proposition:2.5:eqn:8} holds.
            According to \eqref{supp:proposition:2.5:eqn:7}, 
            we have
            \begin{align}
                \rank{\byln{2}(1)}{\svector{\byln{2}}{\lconstant}{\tonumber{\n} \setminus (\bv \setminus \{1\})}} = \rank{\byln{1}(1)}{\svector{\byln{1}}{\lconstant}{\tonumber{\n} \setminus (\bv \setminus \{1\})}} =^{\eqref{supp:proposition:2.5:eqn:4}} \alpha. \label{supp:proposition:2.5:eqn:9}
            \end{align}
            Since $\bv = \{1,\ldots,\tconstant \}$, we have $\tonumber{\n} \setminus \bv = \{\tconstant + 1, \ldots, \n\}$, and $\tonumber{\n} \setminus (\bv \setminus \{1\}) = \{1, \tconstant+1,\ldots, \n\}$. Then, since $\bu = \{\tconstant+1, \ldots, \tconstant + \m\}$, we have
            $\bu \subset \tonumber{\n} \setminus \bv$, and
            $\bu \subset \tonumber{\n} \setminus (\bv \setminus \{1\})$. 
            Hence, according to \eqref{supp:proposition:2.5:eqn:7}, we have for any $i \in \bu$,
            \begin{align*}
                \rank{\byln{2}(i)}{\svector{\byln{2}}{\lconstant}{\tonumber{\n} \setminus (\bv \setminus \{1\})}} = \rank{\byln{1}(i)}{\svector{\byln{1}}{\lconstant}{\tonumber{\n} \setminus (\bv \setminus \{1\})}} >^{\eqref{supp:proposition:2.5:eqn:5}} \alpha.
            \end{align*}
            According to \eqref{supp:proposition:2.5:eqn:6}, we have
            $\bxln{2}(1) = \bx(1)$, and 
            \begin{align*}
                \rank{\bxln{2}(1)}{\svector{\bxln{2}}{\lconstant}{\tonumber{\n} \setminus \bu}} = \rank{\bx(1)}{\svector{\bx}{\lconstant}{\tonumber{\n} \setminus \bu}} = \alpha.
            \end{align*}
            Hence, for any $i \in \bu$
            \begin{align*}
                \rank{\byln{2}(i)}{\svector{\byln{2}}{\lconstant}{\tonumber{\n} \setminus (\bv \setminus \{1\})}} > \alpha = \rank{\bxln{2}(1)}{\svector{\bxln{2}}{\lconstant}{\tonumber{\n} \setminus \bu}}.
            \end{align*}
            According to Lemma \ref{supp:lemma:2}, for any $i \in \bu$, we have
            \begin{align*}
                \rank{\byln{2}(i)}{\byln{2}} \ge \rank{\byln{2}(i)}{\svector{\byln{2}}{\lconstant}{\tonumber{\n} \setminus (\bv \setminus \{1\})}}, 
            \end{align*}
            which implies that
            \begin{align*}
                \rank{\byln{2}(i)}{\byln{2}} >  \alpha = \rank{\bxln{2}(1)}{\svector{\bxln{2}}{\lconstant}{\tonumber{\n} \setminus \bu}}. 
            \end{align*}
            Subsequently, for any $i \in \bu$, we have
            \begin{align*}
                \rank{\bxln{2}(i)}{\bxln{2}} =^{\eqref{supp:proposition:2.5:eqn:8}} \rank{\byln{2}(i)}{\byln{2}} > \alpha = \rank{\bxln{2}(1)}{\svector{\bxln{2}}{\lconstant}{\tonumber{\n} \setminus \bu}}. 
            \end{align*}	
            According to Lemma \ref{supp:lemma:3}, we then have
            \begin{align} \label{supp:proposition:2.5:eqn:10}
                \begin{split}
                    \rank{\bxln{2}(1)}{\bxln{2}} &= \rank{\bxln{2}(1)}{\svector{\bxln{2}}{\lconstant}{\tonumber{\n} \setminus \bu}} = \alpha.
                \end{split}
            \end{align}

            Since $\bv \subset \tonumber{\n} \setminus \bu$, \eqref{supp:proposition:2.5:eqn:6} gives $\bxln{2}(i) = \bx(i)$ for any $i \in \bv$, and
            \begin{align*}
                \rank{\bxln{2}(i)}{\svector{\bxln{2}}{\lconstant}{\tonumber{\n} \setminus \bu}} = \rank{\bx(i)}{\svector{\bx}{\lconstant}{\tonumber{\n} \setminus \bu}}, \quad i \in \bv.
            \end{align*}
            Recall that $\bv = \{1,\ldots, \tconstant\}$. Let us assume $\bv \setminus \{1\} \neq \emptyset$. Then, for any $i \in \bv \setminus \{1\}$, we have
            \begin{align*}
                \rank{\bxln{2}(i)}{\svector{\bxln{2}}{\lconstant}{\tonumber{\n} \setminus \bu}} = \rank{\bx(i)}{\svector{\bx}{\lconstant}{\tonumber{\n} \setminus \bu}}.
            \end{align*}
            Since $\bx(\tconstant) > \ldots > \bx(1)$, we have
            \begin{align*}
                \rank{\bxln{2}(i)}{\svector{\bxln{2}}{\lconstant}{\tonumber{\n} \setminus \bu}} = \rank{\bx(i)}{\svector{\bx}{\lconstant}{\tonumber{\n} \setminus \bu}} > \rank{\bx(1)}{\svector{\bx}{\lconstant}{\tonumber{\n} \setminus \bu}} =^{\eqref{supp:proposition:2.5:eqn:0}} \alpha.
            \end{align*}
            According to Lemma \ref{supp:lemma:2}, for any $i \in \bv \setminus \{1\}$, we have
            \begin{align*}
                \rank{\bxln{2}(i)}{\bxln{2}} \ge \rank{\bxln{2}(i)}{\svector{\bxln{2}}{\lconstant}{\tonumber{\n} \setminus \bu}} > \alpha. 
            \end{align*}
            Then, for any $i \in \bv \setminus \{1\}$, we have
            \begin{align*}
                \rank{\byln{2}(i)}{\byln{2}} =^{\eqref{supp:proposition:2.5:eqn:8}} \rank{\bxln{2}(i)}{\bxln{2}} > \alpha =^{\eqref{supp:proposition:2.5:eqn:9}} \rank{\byln{2}(1)}{\svector{\byln{2}}{\lconstant}{\tonumber{\n} \setminus (\bv \setminus \{1\})}}.
            \end{align*}
            According to Lemma~\ref{supp:lemma:3}, we have
            \begin{align*}
                \begin{split}
                    \rank{\byln{2}(1)}{\byln{2}} &= \rank{\byln{2}(1)}{\svector{\byln{2}}{\lconstant}{\tonumber{\n} \setminus (\bv \setminus \{1\})}}.
                \end{split}
            \end{align*}
            Notice that the above equation still holds when $\bv \setminus \{1\} = \emptyset$. 

            Subsequently, we have
            \begin{align*}
                \rank{\byln{2}(1)}{\byln{2}} = \rank{\byln{2}(1)}{\svector{\byln{2}}{\lconstant}{\tonumber{\n} \setminus (\bv \setminus \{1\})}} =^{\eqref{supp:proposition:2.5:eqn:9}} \alpha =^{\eqref{supp:proposition:2.5:eqn:10}} \rank{\bxln{2}(1)}{\bxln{2}}.
            \end{align*}
            Combining this result with \eqref{supp:proposition:2.5:eqn:8}, we obtain
            \begin{align} \label{supp:proposition:2.5:eqn:11}
                \rank{\bxln{2}(i)}{\bxln{2}} = \rank{\byln{2}(i)}{\byln{2}}, \text{ for any } i \in \bu \cup \bv.
            \end{align}
            Recall that we have
            \begin{align*}
                \byln{2}(i) =^{\eqref{supp:proposition:2.5:eqn:7} } \byln{1}(i) =^{\eqref{supp:proposition:2.5:eqn:3} } \by(i), \text{ for any } i \in \bv,
            \end{align*}
            and
            \begin{align*}
                \bxln{2}(i) =^{\eqref{supp:proposition:2.5:eqn:6}} \bx(i), \text{ for any } i \in \tonumber{\n} \setminus \bu.
            \end{align*}
            Hence, we have found imputations $\bxln{2}, \byln{2} \in \vndistinct$ 
            of $\bx$ and $\by$ for $\bu$ and $\bv$, respectively, such that
            \eqref{supp:proposition:2.5:eqn:11} holds. Therefore, we have
            shown $\bpn{\kk + 1}$ is true. This proves our results when the 
            case $(\rn{ii})$ holds and
            completes our proof.

        \end{proof}

        \subsection{Proof of Proposition 2.6}	

        This subsection proves Proposition 2.6. First, we prove several lemmas
        which will be useful for showing Proposition 2.6.

        \begin{lemma} \label{supp:proposition:equivalence:lemma:1}
            Suppose $\bxln{1}, \bxln{2} \in \vndistinct$, and $\bo \subset \tonumber{\n}$ is 
            a non-empty subset of 
            indices such that $\bo \neq \tonumber{\n}$.	
            Suppose there exist an index $\kk \in \tonumber{\n} \setminus \bo$ such that
            \begin{align*}
                \rank{\bxln{1}(\kk)}{\svector{\bxln{1}}{\lconstant}{\{\kk\} \cup \bo}} = 
                \rank{\bxln{2}(\kk)}{\svector{\bxln{2}}{\lconstant}{\{\kk\} \cup \bo}},
            \end{align*}
            Then, for any $\kln{1}, \kln{2} \in \bo$, if
            \begin{align*}
                \rank{\bxln{1}(\kln{1})}{\svector{\bxln{1}}{\lconstant}{\bo}} = 
                \rank{\bxln{2}(\kln{2})}{\svector{\bxln{2}}{\lconstant}{\bo}},
            \end{align*}
            we have
            \begin{align*}
                \rank{\bxln{1}(\kln{1})}{\svector{\bxln{1}}{\lconstant}{\{\kk\} \cup \bo}} = 
                \rank{\bxln{2}(\kln{2})}{\svector{\bxln{2}}{\lconstant}{\{\kk\} \cup \bo}}.
            \end{align*}
        \end{lemma}

        \begin{proof}	
            To start, let us denote  
            \begin{align*}
                \alpha = \rank{\bxln{1}(\kk)}{\svector{\bxln{1}}{\lconstant}{\{\kk\} \cup \bo}} = 
                \rank{\bxln{2}(\kk)}{\svector{\bxln{2}}{\lconstant}{\{\kk\} \cup \bo}}.
            \end{align*}
            Suppose $\kln{1}, \kln{2} \in \bo$ are such that 
            \begin{align*}
                \rank{\bxln{1}(\kln{1})}{\svector{\bxln{1}}{\lconstant}{\bo}} = 
                \rank{\bxln{2}(\kln{2})}{\svector{\bxln{2}}{\lconstant}{\bo}}.
            \end{align*}
            Then, we have either
            \begin{align*}
                &\text{case } (\rn{i}): \alpha > \rank{\bxlnn{1} {\kln{1}}}{\svector{\bxln{1}}{\lconstant}{\boconstant}} = 
                \rank{\bxlnn{2}{\kln{2}}}{\svector{\bxln{2}}{\lconstant}{\boconstant}},\\
                \text{ or }&\text{case } (\rn{ii}):  \alpha \le \rank{\bxlnn{1}{\kln{1}}}{\svector{\bxln{1}}{\lconstant}{\boconstant}} = 
                \rank{\bxlnn{2}{\kln{2}}}{\svector{\bxln{2}}{\lconstant}{\boconstant}},
            \end{align*}
            is true. In the following, we prove our result when
            the case $(\rn{i})$ and the case $(\rn{ii})$ are true separately.

            Suppose the case
            \begin{align*}
                (\rn{i}): \alpha > \rank{\bxlnn{1}{\kln{1}}}{\svector{\bxln{1}}{\lconstant}{\boconstant}} = 
                \rank{\bxlnn{2}{\kln{2}}}{\svector{\bxln{2}}{\lconstant}{\boconstant}}
            \end{align*}
            is true. Then it follows
            \begin{align*}
                &\alpha = \rank{\bxlnn{1}{\kk}}{\svector{\bxln{1}}{\lconstant}{\{\kk\} \cup \bo}} > \rank{\bxlnn{1}{\kln{1}}}{\svector{\bxln{1}}{\lconstant}{\boconstant}}, \\
                \text{and }&\alpha = \rank{\bxlnn{2}{\kk}}{\svector{\bxln{2}}{\lconstant}{\{\kk\} \cup \bo}} > \rank{\bxlnn{2}{\kln{2}}}{\svector{\bxln{2}}{\lconstant}{\boconstant}}.
            \end{align*}
            Applying Lemma \ref{supp:lemma:3} to $\svector{\bxln{1}}{\lconstant}{\{\kk\} \cup \bo}$ 
            and $\svector{\bxln{2}}{\lconstant}{\{\kk\} \cup \bo}$ separately,
            we have
            \begin{align*}
                &\rank{\bxlnn{1}{\kln{1}}}{\svector{\bxln{1}}{\lconstant}{\{\kk\} \cup \bo}} = \rank{\bxlnn{1}{\kln{1}}}{\svector{\bxln{1}}{\lconstant}{\boconstant}},\\
                \text{and } &\rank{\bxlnn{2}{\kln{2}}}{\svector{\bxln{2}}{\lconstant}{\{\kk\} \cup \bo} } = \rank{\bxlnn{2}{\kln{2}}}{\svector{\bxln{2}}{\lconstant}{\boconstant}},
            \end{align*}
            respectively. Since $\rank{\bxlnn{1}{\kln{1}}}{\svector{\bxln{1}}{\lconstant}{\boconstant}} = \rank{\bxlnn{2}{\kln{2}}}{\svector{\bxln{2}}{\lconstant}{\boconstant}}$, then we have
            \begin{align*}
                \rank{\bxlnn{1}{\kk}}{\svector{\bxln{1}}{\lconstant}{\{\kk\} \cup \bo} } = \rank{\bxlnn{2}{\kk}}{\svector{\bxln{2}}{\lconstant}{\{\kk\} \cup \bo} },
            \end{align*}
            which proves our result when the case $(\rn{i})$ holds.

            If, however, the case
            \begin{align*}
                (\rn{ii}): \alpha \le \rank{\bxlnn{1}{\kln{1}}}{\svector{\bxln{1}}{\lconstant}{\boconstant}} = 
                \rank{\bxlnn{2}{\kln{2}}}{\svector{\bxln{2}}{\lconstant}{\boconstant}}
            \end{align*}
            is true, we can apply Lemma \ref{supp:lemma:2} and get
            \begin{align*}
                \rank{\bxlnn{1}{\kln{1}}}{\svector{\bxln{1}}{\lconstant}{\{\kk\} \cup \bo} } \ge \rank{\bxlnn{1}{\kln{1}}}{\svector{\bxln{1}}{\lconstant}{\boconstant}} \ge \alpha = \rank{\bxlnn{1}{\kk}}{\svector{\bxln{1}}{\lconstant}{\{\kk\} \cup \bo} }, \\
                \rank{\bxlnn{2}{\kln{2}}}{\svector{\bxln{2}}{\lconstant}{\{\kk\} \cup \bo} } \ge \rank{\bxlnn{2}{\kln{2}}}{\svector{\bxln{2}}{\lconstant}{\boconstant}} \ge \alpha = \rank{\bxlnn{2}{\kk}}{\svector{\bxln{2}}{\lconstant}{\{\kk\} \cup \bo} }.
            \end{align*}
            Since $\bxln{1}$ and $\bxln{2}$ are both vectors of distinct real values and $\kk \neq \kln{1}, \kln{2}$, we have
            \begin{align*}
                &\rank{\bxlnn{1}{\kln{1}}}{\svector{\bxln{1}}{\lconstant}{\{\kk\} \cup \bo} } \neq \rank{\bxlnn{1}{\kk}}{\svector{\bxln{1}}{\lconstant}{\{\kk\} \cup \bo} }\\
                \Rightarrow  &\rank{\bxlnn{1}{\kln{1}}}{\svector{\bxln{1}}{\lconstant}{\{\kk\} \cup \bo} } > \rank{\bxlnn{1}{\kk}}{\svector{\bxln{1}}{\lconstant}{\{\kk\} \cup \bo} } \\
                \Rightarrow  &\bxlnn{1}{\kln{1}} > \bxlnn{1}{\kk},
            \end{align*}
            and
            \begin{align*}
                &\rank{\bxlnn{2}{\kln{2}}}{\svector{\bxln{2}}{\lconstant}{\{\kk\} \cup \bo} } \neq \rank{\bxlnn{2}{\kk}}{\svector{\bxln{2}}{\lconstant}{\{\kk\} \cup \bo} }\\
                \Rightarrow  &\rank{\bxlnn{2}{\kln{2}}}{\svector{\bxln{2}}{\lconstant}{\{\kk\} \cup \bo} } > \rank{\bxlnn{2}{\kk}}{\svector{\bxln{1}}{\lconstant}{\{\kk\} \cup \bo} } \\
                \Rightarrow  &\bxlnn{2}{\kln{2}} > \bxlnn{2}{\kk}.
            \end{align*}
            Subsequently, according to the definition of rank and notice that
            $\bxln{1}, \bxln{2} \in \vndistinct$ are vectors of distinct real values, we 
            can show that
            \begin{align*}
                \rank{\bxlnn{1}{\kln{1}}}{\svector{\bxln{1}}{\lconstant}{\{\kk\} \cup \bo}  } &= \rank{\bxlnn{1}{\kln{1}}}{\svector{\bxln{1}}{\lconstant}{\boconstant}} + \indicator{\bxlnn{1}{\kln{1}} > \bxlnn{1}{\kk}}\\
                &= \rank{\bxlnn{1}{\kln{1}}}{\svector{\bxln{1}}{\lconstant}{\boconstant}} +  1,
            \end{align*}
            and
            \begin{align*}
                \rank{\bxlnn{2}{\kln{2}}}{\svector{\bxln{2}}{\lconstant}{\{\kk\} \cup \bo}  } &= \rank{\bxlnn{2}{\kln{2}}}{\svector{\bxln{2}}{\lconstant}{\boconstant}} + \indicator{\bxlnn{2}{\kln{2}} > \bxlnn{2}{\kk}}\\
                &= \rank{\bxlnn{2}{\kln{2}}}{\svector{\bxln{2}}{\lconstant}{\boconstant}} +  1.
            \end{align*}
            Recall that $\rank{\bxlnn{1}{\kln{1}}}{\svector{\bxln{1}}{\lconstant}{\boconstant}} = \rank{\bxlnn{2}{\kln{2}}}{\svector{\bxln{2}}{\lconstant}{\boconstant}}$. Hence, we have
            \begin{align*}
                \rank{\bxlnn{1}{\kk}}{\svector{\bxln{1}}{\lconstant}{\{\kk\} \cup \bo} } = \rank{\bxlnn{2}{\kk}}{\svector{\bxln{2}}{\lconstant}{\{\kk\} \cup \bo} }.
            \end{align*}
            This proves our results when the case $(\rn{ii})$ is true, and completes our proof.

        \end{proof}

        \begin{lemma} \label{supp:proposition:equivalence:lemma:1.0}
            Suppose $\bxln{1}, \bxln{2} \in \vndistinct$, and $\boln{1}, \boln{2} \subset \tonumber{\n}$
            are non-empty disjoint subset of indices such that $\boln{1}, \boln{2} \neq \emptyset$ 
            and $\boln{1} \cap \boln{2} = \emptyset$.	
            Suppose
            \begin{align*}
                \rank{\bxln{1}(i)}{\svector{\bxln{1}}{\lconstant}{\boln{1} \cup \boln{2}}} = 
                \rank{\bxln{2}(i)}{\svector{\bxln{2}}{\lconstant}{\boln{1} \cup \boln{2}}}, \text{ for any } \iconstant \in \boln{2}
            \end{align*}
            Then, for any $\kln{1}, \kln{2} \in \boln{1}$, if
            \begin{align*}
                \rank{\bxln{1}(\kln{1})}{\svector{\bxln{1}}{\lconstant}{\boln{1}}} = 
                \rank{\bxln{2}(\kln{2})}{\svector{\bxln{2}}{\lconstant}{\boln{1}}},
            \end{align*}
            we have
            \begin{align*}
                \rank{\bxln{1}(\kln{1})}{\svector{\bxln{1}}{\lconstant}{\boln{1} \cup \boln{2}}} = 
                \rank{\bxln{2}(\kln{2})}{\svector{\bxln{2}}{\lconstant}{\boln{1} \cup \boln{2}}}.
            \end{align*}
        \end{lemma}

        \begin{proof}
            For any given $\n \in \mathbb{N}$, let $\bpn{\kk}$ be the statement of Lemma~\ref{supp:proposition:equivalence:lemma:1.0} when $|\boln{2}|= \kk$.
            We prove $\bpn{\kk}$ is true for any $\kk \in \{1,\ldots, \n-1\}$ by
            induction on $\kk$.

            \textit{Base Case:} The case $\bpn{1}$ is shown in Lemma~\ref{supp:proposition:equivalence:lemma:1}.   

            \textit{Induction Step:} We show the implication
            $\bpn{1}, \bpn{\kk} \Rightarrow \bpn{\kk + 1}$ for any 
            $\kk \in \{1,\ldots, \n-2\}$.

            Let us now consider the case when $|\boln{2}| = \kk + 1$. 
            Without loss of generality, let us assume (after relabeling)
            $\boln{2} = \{1, \ldots, \kk + 1\}$.
            Then according to the definition of ranks and notice that
            $\bxln{1}, \bxln{2} \in \vndistinct$ are vectors of distinct real values,
            for any $\iconstant \in \boln{2}$,
            and $\jconstant = 1,2$, we can show that
            \begin{align*}
                \rank{\bxln{\jconstant}(i)}{\svector{\bxln{\jconstant}}{\lconstant}{\boln{1} \cup \big(\boln{2} \setminus \{\kk\} \big) }}
                = \rank{\bxln{\jconstant}(i)}{\svector{\bxln{\jconstant}}{\lconstant}{\boln{1} \cup \boln{2} }} - \indicator{\bxlnn{\jconstant}{\iconstant} > \bxlnn{\jconstant}{\kk}}.
            \end{align*}
            Notice that for any $\iconstant \in \boln{2}$,
            and $\jconstant = 1,2$, we have
            \begin{align*}
                \indicator{\bxlnn{\jconstant}{\iconstant} > \bxlnn{\jconstant}{\kk}} = \indicator{\rank{\bxlnn{\jconstant}{\iconstant}}{\svector{\bxln{\jconstant}}{\lconstant}{\boln{1} \cup \boln{2}}} > \rank{\bxlnn{\jconstant}{\kk}}{\svector{\bxln{\jconstant}}{\lconstant}{\boln{1} \cup \boln{2}}}}.
            \end{align*}
            Hence, for any $\iconstant \in \boln{2}$,
            and $\jconstant = 1,2$, we have
            \begin{align*}
                &\rank{\bxln{\jconstant}(i)}{\svector{\bxln{\jconstant}}{\lconstant}{\boln{1} \cup \big(\boln{2} \setminus \{\kk\} \big) }} \\
                &=  \rank{\bxln{\jconstant}(i)}{\svector{\bxln{\jconstant}}{\lconstant}{\boln{1} \cup \boln{2} }}\\
                & - \indicator{\rank{\bxlnn{\jconstant}{\iconstant}}{\svector{\bxln{\jconstant}}{\lconstant}{\boln{1} \cup \boln{2}}} > \rank{\bxlnn{\jconstant}{\kk}}{\svector{\bxln{\jconstant}}{\lconstant}{\boln{1} \cup \boln{2}}}}.
            \end{align*}
            Since 
            \begin{align*}
                \rank{\bxln{1}(i)}{\svector{\bxln{1}}{\lconstant}{\boln{1} \cup \boln{2}}} = 
                \rank{\bxln{2}(i)}{\svector{\bxln{2}}{\lconstant}{\boln{1} \cup \boln{2}}}, \text{ for any } \iconstant \in \boln{2},
            \end{align*}
            then for any $\iconstant \in \boln{2} \setminus \{\kk\}$, we have
            \begin{align*}
                \rank{\bxln{1}(i)}{\svector{\bxln{1}}{\lconstant}{\boln{1} \cup \big(\boln{2} \setminus \{\kk\} \big) }} = 
                \rank{\bxln{2}(i)}{\svector{\bxln{2}}{\lconstant}{\boln{1} \cup  \big(\boln{2} \setminus \{\kk\} \big) }}.
            \end{align*} 
            Notice that $|\big(\boln{2} \setminus \{\kk\} \big)| = \kk$, and we have 
            $\rank{\bxln{1}(\kln{1})}{\svector{\bxln{1}}{\lconstant}{\boln{1}}} = 
            \rank{\bxln{2}(\kln{2})}{\svector{\bxln{2}}{\lconstant}{\boln{1}}}$. 
            Then, since $\bpn{\kk}$ is true, we have
            \begin{align} \label{supp:proposition:equivalence:lemma:1.0:eqn:1}
                \rank{\bxln{1}(\kln{1})}{\svector{\bxln{1}}{\lconstant}{\boln{1} \cup \big(\boln{2} \setminus \{\kk\} \big) }} = 
                \rank{\bxln{2}(\kln{2})}{\svector{\bxln{2}}{\lconstant}{\boln{1} \cup  \big(\boln{2} \setminus \{\kk\} \big) }}.
            \end{align}

            Next, since 
            \begin{align*}
                \rank{\bxln{1}(i)}{\svector{\bxln{1}}{\lconstant}{\boln{1} \cup \boln{2}}} = 
                \rank{\bxln{2}(i)}{\svector{\bxln{2}}{\lconstant}{\boln{1} \cup \boln{2}}}, \text{ for any } \iconstant \in \boln{2},
            \end{align*}
            we have 
            \begin{align*}
                \rank{\bxln{1}(\kk)}{\svector{\bxln{1}}{\lconstant}{ \left\{\boln{1} \cup \big(\boln{2} \setminus \{\kk\} \big) \right\} \cup \{\kk\} }} = 
                \rank{\bxln{2}(\kk)}{\svector{\bxln{2}}{\lconstant}{\left\{\boln{1} \cup \big(\boln{2} \setminus \{\kk\} \big) \right\} \cup \{\kk\}  }}.
            \end{align*}
            Notice that \eqref{supp:proposition:equivalence:lemma:1.0:eqn:1} is true,
            and $\bpn{1}$ is true. Then, we have
            \begin{align*}
                \rank{\bxln{1}(\kln{1})}{\svector{\bxln{1}}{\lconstant}{\boln{1} \cup \boln{2} }} = 
                \rank{\bxln{2}(\kln{2})}{\svector{\bxln{2}}{\lconstant}{\boln{1} \cup \boln{2} }}.
            \end{align*}
            Hence, we have shown $\bpn{\kk + 1}$ is true. Thus, we complete our proof.

        \end{proof}

        \begin{lemma} \label{supp:proposition:equivalence:lemma:2}
            Suppose $\bxln{1}, \bxln{2} \in \vndistinct$ and 
            $\bo \subset \tonumber{\n}$ is a non-empty subset of indices.
            Let $\boln{1}, \boln{2} \subset \tonumber{\n}$ be 
            non-empty subsets of indices such that 
            $\boln{1} \cup \boln{2} = \bo$, $\boln{1} \cap \boln{2} \neq \emptyset$, and $\boln{1} \setminus \boln{2} \neq \emptyset$. Suppose
            \begin{align} \label{supp:proposition:equivalence:lemma:2:eqn:1}
                \rank{\bxlnn{1}{\iconstant}}{\svector{\bxln{1}}{\lconstant}{\boln{1}}} = \rank{\bxlnn{2}{\iconstant}}{\svector{\bxln{2}}{\lconstant}{\boln{1}}}, \text{ for any } \iconstant \in \boln{1} \cap \boln{2},
            \end{align}
            and $\big(\rank{\bxlnn{1}{\iconstant}}{\svector{\bxln{1}}{\lconstant}{\bo}}\big)_{\iconstant \in \boln{2}}$ is a permutation of $\big(\rank{\bxlnn{2}{\iconstant}}{\svector{\bxln{2}}{\lconstant}{\bo}}\big)_{\iconstant \in \boln{2}}$. 
            Then, for any $\kk \in \boln{1} \setminus \boln{2}$, if
            \begin{align} \label{supp:proposition:equivalence:lemma:2:eqn:3}
                \rank{\bxlnn{1}{\kk}}{\svector{\bxln{1}}{\lconstant}{\boln{1}}} = \rank{\bxlnn{2}{\kk}}{\svector{\bxln{2}}{\lconstant}{\boln{1}}},
            \end{align}
            we have
            \begin{align*}
                \rank{\bxlnn{1}{\kk}}{\svector{\bxln{1}}{\lconstant}{\bo}} = \rank{\bxlnn{2}{\kk}}{\svector{\bxln{2}}{\lconstant}{\bo}}.
            \end{align*}
        \end{lemma}

        \begin{proof}
            To start, since $\rank{\bxlnn{1}{\iconstant}}{\svector{\bxln{1}}{\lconstant}{\boln{1}}} = \rank{\bxlnn{2}{\iconstant}}{\svector{\bxln{2}}{\lconstant}{\boln{1}}}$ 
            for any $\iconstant \in \boln{1} \cap \boln{2}$, and $\rank{\bxlnn{1}{\kk}}{\svector{\bxln{1}}{\lconstant}{\boln{1}}} = \rank{\bxlnn{2}{\kk}}{\svector{\bxln{2}}{\lconstant}{\boln{1}}}$, 
            we have
            \begin{align*}
                &\sum_{\iconstant \in \boln{1} \cap \boln{2}} \indicator{\rank{\bxlnn{1}{\iconstant}}{\svector{\bxln{1}}{\lconstant}{\boln{1}}} < \rank{\bxlnn{1}{\kk}}{\svector{\bxln{1}}{\lconstant}{\boln{1}}}} \\
                &= \sum_{\iconstant \in \boln{1} \cap \boln{2}} \indicator{\rank{\bxlnn{2}{\iconstant}}{\svector{\bxln{2}}{\lconstant}{\boln{1}}} < \rank{\bxlnn{2}{\kk}}{\svector{\bxln{2}}{\lconstant}{\boln{1}}}}.
            \end{align*}
            Notice that for any $\iconstant \in \boln{1} \cap \boln{2}$, we have
            \begin{align*}
                &\indicator{\rank{\bxlnn{1}{\iconstant}}{\svector{\bxln{1}}{\lconstant}{\boln{1}}} < \rank{\bxlnn{1}{\kln{1}}}{\svector{\bxln{1}}{\lconstant}{\boln{1}}}} = \indicator{\bxlnn{1}{\iconstant} < \bxlnn{1}{\kln{1}}}, \\
                \text{ and }&\indicator{\rank{\bxlnn{2}{\iconstant}}{\svector{\bxln{2}}{\lconstant}{\boln{1}}} < \rank{\bxlnn{2}{\kln{1}}}{\svector{\bxln{2}}{\lconstant}{\boln{1}}}} = \indicator{\bxlnn{2}{\iconstant} < \bxlnn{2}{\kln{1}}}.
            \end{align*}
            Thus, we have
            \begin{align*}
                \sum_{\iconstant \in \boln{1} \cap \boln{2}} \indicator{\bxlnn{1}{\iconstant} < \bxlnn{1}{\kk}} = \sum_{\iconstant \in \boln{1} \cap \boln{2}} \indicator{\bxlnn{2}{\iconstant} < \bxlnn{2}{\kk}}.
            \end{align*}
            Then, since $\bxln{1}, \bxln{2} \in \vndistinct$ are both vectors of distinct real values and $\kk \notin \boln{1} \cap \boln{2}$, we have
            \begin{align} \label{supp:proposition:equivalence:lemma:2:eqn:4}
                \sum_{\iconstant \in \boln{1} \cap \boln{2}} \indicator{\bxlnn{1}{\iconstant} \le \bxlnn{1}{\kk}} = \sum_{\iconstant \in \boln{1} \cap \boln{2}} \indicator{\bxlnn{2}{\iconstant} \le \bxlnn{2}{\kk}}.
            \end{align}    
            Subsequently, according to the definition of rank, we have
            \begin{align*}
                \rank{\bxlnn{1}{\kk}}{\svector{\bxln{1}}{\lconstant}{\boln{1} \setminus \boln{2}}} &= \sum_{\lconstant \in \boln{1} \setminus \boln{2}} \indicator{\bxlnn{1}{\lconstant} \le \bxlnn{1}{\kk }} \\
                &= \sum_{\lconstant \in \boln{1}} \indicator{\bxlnn{1}{\lconstant} \le \bxlnn{1}{\kk}} - \sum_{\lconstant \in \boln{1} \cap \boln{2}} \indicator{\bxlnn{1}{\lconstant} \le \bxlnn{1}{\kk}} \\
                &= \rank{\bxlnn{1}{\kk}}{\svector{\bxln{1}}{\lconstant}{\boln{1}}} - \sum_{\lconstant \in \boln{1} \cap \boln{2}} \indicator{\bxlnn{1}{\lconstant} \le \bxlnn{1}{\kk}}.
            \end{align*}
            According to \eqref{supp:proposition:equivalence:lemma:2:eqn:3} and \eqref{supp:proposition:equivalence:lemma:2:eqn:4}, we further have
            \begin{align}
                \rank{\bxlnn{1}{\kk}}{\svector{\bxln{1}}{\lconstant}{\boln{1} \setminus \boln{2}}} &=^{\eqref{supp:proposition:equivalence:lemma:2:eqn:3}} \rank{\bxlnn{2}{\kk}}{\svector{\bxln{2}}{\lconstant}{\boln{1}}} - \sum_{\lconstant \in \boln{1} \cap \boln{2}} \indicator{\bxlnn{1}{\lconstant} \le \bxlnn{1}{\kk}} \nonumber \\
                &=^{\eqref{supp:proposition:equivalence:lemma:2:eqn:4}} \rank{\bxlnn{2}{\kk}}{\svector{\bxln{2}}{\lconstant}{\boln{1}}} - \sum_{\lconstant \in \boln{1} \cap \boln{2}} \indicator{\bxlnn{2}{\lconstant} \le \bxlnn{2}{\kk}} \nonumber \\
                &= \sum_{\lconstant \in \boln{1}} \indicator{\bxlnn{2}{\lconstant} \le \bxlnn{2}{\kk}} - \sum_{\lconstant \in \boln{1} \cap \boln{2}} \indicator{\bxlnn{2}{\lconstant} \le \bxlnn{2}{\kk}} \nonumber \\
                &= \sum_{\lconstant \in \boln{1} \setminus \boln{2}} \indicator{\bxlnn{2}{\lconstant} \le \bxlnn{2}{\kk}} \nonumber \\
                &= \rank{\bxlnn{2}{\kk}}{\svector{\bxln{2}}{\lconstant}{\boln{1} \setminus \boln{2}}}. \label{supp:proposition:equivalence:lemma:2:eqn:4.0}
            \end{align}		

            Since $\big(\rank{\bxlnn{1}{\iconstant}}{\svector{\bxln{1}}{\lconstant}{\bo}}\big)_{\iconstant \in \boln{2}}$
            is a permutation of 
            $\big(\rank{\bxlnn{2}{\iconstant}}{\svector{\bxln{2}}{\lconstant}{\bo}}\big)_{\iconstant \in \boln{2}}$,
            then there exist a permutation $\svector{\sigma}{i}{\boln{2}}$ of $\boln{2}$
            such that
            \begin{align} \label{supp:proposition:equivalence:lemma:2:eqn:5}
                \rank{\bxlnn{1}{\sigma(\iconstant)}}{\svector{\bxln{1}}{\lconstant}{\bo}} = \rank{\bxlnn{2}{i}}{\svector{\bxln{2}}{\lconstant}{\bo}}, \text{ for any } i \in \boln{2}.
            \end{align}
            Let $\bxln{3} \in \vndistinct$ be a permutation of $\bxln{1}$ such that
            \begin{align*}
                &\bxlnn{3}{\iconstant} = \bxlnn{1}{\iconstant}, \text{ for any } i \in \tonumber{\n} \setminus \boln{2},\\
                \text{and }&\bxlnn{3}{\iconstant} = \bxlnn{1}{\sigma(\iconstant)}, \text{ for any } i \in \boln{2}.
            \end{align*}
            Notice that $\boln{1} \setminus \boln{2} \subset \tonumber{\n} \setminus \boln{2}$, hence we have
            $\svector{\bxln{1}}{\lconstant}{\boln{1} \setminus \boln{2}} = \svector{\bxln{3}}{\lconstant}{\boln{1} \setminus \boln{2}}$.
            Next, since $\kk \in \boln{1} \setminus \boln{2}$, we have $\bxlnn{1}{\kk} = \bxlnn{3}{\kk}$.
            Hence, we have
            \begin{align}  \label{supp:proposition:equivalence:lemma:2:eqn:5.0}
                \rank{\bxlnn{3}{\kk}}{\svector{\bxln{3}}{\lconstant}{\boln{1} \setminus \boln{2}}} = 	\rank{\bxlnn{1}{\kk}}{\svector{\bxln{1}}{\lconstant}{\boln{1} \setminus \boln{2}}}.
            \end{align}
            According to \eqref{supp:proposition:equivalence:lemma:2:eqn:4.0}, we further have
            \begin{align} \label{supp:proposition:equivalence:lemma:2:eqn:6}
                \rank{\bxlnn{3}{\kk}}{\svector{\bxln{3}}{\lconstant}{\boln{1} \setminus \boln{2}}} = 	\rank{\bxlnn{2}{\kk}}{\svector{\bxln{2}}{\lconstant}{\boln{1} \setminus \boln{2}}}.
            \end{align}

            Next, notice that $\bo = \boln{2} \cup (\bo \setminus \boln{2})$, 
            where $\bo \setminus \boln{2} \subset \tonumber{\n} \setminus \boln{2}$.
            Then, since $\bxlnn{3}{\iconstant} = \bxlnn{1}{\sigma(\iconstant)}, \text{ for any } i \in \boln{2}, \text{ and }\bxlnn{3}{\iconstant} = \bxlnn{1}{\iconstant}, \text{ for any } i \in \tonumber{\n} \setminus \boln{2}$, we have 
            $\svector{\bxln{1}}{\lconstant}{\bo}$ 
            is a permutation of $\svector{\bxln{3}}{\lconstant}{\bo}$. 
            Hence, for any $i \in \boln{2}$, we have
            \begin{align*}
                \rank{\bxlnn{3}{\iconstant}}{\svector{\bxln{3}}{\lconstant}{\bo}} = \rank{\bxlnn{1}{\sigma(i)}}{\svector{\bxln{1}}{\lconstant}{\bo}}.
            \end{align*}
            According to \eqref{supp:proposition:equivalence:lemma:2:eqn:3}, we further have
            \begin{align} \label{supp:proposition:equivalence:lemma:2:eqn:7}
                \rank{\bxlnn{3}{\iconstant}}{\svector{\bxln{1}}{\lconstant}{\bo}} = \rank{\bxlnn{2}{i}}{\svector{\bxln{2}}{\lconstant}{\bo}}, \text{ for any } i \in \boln{2}.
            \end{align}

            Notice that \eqref{supp:proposition:equivalence:lemma:2:eqn:6} 
            and \eqref{supp:proposition:equivalence:lemma:2:eqn:7} is true.
            Then, according to Lemma~\ref{supp:proposition:equivalence:lemma:1.0},
            we have
            \begin{align*}
                \rank{\bxlnn{3}{\kk}}{\svector{\bxln{3}}{\lconstant}{\bo}} = 	\rank{\bxlnn{2}{\kk}}{\svector{\bxln{2}}{\lconstant}{\bo}}.
            \end{align*}
            Since $\svector{\bxln{1}}{\lconstant}{\bo}$ 
            is a permutation of $\svector{\bxln{3}}{\lconstant}{\bo}$, and 
            $\bxlnn{1}{\kk} = \bxlnn{3}{\kk}$, we have 
            \begin{align*}
                \rank{\bxlnn{3}{\kk}}{\svector{\bxln{3}}{\lconstant}{\bo}} = 	\rank{\bxlnn{1}{\kk}}{\svector{\bxln{1}}{\lconstant}{\bo}}.
            \end{align*}
            Thus,
            \begin{align*}
                \rank{\bxlnn{1}{\kk}}{\svector{\bxln{1}}{\lconstant}{\bo}} =^{\eqref{supp:proposition:equivalence:lemma:2:eqn:5.0}}
                \rank{\bxlnn{2}{\kk}}{\svector{\bxln{2}}{\lconstant}{\bo}}.
            \end{align*}
            This completes our proof.

        \end{proof}

        \begin{lemma} \label{supp:proposition:equivalence:lemma:3.0}
            Suppose $\bx, \by \in \vndistinct$, and let 
            $\bv = \{1,\cdots,\tconstant\}$, 
            $\bu = \{\tconstant+1,\cdots,\tconstant+\m\}$
            be subsets of indices, 
            where $\tconstant,\m \ge 1$ and $\tconstant + \m \le \n-1$. 
            Suppose $\bxn{\tconstant} > \ldots > \bxn{1}$, $\byn{\tconstant+\m} > \ldots > \byn{\tconstant+1}$ and
            $\rank{\byn{\tconstant+1}}{\svector{\by}{\lconstant}{\tonumber{\n}\setminus \bv}} > \rank{\bxn{1}}{\svector{\bx}{\lconstant}{\tonumber{\n}\setminus \bu}}$
            Then, if
            $\rank{\bxn{\iconstant}}{\bx} = \rank{\byn{\iconstant}}{\by}$ for any $\iconstant \in \bu \cup \bv$, we have
            $\rank{\bxn{1}}{\svector{\bx}{\lconstant}{\tonumber{\n}\setminus \bu}} = \rank{\bxn{1}}{\bx} = \rank{\byn{1}}{\by}$.
        \end{lemma}

        \begin{proof}
            To start, according to Lemma~\ref{supp:lemma:2}, 
            for any $\iconstant \in \bu \cup \bv$, we have
            \begin{align*}
                \rank{\bxn{\iconstant}}{\bx} =  \rank{\byn{\iconstant}}{\by} \ge \rank{\byn{\iconstant}}{\svector{\by}{\lconstant}{\tonumber{\n} \setminus \bv}}.
            \end{align*}
            Since it is assumed that $\byn{\tconstant+\m} > \ldots > \byn{\tconstant+1}$, 
            then for any $\iconstant \in \bu = \{\tconstant+1,\ldots,\tconstant+\m\}$, 
            we have
            \begin{align*}
                \rank{\byn{\tconstant+\m}}{\svector{\by}{\lconstant}{\tonumber{\n} \setminus \bv}} > \ldots > \rank{\byn{\tconstant + 1}}{\svector{\by}{\lconstant}{\tonumber{\n} \setminus \bv}}.
            \end{align*}
            Hence, for any $i \in \bu$, we have
            \begin{align*}
                \rank{\bxn{\iconstant}}{\bx} =  \rank{\byn{\iconstant}}{\by} \ge \rank{\byn{\iconstant}}{\svector{\by}{\lconstant}{\tonumber{\n} \setminus \bv}} \ge \rank{\byn{\tconstant+1}}{\svector{\by}{\lconstant}{\tonumber{\n} \setminus \bv}}.
            \end{align*}
            Since 
            \begin{align*}
                \rank{\byn{\tconstant+1}}{\svector{\by}{\lconstant}{\tonumber{\n}\setminus \bv}} > \rank{\bxn{1}}{\svector{\bx}{\lconstant}{\tonumber{\n}\setminus \bu}},
            \end{align*}
            then for any $\iconstant \in \bu$, we have
            \begin{align*}
                \rank{\bxn{\iconstant}}{\bx} > \rank{\bxn{1}}{\svector{\bx}{\lconstant}{\tonumber{\n} \setminus \bu}}.
            \end{align*}
            According to Lemma~\ref{supp:lemma:3}, we further have
            \begin{align*}
                \rank{\bxn{1}}{\svector{\bx}{\lconstant}{\tonumber{\n} \setminus \bu}} = \rank{\bxn{1}}{\bx}.
            \end{align*}
            Since for any $\iconstant \in \bu \cup \bv$, $\rank{\bxn{\iconstant}}{\bx} = \rank{\byn{\iconstant}}{\by}$, we have 
            \begin{align*}
                \rank{\bxn{1}}{\svector{\bx}{\lconstant}{\tonumber{\n} \setminus \bu}} = \rank{\bxn{1}}{\bx} = \rank{\byn{1}}{\by}.
            \end{align*}  
            This completes our proof.
        \end{proof}

        \begin{lemma} \label{supp:proposition:equivalence:lemma:3}
            Suppose $\bx, \by \in \vndistinct$, and let 
            $\bv = \{1,\cdots,\tconstant\}$, 
            $\bu = \{\tconstant+1,\cdots,\tconstant+\m\}$
            be subsets of indices, 
            where $\tconstant,\m \ge 1$ and $\tconstant + \m \le \n-1$. 
            Suppose $\bxn{\tconstant} > \ldots > \bxn{1}$, $\byn{\tconstant+\m} > \ldots > \byn{\tconstant+1}$ and
            $\rank{\byn{\tconstant+1}}{\svector{\by}{\lconstant}{\tonumber{\n}\setminus \bv}} = \rank{\bxn{1}}{\svector{\bx}{\lconstant}{\tonumber{\n}\setminus \bu}}$.
            Denote 
            \begin{align} \label{supp:proposition:equivalence:lemma:3:eqn:0}
                \alpha = \rank{\byn{\tconstant+1}}{\svector{\by}{\lconstant}{\tonumber{\n}\setminus \bv}} = \rank{\bxn{1}}{\svector{\bx}{\lconstant}{\tonumber{\n}\setminus \bu}}.
            \end{align}	
            Define $\aconstant$ and $\bconstant$ as
            \begin{align*}
                &\aconstant = \max \{\iconstant \in \{0,\cdots,\tconstant-1\} \mid \rank{\bxn{1+\iconstant}}{\svector{\bx}{\lconstant}{\tonumber{\n}\setminus \bu}} = \alpha + \iconstant\},\\
                \text{and }&\bconstant = \max \{\iconstant \in \{0,\cdots,\m-1\} \mid \rank{\byn{\tconstant+1+\iconstant}}{\svector{\by}{\lconstant}{\tonumber{\n}\setminus \bv}} = \alpha + \iconstant\},
            \end{align*}
            respectively. Then, if $\rank{\bxn{\iconstant}}{\bx} = \rank{\byn{\iconstant}}{\by}$, for any  $\iconstant \in \bu \cup \bv$,
            we have
            \begin{align}
                &\alpha + \iconstant \le \rank{\bxn{1+\iconstant}}{\bx} \le \alpha + \bconstant + \iconstant + 1, \quad \iconstant \in \{0,\cdots,\aconstant\}, \label{supp:proposition:equivalence:lemma:3:eqn:2}\\
                \text{and }&\alpha + \iconstant \le \rank{\byn{\tconstant+1+\iconstant}}{\by} \le \alpha + \aconstant + \iconstant + 1, \quad \iconstant \in \{0,\cdots,\bconstant\}. \label{supp:proposition:equivalence:lemma:3:eqn:3}
            \end{align}
        \end{lemma}
        \begin{proof}
            We only prove inequality \eqref{supp:proposition:equivalence:lemma:3:eqn:3}
            is true, since inequality \eqref{supp:proposition:equivalence:lemma:3:eqn:2} 
            can be proved similarly, or be proved in the same way after switching the labels between  
            $\bx$ and $\by$, and relabeling the relevant components of the data.

            According to the definitions of $\aconstant$ and $\bconstant$, we have 
            \begin{align}
                &\rank{\bxn{1+\iconstant}}{\svector{\bx}{\lconstant}{\tonumber{\n}\setminus \bu}} = \alpha + \iconstant, \text{ for any } \iconstant \in \{0,\cdots,\aconstant\}, \label{supp:proposition:equivalence:lemma:3:eqn:4}\\
                \text{and }&\rank{\byn{\tconstant+1+\iconstant}}{\svector{\by}{\lconstant}{\tonumber{\n}\setminus \bv}} = \alpha + \iconstant, \text{ for any } \iconstant \in \{0,\cdots,\bconstant\}. \label{supp:proposition:equivalence:lemma:3:eqn:5}
            \end{align}
            Subsequently, according to Lemma~\ref{supp:lemma:2},
            we have
            \begin{align*}
                &\rank{\byn{\tconstant+1+\iconstant}}{\by} \ge \rank{\byn{\tconstant+1+\iconstant}}{\svector{\by}{\lconstant}{\tonumber{\n}\setminus \bv}} = \alpha +\iconstant, \text{ for any } \iconstant \in \{0, \cdots,\bconstant\},
            \end{align*}
            which proves the left-hand side of \eqref{supp:proposition:equivalence:lemma:3:eqn:3}.

            In the following, we prove the right-hand side of \eqref{supp:proposition:equivalence:lemma:3:eqn:3} is true. 

            To start, let us consider the case when 
            $\aconstant = \tconstant-1$. 
            According to Lemma~\ref{supp:lemma:2}, for any $\iconstant \in \{0,\cdots,\bconstant\}$, we have
            \begin{align*}
                \begin{split}
                    \rank{\byn{\tconstant+1+\iconstant}}{\by} &\le  \tconstant + \rank{\byn{\tconstant+1+\iconstant}}{\svector{\by}{\lconstant}{\tonumber{\n}\setminus \bv}}.
                \end{split}
            \end{align*}
            Then, we have
            \begin{align*}
                \rank{\byn{\tconstant+1+\iconstant}}{\by} \le^{ \eqref{supp:proposition:equivalence:lemma:3:eqn:5}} \tconstant + \alpha + \iconstant = \alpha + \iconstant + \aconstant + 1.
            \end{align*}
            Hence, we have shown that the right-hand side of \eqref{supp:proposition:equivalence:lemma:3:eqn:3} 
            is true when $\aconstant = \tconstant - 1$.

            Now we consider the cases when $\aconstant < \tconstant-1$. 
            For any fixed non-negative integer $b \ge 0$, let $\bpn{\kk}$ be the statement that 
            \begin{align*}
                \bpn{\kk}: \rank{\byn{\tconstant+1+\iconstant}}{\by} \le \alpha + \aconstant + \iconstant + 1 \text{ for any } \iconstant \le \kk.
            \end{align*}
            We prove $\bpn{\kk}$ holds 
            for any $\kk \in \{0,\ldots, \bconstant\}$ by induction 
            on $\kk$.

            \textit{Base Case:} We prove $\bpn{0}$ holds.
            Since $\aconstant < \tconstant - 1$, it follows 
            $\aconstant + 1 < \tconstant$ and 
            $\aconstant+2 \le \tconstant$. 
            According to the definition of $\aconstant$,
            we have
            \begin{align*}
                \rank{\bxn{\aconstant+1}}{\svector{\bx}{\lconstant}{\tonumber{\n}\setminus \bu}} = \alpha + \aconstant,
            \end{align*}
            and 
            \begin{align*}
                \rank{\bxn{\aconstant+2}}{\svector{\bx}{\lconstant}{\tonumber{\n}\setminus \bu}} \neq \alpha + \aconstant + 1.
            \end{align*}
            Since $\aconstant + 2 \le \tconstant$ and $\bxn{\tconstant} > \ldots > \bxn{1}$, we have
            \begin{align*}
                \rank{\bxn{\aconstant+2}}{\svector{\bx}{\lconstant}{\tonumber{\n}\setminus \bu}} > \alpha + \aconstant.
            \end{align*}
            Thus, we have
            \begin{align} \label{supp:proposition:equivalence:lemma:3:eqn:6}
                \rank{\bxn{\aconstant+2}}{\svector{\bx}{\lconstant}{\tonumber{\n}\setminus \bu}} > \alpha + \aconstant + 1.
            \end{align}

            Since $\aconstant+2 \le \tconstant$, we have $\{\aconstant+2,\cdots,\tconstant\} \neq \emptyset$, and $\{\aconstant+2,\cdots,\tconstant\} \subset \bv$.
            Then, since $\rank{\bxn{\iconstant}}{\bx} = \rank{\byn{\iconstant}}{\by}$, 
            for any  $\iconstant \in \bu \cup \bv$,
            we have
            \begin{align*}
                \rank{\bxn{\iconstant}}{\bx} = \rank{\byn{\iconstant}}{\by}, \text{ for any } \iconstant \in \{\aconstant+2,\cdots,\tconstant\}.
            \end{align*} 
            According to Lemma~\ref{supp:lemma:2}, 
            for any $\iconstant \in \{\aconstant+2,\cdots,\tconstant\}$,
            we have
            \begin{align*}
                \begin{split}
                    \rank{\bxn{\iconstant}}{\bx} \ge \rank{\bxn{\iconstant}}{\svector{\bx}{\lconstant}{\tonumber{\n}\setminus \bu}}.
                \end{split}
            \end{align*}
            Hence,  for any $\iconstant \in \{\aconstant+2,\cdots,\tconstant\}$, 
            it follows 
            \begin{align*}
                \begin{split}
                    \rank{\byn{\iconstant}}{\by} = \rank{\bxn{\iconstant}}{\bx} \ge \rank{\bxn{\iconstant}}{\svector{\bx}{\lconstant}{\tonumber{\n}\setminus \bu}}.
                \end{split}
            \end{align*}
            Since $ \bxn{\tconstant} > \ldots > \bxn{1}$, we further have 
            for any $\iconstant \in \{\aconstant+2,\cdots,\tconstant\}$, 
            \begin{align*}
                \rank{\bxn{\iconstant}}{\svector{\bx}{\lconstant}{\tonumber{\n}\setminus \bu}} \ge \rank{\bxn{\aconstant+2}}{\svector{\bx}{\lconstant}{\tonumber{\n}\setminus \bu}}.
            \end{align*}
            Therefore, for any $\iconstant \in \{\aconstant+2,\cdots,\tconstant\}$, 
            we have
            \begin{align*}
                \rank{\byn{\iconstant}}{\by} \ge \rank{\bxn{\iconstant}}{\svector{\bx}{\lconstant}{\tonumber{\n}\setminus \bu}} \ge \rank{\bxn{\aconstant+2}}{\svector{\bx}{\lconstant}{\tonumber{\n}\setminus \bu}} >^{\eqref{supp:proposition:equivalence:lemma:3:eqn:6}} \alpha + \aconstant + 1.
            \end{align*}
            According to \eqref{supp:proposition:equivalence:lemma:3:eqn:2},
            we have
            \begin{align*}
                \alpha + \aconstant + 1
                = \rank{\byn{\tconstant+1}}{\svector{\by}{\lconstant}{\tonumber{\n}\setminus \bv}} + \aconstant + 1. 
            \end{align*}
            According to Lemma~\ref{supp:lemma:2}, we have
            \begin{align*}
                \begin{split}
                    \rank{\byn{\tconstant+1}}{\svector{\by}{\lconstant}{\tonumber{\n}\setminus \bv}} + \aconstant + 1
                    \ge \rank{\byn{\tconstant+1}}{\svector{\by}{\lconstant}{ \{1,\ldots,\aconstant+1\}  \cup \{\tconstant+1,\ldots,\n\}} }  .
                \end{split}
            \end{align*}
            Hence, for any $\iconstant \in \{\aconstant+2,\cdots,\tconstant\}$,
            \begin{align*}
                \rank{\byn{\iconstant}}{\by} > \alpha + \aconstant + 1 \ge  \rank{\byn{\tconstant+1}}{\svector{\by}{\lconstant}{ \{1,\ldots,\aconstant+1\}  \cup \{\tconstant+1,\ldots,\n\}} }  .
            \end{align*}
            Subsequently, according to Lemma~\ref{supp:lemma:3}, 
            we have
            \begin{align*}
                \rank{\byn{\tconstant+1}}{\by} = \rank{\byn{\tconstant+1}}{\svector{\by}{\lconstant}{ \{1,\ldots,\aconstant+1\}  \cup \{\tconstant+1,\ldots,\n\}} } \le \alpha + \aconstant + 1,
            \end{align*}
            which proves $\bpn{0}$ holds.

            \textit{Induction Step:} We show the implication
            $\bpn{\kk} \Rightarrow \bpn{\kk + 1}$ for any 
            $\kk \in \{0,\ldots, \bconstant-1\}$.

            Since $\bpn{\kk}$ is true, we have
            \begin{align*}
                \rank{\byn{\tconstant+1+\iconstant}}{\by} \le \alpha + \iconstant + \aconstant + 1, \quad \iconstant \in \{0,\cdots,\kk\}.
            \end{align*}
            Notice that $0 \le \kk \le \bconstant - 1 \le \m - 2$. We have 
            $\{\tconstant+1, \ldots, \tconstant +\kk + 1\} \subset \{\tconstant+1, \ldots, \tconstant +\m - 1\} \subset \bu$.
            Since $\rank{\bxn{\iconstant}}{\bx} = \rank{\byn{\iconstant}}{\by}$ for any  $\iconstant \in \bu \cup \bv$,  
            we have for any $\iconstant \in \{0,\cdots,\kk\}$,
            \begin{align*}
                \rank{\bxn{\tconstant+1+\iconstant}}{\bx} = \rank{\byn{\tconstant+1+\iconstant}}{\by}.
            \end{align*}
            Hence, for any $\iconstant \in \{0,\cdots,\kk\}$, we have
            \begin{align}
                &\alpha + \iconstant +  \aconstant + 1 \nonumber \\
                & \ge \rank{\byn{\tconstant+1+\iconstant}}{\by}  \nonumber \\
                & = \rank{\bxn{\tconstant+1+\iconstant}}{\bx} \nonumber  \\
                &= \sum_{l \in \tonumber{\n} \setminus \bu} \indicator{\bxn{l} \le \bxn{\tconstant+1+\iconstant}} + \sum_{l \in \bu} \indicator{\bxn{l} \le \bxn{\tconstant+1+\iconstant}}  \nonumber \\
                & = \rank{\bxn{\tconstant+1+\iconstant}}{\svector{\bx}{\lconstant}{\tonumber{\n}\setminus \bu}} + \sum_{l \in\{\tconstant+1,\cdots,\tconstant+\m\}} \indicator{\bxn{l} \le \bxn{\tconstant+1+\iconstant}}.  \label{supp:proposition:equivalence:lemma:3:eqn:7}
            \end{align}
            Since $\byn{\tconstant+1} < \cdots < \byn{\tconstant+\m}$, we have
            \begin{align*}
                \rank{\byn{\tconstant+1}}{\by}  < \cdots < \rank{\byn{\tconstant+\m}}{\by}.
            \end{align*}
            Then, since $\rank{\bxn{\iconstant}}{\bx} = \rank{\byn{\iconstant}}{\by}$ for any  $\iconstant \in \bu \cup \bv$,  we have
            \begin{align*}
                & \rank{\bxn{\tconstant+1}}{\bx} < \cdots < \rank{\bxn{\tconstant+\m}}{\bx} \\
                &\Rightarrow \bxn{\tconstant+1}  < \cdots <  \bxn{\tconstant+\m} \\
                &\Rightarrow \sum_{l \in\{\tconstant+1,\cdots,\tconstant+\m\}} \indicator{\bxn{l} \le \bxn{\tconstant+1+\iconstant}} =  \iconstant + 1, \quad \iconstant \in \{0,\cdots,\m-1\}.
            \end{align*}
            Since $\kk \le \bconstant-1 \le \m -2$, for any $\iconstant \in \{0,1,\cdots,\kk\}$, we have
            \begin{align*}
                \sum_{l \in\{\tconstant+1,\cdots,\tconstant+\m\}} \indicator{\bxn{l} \le \bxn{\tconstant+1+\iconstant}} =  \iconstant + 1.
            \end{align*}
            Thus, for any $\iconstant \in \{0,\cdots,\kk\}$, we have
            \begin{align*}
                &\alpha + \iconstant + \aconstant + 1  \ge^{\eqref{supp:proposition:equivalence:lemma:3:eqn:7}}  \rank{\bxn{\tconstant+1+\iconstant}}{\svector{\bx}{\lconstant}{\tonumber{\n}\setminus \bu}} + \iconstant + 1 \\
                &\Rightarrow \rank{\bxn{\tconstant+1+\iconstant}}{\svector{\bx}{\lconstant}{\tonumber{\n}\setminus \bu}} \le \alpha + \aconstant, \nonumber \\
                &\Rightarrow \rank{\bxn{\tconstant+1+\iconstant}}{\svector{\bx}{\lconstant}{\tonumber{\n}\setminus \bu}} < \alpha + \aconstant + 1 <^{\eqref{supp:proposition:equivalence:lemma:3:eqn:6}} \rank{\bxn{\aconstant+2}}{\svector{\bx}{\lconstant}{\tonumber{\n}\setminus \bu}}.
            \end{align*}
            Hence, we have $\bxn{\tconstant+1+\iconstant} < \bxn{\aconstant+2}$ for any $\iconstant \in \{0,\cdots,\kk\}$, 
            which implies that
            \begin{align*}
                \sum_{\iconstant \in \{\tconstant+1, \cdots, \tconstant+1+\kk\}}\indicator{\bxn{\iconstant} \le \bxn{\aconstant+2}} = \kk+1.
            \end{align*}
            Since $\bu = \{\tconstant+1,\cdots,\tconstant+\m\}$ and $\kk < \m$, we have
            \begin{align} \label{supp:proposition:equivalence:lemma:3:eqn:8}
                \begin{split}
                    &\rank{\bxn{\aconstant + 2}}{\svector{\bx}{\lconstant}{\{\tconstant+1, \cdots, \tconstant+1+\kk\} \cup (\tonumber{\n} \setminus \bu)} } \\
                    &= \sum_{\iconstant \in \{\tconstant+1, \cdots, \tconstant+1+\kk\}}\indicator{\bxn{\iconstant} \le \bxn{\aconstant+2}} + \sum_{\iconstant \in \tonumber{\n} \setminus \bu }\indicator{\bxn{\iconstant} \le \bxn{\aconstant+2}}  \\
                    &= \kk+1 + \rank{\bxn{\aconstant+2}}{\svector{\bx}{\lconstant}{\tonumber{\n} \setminus \bu}}\\
                    & >^{\eqref{supp:proposition:equivalence:lemma:3:eqn:6}} \alpha + \aconstant + 2 + \kk. 
                \end{split}
            \end{align}

            Since $\bxn{1} < \cdots < \bxn{\tconstant}$, we have
            \begin{align*}
                &\rank{\bxn{1}}{\bx} <  \cdots < \rank{\bxn{\tconstant}}{\bx}.
            \end{align*}
            Then, since $\rank{\bxn{\iconstant}}{\bx} = \rank{\byn{\iconstant}}{\by}$ for any  $\iconstant \in \bu \cup \bv$,  we have
            \begin{align*}
                \rank{\byn{1}}{\by}  < \cdots < \rank{\byn{\tconstant}}{\by}  \Rightarrow \byn{1}  < \cdots \byn{\tconstant}.
            \end{align*}
            Then, for any $\iconstant \in \{\aconstant+2, \cdots, \tconstant\}$,
            we have
            \begin{align*}
                \rank{\byn{\iconstant}}{\by} &\ge \rank{\byn{\aconstant+2}}{\by}.
            \end{align*}
            Since $\rank{\bxn{\iconstant}}{\bx} = \rank{\byn{\iconstant}}{\by}$ for any  $\iconstant \in \bu \cup \bv$, 
            we have
            \begin{align*}
                \rank{\byn{\iconstant}}{\by} \ge \rank{\byn{\aconstant+2}}{\by} 
                = \rank{\bxn{\aconstant+2}}{\bx}.
            \end{align*}
            According to Lemma~\ref{supp:lemma:2}, we have
            \begin{align*}
                \rank{\bxn{\aconstant+2}}{\bx}
                &\ge \rank{\bxn{\aconstant + 2}}{\svector{\bx}{\lconstant}{\{\tconstant+1, \cdots, \tconstant+1+\kk\} \cup (\tonumber{\n} \setminus \bu)}} \\
                &>^{\eqref{supp:proposition:equivalence:lemma:3:eqn:8}} \alpha + \aconstant + 2 + \kk \\
                & =^{\eqref{supp:proposition:equivalence:lemma:3:eqn:5} } \rank{\byn{\tconstant+\kk+2}}{\svector{\by}{\lconstant}{\tonumber{\n} \setminus \bv}} + \aconstant + 1. 
            \end{align*}
            Applying Lemma~\ref{supp:lemma:2} again, we have
            \begin{align*}
                \rank{\byn{\tconstant+\kk+2}}{\svector{\by}{\lconstant}{\tonumber{\n} \setminus \bv}} + \aconstant + 1 \ge \rank{\byn{\tconstant+\kk+2}}{\svector{\by}{\lconstant}{\{1,\cdots, \aconstant+1\} \cup \{\tconstant+1,\cdots,\n\}}}.
            \end{align*}
            Hence, for any  $\iconstant \in \{\aconstant+2, \cdots, \tconstant\}$, it follows ithat
            \begin{align*}
                \rank{\byn{\iconstant}}{\by}  \ge \rank{\byn{\aconstant+2}}{\by}  >  \rank{\byn{\tconstant+\kk+2}}{\svector{\by}{\lconstant}{\{1,\cdots, \aconstant+1\} \cup \{\tconstant+1,\cdots,\n\}}}.
            \end{align*}
            Thus, according to Lemma~\ref{supp:lemma:3}, we have
            \begin{align*}
                \begin{split}
                    \rank{\byn{\tconstant+\kk+2}}{\by} &= \rank{\byn{\tconstant+\kk+2}}{\svector{\by}{\lconstant}{\{1,\cdots, \aconstant+1\} \cup \{\tconstant+1,\cdots,\n\}}} \\
                    &\le \alpha + \aconstant + \kk + 2.
                \end{split}
            \end{align*}	
            That is, 
            \begin{align*}
                \begin{split}
                    \rank{\byn{\tconstant+1 + \kk + 1}}{\by} \le \alpha + \kk + 1 + \aconstant + 1,
                \end{split}
            \end{align*}	
            which proves the right-hand side of \eqref{supp:proposition:equivalence:lemma:3:eqn:3} when $\iconstant = \kk + 1$. Hence, we have shown $\bpn{\kk+1}$ is true.
            Overall, we have shown that the right-hand side of \eqref{supp:proposition:equivalence:lemma:3:eqn:3}
            is true. This completes our proof.

        \end{proof}

        \begin{lemma} \label{supp:proposition:equivalence:lemma:4}
            Following Lemma~\ref{supp:proposition:equivalence:lemma:3}, let $\mathbs, \mathban{1}, \mathban{2} \subset \tonumber{\n}$ be subsets of indices such that
            \begin{align*}
                &\mathbs = \{1, \cdots, \aconstant+1\} \cup \{\tconstant+1, \cdots, \tconstant+\bconstant+1\},\\
                &\mathban{1} = \bv \setminus \{1, \cdots, \aconstant+1\} = \{1, \cdots, \tconstant\} \setminus \{1, \cdots, \aconstant+1\},\\
                \text{and }&\mathban{2} = \bu \setminus \{\tconstant+1, \cdots, \tconstant+\bconstant+1\} = \{\tconstant+1, \cdots, \tconstant+\m\} \setminus \{\tconstant+1, \cdots, \tconstant+\bconstant+1\},
            \end{align*} 
            respectively. Then, we have $\big(\rank{\bxn{\iconstant}}{\svector{\bx}{\lconstant}{\tonumber{\n} \setminus \mathban{2} }}\big)_{\iconstant \in \mathbs}$ and $\big(\rank{\byn{\iconstant}}{\svector{\by}{\lconstant}{\tonumber{\n} \setminus \mathban{1}}}\big)_{\iconstant \in \mathbs}$ are both permutations of $\{\alpha ,\cdots, \alpha + \aconstant + \bconstant + 1\}$. Further, for any $\iconstant \in \mathbs$, we have
            \begin{align*}
                \rank{\bxn{\iconstant}}{\svector{\bx}{\lconstant}{\tonumber{\n} \setminus \mathban{2} } } = \rank{\byn{\iconstant}}{\svector{\by}{\lconstant}{\tonumber{\n} \setminus \mathban{1} } }.
            \end{align*}
        \end{lemma}

        \begin{proof}
            According to Lemma~\ref{supp:proposition:equivalence:lemma:3}, we have
            \begin{align}
                &\alpha + \iconstant \le \rank{\bxn{1+\iconstant}}{\bx} \le \alpha + \bconstant + \iconstant + 1, \text{ for any }\iconstant \in \{0,\cdots,\aconstant\}  \nonumber\\
                \Rightarrow &\alpha \le \rank{\bxn{1+\iconstant}}{\bx} \le \alpha + \bconstant + \aconstant + 1, \text{ for any }\iconstant \in \{0,\cdots,\aconstant\} \nonumber \\
                \Rightarrow &\alpha \le \rank{\bxn{\iconstant}}{\bx} \le \alpha + \bconstant + \aconstant + 1, \text{ for any }\iconstant \in \{1,\cdots,\aconstant+1\}, \label{supp:proposition:equivalence:lemma:4:eqn:0.0}
            \end{align}    
            and
            \begin{align}
                &\alpha + \iconstant \le \rank{\byn{\tconstant+1+\iconstant}}{\by} \le \alpha + \aconstant + \iconstant + 1, 
                \text{ for any } \iconstant \in \{0,1,2,\cdots,\bconstant\}\nonumber \\
                \Rightarrow &\alpha \le \rank{\byn{\tconstant+1+\iconstant}}{\by} \le \alpha + \aconstant + \bconstant + 1, 
                \text{ for any } \iconstant \in \{0,1,2,\cdots,\bconstant\} \nonumber\\
                \Rightarrow &\alpha \le \rank{\byn{\iconstant}}{\by} \le \alpha + \aconstant + \bconstant + 1, \text{ for any } \iconstant \in \{\tconstant+1,\cdots,\tconstant+1+\bconstant\}. \label{supp:proposition:equivalence:lemma:4:eqn:0.1}
            \end{align}
            Then, according to the definition of $\bconstant$, we have
            $\bconstant \le \m - 1$. Hence, we have $\{\tconstant+1,\cdots,\tconstant+1+\bconstant\} \subset \bu$.
            Since $\rank{\bxn{\iconstant}}{\bx} = \rank{\byn{\iconstant}}{\by}$ 
            for any $\iconstant \in \bu \cup \bv$, then according to 
            \eqref{supp:proposition:equivalence:lemma:4:eqn:0.1}, we have
            \begin{align*}
                \alpha \le \rank{\bxn{\iconstant}}{\bx} \le \alpha + \aconstant + \bconstant + 1 \text{ for any } \iconstant \in \{\tconstant+1,\cdots,\tconstant+1+\bconstant\}.
            \end{align*}
            Combining this with \eqref{supp:proposition:equivalence:lemma:4:eqn:0.0}, then 
            for any $\iconstant \in \{1,\cdots, \aconstant+1\} \cup \{\tconstant+1,\cdots,\tconstant+1+\bconstant\}$,
            we have
            \begin{align} \label{supp:proposition:equivalence:lemma:4:eqn:1}
                \alpha \le \rank{\bxn{\iconstant}}{\bx} \le \alpha + \aconstant + \bconstant + 1.
            \end{align}
            That is, 
            \begin{align*}
                \alpha \le \rank{\bxn{\iconstant}}{\bx} \le \alpha + \aconstant + \bconstant + 1, \text{ for any }\iconstant \in \mathbs.
            \end{align*}
            Notice that
            \begin{align*}
                |&\mathbs| = |\{1,\cdots, \aconstant+1\} \cup \{\tconstant+1,\cdots,\tconstant+\bconstant+1\}| =  \aconstant + \bconstant + 2,\\
                \text{and }& \{\alpha,\cdots, \alpha+\aconstant+\bconstant+1\} = \aconstant+\bconstant+2.	
            \end{align*}
            Since $\bx \in \vndistinct$ is a vector of distinct real values, $\big(\rank{\bxn{\iconstant}}{\bx}\big)_{\iconstant \in \mathbs}$ is a permutation of $\aconstant + \bconstant + 2$ distinct integers. Hence, $\big(\rank{\bxn{\iconstant}}{\bx}\big)_{\iconstant \in \mathbs}$ 
            is a permutation of $\{\alpha,\cdots, \alpha + \aconstant + \bconstant + 1\}$.

            Since $\byn{\tconstant+1} < \cdots < \byn{\tconstant+\m}$,
            we have
            \begin{align*}
                \rank{\byn{\tconstant+1}}{\by} < \ldots < \rank{\byn{\tconstant+\m}}{\by}.
            \end{align*}
            Recall that $\rank{\bxn{\iconstant}}{\bx} = \rank{\byn{\iconstant}}{\by}$ 
            for any $\iconstant \in \bu \cup \bv$. Thus,
            \begin{align*}
                &\rank{\bxn{\tconstant+1}}{\bx} < \ldots < \rank{\bxn{\tconstant+\m}}{\bx} \nonumber\\
                \Rightarrow &\bxn{\tconstant+1} < \ldots < \bxn{\tconstant+\m} \nonumber.
            \end{align*}
            Since $\mathban{2} = \bu \setminus \{\tconstant+1,\cdots,\tconstant+\bconstant+1\} = \{\tconstant+1,\cdots,\tconstant+\m\} \setminus \{\tconstant+1,\cdots,\tconstant+\bconstant+1\}$, 
            then if $\mathban{2} \neq \emptyset$, we have for any $\iconstant \in \mathban{2}$, 
            \begin{align} \label{supp:proposition:equivalence:lemma:4:eqn:2}
                \bxn{\iconstant} > \bxn{\tconstant+\bconstant+1}  
                \Rightarrow \rank{\bxn{\iconstant}}{\bx} > \rank{\bxn{\tconstant+\bconstant+1}}{\bx} \ge^{\eqref{supp:proposition:equivalence:lemma:4:eqn:1}} \alpha.
            \end{align}
            Since $\big(\rank{\bxn{\iconstant}}{\bx}\big)_{\iconstant \in \mathbs}$ is a permutation of $\{\alpha,\cdots, \alpha + \aconstant + \bconstant + 1\}$, we have
            \begin{align*}
                \rank{\bxn{\iconstant}}{\bx} \notin \{\alpha, \alpha+1,\cdots, \alpha + \aconstant + \bconstant + 1\}, \text{ for any } \iconstant \in \tonumber{\n} \setminus \mathbs.
            \end{align*}
            Notice that $\mathban{2} = \bu \setminus \{\tconstant+1,\cdots,\tconstant+\bconstant+1\}$,
            and $ \tonumber{\n}\setminus \mathbs = \tonumber{\n} \setminus \big(\{1, \cdots, \aconstant+1\} \cup \{\tconstant+1, \cdots, \tconstant+\bconstant+1\} \big) = \big(\bv \setminus \{1, \cdots, \aconstant+1\}\big) \cup \big(\bu \setminus \{\tconstant+1, \cdots, \tconstant+\bconstant+1\}  \big)
            \cup \big(\tonumber{\n} \setminus (\bv \cup \bu) \big)$. 
            Hence, $\mathban{2} \subset \tonumber{\n}\setminus \mathbs$.
            Therefore, we have
            \begin{align} \label{supp:proposition:equivalence:lemma:4:eqn:3}
                \rank{\bxn{\iconstant}}{\bx} \notin \{\alpha, \alpha+1,\cdots, \alpha + \aconstant + \bconstant + 1\}, \text{ for any } \iconstant \in \mathban{2}.
            \end{align}
            According to \eqref{supp:proposition:equivalence:lemma:4:eqn:2},
            when $\mathban{2} \neq \emptyset$, we have 
            $\rank{\bxn{\iconstant}}{\bx} \ge \alpha$ for any $\iconstant \in \mathban{2}$.
            Combining with \eqref{supp:proposition:equivalence:lemma:4:eqn:3},
            we further have 
            \begin{align*}
                &\rank{\bxn{\iconstant}}{\bx} > \alpha + \aconstant + \bconstant + 1, \text{ for any } \iconstant \in \mathban{2}.
            \end{align*}
            According to \eqref{supp:proposition:equivalence:lemma:4:eqn:1}, we then have
            \begin{align*}
                & \rank{\bxn{\iconstant}}{\bx} > \alpha + \aconstant + \bconstant + 1 \ge  \rank{\bxn{\lconstant}}{\bx} , \text{ for any } \iconstant \in \mathban{2}, \lconstant \in \mathbs \\
                \Rightarrow & \bxn{\iconstant} > \bxn{\lconstant} \text{ for any } \iconstant \in \mathban{2}, \lconstant \in \mathbs.
            \end{align*}
            Subsequently, according to Lemma~\ref{supp:lemma:3}, for any $\iconstant \in \mathbs$,
            \begin{align*}
                \rank{\bxn{\iconstant}}{\svector{\bx}{\lconstant}{\tonumber{\n} \setminus \mathban{2}} } = \rank{\bxn{\iconstant}}{\bx}.
            \end{align*}
            Notice that the above equation still holds when $\mathban{2} = \emptyset$.
            Then, since $\big(\rank{\bxn{\iconstant}}{\bx}\big)_{\iconstant \in \mathbs}$ is a permutation of $\{\alpha,\cdots, \alpha + \aconstant + \bconstant + 1\}$, $\big(\rank{\bxn{\iconstant}}{\svector{\bx}{\lconstant}{\tonumber{\n} \setminus \mathban{2}}}\big)_{\iconstant \in \mathbs}$ is also a permutation of $\{\alpha,\cdots, \alpha + \aconstant + \bconstant + 1\}$.

            Similarly, we can show that $\big(\rank{\byn{\iconstant}}{\by}\big)_{\iconstant \in \mathbs}$ is a permutation of $\{\alpha,\cdots, \alpha + \aconstant + \bconstant + 1\}$ and
            \begin{align*}
                \rank{\byn{\iconstant}}{\svector{\by}{\lconstant}{\tonumber{\n} \setminus \mathban{1}}} = \rank{\byn{\iconstant}}{\by} \text{ for any } \iconstant \in \mathbs.
            \end{align*}
            Since $\big(\rank{\byln{\iconstant}}{\by}\big)_{\iconstant \in \mathbs}$ is a permutation of $\{\alpha,\cdots, \alpha + \aconstant + \bconstant + 1\}$, $\big(\rank{\byln{\iconstant}}{\svector{\by}{\lconstant}{\tonumber{\n} \setminus \mathban{1}}}\big)_{\iconstant \in \mathbs}$ is also a permutation of $\{\alpha,\cdots, \alpha + \aconstant + \bconstant + 1\}$. Further, since for any $\iconstant \in \mathbs \subset \bu \cup \bv$, we have $\rank{\bxln{\iconstant}}{\bx} = \rank{\byln{\iconstant}}{\by}$, it follows that for any $\iconstant \in \mathbs$, we have
            \begin{align*}
                \rank{\bxln{\iconstant}}{\svector{\bx}{\lconstant}{\tonumber{\n} \setminus \mathban{2}}} = \rank{\byln{\iconstant}}{\svector{\by}{\lconstant}{\tonumber{\n} \setminus \mathban{1}}}.
            \end{align*}
            This completes our proof.
        \end{proof}

        \begin{lemma}  \label{supp:proposition:equivalence:lemma:5}
            Suppose $\n \ge 2$ is a positive integer.
            Let $\bv = \{1,\ldots,\tconstant\}, 
            \bu = \{\tconstant+1,\ldots,\tconstant+\m\} \subset \tonumber{\n}$ 
            be subsets of $\tonumber{\n}$
            such that $\tconstant,\m \ge 1$ and $\tconstant + \m \le \n$. 
            Consider two integers $\aconstant$ and $\bconstant$ 
            such that $\tconstant - 1 \ge \aconstant \ge 0$,
            and $\m - 1 \ge \bconstant \ge 0$, and define 
            \begin{align*}
                &\mathbs = \{1, \ldots, \aconstant+1\} \cup \{\tconstant+1, \ldots, \tconstant+\bconstant+1\},\\
                &\mathban{1} = \bv \setminus \{1, \ldots, \aconstant+1\} = \{1, \ldots, \tconstant\} \setminus \{1, \ldots, \aconstant+1\},\\
                &\mathban{2} = \bu \setminus \{\tconstant+1, \ldots, \tconstant+\bconstant+1\} = \{\tconstant+1, \ldots, \tconstant+\m\} \setminus \{\tconstant+1, \ldots, \tconstant+\bconstant+1\}.
            \end{align*} 
            Denote $\boconstant = \tonumber{\n} \setminus (\bu \cup \bv) 
            = \tonumber{\n} \setminus \{1,\ldots, \tconstant + \m\}$.	
            Then, we have
            \begin{align*}
                &\tonumber{\n} \setminus \mathban{2} = \mathbs \cup \mathban{1} \cup \bo, 
                ~\mathbs, \tonumber{\n} \setminus \bu \subset \tonumber{\n} \setminus \mathban{2}, 
                ~\mathbs \cap (\tonumber{\n} \setminus \bu) \neq \emptyset,  
                ~\mathbs \cup (\tonumber{\n} \setminus \bu) = \tonumber{\n} \setminus \mathban{2},\\
                \text{and } &\tonumber{\n} \setminus \mathban{1} = \mathbs \cup \mathban{2} \cup \boconstant,
                ~\mathbs, \tonumber{\n} \setminus \bv \subset \tonumber{\n} \setminus \mathban{1}, 
                ~\mathbs \cap (\tonumber{\n} \setminus \bv) \neq \emptyset,  
                ~\mathbs \cup (\tonumber{\n} \setminus \bv) = \tonumber{\n} \setminus \mathban{1}.
            \end{align*}
        \end{lemma}

        \begin{proof}
            We first prove that
            $\tonumber{\n} \setminus \mathban{2} = \mathbs \cup \mathban{1} \cup \boconstant$,  
            Suppose $\mathban{2} \neq \emptyset$.
            Then, we have $\mathban{2}  = \{\tconstant+1,\ldots,\tconstant+\m\} 
            \setminus \{\tconstant+1,\ldots,\tconstant+1+\bconstant\} 
            = \{\tconstant+2+\bconstant,\ldots,\tconstant+\m\}$. 
            Hence,
            \begin{align*}
                \tonumber{\n} \setminus \mathban{2} &= \big(\{1,\ldots,\tconstant\} \cup \{\tconstant+1,\ldots,\tconstant+\m\} \cup \boconstant \big) \setminus  \{\tconstant+2+\bconstant,\ldots,\tconstant+\m\}\\
                & = \{1,\ldots,\tconstant\} \cup \big(\{\tconstant+1,\ldots,\tconstant+\m\} \setminus \{\tconstant+\bconstant+2,\ldots,\tconstant+\m\}\big) \cup \bo \\
                & =  \{1,\ldots,\tconstant\} \cup \{\tconstant+1,\ldots,\tconstant+\bconstant+1\} \cup \bo.
            \end{align*}
            Since $\aconstant \le \tconstant - 1$, we have $\{1,\ldots,\aconstant+1\} \subset \{1,\ldots,\tconstant\}$.
            Then, 
            \begin{align*}
                \tonumber{\n} \setminus \mathban{2} &= \{1,\ldots,\aconstant+1\}  \cup \big(\{1,\ldots,\tconstant\} \setminus \{1,\ldots,\aconstant+1\} \big) \cup \{\tconstant+1,\ldots,\tconstant+\bconstant+1\} \cup \boconstant \\
                &=\{1,\ldots,\aconstant+1\}  \cup \mathban{1} \cup \{\tconstant+1,\ldots,\tconstant+\bconstant+1\} \cup \bo \\
                & = \{1,\ldots,\aconstant+1\} \cup \{\tconstant+1,\ldots,\tconstant+\bconstant+1\}  \cup \mathban{1} \cup \bo \\
                & = \mathbs \cup \mathban{1} \cup \bo,
            \end{align*}
            which proves $\tonumber{\n} \setminus \mathban{2} = \mathbs \cup \mathban{1} \cup \bo$ when $\mathban{2} \neq \emptyset$.

            However, when  $\mathban{2} = \emptyset$, we have
            \begin{align*}
                \{\tconstant+1, \ldots, \tconstant+\bconstant+1\} = \bu \Rightarrow \tconstant + \bconstant + 1 = \tconstant + \m \Rightarrow \bconstant = \m-1.
            \end{align*}
            Hence, $\mathbs = \{1,\ldots,\aconstant+1\} \cup \{\tconstant+1,\ldots,\tconstant+\m\}$.
            Then, we have
            \begin{align*}
                \mathbs \cup \mathban{1} \cup \bo &= \{1,\ldots,\aconstant+1\} \cup \{\tconstant+1,\ldots,\tconstant+\m\}  \cup \mathban{1} \cup \bo\\
                & = \{1,\ldots,\aconstant+1\} \cup \mathban{1} \cup \{\tconstant+1,\ldots,\n\}\\
                & = \{1,\ldots,\aconstant+1\}  \cup \big(\{1, \ldots, \tconstant\} \setminus \{1, \ldots, \aconstant+1\} \big) \cup \{\tconstant+1,\ldots,\n\}\\
                & = \tonumber{\n} = \tonumber{\n} \setminus \mathban{2},
            \end{align*}
            which proves $\tonumber{\n} \setminus \mathban{2} = \mathbs \cup \mathban{1} \cup \boconstant$ when $\mathban{2} = \emptyset$.
            Therefore, we have shown $\tonumber{\n} \setminus \mathban{2} = \mathbs \cup \mathban{1} \cup \{\tconstant+\m+1,\ldots,\n\}$.

            Next, we show $\mathbs, \tonumber{\n} \setminus \bu \subset \tonumber{\n} \setminus \mathban{2}$.
            Suppose $\mathban{2} \neq \emptyset$, we have $\mathban{2}  = \{\tconstant+1,\ldots,\tconstant+\m\} 
            \setminus \{\tconstant+1,\ldots,\tconstant+1+\bconstant\} 
            = \{\tconstant+2+\bconstant,\ldots,\tconstant+\m\}$. Then
            \begin{align*}
                \tonumber{\n} \setminus \mathban{2} &= \tonumber{\n} \setminus  \{\tconstant+2+\bconstant,\ldots,\tconstant+\m\} \\
                &= \{1,\ldots,\tconstant\} \cup \{\tconstant+1,\ldots,\tconstant+1+\bconstant\} \cup \bo.
            \end{align*}
            Since $\mathbs = \{1,\ldots,\aconstant+1\} \cup \{\tconstant+1,\ldots,\tconstant+1+\bconstant\}$,
            we have $\mathbs \subset \tonumber{\n} \setminus \mathban{2}$.
            However, suppose $\mathban{2} = \emptyset$. Then we have  
            $\tonumber{\n} \setminus \mathban{2} = \tonumber{\n}$ and $\mathbs \subset \tonumber{\n} \setminus \mathban{2}$. 
            Therefore, we have shown $\mathbs \subset \tonumber{\n} \setminus \mathban{2}$.
            Meanwhile, according to the definition of $\mathban{2}$, 
            we have $\mathban{2} \subset \bu$, which gives $\tonumber{\n} \setminus \bu \subset \tonumber{\n} \setminus \mathban{2}$. 
            Hence, we have shown $\tonumber{\n} \setminus \bu \subset \tonumber{\n} \setminus \mathban{2}$.

            Next, we show $\mathbs \cap (\tonumber{\n} \setminus \bu) \neq \emptyset$. Since
            $\bu = \{\tconstant + 1, \ldots, \tconstant + \m\}$,
            we have
            $\tonumber{\n} \setminus \bu  = (\bu \cup \bv \cup \bo) \setminus\bu
            = 	\bv \cup \bo =
            \{1, \ldots, \tconstant\} \cup \bo$. 
            Notice that $\mathbs = \{1, \ldots, \aconstant+1\} \cup \{\tconstant+1, \ldots, \tconstant+\bconstant+1\}$, where 
            $\tconstant - 1 \ge \aconstant \ge 0$,
            and $\m - 1 \ge \bconstant \ge 0$.
            Since
            $\{1, \ldots, \tconstant\} \cap \{1, \ldots, \aconstant+1\} = \{1, \ldots, \aconstant+1\} \neq \emptyset$,  
            we have
            $\mathbs \cap (\tonumber{\n} \setminus \bu)  \neq \emptyset$.

            Next, we show $\mathbs \cup (\tonumber{\n} \setminus \bu) = \tonumber{\n} \setminus \mathban{2}$.
            Suppose $\mathban{2} \neq \emptyset$.
            Then, we have $\mathban{2}  = \{\tconstant+1,\ldots,\tconstant+\m\} 
            \setminus \{\tconstant+1,\ldots,\tconstant+1+\bconstant\} 
            = \{\tconstant+2+\bconstant,\ldots,\tconstant+\m\}$. 
            Hence,
            \begin{align*}
                \tonumber{\n} \setminus \mathban{2} &= \big(\{1,\ldots,\tconstant\} \cup \{\tconstant+1,\ldots,\tconstant+\m\} \cup \boconstant \big) \setminus  \{\tconstant+2+\bconstant,\ldots,\tconstant+\m\}\\
                & = \{1,\ldots,\tconstant\} \cup \big(\{\tconstant+1,\ldots,\tconstant+\m\} \setminus \{\tconstant+\bconstant+2,\ldots,\tconstant+\m\}\big) \cup \bo \\
                & =  \{1,\ldots,\tconstant\} \cup \{\tconstant+1,\ldots,\tconstant+\bconstant+1\} \cup \bo.
            \end{align*}
            Since $\tconstant -1 \ge \aconstant \ge 1$, we have
            $\{1,\ldots,\tconstant\} = \{1,\ldots,\tconstant\} \cup \{1,\ldots,\aconstant+1\}$.
            Then, we have
            \begin{align*}
                \tonumber{\n} \setminus \mathban{2} &=  \{1,\ldots,\tconstant\} \cup \{1,\ldots,\aconstant+1\} \cup \{\tconstant+1,\ldots,\tconstant+\bconstant+1\} \cup \bo \\
                & = \{1,\ldots,\aconstant+1\} \cup \{\tconstant+1,\ldots,\tconstant+\bconstant+1\} \cup \{1,\ldots,\tconstant\} \cup  \bo \\
                & = \mathbs \cup (\{1,\ldots,\tconstant\} \cup \bo) = \mathbs \cup (\tonumber{\n} \setminus \bu).
            \end{align*} 
            However, when $\mathban{2} = \emptyset$, we have
            \begin{align*}
                \{\tconstant+1, \ldots, \tconstant+\bconstant+1\} = \bu \Rightarrow \tconstant + \bconstant + 1 = \tconstant + \m \Rightarrow \bconstant = \m-1.
            \end{align*}
            Then, we have $\mathbs = \{1,\ldots,\aconstant+1\} \cup \{\tconstant+1,\ldots,\tconstant+\m\}$. Hence, we have
            \begin{align*}
                \mathbs \cup (\tonumber{\n} \setminus \bu) 
                &= \{1,\ldots,\aconstant+1\} \cup \{\tconstant+1,\ldots,\tconstant+\m\} \cup (\tonumber{\n} \setminus \{\tconstant+1,\ldots,\tconstant+\m\}) \\
                &= \tonumber{\n} =  \tonumber{\n} \setminus \mathban{2}.
            \end{align*}
            Therefore, we have shown $\mathbs \cup (\tonumber{\n} \setminus \bu) = \tonumber{\n} \setminus \mathban{2}$.

            Now, we show $\tonumber{\n} \setminus \mathban{1} = \mathbs \cup \mathban{2} \cup \boconstant$. 
            Suppose $\mathban{1} \neq \emptyset$. Then, we have $\mathban{1}  = \{1,\ldots,\tconstant\} \setminus \{1,\ldots,\aconstant+1\} = \{\aconstant+2,\ldots,\tconstant\}$.
            Subsequently, it follows that
            \begin{align*}
                \tonumber{\n} \setminus \mathban{1} &= \big(\{1,\ldots,\tconstant\} \cup \{\tconstant+1,\ldots,\tconstant+\m\} \cup \bo \big) \setminus  \{\aconstant+2,\ldots,\tconstant\}\\
                & = \big(\{1,\ldots,\tconstant\} \setminus \{\aconstant+2,\ldots,\tconstant\} \big)		
                \cup \{\tconstant+1,\ldots,\tconstant+\m\} \cup \bo \\
                & =  \{1,\ldots,\aconstant + 1\} \cup \{\tconstant+1,\ldots,\tconstant+\m\} \cup \bo.
            \end{align*}
            Since $\bconstant \le \m - 1$,
            we have	$\{\tconstant + 1,\ldots,\tconstant+ \bconstant + 1\} \subset \{\tconstant + 1,\ldots,\tconstant + \m\}$. Then, we have
            \begin{align*}
                \tonumber{\n} \setminus \mathban{1} &= \{1,\ldots,\aconstant + 1\} \cup (\{\tconstant + 1,\ldots,\tconstant + \m\} \setminus \{\tconstant + 1,\ldots,\tconstant+ \bconstant + 1\}) \cup\{\tconstant + 1,\ldots,\tconstant + \bconstant+1\} \cup \bo \\  
                & = \{1,\ldots,\aconstant + 1\} \cup  \mathban{2} \cup\{\tconstant + 1,\ldots,\tconstant + \bconstant+1\} \cup \bo\\
                & = \{1,\ldots,\aconstant+1\} \cup \{\tconstant+1,\ldots,\tconstant+\bconstant+1\} \cup \mathban{2} \cup \bo \\
                & = \mathbs \cup \mathban{2} \cup \bo,
            \end{align*} 
            which proves $\tonumber{\n} \setminus \mathban{1} = \mathbs \cup \mathban{2} \cup \{\tconstant+\m+1,\ldots,\n\}$ when $\mathban{1} \neq \emptyset$.	

            However, when $\mathban{1} = \emptyset$, we have
            \begin{align*}
                \{1, \ldots, \aconstant+1\} = \bv \Rightarrow \aconstant + 1 = \tconstant \Rightarrow \aconstant = \tconstant-1.
            \end{align*}
            Hence, $\mathbs = \{1,\ldots,\tconstant\} \cup \{\tconstant+1,\ldots,\tconstant+\bconstant+1\}$.
            Then,
            \begin{align*}
                \mathbs \cup \mathban{2} \cup \bo
                &= \{1,\ldots,\tconstant\} \cup \{\tconstant+1,\ldots,\tconstant+\bconstant + 1\}  \cup \mathban{2} \cup \bo \\
                & = \{1,\ldots,\tconstant\} \cup \{\tconstant+1,\ldots,\tconstant+\bconstant + 1\}   \cup \big( \{\tconstant+1, \ldots, \tconstant+\m\} \setminus \{\tconstant+1, \ldots, \tconstant+\bconstant+1\}\big) \cup \bo \\
                & = \{1,\ldots,\tconstant\} \cup \{\tconstant+1, \ldots, \tconstant+\m\} \cup \bo \\
                & = \tonumber{\n} = \tonumber{\n} \setminus \mathban{1},
            \end{align*}
            which proves $\tonumber{\n} \setminus \mathban{1} = \mathbs \cup \mathban{2} \cup \{\tconstant+\m+1,\ldots,\n\}$ when $\mathban{1} = \emptyset$.
            Therefore, we have shown $\tonumber{\n} \setminus \mathban{1} = \mathbs \cup \mathban{2} \cup \{\tconstant+\m+1,\ldots,\n\}$.

            Next, we show $\mathbs, \tonumber{\n} \setminus \bv \subset \tonumber{\n} \setminus \mathban{1}$.		
            Suppose $\mathban{1} \neq \emptyset$, we have $\mathban{1}  = \{1,\ldots,\tconstant\} 
            \setminus \{1,\ldots,\aconstant+1\} 
            = \{\aconstant+2,\ldots,\tconstant\}$. Then
            \begin{align*}
                \tonumber{\n} \setminus \mathban{1} &= \tonumber{\n} \setminus  \{\aconstant+2,\ldots,\tconstant\} \\
                &= \{1,\ldots,\aconstant+1\} \cup \{\tconstant+1,\ldots, \n \}.
            \end{align*}
            Since $\mathbs = \{1,\ldots,\aconstant+1\} \cup \{\tconstant+1,\ldots,\tconstant+\bconstant+1\}$,
            we have $\mathbs \subset \tonumber{\n} \setminus \mathban{1}$.
            However, suppose $\mathban{1} = \emptyset$. Then, we have  
            $\tonumber{\n} \setminus \mathban{1} = \tonumber{\n}$ and $\mathbs \subset \tonumber{\n} \setminus \mathban{1}$. 
            Therefore, we have shown $\mathbs \subset \tonumber{\n} \setminus \mathban{1}$.

            Then, according to the definition of $\mathban{1}$, 
            we have $\mathban{1} \subset \bv$, which gives $\tonumber{\n} \setminus \bv \subset \tonumber{\n} \setminus \mathban{1}$. 
            Hence, we have shown $\tonumber{\n} \setminus \bv \subset \tonumber{\n} \setminus \mathban{1}$.

            Next, we show $\mathbs \cap (\tonumber{\n} \setminus \bv) \neq \emptyset$.
            Since
            $\bv = \{1, \ldots, \tconstant\}$,
            we have
            $\tonumber{\n} \setminus \bv  =  \{
                \tconstant + 1, \ldots, \n\}$. 
            Notice that $\mathbs = \{1, \ldots, \aconstant+1\} \cup \{\tconstant+1, \ldots, \tconstant+\bconstant+1\}$, where 
            $\tconstant - 1 \ge \aconstant \ge 0$,
            and $\m - 1 \ge \bconstant \ge 0$.
            Then, since
            $\{\tconstant+1, \ldots, \tconstant+\bconstant+1\} \cap \{
                \tconstant + 1, \ldots, \n\} = \{\tconstant+1, \ldots, \tconstant+\bconstant+1\} \neq \emptyset$. 
            We have
            $\mathbs \cap (\tonumber{\n} \setminus \bv) \neq \emptyset$.

            Finally, we show 
            $\mathbs \cup (\tonumber{\n} \setminus \bv) = \tonumber{\n} \setminus \mathban{1}$.		
            When $\mathban{1} \neq \emptyset$, we have $\mathban{1}  = \{1,\ldots,\tconstant\} 
            \setminus \{1,\ldots,\aconstant+1\} 
            = \{\aconstant+2,\ldots,\tconstant\}$. Then
            \begin{align*}
                \tonumber{\n} \setminus \mathban{1} &= \big(\{1,\ldots,\tconstant\} \cup \{\tconstant+1,\ldots,\tconstant+\m\} \cup \bo \big) \setminus \{\aconstant+2,\ldots,\tconstant\}\\
                & = \big(\{1,\ldots,\tconstant\} \setminus \{\aconstant+2,\ldots,\tconstant\} \big)		
                \cup \{\tconstant+1,\ldots,\tconstant+\m\}  \cup \bo \\
                & =  \{1,\ldots,\aconstant + 1\} \cup \{\tconstant+1,\ldots,\tconstant+\m\} \cup \bo.
            \end{align*}
            Since $\bconstant \le \m - 1$,
            we have	$\{\tconstant + 1,\ldots,\tconstant+ \bconstant + 1\} \subset \{\tconstant + 1,\ldots,\tconstant + \m\}$, which
            gives  $\{\tconstant + 1,\ldots,\tconstant + \m\} = \{\tconstant + 1,\ldots,\tconstant + \m\} \cup\{\tconstant + 1,\ldots,\tconstant + \bconstant + 1\}$.
            Then, we have
            \begin{align*}
                \tonumber{\n} \setminus \mathban{1} &=   \{1,\ldots,\aconstant + 1\} \cup   \{\tconstant + 1,\ldots,\tconstant + \m\} \cup\{\tconstant + 1,\ldots,\tconstant + \bconstant + 1\} \cup \bo \\
                & = \{1,\ldots,\aconstant+1\} \cup \{\tconstant+1,\ldots,\tconstant+\bconstant+1\} \cup\{\tconstant + 1,\ldots,\tconstant + \m\} \cup \bo \\
                & = \mathbs \cup (\tonumber{\n} \setminus \bv).
            \end{align*} 
            Hence, we have shown $\mathbs \cup (\tonumber{\n} \setminus \bv) = \tonumber{\n} \setminus \mathban{1}$
            when $ \mathban{1} \neq \emptyset$.
            However, suppose $\mathban{1} = \emptyset$. We have
            \begin{align*}
                \{1, \ldots, \aconstant+1\} = \bv \Rightarrow \aconstant + 1 = \tconstant \Rightarrow \aconstant = \tconstant-1.
            \end{align*}
            Then, we have $\mathbs = \{1,\ldots,\tconstant\} \cup \{\tconstant+1,\ldots,\tconstant+\bconstant+1\}$. Hence, we have
            \begin{align*}
                \mathbs \cup (\tonumber{\n} \setminus \bv) 
                &= \{1,\ldots,\tconstant\} \cup \{\tconstant+1,\ldots,\tconstant+\bconstant+1\} \cup (\tonumber{\n} \setminus \{1,\ldots,\tconstant\}) \\
                &= \tonumber{\n} =  \tonumber{\n} \setminus \mathban{1}.
            \end{align*}
            Therefore, we have shown $\mathbs \cup (\tonumber{\n} \setminus \bv) = \tonumber{\n} \setminus \mathban{1}$. 
            This completes our proof.
        \end{proof}

        The following proposition is crucial for proving Proposition~2.6.

        \begin{proposition} \label{supp:proposition:equivalence}
            Suppose $\bxln{1}, \byln{1}, \bxln{2}, \byln{2} \in \vndistinct$,
            and $\bu, \bv \subset \tonumber{\n}$ are disjoint subsets of 
            indices such that $\bu \cap \bv = \emptyset$. Then, if
            \begin{align}
                &\rank{\bxlnn{1}{\iconstant}}{\svector{\bxln{1}}{\lconstant}{ \tonumber{\n} \setminus \bu}} = \rank{\bxlnn{2}{\iconstant}}{\svector{\bxln{2}}{\lconstant}{\tonumber{\n} \setminus \bu}}, \text{ for any } \iconstant \in \tonumber{\n} \setminus \bu, \label{supp:proposition:equivalence:eqn:1}\\
                &\rank{\bylnn{1}{\iconstant}}{\svector{\byln{1}}{\lconstant}{ \tonumber{\n} \setminus \bv}} = \rank{\bylnn{2}{\iconstant}}{\svector{\byln{2}}{\lconstant}{\tonumber{\n} \setminus \bv}}, \text{ for any } \iconstant \in \tonumber{\n} \setminus \bv, \label{supp:proposition:equivalence:eqn:2}
            \end{align}
            and
            $\rank{\bxlnn{\jconstant}{\iconstant}}{\bxln{\jconstant}} = \rank{\bylnn{\jconstant}{\iconstant}}{\byln{\jconstant}},$ $\text{ for any } \iconstant \in \bu \cup \bv, \jconstant \in \{1,2\}$,
            we have $\sfd{\bxln{1}}{\byln{1}} = \sfd{\bxln{2}}{\byln{2}}$.
        \end{proposition}

        \begin{proof}
            For any given $\n \in \mathbb{N}$, let $\bpn{\kk}$ be the statement of Proposition~\ref{supp:proposition:equivalence} when $|\bu| + |\bv| = \kk$.
            We prove $\bpn{\kk}$ holds for any $\kk \in \{0,\ldots, \n\}$ by
            induction on $\kk$.

            \textit{Base Case:} we prove $\bpn{0}$ is true.
            Since $|\bu| = |\bv| = 0$, we have
            $\bu = \bv = \emptyset$. Then, according to
            \eqref{supp:proposition:equivalence:eqn:1} and
            \eqref{supp:proposition:equivalence:eqn:2},
            we have
            $\rank{\bxln{1}}{\bxln{1}} = \rank{\bxln{2}}{\bxln{2}}$ and $\rank{\byln{1}}{\byln{1}} = \rank{\byln{2}}{\byln{2}}$, respectively.
            Hence, we have $\sfd{\bxln{1}}{\byln{1}} = \sfd{\bxln{2}}{\byln{2}}$.
            This proves $\bpn{0}$ is true.

            \textit{Induction Step:} We show the implication
            \begin{align*}
                \bpn{\iconstant}, \text{ for any } \iconstant \le \kk \Rightarrow \bpn{\kk+1}
            \end{align*}
            for any 
            $\kk \in \{0,\ldots, \n-1\}$. Assume the induction 
            hypothesis: when $|\bu| + |\bv| = \iconstant \le \kk $, the cases 
            $\bpn{\iconstant}$ is true.

            Suppose $|\bu| + |\bv| = \kk+1$. Then we have either
            \begin{align*}
                &\text{case } (\rn{I}):   |\bu| = \kk+1, |\bv| = 0 \text{ or } |\bu| = 0, |\bv| = \kk+1,\\
                \text{or }&\text{case } (\rn{II}):  |\bu| > 0, |\bv| > 0, |\bu| + |\bv| = \kk +1,
            \end{align*}
            is true.

            Suppose the case $(\rn{I})$: $|\bu| = \kk+1, |\bv| = 0$ 
            or $|\bu| = 0, |\bv| = \kk+1$ holds. 
            Without loss of generality, let us 
            assume
            $|\bu| = \kk+1$  and $|\bv| =  0$.
            The other case then can be proved 
            in the same way after switching the labels of $\bx$ and $\by$.

            Let us assume (after relabeling) 
            $\bu = \{1,\cdots,\m\}$, and $\bylnn{1}{1} < \ldots < \bylnn{1}{\m}$.
            According to \eqref{supp:proposition:equivalence:eqn:2},
            the order of $\bylnn{1}{1}, \ldots , \bylnn{1}{\m}$
            equals to the order of $\bylnn{2}{1}, \ldots , \bylnn{2}{\m}$.
            Hence, we have 
            $\bylnn{2}{1} < \cdots < \bylnn{2}{\m}$.
            Thus, for both $\jconstant = 1,2$, we have
            $\bylnn{\jconstant}{1}, \ldots , \bylnn{\jconstant}{\m}$.
            Then, we have
            \begin{align*}
                \rank{\bylnn{\jconstant}{1}}{\byln{\jconstant}} < \cdots < \rank{\bylnn{\jconstant}{\m}}{\byln{\jconstant}}, \text{ for both } \jconstant = 1,2.
            \end{align*}
            Since $\rank{\bxlnn{\jconstant}{\iconstant}}{\bxln{\jconstant}} = \rank{\bylnn{\jconstant}{\iconstant}}{\byln{\jconstant}}, \text{ for any } \iconstant \in \bu \cup \bv, \jconstant \in \{1,2\}$,
            we further have
            \begin{align*}
                \rank{\bxlnn{\jconstant}{1}}{\bxln{\jconstant}} < \cdots < \rank{\bxlnn{\jconstant}{\m}}{\bxln{\jconstant}} 
                \Rightarrow \bxlnn{\jconstant}{1} < \cdots < \bxlnn{\jconstant}{\m}.
            \end{align*}
            Then, according to Lemma~\ref{supp:lemma:3}, we have 
            \begin{align*}
                \rank{\bxlnn{\jconstant}{1}}{\svector{\bxln{\jconstant}}{\lconstant}{\{1\} \cup (\tonumber{\n} \setminus \bu)}} = \rank{\bxlnn{\jconstant}{1}}{\bxln{\jconstant}}, \jconstant \in \{1,2\}.
            \end{align*}
            Since $\rank{\bxlnn{\jconstant}{\iconstant}}{\bxln{\jconstant}} = \rank{\bylnn{\jconstant}{\iconstant}}{\byln{\jconstant}}, \text{ for any } \iconstant \in \bu \cup \bv, \jconstant \in \{1,2\}$, we have
            \begin{align*}
                \rank{\bxlnn{\jconstant}{1}}{\svector{\bxln{\jconstant}}{\lconstant}{\{1\} \cup (\tonumber{\n} \setminus \bu)}} = \rank{\bxlnn{\jconstant}{1}}{\bxln{\jconstant}}
                = \rank{\bylnn{\jconstant}{1}}{\byln{\jconstant}}, \jconstant \in \{1,2\}.
            \end{align*}
            Next, since $\bv = \emptyset$, and $\bylnn{1}{\iconstant} = \bylnn{2}{\iconstant}, \text{ for any } \iconstant \in \tonumber{\n} \setminus \bv,$
            then we have 
            $\byln{1} = \byln{2}$. Hence,
            $\rank{\bylnn{1}{1}}{\byln{1}} = \rank{\bylnn{2}{1}}{\byln{2}}$. 
            Then, it follows that
            \begin{align*}
                \rank{\bxlnn{1}{1}}{\svector{\bxln{1}}{\lconstant}{\{1\} \cup (\tonumber{\n} \setminus \bu)}} = \rank{\bxlnn{2}{1}}{\svector{\bxln{2}}{\lconstant}{\{1\} \cup (\tonumber{\n} \setminus \bu)}}.
            \end{align*}
            According to \eqref{supp:proposition:equivalence:eqn:1}, we have 
            \begin{align*}
                \rank{\bxlnn{1}{\iconstant}}{\svector{\bxln{1}}{\lconstant}{ \tonumber{\n} \setminus \bu}} = \rank{\bxlnn{2}{\iconstant}}{\svector{\bxln{2}}{\lconstant}{\tonumber{\n} \setminus \bu}}, \text{ for any } \iconstant \in \tonumber{\n} \setminus \bu.
            \end{align*}
            Subsequently, according to
            Lemma~\ref{supp:proposition:equivalence:lemma:1}, 
            we have
            \begin{align*}
                \begin{split}
                    \rank{\bxlnn{1}{\iconstant}}{\svector{\bxln{1}}{\lconstant}{\{1\} \cup (\tonumber{\n} \setminus \bu)}} = \rank{\bxlnn{2}{\iconstant}}{\svector{\bxln{2}}{\lconstant}{\{1\} \cup (\tonumber{\n} \setminus \bu)}},  \text{ for any } \iconstant \in \tonumber{\n} \setminus \bu.
                \end{split}
            \end{align*}
            Therefore, we have shown that
            for any $\iconstant \in \{1\} \cup (\tonumber{\n} \setminus \bu)$, we have
            \begin{align*}
                \rank{\bxlnn{1}{\iconstant}}{\svector{\bxln{1}}{\lconstant}{\{1\} \cup (\tonumber{\n} \setminus \bu)}} = \rank{\bxlnn{2}{\iconstant}}{\svector{\bxln{2}}{\lconstant}{\{1\} \cup (\tonumber{\n} \setminus \bu)}}.
            \end{align*}
            Notice that $\bv = \emptyset$. Hence, we have $|\bv| + |\tonumber{\n} \setminus (\{1\} \cup (\tonumber{\n} \setminus \bu))| = 0 + |\bu \setminus \{1\}| = \m - 1 = \kk$. 
            Meanwhile, notice that according to \eqref{supp:proposition:equivalence:eqn:2}, we have
            $\rank{\byln{1}}{\byln{1}} = \rank{\byln{2}}{\byln{2}}$. 
            Also, we have $\rank{\bxlnn{\jconstant}{\iconstant}}{\bxln{\jconstant}} = \rank{\bylnn{\jconstant}{\iconstant}}{\byln{\jconstant}}, \text{ for any } \iconstant \in (\bu \cup \bv) \setminus \{1\} , \jconstant \in \{1,2\}$.		
            Then, since $\bpn{\kk}$ is true, we have $\sfd{\bxln{1}}{\byln{1}} = \sfd{\bxln{2}}{\byln{2}}$.
            This proves $\bpn{\kk+1}$ when case $(\rn{I})$ is true.

            Suppose the case $(\rn{II})$: $|\bu| > 0, |\bv| > 0, |\bu| + |\bv| = \kk+1$ is true. 
            Since $|\bu| > 0, |\bv| > 0$, we have $\bu \neq \emptyset$,
            and $\bv \neq \emptyset$. 
            Without loss of generality, let us assume (after relabeling) 
            $\bv = \{1,\cdots,\tconstant\}$, $\bu = \{\tconstant+1,\cdots,\tconstant+\m\}$, 
            $\bxlnn{1}{1}  < \cdots < \bxlnn{1}{\tconstant}$,
            and $\bylnn{1}{\tconstant+1} < \cdots < \bylnn{1}{\tconstant+\m}$.
            According to \eqref{supp:proposition:equivalence:eqn:1},
            and  \eqref{supp:proposition:equivalence:eqn:2},
            the order of 
            $\bxlnn{1}{1}, \ldots, \bxlnn{1}{\tconstant}$
            and the order of 
            $\bylnn{1}{1}, \ldots , \bylnn{1}{\m}$
            equals to the order of 
            $\bxlnn{2}{1}, \ldots, \bxlnn{2}{\tconstant}$ 
            and the order of
            $\bylnn{2}{1}, \ldots , \bylnn{2}{\m}$,
            respectively. 
            Hence, we have
            $\bxlnn{\jconstant}{1}  < \cdots < \bxlnn{\jconstant}{\tconstant}$,
            and $\bylnn{\jconstant}{\tconstant+1} < \cdots < \bylnn{\jconstant}{\tconstant+\m}$ for both $\jconstant = 1,2$.

            Since $\bu = \{\tconstant+1,\cdots,\tconstant+\m\}$ and $\bv = \{1,\cdots,\tconstant\}$, where $\tconstant, \m \ge 1$, we have
            $1 \in \tonumber{\n} \setminus \bu$ and $\tconstant + 1 \in \tonumber{\n}\setminus \bv$. Then, 
            according to \eqref{supp:proposition:equivalence:eqn:1}
            and \eqref{supp:proposition:equivalence:eqn:2}, we have
            \begin{align*}
                & \rank{\bxlnn{1}{1}}{\svector{\bxln{1}}{\lconstant}{\tonumber{\n}\setminus \bu}} = \rank{\bxlnn{2}{1}}{\svector{\bxln{2}}{\lconstant}{\tonumber{\n}\setminus \bu}}, \\
                \text{and }
                &\rank{\bylnn{1}{\tconstant+1}}{\svector{\byln{1}}{\lconstant}{\tonumber{\n}\setminus \bv}} = \rank{\bylnn{2}{\tconstant+1}}{\svector{\byln{2}}{\lconstant}{\tonumber{\n}\setminus \bv}},
            \end{align*}
            respectively. Let us denote
            \begin{align*}
                &\alpha = \rank{\bxlnn{1}{1}}{\svector{\bxln{1}}{\lconstant}{\tonumber{\n}\setminus \bu}} = \rank{\bxlnn{2}{1}}{\svector{\bxln{2}}{\lconstant}{\tonumber{\n}\setminus \bu}}, \\
                &\beta = \rank{\bylnn{1}{\tconstant+1}}{\svector{\byln{1}}{\lconstant}{\tonumber{\n}\setminus \bv}} = \rank{\bylnn{2}{\tconstant+1}}{\svector{\byln{2}}{\lconstant}{\tonumber{\n}\setminus \bv}}.
            \end{align*}
            Notice that it is either $\alpha \le \beta$ or $\alpha \ge \beta$. 
            Let us assume
            $\alpha \le \beta$. However, if $\alpha \ge \beta$,
            we can switch the labels between  
            $\bxln{\jconstant}$ and $\byln{\jconstant}$ 
            for both $\jconstant \in \{1,2\}$,
            and relabel the relevant components of the data.

            In the following, we are going to consider the 
            following two cases
            \begin{align*}
                &\text{case } (\rn{i}):  \alpha < \beta, \\
                \text{and }&\text{case } (\rn{ii}):  \alpha = \beta,
            \end{align*}
            separately.

            Suppose the case $(\rn{i})$: $\alpha < \beta$ holds. In other words,
            \begin{align*}
                \rank{\bxlnn{\jconstant}{1}}{\svector{\bxln{\jconstant}}{\lconstant}{\tonumber{\n}\setminus \bu}} < \rank{\bylnn{\jconstant}{\tconstant+1}}{\svector{\byln{\jconstant}}{\lconstant}{\tonumber{\n}\setminus \bv}}, \text{ for both } \jconstant = 1,2.
            \end{align*}
            According to Lemma~\ref{supp:proposition:equivalence:lemma:3.0}, we have
            \begin{align}
                \rank{\bxlnn{\jconstant}{1}}{\svector{\bxln{\jconstant}}{\lconstant}{\tonumber{\n}\setminus \bu}} = \rank{\bxlnn{\jconstant}{1}}{\bxln{\jconstant}} = \rank{\bylnn{\jconstant}{1}}{\byln{\jconstant}}, \text{ for both } \jconstant = 1,2. \label{supp:proposition:equivalence:eqn:0}
            \end{align}
            Since $\bxlnn{\jconstant}{1}  < \cdots < \bxlnn{\jconstant}{\tconstant}$ 
            for both $\jconstant = 1,2$, we have
            \begin{align*}
                \rank{\bxlnn{\jconstant}{1}}{\bxln{\jconstant}} < \cdots < \rank{\bxlnn{\jconstant}{\tconstant}}{\bxln{\jconstant}}, \text{ for both } \jconstant  = 1,2.
            \end{align*}
            Then, since $\rank{\bxlnn{\jconstant}{\iconstant}}{\bxln{\jconstant}} = \rank{\bylnn{\jconstant}{\iconstant}}{\byln{\jconstant}}, \text{ for any } \iconstant \in \bu \cup \bv, \jconstant \in \{1,2\}$, we have
            \begin{align*}
                &\rank{\bylnn{\jconstant}{1}}{\byln{\jconstant}} < \cdots < \rank{\bylnn{\jconstant}{\tconstant}}{\byln{\jconstant}}, \text{ for both } \jconstant  = 1,2\\
                \Rightarrow &\bylnn{\jconstant}{1} < \bylnn{\jconstant}{2} < \cdots < \bylnn{\jconstant}{\tconstant} \text{ for both } \jconstant  = 1,2.
            \end{align*}		
            Notice that $\bv = \{1,\ldots, \tconstant\}$. 
            Next, according to Lemma~\ref{supp:lemma:3}, 
            for both $ \jconstant= 1,2$, we have
            \begin{align*}
                \rank{\bylnn{\jconstant}{1}}{\svector{\byln{\jconstant}}{\lconstant}{\{1\} \cup (\tonumber{\n} \setminus \bv)}} = \rank{\bylnn{\jconstant}{1}}{\byln{\jconstant}} =^{\eqref{supp:proposition:equivalence:eqn:0}} \rank{\bxlnn{\jconstant}{1}}{\svector{\bxln{\jconstant}}{\lconstant}{\tonumber{\n}\setminus \bu}}.
            \end{align*}
            According to \eqref{supp:proposition:equivalence:eqn:1},
            we have
            \begin{align*}
                \rank{\bxlnn{1}{1}}{\svector{\bxln{1}}{\lconstant}{\tonumber{\n}\setminus \bu}} = \rank{\bxlnn{2}{1}}{\svector{\bxln{2}}{\lconstant}{\tonumber{\n}\setminus \bu}}.
            \end{align*}
            Then, it follows
            \begin{align*}
                \rank{\bylnn{1}{1}}{\svector{\byln{1}}{\lconstant}{\{1\} \cup (\tonumber{\n} \setminus \bv)}} = \rank{\bylnn{2}{1}}{\svector{\byln{2}}{\lconstant}{\{1\} \cup (\tonumber{\n} \setminus \bv)}}.
            \end{align*}
            According to \eqref{supp:proposition:equivalence:eqn:2}, for any $\iconstant \in \tonumber{\n} \setminus \bv$,
            \begin{align*}
                \rank{\bylnn{1}{\iconstant}}{\svector{\byln{1}}{\lconstant}{\tonumber{\n}\setminus \bv}} = \rank{\bylnn{2}{\iconstant}}{\svector{\byln{2}}{\lconstant}{\tonumber{\n}\setminus \bv}}.
            \end{align*}
            Subsequently, according to Lemma~\ref{supp:proposition:equivalence:lemma:1}, for any $\iconstant \in \tonumber{\n} \setminus \bv$, we have
            \begin{align*}
                \rank{\bylnn{1}{\iconstant}}{\svector{\byln{1}}{\lconstant}{\{1\} \cup (\tonumber{\n} \setminus \bv)}} = \rank{\bylnn{2}{\iconstant}}{\svector{\byln{2}}{\lconstant}{\{1\} \cup (\tonumber{\n} \setminus \bv)}}.
            \end{align*}
            Therefore, we have shown that
            for any $\iconstant \in (\tonumber{\n} \setminus \bv) \cup \{1\}$,
            \begin{align*}
                \rank{\bylnn{1}{\iconstant}}{\svector{\byln{1}}{\lconstant}{\{1\} \cup (\tonumber{\n} \setminus \bv)}} = \rank{\bylnn{2}{\iconstant}}{\svector{\byln{2}}{\lconstant}{\{1\} \cup (\tonumber{\n} \setminus \bv)}}.
            \end{align*}
            Notice that $|\bu| + |\bv \setminus \{1\}| = |\bu| + |\bv| - 1 = \kk$. Meanwhile, notice that according to \eqref{supp:proposition:equivalence:eqn:1}, we have
            $\rank{\bxlnn{1}{\iconstant}}{\svector{\bxln{1}}{\lconstant}{ \tonumber{\n} \setminus \bu}} = \rank{\bxlnn{2}{\iconstant}}{\svector{\bxln{2}}{\lconstant}{\tonumber{\n} \setminus \bu}}, \text{ for any } \iconstant \in \tonumber{\n} \setminus \bu$. 
            Also, we have $\rank{\bxlnn{\jconstant}{\iconstant}}{\bxln{\jconstant}} = \rank{\bylnn{\jconstant}{\iconstant}}{\byln{\jconstant}}, \text{ for any } \iconstant \in (\bu \cup \bv) \setminus \{1\}, \jconstant \in \{1,2\}$.		
            Then, since $\bpn{\kk}$ is true, we have $\sfd{\bxln{1}}{\byln{1}} = \sfd{\bxln{2}}{\byln{2}}$.
            This proves $\bpn{\kk+1}$ when the case $(\rn{i})$ holds.

            Now, suppose the case $(\rn{ii})$: $\alpha = \beta$ holds. In other words,
            \begin{align*}
                \alpha = \beta = \rank{\bxlnn{\jconstant}{1}}{\svector{\bxln{\jconstant}}{\lconstant}{\tonumber{\n} \setminus \bu}} 
                =\rank{\bylnn{\jconstant}{\tconstant+1}}{\svector{\byln{\jconstant}}{\lconstant}{\tonumber{\n} \setminus \bv}}, \text{ for any } \jconstant \in \{1,2\}.
            \end{align*}
            Define
            \begin{align*}
                &\aconstant = \max \{\iconstant \in \{0,\cdots,\tconstant-1\} 
                ~|~ \rank{\bxlnn{1}{1+i}}{\svector{\bxln{1}}{\lconstant}{\tonumber{\n} \setminus \bu}} = \alpha + \iconstant\},  \\
                \text{and }&\bconstant = \max \{\iconstant \in \{0,\cdots,\m-1\} 
                ~|~ \rank{\bylnn{1}{\tconstant+1+i}}{\svector{\byln{1}}{\lconstant}{\tonumber{\n} \setminus \bv}} = \alpha + \iconstant\}.
            \end{align*}
            Next, define
            \begin{align*}
                &\mathbs = \{1,\cdots,\aconstant+1\} \cup \{\tconstant+1,\cdots,\tconstant+1+\bconstant\},\\
                &\mathban{1} = \bv \setminus \{1,\cdots,\aconstant+1\} = \{1,\cdots,\tconstant\}  \setminus \{1,\cdots,\aconstant+1\},\\
                \text{and }&\mathban{2} = \bu \setminus \{\tconstant+1,\cdots,\tconstant+1+\bconstant\} = \{\tconstant+1,\cdots,\tconstant+\m\} \setminus \{\tconstant+1,\cdots,\tconstant+1+\bconstant\}.
            \end{align*}
            Then, according to Lemma~\ref{supp:proposition:equivalence:lemma:4},  
            for both $\jconstant = 1,2$, we have, 
            $\big(\rank{\bxlnn{j}{\iconstant}}{\svector{\bxln{j}}{\lconstant}{\tonumber{\n} \setminus \mathban{2}}}\big)_{\iconstant \in \mathbs}$, and $\big(\rank{\bylnn{j}{\iconstant}}{\svector{\byln{j}}{\lconstant}{\tonumber{\n} \setminus \mathban{1}}}\big)_{\iconstant \in \mathbs}$ 
            are permutations of $\{\alpha ,\cdots, \alpha + \aconstant + \bconstant + 1\}$, and
            \begin{align*}
                \rank{\bxlnn{j}{\iconstant}}{\svector{\bxln{j}}{\lconstant}{\tonumber{\n} \setminus \mathban{2}}} = \rank{\bylnn{j}{\iconstant}}{\svector{\byln{j}}{\lconstant}{\tonumber{\n} \setminus \mathban{1}}}, \text{ for any } \iconstant \in \mathbs.
            \end{align*} 
            Thus, $\big(\rank{\bxlnn{1}{\iconstant}}{\svector{\bxln{1}}{\lconstant}{\tonumber{\n} \setminus \mathban{2}}}\big)_{\iconstant \in \mathbs}$ is a permutation of $\big(\rank{\bxlnn{2}{\iconstant}}{\svector{\bxln{2}}{\lconstant}{\tonumber{\n} \setminus \mathban{2}}}\big)_{\iconstant \in \mathbs}$. 
            Then, there exist a permutation 
            $(\sigma(\lconstant))_{\lconstant \in \mathbs}$
            of $\mathbs$ such that
            \begin{align} \label{supp:proposition:equivalence:eqn:4}
                \rank{\bxlnn{2}{\iconstant}}{\svector{\bxln{2}}{\lconstant}{\tonumber{\n} \setminus \mathban{2}}}
                = \rank{\bxlnn{1}{\sigma(\iconstant)}}{\svector{\bxln{1}}{\lconstant}{\tonumber{\n} \setminus \mathban{2}}},\text{ for any }\iconstant \in \mathbs.
            \end{align}
            Since $\rank{\bxlnn{j}{\iconstant}}{\svector{\bxln{j}}{\lconstant}{\tonumber{\n} \setminus \mathban{2}}} = \rank{\bylnn{j}{\iconstant}}{\svector{\byln{j}}{\lconstant}{\tonumber{\n} \setminus \mathban{1}}}, \text{for any } \iconstant \in \mathbs, \jconstant \in \{1,2\}$, then we have
            \begin{align} \label{supp:proposition:equivalence:eqn:5}
                \begin{split} 		  
                    \rank{\bylnn{2}{\iconstant}}{\svector{\byln{2}}{\lconstant}{\tonumber{\n} \setminus \mathban{1}}} 
                    &= \rank{\bxlnn{2}{\iconstant}}{\svector{\bxln{2}}{\lconstant}{\tonumber{\n} \setminus \mathban{2}}}  \\
                    &=^{\eqref{supp:proposition:equivalence:eqn:4}} \rank{\bxlnn{1}{\sigma(\iconstant)}}{\svector{\bxln{1}}{\lconstant}{\tonumber{\n} \setminus \mathban{2}}}  \\
                    &= \rank{\bylnn{1}{\sigma(\iconstant)}}{\svector{\byln{1}}{\lconstant}{\tonumber{\n} \setminus \mathban{1}}}. 
                \end{split}
            \end{align}

            Next, define $\bo = \tonumber{\n} \setminus (\bu \cup \bv)$.
            Then, according to Lemma~\ref{supp:proposition:equivalence:lemma:5}, we have	
            \begin{align}
                &\tonumber{\n} \setminus \mathban{2} = \mathbs \cup \mathban{1} \cup \bo, \label{supp:proposition:equivalence:eqn:6.0}\\
                &\tonumber{\n} \setminus \mathban{1} = \mathbs \cup \mathban{2} \cup \bo, \label{supp:proposition:equivalence:eqn:6.1}\\
                &\mathbs, \tonumber{\n} \setminus \bu \subset \tonumber{\n} \setminus \mathban{2}, 
                ~\mathbs \cap (\tonumber{\n} \setminus \bu) \neq \emptyset,  
                ~\mathbs \cup (\tonumber{\n} \setminus \bu) = \tonumber{\n} \setminus \mathban{2}, \label{supp:proposition:equivalence:eqn:6}\\
                \text{and } &\mathbs, \tonumber{\n} \setminus \bv \subset \tonumber{\n} \setminus \mathban{1}, 
                ~\mathbs \cap (\tonumber{\n} \setminus \bv) \neq \emptyset,  
                ~\mathbs \cup (\tonumber{\n} \setminus \bv) = \tonumber{\n} \setminus \mathban{1}. \label{supp:proposition:equivalence:eqn:7}
            \end{align}		

            According to \eqref{supp:proposition:equivalence:eqn:1}, we have
            \begin{align}
                \rank{\bxlnn{1}{\iconstant}}{\svector{\bxln{1}}{\lconstant}{\tonumber{n} \setminus \bu}} 
                = \rank{\bxlnn{2}{\iconstant}}{\svector{\bxln{2}}{\lconstant}{\tonumber{n} \setminus \bu}}, \text{ for any } \iconstant \in \mathbs \cap (\tonumber{n} \setminus \bu).
            \end{align}
            Notice that $\big(\rank{\bxlnn{1}{\iconstant}}{\svector{\bxln{1}}{\lconstant}{\tonumber{n} \setminus \mathban{2}}}\big)_{\iconstant \in \mathbs}$ is a permutation of $\big(\rank{\bxlnn{2}{\iconstant}}{\svector{\bxln{2}}{\lconstant}{\tonumber{n} \setminus \mathban{2}}}\big)_{\iconstant \in \mathbs}$,  and according to 
            \eqref{supp:proposition:equivalence:eqn:1}, we have
            \begin{align*}
                \rank{\bxlnn{1}{\iconstant}}{\svector{\bxln{1}}{\lconstant}{\tonumber{n} \setminus  \bu}} 
                = \rank{\bxlnn{2}{\iconstant}}{\svector{\bxln{2}}{\lconstant}{\tonumber{n} \setminus \bu }} \text{ for any } \iconstant \in (\tonumber{n} \setminus \bu) \setminus \mathbs.
            \end{align*}
            According to \eqref{supp:proposition:equivalence:eqn:6}, we have $\mathbs, \tonumber{\n} \setminus \bu \subset \tonumber{\n} \setminus \mathban{2}, 
            ~\mathbs \cap (\tonumber{\n} \setminus \bu) \neq \emptyset$,  
            and $\mathbs \cup (\tonumber{\n} \setminus \bu) = \tonumber{\n} \setminus \mathban{2}$.
            Then, we can apply Lemma~\ref{supp:proposition:equivalence:lemma:2} 
            and get 		
            \begin{align} \label{supp:proposition:equivalence:eqn:9}
                \rank{\bxlnn{1}{\iconstant}}{\svector{\bxln{1}}{\lconstant}{\tonumber{n} \setminus \mathban{2}}} 
                = \rank{\bxlnn{2}{\iconstant}}{\svector{\bxln{2}}{\lconstant}{\tonumber{n} \setminus \mathban{2}}}, \text{ for any } \iconstant \in (\tonumber{n} \setminus \bu) \setminus \mathbs.
            \end{align}

            According to \eqref{supp:proposition:equivalence:eqn:5}, $\rank{\bylnn{2}{\iconstant}}{\svector{\byln{2}}{\lconstant}{\tonumber{\n} \setminus \mathban{1}}}$
            is a permutation of 
            $\rank{\bylnn{1}{\iconstant}}{\svector{\byln{1}}{\lconstant}{\tonumber{\n} \setminus \mathban{1}}}$.
            Meanwhile,		
            according to  \eqref{supp:proposition:equivalence:eqn:2}, we have
            \begin{align*}
                \rank{\bylnn{1}{\iconstant}}{\svector{\byln{1}}{\lconstant}{ \tonumber{\n} \setminus \bv}} = \rank{\bylnn{2}{\iconstant}}{\svector{\byln{2}}{\lconstant}{\tonumber{\n} \setminus \bv}}, \text{ for any } \iconstant \in  (\tonumber{\n} \setminus \bv) \setminus \mathbs.
            \end{align*}
            According to \eqref{supp:proposition:equivalence:eqn:7}, we have $\mathbs, \tonumber{\n} \setminus \bv \subset \tonumber{\n} \setminus \mathban{1}, 
            ~\mathbs \cap (\tonumber{\n} \setminus \bv) \neq \emptyset$,  
            and $\mathbs \cup (\tonumber{\n} \setminus \bv) = \tonumber{\n} \setminus \mathban{1}$.		
            Subsequently, according to		
            Lemma~\ref{supp:proposition:equivalence:lemma:2}, we have
            \begin{align} \label{supp:proposition:equivalence:eqn:10}
                \rank{\bylnn{1}{\iconstant}}{\svector{\byln{1}}{\lconstant}{\tonumber{n} \setminus \mathban{1}}} 
                = \rank{\bylnn{2}{\iconstant}}{\svector{\byln{2}}{\lconstant}{\tonumber{n} \setminus \mathban{1}}},
                \text{ for any } \iconstant \in (\tonumber{n} \setminus \bv) \setminus \mathbs.
            \end{align}

            Next, consider vectors $\bxln{3}, \byln{3} \in \vndistinct$ such that
            \begin{align*}
                &\bxlnn{3}{\iconstant} = \bxlnn{1}{\sigma(\iconstant)}, \text{ for any } \iconstant \in \mathbs,  \text{ and } \bxlnn{3}{\iconstant} = \bxlnn{1}{\iconstant}, \text{ for any } \iconstant \in \tonumber{\n} \setminus \mathbs;\\
                &\bylnn{3}{\iconstant} = \bylnn{1}{\sigma(\iconstant)}, \text{ for any } \iconstant \in \mathbs,  \text{ and } \bylnn{3}{\iconstant} = \bylnn{1}{\iconstant}, \text{ for any } \iconstant \in \tonumber{\n} \setminus \mathbs,
            \end{align*}
            where $(\sigma(\lconstant))_{\lconstant \in \mathbs}$ 
            such that \eqref{supp:proposition:equivalence:eqn:4} is 
            a permutation of $\mathbs$. 
            Since $\bxln{3}$ and $\byln{3}$ are permutations of
            $\bxln{1}$ and $\byln{1}$ according to indices, 
            we have
            \begin{align} \label{supp:proposition:equivalence:eqn:5.0}
                D(\bxln{3}, \byln{3}) = D(\bxln{1}, \byln{1}).
            \end{align}
            Meanwhile, since $\bxlnn{3}{\iconstant} = \bxlnn{1}{\iconstant}, \text{ for any } \iconstant \in \tonumber{\n} \setminus \mathbs$, and $\bxln{3}$ is a permutation of $\bxln{1}$, we have
            \begin{align*}
                \rank{\bxlnn{3}{\iconstant}}{\bxln{3}} = \rank{\bxlnn{1}{\iconstant}}{\bxln{1}}, \text{ for any } \iconstant \in (\bu \cup \bv) \setminus \mathbs.
            \end{align*}
            Similarly, since $\bylnn{3}{\iconstant} = \bylnn{1}{\iconstant}, \text{ for any } \iconstant \in \tonumber{\n} \setminus \mathbs$, and $\byln{3}$ is a permutation of $\byln{1}$, we have
            \begin{align*}
                \rank{\bylnn{3}{\iconstant}}{\byln{3}} = \rank{\bylnn{1}{\iconstant}}{\byln{1}}, \text{ for any } \iconstant \in (\bu \cup \bv) \setminus \mathbs.
            \end{align*}
            Then, since $\rank{\bxlnn{1}{\iconstant}}{\bxln{1}} = \rank{\bylnn{1}{\iconstant}}{\byln{1}}$ for any 
            $\iconstant \in \bu \cup \bv$,
            we have	
            \begin{align*}
                \rank{\bxlnn{3}{\iconstant}}{\bxln{3}} = \rank{\bylnn{3}{\iconstant}}{\byln{3}}, \text{ for any } \iconstant \in (\bu \cup \bv) \setminus \mathbs.
            \end{align*}	
            Since we also have $\rank{\bxlnn{2}{\iconstant}}{\bxln{2}} = \rank{\bylnn{2}{\iconstant}}{\byln{2}}$ for any $\iconstant \in \bu \cup \bv$, then if follows that
            \begin{align*}
                \rank{\bxlnn{j}{\iconstant}}{\bxln{j}} = \rank{\bylnn{j}{\iconstant}}{\byln{j}}, \text{ for any } \iconstant \in (\bu \cup \bv) \setminus \mathbs, \jconstant \in \{2,3\}.
            \end{align*}	
            Notice that according to the definition of $\mathbs, \mathban{1}$ and $\mathban{2}$,
            we have $(\bu \cup \bv) \setminus \mathbs = \mathban{1} \cup \mathban{2}$. Hence, we have
            \begin{align} \label{supp:proposition:equivalence:eqn:8}
                \rank{\bxlnn{j}{\iconstant}}{\bxln{j}} = \rank{\bylnn{j}{\iconstant}}{\byln{j}}, \text{ for any } \iconstant \in \mathban{1} \cup \mathban{2}, \jconstant \in \{2,3\}.
            \end{align}

            According to \eqref{supp:proposition:equivalence:eqn:6.0},
            we have $\tonumber{\n} \setminus \mathban{2} = \mathbs \cup \mathban{1} \cup \bo$. According to 
            the definition of $\mathbs$, $\mathban{1}$, $\bo$, we have $\mathbs \cap \mathban{1} = \emptyset$, 
            and $\mathbs \cap \bo = \emptyset$, which gives $(\mathban{1} \cup \bo) \cap \mathbs = \emptyset$.
            Then, we have $\mathban{1} \cup \bo \subset \tonumber{\n} \setminus \mathbs$.
            Since $\bxlnn{3}{\iconstant} = \bxlnn{1}{\sigma(\iconstant)}, \text{ for any } \iconstant \in \mathbs,  \text{ and } \bxlnn{3}{\iconstant} = \bxlnn{1}{\iconstant}, \text{ for any } \iconstant \in \tonumber{\n} \setminus \mathbs$, then for any $\iconstant \in \tonumber{\n} \setminus \mathban{2}$, we have
            \begin{align*}
                &\bxlnn{3}{\iconstant} = \bxlnn{1}{\sigma(\iconstant)}, \text{ if } \iconstant \in \mathbs, \\
                &\bxlnn{3}{\iconstant} = \bxlnn{1}{\iconstant}, \text{ if } \iconstant \in \tonumber{\n} \setminus \mathbs.
            \end{align*}	
            Hence, $\svector{\bxln{3}}{\lconstant}{\tonumber{\n} \setminus \mathban{2}}$
            is a permutation of $\svector{\bxln{1}}{\lconstant}{\tonumber{\n} \setminus \mathban{2}}$.
            Then, we have 
            \begin{align*}
                \rank{\bxlnn{3}{\iconstant}}{\svector{\bxln{3}}{\lconstant}{\tonumber{\n} \setminus \mathban{2}}} 
                = \rank{\bxlnn{1}{\sigma(\iconstant)}}{\svector{\bxln{1}}{\lconstant}{\tonumber{\n} \setminus \mathban{2}}}, \text{ for any } \iconstant \in \mathbs.
            \end{align*}
            According to \eqref{supp:proposition:equivalence:eqn:4}, we further have
            for any $ \iconstant \in \mathbs$,
            \begin{align*}
                \rank{\bxlnn{3}{\iconstant}}{\svector{\bxln{3}}{\lconstant}{\tonumber{\n} \setminus \mathban{2}}} 
                = \rank{\bxlnn{1}{\sigma(\iconstant)}}{\svector{\bxln{1}}{\lconstant}{\tonumber{\n} \setminus \mathban{2}}}
                = \rank{\bxlnn{2}{\iconstant}}{\svector{\bxln{2}}{\lconstant}{\tonumber{\n} \setminus \mathban{2}}}.
            \end{align*}
            Next, since  $\svector{\bxln{3}}{\lconstant}{\tonumber{\n} \setminus \mathban{2}}$
            is a permutation of $\svector{\bxln{1}}{\lconstant}{\tonumber{\n} \setminus \mathban{2}}$,
            and for any $\iconstant \in (\tonumber{\n} \setminus \bu) \setminus \mathbs$, 
            we have $\bxlnn{3}{\iconstant} = \bxlnn{1}{\iconstant}$. Then, for any $\iconstant \in (\tonumber{n} \setminus \bu) \setminus \mathbs$, we have
            \begin{align*}
                \rank{\bxlnn{3}{\iconstant}}{\svector{\bxln{3}}{\lconstant}{\tonumber{\n} \setminus \mathban{2}}}
                = \rank{\bxlnn{1}{\iconstant}}{\svector{\bxln{1}}{\lconstant}{\tonumber{\n} \setminus \mathban{2}}}
                =^{\eqref{supp:proposition:equivalence:eqn:9} } \rank{\bxlnn{2}{\iconstant}}{\svector{\bxln{2}}{\lconstant}{\tonumber{\n} \setminus \mathban{2}}}.
            \end{align*}
            Hence, we have for any $\iconstant \in (\tonumber{n} \setminus \bu) \cup \mathbs$, we have
            \begin{align*}
                \rank{\bxlnn{3}{\iconstant}}{\svector{\bxln{3}}{\lconstant}{\tonumber{\n} \setminus \mathban{2}}} 
                = \rank{\bxlnn{2}{\iconstant}}{\svector{\bxln{2}}{\lconstant}{\tonumber{\n} \setminus \mathban{2}}}.
            \end{align*}
            According to \eqref{supp:proposition:equivalence:eqn:6}, we have $\mathbs \cup (\tonumber{\n} \setminus \bu) = \tonumber{\n} \setminus \mathban{2}$. Hence, it follows that 
            \begin{align} \label{supp:proposition:equivalence:eqn:11}
                \rank{\bxlnn{3}{\iconstant}}{\svector{\bxln{3}}{\lconstant}{\tonumber{\n} \setminus \mathban{2}}} 
                = \rank{\bxlnn{2}{\iconstant}}{\svector{\bxln{2}}{\lconstant}{\tonumber{\n} \setminus \mathban{2}}}, \text{ for any } \iconstant \in \tonumber{\n} \setminus \mathban{2}.
            \end{align}

            Similarly, according to \eqref{supp:proposition:equivalence:eqn:6.1}, we have $\tonumber{\n} \setminus \mathban{1} = \mathbs \cup \mathban{2} \cup \bo$. According to 
            the definition of $\mathbs$, $\mathban{2}$, $\bo$, we have $\mathbs \cap \mathban{2} = \emptyset$, 
            and $\mathbs \cap \bo = \emptyset$, which gives $(\mathban{2} \cup \bo) \cap \mathbs = \emptyset$.
            Then, we have $\mathban{2} \cup \bo \subset \tonumber{\n} \setminus \mathbs$.
            Since $\bylnn{3}{\iconstant} = \bylnn{1}{\sigma(\iconstant)}, \text{ for any } \iconstant \in \mathbs,  \text{ and } \bylnn{3}{\iconstant} = \bylnn{1}{\iconstant}, \text{ for any } \iconstant \in \tonumber{\n} \setminus \mathbs$, then for any $\iconstant \in \tonumber{\n} \setminus \mathban{1}$, we have
            \begin{align*}
                &\bylnn{3}{\iconstant} = \bylnn{1}{\sigma(\iconstant)}, \text{ if } \iconstant \in \mathbs, \\
                &\bylnn{3}{\iconstant} = \bylnn{1}{\iconstant}, \text{ if } \iconstant \in \tonumber{\n} \setminus \mathbs.
            \end{align*}	
            Hence, $\svector{\byln{3}}{\lconstant}{\tonumber{\n} \setminus \mathban{1}}$
            is a permutation of $\svector{\byln{1}}{\lconstant}{\tonumber{\n} \setminus \mathban{1}}$.
            Then, we have 
            \begin{align*}
                \rank{\bylnn{3}{\iconstant}}{\svector{\byln{3}}{\lconstant}{\tonumber{\n} \setminus \mathban{1}}}
                = \rank{\bylnn{1}{\sigma(\iconstant)}}{\svector{\byln{1}}{\lconstant}{\tonumber{\n} \setminus \mathban{1}}}, \text{ for any } \iconstant \in \mathbs.
            \end{align*}
            According to \eqref{supp:proposition:equivalence:eqn:5}, we further have 
            for any $ \iconstant \in \mathbs$,
            \begin{align*}
                \rank{\bylnn{3}{\iconstant}}{\svector{\byln{3}}{\lconstant}{\tonumber{\n} \setminus \mathban{1}}}
                = \rank{\bylnn{1}{\sigma(\iconstant)}}{\svector{\byln{1}}{\lconstant}{\tonumber{\n} \setminus \mathban{1}}}
                = \rank{\bylnn{2}{\iconstant}}{\svector{\byln{2}}{\lconstant}{\tonumber{\n} \setminus \mathban{1}}}.
            \end{align*}
            Since $\svector{\byln{3}}{\lconstant}{\tonumber{\n} \setminus \mathban{1}}$
            is a permutation of $\svector{\byln{1}}{\lconstant}{\tonumber{\n} \setminus \mathban{1}}$,
            and for any $\iconstant \in (\tonumber{\n} \setminus \bv) \setminus \mathbs$, we have $\bylnn{3}{\iconstant} = \bylnn{1}{\iconstant}$. Then, we have
            \begin{align*}
                \rank{\bylnn{3}{\iconstant}}{\svector{\byln{3}}{\lconstant}{\tonumber{\n} \setminus \mathban{1}}}
                = \rank{\bylnn{1}{\iconstant}}{\svector{\byln{1}}{\lconstant}{\tonumber{\n} \setminus \mathban{1}}}
                =^{\eqref{supp:proposition:equivalence:eqn:10}} \rank{\bylnn{2}{\iconstant}}{\svector{\byln{2}}{\lconstant}{\tonumber{\n} \setminus \mathban{1}}}.
            \end{align*}
            Hence, for any $\iconstant \in (\tonumber{\n} \setminus \bv) \cup \mathbs$, we have
            \begin{align*}
                \rank{\bylnn{3}{\iconstant}}{\svector{\byln{3}}{\lconstant}{\tonumber{\n} \setminus \mathban{1}}}
                = \rank{\bylnn{2}{\iconstant}}{\svector{\byln{2}}{\lconstant}{\tonumber{\n} \setminus \mathban{1}}}.
            \end{align*}
            According to \eqref{supp:proposition:equivalence:eqn:7}, we have	
            $\mathbs \cup (\tonumber{\n} \setminus \bv) = \tonumber{\n} \setminus \mathban{1}$. 
            Hence, it follows that
            \begin{align} \label{supp:proposition:equivalence:eqn:12}
                \rank{\bylnn{3}{\iconstant}}{\svector{\byln{3}}{\lconstant}{\tonumber{\n} \setminus \mathban{1}}}
                = \rank{\bylnn{2}{\iconstant}}{\svector{\byln{2}}{\lconstant}{\tonumber{\n} \setminus \mathban{1}}},
                \text{ for any } \iconstant \in  \tonumber{\n} \setminus \mathban{1}.
            \end{align}

            Notice that \eqref{supp:proposition:equivalence:eqn:10}, \eqref{supp:proposition:equivalence:eqn:11}
            and \eqref{supp:proposition:equivalence:eqn:12} are true, and 
            $| \mathban{1}| + |\mathban{2}| = (|\bv| - 1 - \aconstant) + (|\bu| - 1 - \bconstant) = |\bv| + |\bu| - 2 - \aconstant - \bconstant = \kk -1 - \aconstant - \bconstant \le \kk -1 \le \kk$.
            Then, since  $\bpn{\iconstant}$ is true for any $\iconstant \le \kk$, we have
            \begin{align*}
                \sfd{\bxln{3}}{\byln{3}} = \sfd{\bxln{2}}{\byln{2}} =^{\eqref{supp:proposition:equivalence:eqn:5.0}}  \sfd{\bxln{1}}{\byln{1}}. 
            \end{align*}
            Therefore, we have shown $\bpn{\kk+1}$ is true. This completes our proof.
        \end{proof}    

        Now, we prove Proposition~2.6 using Proposition~\ref{supp:proposition:equivalence}.

        \begin{proposition} \label{supp:proposition:equivalence:1}
            Suppose $\bx, \by \in \vndistinct$ and $\bu, \bv \subset \tonumber{\n}$
            are disjoint subsets of indices such that $\bu \cap \bv = \emptyset$.
            Suppose $\bxln{1}, \bxln{2} \in \vndistinct$ are imputations of $\bx$ for indices $\bu$,
            $\byln{1}, \byln{2} \in \vndistinct$ are imputations of $\by$ for indices $\bv$,
            and	$\rank{\bxlnn{\jconstant}{\iconstant}}{\bxln{\jconstant}} = \rank{\bylnn{\jconstant}{\iconstant}}{\byln{\jconstant}},$ for any $\iconstant \in \bu \cup \bv, \jconstant \in \{1,2\}$. Then, we have	
            $\sfd{\bxln{1}}{\byln{1}} = \sfd{\bxln{2}}{\byln{2}}$. 
        \end{proposition}

        \begin{proof}
            Since $\bxln{1}, \bxln{2} \in \vndistinct$ are imputations of $\bx$ for indices $\bu$,
            then according to the definition of imputations,
            we have $\bxlnn{1}{\iconstant} = \bxlnn{2}{\iconstant}$ for any $\iconstant \in \tonumber{\n} \setminus \bu$.
            Hence, we have
            \begin{align*}
                &\rank{\bxlnn{1}{\iconstant}}{\svector{\bxln{1}}{\lconstant}{ \tonumber{\n} \setminus \bu}} = \rank{\bxlnn{2}{\iconstant}}{\svector{\bxln{2}}{\lconstant}{\tonumber{\n} \setminus \bu}}, \text{ for any } \iconstant \in \tonumber{\n} \setminus \bu.
            \end{align*}

            Similarly, since $\byln{1}, \byln{2} \in \vndistinct$ are imputations of $\by$ for indices $\bv$, then we have 
            $\bylnn{1}{\iconstant} = \bylnn{2}{\iconstant}$ for any $\iconstant \in \bv$, which
            follows
            \begin{align*}
                \rank{\bylnn{1}{\iconstant}}{\svector{\byln{1}}{\lconstant}{ \tonumber{\n} \setminus \bv}} = \rank{\bylnn{2}{\iconstant}}{\svector{\byln{2}}{\lconstant}{\tonumber{\n} \setminus \bv}}, \text{ for any } \iconstant \in \tonumber{\n} \setminus \bv.
            \end{align*}

            Since $\rank{\bxlnn{\jconstant}{\iconstant}}{\bxln{\jconstant}} = \rank{\bylnn{\jconstant}{\iconstant}}{\byln{\jconstant}},$ for any $\iconstant \in \bu \cup \bv, \jconstant \in \{1,2\}$, then according to 
            Proposition~\ref{supp:proposition:equivalence},
            we have $\sfd{\bxln{1}}{\byln{1}} = \sfd{\bxln{2}}{\byln{2}}$. 
            This completes our proof.
        \end{proof}

        \subsection{Proof of Theorem 2.7}	

        This subsection proves Theorem 2.7. First, we prove some lemmas which
        will be useful for proving Theorem 2.7.

        \begin{lemma} \label{supp:theorem:2.7:lemma:1}
            Suppose $\bx, \by \in \vndistinct$. Let $\bv = \{1,\ldots, \tconstant\},
            \bu = \{\tconstant+1,\ldots, \tconstant +\m\} \subset\tonumber{\n}$ 
            be subsets of indices such that $\tconstant, \m \ge 1$.
            Suppose $\bx(1) < \ldots < \bx(\tconstant)$ 
            and $\by(\tconstant + 1) < \ldots < \by(\tconstant + \m)$.		
            Assume $\rank{\bx(1)}{\svector{\bx}{\lconstant}{\tonumber{\n} \setminus \bu}} =
            \rank{\by(\tconstant + 1)}{\svector{\by}{\lconstant}{\tonumber{\n} \setminus \bv}}$,
            and $\by(1) > \by(\tconstant + 1)$.
            Denote 
            \begin{align} \label{supp:theorem:2.7:lemma:1:eqn:0}
                \alpha = \rank{\bx(1)}{\svector{\bx}{\lconstant}{\tonumber{\n} \setminus \bu}} = \rank{\by(\tconstant + 1)}{\svector{\by}{\lconstant}{\tonumber{\n} \setminus \bv}}.
            \end{align}
            Let $\bxln{1}, \byln{1}$ be imputations of $\bx, \by$
            for indices $\bu$ and $\bv \setminus \{1\}$, respectively.
            Then, if $\rank{\bxlnn{1}{\iconstant}}{\bxln{1}} 
            = \rank{\bylnn{1}{\iconstant}}{\byln{1}} 
            \text{ for any } \iconstant \in \bu \cup (\bv \setminus \{1\})$,
            we have
            \begin{align}
                \begin{split}
                    \rank{\byln{1}(\tconstant+1)}{\svector{\byln{1}}{\lconstant}{\tonumber{\n} \setminus \bv}} &= \rank{\bylnn{1}{\tconstant+1}}{\byln{1}} \\
                    &= \rank{\bylnn{1}{\tconstant+1}}{\svector{\byln{1}}{\lconstant}{\tonumber{\n} \setminus (\bv \setminus \{1\})}} = \alpha, \label{supp:theorem:2.7:lemma:1:eqn:0.0}
                \end{split}
            \end{align}	
            and
            \begin{align}
                \begin{split}
                    \rank{\bxln{1}(1)}{\svector{\bxln{1}}{\lconstant}{\tonumber{\n} \setminus \bu}} &= \rank{\bxlnn{1}{\tconstant+1}}{\bxln{1}} \\
                    &= \rank{\bxlnn{1}{\tconstant+1}}{\svector{\bxln{1}}{\lconstant}{\tonumber{\n} \setminus (\bu \setminus \{\tconstant+1\})}} = \alpha. \label{supp:theorem:2.7:lemma:1:eqn:0.1}	
                \end{split}		
            \end{align}
            Additionally, we have 
            $\bxln{1}(1) < \ldots < \bxln{1}(\tconstant)$, and $\byln{1}(\tconstant + 1) < \ldots < \byln{1}(\tconstant + \m)$. 
        \end{lemma}
        \begin{proof}
            Since $\bxln{1}, \byln{1}$ are imputations of $\bx$, and $\by$
            for indices $\bu$ and $\bv \setminus \{1\}$, respectively, we have
            \begin{align}
                &\bxln{1}(\iconstant) = \bx(\iconstant), \text{ for any } \iconstant \in \tonumber{\n} \setminus \bu, \label{supp:theorem:2.7:lemma:1:eqn:2}\\
                \text{ and }&\byln{1}(\iconstant) = \by(\iconstant), \text{ for any } \iconstant \in \tonumber{\n} \setminus (\bv \setminus \{1\}). \label{supp:theorem:2.7:lemma:1:eqn:3}
            \end{align} 

            We first show that  $\bxln{1}(1) < \ldots < \bxln{1}(\tconstant)$, and $\byln{1}(\tconstant + 1) < \ldots < \byln{1}(\tconstant + \m)$. 		
            Notice that $\{1,\ldots, \tconstant\} \subset \tonumber{\n} \setminus \bu$,
            and $\{\tconstant+1, \tconstant+\m\} \subset \tonumber{\n} \setminus \bv \subset \tonumber{\n} \setminus (\bv \setminus \{1\})$.
            Then according to \eqref{supp:theorem:2.7:lemma:1:eqn:2} and \eqref{supp:theorem:2.7:lemma:1:eqn:3},
            we have $\bxlnn{1}{\iconstant} = \bxn{\iconstant}, \text{ for any } \iconstant \in \{1,\ldots,\tconstant\}$ and $\bylnn{1}{\iconstant} = \byn{\iconstant}, \text{ for any } 
            \iconstant \in \{\tconstant+1, \ldots, \tconstant+\m\}$, respectively.
            Then, since $\bx(1) < \ldots < \bx(\tconstant)$ 
            and $\by(\tconstant + 1) < \ldots < \by(\tconstant + \m)$, 
            we have $\bxln{1}(1) < \ldots < \bxln{1}(\tconstant)$, 
            and $\byln{1}(\tconstant + 1) < \ldots < \byln{1}(\tconstant + \m)$. 

            Next, we show that \eqref{supp:theorem:2.7:lemma:1:eqn:0.0} is true. To start,		
            according to the definition of rank, we have
            \begin{align*}
                \rank{\by(\tconstant + 1)}{\svector{\by}{\lconstant}{\{1\} \cup (\tonumber{\n} \setminus \bv)}} &= \sum_{\lconstant \in \{1\} \cup (\tonumber{\n} \setminus \bv) } \indicator{\by(\lconstant) \le \by(\tconstant + 1)} \\
                & = \indicator{\by(1) \le \by(\tconstant + 1)} + \sum_{\lconstant \in \tonumber{\n} \setminus \bv} \indicator{\by(\lconstant) \le \by(\tconstant + 1)}.
            \end{align*}
            Since $\by(1) > \by(\tconstant + 1)$, we have $\indicator{\by(1) \le \by(\tconstant + 1)} = 0$.
            Hence, we have
            \begin{align*}
                \rank{\by(\tconstant + 1)}{\svector{\by}{\lconstant}{\{1\} \cup (\tonumber{\n} \setminus \bv)}} &=
                \sum_{\lconstant \in \tonumber{\n} \setminus \bv} \indicator{\by(\lconstant) \le \by(\tconstant + 1)}\\
                & = \rank{\by(\tconstant + 1)}{\svector{\by}{\lconstant}{\tonumber{\n} \setminus \bv}} =^{\eqref{supp:theorem:2.7:lemma:1:eqn:0}} \alpha.
            \end{align*}

            Since $\{1\} \cup (\tonumber{\n} \setminus \bv) = \{1\} \cup \{\tconstant+1, \ldots, \n\}
            = \tonumber{\n} \setminus (\{1,\ldots,\tconstant\}\setminus \{1\}) = 
            \tonumber{\n} \setminus (\bv \setminus \{1\})$,
            we have $\svector{\by}{\lconstant}{\{1\} \cup (\tonumber{\n} \setminus \bv)} 
            = \svector{\by}{\lconstant}{\tonumber{\n} \setminus (\bv \setminus \{1\})}$.
            Then, it follows that
            \begin{align*}
                \rank{\by(\tconstant + 1)}{\svector{\by}{\lconstant}{\{1\} \cup (\tonumber{\n} \setminus \bv)}}
                = \rank{\by(\tconstant + 1)}{\svector{\by}{\lconstant}{\tonumber{\n} \setminus (\bv \setminus \{1\})}}.
            \end{align*}
            Hence, we have
            \begin{align} \label{supp:theorem:2.7:lemma:1:eqn:1}
                \rank{\by(\tconstant + 1)}{\svector{\by}{\lconstant}{\tonumber{\n} \setminus (\bv \setminus \{1\})}} = \alpha.
            \end{align}

            Since $\bu = \{\tconstant+1,\ldots,\tconstant+\m\}$, where $\tconstant, \m \ge 1$, 
            we have $1 \in \tonumber{\n} \setminus \bu$.
            Then, according to \eqref{supp:theorem:2.7:lemma:1:eqn:2},
            we have $\bxlnn{1}{1} = \bxn{1}$. Notice that 
            \eqref{supp:theorem:2.7:lemma:1:eqn:2} also means that
            $\svector{\bxln{1}}{\lconstant}{\tonumber{\n} \setminus \bu}
            = \svector{\bx}{\lconstant}{\tonumber{\n} \setminus \bu}$.
            Hence, we have
            \begin{align*}
                \rank{\bxlnn{1}{1}}{\svector{\bxln{1}}{\lconstant}{\tonumber{\n} \setminus \bu}} = \rank{\bxn{1}}{\svector{\bx}{\lconstant}{\tonumber{\n} \setminus \bu}} =^{\eqref{supp:theorem:2.7:lemma:1:eqn:0}} \alpha.
            \end{align*}

            Similarly, since $\bv = \{1,2,\ldots,\tconstant\}$, where $\tconstant \ge 1$,
            we have $\tconstant+1 \in \tonumber{\n} \setminus (\bv \setminus \{1\})$. 
            Then, according to \eqref{supp:theorem:2.7:lemma:1:eqn:3},
            we have $\bylnn{1}{\tconstant+1} = \byn{\tconstant+1}$. Notice that 
            \eqref{supp:theorem:2.7:lemma:1:eqn:3} also means that
            $\svector{\byln{1}}{\lconstant}{\tonumber{\n} \setminus (\bv \setminus \{1\})} = \svector{\by}{\lconstant}{\tonumber{\n} \setminus (\bv \setminus \{1\})}$.
            Hence, we have
            \begin{align*}
                \rank{\bylnn{1}{\tconstant+1}}{\svector{\byln{1}}{\lconstant}{\tonumber{\n} \setminus (\bv \setminus \{1\})}} = \rank{\byn{\tconstant+1}}{\svector{\by}{\lconstant}{\tonumber{\n} \setminus (\bv \setminus \{1\})}} =^{\eqref{supp:theorem:2.7:lemma:1:eqn:1}} \alpha.
            \end{align*}
            Notice that $\tonumber{\n} \setminus \bv \subset \tonumber{\n} \setminus (\bv \setminus \{1\})$.
            Then, according to \eqref{supp:theorem:2.7:lemma:1:eqn:3},
            we have $\byln{1}(\iconstant) = \by(\iconstant), \text{ for any } \iconstant \in \tonumber{\n} \setminus \bv$. Hence, we have $\svector{\byln{1}}{\lconstant}{\tonumber{\n} \setminus \bv} = \svector{\by}{\lconstant}{\tonumber{\n} \setminus \bv}$.
            Then, it follows that 		
            \begin{align*}
                &\rank{\bylnn{1}{\tconstant+1}}{\svector{\byln{1}}{\lconstant}{\tonumber{\n} \setminus \bv}} = \rank{\byn{\tconstant+1}}{\svector{\by}{\lconstant}{\tonumber{\n} \setminus \bv}} =^{\eqref{supp:theorem:2.7:lemma:1:eqn:0}} \alpha.
            \end{align*}
            Combining the above results, we have
            \begin{align} 
                \begin{split} \label{supp:theorem:2.7:lemma:1:eqn:4}
                    \rank{\bxlnn{1}{1}}{\svector{\bxln{1}}{\lconstant}{\tonumber{\n} \setminus \bu}}  &= \rank{\bylnn{1}{\tconstant+1}}{\svector{\byln{1}}{\lconstant}{\tonumber{\n} \setminus \bv}}\\
                    &= \rank{\bylnn{1}{\tconstant+1}}{\svector{\byln{1}}{\lconstant}{\tonumber{\n} \setminus (\bv \setminus \{1\})}}  =\alpha.
                \end{split}
            \end{align}

            Next, suppose $\bv \setminus \{1\} \neq \emptyset$.
            Then according to Lemma \ref{supp:lemma:2}, we have
            \begin{align*}
                \rank{\bxlnn{1}{\iconstant}}{\bxln{1}} \ge \rank{\bxlnn{1}{\iconstant}}{\svector{\bxln{1}}{\lconstant}{\tonumber{\n} \setminus \bu}}, \text{ for any } \iconstant \in \bv \setminus \{1\}.
            \end{align*}
            Since $\rank{\bxlnn{1}{\iconstant}}{\bxln{1}} 
            = \rank{\bylnn{1}{\iconstant}}{\byln{1}} 
            \text{ for any } \iconstant \in \bu \cup (\bv \setminus \{1\})$,
            then we have 
            \begin{align*}
                \rank{\bylnn{1}{\iconstant}}{\byln{1}} &= \rank{\bxlnn{1}{\iconstant}}{\bxln{1}} \ge \rank{\bxlnn{1}{\iconstant}}{\svector{\bxln{1}}{\lconstant}{\tonumber{\n} \setminus \bu}}, \text{ for any } \iconstant \in \bv \setminus \{1\}.
            \end{align*}
            Since $\bxln{1}(1) < \ldots < \bxln{1}(\tconstant)$, then
            for any $\iconstant \in \bv \setminus \{1\}$, we further have
            \begin{align*}
                \rank{\bylnn{1}{\iconstant}}{\byln{1}} &\ge \rank{\bxlnn{1}{\iconstant}}{\svector{\bxln{1}}{\lconstant}{\tonumber{\n} \setminus \bu}} \\
                & > \rank{\bxlnn{1}{1}}{\svector{\bxln{1}}{\lconstant}{\tonumber{\n} \setminus \bu}} \\
                & =^{ \eqref{supp:theorem:2.7:lemma:1:eqn:4}} \rank{\bylnn{1}{\tconstant+1}}{\svector{\byln{1}}{\lconstant}{\tonumber{\n} \setminus (\bv \setminus \{1\})}} =^{ \eqref{supp:theorem:2.7:lemma:1:eqn:4}} \alpha.
            \end{align*}
            Hence, for any $\iconstant \in  \bv \setminus \{1\}$, we have
            \begin{align*}
                \rank{\bylnn{1}{\iconstant}}{\byln{1}} > \rank{\bylnn{1}{\tconstant+1}}{\svector{\byln{1}}{\lconstant}{\tonumber{\n} \setminus (\bv \setminus \{1\})}}.
            \end{align*}
            Subsequently, according to Lemma~\ref{supp:lemma:3}, we have
            \begin{align} \label{supp:theorem:2.7:lemma:1:eqn:5}
                \rank{\bylnn{1}{\tconstant+1}}{\byln{1}} = \rank{\bylnn{1}{\tconstant+1}}{\svector{\byln{1}}{\lconstant}{\tonumber{\n} \setminus (\bv \setminus \{1\})}} =^{\eqref{supp:theorem:2.7:lemma:1:eqn:4}} \alpha.
            \end{align}
            In particular, the above equations still hold when $\bv \setminus \{1\} = \emptyset$.
            Combining \eqref{supp:theorem:2.7:lemma:1:eqn:4} and \eqref{supp:theorem:2.7:lemma:1:eqn:5},
            we have 
            \begin{align*}
                \rank{\byln{1}(\tconstant+1)}{\svector{\byln{1}}{\lconstant}{\tonumber{\n} \setminus \bv}}
                &= \rank{\bylnn{1}{\tconstant+1}}{\svector{\byln{1}}{\lconstant}{\tonumber{\n} \setminus (\bv \setminus \{1\})}}\\
                &= \rank{\bylnn{1}{\tconstant+1}}{\byln{1}} = \alpha.
            \end{align*}
            This proves \eqref{supp:theorem:2.7:lemma:1:eqn:0.0}.

            Next, we show that \eqref{supp:theorem:2.7:lemma:1:eqn:0.1} is true.
            Suppose $\bu \setminus \{\tconstant+1\} \neq \emptyset$.
            Since $\rank{\bxlnn{1}{\iconstant}}{\bxln{1}} 
            = \rank{\bylnn{1}{\iconstant}}{\byln{1}} 
            \text{ for any } \iconstant \in \bu \cup (\bv \setminus \{1\})$, 
            we have
            \begin{align*}
                \rank{\bxlnn{1}{\iconstant}}{\bxln{1}} = \rank{\bylnn{1}{\iconstant}}{\byln{1}},\text{ for any }
                \iconstant \in \bu \setminus \{\tconstant+1\}.
            \end{align*}
            According to Lemma \ref{supp:lemma:2}, we have
            \begin{align*}
                \rank{\bylnn{1}{\iconstant}}{\byln{1}} \ge \rank{\bylnn{1}{\iconstant}}{\svector{\byln{1}}{\lconstant}{\tonumber{\n} \setminus \bv}}, \text{ for any }
                \iconstant \in \bu \setminus \{\tconstant+1\}.
            \end{align*}
            Further, since $\bylnn{1}{\tconstant+1} < \ldots < \bylnn{1}{\tconstant+\m}$,
            we have
            \begin{align*}
                \rank{\bylnn{1}{\iconstant}}{\svector{\byln{1}}{\lconstant}{\tonumber{\n} \setminus \bv}}
                &> \rank{\bylnn{1}{\tconstant+1}}{\svector{\byln{1}}{\lconstant}{\tonumber{\n} \setminus \bv}}, \text{ for any } \iconstant \in \bu \setminus \{\tconstant+1\}.
            \end{align*}
            Hence, we have 
            \begin{align*}
                \rank{\bxlnn{1}{\iconstant}}{\bxln{1}} > \rank{\bylnn{1}{\tconstant+1}}{\svector{\byln{1}}{\lconstant}{\tonumber{\n} \setminus \bv}} =^{\eqref{supp:theorem:2.7:lemma:1:eqn:0}} \alpha, \text{ for any } \iconstant \in \bu \setminus \{\tconstant+1\}.
            \end{align*}
            Next, notice that $\rank{\bxlnn{1}{\iconstant}}{\bxln{1}} 
            = \rank{\bylnn{1}{\iconstant}}{\byln{1}} 
            \text{ for any } \iconstant \in \bu \cup (\bv \setminus \{1\})$,
            and $\tconstant+1 \in \bu$. Then we have 
            \begin{align*}
                \rank{\bxlnn{1}{\tconstant+1}}{\bxln{1}} = \rank{\bylnn{1}{\tconstant+1}}{\byln{1}} =^{\eqref{supp:theorem:2.7:lemma:1:eqn:5}} \alpha.
            \end{align*}
            Then, according to Lemma~\ref{supp:lemma:3}, we have
            \begin{align} \label{supp:theorem:2.7:lemma:1:eqn:6}
                \rank{\bxlnn{1}{\tconstant+1}}{\svector{\bxln{1}}{\lconstant}{\tonumber{\n} \setminus (\bu \setminus \{\tconstant+1\})}}  = \rank{\bxlnn{1}{\tconstant+1}}{\bxln{1}} = \alpha.
            \end{align}
            In particular, the above equations still hold when $\bu \setminus \{\tconstant+1\} = \emptyset$.
            Combining \eqref{supp:theorem:2.7:lemma:1:eqn:4} and \eqref{supp:theorem:2.7:lemma:1:eqn:6},
            we have 
            \begin{align*}
                \rank{\bxlnn{1}{1}}{\svector{\bxln{1}}{\lconstant}{\tonumber{\n} \setminus \bu}} &=
                \rank{\bxlnn{1}{\tconstant+1}}{\svector{\bxln{1}}{\lconstant}{\tonumber{\n} \setminus (\bu \setminus \{\tconstant+1\})}} \\
                & = \rank{\bxlnn{1}{\tconstant+1}}{\bxln{1}} = \alpha.
            \end{align*}
            This proves \eqref{supp:theorem:2.7:lemma:1:eqn:0.1}, and completes our proof.

        \end{proof}

        \begin{lemma} \label{supp:theorem:2.7:lemma:2}
            Following Lemma~\ref{supp:theorem:2.7:lemma:1}, let $\bxln{2}$ and $\byln{2}$ 
            be imputations of $\bxln{1}$, $\byln{1}$
            for indices $\bu \setminus \{\tconstant+1\}$ and $\bv$, respectively, such that
            $\rank{\bxlnn{2}{\iconstant}}{\bxln{2}} = \rank{\bylnn{2}{\iconstant}}{\byln{2}} \text{ for any } \iconstant \in (\bu \setminus \{\tconstant+1\}) \cup \bv$.
            Then, we have $ \rank{\bxlnn{2}{\iconstant}}{\bxln{2}} = \rank{\bylnn{2}{\iconstant}}{\byln{2}}, \text{ for any } \iconstant \in \bu \cup \bv.$
        \end{lemma}

        \begin{proof}
            According to the definition of imputations, since $\bxln{2}$ and $\byln{2}$ are imputations of $\bxln{1}$, $\byln{1}$ or indices $\bu \setminus \{\tconstant+1\}$ and $\bv$, respectively,
            we have
            \begin{align}
                &\bxlnn{2}{\iconstant} = \bxlnn{1}{\iconstant}, \text{ for any } \iconstant \in \tonumber{\n} \setminus (\bu \setminus \{\tconstant+1\}), \label{supp:theorem:2.7:lemma:2:eqn:1}\\
                \text{ and }&\bylnn{2}{\iconstant} = \bylnn{1}{\iconstant}, \text{ for any } \iconstant \in \tonumber{\n} \setminus \bv. \label{supp:theorem:2.7:lemma:2:eqn:2}
            \end{align} 

            According to the definition of Lemma~\ref{supp:theorem:2.7:lemma:1},
            we have $\bxlnn{1}{1} < \ldots < \bxlnn{1}{\tconstant}$ 
            and $\bylnn{1}{\tconstant+1} < \ldots < \bylnn{1}{\tconstant+\m}$. 
            Notice that $\{1,\ldots, \tconstant\} \subset \tonumber{\n} \setminus \bu \subset \tonumber{\n} \setminus (\bu \setminus \{\tconstant + 1\})$,
            and $\{\tconstant+1, \tconstant+\m\} \subset \tonumber{\n} \setminus \bv$.
            Then according to \eqref{supp:theorem:2.7:lemma:2:eqn:1} and \eqref{supp:theorem:2.7:lemma:2:eqn:2},
            we have $\bxlnn{2}{\iconstant} = \bxlnn{1}{\iconstant}, \text{ for any } \iconstant \in \{1,\ldots,\tconstant\}$ and $\bylnn{2}{\iconstant} = \bylnn{1}{\iconstant}, \text{ for any } 
            \iconstant \in \{\tconstant+1,\ldots, \tconstant+\m\}$, respectively.
            Then, since $\bxlnn{1}{1} < \ldots < \bxlnn{1}{\tconstant}$ 
            and $\bylnn{1}{\tconstant+1} < \ldots < \bylnn{1}{\tconstant+\m}$,
            we have  $\bxlnn{2}{1} < \ldots < \bxlnn{2}{\tconstant}$ and $\bylnn{2}{\tconstant+1} < \ldots < \bylnn{2}{\tconstant+\m}$. 

            Since $\tconstant+1 \in \tonumber{\n} \setminus (\bu \setminus \{\tconstant+1\})$, 
            then according to \eqref{supp:theorem:2.7:lemma:2:eqn:1}, we have
            $\bxlnn{2}{\tconstant+1} = \bxlnn{1}{\tconstant+1}$.
            Notice that \eqref{supp:theorem:2.7:lemma:2:eqn:1} also gives
            $\svector{\bxln{2}}{\lconstant}{\tonumber{\n} \setminus (\bu \setminus \{\tconstant+1\})}
            = \svector{\bxln{1}}{\lconstant}{\tonumber{\n} \setminus (\bu \setminus \{\tconstant+1\})}$.
            Hence, we have
            \begin{align*}
                \rank{\bxlnn{2}{\tconstant+1}}{\svector{\bxln{2}}{\lconstant}{\tonumber{\n} \setminus (\bu \setminus \{\tconstant+1\})}} = \rank{\bxlnn{1}{\tconstant+1}}{\svector{\bxln{1}}{\lconstant}{\tonumber{\n} \setminus (\bu \setminus \{\tconstant+1\})}} =^{\eqref{supp:theorem:2.7:lemma:1:eqn:0.1}} \alpha.
            \end{align*}

            Next, since $\bv = \{1,\ldots,\tconstant\}$, 
            we have $\tconstant+1 \in \tonumber{\n} \setminus \bv$. 
            Then, according to \eqref{supp:theorem:2.7:lemma:2:eqn:2},
            we have $\bylnn{2}{\tconstant+1} = \bylnn{1}{\tconstant+1}$.
            Notice that \eqref{supp:theorem:2.7:lemma:2:eqn:2} also gives
            $ \svector{\byln{2}}{\lconstant}{\tonumber{\n} \setminus \bv} 
            = \svector{\byln{1}}{\lconstant}{\tonumber{\n} \setminus \bv}$.
            Hence, we have
            \begin{align*}
                \rank{\bylnn{2}{\tconstant+1}}{\svector{\byln{2}}{\lconstant}{\tonumber{\n} \setminus \bv}} = \rank{\bylnn{1}{\tconstant+1}}{\svector{\byln{1}}{\lconstant}{\tonumber{\n} \setminus \bv}} =^{\eqref{supp:theorem:2.7:lemma:1:eqn:0.0}} \alpha.
            \end{align*}
            Combining the above results, we have
            \begin{align} \label{supp:theorem:2.7:lemma:2:eqn:3}
                \rank{\bxlnn{2}{\tconstant+1}}{\svector{\bxln{2}}{\lconstant}{\tonumber{\n} \setminus (\bu \setminus \{\tconstant+1\})}}  = 	\rank{\bylnn{2}{\tconstant+1}}{\svector{\byln{2}}{\lconstant}{\tonumber{\n} \setminus \bv}}  = \alpha.
            \end{align}

            Suppose $\bu \setminus \{\tconstant+1\} \neq \emptyset$.
            Then, according to Lemma~\ref{supp:lemma:2}, we have
            \begin{align*}
                \rank{\bylnn{2}{\iconstant}}{\byln{2}}  \geq \rank{\bylnn{2}{\iconstant}}{\svector{\byln{2}}{\lconstant}{\tonumber{\n} \setminus \bv}}, \text{ for any } \bu \setminus \{\tconstant+1\}.
            \end{align*}
            Since $\rank{\bxlnn{2}{\iconstant}}{\bxln{2}} = \rank{\bylnn{2}{\iconstant}}{\byln{2}}$ for any $\iconstant \in (\bu \setminus \{\tconstant+1\}) \cup \bv$, we have
            $\rank{\bxlnn{2}{\iconstant}}{\bxln{2}} = \rank{\bylnn{2}{\iconstant}}{\byln{2}}$
            for any $\iconstant \in \bu \setminus \{\tconstant+1\}$. Then, it follows that
            \begin{align*}
                \rank{\bxlnn{2}{\iconstant}}{\bxln{2}} &= \rank{\bylnn{2}{\iconstant}}{\byln{2}}  \geq \rank{\bylnn{2}{\iconstant}}{\svector{\byln{2}}{\lconstant}{\tonumber{\n} \setminus \bv}}, \text{ for any } \bu \setminus \{\tconstant+1\}.
            \end{align*}		
            Since $\bylnn{2}{\tconstant+1} < \ldots < \bylnn{2}{\tconstant+\m}$, 
            then for any $\iconstant \in \bu \setminus \{\tconstant+1\} = \{\tconstant+2,\ldots,\tconstant+\m\}$,
            we have
            \begin{align*}
                \rank{\bxlnn{2}{\iconstant}}{\bxln{2}} 
                & \ge \rank{\bylnn{2}{\iconstant}}{\svector{\byln{2}}{\lconstant}{\tonumber{\n} \setminus \bv}}\\ 
                &>  \rank{\bylnn{2}{\tconstant+1}}{\svector{\byln{2}}{\lconstant}{\tonumber{\n} \setminus \bv}} \\
                & =^{\eqref{supp:theorem:2.7:lemma:2:eqn:3}} \rank{\bxlnn{2}{\tconstant+1}}{\svector{\bxln{2}}{\lconstant}{\tonumber{\n} \setminus (\bu \setminus \{\tconstant+1\})}}.
            \end{align*}
            Then, according to Lemma~\ref{supp:lemma:3}, we have
            \begin{align} \label{supp:theorem:2.7:lemma:2:eqn:4}
                \rank{\bxlnn{2}{\tconstant+1}}{\bxln{2}} &= \rank{\bxlnn{2}{\tconstant+1}}{\svector{\bxln{2}}{\lconstant}{\tonumber{\n} \setminus (\bu \setminus \{\tconstant+1\})}} =^{\eqref{supp:theorem:2.7:lemma:2:eqn:3}} \alpha.
            \end{align}
            In particular, the above equation still holds when $\bu \setminus \{\tconstant+1\} = \emptyset$. 

            Next, notice that $\tonumber{\n} \setminus \bu \subset \tonumber{\n} \setminus (\bu \setminus \{\tconstant+1\})$. Then, according to Lemma~\ref{supp:lemma:2}, we have
            \begin{align*}
                \rank{\bxlnn{2}{1}}{\svector{\bxln{2}}{\lconstant}{\tonumber{\n} \setminus (\bu \setminus \{\tconstant+1\})}} \ge \rank{\bxlnn{2}{1}}{\svector{\bxln{2}}{\lconstant}{\tonumber{\n} \setminus \bu}}.
            \end{align*}
            Since $\bu = \{\tconstant+1, \ldots, \tconstant+\m\}$, where $\tconstant, \m \geq 1$, 
            we have $1 \in \tonumber{\n} \setminus (\bu \setminus \{\tconstant+1\})$. 
            Hence, according to \eqref{supp:theorem:2.7:lemma:2:eqn:1}, we have $\bxlnn{2}{1} = \bxlnn{1}{1}$.
            Meanwhile, notice that $\tonumber{\n} \setminus \bu \subset \tonumber{\n} \setminus (\bu \setminus \{\tconstant+1\})$. Then \eqref{supp:theorem:2.7:lemma:2:eqn:1} also gives that
            $\svector{\bxln{2}}{\lconstant}{\tonumber{\n} \setminus \bu}
            = \svector{\bxln{1}}{\lconstant}{\tonumber{\n} \setminus \bu}$. 
            Hence, we have
            \begin{align*}
                \rank{\bxlnn{2}{1}}{\svector{\bxln{2}}{\lconstant}{\tonumber{\n} \setminus \bu}} &= \rank{\bxlnn{1}{1}}{\svector{\bxln{1}}{\lconstant}{\tonumber{\n} \setminus \bu}} \\
                &=^{\eqref{supp:theorem:2.7:lemma:1:eqn:0.0}} \alpha  
                =^{\eqref{supp:theorem:2.7:lemma:2:eqn:4}} \rank{\bxlnn{2}{\tconstant+1}}{\svector{\bxln{2}}{\lconstant}{\tonumber{\n} \setminus (\bu \setminus \{\tconstant+1\})}}.  
            \end{align*}
            Therefore, we have
            \begin{align*}
                \rank{\bxlnn{2}{1}}{\svector{\bxln{2}}{\lconstant}{\tonumber{\n} \setminus (\bu \setminus \{\tconstant+1\})}} \ge \rank{\bxlnn{2}{\tconstant+1}}{\svector{\bxln{2}}{\lconstant}{\tonumber{\n} \setminus (\bu \setminus \{\tconstant+1\})}}.
            \end{align*}
            Since $\bxln{2} \in \vndistinct$ is a vector of distinct real values,
            \begin{align*}
                &\rank{\bxlnn{2}{1}}{\svector{\bxln{2}}{\lconstant}{\tonumber{\n} \setminus (\bu \setminus \{\tconstant+1\})}} 
                \neq \rank{\bxlnn{2}{\tconstant+1}}{\svector{\bxln{2}}{\lconstant}{\tonumber{\n} \setminus (\bu \setminus \{\tconstant+1\})}} \\
                &\Rightarrow \rank{\bxlnn{2}{1}}{\svector{\bxln{2}}{\lconstant}{\tonumber{\n} \setminus (\bu \setminus \{\tconstant+1\})}} > \rank{\bxlnn{2}{\tconstant+1}}{\svector{\bxln{2}}{\lconstant}{\tonumber{\n} \setminus (\bu \setminus \{\tconstant+1\})}} =^{\eqref{supp:theorem:2.7:lemma:2:eqn:4}} \alpha.
            \end{align*}
            Since $\bxlnn{2}{1} < \ldots < \bxlnn{2}{\tconstant}$. For any $\iconstant \in \bv = \{1,\ldots, \tconstant\}$, we have
            \begin{align*}
                \rank{\bxlnn{2}{\iconstant}}{\svector{\bxln{2}}{\lconstant}{\tonumber{\n} \setminus (\bu \setminus \{\tconstant+1\})}} 
                &\geq \rank{\bxlnn{2}{1}}{\svector{\bxln{2}}{\lconstant}{\tonumber{\n} \setminus (\bu \setminus \{\tconstant+1\})}} > \alpha.
            \end{align*}
            Then, according to Lemma~\ref{supp:lemma:2}, 
            for any $\iconstant \in \bv = \{1, \ldots, \tconstant\}$, we have
            \begin{align*}
                \rank{\bxlnn{2}{\iconstant}}{\bxln{2}}  
                &\geq \rank{\bxlnn{2}{\iconstant}}{\svector{\bxln{2}}{\lconstant}{\tonumber{\n} \setminus (\bu \setminus \{\tconstant+1\})}} \\
                &> \alpha =^\eqref{supp:theorem:2.7:lemma:2:eqn:3} \rank{\bylnn{2}{\tconstant+1}}{\svector{\byln{2}}{\lconstant}{\tonumber{\n} \setminus \bv}}.
            \end{align*}
            Since $\rank{\bxlnn{2}{\iconstant}}{\bxln{2}} = \rank{\bylnn{2}{\iconstant}}{\byln{2}}$ for any $\iconstant \in (\bu \setminus \{\tconstant+1\}) \cup \bv$,
            we have
            $\rank{\bylnn{2}{\iconstant}}{\byln{2}} = \rank{\bxlnn{2}{\iconstant}}{\bxln{2}}$
            for any $\iconstant \in \bv$. Then, it follows that
            \begin{align*}
                \rank{\bylnn{2}{\iconstant}}{\byln{2}} = \rank{\bxlnn{2}{\iconstant}}{\bxln{2}}  > \rank{\bylnn{2}{\tconstant+1}}{\svector{\byln{2}}{\lconstant}{\tonumber{\n} \setminus \bv}},
                \text{ for any } \iconstant \in \bv.
            \end{align*}
            Subsequently, according to       
            Lemma~\ref{supp:lemma:3}, we have
            \begin{align*}
                \rank{\bylnn{2}{\tconstant+1}}{\byln{2}} = \rank{\bylnn{2}{\tconstant+1}}{\svector{\byln{2}}{\lconstant}{\tonumber{\n} \setminus \bv}} =^\eqref{supp:theorem:2.7:lemma:2:eqn:2} \alpha =^\eqref{supp:theorem:2.7:lemma:2:eqn:3} \rank{\bxlnn{2}{\tconstant+1}}{\bxln{2}}.
            \end{align*}
            Hence, we obtain $\rank{\bylnn{2}{\tconstant+1}}{\byln{2}} = \rank{\bxlnn{2}{\tconstant+1}}{\bxln{2}}$.
            Since $\rank{\bxlnn{2}{\iconstant}}{\bxln{2}} = \rank{\bylnn{2}{\iconstant}}{\byln{2}}$ for any $\iconstant \in (\bu \setminus \{\tconstant+1\}) \cup \bv$,
            then we have
            \begin{align*}
                \rank{\bxlnn{2}{\iconstant}}{\bxln{2}} = \rank{\bylnn{2}{\iconstant}}{\byln{2}}, \text{for any } \iconstant \in \bu \cup \bv.
            \end{align*}
            This completes our proof.
        \end{proof}

        \begin{lemma} \label{supp:theorem:2.7:lemma:3}
            Suppose $\bx, \by \in \vndistinct$. Let $\bv = \{1,\ldots, \tconstant\},
            \bu = \{\tconstant+1,\ldots, \tconstant +\m\} \subset\tonumber{\n}$ 
            be subsets of indices such that $\tconstant, \m \ge 1$.
            Suppose $\bx(1) < \ldots < \bx(\tconstant)$ 
            and $\by(\tconstant + 1) < \ldots < \by(\tconstant + \m)$,		
            and assume $\rank{\byn{\tconstant+1}}{\svector{\by}{\lconstant}{\{1\} \cup (\tonumber{\n}\setminus \bv)}} 
            > \rank{\bxn{1}}{\svector{\bx}{\lconstant}{\tonumber{\n}\setminus \bu}}$. Let us denote
            \begin{align} \label{supp:theorem:2.7:lemma:3:eqn:1}
                \alpha = \rank{\bxn{1}}{\svector{\bx}{\lconstant}{\tonumber{\n}\setminus \bu}}.
            \end{align}
            Let  $\bxln{1}$ and $\byln{1}$ be imputations of $\bx$ and $\by$
            for $\bu$ and $\bv \setminus \{1\}$ such that
            $\rank{\bxlnn{1}{\iconstant}}{\bxln{1}} = \rank{\bylnn{1}{\iconstant}}{\byln{1}} 
            \text{ for any } \iconstant \in (\bv \setminus \{1\}) \cup \bu$.
            Then, when $ \bv \setminus \{1\} \neq \emptyset$,
            for any $\bv \setminus \{1\}$,
            \begin{align*}
                \rank{\bylnn{1}{\iconstant}}{\svector{\byln{1}}{\lconstant}{\tonumber{\n} \setminus \{1\}}} \geq \rank{\bxlnn{1}{1}}{\svector{\bxln{1}}{\lconstant}{\tonumber{\n} \setminus \bu}} = \rank{\bxlnn{1}{1}}{\bxln{1}} = \alpha.
            \end{align*}
            Additionally, we have $\bxln{1}(1) < \ldots < \bxln{1}(\tconstant)$, 
            and $\byln{1}(\tconstant + 1) < \ldots < \byln{1}(\tconstant + \m)$. 
        \end{lemma}

        \begin{proof}
            Since $\bxln{1}, \byln{1}$ are imputations of $\bx$, and $\by$
            for indices $\bu$ and $\bv \setminus \{1\}$, respectively, we have
            \begin{align}
                &\bxln{1}(\iconstant) = \bx(\iconstant), \text{ for any } \iconstant \in \tonumber{\n} \setminus \bu, \label{supp:theorem:2.7:lemma:3:eqn:2}\\
                \text{ and }&\byln{1}(\iconstant) = \by(\iconstant), \text{ for any } \iconstant \in \tonumber{\n} \setminus (\bv \setminus \{1\}). \label{supp:theorem:2.7:lemma:3:eqn:3}
            \end{align} 
            To start, we first show that  $\bxln{1}(1) < \ldots < \bxln{1}(\tconstant)$, and $\byln{1}(\tconstant + 1) < \ldots < \byln{1}(\tconstant + \m)$. 		
            Notice that $\{1,\ldots, \tconstant\} \subset \tonumber{\n} \setminus \bu$,
            and $\{\tconstant+1, \tconstant+\m\} \subset \tonumber{\n} \setminus \bv \subset \tonumber{\n} \setminus (\bv \setminus \{1\})$.
            Then according to \eqref{supp:theorem:2.7:lemma:3:eqn:2} and \eqref{supp:theorem:2.7:lemma:3:eqn:3},
            we have $\bxlnn{1}{\iconstant} = \bxn{\iconstant}, \text{ for any } \iconstant \in \{1,\ldots,\tconstant\}$ and $\bylnn{1}{\iconstant} = \byn{\iconstant}, \text{ for any } 
            \iconstant \in \{\tconstant+1, \ldots, \tconstant+\m\}$, respectively.
            Then, since $\bx(1) < \ldots < \bx(\tconstant)$ 
            and $\by(\tconstant + 1) < \ldots < \by(\tconstant + \m)$, 
            we have $\bxln{1}(1) < \ldots < \bxln{1}(\tconstant)$, 
            and $\byln{1}(\tconstant + 1) < \ldots < \byln{1}(\tconstant + \m)$. 

            Since $\bu = \{\tconstant+1,\ldots, \tconstant + \m\}$, where $\tconstant \ge 1$, we have $1 \in \tonumber{\n} \setminus \bu $. 
            Then according to \eqref{supp:theorem:2.7:lemma:3:eqn:2}, we have
            $\bxlnn{1}{1} = \bxn{1}$.
            Notice that \eqref{supp:theorem:2.7:lemma:3:eqn:2} also gives
            $\svector{\bxln{1}}{\lconstant}{\tonumber{\n}\setminus \bu}
            = \svector{\bx}{\lconstant}{\tonumber{\n}\setminus \bu}$.
            Hence, we have
            \begin{align} \label{supp:theorem:2.7:lemma:3:eqn:3.1}
                \rank{\bxlnn{1}{1}}{\svector{\bxln{1}}{\lconstant}{\tonumber{\n}\setminus \bu}} 
                = \rank{\bxn{1}}{\svector{\bx}{\lconstant}{\tonumber{\n}\setminus \bu}} =^{\eqref{supp:theorem:2.7:lemma:3:eqn:1}} \alpha.
            \end{align}

            Similarly, since $\bv = \{1,\ldots,\tconstant\}$, where $\tconstant \ge 1$,
            we have $\tconstant + 1 \in \tonumber{\n} \setminus (\bv \setminus \{1\})$. Hence, according to \eqref{supp:theorem:2.7:lemma:3:eqn:3}, we have 
            $\bylnn{1}{\tconstant+1} = \byn{\tconstant+1}$. 
            Since $\tonumber{\n} \setminus (\bv \setminus \{1\}) = \{1\} \cup (\tonumber{\n}\setminus \bv)$,
            then \eqref{supp:theorem:2.7:lemma:3:eqn:3} also gives
            that $\svector{\byln{1}}{\lconstant}{\{1\} \cup (\tonumber{\n}\setminus \bv)} = \svector{\by}{\lconstant}{\{1\} \cup (\tonumber{\n}\setminus \bv)}$.
            Hence, we have
            \begin{align*}
                \rank{\bylnn{1}{\tconstant+1}}{\svector{\byln{1}}{\lconstant}{\{1\} \cup (\tonumber{\n}\setminus \bv)}} 
                = \rank{\byn{\tconstant+1}}{\svector{\by}{\lconstant}{\{1\} \cup (\tonumber{\n}\setminus \bv)}}.
            \end{align*}
            Since $\rank{\byn{\tconstant+1}}{\svector{\by}{\lconstant}{\{1\} \cup (\tonumber{\n}\setminus \bv)}} 
            > \rank{\bxn{1}}{\svector{\bx}{\lconstant}{\tonumber{\n}\setminus \bu}}$, we further have
            \begin{align} \label{supp:theorem:2.7:lemma:3:eqn:4}
                \rank{\bylnn{1}{\tconstant+1}}{\svector{\byln{1}}{\lconstant}{\{1\} \cup (\tonumber{\n}\setminus \bv)}}  > \rank{\bxn{1}}{\svector{\bx}{\lconstant}{\tonumber{\n}\setminus \bu}}	=^{\eqref{supp:theorem:2.7:lemma:3:eqn:1}} \alpha.
            \end{align}

            Next, notice that according to \eqref{supp:theorem:2.7:lemma:3:eqn:3},
            we have 
            \begin{align*}
                \rank{\bxlnn{1}{\iconstant}}{\bxln{1}} = \rank{\bylnn{1}{\iconstant}}{\byln{1}}, \text{ for any } \iconstant \in \bu.
            \end{align*}				
            According to Lemma~\ref{supp:lemma:2},
            for any $\iconstant \in \bu$, we have 
            \begin{align*}
                \rank{\bylnn{1}{\iconstant}}{\byln{1}} 
                \ge \rank{\bylnn{1}{\iconstant}}{\svector{\byln{1}}{\lconstant}{\tonumber{\n} \setminus (\bv \setminus \{1\})}}.
            \end{align*}
            Since $ \bylnn{1}{\tconstant+\m} > \cdots > \bylnn{1}{\tconstant+1}$, for any $\iconstant \in \bu$, 
            we further have
            \begin{align*}
                \rank{\bylnn{1}{\iconstant}}{\svector{\byln{1}}{\lconstant}{\tonumber{\n} \setminus (\bv \setminus \{1\})}}  \ge \rank{\bylnn{1}{\tconstant+1}}{\svector{\byln{1}}{\lconstant}{\tonumber{\n} \setminus (\bv \setminus \{1\})}} >^{\eqref{supp:theorem:2.7:lemma:3:eqn:4}} \alpha.
            \end{align*}
            Hence, we have
            \begin{align*}
                \rank{\bxlnn{3}{\iconstant}}{\bxln{3}} > \alpha, \text{ for any } \iconstant \in \bu.
            \end{align*}
            Notice that according to \eqref{supp:theorem:2.7:lemma:3:eqn:3.1}, we have 
            \begin{align*}
                \rank{\bxlnn{1}{1}}{\svector{\bxln{1}}{\lconstant}{\tonumber{\n} \setminus \bu}} = \alpha.
            \end{align*}
            Subsequently, according to Lemma~\ref{supp:lemma:3}, we have
            \begin{align} \label{supp:theorem:2.7:lemma:3:eqn:5}
                \rank{\bxlnn{1}{1}}{\bxln{1}} = \rank{\bxlnn{1}{1}}{\svector{\bxln{1}}{\lconstant}{\tonumber{\n} \setminus \bu}} = \alpha.
            \end{align}

            Suppose $\bv \setminus \{1\} \neq \emptyset$. Then
            for any $\iconstant \in \bv \setminus \{1\}$, according to
            Lemma~\ref{supp:lemma:2}, we have
            \begin{align*}
                \rank{\bylnn{1}{\iconstant}}{\svector{\byln{1}}{\lconstant}{\tonumber{\n} \setminus \{1\}}} \ge \rank{\bylnn{1}{\iconstant}}{\byln{1}} - 1.
            \end{align*}
            Since $\rank{\bxlnn{1}{\iconstant}}{\bxln{1}} = \rank{\bylnn{1}{\iconstant}}{\byln{1}} 
            \text{ for any } \iconstant \in (\bv \setminus \{1\}) \cup \bu$,
            we have 
            $ \rank{\bylnn{1}{\iconstant}}{\byln{1}} =  \rank{\bxlnn{1}{\iconstant}}{\bxln{1}}$
            for any $\iconstant \in \bv \setminus \{1\}$.
            Then, it follows that
            \begin{align*}
                \rank{\bylnn{1}{\iconstant}}{\svector{\byln{1}}{\lconstant}{\tonumber{\n} \setminus \{1\}}} &\geq \rank{\bylnn{1}{\iconstant}}{\byln{1}} - 1 = \rank{\bxlnn{1}{\iconstant}}{\bxln{1}} - 1, \text{ for any } \iconstant \in \bv \setminus \{1\}.
            \end{align*}
            Next, notice that $\bxlnn{1}{1} < \ldots < \bxlnn{1}{\tconstant}$. Hence, for any $\iconstant \in \bv \setminus \{1\}$, we further have
            \begin{align*} 
                \rank{\bxlnn{1}{\iconstant}}{\bxln{1}} - 1
                > \rank{\bxlnn{1}{1}}{\bxln{1}} - 1 =^{\eqref{supp:theorem:2.7:lemma:3:eqn:5}} \rank{\bxlnn{1}{1}}{\svector{\bxln{1}}{\lconstant}{\tonumber{\n} \setminus \bu}} - 1.
            \end{align*}
            Thus, we have
            \begin{align*}
                \rank{\bylnn{1}{\iconstant}}{\svector{\byln{1}}{\lconstant}{\tonumber{\n} \setminus \{1\}}} > \rank{\bxlnn{1}{1}}{\svector{\bxln{1}}{\lconstant}{\tonumber{\n} \setminus \bu}} - 1, \text{ for any } \iconstant \in \bv \setminus \{1\}.
            \end{align*}
            Then, $\text{ for any } \iconstant \in \bv \setminus \{1\}$,  it follows that 
            \begin{align*} 
                \rank{\bylnn{1}{\iconstant}}{\svector{\byln{1}}{\lconstant}{\tonumber{\n} \setminus \{1\}}} \geq \rank{\bxlnn{1}{1}}{\svector{\bxln{1}}{\lconstant}{\tonumber{\n} \setminus \bu}} =^{\eqref{supp:theorem:2.7:lemma:3:eqn:5}} \rank{\bxlnn{1}{1}}{\bx} = \alpha.
            \end{align*}
            This completes our proof.
        \end{proof}

        \begin{lemma} \label{supp:theorem:2.7:lemma:4}
            Following Lemma~\ref{supp:theorem:2.7:lemma:3}, let $\byln{2}$ be 
            an imputation of $\byln{1}$ for the index 1 such that 
            $\rank{\bylnn{2}{1}}{\byln{2}} 
            = \rank{\bxlnn{1}{1}}{\bxln{1}}$. Then, we have 
            \begin{align*}
                \rank{\bylnn{2}{\iconstant}}{\svector{\byln{2}}{\lconstant}{\{1\} \cup (\tonumber{\n} \setminus \bv)}} > \rank{\bylnn{2}{1}}{\svector{\byln{2}}{\lconstant}{\tonumber{\n} \setminus (\bv \setminus \{1\})}} =\alpha, \text{ for any } \iconstant \in \bu.
            \end{align*}
        \end{lemma}

        \begin{proof}
            To start, since $\bylnn{2}{1}$ is an imputation of $\byln{1}$ for the index 1, 
            we have
            \begin{align}
                \bylnn{2}{\iconstant} = \bylnn{1}{\iconstant}, \text{ for any } \iconstant \in \tonumber{\n} \setminus \{1\}. \label{supp:theorem:2.7:lemma:4:eqn:1}
            \end{align}		
            Suppose $\bv \setminus \{1\} \neq \emptyset$.
            Then for any $\iconstant \in \bv \setminus \{1\}$, 
            according to Lemma~\ref{supp:lemma:2}, we have
            \begin{align*}
                \rank{\bylnn{2}{\iconstant}}{\byln{2}} \geq \rank{\bylnn{2}{\iconstant}}{\svector{\byln{2}}{\lconstant}{\tonumber{\n} \setminus \{1\}}}.
            \end{align*}
            According to \eqref{supp:theorem:2.7:lemma:4:eqn:1}, we have
            $\bylnn{2}{\iconstant} = \bylnn{1}{\iconstant}$ for any
            $\iconstant \in \bv \setminus \{1\}$, and 
            $\svector{\byln{1}}{\lconstant}{\tonumber{\n} \setminus \{1\}}
            = \svector{\byln{2}}{\lconstant}{\tonumber{\n} \setminus \{1\}}$.
            Hence, we have
            \begin{align*}
                \rank{\bylnn{2}{\iconstant}}{\byln{2}} \ge \rank{\bylnn{2}{\iconstant}}{\svector{\byln{2}}{\lconstant}{\tonumber{\n} \setminus \{1\}}} = \rank{\bylnn{1}{\iconstant}}{\svector{\byln{1}}{\lconstant}{\tonumber{\n} \setminus \{1\}}}, \text{ for any } \iconstant \in \bv \setminus \{1\}.
            \end{align*}
            According to Lemma~\ref{supp:theorem:2.7:lemma:3}, we have
            \begin{align*}
                \rank{\bylnn{1}{\iconstant}}{\svector{\byln{1}}{\lconstant}{\tonumber{\n} \setminus \{1\}}} \geq \rank{\bxlnn{1}{1}}{\bxln{1}},   \text{ for any } \iconstant \in \bv \setminus \{1\}.
            \end{align*}
            Since $\rank{\bxlnn{1}{1}}{\bxln{1}} = \rank{\bylnn{2}{1}}{\byln{2}}$,
            we then have 
            \begin{align*}
                \begin{split}
                    \rank{\bylnn{2}{\iconstant}}{\byln{2}} \ge \rank{\bylnn{2}{1}}{\byln{2}}, \text{ for any } \iconstant \in \bv \setminus \{1\}.
                \end{split}
            \end{align*}
            Since $\byln{2} \in \vndistinct$ is a vector of distinct real values, we have for any $\iconstant \in \bv \setminus \{1\}$,
            \begin{align*}
                \rank{\bylnn{2}{\iconstant}}{\byln{2}} \neq \rank{\bylnn{2}{1}}{\byln{2}}.
            \end{align*}
            Thus, it follows that
            \begin{align}
                \rank{\bylnn{2}{\iconstant}}{\byln{2}} > \rank{\bylnn{2}{1}}{\byln{2}} \Rightarrow \bylnn{2}{\iconstant} > \bylnn{2}{1}, \text{ for any } \iconstant \in \bv \setminus \{1\}. \nonumber
            \end{align}
            Subsequently, according to Lemma~\ref{supp:lemma:3}, we have
            \begin{align*}
                \rank{\bylnn{2}{1}}{\byln{2}} = \rank{\bylnn{2}{1}}{\svector{\byln{2}}{\lconstant}{\tonumber{\n} \setminus (\bv \setminus \{1\})}}.
            \end{align*}
            In particular, the above equation still holds when $\bv \setminus \{1\} = \emptyset$. 			
            Further, notice that $\rank{\bylnn{2}{1}}{\byln{2}}  = \rank{\bxlnn{1}{1}}{\bxln{1}}$,
            and according to Lemma~\ref{supp:theorem:2.7:lemma:3}, 
            $\rank{\bxlnn{1}{1}}{\bxln{1}} = \alpha$. Hence, we have
            \begin{align}
                \rank{\bylnn{2}{1}}{\byln{2}} = \rank{\bylnn{2}{1}}{\svector{\byln{2}}{\lconstant}{\tonumber{\n} \setminus (\bv \setminus \{1\})}} = \alpha. \label{supp:theorem:2.7:lemma:4:eqn:2}
            \end{align}

            Next, according to Lemma \ref{supp:lemma:2}, for any $\iconstant \in \bu$,
            \begin{align*}
                \rank{\bylnn{2}{\iconstant}}{\byln{2}} \geq \rank{\bylnn{2}{\iconstant}}{\svector{\byln{2}}{\lconstant}{\tonumber{\n} \setminus \bv}}.
            \end{align*}
            Since $\bv = \{1, \ldots, \tconstant\}, \bu = \{\tconstant+1, \ldots, \tconstant+\m\}$, where $\tconstant, \m \geq 1$, it follows $\tonumber{\n} \setminus \bv \subset \tonumber{\n} \setminus \{1\}$ and $\bu \subset \tonumber{\n} \setminus \{1\}$. Thus,
            according to  \eqref{supp:theorem:2.7:lemma:4:eqn:1}, we have
            $\bylnn{2}{\iconstant} = \bylnn{1}{\iconstant}$ for any
            $\iconstant \in \bu$, and 
            $\svector{\byln{2}}{\lconstant}{\tonumber{\n} \setminus \bv}
            = \svector{\byln{1}}{\lconstant}{\tonumber{\n} \setminus \bv}$.
            Then, we have
            \begin{align*}
                \rank{\bylnn{2}{\iconstant}}{\svector{\byln{2}}{\lconstant}{\tonumber{\n} \setminus \bv}} = \rank{\bylnn{1}{\iconstant}}{\svector{\byln{1}}{\lconstant}{\tonumber{\n} \setminus \bv}}, \text{ for any } \iconstant \in \bu.
            \end{align*}
            Further, since $\tonumber{\n} \setminus \bv \subset \tonumber{\n} \setminus (\bv \setminus \{1\})$ and $\bu \subset \tonumber{\n} \setminus (\bv \setminus \{1\})$, 
            then according to \eqref{supp:theorem:2.7:lemma:3:eqn:3},
            we have  $\bylnn{1}{\iconstant} = \byn{\iconstant}$ for any
            $\iconstant \in \bu$, and 
            $\svector{\byln{1}}{\lconstant}{\tonumber{\n} \setminus \bv}
            = \svector{\by}{\lconstant}{\tonumber{\n} \setminus \bv}$.
            Hence, we have
            \begin{align*}
                \rank{\bylnn{1}{\iconstant}}{\svector{\byln{1}}{\lconstant}{\tonumber{\n} \setminus \bv}} = \rank{\byn{\iconstant}}{\svector{\by}{\lconstant}{\tonumber{\n} \setminus \bv}}, \text{ for any } \iconstant \in \bu.
            \end{align*}
            Hence, we have 
            \begin{align*}
                \rank{\bylnn{2}{\iconstant}}{\byln{2}} \geq \rank{\byn{\iconstant}}{\svector{\by}{\lconstant}{\tonumber{\n} \setminus \bv}}, \text{ for any } \iconstant \in \bu.
            \end{align*}
            Further, since $\byn{\tconstant+1} < \ldots < \byn{\tconstant+\m}$, for any $\iconstant \in \bu$, we have
            \begin{align*}
                \rank{\bylnn{2}{\iconstant}}{\byln{2}} \geq \rank{\byn{\iconstant}}{\svector{\by}{\lconstant}{\tonumber{\n} \setminus \bv}} \geq \rank{\byn{\tconstant+1}}{\svector{\by}{\lconstant}{\tonumber{\n} \setminus \bv}}.
            \end{align*}
            Notice that we have
            \begin{align*}
                \rank{\byn{\tconstant+1}}{\svector{\by}{\lconstant}{\{1\} \cup (\tonumber{\n}\setminus \bv)}} 
                > \rank{\bxn{1}}{\svector{\bx}{\lconstant}{\tonumber{\n}\setminus \bu}} =^{\eqref{supp:theorem:2.7:lemma:3:eqn:1}} \alpha. 
            \end{align*}
            Then, it follows that
            \begin{align*}
                \rank{\byn{\tconstant+1}}{\svector{\by}{\lconstant}{\{1\} \cup (\tonumber{\n}\setminus \bv)}} 
                \ge \alpha + 1 \Rightarrow \rank{\byn{\tconstant+1}}{\svector{\by}{\lconstant}{\{1\} \cup (\tonumber{\n}\setminus \bv)}}  - 1 \ge \alpha.
            \end{align*}
            According to Lemma~\ref{supp:lemma:2}, we have
            \begin{align*}
                \rank{\byn{\tconstant+1}}{\svector{\by}{\lconstant}{\tonumber{\n} \setminus \bv}} \ge \rank{\byn{\tconstant+1}}{\svector{\by}{\lconstant}{\{1\} \cup (\tonumber{\n}\setminus \bv)}}  - 1.
            \end{align*}
            Hence, we have
            \begin{align*}
                \rank{\byn{\tconstant+1}}{\svector{\by}{\lconstant}{\tonumber{\n} \setminus \bv}} \ge \alpha.
            \end{align*}
            Then, we have
            \begin{align*}
                \rank{\bylnn{2}{\iconstant}}{\byln{2}} \geq \alpha, \text{ for any }\iconstant \in \bu
            \end{align*}
            According to \eqref{supp:theorem:2.7:lemma:4:eqn:2}, we have $\rank{\bylnn{2}{1}}{\byln{2}} = \alpha$. Since $\byln{2} \in \vndistinct$ is a vector of distinct real values, then for any $\iconstant \in \bu$, we have
            \begin{align}
                &\rank{\bylnn{2}{\iconstant}}{\byln{2}} \neq \rank{\bylnn{2}{1}}{\byln{2}} = \alpha \nonumber \\
                &\Rightarrow \rank{\bylnn{2}{\iconstant}}{\byln{2}} > \rank{\bylnn{2}{1}}{\byln{2}} \nonumber \\
                &\Rightarrow \bylnn{2}{\iconstant} > \bylnn{2}{1} \nonumber \\
                &\Rightarrow \rank{\bylnn{2}{\iconstant}}{\svector{\byln{2}}{\lconstant}{\{1\} \cup (\tonumber{\n} \setminus \bv)}} > \rank{\bylnn{2}{1}}{\svector{\byln{2}}{\lconstant}{\tonumber{\n} \setminus (\bv \setminus \{1\})}} =^{\eqref{supp:theorem:2.7:lemma:4:eqn:2}} \alpha. \nonumber
            \end{align}
            This completes our proof.
        \end{proof}

        \begin{lemma} \label{supp:theorem:2.7:lemma:5}
            Following Lemma~\ref{supp:theorem:2.7:lemma:3} and Lemma~\ref{supp:theorem:2.7:lemma:4},
            let  $\bxln{3}$ and $\byln{3}$ be imputations of $\bxln{1}, \byln{2}$
            for $\bu$ and $\bv \setminus \{1\}$, respectively, such that 
            $\rank{\bxlnn{3}{\iconstant}}{\bxln{3}} = \rank{\bylnn{3}{\iconstant}}{\byln{3}} \text{ for any } \iconstant \in (\bv \setminus \{1\}) \cup \bu$. Then, we have $			\rank{\bxlnn{3}{\iconstant}}{\bxln{3}} = \rank{\bylnn{3}{\iconstant}}{\byln{3}}, \text{ for any }\iconstant \in \bu \cup \bv$.
        \end{lemma}

        \begin{proof}
            To start, since $\bxln{3}$ and $\byln{3}$ are imputations of $\bxln{1}, \byln{2}$
            for $\bu$ and $\bv \setminus \{1\}$, respectively, we have
            \begin{align}
                &\bxlnn{3}{\iconstant} = \bxlnn{1}{\iconstant}, \text{ for any } \iconstant \in \tonumber{\n} \setminus \bu, \label{supp:theorem:2.7:lemma:5:eqn:1}\\
                &\bylnn{3}{\iconstant} = \bylnn{2}{\iconstant}, \text{ for any } \iconstant \in \tonumber{\n} \setminus (\bv \setminus \{1\}). \label{supp:theorem:2.7:lemma:5:eqn:2}
            \end{align}

            Recall that $\bu = \{\tconstant+1, \ldots, \tconstant+\m\}$, where $\tconstant, \m \geq 1$.
            Hence, $1 \in \tonumber{\n} \setminus \bu$. Then according to \eqref{supp:theorem:2.7:lemma:5:eqn:1},
            we have $\bxlnn{3}{1} = \bxlnn{1}{1}$. Notice that \eqref{supp:theorem:2.7:lemma:5:eqn:1} also
            gives that $\svector{\bxln{3}}{\lconstant}{\tonumber{\n} \setminus \bu}
            = \svector{\bxln{1}}{\lconstant}{\tonumber{\n} \setminus \bu}$. Hence, we have
            \begin{align*}
                \rank{\bxlnn{3}{1}}{\svector{\bxln{3}}{\lconstant}{\tonumber{\n} \setminus \bu}} = \rank{\bxlnn{1}{1}}{\svector{\bxln{1}}{\lconstant}{\tonumber{\n} \setminus \bu}} =^{\eqref{supp:theorem:2.7:lemma:3:eqn:1}} \alpha.
            \end{align*}
            Similarly, since $1 \in \tonumber{\n} \setminus (\bv \setminus \{1\})$, according to
            \eqref{supp:theorem:2.7:lemma:5:eqn:2}, we have $\bylnn{3}{1} = \bylnn{2}{1}$.
            Notice that $\tonumber{\n} \setminus (\bv \setminus \{1\} = 
            \{1\}) \cup (\tonumber{\n} \setminus \bv)$. Then \eqref{supp:theorem:2.7:lemma:5:eqn:2}
            also gives that $\svector{\byln{3}}{\lconstant}{\{1\} \cup (\tonumber{\n} \setminus \bv)}
            = \svector{\byln{2}}{\lconstant}{\{1\} \cup (\tonumber{\n} \setminus \bv)}$.
            Hence, we have
            \begin{align} \label{supp:theorem:2.7:lemma:5:eqn:3}
                \rank{\bylnn{3}{1}}{\svector{\byln{3}}{\lconstant}{\{1\} \cup (\tonumber{\n} \setminus \bv)}} = \rank{\bylnn{2}{1}}{\svector{\byln{2}}{\lconstant}{\{1\} \cup (\tonumber{\n} \setminus \bv)}}.
            \end{align}
            According to Lemma~\ref{supp:theorem:2.7:lemma:4}, we have 
            \begin{align*}
                \rank{\bylnn{2}{1}}{\svector{\byln{2}}{\lconstant}{\{1\} \cup (\tonumber{\n} \setminus \bv)}} = \alpha.
            \end{align*}
            Hence
            \begin{align*}
                \rank{\bylnn{3}{1}}{\svector{\byln{3}}{\lconstant}{\{1\} \cup (\tonumber{\n} \setminus \bv)}} = \alpha.
            \end{align*}
            Combining the above results, we have
            \begin{align} \label{supp:theorem:2.7:lemma:5:eqn:4}
                \rank{\bxlnn{3}{1}}{\svector{\bxln{3}}{\lconstant}{\tonumber{\n} \setminus \bu}}
                =\rank{\bylnn{3}{1}}{\svector{\byln{3}}{\lconstant}{\{1\} \cup (\tonumber{\n} \setminus \bv)}} 
                = \alpha.
            \end{align}

            Suppose $\bv \setminus \{1\} \neq \emptyset$, then since $\rank{\bxlnn{3}{\iconstant}}{\bxln{3}} = \rank{\bylnn{3}{\iconstant}}{\byln{3}} \text{ for any } \iconstant \in (\bv \setminus \{1\}) \cup \bu$,
            we have
            \begin{align*}
                \rank{\bylnn{3}{\iconstant}}{\byln{3}} = \rank{\bxlnn{3}{\iconstant}}{\bxln{3}}, \text{ for any }
                \iconstant \in \bv \setminus \{1\}.
            \end{align*}
            According to Lemma~\ref{supp:theorem:2.7:lemma:3}, we have 
            $\bxlnn{1}{1} < \ldots < \bxlnn{1}{\tconstant}$. Notice that
            from \eqref{supp:theorem:2.7:lemma:5:eqn:1}, 
            we have $\bxlnn{1}{i} = \bxlnn{3}{i}$ for any $\iconstant \in \bv = \{1,\ldots,\tconstant\}$.
            Hence, we further have
            $\bxlnn{3}{1} < \ldots < \bxlnn{3}{\tconstant}$. Subsequently,
            \begin{align*}
                \rank{\bxlnn{3}{\iconstant}}{\bxln{3}} > \rank{\bxlnn{3}{1}}{\bxln{3}}, \text{ for any }
                \iconstant \in \bv \setminus \{1\}.
            \end{align*}
            According to Lemma~\ref{supp:lemma:2}, we further have
            \begin{align*}
                \rank{\bxlnn{3}{1}}{\bxln{3}}
                \ge \rank{\bxlnn{3}{1}}{\svector{\bxln{3}}{\lconstant}{\tonumber{\n} \setminus \bu}} =^{\eqref{supp:theorem:2.7:lemma:5:eqn:4}} \rank{\bylnn{3}{1}}{\svector{\byln{3}}{\lconstant}{\{1\} \cup (\tonumber{\n} \setminus \bv)}}.
            \end{align*}
            Hence, we have 
            \begin{align*}
                \rank{\bylnn{3}{\iconstant}}{\byln{3}} >  \rank{\bylnn{3}{1}}{\svector{\byln{3}}{\lconstant}{\{1\} \cup (\tonumber{\n} \setminus \bv)}}, \text{ for any }
                \iconstant \in \bv \setminus \{1\}.
            \end{align*}
            Subsequently, according to Lemma~\ref{supp:lemma:3}, we have
            \begin{align} \label{supp:theorem:2.7:lemma:5:eqn:5}
                \rank{\bylnn{3}{1}}{\byln{3}} = \rank{\bylnn{3}{1}}{\svector{\byln{3}}{\lconstant}{\tonumber{\n} \setminus (\bv \setminus \{1\})}} = \alpha.
            \end{align} 
            In particular, the above equations still hold when $\bv \setminus \{1\} = \emptyset$.

            Since $\rank{\bxlnn{3}{\iconstant}}{\bxln{3}} = \rank{\bylnn{3}{\iconstant}}{\byln{3}} \text{ for any } \iconstant \in (\bv \setminus \{1\}) \cup \bu$, then for any $\iconstant \in \bu$,
            we have
            \begin{align*}
                \rank{\bxlnn{3}{\iconstant}}{\bxln{3}} = \rank{\bylnn{3}{\iconstant}}{\byln{3}}.
            \end{align*}
            Then, according to Lemma~\ref{supp:lemma:2}, for any $\iconstant \in \bu$, we have
            \begin{align*}
                \rank{\bylnn{3}{\iconstant}}{\byln{3}} &\geq \rank{\bylnn{3}{\iconstant}}{\svector{\byln{3}}{\lconstant}{\{1\} \cup (\tonumber{\n} \setminus \bv)}}.
            \end{align*}
            Hence, for any $\iconstant \in \bu$, it follows that
            \begin{align*}
                \rank{\bxlnn{3}{\iconstant}}{\bxln{3}} &\ge \rank{\bylnn{3}{\iconstant}}{\svector{\byln{3}}{\lconstant}{\{1\} \cup (\tonumber{\n} \setminus \bv)}}.
            \end{align*}
            Since $\bu \subset \tonumber{\n} \setminus \bv \subset \tonumber{\n} \setminus (\bv \setminus \{1\})$,
            then according to \eqref{supp:theorem:2.7:lemma:5:eqn:2}, we have $\bylnn{3}{i} = \bylnn{2}{i}$ 
            for any $\iconstant \in \bu$.
            Notice that $\tonumber{\n} \setminus (\bv \setminus \{1\}) = 
            \{1\} \cup (\tonumber{\n} \setminus \bv)$. Then \eqref{supp:theorem:2.7:lemma:5:eqn:2}
            also gives that $\svector{\byln{3}}{\lconstant}{\{1\} \cup (\tonumber{\n} \setminus \bv)}
            = \svector{\byln{2}}{\lconstant}{\{1\} \cup (\tonumber{\n} \setminus \bv)}$.
            Hence, we have
            \begin{align*}
                \rank{\bylnn{3}{\iconstant}}{\svector{\byln{3}}{\lconstant}{\{1\} \cup (\tonumber{\n} \setminus \bv)}} =\rank{\bylnn{2}{\iconstant}}{\svector{\byln{2}}{\lconstant}{\{1\} \cup (\tonumber{\n} \setminus \bv)}}, \text{ for any } \iconstant \in \bu.
            \end{align*}
            Combining the above results, for any $\iconstant \in \bu$, we have
            \begin{align*}
                \rank{\bxlnn{3}{\iconstant}}{\bxln{3}} &\geq \rank{\bylnn{2}{\iconstant}}{\svector{\byln{2}}{\lconstant}{\{1\} \cup (\tonumber{\n} \setminus \bv)}}.
            \end{align*}
            According to Lemma~\ref{supp:theorem:2.7:lemma:4}, we further have
            \begin{align*}
                \rank{\bylnn{2}{\iconstant}}{\svector{\byln{2}}{\lconstant}{\{1\} \cup (\tonumber{\n} \setminus \bv)}}
                > \alpha =^{\eqref{supp:theorem:2.7:lemma:5:eqn:4}} \rank{\bxlnn{3}{1}}{\svector{\bxln{3}}{\lconstant}{\tonumber{\n} \setminus \bu}}.
            \end{align*}
            Hence, we have
            \begin{align*}
                \rank{\bxlnn{3}{\iconstant}}{\bxln{3}} > \rank{\bxlnn{3}{1}}{\svector{\bxln{3}}{\lconstant}{\tonumber{\n} \setminus \bu}}, \text{ for any } \iconstant \in \bu.
            \end{align*}
            Subsequently, according to Lemma~\ref{supp:lemma:3}, we have
            \begin{align*}
                \rank{\bxlnn{3}{1}}{\bxln{3}} = \rank{\bxlnn{3}{1}}{\svector{\bxln{3}}{\lconstant}{\tonumber{\n} \setminus \bu}} = \alpha = \rank{\bylnn{3}{1}}{\byln{3}}.
            \end{align*}
            Combining this result with \eqref{supp:theorem:2.7:lemma:5:eqn:2}, we have
            \begin{align*}
                \rank{\bxlnn{3}{\iconstant}}{\bxln{3}} = \rank{\bylnn{3}{\iconstant}}{\byln{3}}, \text{ for any }\iconstant \in \bu \cup \bv.
            \end{align*}
            This completes our proof.
        \end{proof}

        Now we are ready to prove Theorem~2.7.

        \begin{theorem} \label{supp:theorem:2.7}
            Suppose $\bx, \by \in \vndistinct$, and $\bu, \bv \subset \tonumber{\n}$ are 
            disjoint subsets of indices such that $\bu \cap \bv = \emptyset$.
            Suppose $\bxs, \bys \in \vndistinct$ are imputations of $\bx$,
            $\by$ for indices $\bu$ and $\bv$, respectively.
            Then, if 
            $\rank{\bxsn{\iconstant}}{\bxs} = \rank{\bysn{\iconstant}}{\bys}, \text{ for any } \iconstant \in \bu \cup \bv$, 
            we have $\sfd{\bxs}{\bys} \le \sfd{\bx}{\by}$. Furthermore,
            for any other imputations $\bxsp, \bysp \in \vndistinct$ of 
            $\bx, \by$ for indices $\bu, \bv$, respectively, we have 
            $\sfd{\bxs}{\bys} \le \sfd{\bxp}{\byp}$.
        \end{theorem}

        \begin{proof}
            Notice that there are two statements in Theorem~\ref{supp:theorem:2.7}:
            \begin{itemize}
                \item[(1)]: Suppose $\bx, \by \in \vndistinct$, and $\bu, \bv \subset \tonumber{\n}$ 
                    are disjoint subsets of indices such that $\bu \cap \bv = \emptyset$. Then
                    if $\bxs, \bys \in \vndistinct$ are imputations of $\bx$,
                    $\by$ for indices $\bu$ and $\bv$, respectively, and
                    $\rank{\bxsn{\iconstant}}{\bxs} = \rank{\bysn{\iconstant}}{\bys}, \text{ for any } \iconstant \in \bu \cup \bv$, we have $\sfd{\bxs}{\bys} \le \sfd{\bx}{\by}$. 
                \item[(2)]: Suppose $\bx, \by \in \vndistinct$, and $\bu, \bv \subset \tonumber{\n}$ 
                    are disjoint subsets of indices such that $\bu \cap \bv = \emptyset$. Suppose
                    $\bxs, \bys \in \vndistinct$ are imputations of $\bx$,
                    $\by$ for indices $\bu$ and $\bv$, respectively, and
                    $\rank{\bxsn{\iconstant}}{\bxs} = \rank{\bysn{\iconstant}}{\bys}, \text{ for any } \iconstant \in \bu \cup \bv$.  Then for any other imputations $\bxsp, \bysp \in \vndistinct$ of 
                    $\bx, \by$ for indices $\bu, \bv$, respectively, we have 
                    $\sfd{\bxs}{\bys} \le \sfd{\bxp}{\byp}$.
            \end{itemize}
            Below, we first show that the statement $(1)$ is true. Then 
            we prove the statement $(2)$
            using the statement $(1)$.

            First, we show the statement $(1)$ is true.
            To begin with, let us consider the case when $|\bu| + |\bv| = \n$. 
            Since $\bu \cap \bv = \emptyset$, then $|\bu| + |\bv| = \n$ means
            $\bu \cup \bv = \tonumber{\n}$.
            Since $\rank{\bxsn{\iconstant}}{\bxs} = \rank{\bysn{\iconstant}}{\bys}, \text{ for any } \iconstant \in \bu \cup \bv$, we have  
            $\rank{\bxs}{\bxs} = \rank{\bys}{\bys}$. 
            Hence $\sfd{\bxs}{\bys} = 0.$ According to the definition 
            of Spearman's footrule, we have $\sfd{\bx}{\by} \ge 0 =\sfd{\bxs}{\bys}$, 
            which proves the statement $(1)$ when $|\bu| + |\bv| = \n$.

            When $|\bu| + |\bv| = 0$, then according to the definition of imputations,
            we have $\bx = \bxs$ and $\by = \bys$. Hence we have $\sfd{\bxs}{\bys} = \sfd{\bx}{\by}$.
            This proves statement $(1)$ when $|\bu| + |\bv| = 0$.

            Then, we only need to prove statement $(1)$ when $\n - 1 \ge |\bu| + |\bv| \ge 1$.
            For any given $\n \in \mathbb{N}$, let $\bpn{\kk}$ be the statement of 
            the statement $(1)$ when $|\bu| + |\bv| = \kk$.
            We prove $\bpn{\kk}$ holds for any $\kk \in \{1,\ldots, \n-1\}$ by
            induction on $\kk$.

            \textit{Base Case:} We show $\bpn{1}$ holds.     
            Suppose $|\bu| + |\bv| = 1$. Then we have either $|\bu| = 0, |\bv| = 1$ 
            or $|\bu| = 1, |\bv| = 0$. For both cases, the statement  
            $\bpn{1}$ is true according to Theorem~\ref{supp:theorem:2.4}.

            \textit{Induction Step:} We show the implication
            $ \bpn{\kk} \Rightarrow \bpn{\kk + 1}$ for any 
            $\kk \in \{1,\ldots, \n-2\}$.

            Suppose $|\bu| + |\bv| = \kk + 1$. Then, we have either
            \begin{align*}
                &\text{case } (\rn{I}):~ |\bu| = \kk + 1, |\bv| = 0 \text{ or } |\bu| = 0, |\bv| = \kk + 1,\\
                \text{or } &\text{case } (\rn{II}):~ |\bu| > 0, |\bv| > 0, |\bu| + |\bv| = \kk + 1,
            \end{align*}
            is true.

            Suppose the $\text{case } (\rn{I}):~ |\bu| = \kk + 1, |\bv| = 0 \text{ or } |\bu| = 0, |\bv| = \kk + 1$ 
            is true. Then the statement $\bpn{\kk + 1}$ is true according to Theorem~\ref{supp:theorem:2.4}.

            Suppose the $\text{case } (\rn{II}):~ |\bu| > 0, |\bv| > 0, |\bu| + |\bv| = \kk + 1$ is true.
            Without loss of generality, let us assume (after relabeling) 
            $\bv = \{1, \ldots, \tconstant\}$, 
            $\bu = \{\tconstant + 1, \ldots, \tconstant + \m\}$,
            $\bx(1) < \ldots < \bx(\tconstant)$, and
            $\by(\tconstant + 1) < \by(\tconstant + 2) < \ldots < \by(\tconstant + \m)$.

            Notice that for $\rank{\bx(1)}{\svector{\bx}{\lconstant}{\tonumber{\n} \setminus \bu}}$ and $\rank{\by(\tconstant + 1)}{\svector{\by}{\lconstant}{\tonumber{\n} \setminus \bv}}$, we have either
            \begin{align*}
                &\rank{\bx(1)}{\svector{\bx}{\lconstant}{\tonumber{\n} \setminus \bu}} \ge \rank{\by(\tconstant + 1)}{\svector{\by}{\lconstant}{\tonumber{\n} \setminus \bv}},\\
                \text{or } &\rank{\bx(1)}{\svector{\bx}{\lconstant}{\tonumber{\n} \setminus \bu}} \le \rank{\by(\tconstant + 1)}{\svector{\by}{\lconstant}{\tonumber{\n} \setminus \bv}},
            \end{align*}
            is true. Let us assume $\rank{\bx(1)}{\svector{\bx}{\lconstant}{\tonumber{\n} \setminus \bu}} \le \rank{\by(\tconstant + 1)}{\svector{\by}{\lconstant}{\tonumber{\n} \setminus \bv}}$ is true.
            However, when $\rank{\bx(1)}{\svector{\bx}{\lconstant}{\tonumber{\n} \setminus \bu}} \ge \rank{\by(\tconstant + 1)}{\svector{\by}{\lconstant}{\tonumber{\n} \setminus \bv}}$ holds,
            we can switch the labels between $\bx$ and $\by$, and relabel the relevant components
            of the data.

            Let us denote
            \begin{align} \label{supp:theorem:2.7:eqn:0}
                \alpha = \rank{\bx(1)}{\svector{\bx}{\lconstant}{\tonumber{\n} \setminus \bu}}.
            \end{align}
            Then, since $\alpha = \rank{\bx(1)}{\svector{\bx}{\lconstant}{\tonumber{\n} \setminus \bu}} \le \rank{\by(\tconstant + 1)}{\svector{\by}{\lconstant}{\tonumber{\n} \setminus \bv}}$, 
            we have either 
            \begin{align*}
                &\text{case } (\rn{i}): \alpha = \rank{\by(\tconstant + 1)}{\svector{\by}{\lconstant}{\tonumber{\n} \setminus \bv}}, \text{ and } \by(1) > \by(\tconstant + 1),\\
                &\text{case } (\rn{ii}): \alpha = \rank{\by(\tconstant + 1)}{\svector{\by}{\lconstant}{\tonumber{\n} \setminus \bv}}, \text{ and } \by(1) < \by(\tconstant + 1),\\
                \text{or } &\text{case } (\rn{iii}): \alpha < \rank{\by(\tconstant + 1)}{\svector{\by}{\lconstant}{\tonumber{\n} \setminus \bv}},
            \end{align*}
            is true.
            In the following, we are going to consider the three cases separately.

            Suppose the case $(\rn{i})$ is true. In other words,
            \begin{align}
                & \alpha = \rank{\by(\tconstant + 1)}{\svector{\by}{\lconstant}{\tonumber{\n} \setminus \bv}}, \label{supp:theorem:2.7:eqn:1}\\
                \text{ and }&\by(1) > \by(\tconstant + 1). \label{supp:theorem:2.7:eqn:2}
            \end{align}

            Then, according to Proposition~\ref{supp:proposition:2.5}, 
            there exist $\bxln{1}, \byln{1} \in \vndistinct$
            such that
            \begin{align}
                &\bxln{1}(\iconstant) = \bx(\iconstant), \text{ for any } \iconstant \in \tonumber{\n} \setminus \bu, \label{supp:theorem:2.7:eqn:4}\\
                &\byln{1}(\iconstant) = \by(\iconstant), \text{ for any } \iconstant \in \tonumber{\n} \setminus (\bv \setminus \{1\}), \label{supp:theorem:2.7:eqn:5}\\
                &\rank{\bxlnn{1}{\iconstant}}{\bxln{1}} = \rank{\bylnn{1}{\iconstant}}{\byln{1}} \text{ for any } \iconstant \in \bu \cup (\bv \setminus \{1\}). \label{supp:theorem:2.7:eqn:6}
            \end{align} 
            In other words, $\bxln{1}, \byln{1}$ are imputations of $\bx$, and $\by$
            for indices $\bu$ and $\bv \setminus \{1\}$, respectively, such that
            the equation \eqref{supp:theorem:2.7:eqn:6} holds.

            Since $\bu$ and $\bv$ are disjoint such that $\bu \cap \bv = \emptyset$, 
            then $\bu \cap (\bv \setminus \{1\}) = \emptyset$, i.e. 
            $\bu$ and $\bv \setminus \{1\}$ are also disjoint. 
            Notice that $|\bu| + |\bv \setminus \{1\}| = \m + \tconstant -1 = \kk$. 
            Then, since $\bpn{\kk}$ is true, we have
            \begin{align} \label{supp:theorem:2.7:eqn:6.0}
                \sfd{\bxln{1}}{\byln{1}} \le \sfd{\bx}{\by}.
            \end{align}	    

            Next, applying Proposition \ref{supp:proposition:2.5} again, we can find $\bxln{2}, \byln{2}\in \vndistinct$ such that
            \begin{align}
                &\bxlnn{2}{\iconstant} = \bxlnn{1}{\iconstant}, \text{ for any } \iconstant \in \tonumber{\n} \setminus (\bu \setminus \{\tconstant+1\}), \label{supp:theorem:2.7:eqn:9}\\
                &\bylnn{2}{\iconstant} = \bylnn{1}{\iconstant}, \text{ for any } \iconstant \in \tonumber{\n} \setminus \bv, \label{supp:theorem:2.7:eqn:10}\\
                &\rank{\bxlnn{2}{\iconstant}}{\bxln{2}} = \rank{\bylnn{2}{\iconstant}}{\byln{2}} \text{ for any } \iconstant \in (\bu \setminus \{\tconstant+1\}) \cup \bv. \label{supp:theorem:2.7:eqn:11}
            \end{align} 
            In other words, $\bxln{2}$ and $\byln{2}$ are imputations of $\bxln{1}$, $\byln{1}$
            for indices $\bu \setminus \{\tconstant+1\}$ and $\bv$, respectively, such that
            \eqref{supp:theorem:2.7:eqn:11} is true.

            Notice that $\bu$ and $\bv$ are disjoint, i.e. $\bu \cap \bv = \emptyset$.
            Hence, $(\bu \setminus \{\tconstant+1\}) \cap \bv = \emptyset$.
            Next, since $|\bu \setminus \{\tconstant+1\}| + |\bv| = \m + \tconstant -1 = \kk$,
            we can apply $\bpn{\kk}$ and get
            \begin{align} \label{supp:theorem:2.7:eqn:12.0}
                \sfd{\bxln{2}}{\byln{2}} \le \sfd{\bxln{1}}{\byln{1}}.
            \end{align}

            According to Lemma~\ref{supp:theorem:2.7:lemma:2}, we have
            \begin{align*}
                \rank{\bxlnn{2}{\iconstant}}{\bxln{2}} = \rank{\bylnn{2}{\iconstant}}{\byln{2}}, \text{for any } \iconstant \in \bu \cup \bv.
            \end{align*}
            Notice that
            \begin{align*}
                &\bxlnn{2}{\iconstant} =^{\eqref{supp:theorem:2.7:eqn:9}} \bxlnn{1}{\iconstant}
                =^{\eqref{supp:theorem:2.7:eqn:4}} \bxn{\iconstant} \quad \text{for any } \iconstant \in \tonumber{\n} \setminus \bu,\\
                \text{ and } &\bylnn{2}{\iconstant} =^{\eqref{supp:theorem:2.7:eqn:10}} \bylnn{1}{\iconstant} 
                =^{\eqref{supp:theorem:2.7:eqn:5}} \byn{\iconstant} \quad \text{for any } \iconstant \in \tonumber{\n} \setminus \bv.
            \end{align*}
            In other words, $\bxln{2}, \byln{2} \in \vndistinct$ are imputations of $\bx$,
            $\by$ for indices $\bu$ and $\bv$, respectively, such that
            $\rank{\bxlnn{2}{\iconstant}}{\bxln{2}} = \rank{\bylnn{2}{\iconstant}}{\byln{2}}, \text{for any } \iconstant \in \bu \cup \bv$. 
            Since $\bxs, \bys \in \vndistinct$ are imputations of $\bx$,
            $\by$ for indices $\bu$ and $\bv$, respectively, such that
            $\rank{\bxsn{\iconstant}}{\bxs} = \rank{\bysn{\iconstant}}{\bys}, \text{ for any } \iconstant \in \bu \cup \bv$, then according to Proposition~\ref{supp:proposition:equivalence}, we have
            $\sfd{\bxln{2}}{\byln{2}} = \sfd{\bxs}{\bys}$.
            Thus,
            \begin{align*}
                \sfd{\bxs}{\bys} = \sfd{\bxln{2}}{\byln{2}} \le^{\eqref{supp:theorem:2.7:eqn:12.0}} \sfd{\bxln{1}}{\byln{1}} \le^{\eqref{supp:theorem:2.7:eqn:6.0}} \sfd{\bx}{\by},
            \end{align*}
            which proves $\bpn{\kk + 1}$ when the case $(\rn{i})$ holds.

            Suppose the case $(\rn{ii})$ or the case $(\rn{iii})$ holds. In other words,
            we have either
            \begin{align}
                &\alpha = \rank{\byn{\tconstant+1}}{\svector{\by}{\lconstant}{\tonumber{\n}\setminus \bv}}, 
                \text{ and }\byn{1} < \byn{\tconstant+1},  \label{supp:theorem:2.7:eqn:14}\\
                \text{ or }
                &\alpha < \rank{\byn{\tconstant+1}}{\svector{\by}{\lconstant}{\tonumber{\n}\setminus \bv}}.  \label{supp:theorem:2.7:eqn:15}
            \end{align}

            According to the definition of rank,
            we have
            \begin{align*}
                \begin{split}
                    \rank{\byn{\tconstant+1}}{\svector{\by}{\lconstant}{\{1\} \cup (\tonumber{\n}\setminus \bv)}} 
                    &= \sum_{\iconstant \in \{1\} \cup (\tonumber{\n}\setminus \bv) }  \indicator{\byn{\iconstant} \leq \byn{\tconstant+1}} \\
                    &= \indicator{\byn{1} \leq \byn{\tconstant+1}} 
                    + \sum_{\iconstant \in \tonumber{\n}\setminus \bv }  \indicator{\byn{\iconstant} \leq \byn{\tconstant+1}} \\
                    & = \indicator{\byn{1} \leq \byn{\tconstant+1}}  +  \rank{\byn{\tconstant+1}}{\svector{\by}{\lconstant}{\tonumber{\n}\setminus \bv}}.
                \end{split}
            \end{align*}
            Notice that $\indicator{\byn{1} \leq \byn{\tconstant+1}} \ge 0$.
            Hence, we have
            \begin{align*}
                \rank{\byn{\tconstant+1}}{\svector{\by}{\lconstant}{\{1\} \cup (\tonumber{\n}\setminus \bv)}} 
                &\geq \rank{\byn{\tconstant+1}}{\svector{\by}{\lconstant}{\tonumber{\n}\setminus \bv}}.
            \end{align*}  
            Next, suppose \eqref{supp:theorem:2.7:eqn:15} is true. We further have
            \begin{align*}
                \rank{\byn{\tconstant+1}}{\svector{\by}{\lconstant}{\{1\} \cup (\tonumber{\n}\setminus \bv)}} 
                &\geq \rank{\byn{\tconstant+1}}{\svector{\by}{\lconstant}{\tonumber{\n}\setminus \bv}} \\
                &> \alpha =^{\eqref{supp:theorem:2.7:eqn:0}} \rank{\bxn{1}}{\svector{\bx}{\lconstant}{\tonumber{\n}\setminus \bu}}.
            \end{align*}
            However, suppose \eqref{supp:theorem:2.7:eqn:14} is true. Then, we have
            \begin{align*}
                \rank{\byn{\tconstant+1}}{\svector{\by}{\lconstant}{\{1\} \cup (\tonumber{\n}\setminus \bv)}} 
                &= \indicator{\byn{1} \leq \byn{\tconstant+1}} 
                + \rank{\byn{\tconstant+1}}{\svector{\by}{\lconstant}{\tonumber{\n}\setminus \bv}} \\
                &= 1 + \alpha > \alpha =^{\eqref{supp:theorem:2.7:eqn:0}} \rank{\bxn{1}}{\svector{\bx}{\lconstant}{\tonumber{\n}\setminus \bu}}. 
            \end{align*}
            Hence, when the case $(\rn{ii})$ or the case $(\rn{iii})$ is true, we have
            \begin{align} \label{supp:theorem:2.7:eqn:16}
                \rank{\byn{\tconstant+1}}{\svector{\by}{\lconstant}{\{1\} \cup (\tonumber{\n}\setminus \bv)}} 
                > \alpha = \rank{\bxn{1}}{\svector{\bx}{\lconstant}{\tonumber{\n}\setminus \bu}}.
            \end{align}

            According to Proposition~\ref{supp:proposition:2.5}, 
            there exist $\bxln{3}, \byln{3} \in \vndistinct$ such that
            \begin{align}
                &\bxlnn{3}{\iconstant} = \bxn{\iconstant}, \text{ for any } \iconstant \in \tonumber{\n}\setminus \bu, \label{supp:theorem:2.7:eqn:17}\\
                &\bylnn{3}{\iconstant} = \byn{\iconstant}, \text{ for any } \iconstant \in \tonumber{\n} \setminus (\bv \setminus \{1\}), \label{supp:theorem:2.7:eqn:18}\\
                &\rank{\bxlnn{3}{\iconstant}}{\bxln{3}} = \rank{\bylnn{3}{\iconstant}}{\byln{3}} 
                \text{ for any } \iconstant \in (\bv \setminus \{1\}) \cup \bu. \label{supp:theorem:2.7:eqn:19} 
            \end{align}
            That is, $\bxln{3}$ and $\byln{3}$ are imputations of $\bx$ and $\by$
            for $\bu$ and $\bv \setminus \{1\}$ such that \eqref{supp:theorem:2.7:eqn:19}
            is true.

            Notice that $\bu$ and $\bv$ are disjoint, i.e. $\bu \cap \bv = \emptyset$.
            Hence, $(\bu \setminus \{\tconstant+1\}) \cap \bv = \emptyset$.
            Next, since $|\bu \setminus \{\tconstant+1\}| + |\bv| = \m + \tconstant -1 = \kk$,
            we can apply $\bpn{\kk}$ and get
            \begin{align} \label{supp:theorem:2.7:eqn:19.0}
                \sfd{\bxln{3}}{\byln{3}} \leq \sfd{\bx}{\by}.
            \end{align}
            Notice that \eqref{supp:theorem:2.7:eqn:16} is true.
            Then, according to Lemma~\ref{supp:theorem:2.7:lemma:3},
            when $\bv \setminus \{1\} \neq \emptyset$, for any $\iconstant \in \bv \setminus \{1\}$, we have 
            \begin{align} \label{supp:theorem:2.7:eqn:21}
                \rank{\bylnn{3}{\iconstant}}{\svector{\byln{3}}{\lconstant}{\tonumber{\n} \setminus \{1\}}} \geq \rank{\bxlnn{3}{1}}{\svector{\bxln{3}}{\lconstant}{\tonumber{\n} \setminus \bu}} = \rank{\bxlnn{3}{1}}{\bxln{3}}= \alpha.
            \end{align}	

            Next, according to Lemma~\ref{supp:theorem:2.2:lemma:1}, there exists $\byln{4} \in \vndistinct$ such that
            \begin{align}
                &\bylnn{4}{\iconstant} = \bylnn{3}{\iconstant}, \text{ for any } \iconstant \in \tonumber{\n} \setminus \{1\}, \label{supp:theorem:2.7:eqn:22} \\
                &\rank{\bylnn{4}{1}}{\byln{4}} 
                = \rank{\bxlnn{3}{1}}{\bxln{3}}. \label{supp:theorem:2.7:eqn:23}
            \end{align}
            In other words, $\byln{4}$ is an imputation of $\byln{3}$
            of the index 1 such that $\eqref{supp:theorem:2.7:eqn:23}$
            is true. Then,
            according to Theorem ~\ref{supp:theorem:2.4}, we have
            \begin{align} \label{supp:theorem:2.7:eqn:23.0}
                \sfd{\bxln{3}}{\byln{4}} \leq \sfd{\bxln{3}}{\byln{3}}.
            \end{align}
            According to Lemma~\ref{supp:theorem:2.7:lemma:4},
            for any $\iconstant \in \bu$, we have
            \begin{align}
                \rank{\bylnn{4}{\iconstant}}{\svector{\byln{4}}{\lconstant}{\{1\} \cup (\tonumber{\n} \setminus \bv)}} > \rank{\bylnn{4}{1}}{\svector{\byln{4}}{\lconstant}{\tonumber{\n} \setminus (\bv \setminus \{1\})}} = \alpha. \label{supp:theorem:2.7:eqn:25}
            \end{align}	

            Next, according to Proposition~\ref{supp:proposition:2.5}, there exist $\bxln{5}, \byln{5} \in \vndistinct$ such that
            \begin{align}
                &\bxlnn{5}{\iconstant} = \bxlnn{3}{\iconstant}, \text{ for any } \iconstant \in \tonumber{\n} \setminus \bu, \label{supp:theorem:2.7:eqn:26}\\
                &\bylnn{5}{\iconstant} = \bylnn{4}{\iconstant}, \text{ for any } \iconstant \in \tonumber{\n} \setminus (\bv \setminus \{1\}), \label{supp:theorem:2.7:eqn:27}\\
                &\rank{\bxlnn{5}{\iconstant}}{\bxln{5}} = \rank{\bylnn{5}{\iconstant}}{\byln{5}} \text{ for any } \iconstant \in (\bv \setminus \{1\}) \cup \bu. \label{supp:theorem:2.7:eqn:28} 
            \end{align} 
            In other words, $\bxln{5}$ and $\byln{5}$ are imputations of $\bxln{3}, \byln{4}$
            for $\bu$ and $\bv \setminus \{1\}$, respectively, such that \eqref{supp:theorem:2.7:eqn:28}  is true.

            Since $\bu$ and $\bv$ are disjoint, 
            $\bu$ and $\bv \setminus \{1\}$ are also disjoint. 
            Notice that $|\bu| + |\bv \setminus \{1\}| = \m + \tconstant - 1 = \kk$. 
            Then, according to $\bpn{\kk}$, we have
            \begin{align} \label{supp:theorem:2.7:eqn:28.0} 
                \sfd{\bxln{5}}{\byln{5}} \leq \sfd{\bxln{3}}{\byln{4}}.
            \end{align}

            According to Lemma~\ref{supp:theorem:2.7:lemma:5}, we have
            \begin{align*}
                \rank{\bxlnn{5}{\iconstant}}{\bxln{5}} = \rank{\bylnn{5}{\iconstant}}{\byln{5}}, \text{ for any }\iconstant \in \bu \cup \bv.
            \end{align*}
            Combining \eqref{supp:theorem:2.7:eqn:26} and \eqref{supp:theorem:2.7:eqn:17}, we have
            \begin{align*}
                \bxlnn{5}{\iconstant} =^{\eqref{supp:theorem:2.7:eqn:26}} \bxlnn{3}{\iconstant} =^{\eqref{supp:theorem:2.7:eqn:17}} \bxn{\iconstant}, \text{ for any }\iconstant \in \tonumber{\n} \setminus \bu.
            \end{align*}
            Combining \eqref{supp:theorem:2.7:eqn:27}, \eqref{supp:theorem:2.7:eqn:22}, and \eqref{supp:theorem:2.7:eqn:18}, we have 
            \begin{align*}
                \bylnn{5}{\iconstant} =^{\eqref{supp:theorem:2.7:eqn:27}} \bylnn{4}{\iconstant} =^{\eqref{supp:theorem:2.7:eqn:22}} \bylnn{3}{\iconstant} =^{\eqref{supp:theorem:2.7:eqn:18}} \byn{\iconstant}, \text{ for any }\iconstant \in \tonumber{\n} \setminus \bv.
            \end{align*}
            In other words, $\bxln{5}, \byln{5} \in \vndistinct$ are imputations of $\bx$,
            $\by$ for indices $\bu$ and $\bv$, respectively, such that
            $\rank{\bxlnn{5}{\iconstant}}{\bxln{5}} = \rank{\bylnn{5}{\iconstant}}{\byln{5}}, \text{for any } \iconstant \in \bu \cup \bv$. 
            Since $\bxs, \bys \in \vndistinct$ are imputations of $\bx$,
            $\by$ for indices $\bu$ and $\bv$, respectively, such that
            $\rank{\bxsn{\iconstant}}{\bxs} = \rank{\bysn{\iconstant}}{\bys}, \text{ for any } \iconstant \in \bu \cup \bv$, then according to Proposition~\ref{supp:proposition:equivalence}, we have
            $\sfd{\bxln{5}}{\byln{5}} = \sfd{\bxs}{\bys}$.
            Thus,
            \begin{align*}
                \sfd{\bxs}{\bys} = \sfd{\bxln{5}}{\byln{5}} \le^{\eqref{supp:theorem:2.7:eqn:28.0}} \sfd{\bxln{3}}{\byln{4}} \le^{\eqref{supp:theorem:2.7:eqn:23.0}} \sfd{\bxln{3}}{\byln{3}} \le^{\eqref{supp:theorem:2.7:eqn:19.0}} \sfd{\bx}{\by}.
            \end{align*}
            which proves $\bpn{\kk + 1}$ when the case $(\rn{ii})$ or the case $(\rn{iii})$ holds.
            This completes our proof for the statement $(1)$.

            Next, we prove the statement $(2)$ using the statement $(1)$.
            According to Proposition~\ref{supp:proposition:2.5}, we can find  
            $\bxsp, \bysp \in \vndistinct$ such that $\bxsp, \bysp$
            are imputations of $\bxp, \byp$ for indices 
            $\bu$, and $\bv$, respectively, and 
            $\rank{\bxspn{i}}{\bxsp} = \rank{\byspn{i}}{\bysp}$ for any $i \in \bu \cup \bv$. 
            Then, according to the statement $(1)$, we have 
            \begin{align*}
                \sfd{\bxsp}{\bysp} \le \sfd{\bxp}{\byp}.
            \end{align*}

            Next, since $\bxs, \bys \in \vndistinct$ are imputation of $\bx, \by$
            for $\bu$ and $\bv$, respectively, then we have 
            \begin{align*}
                &\bxsn{i} = \bxn{i}, \text{ for any } i \in \tonumber{\n}\setminus \bu,\\
                &\bysn{i} = \byn{i}, \text{ for any } i \in \tonumber{\n}\setminus \bv.
            \end{align*}
            Further, since $\bxsp, \bysp \in \vndistinct$ such that $\bxsp, \bysp$
            are imputations of $\bxp, \byp$ for indices 
            $\bu$, and $\bv$, respectively, we have
            \begin{align*}
                &\bxspn{i} = \bxsn{i} = \bxn{i}, \text{ for any } i \in \tonumber{\n}\setminus \bu,\\
                &\byspn{i} = \bysn{i} = \byn{i}, \text{ for any } i \in \tonumber{\n}\setminus \bv.
            \end{align*}
            In other words, $\bxsp, \bysp \in \vndistinct$ 
            are also imputations of $\bx, \by$ for indices 
            $\bu$, and $\bv$, respectively.
            Notice that for $\bxs$ and $\bys$, we have 	$\rank{\bxsn{\iconstant}}{\bxs} = \rank{\bysn{\iconstant}}{\bys}, \text{ for any } \iconstant \in \bu \cup \bv$. 
            Also, for $\bxsp$ and $\bysp$, we have $\rank{\bxspn{i}}{\bxsp} = \rank{\byspn{i}}{\bysp}$ for any $i \in \bu \cup \bv$.		
            Then, according to Proposition~\ref{supp:proposition:equivalence:1}, we have $\sfd{\bxsp}{\bysp} = \sfd{\bxs}{\bys}$. Hence, we have
            \begin{align*}
                \sfd{\bxs}{\bys} = \sfd{\bxsp}{\bysp} \le \sfd{\bxp}{\byp}.
            \end{align*}
            This proves the statement $(2)$ and completes our proof.
        \end{proof}

        \subsection{Proof of Proposition 2.9}		

        This subsection proves Proposition 2.9. That is, we show the following result is true.

        \begin{proposition} \label{supp:proposition:2.9} 
            Suppose $\bx, \by \in \vndistinct$ and $\bw \subset \tonumber{\n}$ is a non-empty subset of indices. Define the subvectors $\bxp = (\bx(\lconstant))_{\lconstant \in \tonumber{\n} \setminus \bw}$ and $\byp = (\by(\lconstant))_{\lconstant \in \tonumber{\n} \setminus \bw}$. If the following three conditions all hold,
            \begin{align*}
                & (\rn{i}):  \rank{\bx(\iconstant)}{\bx} = \rank{\by(\iconstant)}{\by}, \text{ for any } \iconstant \in \bw,\\
                & (\rn{ii}): \min(\bx(\lconstant))_{\lconstant \in \bw} > \max(\bx(\lconstant))_{\lconstant \in \tonumber{\n} \setminus \bw}, \\
                & (\rn{iii}): \min(\by(\lconstant))_{\lconstant \in \bw} > \max(\by(\lconstant))_{\lconstant \in \tonumber{\n} \setminus \bw},
            \end{align*}
            then we have $\sfd{\bx}{\by} = \sfd{\bxp}{\byp}$.
        \end{proposition}

        \begin{proof}
            To start, according to the definition of Spearman's footrule, we have
            \begin{align*}
                \sfd{\bx}{\by} &= \sum_{\iconstant \in \bw} \big|\rank{\bx(\iconstant)}{\bx} - \rank{\by(\iconstant)}{\by}\big| 
                + \sum_{\iconstant \in \tonumber{\n} \setminus \bw} \big|\rank{\bx(\iconstant)}{\bx} - \rank{\by(\iconstant)}{\by}\big|.
            \end{align*}
            According to the condition $(\rn{i})$, for any $\iconstant \in \bw$, we have $\rank{\bx(\iconstant)}{\bx} = \rank{\by(\iconstant)}{\by}$.
            Then, 
            \begin{align*}
                \sum_{\iconstant \in \bw} \big|\rank{\bx(\iconstant)}{\bx} - \rank{\by(\iconstant)}{\by}\big|  = 0,
            \end{align*}
            which follows that
            \begin{align*}
                \sfd{\bx}{\by} &= \sum_{\iconstant \in \tonumber{\n} \setminus \bw} \big|\rank{\bx(\iconstant)}{\bx} - \rank{\by(\iconstant)}{\by}\big|.
            \end{align*}

            Next, according to the condition $(\rn{ii})$, we have $\min(\bx(\lconstant))_{\lconstant \in \bw} > \max(\bx(\lconstant))_{\lconstant \in \tonumber{\n} \setminus \bw}$.
            Then, according to Lemma~\ref{supp:lemma:3}, for any $\iconstant \in \tonumber{\n} \setminus \bw$, we have
            \begin{align*}
                \rank{\bx(\iconstant)}{\bx} = \rank{\bx(\iconstant)}{\bxp}.
            \end{align*}
            Similarly, according to the condition $(\rn{iii})$, 
            we have $\min(\by(\lconstant))_{\lconstant \in \bw} > \max(\by(\lconstant))_{\lconstant \in \tonumber{\n} \setminus \bw}$.
            Then, according to Lemma~\ref{supp:lemma:3}, for any $\iconstant \in \tonumber{\n} \setminus \bw$, we have
            \begin{align*}
                \rank{\by(\iconstant)}{\by} = \rank{\by(\iconstant)}{\byp}.
            \end{align*}
            Subsequently, we have
            \begin{align*}
                \sfd{\bx}{\by} &= \sum_{\iconstant \in \tonumber{\n} \setminus \bw} \big|\rank{\bx(\iconstant)}{\bx} - \rank{\by(\iconstant)}{\by}\big| \\
                &= \sum_{\iconstant \in \tonumber{\n} \setminus \bw} \big|\rank{\bx(\iconstant)}{\bxp} - \rank{\by(\iconstant)}{\byp}\big| = \sfd{\bxp}{\byp}.
            \end{align*}
            This completes our proof.
        \end{proof}

        \subsection{Proof of Theorem 2.10} 

        Now, we are ready to prove Theorem 2.10. 

        \begin{theorem} \label{supp:theorem:2.10}
            Suppose $\bx, \by \in \vndistinct$ and $\bu, \bv, \bw \subset \tonumber{\n}$ 
            are pairwise disjoint subsets of indices.
            Suppose $\bxs, \bys \in \vndistinct$ are imputations of $\bx, \by$
            for indices $\bu \cup \bw$, and $\bv \cup \bw$, respectively.
            If the following three conditions all hold
            \begin{align*}
                &(\rn{i}): \rank{\bxs(\iconstant)}{\bxs} = \rank{\bys(\iconstant)}{\bys}, \text{ for any } \iconstant \in \bu \cup \bv \cup \bw, \\
                &(\rn{ii}): \min (\bxs(\lconstant))_{\lconstant \in \bw} > \max (\bxs(\lconstant))_{\lconstant \in \tonumber{\n} \setminus \bw}, \\
                &(\rn{iii}): \min (\bys(\lconstant))_{\lconstant \in \bw} > \max (\bys(\lconstant))_{\lconstant \in \tonumber{\n} \setminus \bw}.
            \end{align*}
            Then, we have $\sfd{\bxs}{\bys} \le \sfd{\bx}{\by}$. Furthermore,
            for any other imputations $\bxln{1}, \byln{1} \in \vndistinct$
            of $\bx, \by$ for indices $\bu \cup \bw$, $\bv \cup \bw$, respectively,
            we have $\sfd{\bxs}{\bys} \le \sfd{\bxln{1}}{\byln{1}}$.
        \end{theorem}

        \begin{proof}
            Notice that there are two statements in Theorem~\ref{supp:theorem:2.10}:
            \begin{itemize}
                \item[(1)]: Suppose $\bx, \by \in \vndistinct$, and $\bu, \bv, \bw \subset \tonumber{\n}$ 
                    are pairwise disjoint subsets of indices.
                    If $\bxs, \bys \in \vndistinct$ are imputations of $\bx$,
                    $\by$ for indices $\bu$ and $\bv$, respectively,
                    and condition $(\rn{i})$, $(\rn{ii})$ and $(\rn{iii})$ all hold, 
                    then we have $\sfd{\bxs}{\bys} \le \sfd{\bx}{\by}$. 
                \item[(2)]: Suppose $\bx, \by \in \vndistinct$, and $\bu, \bv, \bw \subset \tonumber{\n}$ 
                    are pairwise disjoint subsets of indices.
                    If $\bxs, \bys \in \vndistinct$ are imputations of $\bx$,
                    $\by$ for indices $\bu$ and $\bv$, respectively,
                    and condition $(\rn{i})$, $(\rn{ii})$ and $(\rn{iii})$ all hold, 
                    then for any other imputations $\bxsp, \bysp \in \vndistinct$ of 
                    $\bx, \by$ for indices $\bu, \bv$, respectively, we have 
                    $\sfd{\bxs}{\bys} \le \sfd{\bxp}{\byp}$.
            \end{itemize}
            Below, we first show the statement $(1)$ is true. Then 
            we prove the statement $(2)$
            using the statement $(1)$.

            First, we show the statement $(1)$ is true.
            Suppose $\bw =\emptyset$ and $\bu \cup \bv = \emptyset$.
            Then according to the definition of imputations, we have
            $\bxs = \bx$ and $\bys = \by$. Hence, we have
            $\sfd{\bxs}{\bys} = \sfd{\bx}{\by}$, which proves the statement $(1)$
            when $\bw =\emptyset$ and $\bu \cup \bv = \emptyset$.

            Suppose $\bw = \emptyset$ and $\bu \cup \bv \neq \emptyset$. Then $\bxs, \bys \in \vndistinct$ are imputations of $\bx, \by$ for indices $\bu$, and $\bv$, respectively, such that 
            $\rank{\bxs(\iconstant)}{\bx} = \rank{\bys(\iconstant)}{\by}, \text{ for any } \iconstant \in \bu \cup \bv \cup \bw$.
            Then, according to Theorem~\ref{supp:theorem:2.7}, we have
            $\sfd{\bxs}{\bys} \le \sfd{\bx}{\by}$, which proves the statement $(1)$
            when $\bw =\emptyset$ and $\bu \cup \bv \neq \emptyset$.

            Suppose $\bu \cup \bv = \emptyset$ and $\bw \neq \emptyset$. 
            Then since $(\rn{1})$, $(\rn{2})$ and $(\rn{3})$ all hold, we have
            $\bxs, \bys \in \vndistinct$ are imputations of $\bx, \by$
            for indices $\bw$ such that 
            \begin{align*}
                &\rank{\bxs(\iconstant)}{\bx} = \rank{\bys(\iconstant)}{\by}, \text{ for any } \iconstant \in \bw,  \\
                &\min (\bxs(\lconstant))_{\lconstant \in \bw} > \max (\bxs(\lconstant))_{\lconstant \in \tonumber{\n} \setminus \bw}, \\
                &\min (\bys(\lconstant))_{\lconstant \in \bw} > \max (\bys(\lconstant))_{\lconstant \in \tonumber{\n} \setminus \bw}. 
            \end{align*}
            Denote $\bxp = (\bxn{\lconstant})_{\lconstant \in \tonumber{\n} \setminus \bw},\text{ and } \byp = (\byn{\lconstant})_{\lconstant \in \tonumber{\n} \setminus \bw}$. Then, according to Proposition~\ref{supp:proposition:2.9}, we have
            \begin{align*}
                \sfd{\bxs}{\bys} = \sfd{\bxp}{\byp}.
            \end{align*}
            Notice that according to Theorem~~\ref{supp:proposition:2.9}, we have
            \begin{align*}
                \sfd{\bxp}{\byp} \le \sfd{\bx}{\by}.
            \end{align*}
            Hence, we have $\sfd{\bxs}{\bys} \le \sfd{\bx}{\by}$, which proves 
            the statement $(1)$ when $\bw \neq \emptyset$ and $\bu \cup \bv = \emptyset$.

            Next, suppose $\bw \neq \emptyset$ and $\bu \cup \bv \neq \emptyset$.
            Denote $\aconstant = |\bw|$.   	
            Without loss of generality, let us assume (after relabeling) 
            $\bw = \{\n-\aconstant + 1, \ldots, \n\}$.   	
            Denote $\np = \n - \aconstant$.
            Then we have $\tonumber{\n} \setminus \bw = \tonumber{\np}$.
            Denote $\bxp = (\bxn{\lconstant})_{\lconstant \in \tonumber{\n} \setminus \bw}, \byp = (\byn{\lconstant})_{\lconstant \in \tonumber{\n} \setminus \bw},
            \bxsp = (\bxsn{\lconstant})_{\lconstant \in \tonumber{\n} \setminus \bw}, \text{ and } \bysp = (\bysn{\lconstant})_{\lconstant \in \tonumber{\n} \setminus \bw}$.
            Then, we have $\bxp, \byp, \bxsp, \bysp \in \mathcal{V}^{\np}$. 

            Since $\bxs, \bys \in \vndistinct$ are imputations of $\bx, \by$
            for indices $\bu \cup \bw$, and $\bv \cup \bw$, respectively, then we have
            \begin{align*}
                &\bxsn{\iconstant} = \bxn{\iconstant}, \text{ for any } \iconstant \in \tonumber{ \n} \setminus (\bu \cup \bw),\\
                \text{and }&\bysn{\iconstant} = \byn{\iconstant}, \text{ for any } \iconstant \in \tonumber{ \n} \setminus (\bv \cup \bw).
            \end{align*}
            Notice that $\tonumber{ \n} \setminus (\bu \cup \bw) = (\tonumber{ \n} \setminus \bw) \setminus \bu$, and $\tonumber{ \n} \setminus (\bv \cup \bw) = (\tonumber{ \n} \setminus \bv) \setminus \bu$. Hence, we have
            \begin{align*}
                &\bxsn{\iconstant} = \bxn{\iconstant}, \text{ for any } \iconstant \in (\tonumber{ \n} \setminus \bw) \setminus \bu,\\
                \text{and }&\bysn{\iconstant} = \byn{\iconstant}, \text{ for any } \iconstant \in (\tonumber{ \n} \setminus \bw) \setminus \bv.
            \end{align*}
            Therefore, $\bxsp, \bysp$ are imputations of $\bxp, \byp$
            for induces $\bu, \bv \subset \tonumber{\n} \setminus \bw$, respectively.

            Since the condition $(\rn{ii})$ is true,
            then according to Lemma~\ref{supp:lemma:3}, for any $\iconstant \in \bu \cup \bv$, we have
            \begin{align*}
                \rank{\bxsn{\iconstant}}{\bxs} = \rank{\bxsn{\iconstant}}{(\bxsn{\lconstant})_{\lconstant \in \tonumber{\n} \setminus \bw}} = \rank{\bxsn{\iconstant}}{\bxsp}.
            \end{align*}
            Similarly, since the condition $(\rn{iii})$ is true,
            then according to Lemma~\ref{supp:lemma:3}, for any $\iconstant \in \bu \cup \bv$, we have
            \begin{align*}
                \rank{\bysn{\iconstant}}{\bys} = \rank{\bysn{\iconstant}}{(\bysn{\lconstant})_{\lconstant \in \tonumber{\n} \setminus \bw}} = \rank{\bysn{\iconstant}}{\bysp}.
            \end{align*}
            According to the condition $(\rn{i})$, for any $\iconstant \in \bu \cup \bv$, we have 
            \begin{align*}
                \rank{\bxsn{\iconstant}}{\bxs} = 	\rank{\bysn{\iconstant}}{\bys}.
            \end{align*}
            Hence,  for any $\iconstant \in \bu \cup \bv$, we have       
            \begin{align} \label{supp:theorem:2.10:eqn:6}
                \rank{\bxsn{\iconstant}}{\bxsp} = \rank{\bysn{\iconstant}}{\bysp}.
            \end{align}
            Since $\bxsp, \bysp$ are imputations of $\bxp, \byp$
            for indices $\bu, \bv \subset \tonumber{\np}$, respectively,
            then when $\np \ge 2$,    
            according to Theorem~\ref{supp:theorem:2.7}, we have
            \begin{align*}
                \sfd{\bxsp}{\bysp} \le \sfd{\bxp}{\byp}.
            \end{align*}
            However, when $\np = 1$, since $\bu \cup \bv \neq \emptyset$,
            and $\bu \cup \bv \subset \tonumber{\n} \setminus \bw = \tonumber{\np}$,
            we have $\bu \cup \bv = \tonumber{\np}$.    
            Then according to condition $(\rn{ii})$,
            we have $\sfd{\bxsp}{\bysp}  = 0$. Hence, it also follows 
            that
            \begin{align*}
                \sfd{\bxsp}{\bysp} \le \sfd{\bxp}{\byp}.
            \end{align*}

            Next, notice that according to Proposition~\ref{supp:proposition:2.9}, we have
            \begin{align*}
                \sfd{\bxs}{\bys} = \sfd{\bxsp}{\bysp},
            \end{align*}
            and according to Theorem~~\ref{supp:proposition:2.9}, we have
            \begin{align*}
                \sfd{\bxp}{\byp} \le \sfd{\bx}{\by}.
            \end{align*}
            Hence, we have
            \begin{align*}
                \sfd{\bxs}{\bys} = \sfd{\bxsp}{\bysp} \le \sfd{\bxp}{\byp}  \le \sfd{\bx}{\by}.
            \end{align*}
            This completes our proof for the statement $(1)$.

            Next, we prove the statement $(2)$ is true using the statement $(1)$.
            Suppose $\bw = \emptyset$ and $\bu \cup \bv = \emptyset$.
            Then according to the definition of imputations,
            we have $\bxs = \bxln{1} =\bx$ and $\bys = \byln{1} = \by$.
            Hence, we have $\sfd{\bxs}{\bys} = \sfd{\bxln{1}}{\byln{1}}$.
            This completes statement $(2)$ when $\bw = \emptyset$ 
            and $\bu \cup \bv = \emptyset$.

            Suppose $\bw = \emptyset$ and $\bu \cup \bv \neq \emptyset$.
            Then $\bxs, \bys \in \vndistinct$ are imputations of $\bx, \by$
            for indices $\bu$, and $\bv$, respectively, such that 
            $\rank{\bxs(\iconstant)}{\bxs} = \rank{\bys(\iconstant)}{\bys}, \text{ for any } \iconstant \in \bu \cup \bv$.
            Meanwhile, $\bxln{1}, \byln{1} \in \vndistinct$ are imputations of
            of $\bx, \by$ for indices $\bu$, $\bv$.
            Then, according to Theorem~\ref{supp:theorem:2.7},
            we have $\sfd{\bxs}{\bys} \le \sfd{\bxln{1}}{\byln{1}}$.
            This completes the statement $(2)$ when
            $\bw = \emptyset$ and $\bu \cup \bv \neq \emptyset$.

            Suppose $\bw \neq \emptyset$ and $\bu \cup \bv = \emptyset$.
            Let us denote  $\bxp = (\bxn{\lconstant})_{\lconstant \in \tonumber{\n} \setminus \bw},\text{ and } \byp = (\byn{\lconstant})_{\lconstant \in \tonumber{\n} \setminus \bw}$. Then, according to Proposition~\ref{supp:proposition:2.9}, we have
            \begin{align*}
                \sfd{\bxs}{\bys} = \sfd{\bxp}{\byp}.
            \end{align*}

            Next, let us denote $\bxpln{1} = (\bxplnn{1}{\lconstant})_{\lconstant \in \tonumber{\n} \setminus \bw},\text{ and } \bypln{1} = (\byplnn{1}{\lconstant})_{\lconstant \in \tonumber{\n} \setminus \bw}$. 
            Then, according to Theorem~\ref{supp:proposition:2.9}, we have
            \begin{align*}
                \sfd{\bxpln{1}}{\bypln{1}} \le \sfd{\bxln{1}}{\byln{1}}.
            \end{align*}
            Since $\bxln{1}, \byln{1} \in \vndistinct$
            are imputations of $\bx, \by$ for indices $\bw$, we have 
            $\bxpln{1} = \bxp$ and $\bypln{1} = \byp$. Thus,
            \begin{align*}
                \sfd{\bxp}{\byp} = \sfd{\bxpln{1}}{\bypln{1}}.
            \end{align*}
            Combining the above results, we have 
            \begin{align*}
                \sfd{\bxln{1}}{\byln{1}} \ge 	\sfd{\bxpln{1}}{\bypln{1}} = \sfd{\bxp}{\byp} = \sfd{\bxs}{\bys}.
            \end{align*}
            This completes our proof when  $\bw \neq \emptyset$ and $\bu \cup \bv = \emptyset$.

            Next, suppose $\bw \neq \emptyset $ and $\bu \cup \bv \neq \emptyset$.
            Denote $\aconstant = |\bw|$.   	
            Without loss of generality, let us assume (after relabeling) 
            $\bw = \{\n-\aconstant + 1, \ldots, \n\}$.   	
            Denote $\np = \n - \aconstant$.
            Then we have $\tonumber{\n} \setminus \bw = \tonumber{\np}$.
            Denote $\bxp = (\bxn{\lconstant})_{\lconstant \in \tonumber{\n} \setminus \bw},~\byp = (\byn{\lconstant})_{\lconstant \in \tonumber{\n} \setminus \bw},~
            \bxpln{1} = (\bxplnn{1}{\lconstant})_{\lconstant \in \tonumber{\n} \setminus \bw},~\bypln{1} = (\byplnn{1}{\lconstant})_{\lconstant \in \tonumber{\n} \setminus \bw},
            \bxsp = (\bxsn{\lconstant})_{\lconstant \in \tonumber{\n} \setminus \bw}, \bysp = (\bysn{\lconstant})_{\lconstant \in \tonumber{\n} \setminus \bw}$. 
            Then, we have $\bxp, \byp, \bxpln{1}, \bypln{1}, \bxsp, \bysp \in \mathcal{V}^{\np}$. 

            Let us first show $ \sfd{\bxln{1}}{\byln{1}}  \ge \sfd{\bxs}{\bys}$
            when $\np = 1$.
            Suppose $\np = 1$. Then $\bxsp = \bxsn{1}$ and $\bysp = \bysn{1}$.
            According to the definition of rank, we have
            $\rank{\bxsp}{\bxsp} = \rank{\bysp}{\bysp} = 1$. Hence,
            $\sfd{\bxsp}{\bysp} = 0$.

            Next, since the conditions $(\rn{i})$, $(\rn{ii})$ and $(\rn{iii})$ all
            hold. Then according to Proposition~\ref{supp:proposition:2.9}, we have 
            $\sfd{\bxs}{\bys} = \sfd{\bxsp}{\bysp} = 0$.  Hence, we have 
            $ \sfd{\bxln{1}}{\byln{1}}  \ge 0 = \sfd{\bxs}{\bys}$.

            In the following, we show $ \sfd{\bxln{1}}{\byln{1}}  \ge \sfd{\bxs}{\bys}$
            is true when $\np \ge 2$. 

            First, we show that $\bxsp$ and $\bysp$ are imputations of 
            $\bxpln{1}, \bypln{1}$ for $\bu$ and $\bv$, respectively,
            such that $\rank{\bxspn{\iconstant}}{\bxsp} = \rank{\byspn{\iconstant}}{\bysp}, \text{ for any } \iconstant \in \bu \cup \bv.$

            Notice that $\bxs, \bys \in \vndistinct$ are imputations of $\bx, \by$
            for indices $\bu \cup \bw$, and $\bv \cup \bw$, respectively.
            Then we have 
            $\bxsn{\iconstant} = \bxn{\iconstant}$ for any $\iconstant \in \tonumber{\n} \setminus (\bu \cup \bw)$,
            and $\bysn{\iconstant} = \byn{\iconstant}$ for any $\iconstant \in \tonumber{\n} \setminus (\bv \cup \bw)$.

            Notice that $\tonumber{\n} \setminus (\bu \cup \bw)
            = (\tonumber{\n} \setminus \bw) \setminus \bu = \tonumber{\np} \setminus \bu$
            and
            $\tonumber{\n} \setminus (\bv \cup \bw)
            = (\tonumber{\n} \setminus \bw) \setminus \bv = \tonumber{\np} \setminus \bv$.
            Then, we have 
            $\bxsn{\iconstant} = \bxn{\iconstant}$ for any $\iconstant \in \tonumber{\np} \setminus \bu $,
            and $\bysn{\iconstant} = \byn{\iconstant}$ for any $\iconstant \in \tonumber{\np}  \setminus \bv$.

            According to the definition of $\bxsp, \bysp, \bxp, \byp$, we have    	 
            $\bxspn{\iconstant} = \bxn{\iconstant} = \bxpn{\iconstant}$ for any $\iconstant \in \tonumber{\np} \setminus \bu$ and 
            $\byspn{\iconstant} = \byn{\iconstant} = \bypn{\iconstant}$ for any $\iconstant \in \tonumber{\np} \setminus \bv$. 
            Hence, $\bxsp$ and $\bysp$ are imputations of 
            $\bxpln{1}, \bypln{1}$ for $\bu$ and $\bv$, respectively.

            Since the condition $(\rn{ii})$ is true,
            then according to Lemma~\ref{supp:lemma:3}, for any $\iconstant \in \bu \cup \bv$, we have
            \begin{align*}
                \rank{\bxsn{\iconstant}}{\bxs} = \rank{\bxsn{\iconstant}}{(\bxsn{\lconstant})_{\lconstant \in \tonumber{\n} \setminus \bw}} = \rank{\bxsn{\iconstant}}{\bxsp}.
            \end{align*}
            Similarly, since the condition $(\rn{iii})$ is true,
            then according to Lemma~\ref{supp:lemma:3}, for any $\iconstant \in \bu \cup \bv$, we have
            \begin{align*}
                \rank{\bysn{\iconstant}}{\bys} = \rank{\bysn{\iconstant}}{(\bysn{\lconstant})_{\lconstant \in \tonumber{\n} \setminus \bw}} = \rank{\bysn{\iconstant}}{\bysp}.
            \end{align*}
            According to condition $(\rn{i})$, for any $\iconstant \in \bu \cup \bv$, we have 
            \begin{align*}
                \rank{\bxsn{\iconstant}}{\bxs} = 	\rank{\bysn{\iconstant}}{\bys}.
            \end{align*}
            Hence, for any $\iconstant \in \bu \cup \bv$, we have       
            \begin{align*}
                \rank{\bxsn{\iconstant}}{\bxsp} = \rank{\bysn{\iconstant}}{\bysp}.
            \end{align*}
            According to the definition of $\bxsp, \bysp$, we have 
            \begin{align*}
                \rank{\bxspn{\iconstant}}{\bxsp} = \rank{\byspn{\iconstant}}{\bysp}, \text{ for any } \iconstant \in \bu \cup \bv.
            \end{align*}

            Next, according to Proposition~\ref{supp:proposition:2.5},
            we can find imputations
            $\bxpln{2}, \bypln{2} \in \mathcal{V}^{\np}$
            of $\bxpln{1}, \bypln{1}$ for $\bu$ and $\bv$, respectively, such that
            $\rank{\bxplnn{2}{\iconstant}}{\bxpln{1}} = \rank{\byplnn{2}{\iconstant}}{\bypln{1}} $ for any
            $\iconstant \in \bu \cup \bv$.
            Since $\bxsp$ and $\bysp$ are imputations of 
            $\bxpln{1}, \bypln{1}$ for $\bu$ and $\bv$, respectively,
            such that $\rank{\bxspn{\iconstant}}{\bxsp} = \rank{\byspn{\iconstant}}{\bysp}, \text{ for any } \iconstant \in \bu \cup \bv.$
            Then, according to Proposition~\ref{supp:proposition:equivalence:1},
            we have
            \begin{align*}
                \sfd{\bxpln{2}}{\bypln{2}} = \sfd{\bxsp}{\bysp}.
            \end{align*}
            Notice that the conditions $(\rn{i})$, $(\rn{ii})$, and $(\rn{iii})$ all
            hold. Then according to Proposition~\ref{supp:proposition:2.9}, we have 
            \begin{align*}
                \sfd{\bxsp}{\bysp} = \sfd{\bxs}{\bys}.
            \end{align*}
            Hence, we have
            \begin{align} \label{supp:theorem:2.10:eqn:0}
                \sfd{\bxs}{\bys} = \sfd{\bxpln{2}}{\bypln{2}}.
            \end{align}

            Next, let $\bxln{2}, \byln{2} \in \vndistinct$
            be vectors such that the following
            five conditions all hold
            \begin{align}
                &\bxlnn{2}{i} = \bxplnn{2}{i}, \text{ for any } i \in \tonumber{\n} \setminus \bw, \label{supp:theorem:2.10:1}\\
                &\bylnn{2}{i} = \byplnn{2}{i}, \text{ for any } i \in \tonumber{\n} \setminus \bw,\label{supp:theorem:2.10:2}\\
                &\rank{\bxlnn{2}{i}}{\bxln{2}} = \rank{\bylnn{2}{i}}{\byln{2}},  \text{ for any } i \in \bw, \label{supp:theorem:2.10:3}\\
                &\min (\bxlnn{2}{\lconstant})_{\lconstant \in \bw} > \max (\bxlnn{2}{\lconstant})_{\lconstant \in \tonumber{\n} \setminus \bw},\label{supp:theorem:2.10:4}\\
                \text{and }&\min (\bylnn{2}{\lconstant})_{\lconstant \in \bw} > \max (\bylnn{2}{\lconstant})_{\lconstant \in \tonumber{\n} \setminus \bw}. \label{supp:theorem:2.10:5}
            \end{align}
            Combining \eqref{supp:theorem:2.10:1} and \eqref{supp:theorem:2.10:2},
            we have $\bxlnn{2}{i} = \bxplnn{2}{i}$
            and $\bylnn{2}{i} = \byplnn{2}{i} $ for any $\iconstant \in \tonumber{\np} \setminus \bw = \tonumber{\np}$.
            Further, since $\bxpln{2}, \bypln{2} \in \mathcal{V}^{\np}$ are
            imputations of
            $\bxpln{1}, \bypln{1}$ for $\bu$ and $\bv$, respectively,
            we have $\bxplnn{2}{i} = \bxplnn{1}{i}$ for any $\iconstant \in \tonumber{\np} \setminus \bu$
            and $\byplnn{2}{i} = \byplnn{1}{i}$ for any $\iconstant \in \tonumber{\np} \setminus \bv$.
            Hence, we have
            \begin{align*}
                &\bxlnn{2}{i} = \bxplnn{2}{i} = \bxplnn{1}{i} = \bxlnn{1}{i}, \text{ for any } \iconstant \in \tonumber{\np} \setminus \bu,\\
                \text{and }&   \bylnn{2}{i} = \byplnn{2}{i} = \byplnn{1}{i} = \bylnn{1}{i}, \text{ for any } \iconstant \in \tonumber{\np} \setminus \bv.
            \end{align*}    	
            Notice that $\tonumber{\np} = \tonumber{\n} \setminus \bw$. Hence, we 
            have $ \tonumber{\np} \setminus \bu = \tonumber{\n} \setminus (\bw \cup \bu)$,
            and $ \tonumber{\np} \setminus \bv = \tonumber{\n} \setminus (\bw \cup \bv)$.
            Therefore, it follows that
            \begin{align*}
                &\bxlnn{2}{i} = \bxplnn{2}{i} = \bxplnn{1}{i} = \bxlnn{1}{i}, \text{ for any } \iconstant \in \tonumber{\n} \setminus (\bw \cup \bu),\\
                \text{and }&   \bylnn{2}{i} = \byplnn{2}{i} = \byplnn{1}{i} = \bylnn{1}{i}, \text{ for any } \iconstant \in \tonumber{\n} \setminus (\bw \cup \bv).
            \end{align*}
            In other words, $\bxln{2}$ and $\byln{2}$ are imputations of $\bxln{1}, \byln{1}$
            for indices $\bw \cup \bu$ and $\bw \cup \bv$, respectively. 
            Notice that \eqref{supp:theorem:2.10:3}, \eqref{supp:theorem:2.10:4} and \eqref{supp:theorem:2.10:5} are true. Then according to 
            Theorem~\ref{supp:theorem:2.10}, we have
            \begin{align*}
                \sfd{\bxln{1}}{\byln{1}} \ge \sfd{\bxln{2}}{\byln{2}}.
            \end{align*} 
            Since \eqref{supp:theorem:2.10:3}, \eqref{supp:theorem:2.10:4} and \eqref{supp:theorem:2.10:5} are true, then according to 
            Proposition~\ref{supp:proposition:2.9}, we have 
            \begin{align*}
                \sfd{\bxln{2}}{\byln{2}} = \sfd{\bxpln{2}}{\bypln{2}}.
            \end{align*}
            Hence, we have
            \begin{align*}
                \sfd{\bxln{1}}{\byln{1}} \ge \sfd{\bxpln{2}}{\bypln{2}} =^{\eqref{supp:theorem:2.10:eqn:0}} \sfd{\bxs}{\bys}.
            \end{align*}
            This completes proof for the statement $(2)$, and finishes our proof.

        \end{proof}

        \section{Proof of upper bounds} \label{appC}

        This section provides results for deriving exact upper bounds of Spearman's footrule
        in the presence of missing data.

        \subsection{Proof of Proposition 2.11}
        This subsection proves Proposition~2.11. We start by showing
        the following lemma:

        \begin{lemma} \label{supp:proposition:2.11:lemma:1}
            Suppose $\bx, \by \in \vndistinct$, and let $\bxln{1}$, $\bxln{2}$
            be imputations of $\bx$ for an index $\uu \in \tonumber{\n}$ 
            such that $\rank{\bxlnn{1}{\uu}}{\bxln{1}} = 1, \text{ and } \rank{\bxlnn{2}{\uu}}{\bxln{2}} = \n$. Then, if
            \begin{align*}
                (\rn{i}):~\rank{\bxn{\uu}}{\bx} \ge \rank{\byn{\uu}}{\by}
            \end{align*}
            holds, we have $\sfd{\bx}{\by} \le  \sfd{\bxln{2}}{\by}$.
            However, if
            \begin{align*}
                (\rn{ii}):~\rank{\bxn{\uu}}{\bx} \le \rank{\byn{\uu}}{\by}
            \end{align*}
            holds, we have $\sfd{\bx}{\by} \le  \sfd{\bxln{1}}{\by}$.
        \end{lemma}
        \begin{proof}
            We prove the results for the case $(\rn{i})$ and $(\rn{ii})$ separately.

            Suppose the case $(\rn{i}):~\rank{\bxn{\uu}}{\bx} \ge \rank{\byn{\uu}}{\by}$
            is true.
            Then, we have either 
            $\rank{\bxn{\uu}}{\bx} = \n$ or
            $\rank{\bxn{\uu}}{\bx} < \n$.
            Suppose $\rank{\bxn{\uu}}{\bx} = \n$. Then, we have
            \begin{align*}
                \rank{\bxn{\uu}}{\bx} = \n = \rank{\bxlnn{2}{\uu}}{\bxln{2}}.
            \end{align*}
            Since $\bxln{2}$ is an imputation of $\bx$ for the index $\uu$, 
            then according to Lemma~\ref{supp:lemma:4}, we have
            $\rank{\bxln{2}}{\bxln{2}} = \rank{\bx}{\bx}$.
            Hence, we have $\sfd{\bx}{\by} = \sfd{\bxln{2}}{\by}$.

            Suppose $\rank{\bxn{\uu}}{\bx} < \n$.
            Denote $\rank{\bxn{\uu}}{\bx} = \n - \bconstant$, 
            where $\bconstant \in \natnum$ and $\bconstant > 0$. 
            Then, according to Lemma~\ref{supp:proposition:1:lemma:2},
            there exist $\bvln{1}, \ldots, \bvln{\bconstant} \in \vndistinct$ such that
            for any $\kk \in \{1,\ldots, \bconstant\}$, $\bvln{\kk}$
            is an imputation of $\bx$ for the index $\uu$ and
            \begin{align*}
                \rank{\bvlnn{\kk}{\uu}}{\bvln{\kk}} = \rank{\bxn{\uu}}{\bx} + \kk.
            \end{align*}
            Define $\bvln{0} = \bx$. Then, for any $\kk \in \{1,\ldots,\bconstant\}$,
            we have
            \begin{align*}
                &\rank{\bvlnn{\kk-1}{\uu}}{\bvln{\kk-1}} = \rank{\bxn{\uu}}{\bx} + (\kk - 1) \ge  \rank{\bxn{\uu}}{\bx} \ge  \rank{\byn{\uu}}{\by},\\
                \text{and }& \rank{\bvlnn{\kk}{\uuln{1}}}{\bvln{\kk}} = \rank{\bvlnn{\kk-1}{\uuln{1}}}{\bvln{\kk-1}} + 1.
            \end{align*}
            By applying Lemma~\ref{supp:proposition:1:lemma:3} for each pair $\bvln{\kk}, \by$ and $\bvln{\kk-1}, \by$, where $\kk \in \{1,\ldots,\bconstant\}$,
            we have
            \begin{align} \label{supp:proposition:2.11:lemma:1:eqn:1}
                \sfd{\bvln{0}}{\by} \le \sfd{\bvln{1}}{\by} \le \ldots \le \sfd{\bvln{\bconstant}}{\by}.
            \end{align}
            Notice that 
            \begin{align*}
                \rank{\bvlnn{\bconstant}{\uu}}{\bvln{\bconstant}} = \rank{\bxn{\uu}}{\bx} + \bconstant = \n = \rank{\bxlnn{2}{\uu}}{\bxln{2}}.
            \end{align*}
            Since $\bvln{\bconstant}$ and $\bxln{2}$ are
            both imputation of $\bx$ for the index $\uu$,
            then according to Lemma~\ref{supp:lemma:4},
            we have 
            \begin{align*}
                \sfd{\bxln{2}}{\by} = \sfd{\bvln{\bconstant}}{\by}.
            \end{align*}
            According to \eqref{supp:proposition:2.11:lemma:1:eqn:1}, we further have
            \begin{align*}
                \sfd{\bxln{2}}{\by} \ge \sfd{\bvln{0}}{\by}. 
            \end{align*}
            Since $\bvln{0} = \bx$, then we have $ \sfd{\bx}{\by} \le	\sfd{\bxln{2}}{\by}$.
            This completes our proof when the case $(\rn{i})$ is true.

            When the case $(\rn{ii})$ is true, the results can be proved similarly.   	
            Suppose the case $(\rn{ii})$ is true, 
            then we have either $\rank{\bxn{\uu}}{\bx} = 1$,
            or $\rank{\bxn{\uu}}{\bx} > 1$ is true.
            Suppose $\rank{\bxn{\uu}}{\bx} = 1$. Then, 
            \begin{align*}
                \rank{\bxn{\uu}}{\bx} = 1 = \rank{\bxlnn{1}{\uu}}{\bx}.
            \end{align*}
            Since $\bxln{1}$ is an imputation of $\bx$ for the index $\uu$, then 
            according to Lemma~\ref{supp:lemma:4} we have
            $\rank{\bxln{1}}{\bxln{1}} = \rank{\bx}{\bx}$.
            Hence, we have $\sfd{\bx}{\by} = \sfd{\bxln{1}}{\by}$.

            Suppose $\rank{\bxn{\uu}}{\bx} > 1$. Denote $\rank{\bxn{\uuln{1}}}{\bx} = 1 + \aconstant$, where $\aconstant \in \natnum$ and $\aconstant > 0$. Then, according to Lemma~\ref{supp:proposition:1:lemma:2},
            there exist $\bvln{1}, \ldots, \bvln{\aconstant} \in \vndistinct$ such that
            for any $\kk \in \{1,\ldots, \aconstant\}$, $\bvln{\kk}$
            is an imputation of $\bx$ for the index $\uu$ and
            \begin{align*}
                \rank{\bvlnn{\kk}{\uu}}{\bvln{\kk}} = \rank{\bxn{\uu}}{\bx} - \kk.
            \end{align*}
            Define $\bvln{0} = \bx$. Then, for any $\kk \in \{1,\ldots,\aconstant\}$, we have
            \begin{align*}
                &\rank{\bvlnn{\kk-1}{\uu}}{\bvln{\kk-1}} = \rank{\bxn{\uu}}{\bx} - (\kk - 1) \le  \rank{\bxn{\uu}}{\bx} \le \rank{\byn{\uu}}{\by}, \\
                \text{and }&   		\rank{\bvlnn{\kk}{\uu}}{\bvln{\kk}} = \rank{\bvlnn{\kk-1}{\uu}}{\bvln{\kk-1}} - 1.
            \end{align*}
            By applying Lemma~\ref{supp:proposition:1:lemma:3} 
            for each pair $\bvln{\kk}, \by$ and $\bvln{\kk-1}, \by$, where $\kk \in \{1,\ldots,\aconstant\}$, we have
            \begin{align}  \label{supp:proposition:2.11:lemma:1:eqn:2}
                \sfd{\bvln{0}}{\by} \le \sfd{\bvln{1}}{\by} \le \cdots \le \sfd{\bvln{\aconstant}}{\by}.
            \end{align}
            Notice that 
            \begin{align*}
                \rank{\bvlnn{\aconstant}{\uu}}{\bvln{\aconstant}} = \rank{\bxn{\uu}}{\bx} - \aconstant = 1 = \rank{\bxlnn{1}{\uu}}{\bxln{1}}.
            \end{align*}
            Since $\bvln{\aconstant}$ and $\bxln{1}$ are
            both imputation of $\bx$ for the index $\uu$,
            then according to Lemma~\ref{supp:lemma:4},
            we have 
            \begin{align*}
                \sfd{\bxln{1}}{\by} = \sfd{\bvln{\aconstant}}{\by}.
            \end{align*}
            According to \eqref{supp:proposition:2.11:lemma:1:eqn:2}, we further
            have
            \begin{align*}
                \sfd{\bxln{1}}{\by} \ge \sfd{\bvln{0}}{\by}.
            \end{align*}
            Since $\bvln{0} = \bx$, then we have $\sfd{\bxln{1}}{\by} \ge \sfd{\bx}{\by}$.
            This completes our proof.
        \end{proof}

        Now, we are ready to prove Proposition~2.11.

        \begin{proposition} \label{supp:proposition:2.11}
            Suppose $\bx, \by \in \vndistinct$, and let $\bxln{1}$, $\bxln{2}$
            be imputations of $\bx$ for an index $\uu \in \tonumber{\n}$ 
            such that $\rank{\bxlnn{1}{\uu}}{\bxln{1}} = 1, \text{ and } \rank{\bxlnn{2}{\uu}}{\bxln{2}} = \n$. Then, 
            \begin{align*}
                \sfd{\bx}{\by} \le \max \{ \sfd{\bxln{1}}{\by}, \sfd{\bxln{2}}{\by} \},
            \end{align*}
            and for any imputation $\bxp \in \vndistinct$ of $\bx$ for index $\uu$,
            $\sfd{\bxp}{\by} \le \max \{ \sfd{\bxln{1}}{\by}, \sfd{\bxln{2}}{\by} \}$.
        \end{proposition}

        \begin{proof}
            According to Lemma~\ref{supp:proposition:2.11:lemma:1},
            if $\rank{\bxn{\uu}}{\bx} \ge \rank{\byn{\uu}}{\by}$, we have
            $\sfd{\bx}{\by} \le \sfd{\bxln{1}}{\by}  \le \max \{ \sfd{\bxln{1}}{\by}, \sfd{\bxln{2}}{\by} \}$.
            However, if $\rank{\bxn{\uu}}{\bx} \le \rank{\byn{\uu}}{\by}$, we have
            $\sfd{\bx}{\by} \le \sfd{\bxln{2}}{\by}  \le \max \{ \sfd{\bxln{1}}{\by}, \sfd{\bxln{2}}{\by} \}.$ Hence, we have 
            \begin{align*}
                \sfd{\bx}{\by} \le \max \{ \sfd{\bxln{1}}{\by}, \sfd{\bxln{2}}{\by} \}.
            \end{align*}

            Next, we show that $\sfd{\bxp}{\by} \le \max \{ \sfd{\bxln{1}}{\by}, \sfd{\bxln{2}}{\by} \}$. According to  Lemma~\ref{supp:proposition:1:lemma:2}, there exist imputations $\bxln{3}, \bxln{4} \in \vndistinct$ of $\bxp$ for
            the index $\uu$ such that $\rank{\bxlnn{3}{\uu}}{\bxln{3}} = 1$, and $\rank{\bxlnn{4}{\uu}}{\bxln{4}} = \n$. 

            Then, according to Lemma~\ref{supp:proposition:2.11:lemma:1},
            if $\rank{\bxpn{\uu}}{\bx} \ge \rank{\byn{\uu}}{\by}$, we have
            $\sfd{\bxp}{\by} \le \sfd{\bxln{3}}{\by}  \le \max \{ \sfd{\bxln{3}}{\by}, \sfd{\bxln{4}}{\by} \}$.
            However, if $\rank{\bxpn{\uu}}{\bx} \le \rank{\byn{\uu}}{\by}$, we have
            $\sfd{\bxp}{\by} \le \sfd{\bxln{4}}{\by}  \le \max \{ \sfd{\bxln{3}}{\by}, \sfd{\bxln{4}}{\by} \}.$ Hence, we have $\sfd{\bxp}{\by} \le \max \{ \sfd{\bxln{3}}{\by}, \sfd{\bxln{4}}{\by} \}$.

            Notice that $\bxln{3}, \bxln{4}$ are both imputations of $\bxp$ for the index
            $\uu$, and $\bxp$ is the imputation of $\bx$ for index $\uu$. Hence, according to
            the definition of imputations, we can show that
            $\bxln{3}, \bxln{4}$ are both imputation of $\bx$ for the index $\uu$. Notice that 
            $\bxln{1}, \bxln{2}$ are imputation of $\bx$ for the index $\uu$ such that
            $\rank{\bxlnn{1}{\uu}}{\bxln{1}} = 1, \text{ and } \rank{\bxlnn{2}{\uu}}{\bxln{2}} = \n$. Then, according to Lemma~\ref{supp:lemma:4}, we have
            $\rank{\bxln{3}}{\bxln{3}} =\rank{\bxln{1}}{\bxln{1}}$,
            and
            $\rank{\bxln{4}}{\bxln{4}} = \rank{\bxln{2}}{\bxln{2}}$.
            Hence, we have $\sfd{\bxln{3}}{\by} =\sfd{\bxln{1}}{\by} $
            and $\sfd{\bxln{4}}{\by}=  \sfd{\bxln{2}}{\by}$. Notice that
            $\sfd{\bxp}{\by} \le \max \{ \sfd{\bxln{3}}{\by}, \sfd{\bxln{4}}{\by} \}$.
            Therefore, we have 
            \begin{align*}
                \sfd{\bxp}{\by} \le \max \{ \sfd{\bxln{1}}{\by}, \sfd{\bxln{2}}{\by} \}.
            \end{align*}
            This completes our proof.
        \end{proof}

        \subsection{Proof of Proposition 2.13}   

        This subsection proves Proposition 2.13. First, we make the following definition:

        \begin{definition} \label{def:1}
            Let $\bu \subset \bt \subset \tonumber{\n}$ be a subset of indices. Define
            \begin{align*}
                \myfirstset{\bu}{\bt}{\n} =  \{ \bz \in \vndistinct|
                ~\text{for any}~\iconstant \in \bu,~ 
                \bzn{\iconstant} > \max_{\jconstant \in \bt \setminus \bu} \bzn{\jconstant} \text{ or } \bzn{\iconstant} < \min_{\jconstant \in \bt \setminus \bu} \bzn{\jconstant} \}.
            \end{align*}
        \end{definition}

        Then, we provide the following lemma, which will be useful for proving
        Proposition 2.13.

        \begin{lemma} \label{supp:proposition:2.13:lemma:1}
            Let $\bu \subset \tonumber{\n}$ be a non-empty subset of indices, and suppose $\bxln{1} \in \myfirstset{\bu}{\tonumber{\n}}{\n}$.  Let $\bv \subset \tonumber{\n}$ be 
            a non-empty subset of indices such that $\bv \cap \bu = \emptyset$.
            Then, if $\bxln{2} \in \myfirstset{\bv}{\tonumber{\n}}{\n}$ is an imputation of $\bxln{1}$ for $\bv$, we have $\bxln{2} \in \myfirstset{\bu \cup \bv}{\tonumber{\n}}{\n}$.
        \end{lemma}
        \begin{proof}
            Since $\bxln{1} \in \myfirstset{\bu}{\tonumber{\n}}{\n}$, we have
            \begin{align*}
                \bxln{1}(\iconstant) > \max_{\jconstant \in \tonumber{\n} \setminus \bu} \bxln{1}(\jconstant), \text{ or }
                \bxln{1}(\iconstant) < \min_{\jconstant \in \tonumber{\n} \setminus \bu} \bxln{1}(\jconstant), \text{ for any } \iconstant \in \bu.
            \end{align*}
            Hence, we have
            \begin{align} \label{supp:proposition:2.13:lemma:1:eqn:1}
                \bxln{1}(\iconstant) > \max_{\jconstant \in \tonumber{\n} \setminus (\bu \cup \bv)} \bxln{1}(\jconstant), \text{ or }
                \bxln{1}(\iconstant) < \min_{\jconstant \in \tonumber{\n} \setminus (\bu \cup \bv)} \bxln{1}(\jconstant), \text{ for any } \iconstant \in \bu.
            \end{align}

            Next, since $\bxln{2}$ is an imputation of $\bxln{1}$ for indices $\bv$, we have
            \begin{align*}
                \bxlnn{2}{i} = \bxlnn{1}{i}, \text{ for any } \iconstant \in \tonumber{\n} \setminus\bv.
            \end{align*}
            Notice that $\tonumber{\n} \setminus (\bu \cup \bv) \subset 
            \tonumber{\n} \setminus\bv$. Hence, we have 
            \begin{align*}
                &\bxlnn{2}{i} = \bxlnn{1}{i}, \text{ for any } \iconstant \in \tonumber{\n} \setminus (\bv \cup \bu)\\
                \Rightarrow & \max_{\jconstant \in \tonumber{\n} \setminus (\bu \cup \bv)} \bxln{2}(\jconstant) = \max_{\jconstant \in \tonumber{\n} \setminus (\bu \cup \bv)} \bxln{1}(\jconstant), \text{ and } \min_{\jconstant \in \tonumber{\n} \setminus (\bu \cup \bv)} \bxln{2}(\jconstant)  = \min_{\jconstant \in \tonumber{\n} \setminus (\bu \cup \bv)} \bxln{1}(\jconstant).
            \end{align*}
            Meanwhile, since $\bu \cap \bv = \emptyset$, we have $\bu \subset \tonumber{\n} \setminus \bv$. Hence, we have
            \begin{align*}
                \bxlnn{2}{i} = \bxlnn{1}{i}, \text{ for any } \iconstant \in \bu.
            \end{align*}
            Then, according to \eqref{supp:proposition:2.13:lemma:1:eqn:1}, we have
            \begin{align} \label{supp:proposition:2.13:lemma:1:eqn:2}
                \bxln{2}(\iconstant) > \max_{\jconstant \in \tonumber{\n} \setminus (\bu \cup \bv)} \bxln{2}(\jconstant), \text{ or }
                \bxln{2}(\iconstant) < \min_{\jconstant \in \tonumber{\n} \setminus (\bu \cup \bv)} \bxln{2}(\jconstant), \text{ for any } \iconstant \in \bu.
            \end{align}   	

            Since $\bxln{2} \in \myfirstset{\bv}{\tonumber{\n}}{\n}$, then we have
            \begin{align*}
                \bxln{2}(\iconstant) > \max_{\jconstant \in \tonumber{\n} \setminus \bv} \bxln{2}(\jconstant), \text{ or }
                \bxln{2}(\iconstant) < \min_{\jconstant \in \tonumber{\n} \setminus \bv} \bxln{2}(\jconstant), \text{ for any } \iconstant \in \bv.
            \end{align*}
            Notice that $\bu \cap \bv = \emptyset$.
            Then, we have $\bu \subset \tonumber{\n} \setminus \bv$.
            Hence, we have
            \begin{align*}
                \bxln{2}(\iconstant) > \max_{\jconstant \in \bu} \bxln{2}(\jconstant), \text{ or }
                \bxln{2}(\iconstant) < \min_{\jconstant \in \bu} \bxln{2}(\jconstant), \text{ for any } \iconstant \in \bv.
            \end{align*}
            According to \eqref{supp:proposition:2.13:lemma:1:eqn:2}, 
            we have
            \begin{align*}
                \max_{\jconstant \in \bu} \bxln{2}(\jconstant) > \max_{\jconstant \in \tonumber{\n} \setminus (\bu \cup \bv)} \text{ or } \min_{\jconstant \in \bu} \bxln{2}(\jconstant) < \min_{\jconstant \in \tonumber{\n} \setminus (\bu \cup \bv)} \bxln{2}(\jconstant).
            \end{align*}
            Hence, we further have
            \begin{align*}
                \bxln{2}(\iconstant) > \max_{\jconstant \in \tonumber{\n} \setminus (\bu \cup \bv)} \bxln{2}(\jconstant), \text{ or }
                \bxln{2}(\iconstant) < \min_{\jconstant \in \tonumber{\n} \setminus (\bu \cup \bv)} \bxln{2}(\jconstant), \text{ for any } \iconstant \in \bv.
            \end{align*}
            Combining this result with \eqref{supp:proposition:2.13:lemma:1:eqn:2},
            we have 
            \begin{align*}
                \bxln{2}(\iconstant) > \max_{\jconstant \in \tonumber{\n} \setminus (\bu \cup \bv)} \bxln{2}(\jconstant), \text{ or }
                \bxln{2}(\iconstant) < \min_{\jconstant \in \tonumber{\n} \setminus (\bu \cup \bv)} \bxln{2}(\jconstant), \text{ for any } \iconstant \in \bv \cup \bu.
            \end{align*}   	
            This completes our proof.

        \end{proof}

        Now, we are ready to prove Proposition 2.13.

        \begin{proposition} \label{supp:proposition:2.13}
            Suppose $\bx, \by \in \vndistinct$, and let $\bu \subset \tonumber{\n}$.
            Then, there exists an imputation $\bxs \in \myfirstset{\bu}{\tonumber{\n}}{\n}$ of $\bx$ for indices $\bu$ such that	$\sfd{\bx}{\by} \le \sfd{\bxs}{\by}$, and for any other imputation $\xaa \in \vndistinct$ of $\bx$ for indices $\bu$, 
            $\sfd{\bxp}{\by} \le \sfd{\bxs}{\by}$.
        \end{proposition}

        \begin{proof}
            In order to show Proposition~\ref{supp:proposition:2.13},	
            we first show the following statement is true:
            \begin{itemize}
                \item[$\statement$:] Suppose $\bx, \by \in \vndistinct$, and let $\bu \subset \tonumber{\n}$.
                    Then, there exists an imputation $\bxs \in \myfirstset{\bu}{\tonumber{\n}}{\n}$ of $\bx$ for indices $\bu$ such that	$\sfd{\bx}{\by} \le \sfd{\bxs}{\by}$.
            \end{itemize}
            Then we prove Proposition~\ref{supp:proposition:2.13} using statement $\statement$.

            First, we show the statement $\statement$ is true.
            Suppose $\bu = \emptyset$, then according to 
            the definition of  $\myfirstset{\tonumber{\n}}{\tonumber{\n}}{\n}$
            and imputations, we have $\bxs = \bx$.
            Hence, we have $\sfd{\bxs}{\by} = \sfd{\bx}{\by}$.
            This proves the statement $\statement$ when $\bu = \emptyset$.

            Suppose $\bu = \tonumber{\n}$. Then, according to 
            the definition of  $\myfirstset{\tonumber{\n}}{\tonumber{\n}}{\n}$
            and imputations, 
            any vector $\bxs \in \vndistinct$ is such that
            $\bxs \in \myfirstset{\tonumber{\n}}{\tonumber{\n}}{\n}$
            and $\bxs$ is an imputation of $\bx$ for indices $\bu = \tonumber{\n}$.
            Let $\bxs = \bx$, then we have $\sfd{\bxs}{\by} = \sfd{\bx}{\by}$.
            This proves the statement $\statement$ when $\bu = \tonumber{\n}$.

            In the following, we only prove the statement $\statement$ when $\bu \neq \tonumber{\n}$
            and $\bu \neq \emptyset$.

            For any given $\n \in \mathbb{N}$, let $\bpn{\kk}$ be the statement $\statement$
            when $|\bu| = \kk$.
            We prove $\bpn{\kk}$ is true for any $\kk \in \{1,\ldots, \n-1\}$
            by induction on $\kk$.

            \emph{Base Case: } We show $\bpn{1}$ is true.
            Suppose $|\bu| = 1$. Let us denote $\bu = \{\uu\}$.

            Then, let $\bxln{1}$, $\bxln{2}$
            be imputations of $\bx$ for the index $\uu \in \tonumber{\n}$ 
            such that $\rank{\bxlnn{1}{\uu}}{\bxln{1}} = 1, \text{ and } \rank{\bxlnn{2}{\uu}}{\bxln{2}} = \n$.
            Since $\rank{\bxlnn{1}{\uu}}{\bxln{1}} = 1$, we have
            \begin{align*}
                \bxlnn{1}{\uu} < \min_{\jconstant \in \bt \setminus \bu} \bxlnn{1}{\jconstant}.
            \end{align*}
            Hence, $\bxln{1} \in \myfirstset{\bu}{\tonumber{\n}}{\n}$.
            Similarly, since $\rank{\bxlnn{2}{\uu}}{\bxln{2}} = \n$, we have
            \begin{align*}
                \bxlnn{2}{\uu} > \max_{\jconstant \in \bt \setminus \bu} \bxlnn{2}{\jconstant}.
            \end{align*}
            Hence, $\bxln{2} \in \myfirstset{\bu}{\tonumber{\n}}{\n}$.

            Next, according to Proposition~\ref{supp:proposition:2.11}, we have 
            $\sfd{\bxp}{\by} \le \max \{ \sfd{\bxln{1}}{\by}, \sfd{\bxln{2}}{\by} \}$.
            Hence, we proved that $\bpn{1}$ is true.

            \emph{Induction Steps:} We show the implication 
            $\bpn{1}, \bpn{\kk} \Rightarrow \bpn{\kk+1}$ for any 
            $\kk \in \{1,\ldots, \n-2\}$.

            Denote $\m = \kk + 1$. Without loss of generality,   
            let us assume (after relabeling) $U = \{1,2,\cdots,m\}$. 
            Then, since $\bpn{\kk}$ is true,  there exists an imputation $\bxln{1} \in \myfirstset{\bu \setminus \{1\}}{\tonumber{\n}}{\n}$ of $\bx$ for indices $\bu \setminus \{1\}$ such that $\sfd{\bx}{\by} \le \sfd{\bxln{1}}{\by}$.

            Since $\bpn{1}$ is true, there exists an imputation $\bxln{2} \in \myfirstset{\{1\}}{\tonumber{\n}}{\n}$ of $\bxln{1}$ for the index 
            $\{1\}$ such that $\sfd{\bxln{1}}{\by} \le \sfd{\bxln{2}}{\by}$.

            Then, we have $\sfd{\bx}{\by} \le \sfd{\bxln{2}}{\by}$. Further, according to Lemma~\ref{supp:proposition:2.13:lemma:1}, 
            we have $\bxln{2} \in \myfirstset{\bu}{\tonumber{\n}}{\n}$.

            Next, since $\bxln{1}$ is an imputation of $\bx$ for indices $\bu \setminus \{1\}$,
            we have $\bxlnn{1}{i} = \bxn{\iconstant}$ for any $\iconstant \in \tonumber{\n} \setminus \bu$. Further, since $\bxln{2}$  is an imputation of $\bxln{1}$ for the index $\{1\}$, we have  $\bxlnn{2}{i} = \bxlnn{1}{\iconstant}$ for any $\iconstant \in \tonumber{\n} \setminus \bu$. Hence, we have $\bxlnn{2}{i} = \bxn{\iconstant}$ for any $\iconstant \in \tonumber{\n} \setminus \bu$. In other words, $\bxln{2}$ is an imputation of $\bx$ for indices $\bu$,

            Thus, $\bxln{2}  \in \myfirstset{\bu}{\tonumber{\n}}{\n}$  is an imputation of $\bx$ for indices $\bu$ such that $\sfd{\bx}{\by} \le \sfd{\bxln{2}}{\by}$.
            Hence, we have shown $\bpn{\kk+1}$ is true.
            This completes our proof for the statement $\statement$.

            Next, we prove Proposition~\ref{supp:proposition:2.13} using the Statement $\statement$.
            Let us denote 
            \begin{align*}
                \bs = \{\sfd{\bv}{\by}:  \bv \in \myfirstset{\bu}{\tonumber{\n}}{\n} \text{ is an imputation of } \bx \text{ for indices } \bu\}.
            \end{align*}
            Then, the carnality of $\bs$ is finite and there exist  
            $\bxs \in \myfirstset{\bu}{\tonumber{\n}}{\n}$ of $\bx$ for indices $\bu$ such that $\sfd{\bxs}{\by} = \max \bs$.

            We now show that $\sfd{\bx}{\by} \le \sfd{\bxs}{\by}$.
            According to statement $\statement$,  there exist  imputation
            $\bxln{1} \in \myfirstset{\bu}{\tonumber{\n}}{\n}$ of $\bx$ for indices $\bu$ such that $\sfd{\bx}{\by} \le \sfd{\bxln{1}}{\by}$. Then, according to the definition of 
            $\bs$ and $\bxs$, we have $ \sfd{\bxln{1}}{\by} \le \sfd{\bxs}{\by}$.
            Hence, we have $\sfd{\bx}{\by} \le \sfd{\bxs}{\by}$.

            Next, we show that $\sfd{\bxp}{\by} \le \sfd{\bxs}{\by}$.
            According to statement $\statement$,  there exist  imputation
            $\bxln{2} \in \myfirstset{\bu}{\tonumber{\n}}{\n}$ of $\bxp$ for indices $\bu$ such that $\sfd{\bxp}{\by} \le \sfd{\bxln{2}}{\by}$.
            Since $\bxp$ is an imputation of $\bx$ for indices $\bu$,
            then $\bxln{2}$ is also an imputation of $\bx$ for indices $\bu$.
            Then, according to the definition of $\bxs$, we have
            $\sfd{\bxln{2}}{\by} \le \sfd{\bxs}{\by}$.    	  
            Overall, we have $\sfd{\bxp}{\by} \le \sfd{\bxs}{\by}$.
            This completes our proof.

        \end{proof}

        \subsection{Proof of Lemma 2.14}

        This section proves Lemma 2.14. First, we prove the following result:

        \begin{lemma} \label{supp:lemma:2.14:1}
            Suppose $\bx, \by \in \vndistinct$, and
            assume $\byn{1} < \ldots < \byn{\n}$. 
            Let $\bxln{1}$ be a permutation of $\bx$ 
            such that $\bxlnn{1}{1} > \ldots > \bxlnn{1}{\n}$. Then, we have
            \begin{align} \label{supp:lemma:2.14:1:eqn:1}
                \sum_{\iconstant=1}^{\n} |\bxn{\iconstant} - \byn{\iconstant}| \le \sum_{\iconstant=1}^{\n} |\bxlnn{1}{\iconstant} - \byn{\iconstant}|. 
            \end{align}
        \end{lemma}

        \begin{proof}
            Let $\bpn{\kk}$ be the statement of Lemma~\ref{supp:lemma:2.14:1} when $\n = \kk$.
            We prove that $\bpn{\kk}$ is true for any $\kk \in \mathbb{N}$ such that $\kk \ge 2$
            by induction on $\kk$.

            \emph{Base Case: } We show $\bpn{2}$ is true.
            Suppose $\n = 2$. Then, we have $\bx = (\bxn{1}, \bxn{2})$. 
            If $\bxn{1} > \bxn{2}$, then $\bxln{1} = (\bxn{1}, \bxn{2})$. 
            Hence, we have $\bx = \bxln{1}$ and then \eqref{supp:lemma:2.14:1:eqn:1} holds. 

            Thus, it is sufficient to prove \eqref{supp:lemma:2.14:1:eqn:1} 
            when $\bxn{1} < \bxn{2}$. Suppose $\bxn{1} < \bxn{2}$. 
            Then, we have $\bxln{1} = (\bxn{2}, \bxn{1})$. Therefore, the right-hand side of \eqref{supp:lemma:2.14:1:eqn:1} is such that
            \begin{align} \label{supp:lemma:2.14:1:eqn:2}
                |\bxlnn{1}{1} - \byn{1}| + |\bxlnn{1}{2} - \byn{2}| = |\bxn{2} - \byn{1}| + |\bxn{1} - \byn{2}|.
            \end{align}   

            Since $\byn{1} < \byn{2}$, and $\bxn{1} < \bxn{2}$, then there are 
            following 6 cases of the order between $\byn{1}, \byn{2}, \bxn{1}$, and $\bxn{2}$:
            \begin{align*}
                (1): &\byn{1} < \byn{2} \le \bxn{1} < \bxn{2},\\
                (2): &\byn{1} \le \bxn{1} \le \byn{2} \le \bxn{2},\\
                (3): &\bxn{1} \le \byn{1} < \byn{2} \le \bxn{2},\\
                (4): &\byn{1} \le \bxn{1} < \bxn{2} \le \byn{2},\\
                (5): &\bxn{1} \le \byn{1} \le \bxn{2} \le \byn{2}, \\
                (6): &\bxn{1} < \bxn{2} \le \byn{1} < \byn{2}.  
            \end{align*}

            In the following, we are going to consider the 6 cases separately.

            Suppose the case $(1): \byn{1} < \byn{2} \le \bxn{1} < \bxn{2}$ is true. Then, the left-hand side of \eqref{supp:lemma:2.14:1:eqn:1} is
            \begin{align*}
                |\bxn{1} - \byn{1}| + |\bxn{2} - \byn{2}| = \bxn{1} - \byn{1} + \bxn{2} - \byn{2}.
            \end{align*} 
            The right-hand side of \eqref{supp:lemma:2.14:1:eqn:1} is
            \begin{align*}
                |\bxlnn{1}{1} - \byn{1}| + |\bxlnn{1}{2} - \byn{2}| &=^{\eqref{supp:lemma:2.14:1:eqn:2}} |\bxn{2} - \byn{1}| + |\bxn{1} - \byn{2}| \\
                &= \bxn{2} - \byn{1} + \bxn{1} - \byn{2}\\
                &= \bxn{1} - \byn{1} + \bxn{2} - \byn{2}.
            \end{align*}
            Then, the right-hand side of \eqref{supp:lemma:2.14:1:eqn:1} minus the left-hand side of \eqref{supp:lemma:2.14:1:eqn:1} equals to
            \begin{align*}
                (\bxn{1} - \byn{1} + \bxn{2} - \byn{2}) - (\bxn{1} - \byn{1} + \bxn{2} - \byn{2}) = 0.
            \end{align*}
            Hence, we have shown \eqref{supp:lemma:2.14:1:eqn:1} is true when the case (1) is true.

            Next, suppose the case $(2): \byn{1} \le \bxn{1} \le \byn{2} \le \bxn{2}$ is true. Then, the left-hand side of \eqref{supp:lemma:2.14:1:eqn:1} is
            \begin{align*}
                |\bxn{1} - \byn{1}| + |\bxn{2} - \byn{2}| &= \bxn{1} - \byn{1} + \bxn{2} - \byn{2}\\
                &= (\bxn{2} - \byn{1}) + (\bxn{1} - \byn{2}).
            \end{align*} 
            The right-hand side of \eqref{supp:lemma:2.14:1:eqn:1} is
            \begin{align*}
                |\bxlnn{1}{1} - \byn{1}| + |\bxlnn{1}{2} - \byn{2}| &=^{\eqref{supp:lemma:2.14:1:eqn:2}}
                |\bxn{2} - \byn{1}| + |\bxn{1} - \byn{2}| \\
                & = \bxn{2} - \byn{1} + \byn{2} - \bxn{1} \\
                & = (\bxn{2} - \byn{1}) + (\byn{2} - \bxn{1}).  
            \end{align*}
            Subsequently, the right-hand side of \eqref{supp:lemma:2.14:1:eqn:1} minus the left-hand side of \eqref{supp:lemma:2.14:1:eqn:1} equals to
            \begin{align*}
                &(\bxn{2} - \byn{1}) + (\byn{2} - \bxn{1}) - (\bxn{2} - \byn{1}) - (\bxn{1} - \byn{2}) \\
                & = (\byn{2} - \bxn{1}) - (\bxn{1} - \byn{2}) = 2\byn{2} - 2\bxn{1} \ge 0.
            \end{align*}
            Hence, we have shown \eqref{supp:lemma:2.14:1:eqn:1} is true when the case (2) is true.

            Next, suppose the case $(3): \bxn{1} \le \byn{1} < \byn{2} \le \bxn{2}$ is true. Then, the left-hand side of \eqref{supp:lemma:2.14:1:eqn:1} is
            \begin{align*}
                |\bxn{1} - \byn{1}| + |\bxn{2} - \byn{2}| &= \byn{1} - \bxn{1} + \bxn{2} - \byn{2}\\
                &= (\bxn{2} - \bxn{1}) + (\byn{1} - \byn{2}).
            \end{align*} 
            The right-hand side of \eqref{supp:lemma:2.14:1:eqn:1} is
            \begin{align*}
                |\bxlnn{1}{1} - \byn{1}| + |\bxlnn{1}{2} - \byn{2}| &=^{\eqref{supp:lemma:2.14:1:eqn:2}}
                |\bxn{2} - \byn{1}| + |\bxn{1} - \byn{2}| \\
                &= \bxn{2} - \byn{1} + \byn{2} - \bxn{1}\\
                &= (\bxn{2} - \bxn{1}) + (\byn{2} - \byn{1}).  
            \end{align*}
            Subsequently, the right-hand side of \eqref{supp:lemma:2.14:1:eqn:1} minus the left-hand side of \eqref{supp:lemma:2.14:1:eqn:1} equals to
            \begin{align*}
                &(\bxn{2} - \bxn{1}) + (\byn{2} - \byn{1}) - (\bxn{2} - \bxn{1}) - (\byn{1} - \byn{2})\\
                &= (\byn{2} - \byn{1}) - (\byn{1} - \byn{2}) = 2\byn{2} - 2\byn{1} > 0.
            \end{align*}
            Hence, we have shown \eqref{supp:lemma:2.14:1:eqn:1} is true when the case (3) is true.

            Next, suppose the case $(4): \byn{1} \le \bxn{1} < \bxn{2} \le \byn{2}$ is true. Then, the left-hand side of \eqref{supp:lemma:2.14:1:eqn:1} is
            \begin{align*}
                |\bxn{1} - \byn{1}| + |\bxn{2} - \byn{2}| &= \bxn{1} - \byn{1} + \byn{2} - \bxn{2}\\
                &= (\byn{2} - \byn{1}) + (\bxn{1} - \bxn{2}).
            \end{align*} 
            The right hand side of \eqref{supp:lemma:2.14:1:eqn:1} is,
            \begin{align*}
                |\bxlnn{1}{1} - \byn{1}| + |\bxlnn{1}{2} - \byn{2}| &=^{\eqref{supp:lemma:2.14:1:eqn:2}}
                |\bxn{2} - \byn{1}| + |\bxn{1} - \byn{2}| \\
                & = \bxn{2} - \byn{1} + \byn{2} - \bxn{1}\\
                & = (\byn{2} - \byn{1}) + (\bxn{2} - \bxn{1}).  
            \end{align*}
            Subsequently, the right hand side of \eqref{supp:lemma:2.14:1:eqn:1} minus the left hand side of \eqref{supp:lemma:2.14:1:eqn:1} equals
            \begin{align*}
                &(\byn{2} - \byn{1}) + (\bxn{2} - \bxn{1}) - (\byn{2} - \byn{1}) - (\bxn{1} - \bxn{2})\\
                &=(\bxn{2} - \bxn{1}) - (\bxn{1} - \bxn{2}) = 2\bxn{2} - 2\bxn{1} > 0.
            \end{align*}
            Hence, we have shown \eqref{supp:lemma:2.14:1:eqn:1} is true when the case (4) is true.

            Next, suppose the case $(5): \bxn{1} \le \byn{1} \le \bxn{2} \le \byn{2}$ is true. Then, the left hand side of \eqref{supp:lemma:2.14:1:eqn:1} is
            \begin{align*}
                |\bxn{1} - \byn{1}| + |\bxn{2} - \byn{2}| &= \byn{1} - \bxn{1} + \byn{2} - \bxn{2} \\
                & = (\byn{2} - \bxn{1}) + (\byn{1} - \bxn{2}).
            \end{align*} 
            The right hand side of \eqref{supp:lemma:2.14:1:eqn:1} is,
            \begin{align*}
                |\bxlnn{1}{1} - \byn{1}| + |\bxlnn{1}{2} - \byn{2}| &=^{\eqref{supp:lemma:2.14:1:eqn:2}}
                |\bxn{2} - \byn{1}| + |\bxn{1} - \byn{2}| \\
                & = \bxn{2} - \byn{1} + \byn{2} - \bxn{1}\\
                & = (\byn{2} - \bxn{1}) + (\bxn{2} - \byn{1}).  
            \end{align*}
            Subsequently, the right hand side of \eqref{supp:lemma:2.14:1:eqn:1} minus the left hand side of \eqref{supp:lemma:2.14:1:eqn:1} equals to
            \begin{align*}
                &(\byn{2} - \bxn{1}) + (\bxn{2} - \byn{1}) - (\byn{2} - \bxn{1}) - (\byn{1} - \bxn{2})\\
                &=(\bxn{2} - \byn{1}) - (\byn{1} - \bxn{2}) = 2\bxn{2} - 2\byn{1} \ge 0.
            \end{align*}
            Hence, we have shown \eqref{supp:lemma:2.14:1:eqn:1} is true when the case (5) is true.

            Next, suppose the case $(6): \bxn{1} < \bxn{2} \le \byn{1} < \byn{2}$ is true. Then, the left hand side of \eqref{supp:lemma:2.14:1:eqn:1} is
            \begin{align*}
                |\bxn{1} - \byn{1}| + |\bxn{2} - \byn{2}| = \byn{1} - \bxn{1} + \byn{2} - \bxn{2}.
            \end{align*} 
            The right hand side of \eqref{supp:lemma:2.14:1:eqn:1} is, 
            \begin{align*}
                |\bxlnn{1}{1} - \byn{1}| + |\bxlnn{1}{2} - \byn{2}| &=^{\eqref{supp:lemma:2.14:1:eqn:2}}
                |\bxn{2} - \byn{1}| + |\bxn{1} - \byn{2}| \\
                & = \byn{1} - \bxn{2} + \byn{2} - \bxn{1} \\
                & = \byn{1} - \bxn{1} + \byn{2} - \bxn{2}.
            \end{align*}
            Subsequently, the right hand side of \eqref{supp:lemma:2.14:1:eqn:1} minus the left hand side of \eqref{supp:lemma:2.14:1:eqn:1} equals to
            \begin{align*}
                (\byn{1} - \bxn{1} + \byn{2} - \bxn{2}) - (\byn{1} - \bxn{1} + \byn{2} - \bxn{2}) = 0
            \end{align*}
            Hence, we have shown \eqref{supp:lemma:2.14:1:eqn:1} is true when case (6) is true.
            This completes our proof for $\bpn{2}$.

            \emph{Induction Step: } We show the implication $\bpn{2}, \bpn{\kk} \Rightarrow \bpn{\kk+1}$
            for any integer $\kk \ge 2$.

            Let $\bxln{2}$ be a permutation of $\bx$ 
            such that $\bxlnn{2}{1} > \ldots > \bxlnn{2}{\n-1}$ 
            and $\bxlnn{2}{\n} = \bxn{\n}$. 
            Then, since $\bpn{\kk}$ is true, we have
            \begin{align*}
                \sum_{\iconstant=1}^{\n-1} |\bxn{\iconstant} - \byn{\iconstant}| \le \sum_{\iconstant=1}^{\n-1} |\bxlnn{2}{\iconstant} - \byn{\iconstant}|. 
            \end{align*}
            Since $\bxlnn{2}{\n} = \bxn{\n}$, we have
            \begin{align*}
                |\bxn{\n} - \byn{\n}| = |\bxlnn{2}{\n} - \byn{\n}|. 
            \end{align*}
            Then, we have 
            \begin{align} \label{supp:lemma:2.14:1:eqn:3}
                \sum_{\iconstant=1}^{\n} |\bxn{\iconstant} - \byn{\iconstant}| \le \sum_{\iconstant=1}^{\n} |\bxlnn{2}{\iconstant} - \byn{\iconstant}|. 
            \end{align}

            In the following, we are going to show $\bpn{\kk+1}$ is true
            when  $\bxn{\n} < \min \{\bxn{1} ,\ldots, \bxn{\n-1}\}$
            and  $\bxn{\n} > \min \{\bxn{1} ,\ldots, \bxn{\n-1}\}$,
            separately.

            Suppose $\bxn{\n} < \min \{\bxn{1} ,\ldots, \bxn{\n-1}\}$. 
            Then, since $\bxlnn{2}{1}, \ldots, \bxlnn{2}{\n-1}$ is 
            a permutation of $\{\bxn{1} ,\ldots, \bxn{\n-1}\}$, 
            we have
            \begin{align*}
                \min \{\bxn{1} ,\cdots, \bxn{\n-1}\} 
                = \min \{\bxlnn{2}{1} ,\cdots, \bxlnn{2}{\n-1}\}.
            \end{align*}

            Next, since $\bxlnn{2}{\n} = \bxn{\n}$, then we further have
            $\bxlnn{2}{\n} = \bxn{\n} <  \min \{\bxn{1} ,\cdots, \bxn{\n-1}\} = \min \{\bxlnn{2}{1} ,\cdots, \bxlnn{2}{\n-1}\}$. 
            Since $\bxln{2}$ is a permutation of $\bx$ 
            such that $\bxlnn{2}{1} > \ldots > \bxlnn{2}{\n-1}$,
            and $\bxlnn{2}{\n}  < \min \{\bxlnn{2}{1} ,\cdots, \bxlnn{2}{\n-1}\}$,
            we have $\bxln{2}$ is a permutation of $\bx$ such that 
            $\bxlnn{2}{1} > \ldots > \bxlnn{2}{\n}$.

            Since $\bxln{1}$ is a permutation of $\bx$ 
            such that $\bxlnn{1}{1} > \ldots > \bxlnn{1}{\n}$,  
            then we have $\bxln{2} = \bxln{1}$.
            According to \eqref{supp:lemma:2.14:1:eqn:3}, we have
            \begin{align*}
                \sum_{\iconstant=1}^{\n} |\bxn{\iconstant} - \byn{\iconstant}| \le \sum_{\iconstant=1}^{\n} |\bxlnn{1}{\iconstant} - \byn{\iconstant}|. 
            \end{align*}
            Hence, we have shown $\bpn{\kk+1}$ when  $\bxn{\n} < \text{min} \{\bxn{1} ,\ldots, \bxn{\n-1}\}$.

            However, suppose $\bxn{\n} > \min \{\bxn{1} , \ldots, \bxn{\n-1}\}$. 
            Then, since $\bxlnn{2}{1}, \cdots, \bxlnn{2}{\n-1}$ is a 
            permutation of $\{\bxn{1} ,\cdots, \bxn{\n-1}\}$,
            we have
            \begin{align*}
                \min \{\bxn{1} ,\cdots, \bxn{\n-1}\} 
                = \min \{\bxlnn{2}{1} ,\cdots, \bxlnn{2}{\n-1}\}.
            \end{align*}

            Since $\bxlnn{2}{\n} = \bxn{\n}$, then we further have
            $\bxlnn{2}{\n} = \bxn{\n} > \min \{\bxn{1} ,\ldots, \bxn{\n-1}\} 
            = \min \{\bxlnn{2}{1} ,\ldots, \bxlnn{2}{\n-1}\}$.
            Further, since $\bxlnn{2}{1} > \ldots \bxlnn{2}{\n-1}$, 
            we have $\bxlnn{2}{\n} > \bxlnn{2}{\n-1} = \min \{\bxlnn{2}{1} ,\ldots, \bxlnn{2}{\n-1}\}$,
            and
            \begin{align} \label{supp:lemma:2.14:1:eqn:3.0}
                \bxlnn{2}{\n-1} = \min \{\bxlnn{2}{1} ,\ldots, \bxlnn{2}{\n}\}.
            \end{align}

            Next, let $\bxln{3}$ be a permutation of $\bxln{2}$ 
            such that $\bxlnn{3}{i} = \bxlnn{2}{i}$ 
            for any $i \in \{1,\ldots,\n-2\}$, $\bxlnn{3}{\n-1} = \bxlnn{2}{\n}$, 
            and $\bxlnn{3}{\n} = \bxlnn{2}{\n-1}$. 
            Then, since $\bxlnn{2}{\n} > \bxlnn{2}{\n-1}$, 
            we have $\bxlnn{3}{\n-1} > \bxlnn{3}{\n}$. 

            Since $\bpn{2}$ is true, we have
            \begin{align*}
                |\bxlnn{2}{\n-1} - \byn{\n-1}| + |\bxlnn{2}{\n} - \byn{\n}| \le 	|\bxlnn{3}{\n-1} - \byn{\n-1}| + |\bxlnn{3}{\n} - \byn{\n}|.
            \end{align*} 
            Since $\bxlnn{3}{i} = \bxlnn{2}{i}$ for any $i \in \{1,\ldots,\n-2\}$, we have
            \begin{align*}
                \sum_{\iconstant=1}^{\n-2} |\bxlnn{2}{\iconstant} - \byn{\iconstant}| = \sum_{\iconstant=1}^{\n-2} |\bxlnn{3}{\iconstant} - \byn{\iconstant}|.
            \end{align*}
            Then, we have 
            \begin{align} \label{supp:lemma:2.14:1:eqn:4}
                \sum_{\iconstant=1}^{\n} |\bxlnn{2}{\iconstant} - \byn{\iconstant}| \le \sum_{\iconstant=1}^{\n} |\bxlnn{3}{\iconstant} - \byn{\iconstant}|.
            \end{align}

            Next, let $\bxln{4}$ be a permutation of $\bxln{3}$ 
            such that $\bxlnn{4}{1} > \ldots > \bxlnn{4}{\n-1}$ 
            and $\bxlnn{4}{\n} = \bxlnn{3}{\n}$. 
            Then, since $\bpn{\kk}$ is true, we have
            \begin{align*}
                \sum_{\iconstant=1}^{\n-1} |\bxlnn{3}{\iconstant} - \byn{\iconstant}| \le \sum_{\iconstant=1}^{\n-1} |\bxlnn{4}{\iconstant} - \byn{\iconstant}|.
            \end{align*}
            Since $\bxlnn{4}{\n} = \bxlnn{3}{\n}$, we have
            \begin{align*}
                |\bxlnn{3}{\n} - \byn{\n}| = |\bxlnn{4}{\n} - \byn{\n}|. 
            \end{align*}
            Thus, we have
            \begin{align} \label{supp:lemma:2.14:1:eqn:5}
                \sum_{\iconstant=1}^{\n} |\bxlnn{3}{\iconstant} - \byn{\iconstant}| \le \sum_{\iconstant=1}^{\n} |\bxlnn{4}{\iconstant} - \byn{\iconstant}|.
            \end{align}

            According to \eqref{supp:lemma:2.14:1:eqn:3.0}, we have 
            $\bxlnn{2}{\n-1} = \min \{\bxlnn{2}{1} ,\ldots, \bxlnn{2}{\n}\}.$
            Since $\bxln{3}$ is a permutation of $\bxln{2}$ and 
            $\bxlnn{3}{\n} = \bxlnn{2}{\n-1}$, then we have
            $\bxlnn{3}{\n} = \min \{\bxlnn{3}{1} ,\ldots, \bxlnn{3}{\n}\}.$

            Next, since $\bxln{4}$ is a permutation of $\bxln{3}$, 
            and $\bxlnn{4}{\n} = \bxlnn{3}{\n}$,
            we have $\bxlnn{4}{\n} = \min \{\bxlnn{4}{1} ,\ldots, \bxlnn{4}{\n}\}.$
            Notice that $\bxlnn{4}{1} > \ldots > \bxlnn{4}{\n-1}$.
            Hence, we have $\bxlnn{4}{1} > \ldots > \bxlnn{4}{\n}$. 

            Since $\bxln{4}$ is a permutation of $\bxln{3}$, 
            $\bxln{3}$ is a permutation of $\bxln{2}$, 
            $\bxln{2}$ is a permutation of $\bx$, 
            we have $\bxln{4}$ is a permutation of $\bx$.
            Then, since $\bxln{1}$ is a permutation of $\bx$, 
            and $\bxlnn{1}{1} > \ldots > \bxlnn{1}{\n}$,
            we have $\bxln{4} = \bxln{1}$.
            Hence, we have 
            \begin{align*}
                \sum_{\iconstant=1}^{\n} |\bxlnn{1}{\iconstant} - \byn{\iconstant}| &= \sum_{\iconstant=1}^{\n} |\bxlnn{4}{\iconstant} - \byn{\iconstant}|\\
                &\ge^{\eqref{supp:lemma:2.14:1:eqn:5}} \sum_{\iconstant=1}^{\n} |\bxlnn{3}{\iconstant} - \byn{\iconstant}|\\
                &\ge^{\eqref{supp:lemma:2.14:1:eqn:4}} \sum_{\iconstant=1}^{\n} |\bxlnn{2}{\iconstant} - \byn{\iconstant}| \\
                &\ge^{\eqref{supp:lemma:2.14:1:eqn:3}} \sum_{\iconstant=1}^{\n} |\bxn{\iconstant} - \byn{\iconstant}|.
            \end{align*}
            Thus, we have shown $\bpn{\kk+1}$ is true.
            This completes our proof.

        \end{proof}

        Now, we show Lemma 2.14 is true:

        \begin{lemma} \label{supp:lemma:2.15}
            Suppose $\bx,\by \in \vndistinct$ and for $\m \leq \n$ assume $\bu = \{1,\ldots,\m\}$ and 
            $\byn{1} < \cdots < \byn{\m}$. 
            Let $\bxp$ be an imputation of $\bx$ for indices $\bu$. Suppose 
            $\bxs$ is also an imputation of $\bx$ for indices $\bu$,
            and $\svector{\bxs}{\iconstant}{ \bu}$ is a permutation of 
            $\svector{\bxp}{\iconstant}{ \bu}$ such that 
            $\bxsn{1} > \cdots > \bxsn{\m}$. 
            Then, we have $\sfd{\bxp}{\by} \le \sfd{\bxs}{\by}$.
        \end{lemma}

        \begin{proof}
            Since $\bxp$ is an imputation of $\bx$ for indices $\bu$, we have
            $\bxpn{\iconstant} = \bxn{\iconstant}$ for any $\iconstant \in \tonumber{\n} \setminus \bu$. 
            Similarly, since $\bxsp$ is an imputation of $\bx$ for indices $\bu$, we have $\bxsn{\iconstant} = \bxn{\iconstant}$ for any $\iconstant \in \tonumber{\n} \setminus \bu$. Hence, we have $\bxsn{\iconstant} = \bxpn{\iconstant}$ for any $\iconstant \in \tonumber{\n} \setminus \bu$.

            Further, since $\svector{\bxs}{\iconstant}{ \bu}$ is a permutation of 
            $\svector{\bxp}{\iconstant}{ \bu}$, then we have 
            $\bxs$ is a permutation of $\bxp$.
            Hence, according to the definition of rank, for any $\iconstant \in \tonumber{\n} \setminus \bu$, we have
            \begin{align*}
                \rank{\bxsn{\iconstant}}{\bxs} &= \sum_{\jconstant=1}^{\n} \indicator{\bxsn{\jconstant} \le \bxsn{\iconstant}} \\
                &= \sum_{\jconstant=1}^{\n} \indicator{\bxpn{\jconstant} \le \bxsn{\iconstant}}.
            \end{align*}
            Since $\bxsn{\iconstant} = \bxpn{\iconstant}$ for any $\iconstant \in \tonumber{\n} \setminus \bu$, we further have 
            \begin{align*}
                &\rank{\bxsn{\iconstant}}{\bxs} = \sum_{\jconstant=1}^{\n} \indicator{\bxpn{\jconstant} \le \bxpn{\iconstant}} = \rank{\bxpn{\iconstant}}{\bxp},
                \text{ for any } \iconstant \in \tonumber{\n} \setminus \bu.
            \end{align*}
            Then, we have
            \begin{align}
                \sum_{\iconstant=\m+1}^{\n} \left|\rank{\bxsn{\iconstant}}{\bxs} -  \rank{\byn{\iconstant}}{\by} \right| = \sum_{\iconstant=\m+1}^{\n} \left|\rank{\bxpn{\iconstant}}{\bxp} -  \rank{\byn{\iconstant}}{\by} \right|. \label{supp:lemma:2.15:eqn:1}
            \end{align}

            Next, since $\bxs$ is a permutation of $\bxp$,
            then according to the definition of rank, for any $\iconstant \in \bu$, we have
            \begin{align*}
                \rank{\bxsn{\iconstant}}{\bxs} = \sum_{\jconstant=1}^{\n} \indicator{\bxsn{\jconstant} \le \bxsn{\iconstant}}
                = \sum_{\jconstant=1}^{\n} \indicator{\bxpn{\jconstant} \le \bxsn{\iconstant}}.
            \end{align*}
            Since  $\svector{\bxs}{\iconstant}{ \bu}$ is a permutation of $\svector{\bxp}{\iconstant}{ \bu}$, then there exist
            a permutation $\svector{\sigma}{\lconstant}{\bu}$
            of $\bu$ such that
            \begin{align*}
                \bxsn{\iconstant} = \bxpn{\sigma(\iconstant)}, \text{ for any } \iconstant \in \bu.
            \end{align*}
            Hence, we have 
            \begin{align*}
                \rank{\bxsn{\iconstant}}{\bxs} = \sum_{\jconstant=1}^{\n} \indicator{\bxpn{\jconstant} \le\bxpn{\sigma(\iconstant)} } = \rank{\bxpn{\sigma(\iconstant)}}{\bxp},  \text{ for any } \iconstant \in \bu.
            \end{align*}
            In other words, $\left(\rank{\bxsn{\iconstant}}{\bxs}\right)_{\iconstant \in \bu}$
            is a permutation of $\left(\rank{\bxpn{\iconstant}}{\bxp}\right)_{\iconstant \in \bu}$. Since $\bxsn{1} > \ldots > \bxsn{\m}$, we have
            $\rank{\bxsn{1}}{\bxs} > \ldots > \rank{\bxsn{\m}}{\bxs}$.

            Then, according to Lemma~\ref{supp:lemma:2.14:1}, we have
            \begin{align*}
                \sum_{\iconstant=1}^{\m} \left|\rank{\bxsn{\iconstant}}{\bxs} -  \rank{\byn{\iconstant}}{\by} \right|  \ge  \sum_{\iconstant=1}^{\m} \left|\rank{\bxpn{\iconstant}}{\bxp} -  \rank{\byn{\iconstant}}{\by} \right| .
            \end{align*}
            Combining this result with \eqref{supp:lemma:2.15:eqn:1}, we have
            \begin{align*}
                &	\sum_{\iconstant=1}^{\n} \left|\rank{\bxsn{\iconstant}}{\bxs} -  \rank{\byn{\iconstant}}{\by} \right| \ge \sum_{\iconstant=1}^{\n} \left|\rank{\bxpn{\iconstant}}{\bxp} -  \rank{\byn{\iconstant}}{\by} \right| \\
                \Rightarrow&	\sfd{\bxs}{\by} \ge \sfd{\bxp}{\by}.
            \end{align*}
            This completes our proof.

        \end{proof}

        \subsection{Proof of Theorem 2.16}

        This subsection proves Theorem~2.6. First, we make the following definition:

        \begin{definition} \label{def:2}
            Suppose $\by \in \vndistinct$ and let $\bu \subset \bt \subset \tonumber{\n}$. Define
            \begin{align*}
                \mysecondset{\by}{\bu}{\bt}{\n} 
                = \left\{\bz \in \myfirstset{\bu}{\bt}{\n} |  ~\text{for any}~\iconstant, \jconstant \in \bu,~ \bzn{\iconstant} > \bzn{\jconstant}~\text{if}~\byn{\iconstant} < \byn{\jconstant} \right\}.
            \end{align*}
        \end{definition}

        Then we prove the following lemma, which will be useful for proving Theorem~2.6.

        \begin{lemma} \label{supp:theorem:2.16:lemma:1}
            Suppose $\bx, \by \in \vndistinct$, and for $\m \le \n$ assume $\bu = \{1,\ldots,\m\}$ and $\byn{1} < \ldots < \byn{\m}$. Suppose $\bxln{1} \in \myfirstset{\bu}{\tonumber{\n}}{\n}$ is an imputation of $\bx$ for $\bu$.
            If $\bxln{2}$ is an imputation of $\bx$ for indices $\bu$
            such that $\svector{\bxln{2}}{\iconstant}{ \bu}$ is a permutation of 
            $\svector{\bxln{1}}{\iconstant}{ \bu}$, and 
            $\bxlnn{2}{1} > \ldots > \bxlnn{2}{\m}$,
            then we have  $\bxln{2} \in \mysecondset{\by}{\bu}{\tonumber{\n}}{\n}$.
        \end{lemma}

        \begin{proof}
            Since $\bxln{1} \in \myfirstset{\bu}{\tonumber{\n}}{\n}$, then we have
            \begin{align*}
                \bxlnn{1}{\iconstant} < \min_{\jconstant \in \tonumber{\n} \setminus \bu} \bxlnn{1}{\jconstant}, \text{ or } \bxlnn{1}{\iconstant} > \max_{\jconstant \in \tonumber{\n} \setminus \bu} \bxlnn{1}{\jconstant}, \text{ for any } \iconstant \in \bu.  
            \end{align*}
            Since $\bxln{2}$ is an imputation of $\bx$ for $\bu$,
            we have $\bxlnn{2}{\iconstant} = \bxn{\iconstant}$
            for any $\iconstant \in \tonumber{\n} \setminus \bu$.
            Then, since $\bxln{1}$ is an imputation of $\bx$ for $\bu$,
            we have $\bxlnn{1}{\iconstant} = \bxn{\iconstant}$
            for any $\iconstant \in \tonumber{\n} \setminus \bu$.
            Hence, we have $\bxlnn{1}{\iconstant} = \bxlnn{2}{\iconstant}$
            for any $\iconstant \in \tonumber{\n} \setminus \bu$.
            Thus, we have 
            $\min_{\jconstant \in \tonumber{\n} \setminus \bu} \bxlnn{1}{\jconstant}
            = \min_{\jconstant \in \tonumber{\n} \setminus \bu} \bxlnn{2}{\jconstant}$
            and
            $\max_{\jconstant \in \tonumber{\n} \setminus \bu} \bxlnn{1}{\jconstant}
            = \max_{\jconstant \in \tonumber{\n} \setminus \bu} \bxlnn{2}{\jconstant}$.
            Then, we have
            \begin{align*}
                \bxlnn{1}{\iconstant} < \min_{\jconstant \in \tonumber{\n} \setminus \bu} \bxlnn{2}{\jconstant}, \text{ or } \bxlnn{1}{\iconstant} > \max_{\jconstant \in \tonumber{\n} \setminus \bu} \bxlnn{2}{\jconstant}, \text{ for any } \iconstant \in \bu.  
            \end{align*}  	
            Since $\svector{\bxln{2}}{\iconstant}{ \bu}$ is a permutation of 
            $\svector{\bxln{1}}{\iconstant}{ \bu}$, then we have
            \begin{align*}
                \bxlnn{2}{\iconstant} < \min_{\jconstant \in \tonumber{\n} \setminus \bu} \bxlnn{2}{\jconstant}, \text{ or } \bxlnn{2}{\iconstant} > \max_{\jconstant \in \tonumber{\n} \setminus \bu} \bxlnn{2}{\jconstant}, \text{ for any } \iconstant \in \bu.  
            \end{align*}  	
            Thus, we have $\bxln{2} \in \myfirstset{\bu}{\tonumber{\n}}{\n}$.
            Notice that  $\byn{1} < \ldots < \byn{\m}$ and $\bxlnn{2}{1} > \ldots > \bxlnn{2}{\m}$. Hence, we have $\text{for any}~\iconstant, \jconstant \in \bu,~ \bxlnn{2}{\iconstant} > \bxlnn{2}{\jconstant}~\text{if}~\byn{\iconstant} < \byn{\jconstant}$. Therefore, we have shown $\bxln{2} \in \mysecondset{\by}{\bu}{\tonumber{\n}}{\n}$.
            This completes our proof.
        \end{proof}

        Now, we are ready to prove Theorem~2.6.

        \begin{theorem} \label{supp:theorem:2.16}
            Suppose $\bx,\by \in \vndistinct$ and let $\bu \subset \tonumber{\n}$. 
            Then, there exists an imputation $\bxs \in \mysecondset{\by}{\bu}{\tonumber{\n}}{\n}$ of $\bx$ for indices $\bu$ such that	$\sfd{\bx}{\by} \le \sfd{\bxs}{\by}$, and for any other imputation $\xaa \in \vndistinct$ of $\bx$ for indices $\bu$, 
            $\sfd{\bxp}{\by} \le \sfd{\bxs}{\by}$.	
        \end{theorem}

        \begin{proof}
            Without loss of generality, let us assume (after relabeling)
            $\bu = \{1, \ldots, \m\}$ and $\byn{1} < \ldots < \byn{\m}$.
            Let us denote 
            \begin{align*}
                \bs = \{\sfd{\bv}{\by}:  \bv \in \mysecondset{\by}{\bu}{\tonumber{\n}}{\n} \text{ is an imputation of } \bx \text{ for indices } \bu \}.
            \end{align*}
            The carnality of $\bs$ is finite and there exist  
            imputation $\bxs \in \mysecondset{\by}{\bu}{\tonumber{\n}}{\n}$ 
            of $\bx$ for indices $\bu$ such that $\sfd{\bxs}{\by} = \max \bs$.

            We first show that $\sfd{\bx}{\by} \le \sfd{\bxs}{\by}$.
            According to Proposition~\ref{supp:proposition:2.13},  
            there exist imputation $\bxln{1} \in \myfirstset{\bu}{\tonumber{\n}}{\n}$ 
            of $\bx$ for indices $\bu$ such that
            $\sfd{\bx}{\by} \le \sfd{\bxln{1}}{\by}$.

            Next, let $\bxln{2}$ be an imputation of $\bx$ for indices $\bu$
            such that $\svector{\bxln{2}}{\iconstant}{ \bu}$ is a permutation of 
            $\svector{\bxln{1}}{\iconstant}{ \bu}$, and 
            $\bxlnn{2}{1} > \ldots > \bxlnn{2}{\m}$.
            Then, according to Lemma~\ref{supp:lemma:2.15}, we have 
            $\sfd{\bxln{1}}{\by} \le  \sfd{\bxln{2}}{\by}$.

            According to Lemma~\ref{supp:theorem:2.16:lemma:1}, we have
            $\bxln{2} \in \mysecondset{\by}{\bu}{\tonumber{\n}}{\n}$.
            Then, according to the definition of $\bxs$, we have
            $\sfd{\bxln{2}}{\by} \le \sfd{\bxs}{\by}$.    	  
            Hence, we have $\sfd{\bx}{\by} \le \sfd{\bxs}{\by}$.

            Next, we show that $\sfd{\bxp}{\by} \le \sfd{\bxs}{\by}$.
            According to Proposition~\ref{supp:proposition:2.13},  
            there exist  imputation $\bxln{3} \in \myfirstset{\bu}{\tonumber{\n}}{\n}$ of $\bxp$ for indices $\bu$ such that $\sfd{\bxp}{\by} \le \sfd{\bxln{3}}{\by}$.

            Since $\bxp$ is an imputation of $\bx$ for indices $\bu$,
            then $\bxln{3}$ is also an imputation of $\bx$ for indices $\bu$.
            Let $\bxln{4}$ be a imputation of $\bx$ for indices $\bu$ such that 
            $\svector{\bxln{4}}{\iconstant}{ \bu}$ is a permutation of 
            $\svector{\bxln{3}}{\iconstant}{ \bu}$, and 
            $\bxlnn{4}{1} > \ldots > \bxlnn{4}{\m}$.
            Then, according to Lemma~\ref{supp:lemma:2.15}, we have
            $\sfd{\bxln{3}}{\by} \le \sfd{\bxln{4}}{\by}$.    	  

            According to Lemma~\ref{supp:theorem:2.16:lemma:1}, we have 
            $\bxln{4} \in \mysecondset{\by}{\bu}{\tonumber{\n}}{\n}$.
            Then, according to the definition of $\bxs$, we have
            $\sfd{\bxln{4}}{\by} \le \sfd{\bxs}{\by}$.    	  
            Hence, we have $\sfd{\bxp}{\by} \le \sfd{\bxs}{\by}$.
            This completes our proof.

        \end{proof}

        \subsection{Proof of Theorem 2.17}

        This section proves Theorem~2.17. We start by proving the following lemma:

        \begin{lemma}  \label{supp:theorem:2.17:lemma:1}
            Suppose $\bx,\by \in \vndistinct$ and $\bu, \bv \subset \tonumber{\n}$
            are subsets of indices such that $\bu \cap \bv = \emptyset$.	
            Suppose $\bxln{1}$ is an imputation of $\bx$ for $\bu$. If $\byln{1} \in \mysecondset{\bxln{1}}{\bv}{\tonumber{\n}}{\n}$, we have $\byln{1} \in \mysecondset{\bx}{\bv}{\tonumber{\n}}{\n}$.
        \end{lemma}    

        \begin{proof}

            Since $\byln{1} \in \mysecondset{\bxln{1}}{\bv}{\tonumber{\n}}{\n}$, then
            we have $\byln{1} \in \myfirstset{\bv}{\tonumber{\n}}{\n}$,
            and for any $\iconstant, \jconstant \in \bv$, we have $\byln{\iconstant} > \byln{\jconstant}$ 
            if $\bxlnn{1}{\iconstant} < \bxlnn{1}{\jconstant}$. 

            Next, since $\bxln{1}$ is an imputation of $\bx$ for $\bu$, then
            we have $\bxlnn{1}{\iconstant} = \bxn{\iconstant}$ for any
            $\iconstant \in \tonumber{\n} \setminus \bu$.
            Then, since $\bu \cap \bv = \emptyset$, we have
            $\bv \subset  \tonumber{\n} \setminus \bu$. Hence, we have
            $\bxlnn{1}{\iconstant} = \bxn{\iconstant}$ for any
            $\iconstant \in \bv$.

            Thus, for any $\iconstant, \jconstant \in \bv$, we have 
            $\byln{\iconstant} > \byln{\jconstant}$ if $\bxn{\iconstant} < \bxn{\jconstant}$. 
            Therefore, we have shown $\byln{1} \in \mysecondset{\bx}{\bv}{\tonumber{\n}}{\n}$.
            This completes our proof.
        \end{proof}

        Now, we show Theorem~2.17 is true.

        \begin{theorem} \label{supp:theorem:2.17}
            Suppose $\bx,\by \in \vndistinct$ and $\bu, \bv \subset \tonumber{\n}$
            are disjoint subsets of indices such that $\bu \cap \bv = \emptyset$.	
            Then, there exist imputations $(\bxs,\bys) \in (\mysecondset{\by}{\bu}{\tonumber{\n}}{\n}, \mysecondset{\bx}{\bv}{\tonumber{\n}}{\n})$ of $\bx, \by$ for indices $\bu$ and $\bv$, respectively, such that	$\sfd{\bx}{\by} \le \sfd{\bxs}{\bys}$. Furthermore, consider any imputation $\bxp,\byp \in \vndistinct$ of $\bx, \by$ for indices $\bu$ and $\bv$, respectively, we have 
            $\sfd{\bxp}{\byp} \le \sfd{\bxs}{\bys}$.
        \end{theorem}

        \begin{proof}
            To start, let us denote 
            \begin{align*}
                \bs = \{\sfd{\bv}{\bw}&: (\bv,\bw) \in (\mysecondset{\by}{\bu}{\tonumber{\n}}{\n}, \mysecondset{\bx}{\bv}{\tonumber{\n}}{\n}) \\ 
                &\text{ are imputations of }
                \bx, \by \text{ for indices } \bu \text{ and } \bv, \text {respectively}\}.
            \end{align*}
            Then, the carnality of $\bs$ is finite and there exist  
            imputations
            \begin{align*}
                (\bxs,\bys) \in (\mysecondset{\by}{\bu}{\tonumber{\n}}{\n}, \mysecondset{\bx}{\bv}{\tonumber{\n}}{\n})
            \end{align*}
            of $\bx, \by$ for indices $\bu$ and $\bv$, respectively, such that $\sfd{\bxs}{\bys} = \max \bs$.    	

            We first show that $\sfd{\bx}{\by} \le \sfd{\bxs}{\bys}$.		
            According to Theorem~\ref{supp:theorem:2.16},
            there exist an This completes our proof.imputation $\bxln{1} \in \mysecondset{\by}{\bu}{\tonumber{\n}}{\n}$
            of $\bx$ for indices $\bu$ such that 
            $\sfd{\bx}{\by} \le \sfd{\bxln{1}}{\by}$.
            Applying Theorem~\ref{supp:theorem:2.16} again,
            there exist an imputation $\byln{1} \in \mysecondset{\bxln{1}}{\bv}{\tonumber{\n}}{\n}$
            of $\by$ for indices $\bv$ such that 
            $\sfd{\bxln{1}}{\by} \le \sfd{\bxln{1}}{\byln{1}}$.		
            Hence, we have $\sfd{\bx}{\by} \le \sfd{\bxln{1}}{\byln{1}}$.

            According to Lemma~\ref{supp:theorem:2.17:lemma:1},
            we have $\byln{1} \in \mysecondset{\bx}{\bv}{\tonumber{\n}}{\n}$.
            Notice that $\bxln{1} \in \mysecondset{\by}{\bu}{\tonumber{\n}}{\n}$.		
            Then, according to the definition of $(\bxs,\bys)$,
            we have $\sfd{\bxln{1}}{\byln{1}} \le \sfd{\bxs}{\bys}$.
            Hence, we have $\sfd{\bx}{\by} \le \sfd{\bxs}{\bys}$.

            Next, we show that $\sfd{\bxp}{\byp} \le \sfd{\bxs}{\bys}$.
            According to Theorem~\ref{supp:theorem:2.16},
            there exist an imputation 
            $\bxln{2} \in \mysecondset{\byp}{\bu}{\tonumber{\n}}{\n}$
            of $\bxp$ for indices $\bu$ such that 
            $\sfd{\bxp}{\byp} \le \sfd{\bxln{2}}{\byp}$.

            Since $\bxp$ is an imputation of $\bx$ for $\bu$,
            we have $\bxln{2}$ is also an imputation of $\bx$ for $\bu$.
            Meanwhile, notice that $\byp$ is an imputation of $\by$
            for indices $\bv$, then according to Lemma~\ref{supp:theorem:2.17:lemma:1},
            we have $\bxln{2} \in \mysecondset{\by}{\bu}{\tonumber{\n}}{\n}$.

            Applying Theorem~\ref{supp:theorem:2.16} again,
            there exist an imputation $\byln{2} \in \mysecondset{\bxln{2}}{\bv}{\tonumber{\n}}{\n}$
            of $\byp$ for indices $\bv$ such that 
            $\sfd{\bxln{2}}{\byp} \le \sfd{\bxln{2}}{\byln{2}}$.
            Hence, we have $\sfd{\bxp}{\byp} \le \sfd{\bxln{1}}{\byln{2}}$.

            Since $\byp$ is an imputation of $\by$ for $\bv$, then according to 
            the definition of imputations, we can show that
            $\byln{2}$ is also an imputation of $\by$ for $\bv$.
            Meanwhile, notice that $\bxln{2}$ is an imputation of $\bx$
            for indices $\bu$, then according to Lemma~\ref{supp:theorem:2.17:lemma:1},
            we have $\byln{2} \in \mysecondset{\bx}{\bv}{\tonumber{\n}}{\n}$.

            Then, according to the definition of $(\bxs,\bys)$,
            we have $\sfd{\bxln{2}}{\byln{2}} \le \sfd{\bxs}{\bys}$.
            Hence, we have $\sfd{\bxp}{\byp} \le \sfd{\bxs}{\bys}$.
            This completes our proof.

        \end{proof}

        \subsection{Proof of Proposition 2.18}

        Now, we prove Proposition~2.18 is true.

        \begin{proposition} \label{supp:proposition:2.18}
            Suppose $\bx,\by \in \vndistinct$, and let $\uu \in \tonumber{\n}$ be an index. Suppose $\bxln{1}, \byln{1} \in \vndistinct$ and $\bxln{2}, \byln{2} \in \vndistinct$ are imputations of $\bx$ and $\by$ for the index $\uu$ such that
            \begin{align*}
                &\rank{\bxlnn{1}{\uu}}{\bxln{1}} = 1,~\text{and}~ \rank{\bylnn{1}{\uu}}{\byln{1}} = \n,\\
                &\rank{\bxlnn{2}{\uu}}{\bxln{2}} = \n,~\text{and}~ \rank{\bylnn{2}{\uu}}{\byln{2}} = 1.
            \end{align*}
            Then, $\sfd{\bx}{\by} \le \max \{ \sfd{\bxln{1}}{\byln{1}}, \sfd{\bxln{2}}{\byln{2}}\}$.
        \end{proposition}

        \begin{proof}

            To start, since $\rank{\bylnn{2}{\uu}}{\byln{2}} = 1$, then
            we have $\rank{\bxn{\uu}}{\bx} \ge \rank{\bylnn{2}{\uu}}{\byln{2}}$.
            Hence, according to Lemma~\ref{supp:proposition:2.11:lemma:1},
            we have $\sfd{\bx}{\byln{2}} \le \sfd{\bxln{2}}{\byln{2}}$.

            Similarly, since $\rank{\bylnn{1}{\uu}}{\byln{1}} = \n$, we have
            $\rank{\bxn{\uu}}{\bx} \le \rank{\bylnn{1}{\uu}}{\byln{1}}$.
            Hence, according to Lemma~\ref{supp:proposition:2.11:lemma:1},
            we have $\sfd{\bx}{\byln{1}} \le \sfd{\bxln{1}}{\byln{1}}$.

            Hence, we have $\max \{ \sfd{\bx}{\byln{1}}, \sfd{\bx}{\byln{2}}\} \le \max \{ \sfd{\bxln{1}}{\byln{1}}, \sfd{\bxln{2}}{\byln{2}}\}$.
            According to Proposition~\ref{supp:proposition:2.11},
            we have $\sfd{\bx}{\by} \le \max \{ \sfd{\bx}{\byln{1}}, \sfd{\bx}{\byln{2}}\}$.
            Thus, we further have $\sfd{\bx}{\by} \le \max \{ \sfd{\bxln{1}}{\byln{1}}, \sfd{\bxln{2}}{\byln{2}}\}$.
            This completes our proof.

        \end{proof}

        \subsection{Proof of Theorem 2.20}  This subsection proves Theorem~2.20. First, we make the following definition:

        \begin{definition} \label{def:3}
            Let $\bw \subset \tonumber{\n}$ be a subset of indices.
            Define
            $\mythirdset{\bw}{\n}$ as
            \begin{align*}
                \mythirdset{\bw}{\n}  =
                \left\{
                    (\bzln{1},\bzln{2}) \in (\myfirstset{\bw}{\tonumber{\n}}{\n}, \vndistinct) |\rank{\bzlnn{1}{\iconstant}}{\bzln{1}} + \rank{\bzlnn{2}{\iconstant}}{\bzln{2}} = \n + 1,~\iconstant \in \bw
                    \right\}.
            \end{align*}
        \end{definition}

        Then, we prove the following two lemmas, which will be useful for proving Theorem~2.20.

        \begin{lemma} \label{supp:theorem:2.20:lemma:0}
            Suppose $\bxln{1} \in \vndistinct$, and for $2 \le \m < \n$,
            denote $\bwp = \{\n-\m+1,\ldots,\n-1\}$ as a subset of indices.	
            Suppose $\bxln{1} \in \myfirstset{\bwp}{\tonumber{\n}}{\n}$,
            and define	
            \begin{align*}
                \bwpln{1} = \left\{\iconstant \in \bwp: 	\bxlnn{1}{\iconstant} > \max_{\lconstant \in \tonumber{\n} \setminus \bwp} \bxlnn{1}{\lconstant} \right\}, \text{ and }\bwpln{2} = \left\{\iconstant \in \bwp: 	\bxlnn{1}{\iconstant} < \min_{\lconstant \in \tonumber{\n} \setminus \bwp} \bxlnn{1}{\lconstant} \right\}.
            \end{align*}
            Suppose $\bxln{2}$ is an
            imputation of $\bxln{1}$ for the index $\n$. Then if $\rank{\bxlnn{2}{\n}}{\bxln{2}} =\n$,
            we have $\rank{\bxlnn{2}{\iconstant}}{\bxln{2}} = \rank{\bxlnn{1}{\iconstant}}{\bxln{1}} - 1$,
            for any $\iconstant \in \bwpln{1}$, and $\rank{\bxlnn{2}{\iconstant}}{\bxln{2}} = \rank{\bxlnn{1}{\iconstant}}{\bxln{1}}$,
            for any $\iconstant \in \bwpln{2}$. However, if $\rank{\bxlnn{2}{\n}}{\bxln{2}} =1$,
            we have $\rank{\bxlnn{2}{\iconstant}}{\bxln{2}} = \rank{\bxlnn{1}{\iconstant}}{\bxln{1}}$,
            for any $\iconstant \in \bwpln{1}$, and $\rank{\bxlnn{2}{\iconstant}}{\bxln{2}} = \rank{\bxlnn{1}{\iconstant}}{\bxln{1}} + 1$,	for any $\iconstant \in \bwpln{2}$.
        \end{lemma}

        \begin{proof}
            First, we show that if $\rank{\bxlnn{2}{\n}}{\bxln{2}} =\n$,
            then we have $\rank{\bxlnn{2}{\iconstant}}{\bxln{2}} = \rank{\bxlnn{1}{\iconstant}}{\bxln{1}} - 1$,
            for any $\iconstant \in \bwpln{1}$, and $\rank{\bxlnn{2}{\iconstant}}{\bxln{2}} = \rank{\bxlnn{1}{\iconstant}}{\bxln{1}}$,
            for any $\iconstant \in \bwpln{2}$.

            According to the definition of rank, for any $\iconstant \in \bwp$, we have 
            \begin{align*}
                \rank{\bxlnn{2}{\iconstant}}{\bxln{2}} &= \sum_{\jconstant=1}^{\n} \indicator{\bxlnn{2}{\jconstant} \le \bxlnn{2}{\iconstant}} \\
                &= \sum_{\jconstant=1}^{\n-1} \indicator{\bxlnn{2}{\jconstant} \le \bxlnn{2}{\iconstant}} +  \indicator{\bxlnn{2}{\n} \le \bxlnn{2}{\iconstant}}.
            \end{align*}
            Since $\rank{\bxlnn{2}{\n}}{\bxln{2}} = \n$, we have
            $\indicator{\bxlnn{2}{\n} \le \bxlnn{2}{\iconstant}} = 0$, for any $\iconstant \in \bwp$. 
            Then, for any $\iconstant \in \bwp$, we have
            \begin{align} \label{supp:theorem:2.20:lemma:0:eqn:1}
                \rank{\bxlnn{2}{\iconstant}}{\bxln{2}} = \sum_{\jconstant=1}^{\n-1} \indicator{\bxlnn{2}{\jconstant} \le \bxlnn{2}{\iconstant}} = \rank{\bxlnn{2}{\iconstant}}{(\bxlnn{2}{\lconstant})_{\lconstant \in \tonumber{\n-1}}}.
            \end{align} 
            Since $\bxln{2}$ is an imputation of $\bxln{1}$ for the index $\n$, then we have $\bxlnn{2}{\iconstant} = \bxlnn{1}{\iconstant}$
            for any $\iconstant \in \tonumber{\n} \setminus \{\n\} = \tonumber{\n-1}$.
            Hence, $(\bxlnn{2}{\lconstant})_{\lconstant \in \tonumber{\n-1}}$ is 
            a permutation of $(\bxlnn{1}{\lconstant})_{\lconstant \in \tonumber{\n-1}}$.
            Meanwhile, since $\bwp = \{\n-\m+1,\ldots,\n-1\} \subset \tonumber{\n-1}$,
            we have $\bxlnn{2}{\iconstant} = \bxlnn{1}{\iconstant}$, for any $\iconstant \in \bwp$.
            Then, it follows that
            \begin{align} \label{supp:theorem:2.20:lemma:0:eqn:2}
                &\rank{\bxlnn{2}{\iconstant}}{(\bxlnn{2}{\lconstant})_{\lconstant \in \tonumber{\n-1}}} =  \rank{\bxlnn{1}{\iconstant}}{(\bxlnn{1}{\lconstant})_{\lconstant \in \tonumber{\n-1}}}, \text{ for any }
                \iconstant \in \bwp.
            \end{align}
            Using the definition of ranks and notice that $\bxln{1} \in \vndistinct$ 
            is a vector of distinct real values, we can show that
            \begin{align*}
                \rank{\bxlnn{1}{\iconstant}}{(\bxlnn{1}{\lconstant})_{\lconstant \in \tonumber{\n-1}}} = \rank{\bxlnn{1}{\iconstant}}{\bxln{1}} - \indicator{\bxlnn{1}{\n} < \bxlnn{1}{\iconstant}}, \text{ for any } \iconstant \in \bwp.
            \end{align*} 	
            If $\iconstant \in \bwpln{1}$, then according to the definition of $\bwpln{1}$,  	
            we have $\bxlnn{1}{\iconstant} >  
            \max_{\lconstant \in \tonumber{\n} \setminus \bwp} \bxlnn{1}{\lconstant}$.
            Since $\n \in \tonumber{\n} \setminus \bwp$, we have 
            $\bxlnn{1}{\iconstant} > \bxlnn{1}{\n}$. 	  	
            Hence, we have
            \begin{align} \label{supp:theorem:2.20:lemma:0:eqn:3}
                \rank{\bxlnn{1}{\uuln{1}}}{(\bxlnn{1}{\lconstant})_{\lconstant \in \tonumber{\n-1}}} = \rank{\bxlnn{1}{\uuln{1}}}{\bxln{1}} - 1, \text{ for any } \iconstant \in \bwpln{1}.
            \end{align} 	
            However, if $\iconstant \in \bwpln{2}$, according to the definition of $\bwpln{2}$,  	
            we have $\bxlnn{1}{\iconstant} < 
            \min_{\lconstant \in \tonumber{\n} \setminus \bwp} \bxlnn{1}{\lconstant}$.
            Since $\n \in \tonumber{\n} \setminus \bwp$, we have 
            $\bxlnn{1}{\iconstant} < \bxlnn{1}{\n}$. 	  	
            Hence, we have
            \begin{align} \label{supp:theorem:2.20:lemma:0:eqn:3.0}
                \rank{\bxlnn{1}{\uuln{1}}}{(\bxlnn{1}{\lconstant})_{\lconstant \in \tonumber{\n-1}}} = \rank{\bxlnn{1}{\uuln{1}}}{\bxln{1}}, \text{ for any } \iconstant \in \bwpln{2}.
            \end{align} 	
            Thus, for any $\iconstant \in \bwpln{1}$, we have
            \begin{align*}
                \rank{\bxlnn{2}{\iconstant}}{\bxln{2}} &=^{\eqref{supp:theorem:2.20:lemma:0:eqn:1}}  \rank{\bxlnn{2}{\iconstant}}{(\bxlnn{2}{\lconstant})_{\lconstant \in \tonumber{\n-1}}}\\
                &=^{\eqref{supp:theorem:2.20:lemma:0:eqn:2}} \rank{\bxlnn{1}{\iconstant}}{(\bxlnn{1}{\lconstant})_{\lconstant \in \tonumber{\n-1}}}\\
                &=^{\eqref{supp:theorem:2.20:lemma:0:eqn:3}} \rank{\bxlnn{1}{\iconstant}}{\bxln{1}} - 1.
            \end{align*}
            Similarly, for any $\iconstant \in \bwpln{1}$, we have
            \begin{align*}
                \rank{\bxlnn{2}{\iconstant}}{\bxln{2}} &=^{\eqref{supp:theorem:2.20:lemma:0:eqn:1}}  \rank{\bxlnn{2}{\iconstant}}{(\bxlnn{2}{\lconstant})_{\lconstant \in \tonumber{\n-1}}}\\
                &=^{\eqref{supp:theorem:2.20:lemma:0:eqn:2}} \rank{\bxlnn{1}{\iconstant}}{(\bxlnn{1}{\lconstant})_{\lconstant \in \tonumber{\n-1}}}\\
                &=^{\eqref{supp:theorem:2.20:lemma:0:eqn:3.0}} \rank{\bxlnn{1}{\iconstant}}{\bxln{1}}.
            \end{align*}
            Hence, we have shown that when $\rank{\bxlnn{2}{\n}}{\bxln{2}} =\n$,
            we have $\rank{\bxlnn{2}{\iconstant}}{\bxln{2}} = \rank{\bxlnn{1}{\iconstant}}{\bxln{1}} - 1$,
            for any $\iconstant \in \bwpln{1}$, and $\rank{\bxlnn{2}{\iconstant}}{\bxln{2}} = \rank{\bxlnn{1}{\iconstant}}{\bxln{1}}$,
            for any $\iconstant \in \bwpln{2}$.

            Similarly, we can show that if $\rank{\bxlnn{2}{\n}}{\bxln{2}} =1$,
            we have $\rank{\bxlnn{2}{\iconstant}}{\bxln{2}} = \rank{\bxlnn{1}{\iconstant}}{\bxln{1}}$,
            for any $\iconstant \in \bwpln{1}$, and $\rank{\bxlnn{2}{\iconstant}}{\bxln{2}} = \rank{\bxlnn{1}{\iconstant}}{\bxln{1}} + 1$,
            for any $\iconstant \in \bwpln{2}$.

            According to the definition of rank, for any $\iconstant \in \bw$, we have 
            \begin{align*}
                \rank{\bxlnn{2}{\iconstant}}{\bxln{2}} &= \sum_{\jconstant=1}^{\n} \indicator{\bxlnn{2}{\jconstant} \le \bxlnn{2}{\iconstant}} \\
                &= \sum_{\jconstant=1}^{\n-1} \indicator{\bxlnn{2}{\jconstant} \le \bxlnn{2}{\iconstant}} +  \indicator{\bxlnn{2}{\n} \le \bxlnn{2}{\iconstant}}.
            \end{align*}
            Since $\rank{\bxlnn{2}{\n}}{\bxln{2}} = 1$, we have
            $\indicator{\bxlnn{2}{\n} \le \bxlnn{2}{\iconstant}} = 1$, for any $\iconstant \in \bwp$.
            Then, we have
            \begin{align} \label{supp:theorem:2.20:lemma:0:eqn:4}
                \begin{split}
                    \rank{\bxlnn{2}{\uuln{1}}}{\bxln{2}} &= \sum_{\iconstant=1}^{\n-1} \indicator{\bxlnn{2}{\iconstant} \le \bxlnn{2}{\uuln{1}}} + 1 \\
                    &= \rank{\bxlnn{2}{\uuln{1}}}{(\bxlnn{2}{\lconstant})_{\lconstant \in \tonumber{\n-1}}} + 1, \text{ for any } \iconstant \in \bwp.
                \end{split}
            \end{align} 
            Since $\bxln{2}$ is an imputation of $\bxln{1}$ for the index $\n$, then we have $\bxlnn{2}{\iconstant} = \bxlnn{1}{\iconstant}$
            for any $\iconstant \in \tonumber{\n} \setminus \{\n\} = \tonumber{\n-1}$.
            Hence, $(\bxlnn{2}{\lconstant})_{\lconstant \in \tonumber{\n-1}}$ is 
            a permutation of $(\bxlnn{1}{\lconstant})_{\lconstant \in \tonumber{\n-1}}$.
            Meanwhile, since $\bwp = \{\n-\m+1,\ldots,\n-1\} \subset \tonumber{\n - 1}$,
            we have $\bxlnn{2}{\iconstant} = \bxlnn{1}{\iconstant}$, for any $\iconstant \in \bwp$.
            Then, it follows that
            \begin{align} \label{supp:theorem:2.20:lemma:0:eqn:5}
                &\rank{\bxlnn{2}{\iconstant}}{(\bxlnn{2}{\lconstant})_{\lconstant \in \tonumber{\n-1}}} =  \rank{\bxlnn{1}{\iconstant}}{(\bxlnn{1}{\lconstant})_{\lconstant \in \tonumber{\n-1}}}, \text{ for any }\iconstant \in \bwp.
            \end{align}
            Using the definition of ranks and notice that $\bxln{2} \in \vndistinct$ 
            is a vector of distinct real values, we can show that
            \begin{align*}
                \rank{\bxlnn{1}{\iconstant}}{(\bxlnn{1}{\lconstant})_{\lconstant \in \tonumber{\n-1}}} = \rank{\bxlnn{1}{\iconstant}}{\bxln{1}} - \indicator{\bxlnn{1}{\n} < \bxlnn{1}{\iconstant}}, \text{ for any } \iconstant \in \bwp.
            \end{align*} 	
            If $\iconstant \in \bwpln{1}$, then according to the definition of $\bwpln{1}$,  	
            we have $\bxlnn{1}{\iconstant} >  
            \max_{\lconstant \in \tonumber{\n} \setminus \bwp} \bxlnn{1}{\lconstant}$.
            Since $\n \in \tonumber{\n} \setminus \bwp$, we have 
            $\bxlnn{1}{\iconstant} > \bxlnn{1}{\n}$. 	  	
            Hence, we have
            \begin{align} \label{supp:theorem:2.20:lemma:0:eqn:6}
                \rank{\bxlnn{1}{\uuln{1}}}{(\bxlnn{1}{\lconstant})_{\lconstant \in \tonumber{\n-1}}} = \rank{\bxlnn{1}{\uuln{1}}}{\bxln{1}} - 1, \text{ for any } \iconstant \in \bwpln{1}.
            \end{align} 	
            However, if $\iconstant \in \bwpln{2}$, then according to the definition of $\bwpln{2}$,  	
            we have $\bxlnn{1}{\iconstant} < 
            \min_{\lconstant \in \tonumber{\n} \setminus \bwp} \bxlnn{1}{\lconstant}$.
            Since $\n \in \tonumber{\n} \setminus \bwp$, we have 
            $\bxlnn{1}{\iconstant} < \bxlnn{1}{\n}$. 	  	
            Hence, we have
            \begin{align} \label{supp:theorem:2.20:lemma:0:eqn:7}
                \rank{\bxlnn{1}{\uuln{1}}}{(\bxlnn{1}{\lconstant})_{\lconstant \in \tonumber{\n-1}}} = \rank{\bxlnn{1}{\uuln{1}}}{\bxln{1}}, \text{ for any } \iconstant \in \bwpln{2}.
            \end{align} 	
            Thus, for any $\iconstant \in \bwpln{1}$, we have
            \begin{align*}
                \rank{\bxlnn{2}{\iconstant}}{\bxln{2}} &=^{\eqref{supp:theorem:2.20:lemma:0:eqn:4}}  \rank{\bxlnn{2}{\iconstant}}{(\bxlnn{2}{\lconstant})_{\lconstant \in \tonumber{\n-1}}} + 1\\
                &=^{\eqref{supp:theorem:2.20:lemma:0:eqn:5}} \rank{\bxlnn{1}{\iconstant}}{(\bxlnn{1}{\lconstant})_{\lconstant \in \tonumber{\n-1}}} + 1\\
                &=^{\eqref{supp:theorem:2.20:lemma:0:eqn:6}} \rank{\bxlnn{1}{\iconstant}}{\bxln{1}}.
            \end{align*}
            Similarly, for any $\iconstant \in \bwpln{2}$, we have
            \begin{align*}
                \rank{\bxlnn{2}{\iconstant}}{\bxln{2}} &=^{\eqref{supp:theorem:2.20:lemma:0:eqn:4}}  \rank{\bxlnn{2}{\iconstant}}{(\bxlnn{2}{\lconstant})_{\lconstant \in \tonumber{\n-1}}} + 1\\
                &=^{\eqref{supp:theorem:2.20:lemma:0:eqn:5}} \rank{\bxlnn{1}{\iconstant}}{(\bxlnn{1}{\lconstant})_{\lconstant \in \tonumber{\n-1}}} + 1\\
                &=^{\eqref{supp:theorem:2.20:lemma:0:eqn:6}} \rank{\bxlnn{1}{\iconstant}}{\bxln{1}} + 1.
            \end{align*}
            Hence, we have shown that if $\rank{\bxlnn{2}{\n}}{\bxln{2}} =1$,
            we have $\rank{\bxlnn{2}{\iconstant}}{\bxln{2}} = \rank{\bxlnn{1}{\iconstant}}{\bxln{1}}$,
            for any $\iconstant \in \bwpln{1}$, and $\rank{\bxlnn{2}{\iconstant}}{\bxln{2}} = \rank{\bxlnn{1}{\iconstant}}{\bxln{1}} + 1$, for any $\iconstant \in \bwpln{2}$.
            This completes our proof.
        \end{proof}

        \begin{lemma} \label{supp:theorem:2.20:lemma:1}
            Suppose $\bx, \by \in \vndistinct$, and for $2 \le \m < \n$ assume $\bw = \{\n-\m+1,\ldots,\n\}$,
            and $\bwp = \{\n-\m+1,\ldots,\n-1\}$.
            Suppose $(\bxln{1},\byln{1}) \in \mythirdset{\bwp}{\n}$ are imputations
            of $\bx$ and $\by$ for indices $\bwp$, and $(\bxln{2}, \byln{2}) \in \mythirdset{\{\n\}}{\n} $ 
            are imputations of $\bxln{1}$ and $\byln{1}$ for the index $\n$. Then $\bxln{2},\byln{2}$ 
            are imputations of $\bx$ and $\by$ for indices $\bw$ such that
            $(\bxln{2},\byln{2}) \in \mythirdset{\bw}{\n}$.
        \end{lemma}

        \begin{proof}	
            First, we show that $\bxln{2},\byln{2}$ 
            are imputations of $\bx$ and $\by$ for indices $\bw$.
            Since $\bxln{1},\byln{1}$ are imputations
            of $\bx$ and $\by$ for indices $\bwp$, then according to
            the definitions of imputations, we have
            $\bxlnn{1}{\iconstant} = \bxn{\iconstant}$
            and
            $\bylnn{1}{\iconstant} = \byn{\iconstant}$
            for any $\iconstant \in \tonumber{\n} \setminus \bwp$.

            Since $\bwp \subset \bw$, we have $\tonumber{\n} \setminus \bw \subset \tonumber{\n} \setminus \bwp$.
            Hence, we have
            \begin{align*}
                \bxlnn{1}{\iconstant} = \bxn{\iconstant}, \text{ and } \bylnn{1}{\iconstant} = \byn{\iconstant}, \text{ for any } \iconstant \in \tonumber{\n} \setminus \bw.
            \end{align*}
            Since $\bxln{2}, \byln{2}$ are imputations of $\bxln{1}$ and $\byln{1}$ for the index $\n$,
            then we have
            $\bxlnn{2}{\iconstant} = \bxlnn{1}{\iconstant}$
            and
            $\bylnn{2}{\iconstant} = \bylnn{1}{\iconstant}$
            for any $\iconstant \in \tonumber{\n} \setminus \{\n\}$.
            Since $\{\n\} \subset \bw$, then we have $\tonumber{\n} \setminus \bw \subset  \tonumber{\n} \setminus \{\n\}$. Hence, we have 
            \begin{align*}
                &\bxlnn{2}{\iconstant} = \bxlnn{1}{\iconstant}, \text{ and } \bylnn{2}{\iconstant} = \bylnn{1}{\iconstant}, \text{ for any } \iconstant \in \tonumber{\n} \setminus \bw\\
                \Rightarrow &\bxlnn{2}{\iconstant} = \bxn{\iconstant}, \text{ and } \bylnn{2}{\iconstant} = \byn{\iconstant}, \text{ for any } \iconstant \in \tonumber{\n} \setminus \bw.
            \end{align*}
            This proves that $\bxln{2},\byln{2}$ 
            are imputations of $\bx$ and $\by$ for indices $\bw$.

            Now, we show that $(\bxln{2},\byln{2}) \in \mythirdset{\bw}{\n}$.	
            To start, since $(\bxln{1},\byln{1}) \in \mythirdset{\bwp}{\n}$,
            then we have $\bxln{1} \in \myfirstset{\bwp}{\tonumber{\n}}{\n}$. 
            Similarly, since $(\bxln{2}, \byln{2}) \in \mythirdset{\{\n\}}{\n}$,
            we have $\bxln{2} \in \myfirstset{\{\n\}}{\tonumber{\n}}{\n}$.
            Then, according to Lemma~\ref{supp:proposition:2.13:lemma:1}, 
            we have $\bxln{2} \in \myfirstset{\bw}{\tonumber{\n}}{\n}$.
            Hence, we have 
            $(\bxln{2},\byln{2}) \in (\myfirstset{\bw}{\tonumber{\n}}{\n}, \vndistinct)$.
            Then, in order to show that $(\bxln{2},\byln{2}) \in \mythirdset{\bw}{\n}$,
            it is sufficient to show that
            \begin{align*} 
                \rank{\bxlnn{2}{\iconstant}}{\bxln{2}} + \rank{\bylnn{2}{\iconstant}}{\byln{2}} = \n + 1, \text{ for any } \iconstant \in \bw.
            \end{align*}
            Since $(\bxln{2}, \byln{2}) \in \mythirdset{\{\n\}}{\n}$, we have
            \begin{align} \label{supp:theorem:2.20:lemma:1:eqn:1}
                \rank{\bxlnn{2}{\n}}{\bxln{2}} + \rank{\bylnn{2}{\n}}{\byln{2}} = \n + 1.
            \end{align} 
            Then, we only need to show that
            \begin{align} \label{supp:theorem:2.20:lemma:1:eqn:2}
                \rank{\bxlnn{2}{\iconstant}}{\bxln{2}} + \rank{\bylnn{2}{\iconstant}}{\byln{2}} = \n + 1, \text{ for any } \iconstant \in \bwp.
            \end{align}
            Below, we prove \eqref{supp:theorem:2.20:lemma:1:eqn:2} is true.

            Since $\bxln{2} \in \myfirstset{\{\n\}}{\tonumber{\n}}{\n}$, then we have
            \begin{align*}
                \bxlnn{2}{\n} > \max_{\lconstant \in \tonumber{\n-1}} \bxlnn{2}{\lconstant},~\text{or}~\bxlnn{2}{\n} < \min_{\lconstant \in \tonumber{\n-1}} \bxlnn{2}{\lconstant}.
            \end{align*}
            Hence, we have either
            \begin{align*}
                \rank{\bxlnn{2}{\n}}{\bxln{2}} = \n, \text{ or } \rank{\bxlnn{2}{\n}}{\bxln{2}} = 1.
            \end{align*}
            Combining with \eqref{supp:theorem:2.20:lemma:1:eqn:1}, we
            have either  
            \begin{align*}
                &\text{ case } (1): ~\rank{\bxlnn{2}{\n}}{\bxln{2}} = \n, \text{ and } \rank{\bylnn{2}{\n}}{\byln{2}} = 1,\\
                \text{or }&\text{ case } (2): ~\rank{\bxlnn{2}{\n}}{\bxln{2}} = 1, \text{ and } \rank{\bylnn{2}{\n}}{\byln{2}} = \n,
            \end{align*}
            is true.

            Suppose case $(1)$ is true. In other words, 
            $~\rank{\bxlnn{2}{\n}}{\bxln{2}} = \n$,
            and $\rank{\bylnn{2}{\n}}{\byln{2}} = 1$.

            Since $\bxln{1} \in \myfirstset{\bwp}{\tonumber{\n}}{\n}$, then we have
            \begin{align*}
                \bxlnn{1}{\iconstant} > \max_{\lconstant \in \tonumber{\n} \setminus \bwp} \bxlnn{1}{\lconstant},~\text{or}~ \bxlnn{1}{\iconstant} < \min_{\lconstant \in \tonumber{\n} \setminus \bwp} \bxlnn{1}{\lconstant}, \text{ for any } \iconstant \in \bwp.
            \end{align*} 
            Let us define
            \begin{align*}
                &\bwpln{1} = \left\{\iconstant \in \bwp: 	\bxlnn{1}{\iconstant} > \max_{\lconstant \in \tonumber{\n} \setminus \bwp} \bxlnn{1}{\lconstant} \right\}, \\
                \text{and }&\bwpln{2} = \left\{\iconstant \in \bwp: 	\bxlnn{1}{\iconstant} < \min_{\lconstant \in \tonumber{\n} \setminus \bwp} \bxlnn{1}{\lconstant} \right\}.
            \end{align*}
            Then, we have $\bwpln{1} \cup \bwpln{2} = \bwp$.
            Denote $\npln{1} = |\bwpln{1}|$ and $\npln{2} = |\bwpln{2}|$.

            Next, since $\bxln{2}$ is an imputation of $\bxln{1} \in \myfirstset{\bwp}{\tonumber{\n}}{\n}$
            such that $\rank{\bxlnn{2}{\n}}{\bxln{2}} = \n$. Then	
            according to Lemma~\ref{supp:theorem:2.20:lemma:0}, we have 
            \begin{align}
                &\rank{\bxlnn{2}{\iconstant}}{\bxln{2}} = \rank{\bxlnn{1}{\iconstant}}{\bxln{1}} - 1, \text{ for any } \iconstant \in \bwpln{1}, \label{supp:theorem:2.20:lemma:1:eqn:3}\\
                \text{and }& \rank{\bxlnn{2}{\iconstant}}{\bxln{2}} = \rank{\bxlnn{1}{\iconstant}}{\bxln{1}}, \text{ for any } \iconstant \in \bwpln{2}. \label{supp:theorem:2.20:lemma:1:eqn:4}
            \end{align}

            For any $\iconstant \in \bwpln{1}$, 
            since $\bxlnn{1}{\iconstant} > 
            \max_{\lconstant \in \tonumber{\n} \setminus \bwp} \bxlnn{1}{\lconstant}$, 
            then we have
            \begin{align*}
                \rank{\bxlnn{1}{\iconstant}}{\bxln{1}} \in \{\n, \ldots, \n- \npln{1} +1\}, 
                \text{ for any } \iconstant \in \bwpln{1}.
            \end{align*}
            Further, since $(\bxln{1},\byln{1}) \in \mythirdset{\bwp}{\n}$, then	
            $\rank{\bxlnn{1}{\iconstant}}{\bxln{1}} + \rank{\bylnn{1}{\iconstant}}{\byln{1}} = \n+1$ 
            for any $\iconstant \in \bwp$.
            Hence, we have 
            \begin{align}
                &\rank{\bylnn{1}{\iconstant}}{\bxln{1}} \in \{1, \ldots, \npln{1}\}, 
                \text{ for any } \iconstant \in \bwpln{1} \nonumber\\
                \Rightarrow & \bylnn{1}{\iconstant} < \min_{\lconstant \in \tonumber{\n} \setminus \bwp} \bylnn{1}{\lconstant}, \text{ for any } \iconstant \in \bwpln{1}.  \label{supp:theorem:2.20:lemma:1:eqn:4.0}
            \end{align}
            Similarly, for any $\iconstant \in \bwpln{2}$, 
            since $\bxlnn{1}{\iconstant} < 
            \min_{\lconstant \in \tonumber{\n} \setminus \bwp} \bxlnn{1}{\lconstant}$, 
            then we have
            \begin{align*}
                \rank{\bxlnn{1}{\iconstant}}{\bxln{1}} \in \{1, \ldots, \npln{2}\}, 
                \text{ for any } \iconstant \in \bwpln{2}.
            \end{align*}
            Further, since $(\bxln{1},\byln{1}) \in \mythirdset{\bwp}{\n}$, then	
            $\rank{\bxlnn{1}{\iconstant}}{\bxln{1}} + \rank{\bylnn{1}{\iconstant}}{\byln{1}} = \n+1$ 
            for any $\iconstant \in \bwp$.
            Hence, we have 
            \begin{align}
                &\rank{\bylnn{1}{\iconstant}}{\bxln{1}} \in \{\n, \ldots, \n-\npln{2}+1\}, 
                \text{ for any } \iconstant \in \bwpln{2} \nonumber\\
                \Rightarrow & \bylnn{1}{\iconstant} > \max_{\lconstant \in \tonumber{\n} \setminus \bwp} \bylnn{1}{\lconstant}, \text{ for any } \iconstant \in \bwpln{2}.  \label{supp:theorem:2.20:lemma:1:eqn:4.1}
            \end{align}

            Combining \eqref{supp:theorem:2.20:lemma:1:eqn:4.0} and \eqref{supp:theorem:2.20:lemma:1:eqn:4.1},
            we have $\byln{1} \in \myfirstset{\bwp}{\tonumber{\n}}{\n}$. 	
            Notice that $\byln{2}$ is an imputation of $\byln{1} \in \myfirstset{\bwp}{\tonumber{\n}}{\n}$
            for the index $\n$ such that $\rank{\bylnn{2}{\n}}{\bxln{2}} = 1$.	 
            Then, according to Lemma~\ref{supp:theorem:2.20:lemma:0}, we have 
            \begin{align}
                &\rank{\bylnn{2}{\iconstant}}{\byln{2}} =  \rank{\bylnn{1}{\iconstant}}{\byln{1}} + 1, \text{ for any } \iconstant \in \bwpln{1}, \label{supp:theorem:2.20:lemma:1:eqn:5}\\
                \text{and }&\rank{\bylnn{2}{\iconstant}}{\byln{2}} =  \rank{\bylnn{1}{\iconstant}}{\byln{1}}, \text{ for any } \iconstant \in \bwpln{2}. \label{supp:theorem:2.20:lemma:1:eqn:6}
            \end{align}	
            Combining \eqref{supp:theorem:2.20:lemma:1:eqn:3}, \eqref{supp:theorem:2.20:lemma:1:eqn:4},
            \eqref{supp:theorem:2.20:lemma:1:eqn:5} and \eqref{supp:theorem:2.20:lemma:1:eqn:5}, for any
            $\iconstant \in \bwp$, we have
            \begin{align*}
                \rank{\bxlnn{2}{\iconstant}}{\bxln{2}} + \rank{\bylnn{2}{\iconstant}}{\byln{2}} &= \rank{\bxlnn{1}{\iconstant}}{\bxln{1}} + \rank{\bylnn{1}{\iconstant}}{\byln{1}}.
            \end{align*}    	 
            Since  $(\bxln{1},\byln{1}) \in \mythirdset{\bwp}{\n}$, then  for any
            $\iconstant \in \bwp$, we have
            \begin{align*}
                \rank{\bxlnn{1}{\iconstant}}{\bxln{1}} + \rank{\bylnn{1}{\iconstant}}{\byln{1}} = \n +1 .
            \end{align*}
            Hence, for any 	$\iconstant \in \bwp$, we have
            \begin{align*}
                \rank{\bxlnn{2}{\iconstant}}{\bxln{2}} + \rank{\bylnn{1}{\iconstant}}{\byln{1}} = \n +1.
            \end{align*}
            This proves \eqref{supp:theorem:2.20:lemma:1:eqn:2} when the case $(1)$ is true.

            When the case $(2)$ is true, \eqref{supp:theorem:2.20:lemma:1:eqn:2} can be proved similarly.
            Suppose the case $(2)$ is true. In other words, suppose
            $~\rank{\bxlnn{2}{\n}}{\bxln{2}} = 1$,
            and $\rank{\bylnn{2}{\n}}{\byln{2}} = \n$.

            Since $\bxln{1} \in \myfirstset{\bwp}{\tonumber{\n}}{\n}$, then we have
            \begin{align*}
                \bxlnn{1}{\iconstant} > \max_{\lconstant \in \tonumber{\n} \setminus \bwp} \bxlnn{1}{\lconstant},~\text{or}~ \bxlnn{1}{\iconstant} < \min_{\lconstant \in \tonumber{\n} \setminus \bwp} \bxlnn{1}{\lconstant}, \text{ for any } \iconstant \in \bwp.
            \end{align*} 
            Let us define
            \begin{align*}
                &\bwpln{1} = \left\{\iconstant \in \bwp: 	\bxlnn{1}{\iconstant} > \max_{\lconstant \in \tonumber{\n} \setminus \bwp} \bxlnn{1}{\lconstant} \right\}, \\
                \text{and }&\bwpln{2} = \left\{\iconstant \in \bwp: 	\bxlnn{1}{\iconstant} < \min_{\lconstant \in \tonumber{\n} \setminus \bwp} \bxlnn{1}{\lconstant} \right\}.
            \end{align*}
            Then, we have $\bwpln{1} \cup \bwpln{2} = \bwp$.
            Denote $\npln{1} = |\bwpln{1}|$ and $\npln{2} = |\bwpln{2}|$.

            Next, since $\bxln{2}$ is an imputation of $\bxln{1} \in \myfirstset{\bwp}{\tonumber{\n}}{\n}$
            such that $\rank{\bxlnn{2}{\n}}{\bxln{2}} = 1$. Then	
            according to Lemma~\ref{supp:theorem:2.20:lemma:0}, we have 
            \begin{align}
                &\rank{\bxlnn{2}{\iconstant}}{\bxln{2}} = \rank{\bxlnn{1}{\iconstant}}{\bxln{1}}, \text{ for any } \iconstant \in \bwpln{1}, \label{supp:theorem:2.20:lemma:1:eqn:7}\\
                \text{and }& \rank{\bxlnn{2}{\iconstant}}{\bxln{2}} = \rank{\bxlnn{1}{\iconstant}}{\bxln{1}} + 1, \text{ for any } \iconstant \in \bwpln{2}. \label{supp:theorem:2.20:lemma:1:eqn:8}
            \end{align}

            For any $\iconstant \in \bwpln{1}$, 
            since $\bxlnn{1}{\iconstant} > 
            \max_{\lconstant \in \tonumber{\n} \setminus \bwp} \bxlnn{1}{\lconstant}$, 
            then we have
            \begin{align*}
                \rank{\bxlnn{1}{\iconstant}}{\bxln{1}} \in \{\n, \ldots, \n- \npln{1} +1\}, 
                \text{ for any } \iconstant \in \bwpln{1}.
            \end{align*}
            Further, since $(\bxln{1},\byln{1}) \in \mythirdset{\bwp}{\n}$, then	
            $\rank{\bxlnn{1}{\iconstant}}{\bxln{1}} + \rank{\bylnn{1}{\iconstant}}{\byln{1}} = \n+1$ 
            for any $\iconstant \in \bwp$.
            Hence, we have 
            \begin{align}
                &\rank{\bylnn{1}{\iconstant}}{\bxln{1}} \in \{1, \ldots, \npln{1}\}, 
                \text{ for any } \iconstant \in \bwpln{1} \nonumber\\
                \Rightarrow & \bylnn{1}{\iconstant} < \min_{\lconstant \in \tonumber{\n} \setminus \bwp} \bylnn{1}{\lconstant}, \text{ for any } \iconstant \in \bwpln{1}.  \label{supp:theorem:2.20:lemma:1:eqn:9}
            \end{align}
            Similarly, for any $\iconstant \in \bwpln{2}$, 
            since $\bxlnn{1}{\iconstant} < 
            \min_{\lconstant \in \tonumber{\n} \setminus \bwp} \bxlnn{1}{\lconstant}$, 
            then we have
            \begin{align*}
                \rank{\bxlnn{1}{\iconstant}}{\bxln{1}} \in \{1, \ldots, \npln{2}\}, 
                \text{ for any } \iconstant \in \bwpln{2}.
            \end{align*}
            Further, since $(\bxln{1},\byln{1}) \in \mythirdset{\bwp}{\n}$, then	
            $\rank{\bxlnn{1}{\iconstant}}{\bxln{1}} + \rank{\bylnn{1}{\iconstant}}{\byln{1}} = \n+1$ 
            for any $\iconstant \in \bwp$.
            Hence, we have 
            \begin{align}
                &\rank{\bylnn{1}{\iconstant}}{\bxln{1}} \in \{\n, \ldots, \n-\npln{2}+1\}, 
                \text{ for any } \iconstant \in \bwpln{2} \nonumber\\
                \Rightarrow & \bylnn{1}{\iconstant} > \max_{\lconstant \in \tonumber{\n} \setminus \bwp} \bylnn{1}{\lconstant}, \text{ for any } \iconstant \in \bwpln{2}.  \label{supp:theorem:2.20:lemma:1:eqn:10}
            \end{align}

            Combining \eqref{supp:theorem:2.20:lemma:1:eqn:9} and \eqref{supp:theorem:2.20:lemma:1:eqn:10},
            we have $\byln{1} \in \myfirstset{\bwp}{\tonumber{\n}}{\n}$.	
            Notice that $\byln{2}$ is an imputation of $\byln{1} \in \myfirstset{\bwp}{\tonumber{\n}}{\n}$
            such that $\rank{\bylnn{2}{\n}}{\bxln{2}} = \n$.	 
            Then, according to Lemma~\ref{supp:theorem:2.20:lemma:0}, we have 
            \begin{align}
                &\rank{\bylnn{2}{\iconstant}}{\byln{2}} =  \rank{\bylnn{1}{\iconstant}}{\byln{1}}, \text{ for any } \iconstant \in \bwpln{1}, \label{supp:theorem:2.20:lemma:1:eqn:11}\\
                \text{and }&\rank{\bylnn{2}{\iconstant}}{\byln{2}} =  \rank{\bylnn{1}{\iconstant}}{\byln{1}} - 1, \text{ for any } \iconstant \in \bwpln{2}. \label{supp:theorem:2.20:lemma:1:eqn:12}
            \end{align}	
            Combining \eqref{supp:theorem:2.20:lemma:1:eqn:7}, \eqref{supp:theorem:2.20:lemma:1:eqn:8},
            \eqref{supp:theorem:2.20:lemma:1:eqn:11} and \eqref{supp:theorem:2.20:lemma:1:eqn:12}, for any
            $\iconstant \in \bwp$, we have
            \begin{align*}
                \rank{\bxlnn{2}{\iconstant}}{\bxln{2}} + \rank{\bylnn{2}{\iconstant}}{\byln{2}} &= \rank{\bxlnn{1}{\iconstant}}{\bxln{1}} + \rank{\bylnn{1}{\iconstant}}{\byln{1}}.
            \end{align*}    	 
            Since  $(\bxln{1},\byln{1}) \in \mythirdset{\bwp}{\n}$, then  for any
            $\iconstant \in \bwp$, we have
            \begin{align*}
                \rank{\bxlnn{1}{\iconstant}}{\bxln{1}} + \rank{\bylnn{1}{\iconstant}}{\byln{1}} = \n +1 .
            \end{align*}
            Hence, for any 	$\iconstant \in \bwp$, we have
            \begin{align*}
                \rank{\bxlnn{2}{\iconstant}}{\bxln{2}} + \rank{\bylnn{1}{\iconstant}}{\byln{1}} = \n +1.
            \end{align*}
            This proves \eqref{supp:theorem:2.20:lemma:1:eqn:2} when the case $(2)$ is true.
            Hence, we complete our proof.

        \end{proof}

        Now, we are ready to prove Theorem~2.20:

        \begin{theorem} \label{supp:theorem:2.20}
            Suppose $\bx,\by \in \vndistinct$ and $\bw \subset \tonumber{\n}$.
            Then there exist imputations $(\bxs,\bys) \in \mythirdset{\bw}{\n}$ 
            of $\bx$ and $\by$ for indices $\bw$ such that $\sfd{\bx}{\by} \le \sfd{\bxs}{\bys}$.
            Furthermore, consider any other imputations $\bxp,\byp \in \vndistinct$ of $\bx, \by$ for indices $\bw$. 
            Then $\sfd{\bxp}{\byp} \le \sfd{\bxs}{\bys}$.
        \end{theorem}

        \begin{proof}
            To prove Theorem~\ref{supp:theorem:2.20}, we first prove the following
            statement
            \begin{itemize}
                \item[$\statement$]: Suppose $\bx,\by \in \vndistinct$ and $\bw \subset \tonumber{\n}$.
                    Then there exist imputations $(\bxs,\bys) \in \mythirdset{\bw}{\n}$ 
                    of $\bx$ and $\by$ for indices $\bw$ such that $\sfd{\bx}{\by} \le \sfd{\bxs}{\bys}$.
            \end{itemize}
            Then we will prove Theorem~\ref{supp:theorem:2.20} using statement $\statement$. 

            First, we prove the statement $\statement$ is true.
            Suppose $\bw = \emptyset$. Then according to the definition of imputations,
            we have $\bxs = \bx$ and $\bys = \by$. Hence, we have 
            $\sfd{\bx}{\by} = \sfd{\bxs}{\bys}$. This proves the statement $\statement$ 
            when $\bw = \emptyset$.

            Suppose $\bw = \tonumber{\n}$. Then, let $\bxs \in \vndistinct$
            be a vector such that $\bxsn{1} > \ldots > \bxsn{\n}$.
            In other words,  $\bxs \in \vndistinct$ is a vector such that
            \begin{align*}
                \rank{\bxsn{\iconstant}}{\bxs} = \n + 1 - \iconstant, \text{ for any } \iconstant \in \tonumber{\n}.
            \end{align*}
            Let $\bys \in \vndistinct$
            be a vector such that $\bysn{1} < \ldots < \bysn{\n}$.
            In other words,  $\bys \in \vndistinct$ is a vector such that
            \begin{align*}
                \rank{\bysn{\iconstant}}{\bys} = \iconstant, \text{ for any } \iconstant \in \tonumber{\n}.
            \end{align*}
            Hence, we have 
            \begin{align*}
                \rank{\bxsn{\iconstant}}{\bxs} + \rank{\bysn{\iconstant}}{\bys} = \n + 1, \text{ for any } \iconstant \in \tonumber{\n}.
            \end{align*}
            Thus, $\bxs,\bys$ are imputations of $\bx, \by$ for $\bw = \tonumber{\n}$,
            and $(\bxs,\bys) \in \mythirdset{\bw}{\n}$.

            Next, let us assume (after relabeling) $\byln{1} < \ldots < \byln{\n}$.
            Then according to the definition of Spearman's footrule, we have
            \begin{align*}
                \sfd{\bx}{\by} &= \sum_{\iconstant=1}^{\n} |\rank{\bxn{\iconstant}}{\bx} -  \rank{\byn{\iconstant}}{\by}| \\
                & =\sum_{\iconstant=1}^{\n} |\rank{\bxn{\iconstant}}{\bx} -  \iconstant|.
            \end{align*}
            Since  $\bysn{1} < \ldots < \bysn{\n}$, we have 
            \begin{align*}
                \sfd{\bxs}{\bys} &= \sum_{\iconstant=1}^{\n} |\rank{\bxsn{\iconstant}}{\bxs} -  \rank{\bysn{\iconstant}}{\bys}| \\
                & =\sum_{\iconstant=1}^{\n} |\rank{\bxsn{\iconstant}}{\bxs} -  \iconstant|.
            \end{align*}
            Notice that $(\rank{\bxn{\iconstant}}{\bx})_{\iconstant \in \tonumber{\n}}$
            and $(\rank{\bxsn{\iconstant}}{\bxs})_{\iconstant \in \tonumber{\n}}$ are 
            both permutations of $\tonumber{\n}$. Then 
            $(\rank{\bxsn{\iconstant}}{\bxs})_{\iconstant \in \tonumber{\n}}$ is a permutation
            of $(\rank{\bxn{\iconstant}}{\bx})_{\iconstant \in \tonumber{\n}}$.
            Since $\bxsn{1} > \ldots > \bxsn{\n}$, then we have
            $\rank{\bxsn{1}}{\bxs} > \ldots > \rank{\bxsn{\n}}{\bxs}$.
            Hence, according to Lemma~\ref{supp:lemma:2.14:1}, we have
            \begin{align*}
                \sfd{\bxs}{\bys} \ge \sfd{\bx}{\by}.
            \end{align*}
            This proves the statement $\statement$ is true when $\bw = \tonumber{\n}$.

            In the following, we prove the statement $\statement$ 
            is true when $\bw \neq \emptyset$ 
            and $\bw \neq \tonumber{\n}$.

            For any given $\n \in \mathbb{N}$, let $\bpn{\kk}$ be the statement $\statement$
            when $|\bw| = \kk$.
            We prove $\bpn{\kk}$ is true for any $\kk \in \{1,\ldots,\n-1\}$
            by induction on $\kk$.

            \emph{Base Case: } We prove $\bpn{1}$ is true.
            When $|\bw| = 1$, let us denote $\bw = \uu$.
            Suppose $\bxln{1}, \byln{1} \in \vndistinct$ and 
            $\bxln{2}, \byln{2} \in \vndistinct$ are imputations 
            of $\bx$ and $\by$ for the index $\uu$ such that
            \begin{align*}
                &\rank{\bxlnn{1}{\uu}}{\bxln{1}} = 1,~\text{and}~ \rank{\bylnn{1}{\uu}}{\byln{1}} = \n,\\
                &\rank{\bxlnn{2}{\uu}}{\bxln{2}} = \n,~\text{and}~ \rank{\bylnn{2}{\uu}}{\byln{2}} = 1.
            \end{align*}
            Then, according to Proposition~\ref{supp:proposition:2.18}, we have    	
            $\sfd{\bx}{\by} \le \max \{ \sfd{\bxln{1}}{\byln{1}}, \sfd{\bxln{2}}{\byln{2}}\}$.

            Since $\rank{\bxlnn{1}{\uu}}{\bxln{1}} = 1$, we have $\bxlnn{1}{\uu} < \bxlnn{1}{\iconstant}$
            for any $\iconstant \in \tonumber{\n} \setminus \{\uu\}$. 
            Notice that $\bw = \{\uu\}$. Then, we have $\bxln{1} \in \myfirstset{\bw}{\tonumber{\n}}{\n}$.
            Since $\rank{\bxlnn{1}{\uu}}{\bxln{1}} + \rank{\bylnn{1}{\uu}}{\byln{1}} = \n + 1$,
            we have $(\bxln{1}, \byln{1}) \in \mythirdset{\bw}{\n}$.

            Similarly, since $\rank{\bxlnn{2}{\uu}}{\bxln{2}} = \n$, we have $\bxlnn{2}{\uu} > \bxlnn{2}{\iconstant}$
            for any $\iconstant \in \tonumber{\n} \setminus \{\uu\}$. 
            Notice that $\bw = \{\uu\}$. Then, we have $\bxln{2} \in \myfirstset{\bw}{\tonumber{\n}}{\n}$.
            Since $\rank{\bxlnn{2}{\uu}}{\bxln{2}} + \rank{\bylnn{2}{\uu}}{\byln{2}} = \n + 1$,
            we have $(\bxln{2}, \byln{2}) \in \mythirdset{\bw}{\n}$.

            Notice that $\sfd{\bx}{\by} \le \max \{ \sfd{\bxln{1}}{\byln{1}}, \sfd{\bxln{2}}{\byln{2}}\}$,
            $(\bxln{1}, \byln{1}) \in \mythirdset{\bw}{\n}$ and $(\bxln{2}, \byln{2}) \in \mythirdset{\bw}{\n}$.
            Hence, we have shown $\bpn{1}$ is true.

            \emph{Induction Step:} We show the implication $\bpn{1}, \bpn{\kk} \Rightarrow \bpn{\kk+1}$
            for any $\kk \in \{1,\ldots, \n-2\}$.

            Denote $\m = \kk+1$. 
            Without loss of generality, let us assume 
            (after relabeling) $\bw = \{\n-\m+1, \cdots, \n\}$.
            Define $\bwp = \{\n-\m+1,\cdots,\n-1\}$. 

            Notice that $|\bwp| = |\bw| - 1 = \kk$. 
            Then, since $\bpn{\kk}$ is true, there exist 
            imputations $(\bxln{1},\byln{1}) \in \mythirdset{\bwp}{\n}$ 
            of $\bx$ and $\by$ for indices $\bwp$ 
            such that $\sfd{\bx}{\by} \le \sfd{\bxln{1}}{\byln{1}}$.

            Next, since $\bpn{1}$ is true, then there exist imputations $(\bxln{2}, \byln{2}) \in \mythirdset{\{\n\}}{\n} $ of $\bxln{1}$ and $\byln{1}$ for the index $\n$ 
            such that $\sfd{\bxln{1}}{\byln{1}} \le \sfd{\bxln{2}}{\byln{2}}$.

            Hence, we have $\sfd{\bx}{\by} \le \sfd{\bxln{2}}{\byln{2}}$.
            According to Lemma~\ref{supp:theorem:2.20:lemma:1}, 
            $\bxln{2},\byln{2}$ are imputations of $\bx$ and $\by$ for indices $\bw$
            such that
            $(\bxln{2},\byln{2}) \in \mythirdset{\bw}{\n}$.
            Hence, we have shown $\bpn{\kk+1}$ is true.
            This completes our proof for the statement $\statement$.

            Next, we prove Theorem~\ref{supp:theorem:2.20} is true using the statement $\statement$. 
            let us denote 
            \begin{align*}
                \bs = \{\sfd{\bvln{1}}{\bvln{2}}:  (\bvln{1}, \bvln{2}) \in \mythirdset{\bw}{\n} \text{ are imputations of } \bx, \by \text{ for indices } \bw \}.
            \end{align*}
            Then, the carnality of $\bs$ is finite and there exist  
            imputations $(\bxs, \bys) \in \mythirdset{\bw}{\n}$ of $\bx, \by$ for indices $\bw$ such that $\sfd{\bxs}{\bys} = \max \bs$.

            We now show that $\sfd{\bx}{\by} \le \sfd{\bxs}{\bys}$.
            According to statement $\statement$,  there exist imputations
            $(\bxln{1}, \byln{1}) \in \mythirdset{\bw}{\n}$ of $\bx, \by$ for indices $\bw$
            such that $\sfd{\bx}{\by} \le \sfd{\bxln{1}}{\byln{1}}$. 
            According to the definition of $\bs$, we have
            $\sfd{\bxln{1}}{\byln{1}} \in \bs$.
            Hence, we have $\sfd{\bxln{1}}{\byln{1}} \le  \max \bs = \sfd{\bxs}{\bys}$.
            Thus, we have $\sfd{\bx}{\by} \le \sfd{\bxs}{\bys}$.

            Next, we show that $\sfd{\bxp}{\byp} \le \sfd{\bxs}{\bys}$.
            According to the statement $\statement$, there exist imputation
            $(\bxln{2}, \byln{2}) \in \mythirdset{\bw}{\n}$ of $\bxp, \byp$ for indices $\bw$
            such that $\sfd{\bxp}{\byp} \le \sfd{\bxln{2}}{\byln{2}}$.
            Since $\bxp$, $\byp$ are imputations of $\bx$ and $\by$ for indices $\bw$,
            then $\bxln{2}$, $\byln{2}$ are imputations of $\bx, \by$ for indices $\bw$.
            Then, according to the definition of $\bs$, we have $\sfd{\bxln{2}}{\byln{2}} \in \bs$.
            Hence, we have $\sfd{\bxln{2}}{\byln{2}} \le  \max \bs = \sfd{\bxs}{\bys}$.
            Thus, we have $\sfd{\bxp}{\byp} \le \sfd{\bxs}{\bys}$.
            This completes our proof.
        \end{proof}

        \subsection{Proof of Theorem 2.21}

        Now, we are ready to prove Theorem~2.21.

        \begin{theorem} \label{supp:theorem:2.21}
            Suppose $\bx, \by \in \vndistinct$, and let $\bu$, $\bv$, $\bw \subset \tonumber{\n}$ be pairwise disjoint subsets. 
            Then, there exist $(\bxs,\bys) \in (\mysecondset{\by}{\bu}{\tonumber{\n} \setminus \bw}{\n},  
            \mysecondset{\bx}{\bv}{\tonumber{\n} \setminus \bw}{\n}) \cap \mythirdset{\bw}{\n}$ of imputations $\bx$ and $\by$ for indices $\bu\cup\bw$ and $\bv \cup \bw$, respectively, such that $\sfd{\bx}{\by} \le \sfd{\bxs}{\bys}$.  Furthermore, 
            consider any imputation $\bxp,\byp \in \vndistinct$ of $\bx, \by$ for indices $\bu\cup\bw$ and $\bv \cup \bw$, respectively, we have 
            $\sfd{\bxp}{\byp} \le \sfd{\bxs}{\bys}$.
        \end{theorem}

        \begin{proof}
            To prove Theorem~\ref{supp:theorem:2.21}, we first prove the following
            statement
            \begin{itemize}
                \item[$\statement$]: Suppose $\bx, \by \in \vndistinct$, and let $\bu$, $\bv$, $\bw \subset \tonumber{\n}$ be pairwise disjoint subsets. 
                    Then, there exist $(\bxs,\bys) \in (\mysecondset{\by}{\bu}{\tonumber{\n} \setminus \bw}{\n},  
                    \mysecondset{\bx}{\bv}{\tonumber{\n} \setminus \bw}{\n}) \cap \mythirdset{\bw}{\n}$ of imputations $\bx$ and $\by$ for indices $\bu\cup\bw$ and $\bv \cup \bw$, respectively, such that $\sfd{\bx}{\by} \le \sfd{\bxs}{\bys}$. 
            \end{itemize}
            Then we will prove Theorem~\ref{supp:theorem:2.21} using the statement $\statement$. 

            According to Theorem~\ref{supp:theorem:2.17}, there exist imputations 
            \begin{align*}
                (\bxln{1},\byln{1}) \in (\mysecondset{\by}{\bu}{\tonumber{\n}}{\n}, \mysecondset{\bx}{\bv}{\tonumber{\n}}{\n})
            \end{align*}
            of $\bx, \by$ for indices $\bu$ and $\bv$, respectively, such that $\sfd{\bx}{\by} \le \sfd{\bxln{1}}{\byln{1}}$.

            Further, according to Theorem~\ref{supp:theorem:2.20}, there exist imputations $(\bxln{2},\byln{2}) \in \mythirdset{\bw}{\n}$ of $\bxln{1}$ and $\byln{1}$ for indices $\bw$ such that $\sfd{\bxln{1}}{\byln{1}} \le \sfd{\bxln{2}}{\byln{2}}$.
            Hence, we have $\sfd{\bx}{\by} \le \sfd{\bxln{2}}{\byln{2}}$.

            Then, in order to prove our result, it is sufficient to show that 	
            $\bxln{2},\byln{2}$ are imputations of $\bx$ and $\by$ for indices $\bu\cup\bw$ 
            and $\bv \cup \bw$, respectively, and are such that 
            $(\bxln{2},\byln{2}) \in (\mysecondset{\by}{\bu}{\tonumber{\n} \setminus \bw}{\n},  
            \mysecondset{\bx}{\bv}{\tonumber{\n} \setminus \bw}{\n}) \cap \mythirdset{\bw}{\n}$.

            First, we show that $\bxln{2},\byln{2}$ are imputations of $\bx$ and $\by$ for indices $\bu\cup\bw$ 
            and $\bv \cup \bw$, respectively. Since $\bxln{1},\byln{1}$ are imputations 
            of $\bx, \by$ for indices $\bu$ and $\bv$, respectively, we have
            \begin{align*}
                &\bxlnn{1}{\iconstant} = \bxn{\iconstant}, \text{ for any } \tonumber{\n} \setminus \bu,
                \text{ and } \bylnn{1}{\iconstant} = \byn{\iconstant}, \text{ for any } \tonumber{\n} \setminus \bv\\
                \Rightarrow &\bxlnn{1}{\iconstant} = \bxn{\iconstant}, \text{ for any } \tonumber{\n} \setminus (\bu \cup \bw),
                \text{ and } \bylnn{1}{\iconstant} = \byn{\iconstant}, \text{ for any } \tonumber{\n} \setminus (\bv \cup \bw).
            \end{align*}
            Further, since $\bxln{2},\byln{2}$ are imputations of $\bxln{1}$ and $\byln{1}$ for indices $\bw$,
            we have
            \begin{align*}
                &\bxlnn{2}{\iconstant} = \bxlnn{1}{\iconstant}, \text{ for any } \tonumber{\n} \setminus \bw,
                \text{ and } \bylnn{2}{\iconstant} = \bylnn{1}{\iconstant}, \text{ for any } \tonumber{\n} \setminus \bw\\
                \Rightarrow &\bxlnn{2}{\iconstant} = \bxlnn{1}{\iconstant}, \text{ for any } \tonumber{\n} \setminus (\bu\cup\bw),
                \text{ and } \bylnn{2}{\iconstant} = \bylnn{1}{\iconstant}, \text{ for any } \tonumber{\n} \setminus (\bv\cup\bw).
            \end{align*}
            Hence, we have 
            \begin{align*}
                \bxlnn{2}{\iconstant} = \bxn{\iconstant}, \text{ for any } \tonumber{\n} \setminus (\bu \cup \bw),
                \text{ and } \bylnn{2}{\iconstant} = \byn{\iconstant}, \text{ for any } \tonumber{\n} \setminus (\bv \cup \bw).
            \end{align*} 
            In other words, $\bxln{2},\byln{2}$ are imputations of $\bx$ and $\by$ for indices $\bu\cup\bw$ 
            and $\bv \cup \bw$, respectively.

            Next, we show $(\bxln{2},\byln{2}) \in (\mysecondset{\by}{\bu}{\tonumber{\n} \setminus \bw}{\n},  
            \mysecondset{\bx}{\bv}{\tonumber{\n} \setminus \bw}{\n}) \cap \mythirdset{\bw}{\n}$.
            Since $(\bxln{2},\byln{2}) \in \mythirdset{\bw}{\n}$, we only need to show that 
            $(\bxln{2},\byln{2}) \in (\mysecondset{\by}{\bu}{\tonumber{\n} \setminus \bw}{\n},  
            \mysecondset{\bx}{\bv}{\tonumber{\n} \setminus \bw}{\n})$.

            We now show $\bxln{2} \in \mysecondset{\by}{\bu}{\tonumber{\n} \setminus \bw}{\n}$.
            Since $\bxln{1} \in \mysecondset{\by}{\bu}{\tonumber{\n}}{\n}$, then according
            to the definition of $\mysecondset{\by}{\bu}{\tonumber{\n}}{\n}$, we have
            \begin{align*}
                &\bxlnn{1}{\iconstant} > \max_{\jconstant \in \tonumber{\n} \setminus \bu} \bxlnn{1}{\jconstant}, 
                \text{ or }
                \bxlnn{1}{\iconstant} < \min_{\jconstant \in \tonumber{\n} \setminus \bu} \bxlnn{1}{\jconstant}, 
                \text{ for any } \iconstant \in \bu \\
                \Rightarrow &\bxlnn{1}{\iconstant} > \max_{\jconstant \in \tonumber{\n} \setminus (\bu \cup \bw)} \bxlnn{1}{\jconstant}, 
                \text{ or }
                \bxlnn{1}{\iconstant} < \min_{\jconstant \in \tonumber{\n} \setminus (\bu \cup \bw) } \bxlnn{1}{\jconstant}, 
                \text{ for any } \iconstant \in \bu.
            \end{align*}
            Since $\bxln{2}$ is an imputation of $\bxln{1}$ for $\bw$, we have $\bxlnn{2}{\iconstant} = \bxlnn{1}{\iconstant}$ for any $\iconstant \in \tonumber{\n} \setminus \bw$.
            Since $\tonumber{\n} \setminus (\bu \cup \bw) \subset  \tonumber{\n} \setminus \bw$,
            we have $\max_{\jconstant \in \tonumber{\n} \setminus (\bu \cup \bw)} \bxlnn{1}{\jconstant}
            = \max_{\jconstant \in \tonumber{\n} \setminus (\bu \cup \bw)} \bxlnn{2}{\jconstant}$
            and
            $\min_{\jconstant \in \tonumber{\n} \setminus (\bu \cup \bw)} \bxlnn{1}{\jconstant}
            = \min_{\jconstant \in \tonumber{\n} \setminus (\bu \cup \bw)} \bxlnn{2}{\jconstant}$.
            Meanwhile, since $\bu \in \tonumber{\n} \setminus \bw$, we have
            $\bxlnn{2}{\iconstant} = \bxlnn{1}{\iconstant}$ for any $\iconstant \in \bu$.
            Hence, we have
            \begin{align*}
                \bxlnn{2}{\iconstant} > \max_{\jconstant \in \tonumber{\n} \setminus (\bu \cup \bw)} \bxlnn{2}{\jconstant}, 
                \text{ or }
                \bxlnn{2}{\iconstant} < \min_{\jconstant \in \tonumber{\n} \setminus (\bu \cup \bw) } \bxlnn{2}{\jconstant}, 
                \text{ for any } \iconstant \in \bu.
            \end{align*}
            Thus, we have shown $\bxln{2} \in \myfirstset{\bu}{\tonumber{\n} \setminus \bw}{\n}$.

            Next, since $\bxln{1} \in \mysecondset{\by}{\bu}{\tonumber{\n}}{\n}$, then according
            to the definition of $\mysecondset{\by}{\bu}{\tonumber{\n}}{\n}$, we have
            \begin{align*}
                \bxlnn{1}{\iconstant} > \bxlnn{1}{\jconstant}, \text{ if } \byn{\iconstant} < \byn{\jconstant}, 
                \text{ for any } \iconstant, \jconstant \in \bu. 
            \end{align*}
            Since $\bxlnn{2}{\iconstant} = \bxlnn{1}{\iconstant}$ for any $\iconstant \in \bu$,
            then we have 
            \begin{align*}
                \bxlnn{2}{\iconstant} > \bxlnn{2}{\jconstant}, \text{ if } \byn{\iconstant} < \byn{\jconstant}, 
                \text{ for any } \iconstant, \jconstant \in \bu. 
            \end{align*}
            Hence, we have $\bxln{2} \in \mysecondset{\by}{\bu}{\tonumber{\n} \setminus \bw}{\n}$.

            Similarly, we can show that 
            $\byln{2} \in \mysecondset{\by}{\bv}{\tonumber{\n} \setminus \bw}{\n}$.
            Since $\byln{1} \in \mysecondset{\bx}{\bv}{\tonumber{\n}}{\n}$, then according
            to the definition of $\mysecondset{\bx}{\bv}{\tonumber{\n}}{\n}$, we have
            \begin{align*}
                &\bylnn{1}{\iconstant} > \max_{\jconstant \in \tonumber{\n} \setminus \bv} \bylnn{1}{\jconstant}, 
                \text{ or }
                \bylnn{1}{\iconstant} < \min_{\jconstant \in \tonumber{\n} \setminus \bv} \bylnn{1}{\jconstant}, 
                \text{ for any } \iconstant \in \bv \\
                \Rightarrow &\bylnn{1}{\iconstant} > \max_{\jconstant \in \tonumber{\n} \setminus (\bv \cup \bw)} \bxlnn{1}{\jconstant}, 
                \text{ or }
                \bylnn{1}{\iconstant} < \min_{\jconstant \in \tonumber{\n} \setminus (\bv \cup \bw) } \bylnn{1}{\jconstant}, 
                \text{ for any } \iconstant \in \bv.
            \end{align*}
            Since $\byln{2}$ is an imputation of $\byln{1}$ for $\bw$, we have $\bylnn{2}{\iconstant} = \bylnn{1}{\iconstant}$ for any $\iconstant \in \tonumber{\n} \setminus \bw$.
            Since $\tonumber{\n} \setminus (\bv \cup \bw) \subset  \tonumber{\n} \setminus \bw$,
            we have $\max_{\jconstant \in \tonumber{\n} \setminus (\bv \cup \bw)} \bylnn{1}{\jconstant}
            = \max_{\jconstant \in \tonumber{\n} \setminus (\bv \cup \bw)} \bylnn{2}{\jconstant}$
            and
            $\min_{\jconstant \in \tonumber{\n} \setminus (\bv \cup \bw)} \bylnn{1}{\jconstant}
            = \min_{\jconstant \in \tonumber{\n} \setminus (\bv \cup \bw)} \bylnn{2}{\jconstant}$.
            Meanwhile, since $\bv \in \tonumber{\n} \setminus \bw$, we have
            $\bylnn{2}{\iconstant} = \bylnn{1}{\iconstant}$ for any $\iconstant \in \bv$.
            Hence, we have
            \begin{align*}
                \bylnn{2}{\iconstant} > \max_{\jconstant \in \tonumber{\n} \setminus (\bv \cup \bw)} \bylnn{2}{\jconstant}, 
                \text{ or }
                \bylnn{2}{\iconstant} < \min_{\jconstant \in \tonumber{\n} \setminus (\bv \cup \bw) } \bylnn{2}{\jconstant}, 
                \text{ for any } \iconstant \in \bv.
            \end{align*}
            Thus, we have shown $\byln{2} \in \myfirstset{\bv}{\tonumber{\n} \setminus \bw}{\n}$.

            Next, since $\byln{1} \in \mysecondset{\bx}{\bv}{\tonumber{\n}}{\n}$, then according
            to the definition of $\mysecondset{\bx}{\bv}{\tonumber{\n}}{\n}$, we have
            \begin{align*}
                \bylnn{1}{\iconstant} > \bylnn{1}{\jconstant}, \text{ if } \bxn{\iconstant} < \bxn{\jconstant}, 
                \text{ for any } \iconstant, \jconstant \in \bv. 
            \end{align*}
            Since $\bylnn{2}{\iconstant} = \bylnn{1}{\iconstant}$ for any $\iconstant \in \bv$,
            then we have 
            \begin{align*}
                \bylnn{2}{\iconstant} > \bylnn{2}{\jconstant}, \text{ if } \bxn{\iconstant} < \bxn{\jconstant}, 
                \text{ for any } \iconstant, \jconstant \in \bv. 
            \end{align*}
            Hence, we have $\byln{2} \in \mysecondset{\bx}{\bv}{\tonumber{\n} \setminus \bw}{\n}$.
            This completes our proof for statement $\statement$.

            Next, we prove Theorem~\ref{supp:theorem:2.21} is true using statement $\statement$. 
            Let us denote 
            \begin{align*}
                \bs = \{\sfd{Z_1}{Z_2}: (Z_1,Z_2) \in (\mysecondset{\by}{\bu}{\tonumber{\n} \setminus \bw}{\n}, \mysecondset{\bx}{\bv}{\tonumber{\n} \setminus \bw}{\n}) \cap \mythirdset{\bw}{\n}\\
                \text{ are imputations of } \bx, \by \text{ for indices } \bu\cup\bw \text{ and } \bv \cup \bw, \text{respectively} \}.
            \end{align*}
            The carnality of $\bs$ is finite and there exist  
            imputations 
            \begin{align*}
                (\bxs, \bys) \in (\mysecondset{\by}{\bu}{\tonumber{\n} \setminus \bw}{\n}, \mysecondset{\bx}{\bv}{\tonumber{\n} \setminus \bw}{\n}) \cap \mythirdset{\bw}{\n}
            \end{align*}
            of $\bx, \by$ for indices $\bu\cup\bw$ and $\bv \cup \bw$, respectively, such
            that $ \max \bs = \sfd{\bxs}{\bys}$.

            We now show that $\sfd{\bx}{\by} \le \sfd{\bxs}{\bys}$.
            According to the statement $\statement$, there exist imputations
            \begin{align*}
                (\bxln{1}, \byln{1}) \in (\mysecondset{\by}{\bu}{\tonumber{\n} \setminus \bw}{\n}, \mysecondset{\bx}{\bv}{\tonumber{\n} \setminus \bw}{\n}) \cap \mythirdset{\bw}{\n}
            \end{align*}
            of $\bx, \by$ for indices $\bu\cup\bw$ and $\bv \cup \bw$, respectively
            such that $\sfd{\bx}{\by} \le \sfd{\bxln{1}}{\byln{1}}$. 
            According to the definition of $\bs$, we have
            $\sfd{\bxln{1}}{\byln{1}} \in \bs$.
            Hence, we have $\sfd{\bxln{1}}{\byln{1}} \le  \max \bs = \sfd{\bxs}{\bys}$.
            Thus, we have $\sfd{\bx}{\by} \le \sfd{\bxs}{\bys}$.

            Next, we show that $\sfd{\bxp}{\byp} \le \sfd{\bxs}{\bys}$.
            According to statement $\statement$, there exist imputations
            \begin{align*}
                (\bxln{2}, \byln{2}) \in (\mysecondset{\byp}{\bu}{\tonumber{\n} \setminus \bw}{\n}, \mysecondset{\bxp}{\bv}{\tonumber{\n} \setminus \bw}{\n}) \cap \mythirdset{\bw}{\n}
            \end{align*}
            of $\bxp, \byp$ for indices $\bu\cup\bw$ and $\bv \cup \bw$, respectively
            such that
            \begin{align} \label{supp:theorem:2.21:eqn:1}
                \sfd{\bxp}{\byp} \le \sfd{\bxln{2}}{\byln{2}}.
            \end{align}

            Below, we show that $\sfd{\bxln{2}}{\byln{2}} \in \bs$. In other words,
            we show that $\bxln{2}$, $\byln{2}$ are 
            imputations of $\bx$ and $\by$ for indices $\bu \cup \bw$,
            and $\bv \cup \bw$, respectively such that 
            $(\bxln{2}, \byln{2}) \in (\mysecondset{\by}{\bu}{\tonumber{\n} \setminus \bw}{\n}, \mysecondset{\bx}{\bv}{\tonumber{\n} \setminus \bw}{\n}) \cap \mythirdset{\bw}{\n}$.

            Since $\bxp$, $\byp$ are imputations of $\bx$ and $\by$ for indices $\bu \cup \bw$,
            and $\bv \cup \bw$, respectively, then $\bxln{2}$, $\byln{2}$ are 
            also imputations of $\bx$ and $\by$ for indices $\bu \cup \bw$,
            and $\bv \cup \bw$, respectively.

            Since $\bxln{2} \in \mysecondset{\byp}{\bu}{\tonumber{\n} \setminus \bw}{\n}$,
            then according to the definition of $\mysecondset{\byp}{\bu}{\tonumber{\n} \setminus \bw}{\n}$,
            we have
            \begin{align*}
                \bxlnn{2}{\iconstant} > \bxlnn{2}{\jconstant}, \text{ if } \bypn{\iconstant} < \bypn{\jconstant}, 
                \text{ for any } \iconstant, \jconstant \in \bu. 
            \end{align*}
            Since $\byp$ is an imputation of $\by$ for indices $\bv \cup \bw$, we have
            \begin{align*}
                \bypn{\iconstant} = \byn{\iconstant}, \text{ for any } \iconstant \in \tonumber{\n} \setminus (\bv \cup \bw).
            \end{align*}
            Then, since $\bu \subset \tonumber{\n} \setminus (\bv \cup \bw)$, we have 
            \begin{align*}
                \bypn{\iconstant} = \byn{\iconstant}, \text{ for any } \iconstant \in \bu.
            \end{align*}
            Hence, we have 
            \begin{align*}
                \bxlnn{2}{\iconstant} > \bxlnn{2}{\jconstant}, \text{ if } \byn{\iconstant} < \byn{\jconstant}, 
                \text{ for any } \iconstant, \jconstant \in \bu. 
            \end{align*}
            Since $\bxln{2} \in \mysecondset{\byp}{\bu}{\tonumber{\n} \setminus \bw}{\n}$,
            then according to the definition of $\mysecondset{\byp}{\bu}{\tonumber{\n} \setminus \bw}{\n}$,
            we have $\bxln{2} \in \myfirstset{\bu}{\tonumber{\n} \setminus \bw}{\n}$.
            Hence, we have
            $\bxln{2} \in \mysecondset{\by}{\bu}{\tonumber{\n} \setminus \bw}{\n}$.

            Similarly, we can show $\byln{2} \in \mysecondset{\bx}{\bv}{\tonumber{\n} \setminus \bw}{\n}$.
            Since $\byln{2} \in \mysecondset{\bxp}{\bv}{\tonumber{\n} \setminus \bw}{\n}$,
            then according to the definition of $\mysecondset{\bxp}{\bv}{\tonumber{\n} \setminus \bw}{\n}$,
            we have
            \begin{align*}
                \bylnn{2}{\iconstant} > \bylnn{2}{\jconstant}, \text{ if } \bxpn{\iconstant} < \bxpn{\jconstant}, 
                \text{ for any } \iconstant, \jconstant \in \bv. 
            \end{align*}
            Since $\bxp$ is an imputation of $\bx$ for indices $\bu \cup \bw$, we have
            \begin{align*}
                \bxpn{\iconstant} = \bxn{\iconstant}, \text{ for any } \iconstant \in \tonumber{\n} \setminus (\bu \cup \bw).
            \end{align*}
            Then, since $\bv \subset \subset \tonumber{\n} \setminus (\bu \cup \bw)$, we have 
            \begin{align*}
                \bxpn{\iconstant} = \bxn{\iconstant}, \text{ for any } \iconstant \in \bv.
            \end{align*}
            Hence, we have 
            \begin{align*}
                \bylnn{2}{\iconstant} > \bylnn{2}{\jconstant}, \text{ if } \bxn{\iconstant} < \bxn{\jconstant}, 
                \text{ for any } \iconstant, \jconstant \in \bv. 
            \end{align*}
            Since $\byln{2} \in \mysecondset{\bxp}{\bv}{\tonumber{\n} \setminus \bw}{\n}$,
            then according to the definition of $\mysecondset{\bxp}{\bv}{\tonumber{\n} \setminus \bw}{\n}$,
            we have $\byln{2} \in \myfirstset{\bv}{\tonumber{\n} \setminus \bw}{\n}$.
            Hence, we have
            $\byln{2} \in \mysecondset{\bx}{\bv}{\tonumber{\n} \setminus \bw}{\n}$.

            Next, since $(\bxln{2}, \byln{2}) \in  \mythirdset{\bw}{\n}$,
            we have $(\bxln{2}, \byln{2}) \in (\mysecondset{\by}{\bu}{\tonumber{\n} \setminus \bw}{\n}, \mysecondset{\bx}{\bv}{\tonumber{\n} \setminus \bw}{\n}) \cap \mythirdset{\bw}{\n}$
            Then, according to the definition of $\bs$, we have $\sfd{\bxln{2}}{\byln{2}} \in \bs$.
            Hence, we have $\sfd{\bxln{2}}{\byln{2}} \le  \max \bs = \sfd{\bxs}{\bys}$.
            Thus, we have
            \begin{align*}
                \sfd{\bxp}{\byp} \le^{\eqref{supp:theorem:2.21:eqn:1}} \sfd{\bxln{2}}{\byln{2}}  \le \sfd{\bxs}{\bys}.
            \end{align*}
            This completes our proof.

        \end{proof}

        \section{Proof of bounds of $p$-values} \label{appD}

        \begin{definition} 
            Suppose $\bx, \by \in \vndistinct$, and $F_{n}$ is the cumulative distribution function of a normal distribution with mean, variance equal to $n^2/3$ and $2n^3/45$, respectively. Then when $\n$
            is sufficiently large, the $p$-value
            of Spearman's footrule is defined as
            \begin{align} \label{pvalue}
                p(D(X,Y)) = 2{\min}\{F_{n}(D(X,Y)), 1 - F_{n}(D(X,Y))\}.
            \end{align}
        \end{definition}

        \begin{proposition} \label{supp:proposition:3.1}
            Suppose $\bx,\by \in \vndistinct$ are partially observed. Assume $\n$ is sufficiently large. Let $D_{\min}$ and $D_{\max}$ be the minimum and maximum possible values of Spearman's footrule between $\bx$ and $\by$. Denote $p_1 = p(D_{\min})$ and $p_2 = p(D_{\max})$, where $p(\cdot)$ is defined in \eqref{pvalue}. Define $p_{\min} = \min \{p_1,p_2\}$, and
            \begin{align*}
                p_{\max} = \left\{ \begin{array}{ll}
                    \max \{p_1,p_2\}, & \text{ if } (D_{\min} - n^2/3)(D_{\max} - n^2/3) \ge 0, \\
                    0, & \text{otherwise}.
                \end{array}\right.
            \end{align*}
            Then, the $p$-value of $\sfd{\bx}{\by}$ is bounded such that $p(D(X,Y)) \in [p_{\min}, p_{\max}]$.
        \end{proposition} 

        \begin{proof}
            According to the definition $p(\cdot)$ of $p$-values in \eqref{pvalue},
            we have  
            \begin{align*}
                &p(D(X,Y)) = 2\min\{F_{n}(D(X,Y)), 1 - F_{n}(D(X,Y))\},\\
                &p_1 = 2\min\{F_{n}(D_{\min}), 1 - F_{n}(D_{\min})\},\\
                \text{and } &p_2 = 2\min\{F_{n}(D_{\max}), 1 - F_{n}(D_{\max})\},
            \end{align*}
            separately.

            In the following, we prove $p(D(X,Y)) \in [p_{\min}, p_{\max}]$ 
            when $F_{n}(D(X,Y)) \le 1/2$ and $F_{n}(D(X,Y)) > 1/2$ separately.

            Suppose $F_{\n}(D(X,Y)) \le 1/2$. 
            We first show that $p_{\min} \le p(D(X,Y)).$

            Since $F_{\n}(D(X,Y)) \le 1/2$, then we have $1 - F_{n}(D(X,Y)) \ge 1/2$. Hence,
            \begin{align} \label{supp:proposition:3.1:eqn:0}
                p(D(X,Y)) = 2\min\{F_{n}(D(X,Y)), 1 - F_{n}(D(X,Y))\} = 2F_{n}(D(X,Y)).
            \end{align}
            Then, since $D_{\min} \le D(X,Y)$, we have
            \begin{align*}
                &F_{n}(D_{\min}) \le F_{n}(D(X,Y)) \le 1/2 \\
                \Rightarrow & 1 - F_{n}(D_{\min}) \ge 1/2 \\
                \Rightarrow & p_1 = 2\min\{F_{n}(D_{\min}), 1 - F_{n}(D_{\min})\} = 2F_{n}(D_{\min}).
            \end{align*}
            Since $F_{n}(D_{\min}) \le F_{n}(D(X,Y))$, we then have
            \begin{align*}
                &p_1 = 2F_{n}(D_{\min}) \le 2 F_{n}(D(X,Y)) = p(D(X,Y))\\
                \Rightarrow& 		p_{\min} = \min \{p_1,p_2\} \le p(D(X,Y)).
            \end{align*}

            Now, we show $p(D(X,Y)) \le p_{\max}.$

            If $(D_{\min} - n^2/3)(D_{\max} - n^2/3) < 0$, then we have $p_{\max} = 1$. According
            to the definition of $p$-values in \eqref{pvalue}, we have 
            $p(D(X,Y)) \le 1 \le p_{\max} = 1$.

            However, if $(D_{\min} - n^2/3)(D_{\max} - n^2/3) \ge 0$, then  
            since $F_{n}$ is the cumulative distribution function of a normal 
            distribution with mean equal to $n^2/3$, we have	
            \begin{align*}
                &F_{n}(D_{\min}) \le 1/2 \Rightarrow D_{\min} \le n^2/3\\
                \Rightarrow & (D_{\min} - n^2/3) \le 0.
            \end{align*}
            Since $(D_{\min} - n^2/3)(D_{\max} - n^2/3) \ge 0$, then we have $D_{\max} - n^2/3 \le 0$.
            Hence,
            \begin{align}
                &D_{\min} \le D(X,Y) \le D_{\max} \le n^2/3  \nonumber \\
                \Rightarrow &F_{n}(D_{\min}) \le F_{n}(D(X,Y)) \le  F_{n}(D_{\max}) \nonumber\\
                \Rightarrow^{\eqref{supp:proposition:3.1:eqn:0}}&  F_{n}(D_{\min}) \le p(D(X,Y))/2 \le  F_{n}(D_{\max}). \label{supp:proposition:3.1:eqn:1}
            \end{align}

            Notice that since $D_{\max} - n^2/3 \le 0$, we also have
            \begin{align*}
                &F_{n}(D_{\max}) \le 1/2 \\
                \Rightarrow & 1 - F_{n}(D_{\max}) \ge 1/2 \\
                \Rightarrow & p_2 = 2\min\{F_{n}(D_{\max}), 1 - F_{n}(D_{\max})\} = 2F_{n}(D_{\max}).
            \end{align*}
            Combining this result with \eqref{supp:proposition:3.1:eqn:1}, we have
            \begin{align*}
                p(D(X,Y)) \le p_2 \le  \max \{p_1, p_2\} = p_{\max}.
            \end{align*}
            Hence, we have shown $p(D(X,Y)) \in [p_{\min}, p_{\max}]$ 
            when $F_{n}(D(X,Y)) \le 1/2$.

            Similarly, we can show $p(D(X,Y)) \in [p_{\min}, p_{\max}]$ 
            when $F_{n}(D(X,Y)) > 1/2$.
            Suppose $F_{\n}(D(X,Y)) > 1/2$. 
            We first show that $p_{\min} \le p(D(X,Y)).$

            Since $F_{\n}(D(X,Y)) > 1/2$, then we have $1 - F_{n}(D(X,Y)) < 1/2$. Hence,
            \begin{align} \label{supp:proposition:3.1:eqn:2}
                p(D(X,Y)) = 2\min\{F_{n}(D(X,Y)), 1 - F_{n}(D(X,Y))\} = 2 - 2F_{n}(D(X,Y)).
            \end{align}
            Then, since $D_{\max} \ge D(X,Y)$, we have
            \begin{align*}
                & F_{n}(D_{\max}) \ge F_{n}(D(X,Y)) > 1/2\\
                \Rightarrow & 1 - F_{n}(D_{\max}) < 1/2 \\
                \Rightarrow & p_2 = 2\min\{F_{n}(D_{\max}), 1 - F_{n}(D_{\max})\} = 2 - 2F_{n}(D_{\max}).
            \end{align*}
            Since $F_{n}(D_{\max}) \ge F_{n}(D(X,Y))$, then we have
            \begin{align*}
                &p_2 =  2 - 2F_{n}(D_{\max}) \le 2 - 2F_{n}(D(X,Y)) = p(D(X,Y))\\
                \Rightarrow&p_{\min} = \min \{p_1,p_2\} \le p(D(X,Y)).
            \end{align*}

            Now, we show $p(D(X,Y)) \le p_{\max}.$

            If $(D_{\min} - n^2/3)(D_{\max} - n^2/3) < 0$, then we have $p_{\max} = 1$. According
            to the definition of $p$-values in \eqref{pvalue}, we have 
            $p(D(X,Y)) \le 1 \le p_{\max} = 1$.

            However, if $(D_{\min} - n^2/3)(D_{\max} - n^2/3) \ge 0$, then  
            since $F_{n}$ is the cumulative distribution function of a normal 
            distribution with mean equal to $n^2/3$, we have	
            \begin{align*}
                &F_{n}(D_{\min}) > 1/2 \Rightarrow D_{\min} > n^2/3\\
                \Rightarrow & (D_{\min} - n^2/3) > 0.
            \end{align*}
            Since $(D_{\min} - n^2/3)(D_{\max} - n^2/3) \ge 0$, then we have $D_{\max} - n^2/3 \ge 0$.
            Hence,
            \begin{align}
                &D_{\max} \ge D(X,Y) \ge D_{\min} > n^2/3  \nonumber \\
                \Rightarrow &F_{n}(D_{\max}) \ge F_{n}(D(X,Y)) \ge F_{n}(D_{\min}) \nonumber\\
                \Rightarrow^{\eqref{supp:proposition:3.1:eqn:0}}&  F_{n}(D_{\max}) \ge 1 - p(D(X,Y))/2 \ge  F_{n}(D_{\min}) \nonumber\\
                \Rightarrow& p(D(X,Y)) \le 2 - 2F_{n}(D_{\min}). \label{supp:proposition:3.1:eqn:4}
            \end{align}

            Notice that since $D_{\min} - n^2/3 > 0$, we also have
            \begin{align*}
                &F_{n}(D_{\min}) > 1/2 \\
                \Rightarrow & 1 - F_{n}(D_{\min}) < 1/2 \\
                \Rightarrow & p_1 = 2\min\{F_{n}(D_{\min}), 1 - F_{n}(D_{\min})\} = 2 - 2F_{n}(D_{\min}).
            \end{align*}
            Combining this result with \eqref{supp:proposition:3.1:eqn:4}, we have
            \begin{align*}
                p(D(X,Y)) \le p_1 \le  \max \{p_1, p_2\} = p_{\max}.
            \end{align*}
            Hence, we have shown $p(D(X,Y)) \in [p_{\min}, p_{\max}]$ 
            when $F_{n}(D(X,Y)) > 1/2$.
            This completes our proof.
        \end{proof}

        \section{Efficient algorithms for calculating exact upper bounds} \label{appE}

        This section gives efficient algorithms for calculating exact upper bounds
        of Spearman's footrule under Missing Case I, Missing Case II, Missing
        Case III and General Missing Case.

        \subsection{Missing Case I}
        This subsection provides an efficient algorithm for calculating exact upper bounds
        of Spearman's footrule under Missing Case I. To start, we prove the 
        following lemma:

        \begin{lemma} \label{supp:proposition:alg:1:lemma:0}
            Suppose $\bx, \by \in \vndistinct$ and for $\mln{1} \in \tonumber{n}$, 
            let $\bu = \{\uuln{1},\ldots,\uuln{\mln{1}}\} \subset \tonumber{\n}$
            be a subset of indices. Suppose
            $\byn{\uuln{1}} < \ldots < \byn{\uuln{\mln{1}}}$, and assume
            $\bx \in \mysecondset{\by}{\bu}{\tonumber{\n}}{\n}$ 
            is a vector such that $\sum_{\iconstant \in \bu} \indicator{ \bxn{\iconstant} < \min \svector{\bx}{\lconstant}{\tonumber{\n} \setminus \bu} } = \rr$, where $\rr \in \{0,\ldots, \mln{1}\}$.
            Then, for any $\iconstant \in \{1,\ldots, \mln{1}\}$, we have
            \begin{align*}
                \rank{\bxn{\uuln{\iconstant}}}{\bx} = \left\{ \begin{array}{ll}
                    \n - \iconstant + 1, & \text{ if } \rr = 0, \\
                    \indicator{\uuln{\iconstant} \le \uuln{\mln{1} - \rr}} (\n - \iconstant + 1) 
                    + \indicator{\uuln{\iconstant} > \uuln{\mln{1} - \rr}} (\mln{1} - \iconstant + 1) , & \text{ if } \mln{1} > \rr > 0, \\
                    \mln{1} - \iconstant + 1, & \text{ if } \rr = \mln{1}.
                \end{array}\right.
            \end{align*}
        \end{lemma}

        \begin{proof}
            To start, since $\bx \in \mysecondset{\by}{\bu}{\tonumber{\n}}{\n}$, then
            according to the definition of $\mysecondset{\by}{\bu}{\tonumber{\n}}{\n}$,
            we have
            \begin{align}
                \bxn{\iconstant} > \max_{\jconstant \in \tonumber{\n} \setminus \bu} \bxn{\jconstant},
                \text{ or }
                \bxn{\iconstant} < \min_{\jconstant \in \tonumber{\n} \setminus \bu} \bxn{\jconstant},
                \text{ for any } \iconstant \in \bu. \label{supp:proposition:alg:1:lemma:0:eqn:1}
            \end{align} 
            Since $\byn{\uuln{1}} < \ldots < \byn{\uuln{\mln{1}}}$, then according to
            the definition of $\mysecondset{\by}{\bu}{\tonumber{\n}}{\n}$,
            we also have
            \begin{align}
                \rank{\bxn{\uuln{1}}}{\bx} > \ldots > \rank{\bxn{\uuln{\mln{1}}}}{\bx}. \label{supp:proposition:alg:1:lemma:0:eqn:2}
            \end{align}

            Below, we prove our result when $\rr = 0$, $ \mln{1} > \rr > 0$ and $\rr = \mln{1}$,
            separately.

            Suppose $\rr = 0$, then according to \eqref{supp:proposition:alg:1:lemma:0:eqn:1}, 
            we have
            \begin{align*} 
                &\bxn{\iconstant} > \max_{\jconstant \in \tonumber{\n} \setminus \bu} \bxn{\jconstant}, \text{ for any } \iconstant \in \bu \\
                \Rightarrow & \rank{\bxn{\iconstant}}{\bx} \in \{\n, \ldots, \n - \mln{1} + 1\}, \text{ for any } \iconstant \in \bu\\
                \Rightarrow^{\eqref{supp:proposition:alg:1:lemma:0:eqn:2}}  & \rank{\bxn{\uuln{\iconstant}}}{\bx} = \n - \iconstant + 1~\text{for any}~\iconstant \in \{1,\ldots, \mln{1}\}.
            \end{align*}
            This proves our result when $\rr = 0$.

            Similarly, suppose $\mln{1} > \rr > 0$, then according to \eqref{supp:proposition:alg:1:lemma:0:eqn:1}, 
            we have
            \begin{align*} 
                &\sum_{\iconstant \in \bu} \indicator{\bxn{\iconstant} < \min_{\jconstant \in \tonumber{\n} \setminus \bu} \bxn{\jconstant}} = \rr, \text{ and }\sum_{\iconstant \in \bu} \indicator{\bxn{\iconstant} > \max_{\jconstant \in \tonumber{\n} \setminus \bu} \bxn{\jconstant}} = \mln{1} - \rr \\
                \Rightarrow & \rank{\bxn{\iconstant}}{\bx} \in \{1, \ldots, \rr\} \cup \{\n, \ldots, \n - \mln{1} + \rr + 1\}, \text{ for any } \iconstant \in \bu\\
                \Rightarrow^{\eqref{supp:proposition:alg:1:lemma:0:eqn:2}} & \rank{\bxn{\uuln{\iconstant}}}{\bx} = \n - \iconstant + 1~\text{for any }~\iconstant \in \{1,\ldots, \mln{1} - \rr\}, \\
                &\text{ and } \rank{\bxn{\uuln{\iconstant}}}{\bx} = \mln{1} - \iconstant + 1~\text{for any }~\iconstant \in \{\mln{1}-\rr+1,\ldots, \mln{1}\} \\
                \Rightarrow & \rank{\bxn{\uuln{\iconstant}}}{\bx} = 	\indicator{\uuln{\iconstant} \le \uuln{\mln{1} - \rr}} (\n - \iconstant + 1) 
                + \indicator{\uuln{\iconstant} > \uuln{\mln{1} - \rr}} (\mln{1} - \iconstant + 1).
            \end{align*}
            This proves our result when $\mln{1} > \rr > 0$.

            Finally, suppose $\rr = \mln{1}$, then according to \eqref{supp:proposition:alg:1:lemma:0:eqn:1}, 
            we have
            \begin{align*} 
                &\bxn{\iconstant} < \min_{\jconstant \in \tonumber{\n} \setminus \bu} \bxn{\jconstant}, \text{ for any } \iconstant \in \bu \\
                \Rightarrow & \rank{\bxn{\iconstant}}{\bx} \in \{1, \ldots, \rr\}, \text{ for any } \iconstant \in \bu \\
                \Rightarrow^{\eqref{supp:proposition:alg:1:lemma:0:eqn:2}}  & \rank{\bxn{\uuln{\iconstant}}}{\bx} = \mln{1} - \iconstant + 1~\text{for any}~\iconstant \in \{1,\ldots, \mln{1}\}.
            \end{align*}
            This proves our result when $\rr = \mln{1}$ and completes our proof.		
        \end{proof}

        Then, we prove the following lemma:

        \begin{lemma} \label{supp:proposition:alg:1:lemma:2}
            Suppose $\bx \in \vndistinct$, and $\bu \subset \tonumber{\n}$
            is a subset of indices. Let $\bxs \in \myfirstset{\bu}{\tonumber{\n}}{\n}$ be an imputation
            of $\bx$ for indices $\bu$ such that $\sum_{\iconstant \in \bu} \indicator{ \bxsn{\iconstant} < \min \svector{\bxs}{\lconstant}{\tonumber{\n} \setminus \bu} } = \rr$.
            Then, we have
            \begin{align*}
                \rank{\bxsn{\iconstant}}{\bxs} = \rank{\bxn{\iconstant}}{\svector{\bx}{\lconstant}{\tonumber{\n} \setminus \bu}} + \rr, \text{ for any } \iconstant \in \tonumber{\n} \setminus \bu.
            \end{align*}
        \end{lemma}

        \begin{proof}
            According to the definition of $\myfirstset{\bu}{\tonumber{\n}}{\n}$,
            we have
            \begin{align*}
                \bxsn{\iconstant} > \max_{\jconstant \in \tonumber{\n} \setminus \bu} \bxsn{\jconstant},
                \text{ or }
                \bxsn{\iconstant} < \min_{\jconstant \in \tonumber{\n} \setminus \bu} \bxsn{\jconstant},
                \text{ for any } \iconstant \in \bu. 
            \end{align*}
            Then, since $\sum_{\iconstant \in \bu} \indicator{ \bxsn{\iconstant}   < \min \svector{\bxs}{\lconstant}{\tonumber{\n} \setminus \bu} } = \rr$, we have
            $\rr$ components of $\bxs$ with indices $\bu$ smaller than
            any components of $\bxs$ with indices $\tonumber{\n} \setminus \bu$, 
            and other components of $\bxs$ with indices $\bu$ larger than
            any components of $\bxs$ with indices $\tonumber{\n} \setminus \bu$.	
            Hence, we have
            \begin{align} \label{supp:proposition:alg:1:lemma:2:eqn:1}
                \sum_{\jconstant \in \bu} \indicator{\bxsn{\jconstant} \le \bxsn{\iconstant}} = \rr, \text{ for any } \iconstant \in \tonumber{\n} \setminus \bu. 
            \end{align}	

            Next, according to the definition of rank, for any 
            $\iconstant \in \tonumber{\n} \setminus \bu$, we have
            \begin{align*}
                \rank{\bxsn{\iconstant}}{\bxs} &= \sum_{\jconstant=1}^{\n} \indicator{\bxsn{\jconstant} \le \bxsn{\iconstant}}  \\
                & = \sum_{\jconstant \in \bu} \indicator{\bxsn{\jconstant} \le \bxsn{\iconstant}} + \sum_{\jconstant \in \tonumber{\n} \setminus \bu} \indicator{\bxsn{\jconstant} \le \bxsn{\iconstant}} \\
                & =^{\eqref{supp:proposition:alg:1:lemma:2:eqn:1}} \rr +  \sum_{\jconstant \in \tonumber{\n} \setminus \bu} \indicator{\bxsn{\jconstant} \le \bxsn{\iconstant}}.
            \end{align*}
            Since $\bxs$ is an imputation of $\bx$ for $\bu$, we have
            $\bxsn{\iconstant} = \bxn{\iconstant}$ for any $\iconstant \in \tonumber{\n} \setminus \bu$. Hence, we have 
            \begin{align*}
                \sum_{\jconstant \in \tonumber{\n} \setminus \bu} \indicator{\bxn{\jconstant} \le \bxn{\iconstant}} = \sum_{\jconstant \in \tonumber{\n} \setminus \bu} \indicator{\bxn{\jconstant} \le \bxn{\iconstant}}, \text{ for any } \iconstant \in \tonumber{\n} \setminus \bu. 
            \end{align*}
            Then, for any $\iconstant \in \tonumber{\n} \setminus \bu$,
            we have
            \begin{align*}
                \rank{\bxsn{\iconstant}}{\bxs} &= \rr + \sum_{\jconstant \in \tonumber{\n} \setminus \bu} \indicator{\bxn{\jconstant} \le \bxn{\iconstant}} \\
                & = \rr +  \rank{\bxn{\iconstant}}{\svector{\bx}{\lconstant}{\tonumber{\n} \setminus \bu}}. 
            \end{align*}
            This completes our proof.
        \end{proof}

        We are now ready to prove the first main results for Missing Case I:

        \begin{proposition} \label{supp:proposition:alg:0}
            Suppose $\bx, \by \in \vndistinct$ and for $\mln{1} \in \tonumber{\n}$, 
            let $\bu = \{1, \ldots, {\mln{1}} \} \subset  \tonumber{\n}$ 
            be a subset of indices. Suppose
            $\byn{{1}} < \ldots < \byn{{\mln{1}}}$, and 
            let $\bxs^{(\rr)} \in \mysecondset{\by}{\bu}{\tonumber{\n}}{\n}$ 
            be an imputation of $\bx$ for indices in $\bu$ such that 
            $\sum_{\iconstant \in \bu} \indicator{ \bxs^{(\rr)}(\iconstant)   < \min \svector{\bxs^{(\rr)}}{\lconstant}{\tonumber{\n} \setminus \bu} } = \rr$.
            For any $\iconstant \in \tonumber{\n} \setminus \bu$, denote $d_i = \rank{\byn{\iconstant}}{\by} - \rank{\bxn{\iconstant}}{\svector{\bx}{\jconstant}{\tonumber{\n}\setminus \bu}}$,
            and for any $\rr \in \{0, \ldots, \mln{1}\}$, $\iconstant \in \{1, \ldots, \mln{1}\}$, denote $\qnn{\rr}{\iconstant} = \indicator{{\iconstant} \le {\mln{1} - \rr}} (\n - \iconstant + 1) 
            + \indicator{{\iconstant} > {\mln{1} - \rr}} (\mln{1} - \iconstant + 1)$. Then, for any $\rr \in \{0,\ldots, \mln{1}-1\}$, we have
            \begin{align*}
                \sfdsb{\bxrs{\rr} }{\by} =  \sum_{\iconstant \in \bu} \left|  \qnn{\rr}{\iconstant} - \rank{\byn{\iconstant}}{\by} \right| + \sum_{\iconstant \in \tonumber{\n} \setminus \bu} \left| \rr -  d_i\right|.
            \end{align*}
        \end{proposition}

        \begin{proof}
            To start, according to the definition of Spearman's footrule,
            we have
            \begin{align*}
                &\sfdsb{\bxrs{\rr} }{\by} = \sum_{\iconstant = 1}^{\n} \left|\rank{\bxrsn{\rr}{\iconstant}}{\bxrs{\rr}} - \rank{\byn{\iconstant}}{\by}\right| \\
                & = \sum_{\iconstant=1}^{\mln{1}} \left|\rank{\bxrsn{\rr}{{\iconstant}}}{\bxrs{\rr}} - \rank{\byn{{\iconstant}}}{\by}\right| + \sum_{\iconstant \in \tonumber{\n} \setminus \bu} \left|\rank{\bxrsn{\rr}{\iconstant}}{\bxrs{\rr}} - \rank{\byn{\iconstant}}{\by}\right|.
            \end{align*}
            In order to prove our result, it is then sufficient to show the following two equations hold:
            \begin{align*}
                (1): &\sum_{\iconstant=1}^{\mln{1}} \left|\rank{\bxrsn{\rr}{{\iconstant}}}{\bxrs{\rr}} 
                - \rank{\byn{{\iconstant}}}{\by}\right| 
                = \sum_{\iconstant=1}^{\mln{1}} \left|  \qnn{\rr}{\iconstant} - \rank{\byn{\iconstant}}{\by} \right|, \\
                \text{and }(2):	&\sum_{\iconstant \in \tonumber{\n} \setminus \bu} \left|\rank{\bxrsn{\rr}{\iconstant}}{\bxrs{\rr}} - \rank{\byn{\iconstant}}{\by}\right|
                =  \sum_{\iconstant \in \tonumber{\n} \setminus \bu} \left| \rr -  d_i\right|.
            \end{align*}
            Below, we show equations (1) and (2) are true separately.

            First, we show that equation (1) is true.

            Since $\bxs^{(\rr)} \in \mysecondset{\by}{\bu}{\tonumber{\n}}{\n}$,
            and $\byn{{1}} < \ldots < \byn{{\mln{1}}}$, then
            according to Lemma~\ref{supp:proposition:alg:1:lemma:0}, 
            for any $\iconstant \in \{1, \ldots, \mln{1}\}$, and
            $\rr \in \{0, \ldots, \mln{1}\}$, we have	
            \begin{align*}
                \rank{\bxrsn{\rr}{{\iconstant}}}{\bxrs{\rr}} = \left\{ \begin{array}{ll}
                    \n - \iconstant + 1, & \text{ if } \rr = 0, \\
                    \qnn{\rr}{\iconstant}, & \text{ if } \mln{1} > \rr > 0, \\
                    \mln{1} - \iconstant + 1, & \text{ if } \rr = \mln{1},
                \end{array}\right.
            \end{align*}
            where $\qnn{\rr}{\iconstant} = \indicator{{\iconstant} \le {\mln{1} - \rr}} (\n - \iconstant + 1) + \indicator{{\iconstant} > {\mln{1} - \rr}} (\mln{1} - \iconstant + 1)$.
            Notice that when $\rr = 0$ and $\rr = \mln{1}$, 
            $\rank{\bxn{{\iconstant}}}{\bx}  = \qnn{\rr}{\iconstant}$
            is still true for any
            $\iconstant = 1, \ldots, \mln{1}$. Hence, we have
            \begin{align*}
                \rank{\bxn{{\iconstant}}}{\bx} = \qnn{\rr}{\iconstant} , \text{ for any } \iconstant \in \bu, \rr \in \{0, \ldots, \mln{1}\}.
            \end{align*}
            Thus, for any $\iconstant \in \bu, \rr \in \{0, \ldots, \mln{1}\}$, we have
            \begin{align*}
                &\sum_{\iconstant=1}^{\mln{1}} \left|\rank{\bxrsn{\rr}{{\iconstant}}}{\bxrs{\rr}} 
                - \rank{\byn{{\iconstant}}}{\by}\right|  = \left|  \qnn{\rr}{\iconstant} - \rank{\byn{\iconstant}}{\by} \right|.
            \end{align*}
            This proves equation (1).

            Next, we show that equation (2) is true. 

            Since $\bxrs{\rr} \in \mysecondset{\by}{\bu}{\tonumber{\n}}{\n}$,
            then according to the definition of $\mysecondset{\by}{\bu}{\tonumber{\n}}{\n}$,
            we also have $\bxrs{\rr} \in \myfirstset {\bu}{\tonumber{\n}}{\n}$.
            Notice that $\bxrs{\rr}$ is an imputation of $\bx$ for $\bu$, and $\sum_{\iconstant \in \bu} \indicator{ \bxs^{(\rr)}(\iconstant)   < \min \svector{\bxs^{(\rr)}}{\lconstant}{\tonumber{\n} \setminus \bu} } = \rr$. Then, according to Lemma~\ref{supp:proposition:alg:1:lemma:2}, for any $ \iconstant \in \tonumber{\n} \setminus \bu$, we have
            \begin{align}
                \rank{\bxrsn{\rr}{\iconstant}}{\bxrs{\rr}} = \rr + \rank{\bxn{\iconstant}}{\svector{\bx}{\lconstant}{\tonumber{\n} \setminus \bu}} , \text{ for any } \iconstant \in \tonumber{\n} \setminus \bu.
            \end{align}
            Hence, we have
            \begin{align*}
                &\sum_{\iconstant \in \tonumber{\n} \setminus \bu} \left|\rank{\bxrsn{\rr}{\iconstant}}{\bxrs{\rr}} - \rank{\byn{\iconstant}}{\by}\right|\\
                &= \sum_{\iconstant \in \tonumber{\n} \setminus \bu} \left|\rank{\bxn{\iconstant}}{\svector{\bx}{\lconstant}{\tonumber{\n} \setminus \bu}} + \rr -  \rank{\byn{\iconstant}}{\by} \right| \\
                & = \sum_{\iconstant \in \tonumber{\n} \setminus \bu} \left|\rr -  d_i\right|. 
            \end{align*}
            This proves equation (2), and completes our proof.
        \end{proof}

        Before showing the second main result for Missing Case I, 
        we prove the following two lemmas.

        \begin{lemma} \label{supp:proposition:alg:1:lemma:0.0}
            Suppose $\bx = (\bxn{\uuln{1}}, \ldots, \bxn{\uuln{\n}})$ is a vector of integers,
            where $\bu = \{\uuln{1}, \ldots, \uuln{\n}\}$ is a set of indices such that $\uuln{1} < \ldots < \uuln{\n}$. Then, we have 
            \begin{align*}
                \sum_{\iconstant \in \bu} |\bxn{\iconstant} + 1|
                - \sum_{\iconstant \in \bu} |\bxn{\iconstant}| = 2 \sum_{\iconstant \in \bu} \indicator{\bxn{\iconstant} \ge 0} - \n. 		
            \end{align*}
        \end{lemma}

        \begin{proof}
            To start, for any $\iconstant \in \bu$, if $\bxn{\iconstant} \ge 0$, we have
            \begin{align*}
                |\bxn{\iconstant} + 1| - |\bxn{\iconstant}| & = 1.
            \end{align*}
            If, however $\bxn{\iconstant} < 0$, then since 
            $\bx$ is a vector of integers, we
            have $\bxn{\iconstant} + 1 \le 0$.
            Hence, we have
            \begin{align*}
                |\bxn{\iconstant} + 1| - |\bxn{\iconstant}| &= -1.
            \end{align*}
            Therefore, we have
            \begin{align*}
                \sum_{\iconstant \in \bu} |\bxn{\iconstant} + 1| - \sum_{\iconstant \in \bu} |\bxn{\iconstant}| 
                &= \sum_{\iconstant \in \bu}  \left(|\bxn{\iconstant} + 1| - |\bxn{\iconstant}| \right) \\
                & = \sum_{\iconstant \in \bu} \indicator{\bxn{\iconstant} \ge 0} - \sum_{\iconstant \in \bu} \indicator{\bxn{\iconstant} < 0}\\
                & = \sum_{\iconstant \in \bu} \indicator{\bxn{\iconstant} \ge 0}  - \left(\n -  \sum_{\iconstant \in \bu} \indicator{\bxn{\iconstant} \ge 0}   \right)\\
                & = 2\sum_{\iconstant \in \bu} \indicator{\bxn{\iconstant} \ge 0} - \n.
            \end{align*}
            This completes our proof.	
        \end{proof}

        \begin{lemma} \label{supp:proposition:alg:2:lemma:3}		
            Suppose $\bu = \{1, \ldots, \mln{1}\}$, and
            $\n \in \mathbb{N}$ is a number such that 
            $\n > \mln{1}$.
            For any $\iconstant \in \{1, \ldots, \mln{1} \}$, $\rr \in \{0, \ldots, \mln{1}\}$, define
            \begin{align*}
                \qnn{\rr}{{\iconstant}} = \indicator{{\iconstant} \le {\mln{1} - \rr}} (\n - \iconstant + 1) + \indicator{{\iconstant} > {\mln{1} - \rr}} (\mln{1} - \iconstant + 1).
            \end{align*}		
            Suppose $\byn{{1}}, \ldots, \byn{{{\mln{1}}}}$ are real numbers.
            Then, for any $\rr \in \{0, \ldots, \mln{1} - 1\}$, we have
            \begin{align*}
                & \sum_{\iconstant = 1}^{\mln{1}} \left|\qnn{\rr + 1}{\iconstant}  - \byn{\iconstant}  \right| \\
                & = \sum_{\iconstant = 1}^{\mln{1}} \left|\qnn{\rr}{\iconstant}  - \byn{\iconstant}  \right| + \left|\rr + 1 -  \byn{{\mln{1} -\rr} } \right|  
                - \left|\n - \mln{1} + \rr + 1 - \byn{{\mln{1} -\rr}} \right|.
            \end{align*}
        \end{lemma}

        \begin{proof}
            We show our result for $\rr = 0$, $\mln{1} - 1 > \rr > 0$,
            and $\rr = \mln{1} - 1$ separately.

            When $\rr = 0$, we have
            \begin{align*}
                & \sum_{\iconstant \in \bu} \left|\qnn{\rr}{\iconstant}  - \byn{\iconstant}  \right| = \sum_{\iconstant=1}^{\mln{1}}|\n - \iconstant + 1 - \byn{{\iconstant}}|,\\
                & \sum_{\iconstant \in \bu} \left|\qnn{\rr+1}{\iconstant}  - \byn{\iconstant}  \right| = \sum_{\iconstant=1}^{\mln{1}-1}|\n - \iconstant + 1 - \byn{{\iconstant}} |  + |1 - \byn{{\iconstant}}|.
            \end{align*}
            Hence, we have
            \begin{align*}
                \begin{split}
                    & \sum_{\iconstant \in \bu} \left|\qnn{\rr+1}{\iconstant}  - \byn{\iconstant}  \right| - \sum_{\iconstant \in \bu} \left|\qnn{\rr}{\iconstant}  - \byn{\iconstant}  \right| \\
                    & = |1 -\byn{{\mln{1}}}| - |\n - \mln{1} + 1 - \byn{{\mln{1}}}| \\
                    & = \left|\rr + 1 -  \byn{{\mln{1} -\rr} } \right|  
                    - \left|\n - \mln{1} + \rr + 1 - \byn{{\mln{1} -\rr}} \right|.
                \end{split}
            \end{align*}
            This completes our proof when $\rr = 0$.

            Similarly, when $\mln{1} - 1 > \rr > 0$,
            we have
            \begin{align*}
                \sum_{\iconstant \in \bu} \left|\qnn{\rr}{\iconstant}  - \byn{\iconstant}  \right| 
                = \sum_{\iconstant=1}^{\mln{1} - \rr} |\n - \iconstant + 1 - {\byn{{\iconstant}}}| 
                + \sum_{\iconstant = \mln{1} - \rr + 1}^{\mln{1}}|\mln{1} - \iconstant + 1 - {\byn{{\iconstant}}}|,
            \end{align*}
            and
            \begin{align*}
                \sum_{\iconstant \in \bu} \left|\qnn{\rr+1}{\iconstant}  - \byn{\iconstant}  \right|  = \sum_{\iconstant=1}^{\mln{1} - \rr - 1} |\n - \iconstant + 1 - {\byn{{\iconstant}}} | 
                + \sum_{\iconstant = \mln{1} - \rr}^{\mln{1}}|\mln{1} - \iconstant + 1 - {\byn{{\iconstant}}} |.
            \end{align*}
            Hence, we have
            \begin{align*}
                \begin{split}
                    &\sum_{\iconstant \in \bu} \left|\qnn{\rr+1}{\iconstant}  - \byn{\iconstant}  \right|  -  \sum_{\iconstant \in \bu} \left|\qnn{\rr}{\iconstant}  - \byn{\iconstant}  \right|  \\
                    & = |\mln{1} - (\mln{1} - \rr) + 1 - {\byn{{\mln{1} - \rr}}}| - |\n - (\mln{1} - \rr) + 1 - {\byn{{\mln{1} - \rr}}}| \\
                    & = \left|\rr + 1 -  {\byn{{\mln{1} - \rr}}} \right|  
                    - \left|\n - \mln{1} + \rr + 1 -  {\byn{{\mln{1} - \rr}}} \right|.
                \end{split}
            \end{align*}
            This completes our proof when $\mln{1} - 1 > \rr > 0$.

            Similarly,  when $\rr = \mln{1} - 1$, we have
            \begin{align*}
                &\sum_{\iconstant \in \bu} \left|\qnn{\rr+1}{\iconstant}  - \byn{\iconstant}  \right|  = |\n - 1 + 1 - {\byn{{1}}}| 
                + \sum_{\iconstant = 2}^{\mln{1}}|\mln{1} - \iconstant + 1 - {\byn{{\iconstant}}}|,\\
                &\sum_{\iconstant \in \bu} \left|\qnn{\rr}{\iconstant}  - \byn{\iconstant}  \right| = \sum_{\iconstant = 1}^{\mln{1}}|\mln{1} - \iconstant + 1 - {\byn{{\iconstant}}}|.
            \end{align*}
            Hence, we have
            \begin{align*} 
                \begin{split}
                    & \sum_{\iconstant \in \bu} \left|\qnn{\rr+1}{\iconstant}  - \byn{\iconstant}  \right|  + \sum_{\iconstant \in \bu} \left|\qnn{\rr}{\iconstant}  - \byn{\iconstant}  \right|  \\
                    & = |\mln{1} - 1 + 1 - {\byn{{1}}}| -  |\n - 1 + 1 - {\byn{{1}}}|  \\
                    & = \left|\rr + 1 - {\byn{{\mln{1} - \rr}}} \right|  
                    - \left|\n - \mln{1} + \rr + 1 -  {\byn{{\mln{1} - \rr}}} \right|.
                \end{split}
            \end{align*}
            This completes our proof when $\rr  = \mln{1} - 1$.
            Hence, we have shown our result.
        \end{proof}

        Now, we are ready to prove the second main result for Missing Case I.

        \begin{proposition} \label{supp:proposition:alg:1}
            Suppose $\bx, \by \in \vndistinct$ and for $\mln{1} \in \tonumber{\n}$, 
            let $\bu = \{{1},\ldots,{\mln{1}} \} \subset \tonumber{\n}$ 
            be a subset of indices. Suppose
            $\byn{{1}} < \ldots < \byn{{\mln{1}}}$, and 
            let $\bxs^{(\rr)} \in \mysecondset{\by}{\bu}{\tonumber{\n}}{\n}$ 
            be an imputation of $\bx$ for indices in $\bu$ such that 
            $\sum_{\iconstant \in \bu} \indicator{ \bxs^{(\rr)}(\iconstant)   < \min \svector{\bxs^{(\rr)}}{\lconstant}{\tonumber{\n} \setminus \bu} } = \rr$.
            For any $\iconstant \in \tonumber{\n} \setminus \bu$, denote $d_i = \rank{\byn{\iconstant}}{\by} - \rank{\bxn{\iconstant}}{\svector{\bx}{\jconstant}{\tonumber{\n}\setminus \bu}}$,
            and for any $\rr \in \{0, \ldots, \mln{1} - 1\}$, denote $\ssln{\rr} = \sum_{\iconstant \in \tonumber{\n} 
            \setminus \bu} \indicator{d_i \le \rr}$. Then, for any $\rr \in \{0,\ldots, \mln{1}-1\}$, we have
            \begin{align*}
                \sfdsb{\bxrs{\rr+1} }{\by} &= \sfdsb{\bxrs{\rr} }{\by} +  2\ssln{\rr} - \n + \mln{1} + C^{(\rr)},
            \end{align*}
            where $C^{(\rr)} = |\rr + 1 - \rank{\byn{\mln{1} - \rr}}{\by}| - |\n - \mln{1} + \rr + 1 - \rank{\byn{\mln{1} - \rr}}{\by}|$.
        \end{proposition}

        \begin{proof}
            To start, according to Proposition~\ref{supp:proposition:alg:0},
            we have
            \begin{align*}
                &\sfdsb{\bxrs{\rr} }{\by} =  \sum_{\iconstant \in \bu} \left|  \qnn{\rr}{\iconstant} - \rank{\byn{\iconstant}}{\by} \right| + \sum_{\iconstant \in \tonumber{\n} \setminus \bu} \left| \rr -  d_i\right|, \\
                \text{and }&\sfdsb{\bxrs{\rr+1} }{\by} =  \sum_{\iconstant \in \bu} \left|  \qnn{\rr+1}{\iconstant} - \rank{\byn{\iconstant}}{\by} \right| + \sum_{\iconstant \in \tonumber{\n} \setminus \bu} \left| \rr + 1 -  d_i\right|,
            \end{align*}
            where $\qnn{\rr}{\iconstant} = \indicator{{\iconstant} \le {\mln{1} - \rr}} (\n - \iconstant + 1) 
            + \indicator{{\iconstant} > {\mln{1} - \rr}} (\mln{1} - \iconstant + 1)$ for 
            any $\iconstant \in \{1, \ldots, \mln{1}\}$ and $\rr \in \{0, \ldots, \mln{1}\}$.
            Then, in order to prove our result, it is sufficient to show that
            \begin{align*}
                \sum_{\iconstant \in \tonumber{\n} \setminus \bu} \left| \rr + 1 -  d_i\right| - \sum_{\iconstant \in \tonumber{\n} \setminus \bu} \left| \rr -  d_i\right| = 2\ssln{\rr} - \n + \mln{1},
            \end{align*}
            and
            \begin{align*}
                &\sum_{\iconstant \in \bu} \left|  \qnn{\rr+1}{\iconstant} - \rank{\byn{\iconstant}}{\by} \right|  - \sum_{\iconstant \in \bu} \left|  \qnn{\rr}{\iconstant} - \rank{\byn{\iconstant}}{\by} \right| \\
                &= |\rr + 1 - \rank{\byn{\mln{1} - \rr}}{\by}| - |\n - \mln{1} + \rrln{1} + 1 - \rank{\byn{\mln{1} - \rr}}{\by}|.
            \end{align*}

            According to Lemma~\ref{supp:proposition:alg:1:lemma:0.0},
            we have
            \begin{align*}
                \sum_{\iconstant \in \tonumber{\n} \setminus \bu} \left| \rr + 1 -  d_i\right| - \sum_{\iconstant \in \tonumber{\n} \setminus \bu} \left| \rr -  d_i\right| = 2 \sum_{\iconstant \in \tonumber{\n} \setminus \bu} \indicator{\rr - \dln{\iconstant} \ge 0} - (\n - \mln{1}) = 2\ssln{\rr} - \n + \mln{1}.
            \end{align*}

            Next, notice that $\qnn{\rr}{\iconstant} = \indicator{{\iconstant} \le {\mln{1} - \rr}} (\n - \iconstant + 1) 
            + \indicator{{\iconstant} > {\mln{1} - \rr}} (\mln{1} - \iconstant + 1)$, for any $\iconstant \in \{1, \ldots, \mln{1}\}$ and $\rr \in \{0, \ldots, \mln{1}\}$.
            Then according to Lemma~\ref{supp:proposition:alg:2:lemma:3},
            we have
            \begin{align*}
                &\sum_{\iconstant \in \bu} \left|  \qnn{\rr+1}{\iconstant} - \rank{\byn{\iconstant}}{\by} \right|  - \sum_{\iconstant \in \bu} \left|  \qnn{\rr}{\iconstant} - \rank{\byn{\iconstant}}{\by} \right| \\
                &= |\rr + 1 - \byn{\mln{1} - \rr}| - |\n - \mln{1} + \rr + 1 - \byn{\mln{1} - \rr}|\\
                & = C^{(\rr)}.
            \end{align*}
            This completes our proof.
        \end{proof}

        \begin{algorithm} 
            \caption{An efficient algorithm for computing cumulative counts.}  \label{supp:alg:0}
            \begin{algorithmic}[1]
                \Require{A sequence of integers $\bz = (\bzn{1}, \ldots, \bzn{\n})$, two integer number $\mln{1} \le \mln{2}$.} 
                \Ensure{$\ssln{\rr} = \sum_{\iconstant \in \tonumber{\n}} \indicator{\bzn{\iconstant} \le \rr}$ for any $\rr \in \{\mln{1}, \ldots, \mln{2}\}$.}
                \State Initialize $\ssln{\mln{1}}  = 0$. If $\mln{2} > \mln{1}$, initialize $\tln{\iconstant} = 0$ for any $\iconstant \in \{\mln{1} + 1, \ldots, \mln{2}\}$.
                \For{$\iconstant = 1, \ldots, \n$}
                \If{$\bzn{\iconstant} \le \mln{1}$} \label{sup:alg:0:line:1}
                \State $\ssln{0} = \ssln{0} + 1$.
                \EndIf
                \If{$\mln{1} < \bzn{\iconstant} \le \mln{2}$}
                \State $\tln{\bzn{\iconstant}} = \tln{\bzn{\iconstant}} + 1$.
                \EndIf		 \label{sup:alg:0:line:2}
                \EndFor
                \State If $\mln{2} > \mln{1}$, compute $\ssln{\rr + 1} = \ssln{\rr} + \tln{\rr+1}$ for any $\rr \in \{\mln{1}, \ldots, \mln{2} - 1\}$. \label{sup:alg:0:line:3}
                \State Return $\ssln{\rr}$ for any $\rr \in \{\mln{1}, \ldots, \mln{2}\}$.
            \end{algorithmic}
        \end{algorithm}

        \begin{remark} \label{supp:remark:alg:0}
            The computational complexity of Algorithm~\ref{supp:alg:0} is analyzed as follows. 
            Each step between line~\ref{sup:alg:0:line:1} and line~\ref{sup:alg:0:line:2}
            takes constant time $\bmo(1)$, and since the loop runs $\n$ times,
            the computational complexity for the loop is $\bmo(\n)$.
            The computation complexity for line~\ref{sup:alg:0:line:3} 
            is $\bmo(\mln{2} - \mln{1})$.
            Therefore, the overall computational complexity for Algorithm~\ref{supp:alg:0} is  $\bmo(\n + \mln{2} - \mln{1})$.
        \end{remark}

        \begin{algorithm} 
            \caption{An efficient algorithm for computing exact upper bounds of Spearman's footrule under Missing Case I.}  \label{supp:alg:1}
            \begin{algorithmic}[1]
                \Require{$\bx, \by \in \vndistinct$, where either $\bx$ or $\by$ is fully observed,
                while the other is partially observed. } 
                \Ensure{Maximum possible Spearman's footrule distance between $X$ and $Y$.}
                \State If $\bx$ is fully observed while $\by$ is partially observed, then change the label of $\bx$ and $\by$.
                \State Denote $\mln{1}$ as the number of missing components in $\bx$.
                If $\mln{1} = \n$, return $\sum_{\iconstant=1}^{\n} |\iconstant - (\n -\iconstant + 1)|$.
                \State Rank all observed components in $X$ and $Y$.  \label{supp:alg:1:line:1}
                \State	Relabel $\bx$ and $\by$ such that $(\bxn{1}, \ldots, \bxn{\mln{1}})$ are unobserved and $\byn{1} < \ldots < \byn{\mln{1}}$.
                \State Let $\bu = \{1,2, \dots,\mln{1}\}.$ For $\iconstant \in \tonumber{\n} \setminus \bu$, let $\dln{\iconstant} = \rank{\byn{\iconstant}}{\by} - \rank{\bxn{\iconstant}}{\svector{\bx}{\lconstant}{\tonumber{\n} \setminus \bu}}.$ \label{supp:alg:1:line:2}
                \State Run Algorithm~\ref{supp:alg:0} computing $\ssln{\rr} =  \sum_{\iconstant \in \tonumber{\n} \setminus \bu} \indicator{\dln{\iconstant} \le \rr}$ for any $\rr = \{0, \ldots, \mln{1} - 1\}$. \label{supp:alg:1:line:3}
                \State For any $\rr \in \{0, \ldots, \mln{1} - 1\}$, compute \label{supp:alg:1:line:4}
                \begin{align*}
                    C^{(\rr)} = |\rr + 1 - \rank{\byn{\mln{1} - \rr}}{\by}| - |\n - \mln{1} + \rr + 1 - \rank{\byn{\mln{1} - \rr}}{\by}|.
                \end{align*}
                \State Initialize $\bdln{0}$ = $\sum_{\iconstant = 1}^{\mln{1}} \left| (n - \iconstant + 1) - \rank{\byn{\iconstant}}{\svector{\by}{\lconstant}{\tonumber{\n}}}\right| + \sum_{i \in \tonumber{\n} \setminus \bu} |d_i|$. \label{supp:alg:1:line:5}
                \State For $\rr \in \{0,\ldots,\mln{1}-1\}$, compute $D_{\rr+1} =  D_{\rr}  + 2\ssln{\rr} - \n + \mln{1} + C^{(\rr)}$. \label{supp:alg:1:line:6}
                \State Return $\max \{\bdln{0}, \ldots, \bdln{\mln{1}}\}$.
            \end{algorithmic}
        \end{algorithm}

        \begin{remark}	
            A few comments of Algorithm~\ref{supp:alg:1} are made below:

            In Algorithm~\ref{supp:alg:1}, the initialization $\bdln{0}$ computed in
            line~\ref{supp:alg:1:line:5} equals to $\sfdsb{\bxrs{0}}{\by}$ defined in 
            Proposition~\ref{supp:proposition:alg:0}.
            Spearman's footrule $D_{\rr+1}$ and $D_{\rr}$ 
            equal to $\sfdsb{\bxrs{\rr + 1}}{\by}$
            and $\sfdsb{\bxrs{\rr}}{\by}$, respectively,
            and are updated according to Proposition \ref{supp:proposition:alg:1}.

            By computing $\{\bdln{0}, \ldots, \bdln{\mln{1}}\}$, Algorithm~\ref{supp:alg:1} 
            finds all possible Spearman's footrule
            values $\sfd{\bxs}{\by}$, where $\bxs \in \mysecondset{\by}{\bu}{\tonumber{\n}}{\n}$ 
            is imputation of $\bx$ for indices $\bu$. See Remark~3 in the main
            paper for explanation.
            Hence, according to Theorem~\ref{supp:theorem:2.16},
            the algorithm guarantees to find the maximum possible Spearman's footrule
            between $\bx$ and $\by$.

            The computational complexity of Algorithm~\ref{supp:alg:1} is analyzed as follows. Ranking and relabeling 
            all observed components in $\bx$ and $\by$ in line~\ref{supp:alg:1:line:1} requires $\bmo (\n \log \n)$ steps. 
            Using these rankings, in line~\ref{supp:alg:1:line:2} computing each $\dln{\iconstant}$ is $\bmo(1)$, and so overall
            line~\ref{supp:alg:1:line:2} is $\bmo(\n - \mln{1})$. 
            According to Remark~\ref{supp:remark:alg:0}, the computational
            complexity of running Algorithm~\ref{supp:alg:0} in line~\ref{supp:alg:1:line:3}
            is $\bmo(\n + \mln{1}) = \bmo(\n)$. In line~\ref{supp:alg:1:line:4}, each iteration of the for loop is $\bmo(1)$, and since the loop runs $\mln{1}$ times, the computational complexity for the loop is $\bmo(\mln{1})$. Line~\ref{supp:alg:1:line:5} and \ref{supp:alg:1:line:6}
            takes $\bmo(1)$ and $\bmo(\mln{1})$ steps, respectively.
            Therefore, the overall computational complexity for Algorithm~\ref{supp:alg:1} is $\bmo( \n\log \n)$.
        \end{remark}

        \subsection{Missing Case II}
        This subsection provides an efficient algorithm for computing 
        exact upper bounds of Spearman's footrule under missing case II,
        where $\bx$ and $\by$ might be
        partially observed, but for any pair $(\bxn{\iconstant}, \byn{\iconstant})$,
        where $\iconstant \in \tonumber{\n}$, at least one value is observed.
        First, we show the following result:
        \begin{proposition} \label{supp:proposition:alg:2:1}
            Suppose $\bx, \by \in \vndistinct$ and for $\mln{1}, \mln{2} \in \tonumber{\n}$ such
            that $\mln{1} + \mln{2} < \n$, 
            let $\bv = \{1, \ldots, \mln{2}\}$ and $\bu = \{\mln{2} + 1,\ldots, \mln{2} + \mln{1}\}$ be subsets of indices. Suppose
            $\bxn{1} < \ldots < \bxn{\mln{2}}$, and
            $\byn{\mln{2} + 1} < \ldots < \byn{\mln{2} + \mln{1}}$.
            Let $(\bxs^{(\rrln{1})}, \bys^{(\rrln{2})}) \in (\mysecondset{\by}{\bu}{\tonumber{\n}}{\n},  \mysecondset{\bx}{\bv}{\tonumber{\n}}{\n})$ 
            be imputations of $\bx$, $\by$ for indices $\bu$ and $\bv$, 
            respectively, such that 
            \begin{align*}
                &\sum_{\iconstant \in \bu} \indicator{ \bxs^{(\rrln{1})}(\iconstant)   < \min \svector{\bxs^{(\rrln{1})}}{\lconstant}{\tonumber{\n} \setminus \bu} } = \rrln{1} ,\\
                \text{and } &\sum_{\iconstant \in \bv} \indicator{ \bys^{(\rrln{2})}(\iconstant)   < \min \svector{\bys^{(\rrln{2})}}{\lconstant}{\tonumber{\n} \setminus \bv} } = \rrln{2}.
            \end{align*}
            For any $\iconstant \in \tonumber{\n} \setminus (\bu \cup \bv)$, denote
            $d_i = \rank{\byn{\iconstant}}{\svector{\by}{\lconstant}{\tonumber{\n} \setminus \bv}} - \rank{\bxn{\iconstant}}{\svector{\bx}{\lconstant}{\tonumber{\n} \setminus \bu}}$.Then, for any $\rrln{1} \in \{0, \ldots, \mln{1}\}$, and $\rrln{2} \in \{0, \ldots, \mln{2}\}$, we have
            \begin{align*} 
                \begin{split}
                    \sfdsb{\bxrs{\rrln{1}} }{\byrs{\rrln{2}}} &= \sum_{\iconstant\in\bv} \left| \rank{\bxn{\iconstant}}{\svector{\bx}{\lconstant}{\tonumber{\n} \setminus \bu}} + \rrln{1} -  \qnn{\rrln{2}}{\iconstant} \right| \\
                    &+ \sum_{\iconstant\in\bu} \left| \pnn{\rrln{1}}{\iconstant - \mln{2}} - \rank{\byn{\iconstant}}{\svector{\by}{\lconstant}{\tonumber{\n} \setminus \bv}} - \rrln{2} \right| +  \sum_{\iconstant \in \tonumber{\n} \setminus (\bu \cup \bv)} \left|\rrln{1} - \rrln{2} -  d_i \right|. 
                \end{split}
            \end{align*}
            where $\qnn{\rrln{2}}{\iconstant}  = \indicator{{\iconstant} \le {\mln{2} - \rrln{2}}} (\n - \iconstant + 1) 
            + \indicator{{\iconstant} > {\mln{2} - \rrln{2}}} (\mln{2} - \iconstant + 1)$,
            and $\pnn{\rrln{1}}{\iconstant} = \indicator{\iconstant \le \mln{1} - \rrln{1}} (\n - \iconstant + 1) + \indicator{\iconstant > \mln{1} - \rrln{1}} (\mln{1} - \iconstant + 1)$.
        \end{proposition}
        \begin{proof}
            To start, according to the definition of Spearman's footrule,
            we have
            \begin{align*}
                \sfdsb{\bxrs{\rrln{1}} }{\byrs{\rrln{2}}}  &= \sum_{\iconstant \in \bv} \left|\rank{\bxrsn{\rrln{1}}{\iconstant}}{\bxrs{\rrln{1}}}  - \rank{\byrsn{\rrln{2}}{\iconstant}}{\byrs{\rrln{2}}}\right| \\
                &+ \sum_{\iconstant \in \bu} \left|\rank{\bxrsn{\rrln{1}}{\iconstant}}{\bxrs{\rrln{1}}}  - \rank{\byrsn{\rrln{2}}{\iconstant}}{\byrs{\rrln{2}}}\right| \\
                &+ \sum_{\iconstant \in \tonumber{\n} \setminus (\bu \cup \bv)} \left|\rank{\bxrsn{\rrln{1}}{\iconstant}}{\bxrs{\rrln{1}}}  - \rank{\byrsn{\rrln{2}}{\iconstant}}{\byrs{\rrln{2}}}\right|. 
            \end{align*}
            Hence, in order to prove our results, it is then sufficient to show
            the following three equations all hold:
            \begin{align*}
                (1): &\sum_{\iconstant \in \bv} \left|\rank{\bxrsn{\rrln{1}}{\iconstant}}{\bxrs{\rrln{1}}}  - \rank{\byrsn{\rrln{2}}{\iconstant}}{\byrs{\rrln{2}}}\right| \\
                &= \sum_{\iconstant\in\bv} \left| \rank{\bxn{\iconstant}}{\svector{\bx}{\lconstant}{\tonumber{\n} \setminus \bu}} + \rrln{1} -  \qnn{\rrln{2}}{\iconstant} \right|, \\
                (2): &\sum_{\iconstant \in \bu} \left|\rank{\bxrsn{\rrln{1}}{\iconstant}}{\bxrs{\rrln{1}}}  - \rank{\byrsn{\rrln{2}}{\iconstant}}{\byrs{\rrln{2}}}\right| \\
                &= \sum_{\iconstant\in\bu} \left| \pnn{\rrln{1}}{\iconstant} - \rank{\byn{\iconstant}}{\svector{\by}{\lconstant}{\tonumber{\n} \setminus \bv}} - \rrln{2} \right|, \\
                \text{and }(3): &\sum_{\iconstant \in \tonumber{\n} \setminus (\bu \cup \bv)} \left|\rank{\bxrsn{\rrln{1}}{\iconstant}}{\bxrs{\rrln{1}}}  - \rank{\byrsn{\rrln{2}}{\iconstant}}{\byrs{\rrln{2}}}\right| \\
                &=  \sum_{\iconstant \in \tonumber{\n} \setminus (\bu \cup \bv)} \left|\rrln{1} - \rrln{2} -  d_i \right|. 
            \end{align*}
            Below, we show equations (1), (2) and (3) are true seperately.

            We first show equation (1) is true. 
            Since $\bxrs{\rrln{1}} \in \mysecondset{\by}{\bu}{\tonumber{\n}}{\n}$,
            then according to the definition of 
            $\mysecondset{\by}{\bu}{\tonumber{\n}}{\n}$,
            we have $\bxrs{\rrln{1}} 
            \in \myfirstset{\bu}{\tonumber{\n}}{\n}$.
            Since 
            \begin{align*}
                &\sum_{\iconstant \in \bu} \indicator{ \bxs^{(\rrln{1})}(\iconstant)   < \min \svector{\bxs^{(\rrln{1})}}{\lconstant}{\tonumber{\n} \setminus \bu} } = \rrln{1},
            \end{align*}
            and $\bxrs{\rrln{1}}$ is an imputation 
            of $\bx$ for $\bu$, then according to Lemma~\ref{supp:proposition:alg:1:lemma:2},
            we have 
            \begin{align}
                &\rank{\bxrsn{\rrln{1}}{\iconstant}}{\bxrs{\rrln{1}}} = \rank{\bxn{\iconstant}}{\svector{\bx}{\lconstant}{\tonumber{\n} \setminus \bu}} + \rrln{1}, \text{ for any } \iconstant \in \tonumber{\n} \setminus \bu. \label{supp:proposition:alg:2:1:eqn:1}
            \end{align}

            Next, since $\byrs{\rrln{2}} \in \mysecondset{\bx}{\bv}{\tonumber{\n}}{\n}$,
            and $\bxn{1} < \ldots < \bxn{\mln{2}}$, 
            then according to Lemma~\ref{supp:proposition:alg:1:lemma:0},
            for any $\iconstant \in \{1,\ldots, \mln{2}\}$, we have
            \begin{align*}
                \rank{\byrsn{\rrln{2}}{\iconstant}}{\byrs{\rrln{2}}} = \left\{ \begin{array}{ll}
                    \n - \iconstant + 1, & \text{ if } \rrln{2} = 0, \\
                    \qnn{\rrln{2}}{\iconstant} , & \text{ if } \mln{2} > \rrln{2} > 0, \\
                    \mln{2} - \iconstant + 1, & \text{ if } \rr = \mln{2},
                \end{array}\right.
            \end{align*}
            where $\qnn{\rrln{2}}{\iconstant}  = \indicator{{\iconstant} \le {\mln{2} - \rrln{2}}} (\n - \iconstant + 1) 
            + \indicator{{\iconstant} > {\mln{2} - \rrln{2}}} (\mln{2} - \iconstant + 1)$.
            Notice that when $\rrln{2} = 0$ and $\rrln{2} = \mln{2}$,
            $\rank{\byrsn{\rrln{2}}{\iconstant}}{\byrs{\rrln{2}}}  = \qnn{\rrln{2}}{\iconstant}$ is still true
            for any $\iconstant \in \{1,\ldots, \mln{2}\}$. Hence, we have
            \begin{align} \label{supp:proposition:alg:2:1:eqn:2}
                \rank{\byrsn{\rrln{2}}{\iconstant}}{\byrs{\rrln{2}}} = \qnn{\rrln{2}}{\iconstant}, \text{ for any } \iconstant \in \bv, \rrln{2} \in \{0, \ldots, \mln{2}\}.
            \end{align}
            Then, combining \eqref{supp:proposition:alg:2:1:eqn:1}, and \eqref{supp:proposition:alg:2:1:eqn:2},
            we have
            \begin{align*}
                &\sum_{\iconstant \in \bv} \left|\rank{\bxrsn{\rrln{1}}{\iconstant}}{\bxrs{\rrln{1}}}  - \rank{\byrsn{\rrln{2}}{\iconstant}}{\byrs{\rrln{2}}}\right| \\
                &= \sum_{\iconstant\in\bv} \left| \rank{\bxn{\iconstant}}{\svector{\bx}{\lconstant}{\tonumber{\n} \setminus \bu}} + \rrln{1} -  \qnn{\rrln{2}}{\iconstant} \right|, 
            \end{align*}
            which proves equation (1).

            Similarly, we can show equation (2) is true. 
            Since $\byrs{\rrln{2}} \in \mysecondset{\bx}{\bv}{\tonumber{\n}}{\n}$,
            then according to the definition of 
            $\mysecondset{\bx}{\bv}{\tonumber{\n}}{\n}$,
            we have $\byrs{\rrln{2}} 
            \in \myfirstset{\bv}{\tonumber{\n}}{\n}$.
            Since 
            \begin{align*}
                &\sum_{\iconstant \in \bv} \indicator{ \byrsn{\rrln{2}}{\iconstant}   < \min \svector{\byrs{\rrln{2}}}{\lconstant}{\tonumber{\n} \setminus \bv} } = \rrln{2},
            \end{align*}
            and $\byrs{\rrln{2}}$ is an imputation 
            of $\by$ for $\bv$, then according to Lemma~\ref{supp:proposition:alg:1:lemma:2},
            we have 
            \begin{align}
                &\rank{\byrsn{\rrln{2}}{\iconstant}}{\byrs{\rrln{2}}} = \rank{\byn{\iconstant}}{\svector{\by}{\lconstant}{\tonumber{\n} \setminus \bv}} + \rrln{2}, \text{ for any } \iconstant \in \tonumber{\n} \setminus \bv. \label{supp:proposition:alg:2:1:eqn:1.1}
            \end{align}

            Next, since $\bxrs{\rrln{1}} \in \mysecondset{\by}{\bu}{\tonumber{\n}}{\n}$,
            and $\byn{\mln{2} + 1} < \ldots < \byn{\mln{2} + \mln{1}}$, 
            then according to Lemma~\ref{supp:proposition:alg:1:lemma:0},
            for any $\iconstant \in \{1,\ldots, \mln{2}\}$, we have
            \begin{align*}
                \rank{\bxrsn{\rrln{1}}{\mln{2} + \iconstant}}{\bxrs{\rrln{1}}} = \left\{ \begin{array}{ll}
                    \n - \iconstant + 1, & \text{ if } \rrln{1} = 0, \\
                    \pnn{\rrln{1}}{\iconstant} , & \text{ if } \mln{1} > \rrln{1} > 0, \\
                    \mln{1} - \iconstant + 1, & \text{ if } \rr = \mln{1},
                \end{array}\right.
            \end{align*}
            where $\pnn{\rrln{1}}{\iconstant}  = \indicator{\mln{2} + {\iconstant} \le \mln{2} + {\mln{1} - \rrln{1}}} (\n - \iconstant + 1) 
            +  \indicator{\mln{2} + {\iconstant} > \mln{2} + {\mln{1} - \rrln{1}}} (\mln{1} - \iconstant + 1) = \indicator{{\iconstant} \le {\mln{1} - \rrln{1}}} (\n - \iconstant + 1) 
            +  \indicator{ {\iconstant} >  {\mln{1} - \rrln{1}}} (\mln{1} - \iconstant + 1) $.
            Notice that when $\rrln{1} = 0$ and $\rrln{1} = \mln{1}$,
            $\rank{\bxrsn{\rrln{1}}{\mln{2} + \iconstant}}{\bxrs{\rrln{1}}} = \pnn{\rrln{1}}{\iconstant}$ is still true
            for any $\iconstant \in \{1,\ldots, \mln{2}\}$. Hence, we have
            \begin{align*}
                &\rank{\bxrsn{\rrln{1}}{\mln{2} + \iconstant}}{\bxrs{\rrln{1}}} = \pnn{\rrln{1}}{\iconstant}, \text{ for any } \iconstant \in \{1,\ldots, \mln{1}\}, \rrln{1} \in \{0, \ldots, \mln{1}\}.
            \end{align*}
            In other words, for any $\iconstant \in \{\mln{2} + 1,\ldots, \mln{2} + \mln{1}\}, \rrln{1} \in \{0, \ldots, \mln{1}\}$, we have
            \begin{align}  \label{supp:proposition:alg:2:1:eqn:2.1}
                \rank{\bxrsn{\rrln{1}}{\iconstant}}{\bxrs{\rrln{1}}} = \pnn{\rrln{1}}{\iconstant - \mln{2}}. 
            \end{align}
            Then, combining \eqref{supp:proposition:alg:2:1:eqn:1.1}, and \eqref{supp:proposition:alg:2:1:eqn:2.1},
            we have
            \begin{align*}
                &\sum_{\iconstant \in \bu} \left|\rank{\bxrsn{\rrln{1}}{\iconstant}}{\bxrs{\rrln{1}}}  - \rank{\byrsn{\rrln{2}}{\iconstant}}{\byrs{\rrln{2}}}\right|\\
                & = \sum_{\iconstant\in\bv} \left| \pnn{\rrln{1}}{\iconstant - \mln{2}} - \rank{\byn{\iconstant}}{\svector{\by}{\lconstant}{\tonumber{\n} \setminus \bv}} - \rrln{2} \right|, 
            \end{align*}
            which proves equation (2).

            We now show equation (3) is true.
            Since $\byrs{\rrln{2}} \in \mysecondset{\bx}{\bv}{\tonumber{\n}}{\n}$,
            then according to the definition of 
            $\mysecondset{\bx}{\bv}{\tonumber{\n}}{\n}$,
            we have $\byrs{\rrln{2}} 
            \in \myfirstset{\bv}{\tonumber{\n}}{\n}$.
            Since 
            \begin{align*}
                &\sum_{\iconstant \in \bv} \indicator{ \bys^{(\rrln{2})}(\iconstant)   < \min \svector{\bys^{(\rrln{2})}}{\lconstant}{\tonumber{\n} \setminus \bv} } = \rrln{2},
            \end{align*}
            and $\byrs{\rrln{2}}$ is an imputation 
            of $\by$ for $\bv$, then according to Lemma~\ref{supp:proposition:alg:1:lemma:2},
            we have 
            \begin{align}
                &\rank{\byrsn{\rrln{2}}{\iconstant}}{\byrs{\rrln{2}}} = \rank{\byn{\iconstant}}{\svector{\by}{\lconstant}{\tonumber{\n} \setminus \bv}} + \rrln{2}, \text{ for any } \iconstant \in \tonumber{\n} \setminus \bv. \label{supp:proposition:alg:2:1:eqn:3}
            \end{align}
            Then, combining \eqref{supp:proposition:alg:2:1:eqn:1} and \eqref{supp:proposition:alg:2:1:eqn:3}, we have
            \begin{align*}
                &\sum_{\iconstant \in \tonumber{\n} \setminus (\bu \cup \bv)} \left|\rank{\bxrsn{\rrln{1}}{\iconstant}}{\bxrs{\rrln{1}}}  - \rank{\byrsn{\rrln{2}}{\iconstant}}{\byrs{\rrln{2}}}\right| \\
                &= \sum_{\iconstant \in \tonumber{\n} \setminus (\bu \cup \bv)} \left|\rank{\bxn{\iconstant}}{\svector{\bx}{\lconstant}{\tonumber{\n} \setminus \bu}} + \rrln{1} -  \rank{\byn{\iconstant}}{\svector{\by}{\lconstant}{\tonumber{\n} \setminus \bv}} - \rrln{2} \right| \\
                & =  \sum_{\iconstant \in \tonumber{\n} \setminus (\bu \cup \bv)} \left|\rrln{1} - \rrln{2} -d_i \right|.
            \end{align*}
            This proves equation (3), and completes our proof.
        \end{proof}

        Next, we show the following result.
        \begin{proposition} \label{supp:proposition:alg:2}
            Suppose $\bx, \by \in \vndistinct$ and for $\mln{1}, \mln{2} \in \tonumber{\n}$ such
            that $\mln{1} + \mln{2} < \n$, 
            let $\bv = \{1, \ldots, \mln{2}\}$ and $\bu = \{\mln{2} + 1,\ldots, \mln{2} + \mln{1}\}$ be subsets of indices. Suppose
            $\bxn{1} < \ldots < \bxn{\mln{2}}$, and
            $\byn{\mln{2} + 1} < \ldots < \byn{\mln{2} + \mln{1}}$.
            Let $(\bxs^{(\rrln{1})}, \bys^{(\rrln{2})}) \in (\mysecondset{\by}{\bu}{\tonumber{\n}}{\n},  \mysecondset{\bx}{\bv}{\tonumber{\n}}{\n})$ 
            be imputations of $\bx$, $\by$ for indices $\bu$ and $\bv$, 
            respectively, such that 
            \begin{align*}
                &\sum_{\iconstant \in \bu} \indicator{ \bxs^{(\rrln{1})}(\iconstant)   < \min \svector{\bxs^{(\rrln{1})}}{\lconstant}{\tonumber{\n} \setminus \bu} } = \rrln{1} ,\\
                \text{ and } &\sum_{\iconstant \in \bv} \indicator{ \bys^{(\rrln{2})}(\iconstant)   < \min \svector{\bys^{(\rrln{2})}}{\lconstant}{\tonumber{\n} \setminus \bv} } = \rrln{2}.
            \end{align*}
            For any $\rrln{1} \in \{0, \ldots, \mln{1}\}$ and 
            $\rrln{2} \in \{0,\ldots,\mln{2}\}$, define
            \begin{align*}
                &S^{(\rrln{1}, \rrln{2})} = \sum_{\iconstant \in \bv} \indicator{\rank{\bxn{\iconstant}}{\svector{\bx}{\lconstant}{\tonumber{\n} \setminus \bu}} + \rrln{1} -  \qnn{\rrln{2}}{\iconstant} \ge 0},\\
                \text{and } &R^{(\rrln{1}, \rrln{2})} = \sum_{\iconstant \in \bu} \indicator{\pnn{\rrln{1}}{\iconstant - \mln{2}} - \rank{\byn{\iconstant}}{\svector{\by}{\lconstant}{\tonumber{\n} \setminus \bv}} - \rrln{2} -1 \ge 0},
            \end{align*}
            where $\qnn{\rrln{2}}{\iconstant}  = \indicator{{\iconstant} \le {\mln{2} - \rrln{2}}} (\n - \iconstant + 1) 
            + \indicator{{\iconstant} > {\mln{2} - \rrln{2}}} (\mln{2} - \iconstant + 1)$
            and 
            $\pnn{\rrln{1}}{\iconstant} = \indicator{\iconstant \le \mln{1} - \rrln{1}} (\n - \iconstant + 1) + \indicator{\iconstant > \mln{1} - \rrln{1}} (\mln{1} - \iconstant + 1)$.
            For any $\iconstant \in \tonumber{\n} \setminus (\bu \cup \bv)$, denote
            $d_i = \rank{\byn{\iconstant}}{\svector{\by}{\lconstant}{\tonumber{\n} \setminus \bv}} - \rank{\bxn{\iconstant}}{\svector{\bx}{\lconstant}{\tonumber{\n} \setminus \bu}}$, and for any $\iconstant \in \{-\mln{2}, \ldots, \mln{1} - 1\}$, denote $\ssln{\iconstant} = \sum_{\lconstant \in \tonumber{\n} \setminus (\bu \cup \bv)} \indicator{d_l \le \iconstant} $. Denote $\np = \n - \mln{1} - \mln{2}$.
            Then, for any $\rrln{1} \in \{0, \ldots, \mln{1} - 1\}$, $\rrln{2} \in \{0, \ldots, \mln{2}\}$, we have
            \begin{align} 
                &\sfdsb{\bxrs{\rrln{1}+1} }{\byrs{\rrln{2}}} = \sfdsb{\bxrs{\rrln{1}} }{\byrs{\rrln{2}}} + 2\ssln{\rrln{1} - \rrln{2}} + \np + 2S^{(\rrln{1}, \rrln{2})}  + \mln{2} + 	\bcnn{1}{\rrln{1}, \rrln{2}}, \label{supp:proposition:alg:2:eqn:1}
            \end{align}
            and for any $\rrln{1} \in \{0, \ldots, \mln{1} \}$, $\rrln{2} \in \{0, \ldots, \mln{2} - 1\}$, we have
            \begin{align}
                \begin{split}  \label{supp:proposition:alg:2:eqn:2}
                    \sfdsb{\bxrs{\rrln{1}}}{\byrs{\rrln{2}+1}} &= \sfdsb{\bxrs{\rrln{1}} }{\byrs{\rrln{2}}} \\
                    &- 2\ssln{\rrln{1} - \rrln{2} - 1} + \np - 2R^{(\rrln{1}, \rrln{2})}  + \mln{1} + \bcnn{2}{\rrln{1}, \rrln{2}},
                \end{split}
            \end{align}
            where 
            \begin{align*}
                \bcnn{1}{\rrln{1}, \rrln{2}} &= |\rrln{1} + 1 - \rank{\byn{\mln{2} + \mln{1} - \rrln{1}}}{\svector{\by}{\lconstant}{\tonumber{\n} \setminus \bv}} - \rrln{2}| \\
                &- |\n - \mln{1} + \rrln{1} + 1 - \rank{\byn{\mln{2} + \mln{1} - \rrln{1}}}{\svector{\by}{\lconstant}{\tonumber{\n} \setminus \bv}} -\rrln{2}|,  \\
                \bcnn{2}{\rrln{1}, \rrln{2}} &= |\rrln{2} + 1 - \rank{\bxn{\mln{2} - \rrln{2}}}{\svector{\bx}{\lconstant}{\tonumber{\n} \setminus \bu}} - \rrln{1}| \\
                & - \left|\n - \mln{2} + \rrln{2} + 1 - \rank{\bxn{\mln{2} - \rrln{2}}}{\svector{\bx}{\lconstant}{\tonumber{\n} \setminus \bu}} - \rrln{1} \right|.
            \end{align*}

        \end{proposition}

        \begin{proof}
            We first show \eqref{supp:proposition:alg:2:eqn:1} is true. 
            According to Proposition~\ref{supp:proposition:alg:2:1},
            we have
            \begin{align*} 
                \begin{split}
                    \sfdsb{\bxrs{\rrln{1}} }{\byrs{\rrln{2}}} &= \sum_{\iconstant\in\bv} \left| \rank{\bxn{\iconstant}}{\svector{\bx}{\lconstant}{\tonumber{\n} \setminus \bu}} + \rrln{1} -  \qnn{\rrln{2}}{\iconstant} \right| \\
                    &+ \sum_{\iconstant\in\bu} \left| \pnn{\rrln{1}}{\iconstant - \mln{2}} - \rank{\byn{\iconstant}}{\svector{\by}{\lconstant}{\tonumber{\n} \setminus \bv}} - \rrln{2} \right| +  \sum_{\iconstant \in \tonumber{\n} \setminus (\bu \cup \bv)} \left|\rrln{1} - \rrln{2} -  d_i \right|,
                \end{split}
            \end{align*}
            and
            \begin{align*} 
                \begin{split}
                    \sfdsb{\bxrs{\rrln{1} + 1} }{\byrs{\rrln{2}}} &= \sum_{\iconstant\in\bv} \left| \rank{\bxn{\iconstant}}{\svector{\bx}{\lconstant}{\tonumber{\n} \setminus \bu}} + \rrln{1} + 1 -  \qnn{\rrln{2}}{\iconstant} \right| \\
                    &+ \sum_{\iconstant\in\bu} \left| \pnn{\rrln{1} + 1}{\iconstant - \mln{2}} - \rank{\byn{\iconstant}}{\svector{\by}{\lconstant}{\tonumber{\n} \setminus \bv}} - \rrln{2} \right| \\
                    &+  \sum_{\iconstant \in \tonumber{\n} \setminus (\bu \cup \bv)} \left|\rrln{1} + 1 - \rrln{2} -  d_i \right|.
                \end{split}
            \end{align*}
            In order to prove \eqref{supp:proposition:alg:2:eqn:1},
            it is then sufficient to show the following three equations all hold:
            \begin{align}
                \begin{split} \label{supp:proposition:alg:2:eqn:3}
                    & \sum_{\iconstant\in\bv} \left| \rank{\bxn{\iconstant}}{\svector{\bx}{\lconstant}{\tonumber{\n} \setminus \bu}} + \rrln{1} + 1 -  \qnn{\rrln{2}}{\iconstant} \right|  \\
                    & - \sum_{\iconstant\in\bv} \left| \rank{\bxn{\iconstant}}{\svector{\bx}{\lconstant}{\tonumber{\n} \setminus \bu}} + \rrln{1} -  \qnn{\rrln{2}}{\iconstant} \right| = 2S^{(\rrln{1}, \rrln{2})} + \mln{2},
                \end{split}
            \end{align}
            \begin{align}
                \begin{split} \label{supp:proposition:alg:2:eqn:4}
                    &\sum_{\iconstant\in\bu} \left| \pnn{\rrln{1} + 1}{\iconstant - \mln{2}} - \rank{\byn{\iconstant}}{\svector{\by}{\lconstant}{\tonumber{\n} \setminus \bv}} - \rrln{2} \right| \\
                    & - \sum_{\iconstant\in\bu} \left| \pnn{\rrln{1}}{\iconstant - \mln{2}} - \rank{\byn{\iconstant}}{\svector{\by}{\lconstant}{\tonumber{\n} \setminus \bv}} - \rrln{2} \right|  = \bcnn{1}{\rrln{1}, \rrln{2}},
                \end{split}
            \end{align}
            and
            \begin{align}
                \begin{split} \label{supp:proposition:alg:2:eqn:5}
                    \sum_{\iconstant \in \tonumber{\n} \setminus (\bu \cup \bv)} \left|\rrln{1} + 1 - \rrln{2} -  d_i \right| -  \sum_{\iconstant \in \tonumber{\n} \setminus (\bu \cup \bv)} \left|\rrln{1} - \rrln{2} -  d_i \right| = 2\ssln{\rrln{1} - \rrln{2}} + \np.
                \end{split}
            \end{align}

            First, we show \eqref{supp:proposition:alg:2:eqn:3} is true.
            According to Lemma~\ref{supp:proposition:alg:1:lemma:0.0},
            we have
            \begin{align*}
                &\sum_{\iconstant\in\bv} \left| \rank{\bxn{\iconstant}}{\svector{\bx}{\lconstant}{\tonumber{\n} \setminus \bu}} + \rrln{1} + 1 -  \qnn{\rrln{2}}{\iconstant} \right|  - \sum_{\iconstant\in\bv} \left| \rank{\bxn{\iconstant}}{\svector{\bx}{\lconstant}{\tonumber{\n} \setminus \bu}} + \rrln{1} -  \qnn{\rrln{2}}{\iconstant} \right| \\
                & = 2\sum_{\iconstant \in \bv} \indicator{\rank{\bxn{\iconstant}}{\svector{\bx}{\lconstant}{\tonumber{\n} \setminus \bu}} + \rrln{1} -  \qnn{\rrln{2}}{\iconstant} \ge 0} + \mln{2} \\
                & = 2S^{(\rrln{1}, \rrln{2})} + \mln{2}.
            \end{align*}
            This proves \eqref{supp:proposition:alg:2:eqn:3}.

            Next, we show \eqref{supp:proposition:alg:2:eqn:4} is true.
            Notice that
            \begin{align*}
                &\sum_{\iconstant\in\bu} \left| \pnn{\rrln{1} + 1}{\iconstant - \mln{2}} - \rank{\byn{\iconstant}}{\svector{\by}{\lconstant}{\tonumber{\n} \setminus \bv}} - \rrln{2} \right|  - \sum_{\iconstant\in\bu} \left| \pnn{\rrln{1}}{\iconstant - \mln{2}} - \rank{\byn{\iconstant}}{\svector{\by}{\lconstant}{\tonumber{\n} \setminus \bv}} - \rrln{2} \right| \\
                & = \sum_{\iconstant = \mln{2}+1}^{\mln{2} + \mln{1}} \left| \pnn{\rrln{1} + 1}{\iconstant - \mln{2}} - \rank{\byn{\iconstant}}{\svector{\by}{\lconstant}{\tonumber{\n} \setminus \bv}} - \rrln{2} \right| \\ &-\sum_{\iconstant = \mln{2}+1}^{\mln{2} + \mln{1}} \left| \pnn{\rrln{1}}{\iconstant - \mln{2}} - \rank{\byn{\iconstant}}{\svector{\by}{\lconstant}{\tonumber{\n} \setminus \bv}} - \rrln{2} \right| \\
                & = \sum_{\iconstant = 1}^{\mln{1}} \left| \pnn{\rrln{1} + 1}{\iconstant} - \rank{\byn{\mln{2} + \iconstant}}{\svector{\by}{\lconstant}{\tonumber{\n} \setminus \bv}} - \rrln{2} \right| \\ &-\sum_{\iconstant = 1}^{\mln{1}} \left| \pnn{\rrln{1}}{\iconstant } - \rank{\byn{\mln{2} + \iconstant}}{\svector{\by}{\lconstant}{\tonumber{\n} \setminus \bv}} - \rrln{2} \right|. 
            \end{align*}
            Then, according to Lemma~\ref{supp:proposition:alg:2:lemma:3}, 
            we have 
            \begin{align*}
                &\sum_{\iconstant\in\bu} \left| \pnn{\rrln{1} + 1}{\iconstant} - \rank{\byn{\iconstant}}{\svector{\by}{\lconstant}{\tonumber{\n} \setminus \bv}} - \rrln{2} \right|  - \sum_{\iconstant\in\bu} \left| \pnn{\rrln{1}}{\iconstant} - \rank{\byn{\iconstant}}{\svector{\by}{\lconstant}{\tonumber{\n} \setminus \bv}} - \rrln{2} \right| \\
                & = |\rrln{1} + 1 - \rank{\byn{\mln{2} + \mln{1} - \rrln{1}}}{\svector{\by}{\lconstant}{\tonumber{\n} \setminus \bv}} -\rrln{2}| \\
                &- |\n - \mln{1} + \rrln{1} + 1 - \rank{\byn{\mln{2} + \mln{1} - \rrln{1}}}{\svector{\by}{\lconstant}{\tonumber{\n} \setminus \bv}} -\rrln{2}| \\
                & = \bcnn{1}{\rrln{1}, \rrln{2}}. 
            \end{align*}
            This proves \eqref{supp:proposition:alg:2:eqn:4}.

            Next, we show \eqref{supp:proposition:alg:2:eqn:5} is true.
            According to Lemma~\ref{supp:proposition:alg:1:lemma:0.0},
            we have
            \begin{align*}
                &\sum_{\iconstant \in \tonumber{\n} \setminus (\bu \cup \bv)} \left|\rrln{1} + 1 - \rrln{2} -  d_i \right| -  \sum_{\iconstant \in \tonumber{\n} \setminus (\bu \cup \bv)} \left|\rrln{1} - \rrln{2} -  d_i \right|  \\
                & = 2  \sum_{\iconstant \in \tonumber{\n} \setminus (\bu \cup \bv)} \indicator{\rrln{1} - \rrln{2} -  d_i \ge 0} +  \n - \mln{1} - \mln{2}\\
                & = 2\sum_{\iconstant \in \tonumber{\n} \setminus (\bu \cup \bv)} \indicator{d_i \le \rrln{1} - \rrln{2}} + \np \\
                & = 2\tln{\rrln{1} - \rrln{2}} + \np.
            \end{align*}
            This proves \eqref{supp:proposition:alg:2:eqn:5}
            and completes our proof for \eqref{supp:proposition:alg:2:eqn:1}.

            Similarly, we can prove \eqref{supp:proposition:alg:2:eqn:2} is true.
            According to Proposition~\ref{supp:proposition:alg:2:1},
            we have
            \begin{align*} 
                \begin{split}
                    \sfdsb{\bxrs{\rrln{1}} }{\byrs{\rrln{2}}} &= \sum_{\iconstant\in\bv} \left| \rank{\bxn{\iconstant}}{\svector{\bx}{\lconstant}{\tonumber{\n} \setminus \bu}} + \rrln{1} -  \qnn{\rrln{2}}{\iconstant} \right| \\
                    &+ \sum_{\iconstant\in\bu} \left| \pnn{\rrln{1}}{\iconstant} - \rank{\byn{\iconstant}}{\svector{\by}{\lconstant}{\tonumber{\n} \setminus \bv}} - \rrln{2} \right| +  \sum_{\iconstant \in \tonumber{\n} \setminus (\bu \cup \bv)} \left|\rrln{1} - \rrln{2} -  d_i \right|,
                \end{split}
            \end{align*}
            and
            \begin{align*} 
                \begin{split}
                    \sfdsb{\bxrs{\rrln{1}} }{\byrs{\rrln{2} + 1}} &= \sum_{\iconstant\in\bv} \left| \rank{\bxn{\iconstant}}{\svector{\bx}{\lconstant}{\tonumber{\n} \setminus \bu}} + \rrln{1} -  \qnn{\rrln{2} + 1}{\iconstant} \right| \\
                    &+ \sum_{\iconstant\in\bu} \left| \pnn{\rrln{1}}{\iconstant} - \rank{\byn{\iconstant}}{\svector{\by}{\lconstant}{\tonumber{\n} \setminus \bv}} - \rrln{2} - 1 \right|\\
                    &+  \sum_{\iconstant \in \tonumber{\n} \setminus (\bu \cup \bv)} \left|\rrln{1} - \rrln{2} - 1 -  d_i \right|.
                \end{split}
            \end{align*}
            Then, in order to prove \eqref{supp:proposition:alg:2:eqn:2},
            it is sufficient to show the following three equations all hold:
            \begin{align}
                \begin{split} \label{supp:proposition:alg:2:eqn:6}
                    & \sum_{\iconstant\in\bv} \left| \rank{\bxn{\iconstant}}{\svector{\bx}{\lconstant}{\tonumber{\n} \setminus \bu}} + \rrln{1} -  \qnn{\rrln{2}+1}{\iconstant} \right|  \\
                    & - \sum_{\iconstant\in\bv} \left| \rank{\bxn{\iconstant}}{\svector{\bx}{\lconstant}{\tonumber{\n} \setminus \bu}} + \rrln{1} -  \qnn{\rrln{2}}{\iconstant} \right| =  \bcnn{2}{\rrln{1}, \rrln{2}},
                \end{split}
            \end{align}
            \begin{align}
                \begin{split} \label{supp:proposition:alg:2:eqn:7}
                    &\sum_{\iconstant\in\bu} \left| \pnn{\rrln{1}}{\iconstant} - \rank{\byn{\iconstant}}{\svector{\by}{\lconstant}{\tonumber{\n} \setminus \bv}} - \rrln{2} -1 \right| \\
                    & - \sum_{\iconstant\in\bu} \left| \pnn{\rrln{1}}{\iconstant} - \rank{\byn{\iconstant}}{\svector{\by}{\lconstant}{\tonumber{\n} \setminus \bv}} - \rrln{2} \right|  =  \mln{1} - 2R^{(\rrln{1}, \rrln{2})},
                \end{split}
            \end{align}
            and
            \begin{align}
                \begin{split} \label{supp:proposition:alg:2:eqn:8}
                    \sum_{\iconstant \in \tonumber{\n} \setminus (\bu \cup \bv)} \left|\rrln{1}  - \rrln{2} -  d_i \right| -  \sum_{\iconstant \in \tonumber{\n} \setminus (\bu \cup \bv)} \left|\rrln{1} - \rrln{2} -  d_i - 1\right| = \np - 2\tln{\rrln{1} - \rrln{2} - 1}.
                \end{split}
            \end{align}

            First, we show \eqref{supp:proposition:alg:2:eqn:6} is true.
            Notice that
            \begin{align*}
                \qnn{\rrln{2}}{\iconstant}  = \indicator{{\iconstant} \le {\mln{2} - \rrln{2}}} (\n - \iconstant + 1) 
                + \indicator{{\iconstant} > {\mln{2} - \rrln{2}}} (\mln{1} - \iconstant + 1),
            \end{align*}
            for any $\iconstant \in \{1, \ldots, \mln{2} \}$ and $\rrln{2} \in \{0, \ldots, \mln{2}\}$.
            Then, denote $\uuln{\iconstant} = \iconstant$ for any $\iconstant = 0,\ldots,\mln{2}$. We have
            \begin{align*}
                \qnn{\rrln{1}}{\uuln{\iconstant}} = \indicator{\uuln{\iconstant} \le \uuln{\mln{2} - \rrln{2}}} (\n - \iconstant + 1) + \indicator{\uuln{\iconstant} > \uuln{\mln{2} - \rrln{2}}} (\mln{2} - \iconstant + 1).
            \end{align*}
            Then, according to Lemma~\ref{supp:proposition:alg:2:lemma:3}, we have
            \begin{align*}
                \begin{split}
                    &\sum_{\iconstant \in \bv} \left|\rank{\bxn{\iconstant}}{\svector{\bx}{\lconstant}{\tonumber{\n} \setminus \bu}} + \rrln{1} -  \qnn{\rrln{2} + 1}{\iconstant} \right| - \sum_{\iconstant \in \bv} \left|\rank{\bxn{\iconstant}}{\svector{\bx}{\lconstant}{\tonumber{\n} \setminus \bu}} + \rrln{1} -  \qnn{\rrln{2}}{\iconstant} \right| \\
                    & = \sum_{\iconstant \in \bv} \left|\qnn{\rrln{2} + 1}{\iconstant} - \rank{\bxn{\iconstant}}{\svector{\bx}{\lconstant}{\tonumber{\n} \setminus \bu}} - \rrln{1} \right| - \sum_{\iconstant \in \bv} \left|\qnn{\rrln{2}}{\iconstant} - \rank{\bxn{\iconstant}}{\svector{\bx}{\lconstant}{\tonumber{\n} \setminus \bu}} - \rrln{1} \right| \\
                    & = |\rrln{2} + 1 - \rank{\bxn{\mln{2} - \rrln{2}}}{\svector{\bx}{\lconstant}{\tonumber{\n} \setminus \bu}} - \rrln{1}| \\
                    & - \left|\n - \mln{2} + \rrln{2} + 1 - \rank{\bxn{\mln{2} - \rrln{2}}}{\svector{\bx}{\lconstant}{\tonumber{\n} \setminus \bu}} - \rrln{1} \right|\\
                    & = \bcnn{2}{\rrln{1}, \rrln{2}}.
                \end{split}
            \end{align*}
            This proves \eqref{supp:proposition:alg:2:eqn:6}.

            Next, we show \eqref{supp:proposition:alg:2:eqn:7} is true.
            According to Lemma~\ref{supp:proposition:alg:1:lemma:0.0}, we have
            \begin{align*}
                &\sum_{\iconstant\in\bu} \left| \pnn{\rrln{1}}{\iconstant} - \rank{\byn{\iconstant}}{\svector{\by}{\lconstant}{\tonumber{\n} \setminus \bv}} - \rrln{2} -1 \right| \\
                & - \sum_{\iconstant\in\bu} \left| \pnn{\rrln{1}}{\iconstant} - \rank{\byn{\iconstant}}{\svector{\by}{\lconstant}{\tonumber{\n} \setminus \bv}} - \rrln{2} \right| \\
                & = \mln{1} - 2\sum_{\iconstant \in \bu} \indicator{\pnn{\rrln{1}}{\iconstant} - \rank{\byn{\iconstant}}{\svector{\by}{\lconstant}{\tonumber{\n} \setminus \bv}} - \rrln{2} -1 \ge 0}\\
                & =  \mln{1} - 2R^{(\rrln{1}, \rrln{2})}.
            \end{align*}
            This proves \eqref{supp:proposition:alg:2:eqn:7}.

            Next, we show \eqref{supp:proposition:alg:2:eqn:8} is true.
            According to Lemma~\ref{supp:proposition:alg:1:lemma:0.0}, we have
            \begin{align*}
                &\sum_{\iconstant \in \tonumber{\n} \setminus (\bu \cup \bv)} \left|\rrln{1}  - \rrln{2} -  d_i \right| -  \sum_{\iconstant \in \tonumber{\n} \setminus (\bu \cup \bv)} \left|\rrln{1} - \rrln{2} -  d_i - 1\right| \\
                & = \n - \mln{1} - \mln{2} - 2  \sum_{\iconstant \in \tonumber{\n} \setminus (\bu \cup \bv)} \indicator{\rrln{1} - \rrln{2} -  d_i - 1\ge 0}\\
                & = \np - 2\sum_{\iconstant \in \tonumber{\n} \setminus (\bu \cup \bv)} \indicator{d_i \le \rrln{1} - \rrln{2} - 1} \\
                & = \np - 2\tln{\rrln{1} - \rrln{2} - 1}.
            \end{align*}
            This proves \eqref{supp:proposition:alg:2:eqn:8}
            and completes our proof for \eqref{supp:proposition:alg:2:eqn:2}.
            Hence, we finish our proof.

        \end{proof}

        \begin{algorithm}
            \caption{An efficient algorithm for computing exact upper bounds of Spearman's Footrule under  Missing Case II}  \label{supp:alg:2}
            \begin{algorithmic}[1]
                \Require{$\bx, \by \in \vndistinct$, where $\bx$ and $\by$ might be
                partially observed, but for any pair $(\bxn{\iconstant}, \byn{\iconstant})$,
                where $\iconstant \in \tonumber{\n}$, at least one value is observed.} 
                \Ensure{Maximum possible Spearman's footrule distance between $X$ and $Y$.}
                \State Denote the number of unobserved components in $\bx$, $\by$ as $\mln{1}$
                and $\mln{2}$, respectively. If at least $\mln{1} = 0$ or $\mln{2} = 0$ is true, run  Algorithm~\ref{supp:alg:1}. If $\mln{1} + \mln{2} = \n$, then return $\sum_{\iconstant=1}^{\n} |\iconstant - (\n -\iconstant + 1)|$.
                \State Rank all observed in $\bx$ and $\by$. 
                Relabel $\bx$ and $\by$ such that $(\byn{1}, \ldots, \byn{\mln{2}})$, and $(\bxn{\mln{2}+1},\ldots, \bxn{\mln{2}+\mln{1}})$ are unobserved, and $\bxn{1} < \ldots <\bxn{\mln{2}}$, $\byn{\mln{1}} < \ldots <\byn{\mln{2} + \mln{1}}$.\label{supp:alg:2:line:1}
                \State Let $\bv = \{1, \ldots, \mln{2}\}$, and $\bu = \{\mln{2}+1, \ldots, \mln{2} + \mln{1}\}$.
                \State For $\iconstant \in \tonumber{\n} \setminus (\bu \cup \bv)$, let $\dln{\iconstant} = \rank{\byn{\iconstant}}{\svector{\by}{\lconstant}{\tonumber{\n} \setminus \bv}} - \rank{\bxn{\iconstant}}{\svector{\bx}{\lconstant}{\tonumber{\n} \setminus \bu}}.$ \label{supp:alg:2:line:2}
                \State For any $\iconstant \in \{1, \ldots, \mln{2}\}$ and $\rrln{2} \in \{0, \ldots, \mln{2}\}$, let  $\qnn{\rrln{2}}{\iconstant}  = \indicator{{\iconstant} \le {\mln{2} - \rrln{2}}} (\n - \iconstant + 1) + \indicator{{\iconstant} > {\mln{2} - \rrln{2}}} (\mln{2} - \iconstant + 1)$. \label{supp:alg:2:line:3}
                \State  For any $\iconstant \in \{1, \ldots, \mln{1}\}$, let $\pnn{0}{\iconstant} = \indicator{\iconstant \le \mln{1}} (\n - \iconstant + 1) + \indicator{\iconstant > \mln{1}} (\mln{1} - \iconstant + 1)$. \label{supp:alg:2:line:4}
                \State Run Algorithm~\ref{supp:alg:0} for computing $\ssln{\iconstant} = \sum_{\lconstant \in \tonumber{\n} \setminus (\bu \cup \bv)} \indicator{d_l \le \iconstant}$ for any $\iconstant \in \{-\mln{2}, \ldots, \mln{1} -1\}$.\label{supp:alg:2:line:5}
                \State Run Algorithm~\ref{supp:alg:0} for computing \label{supp:alg:2:line:6}
                \begin{align*}
                    R^{(0, \rrln{2})} = \sum_{\iconstant \in \bu} \indicator{- \pnn{0}{\iconstant - \mln{2}} + \rank{\byn{\iconstant}}{\svector{\by}{\lconstant}{\tonumber{\n} \setminus \bv}} + 1 \le - \rrln{2}}, \text{ for any } \rrln{2} \in \{0, \ldots, \mln{2}\}.
                \end{align*}
                \State For any given $\rrln{2} \in \{0, \ldots, \mln{2}\}$, run Algorithm~\ref{supp:alg:0} for computing \label{supp:alg:2:line:7}
                \begin{align*}
                    S^{(\rrln{1}, \rrln{2})} = \sum_{\iconstant \in \bv} \indicator{ \qnn{\rrln{2}}{\iconstant} - \rank{\bxn{\iconstant}}{\svector{\bx}{\lconstant}{\tonumber{\n} \setminus \bu}} \le \rrln{1}}, \text{ for any } \rrln{1} \in \{0, \ldots, \mln{1}\}.
                \end{align*}
                \State Initialize  \label{supp:alg:2:line:8}
                \begin{align*}
                    \bdln{0,0} &= \sum_{\iconstant\in\bv} \left| \rank{\bxn{\iconstant}}{\svector{\bx}{\lconstant}{\tonumber{\n} \setminus \bu}} + \rrln{1} -  (\n - \iconstant + 1) \right| \\
                    &+ \sum_{\iconstant\in\bu} \left| (\n - \iconstant + \mln{2} +  1)  - \rank{\byn{\iconstant}}{\svector{\by}{\lconstant}{\tonumber{\n} \setminus \bv}} - \rrln{2} \right| +  \sum_{\iconstant \in \tonumber{\n} \setminus (\bu \cup \bv)} \left|\rrln{1} - \rrln{2} -  d_i \right|. 
                \end{align*}

                \algstore{alg:33:part:2}
                \end{algorithmic}
                
            \end{algorithm}

\begin{algorithm} 
\begin{algorithmic}[1]
\algrestore{alg:33:part:2}

                \For{$\rrln{2} = 0,\ldots,\mln{2}$}		 \label{supp:alg:2:line:8.1}		
                \For{$\rrln{1} = 0,\ldots,\mln{1} - 1$} \label{supp:alg:2:line:8.2}
                \State Compute  \label{supp:alg:2:line:9}
                \begin{align*}
                    \bcnn{1}{\rrln{1}, \rrln{2}} &= |\rrln{1} + 1 - \rank{\byn{\mln{2} + \mln{1} - \rrln{1}}}{\svector{\by}{\lconstant}{\tonumber{\n} \setminus \bv}} - \rrln{2}| \\
                    &- |\n - \mln{1} + \rrln{1} + 1 - \rank{\byn{\mln{2} + \mln{1} - \rrln{1}}}{\svector{\by}{\lconstant}{\tonumber{\n} \setminus \bv}} -\rrln{2}|.
                \end{align*}
                \State Compute $\bdln{\rrln{1}+1, \rrln{2}} = \bdln{\rrln{1}, \rrln{2}} + 2\ssln{\rrln{1} - \rrln{2}} + \np + 2S^{(\rrln{1}, \rrln{2})}  + \mln{2} + 	\bcnn{1}{\rrln{1}, \rrln{2}}$.  \label{supp:alg:2:line:10}

                \EndFor
                \If{$\rrln{2} < \mln{2}$}
                \State Let $\rrln{1} = 0$ and compute \label{supp:alg:2:line:11}
                \begin{align*}
                    \bcnn{2}{\rrln{1}, \rrln{2}} &= |\rrln{2} + 1 - \rank{\bxn{\mln{2} - \rrln{2}}}{\svector{\bx}{\lconstant}{\tonumber{\n} \setminus \bu}} - \rrln{1}| \\
                    & - \left|\n - \mln{2} + \rrln{2} + 1 - \rank{\bxn{\mln{2} - \rrln{2}}}{\svector{\bx}{\lconstant}{\tonumber{\n} \setminus \bu}} - \rrln{1} \right|.
                \end{align*}
                \State Compute $\bdln{\rrln{1}, \rrln{2} + 1} = \bdln{\rrln{1}, \rrln{2}} - 2\ssln{\rrln{1} - \rrln{2} - 1} + \np - 2R^{(\rrln{1}, \rrln{2})}  + \mln{1} + \bcnn{2}{\rrln{1}, \rrln{2}}$.  \label{supp:alg:2:line:12}

                \EndIf
                \EndFor
                \State Return $\max \{\bdln{0,0}, \ldots, \bdln{\mln{1},\mln{2}}\}$.
            \end{algorithmic}
        \end{algorithm}	

        \begin{remark}
            A few comments of Algorithm~\ref{supp:alg:2} are made below:

            In Algorithm~\ref{supp:alg:2}, the initialization $\bdln{0, 0}$ computed in
            line~\ref{supp:alg:2:line:8} equals to $\sfdsb{\bxrs{0}}{\byrs{0}}$ defined in 
            Proposition~\ref{supp:proposition:alg:2:1}.
            Spearman's footrule
            $\bdln{\rrln{1} + 1, \rrln{2}}$, $\bdln{\rrln{1}, \rrln{2} + 1}$,
            and $\bdln{\rrln{1}, \rrln{2}}$ equal to 
            $\sfdsb{\bxrs{\rrln{1} + 1}}{\byrs{\rrln{2}}}$, 
            $\sfdsb{\bxrs{\rrln{1}}}{\byrs{\rrln{2} + 1}}$,
            and $\sfdsb{\bxrs{\rrln{1}}}{\byrs{\rrln{2}}}$,
            respectively, and are updated according to
            Proposition~\ref{supp:proposition:alg:2}.

            By computing $\{\bdln{0,0}, \ldots, \bdln{\mln{1}, \mln{2}}\}$, Algorithm~\ref{supp:alg:2} 
            finds all possible Spearman's footrule
            values $\sfd{\bxs}{\bys}$, where $\bxs \in \mysecondset{\by}{\bu}{\tonumber{\n}}{\n}$,
            $\bys \in \mysecondset{\bx}{\bv}{\tonumber{\n}}{\n}$ are
            imputations of $\bx, \by$ for indices $\bu$, and $\bv$, respectively.
            See discussions after Theorem 2.17 for explanations.
            Hence, according to Theorem~\ref{supp:theorem:2.17},
            the algorithm guarantees to find the maximum possible Spearman's footrule
            between $\bx$ and $\by$.

            The computational complexity of Algorithm~\ref{supp:alg:2} is analyzed as follows. 
            Ranking and relabeling all observed components in $\bx$ and $\by$ in line~\ref{supp:alg:2:line:1} 
            requires $\bmo (\n \log \n)$ steps. 
            Using these rankings, in line~\ref{supp:alg:2:line:2} computing each $\dln{\iconstant}$ is $\bmo(1)$, 
            and so overall line~\ref{supp:alg:2:line:2} is $\bmo(\n - \mln{1} - \mln{2})$.
            Line~\ref{supp:alg:2:line:3} and line~\ref{supp:alg:2:line:4} takes $\bmo (\mln{2}^2)$
            and $\bmo (\mln{1})$ steps, respectively.

            According to Remark~\ref{supp:remark:alg:0}, the computational
            complexity of running Algorithm~\ref{supp:alg:0} in 
            line~\ref{supp:alg:2:line:5} and line~\ref{supp:alg:2:line:6} is $\bmo(\mln{1} + \mln{2})$,
            and $\bmo(\mln{2})$, respectively. 		
            In line~\ref{supp:alg:2:line:7}, each iteration of running Algorithm~\ref{supp:alg:0} 
            within the for loop is $\bmo(\mln{1})$. Since the loop runs $\mln{2}$ times, 
            the computational complexity for the loop is $\bmo(\mln{1}\mln{2} )$. 

            Line~\ref{supp:alg:2:line:8} requires $\bmo (\n)$ steps using the ranks of observed components.
            Line~\ref{supp:alg:2:line:9}, line~\ref{supp:alg:2:line:10},
            line~\ref{supp:alg:2:line:11}, and line~\ref{supp:alg:2:line:12}
            all take $\bmo(1)$ steps, and since the loop in \ref{supp:alg:2:line:8.2}
            runs $\mln{1}$ times, 
            the computational complexity for the loop is $\bmo(\mln{1})$. 
            Further, since the loop in \ref{supp:alg:2:line:8.1}
            runs $\mln{2} + 1$ times, 
            the computational complexity for the loop is $\bmo(\mln{1}\mln{2})$.		
            Therefore, the overall computational complexity for Algorithm~\ref{supp:alg:2} is $\bmo(\n\log \n + \mln{1}\mln{2} + \mln{2}^2)$.
        \end{remark}

        \subsection{Missing Case III}

        This section provides an efficient algorithm for computing
        exact upper bounds of Spearman's footrule under 
        missing case III, where $\bx$ and $\by$ might be
        partially observed, and for any pair $(\bxn{\iconstant}, \byn{\iconstant})$,
        where $\iconstant \in \tonumber{\n}$, the two value
        are either both observed, or both missing.

        To start, we prove the following lemma:
        \begin{lemma} \label{supp:proposition:alg:3:lemma:1}
            Suppose $\bx, \by \in \vndistinct$ and for $0 < \mln{3} < \n$, 
            let $\bw = \{\n - \mln{3} + 1, \ldots ,\n\} \subset  \tonumber{\n}$ 
            be a subset of indices. If $(\bx, \by) \in \mythirdset{\bw}{\n}$ 
            is such that
            \begin{align*}
                \sum_{\iconstant \in \bw} \indicator{ \bx^{(\rr)}(\iconstant)   < \min \svector{\bx^{(\rr)}}{\lconstant}{\tonumber{\n} \setminus \bw} } = \rr,
            \end{align*}
            then we have $\by \in \myfirstset{\bw}{\tonumber{\n}}{\n}$,
            $\sum_{\iconstant \in \bw} \indicator{ \by(\iconstant)   < \min \svector{\by}{\lconstant}{\tonumber{\n} \setminus \bw} } = \mln{3} - \rr$,
            and
            \begin{align} \label{supp:proposition:alg:3:lemma:1:eqn:0.0}
                \begin{split}
                    \sum_{\iconstant \in \bw} \left|\rank{\bxn{{\iconstant}}}{\bx} - \rank{\byn{{\iconstant}}}{\by} \right| &= \sum_{\iconstant=1}^{\mln{3}} \left(\indicator{\iconstant \le \mln{3} - \rr} |\n + 1 - 2\iconstant| \right)\\
                    &+ \sum_{\iconstant=1}^{\mln{3}}\left(\indicator{\iconstant > \mln{3} - \rr} |2\mln{3} - 2\iconstant + 1 -\n|\right).
                \end{split}
            \end{align}
        \end{lemma}

        \begin{proof}
            Without loss of generality, let us assume (after relabeling)
            $\byn{\n - \mln{3} + 1} < \ldots < \byn{\n}$.

            Since $(\bx, \by) \in \mythirdset{\bw}{\n}$, then 
            we have $\bx \in \myfirstset{\bw}{\tonumber{\n}}{\n}$,
            and 
            \begin{align} \label{supp:proposition:alg:3:lemma:1:eqn:0}
                &\rank{\bxn{\iconstant}}{\bx} + \rank{\byn{\iconstant}}{\by}= \n + 1, \text{ for any } \iconstant \in \bw.
            \end{align}
            Then, for any $\iconstant, \jconstant \in \bw $, we have 
            \begin{align*}
                \text{ if } \rank{\bxn{\iconstant}}{\bx} > \rank{\bxn{\jconstant}}{\bx}, \text{ then } 
                \rank{\byn{\iconstant}}{\by} < \rank{\byn{\jconstant}}{\by}.
            \end{align*}
            Thus, we have
            \begin{align*}
                \bxn{\iconstant} > \bxn{\jconstant} \Rightarrow \byn{\iconstant} > \byn{\jconstant}, \text{ for any } \iconstant, \jconstant \in \bw.
            \end{align*}
            Hence, we have shown that $\bx \in \mysecondset{\by}{\bw}{\tonumber{\n}}{\n}$.
            Then, according to Lemma~\ref{supp:proposition:alg:1:lemma:0}, 
            for any $\iconstant \in \{1,\ldots, \mln{3}\}$, we have		
            \begin{align} \label{supp:proposition:alg:3:lemma:1:eqn:1}
                \rank{\bxrsn{\rr}{{\n - \mln{3} + \iconstant}}}{\bxrs{\rr}}
                = \left\{ \begin{array}{ll}
                    \n - \iconstant + 1, & \text{ if } \rr = 0, \\
                    q^{(\rr)}  , & \text{ if } \mln{3} > \rr > 0, \\
                    \mln{3} - \iconstant + 1, & \text{ if } \rr = \mln{3},
                \end{array}\right.
            \end{align}
            where $q^{(\rr)} = \indicator{{\n - \mln{3} + \iconstant} \le {\n - \rr}} (\n - \iconstant + 1) 
            + \indicator{\n -\mln{3} + \iconstant > \n - \rr} (\mln{3} - \iconstant + 1)$.

            In the following, we prove that the following three statements:
            \begin{align*}
                &(\rn{i}): \by \in \myfirstset{\bw}{\tonumber{\n}}{\n},\\
                &(\rn{ii}): \sum_{\iconstant \in \bw} \indicator{ \by(\iconstant)   < \min \svector{\by}{\lconstant}{\tonumber{\n} \setminus \bw} } = \mln{3} - \rr,\\
                &(\rn{iii}): \text{ equation } \eqref{supp:proposition:alg:3:lemma:1:eqn:0.0} \text{ holds,}
            \end{align*} 
            are true when $\rr = 0$, $\mln{3} > \rr > 0$ and $\rr =\mln{3}$, separately.

            Suppose $\rr = 0$.  Then according to \eqref{supp:proposition:alg:3:lemma:1:eqn:1},
            we have 
            \begin{align*}
                \rank{\bxn{{\n - \mln{3} + \iconstant}}}{\bx} = \n - \iconstant + 1, \text{ for any }\iconstant \in \{1,\ldots, \mln{3}\}.
            \end{align*}
            According to \eqref{supp:proposition:alg:3:lemma:1:eqn:0}, we further have 
            \begin{align*}
                &\rank{\byn{{\n - \mln{3} + \iconstant}}}{\by} = \iconstant, \text{ for any } \iconstant \in \{1,\ldots, \mln{3}\}.
            \end{align*}	
            Since the $\mln{3}$ components of $\bw$ takes the ranks from $1$ to $\mln{3}$,
            then we have
            \begin{align*}
                \by(\iconstant)   < \min \svector{\by}{\lconstant}{\tonumber{\n} \setminus \bw}, \text{ for any } \iconstant \in \bw.
            \end{align*}
            Hence, we have $\by \in \myfirstset{\bw}{\tonumber{\n}}{\n}$
            and $\sum_{\iconstant \in \bw} \indicator{ \by(\iconstant)   < \min \svector{\by}{\lconstant}{\tonumber{\n} \setminus \bw} } = \mln{3} = \mln{3} - \rr$.
            Further, for any $ \iconstant \in \{1,\ldots, \mln{3}\}$, we have
            \begin{align*}
                \rank{\bxn{{\n - \mln{3} + \iconstant}}}{\bx} - \rank{\byn{{\n - \mln{3} + \iconstant}}}{\by} = \n - \iconstant + 1 - \iconstant = \n + 1 - 2\iconstant,
            \end{align*}
            which gives
            \begin{align*}
                \sum_{\iconstant \in \bw} \left|\rank{\bxn{{\iconstant}}}{\bx} - \rank{\byn{{\iconstant}}}{\by} \right| &= \sum_{\iconstant=1}^{\mln{3}} |\n + 1 - 2\iconstant|.
            \end{align*}		
            This proves our result when $\rr = 0$.

            Suppose $\mln{3} > \rr > 0$.  Then according to \eqref{supp:proposition:alg:3:lemma:1:eqn:1},
            we have 
            \begin{align*}
                &\rank{\bxn{{\n - \mln{3} + \iconstant}}}{\bx} = \n - \iconstant + 1, \text{ for any } \iconstant \in \{1, \ldots, \mln{3} - \rr\},\\
                \text{and }&\rank{\bxn{{\n - \mln{3} + \iconstant}}}{\bx} = \mln{3} - \iconstant + 1, \text{ for any } \iconstant \in \{\mln{3} - \rr + 1, \ldots, \mln{3}\}.
            \end{align*}	
            Then according to \eqref{supp:proposition:alg:3:lemma:1:eqn:0}, we have
            \begin{align*}
                &\rank{\byn{{\n - \mln{3} + \iconstant}}}{\by} = \iconstant, \text{ for any } \iconstant \in \{1, \ldots, \mln{3} - \rr\},\\
                \text{and }&\rank{\byn{{\n - \mln{3} + \iconstant}}}{\by} = \n - \mln{3} + \iconstant, \text{ for any } \iconstant \in \{\mln{3} - \rr + 1, \ldots, \mln{3}\},
            \end{align*}
            Since the $\mln{3}$ components of $\bw$ takes either the ranks from $1$ to $\mln{3} - \rr$,
            or the ranks from $\n - \rr + 1$ to $\n$,
            then we have
            \begin{align*}
                \by(\iconstant)   < \min \svector{\by}{\lconstant}{\tonumber{\n} \setminus \bw}, \text{ or } \by(\iconstant)   > \min \svector{\by}{\lconstant}{\tonumber{\n} \setminus \bw}, \text{ for any } \iconstant \in \bw.
            \end{align*}
            Hence, we have $\by \in \myfirstset{\bw}{\tonumber{\n}}{\n}$
            and $\sum_{\iconstant \in \bw} \indicator{ \by(\iconstant)   < \min \svector{\by}{\lconstant}{\tonumber{\n} \setminus \bw} } = \mln{3} - \rr$.
            Further, for any $\iconstant \in \{1, \ldots, \mln{3} - \rr\}$, we have
            \begin{align*}
                &\rank{\bxn{{\n - \mln{3} + \iconstant}}}{\bx} - \rank{\byn{{\n - \mln{3} + \iconstant}}}{\by} = \n + 1 - 2\iconstant,
            \end{align*}
            and for any $\iconstant \in \{\mln{3} - \rr + 1, \ldots, \mln{3}\}$, we have
            \begin{align*}
                \rank{\bxn{{\n - \mln{3} + \iconstant}}}{\bx} - \rank{\byn{{\n - \mln{3} + \iconstant}}}{\by} = 2\mln{3} - 2\iconstant + 1 - \n.
            \end{align*}
            Hence, we have
            \begin{align*}
                &\sum_{\iconstant \in \bw} \left|\rank{\bxn{{\iconstant}}}{\bx} - \rank{\byn{{\iconstant}}}{\by} \right| \\
                & = \sum_{\iconstant = 1}^{\mln{3} - \rr} |\n + 1 - 2\iconstant| + \sum_{\iconstant = \mln{3} - \rr + 1}^{\mln{3} } |2\mln{3} - 2\iconstant + 1 -\n|.
            \end{align*}
            This proves our result when $\mln{3} > \rr > 0$.

            Suppose $\rr = \mln{3}$.  Then according to \eqref{supp:proposition:alg:3:lemma:1:eqn:1},
            we have 
            \begin{align*}
                \rank{\bxn{{\n - \mln{3} + \iconstant}}}{\bx} = \mln{3} - \iconstant + 1, \text{ for any } \iconstant \in \{1,\ldots, \mln{3}\}.
            \end{align*}
            Then according to \eqref{supp:proposition:alg:3:lemma:1:eqn:0}, we have 
            \begin{align*}
                &\rank{\byn{{\n - \mln{3} + \iconstant}}}{\by} = \n - \mln{3} + \iconstant, \text{ for any } \iconstant \in \{1,\ldots, \mln{3}\}.
            \end{align*}	
            Since the $\mln{3}$ components of $\bw$ takes the ranks from $\n - \mln{3} + 1$ to $\n$,
            then we have
            \begin{align*}
                \by(\iconstant) > \max \svector{\by}{\lconstant}{\tonumber{\n} \setminus \bw}, \text{ for any } \iconstant \in \bw.
            \end{align*}
            Hence, we have $\by \in \myfirstset{\bw}{\tonumber{\n}}{\n}$
            and $\sum_{\iconstant \in \bw} \indicator{ \by(\iconstant)   < \min \svector{\bys}{\lconstant}{\tonumber{\n} \setminus \bw} } = \mln{3} - \rr$.
            Further, for any $ \iconstant \in \{1,\ldots, \mln{3}\}$, we have
            \begin{align*}
                \rank{\bxn{{\n - \mln{3} + \iconstant}}}{\bx} - \rank{\byn{{\n - \mln{3} + \iconstant}}}{\by} = 2\mln{3} - 2\iconstant + 1 - \n,
            \end{align*}
            which gives
            \begin{align*}
                \sum_{\iconstant \in \bw} \left(\rank{\bxn{{\iconstant}}}{\bx} - \rank{\byn{{\iconstant}}}{\by} \right) &= \sum_{\iconstant=1}^{\mln{3}} |2\mln{3} - 2\iconstant + 1 - \n|.
            \end{align*}
            This proves our result when $\rr = \mln{3}$, and completes our proof.
        \end{proof}

        \begin{proposition} \label{supp:proposition:alg:3:1}
            Suppose $\bx, \by \in \vndistinct$ and for $0 < \mln{3} < \n$, 
            let $\bw = \{\n - \mln{3} + 1, \ldots ,\n\} \subset  \tonumber{\n}$ 
            be a subset of indices. Let $(\bxs^{(\rr)}, \bys^{(\rr)}) \in \mythirdset{\bw}{\n}$ 
            be imputations of $\bx$ and $\by$ for indices in $\bw$ such that 
            $\sum_{\iconstant \in \bw} \indicator{ \bxs^{(\rr)}(\iconstant)   < \min \svector{\bxs^{(\rr)}}{\lconstant}{\tonumber{\n} \setminus \bw} } = \rr$. For any $\iconstant \in \tonumber{\n} \setminus \bw$, denote $\dln{\iconstant} = \rank{\byn{\iconstant}}{\svector{\by}{\jconstant}{\tonumber{\n} \setminus \bw}} - \rank{\bxn{\iconstant}}{\svector{\bx}{\jconstant}{\tonumber{\n}\setminus \bw}}$. 
            Then, for any $\rr \in \{0,\ldots, \mln{3}\}$, we have
            \begin{align*}
                \sfdsb{\bxrs{\rr}}{\byrs{\rr}} &=  \sum_{\iconstant=1}^{\n - \mln{3}} |\dln{\iconstant} + \mln{3} - 2\rr| + \grn{\rr},
            \end{align*}
            where $\grn{\rr} = \sum_{\iconstant=1}^{\mln{3}} \left(\indicator{\iconstant \le \mln{3} - \rr} |\n + 1 - 2\iconstant| \right)
            + \sum_{\iconstant=1}^{\mln{3}}\left(\indicator{\iconstant > \mln{3} - \rr} |2\mln{3} - 2\iconstant + 1 -\n|\right)$.
        \end{proposition}

        \begin{proof}
            To start, according to the definition of Spearman's footrule,
            we have
            \begin{align*}
                \sfdsb{\bxrs{\rr}}{\byrs{\rr}} &= \sum_{\iconstant \in \bw} \left|\rank{\byrsn{\rr}{\iconstant}}{\byrs{\rr}}  - \rank{\bxrsn{\rr}{\iconstant}}{\bxrs{\rr}}\right| \\
                & + \sum_{\iconstant \in \tonumber{\n} \setminus \bw} \left|\rank{\byrsn{\rr}{\iconstant}}{\byrs{\rr}}  - \rank{\bxrsn{\rr}{\iconstant}}{\bxrs{\rr}}\right|.
            \end{align*}
            Then, in order to prove our results, it is sufficient
            to show that
            \begin{align*}
                \sum_{\iconstant \in \bw} \left|\rank{\byrsn{\rr}{\iconstant}}{\byrs{\rr}}  - \rank{\bxrsn{\rr}{\iconstant}}{\bxrs{\rr}}\right|  = \grn{\rr},
            \end{align*}
            and
            \begin{align*}
                \sum_{\iconstant \in \tonumber{\n} \setminus \bw} \left|\rank{\byrsn{\rr}{\iconstant}}{\byrs{\rr}}  - \rank{\bxrsn{\rr}{\iconstant}}{\bxrs{\rr}}\right| =  \sum_{\iconstant=1}^{\n - \mln{3}} |d_i + \mln{3} - 2\rr|.
            \end{align*}
            Notice that
            \begin{align*}
                \sum_{\iconstant \in \bw} \left|\rank{\byrsn{\rr}{\iconstant}}{\byrs{\rr}}  - \rank{\bxrsn{\rr}{\iconstant}}{\bxrs{\rr}}\right|  = \grn{\rr}
            \end{align*}
            is true according to Lemma~\ref{supp:proposition:alg:3:lemma:1}.
            Hence, we only need to show that
            \begin{align*}
                \sum_{\iconstant \in \tonumber{\n} \setminus \bw} \left|\rank{\byrsn{\rr}{\iconstant}}{\byrs{\rr}}  - \rank{\bxrsn{\rr}{\iconstant}}{\bxrs{\rr}}\right| =  \sum_{\iconstant=1}^{\n - \mln{3}} |d_i + \mln{3} - 2\rr|.
            \end{align*}

            Since $(\bxs^{(\rr)}, \bys^{(\rr)}) \in \mythirdset{\bw}{\n}$, then 
            we have $\bxs^{(\rr)} \in \myfirstset{\bw}{\tonumber{\n}}{\n}$. 
            Notice that
            \begin{align*}
                \sum_{\iconstant \in \bw} \indicator{ \bxs^{(\rr)}(\iconstant)   < \min \svector{\bxs^{(\rr)}}{\lconstant}{\tonumber{\n} \setminus \bw} } = \rr,
            \end{align*}
            then according to Lemma~\ref{supp:proposition:alg:1:lemma:2}, we have
            \begin{align*}
                \rank{\bxrsn{\rr}{\iconstant}}{\bxrs{\rr}} = \rank{\bxn{\iconstant}}{\svector{\bx}{\lconstant}{\tonumber{\n} \setminus \bw}} + \rr, \text{ for any } \iconstant \in \tonumber{\n} \setminus \bw. 
            \end{align*}
            Next, according to Lemma~\ref{supp:proposition:alg:3:lemma:1}, we have $\bys^{(\rr)} \in \myfirstset{\bw}{\tonumber{\n}}{\n}$, and
            \begin{align*}
                \sum_{\iconstant \in \bw} \indicator{ \bys^{(\rr)}(\iconstant)   < \min \svector{\bys^{(\rr)}}{\lconstant}{\tonumber{\n} \setminus \bw} } = \mln{3} - \rr.
            \end{align*}
            Then, according to Lemma~\ref{supp:proposition:alg:1:lemma:2}, we have
            \begin{align*}
                \rank{\byrsn{\rr}{\iconstant}}{\byrs{\rr}} = \rank{\byn{\iconstant}}{\svector{\by}{\lconstant}{\tonumber{\n} \setminus \bw}} + \mln{3} - \rr, \text{ for any } \iconstant \in \tonumber{\n} \setminus \bw. 
            \end{align*}
            Hence, we have
            \begin{align*}
                &\sum_{\iconstant \in \tonumber{\n} \setminus \bw} \left|\rank{\byrsn{\rr}{\iconstant}}{\byrs{\rr}}  - \rank{\bxrsn{\rr}{\iconstant}}{\bxrs{\rr}}\right|\\
                &=\sum_{\iconstant \in \tonumber{\n} \setminus \bw} \left|\rank{\byn{\iconstant}}{\svector{\by}{\lconstant}{\tonumber{\n} \setminus \bw}} - \rank{\bxn{\iconstant}}{\svector{\bx}{\lconstant}{\tonumber{\n} \setminus \bw}} + \mln{3} - 2\rr \right| \\
                &= \sum_{\iconstant=1}^{\n - \mln{3}} |d_i + \mln{3} - 2\rr|. 
            \end{align*}
            This completes our proof.
        \end{proof}

        \begin{lemma} \label{supp:proposition:alg:3:lemma:1.1}
            Suppose $\n \in \mathbb{N}$ such that $\n > 2$,		
            and $\mln{3} \in \tonumber{\n}$. For any $\rr \in \{0, \ldots, \mln{3}\}$,
            define
            \begin{align*}
                \grn{\rr} = \sum_{\iconstant=1}^{\mln{3}} \left(\indicator{\iconstant \le \mln{3} - \rr} |\n + 1 - 2\iconstant| \right)
                + \sum_{\iconstant=1}^{\mln{3}}\left(\indicator{\iconstant > \mln{3} - \rr} |2\mln{3} - 2\iconstant + 1 -\n|\right).
            \end{align*}
            Then for any $\rr \in \{0, \ldots, \mln{3} - 1\}$, we have		
            \begin{align*}
                \grn{\rr + 1} - \grn{\rr}  = |2\rr + 1 - \n| - |\n + 1 - 2(\mln{3} - \rr)|.
            \end{align*}
        \end{lemma}

        \begin{proof}
            We prove our results when $\rr = 0$, $\mln{3} > \rr > 0$,
            and $\rr = \mln{3}$, separately.

            First, suppose $\rr = 0$. Then we have
            \begin{align*}
                \grn{\rr}  =  \sum_{\iconstant = 1}^{\mln{3}}|\n + 1 - 2\iconstant|
            \end{align*}
            and
            \begin{align*}
                \grn{\rr + 1} =  \sum_{\iconstant = 1}^{\mln{3} - 1}|\n + 1 - 2\iconstant| + |1 - \n|.
            \end{align*}
            Hence, we have
            \begin{align*}
                \grn{\rr + 1}   - \grn{\rr}  = |1 - \n| - |\n + 1 - 2\mln{3}|.
            \end{align*}
            This proves our results when $\rr = 0$.

            Next, suppose $\mln{3} - 1 > \rr > 0$. Then  we have
            \begin{align*}
                \grn{\rr}  = \sum_{\iconstant = 1}^{\mln{3} - \rr} |\n + 1 - 2\iconstant| + \sum_{\iconstant = \mln{3} - \rr + 1}^{\mln{3} } |2\mln{3} - 2\iconstant + 1 -\n|, 
            \end{align*}
            and
            \begin{align*}
                \grn{\rr + 1}  =  \sum_{\iconstant = 1}^{\mln{3} - \rr - 1} |\n + 1 - 2\iconstant| + \sum_{\iconstant = \mln{3} - \rr}^{\mln{3} } |2\mln{3} - 2\iconstant + 1 -\n| .
            \end{align*}
            Hence, we have
            \begin{align*}
                \grn{\rr + 1}   - \grn{\rr}  & = |2\mln{3} - 2(\mln{3} - \rr) + 1 - \n| - |\n + 1 - 2(\mln{3} - \rr)|\\
                & = |2\rr + 1 - \n| - |\n + 1 - 2(\mln{3} - \rr)|.
            \end{align*}
            This proves our results when $\mln{3} - 1 > \rr > 0$.

            Finally, suppose $\rr = \mln{3} - 1$. Then  we have
            \begin{align*}
                \grn{\rr} = |\n - 1| + \sum_{\iconstant = 2}^{\mln{3} } |2\mln{3} - 2\iconstant + 1 -\n|, 
            \end{align*}
            and
            \begin{align*}
                \grn{\rr + 1}  = \sum_{\iconstant = 1}^{\mln{3} } |2\mln{3} - 2\iconstant + 1 -\n|.
            \end{align*}
            Hence, we have
            \begin{align*}
                \grn{\rr + 1}   - \grn{\rr} = |2\mln{3} -1 - \n| - |\n - 1|.
            \end{align*}
            This proves our results when $\rr = \mln{3} -1$, and completes our proof.
        \end{proof}

        \begin{lemma} \label{supp:proposition:alg:3:lemma:2}
            Suppose $\bx = (\bxn{\uuln{1}}, \ldots, \bxn{\uuln{\n}})$ is a vector of integers,
            where $\bu = \{\uuln{1}, \ldots, \uuln{\n}\}$ is a set of indices such that $\uuln{1} < \ldots < \uuln{\n}$. Then, we have 
            \begin{align*}
                \sum_{\iconstant \in \bu} |\bxn{\iconstant} + 2|
                - \sum_{\iconstant \in \bu} |\bxn{\iconstant} | 
                = 2\sum_{\iconstant \in \bu} \indicator{\bxn{\iconstant}  \ge 0} + 2\sum_{\iconstant \in \bu} \indicator{\bxn{\iconstant} + 1 \ge 0} - 2\n.
            \end{align*}
        \end{lemma}

        \begin{proof}
            To start,
            for any $\iconstant \in \bu$, we have
            \begin{align*}
                |\bxn{\iconstant}  + 2| - |\bxn{\iconstant}| &= 2, \text{ if } \bxn{\iconstant} \ge 0,\\
                |\bxn{\iconstant} + 2| - |\bxn{\iconstant}| &= 0, \text{ if } \bxn{\iconstant} = -1.
            \end{align*}
            If, however $\bxn{\iconstant}  < -1$, then since 
            $\bx$ is a vector of integers, we
            have $\bxn{\iconstant} + 2 \le 0$.
            Hence, we have
            \begin{align*}
                |\bxn{\iconstant} + 2| - |\bxn{\iconstant}| &= -2, \text{ if } \bxn{\iconstant} < -1.
            \end{align*}
            Therefore, we have
            \begin{align*}
                \sum_{\iconstant \in \bu} |\bxn{\iconstant}  + 2| - \sum_{\iconstant \in \bu} |\bxn{\iconstant}| 
                &= \sum_{\iconstant \in \bu}  \left(|\bxn{\iconstant} + 2| - |\bxn{\iconstant} - \byn{\iconstant}|\right) \\
                & = 2\sum_{\iconstant \in \bu} \indicator{\bxn{\iconstant}  \ge 0} - 2\sum_{\iconstant \in \bu} \indicator{\bxn{\iconstant}  < -1} \\
                & =  2\sum_{\iconstant \in \bu} \indicator{\bxn{\iconstant}  \ge 0} - 2\left(\n - \sum_{\iconstant \in \bu} \indicator{\bxn{\iconstant}  \ge -1}  \right)\\
                & =  2\sum_{\iconstant \in \bu} \indicator{\bxn{\iconstant}  \ge 0}  + 2\sum_{\iconstant \in \bu} \indicator{\bxn{\iconstant} + 1 \ge 0}  - 2\n.
            \end{align*}
            This completes our proof.
        \end{proof}

        \begin{proposition} \label{supp:proposition:alg:3}
            Suppose $\bx, \by \in \vndistinct$ and for $0 < \mln{3} < \n$, 
            let $\bw = \{\n - \mln{3} + 1, \ldots ,\n\} \subset  \tonumber{\n}$ 
            be a subset of indices. Let $(\bxs^{(\rr)}, \bys^{(\rr)}) \in \mythirdset{\bw}{\n}$ 
            be imputations of $\bx$ and $\by$ for indices in $\bw$ such that 
            $\sum_{\iconstant \in \bw} \indicator{ \bxs^{(\rr)}(\iconstant)   < \min \svector{\bxs^{(\rr)}}{\lconstant}{\tonumber{\n} \setminus \bw} } = \rr$. For any $\iconstant \in \tonumber{\n} \setminus \bw$, let $\dln{\iconstant} = \rank{\byn{\iconstant}}{\svector{\by}{\jconstant}{\tonumber{\n} \setminus \bw}} - \rank{\bxn{\iconstant}}{\svector{\bx}{\jconstant}{\tonumber{\n}\setminus \bw}}$.
            For any $\iconstant \in \{-\mln{3}, \ldots, \mln{3} - 1\}$, let $\ssln{\iconstant} = \sum_{\lconstant \in \tonumber{\n} \setminus \bw} \indicator{\dln{\lconstant} \le \iconstant}$. 
            Then, for any $\rr \in \{0,\ldots, \mln{3}-1\}$, we have
            \begin{align*}
                \sfdsb{\bxrs{\rr+1}}{\byrs{\rr+1}} &= \sfdsb{\bxrs{\rr}}{\byrs{\rr}} + 2(\ssln{ 2\rr + 1 - \mln{3}} + \ssln{2\rr - \mln{3}} - \n + \mln{3}) \\
                & + |2\rr + 1 - \n| - |\n + 1 - 2(\mln{3} - \rr)|.
            \end{align*}
        \end{proposition}

        \begin{proof}
            To start, 
            according to Proposition~\ref{supp:proposition:alg:3:1}, we have
            \begin{align*}
                \sfdsb{\bxrs{\rr}}{\byrs{\rr}} &=  \sum_{\iconstant=1}^{\n - \mln{3}} |d_i + \mln{3} - 2\rr| + \grn{\rr},\\
                \text{and }			\sfdsb{\bxrs{\rr+1}}{\byrs{\rr+1}} &=  \sum_{\iconstant=1}^{\n - \mln{3}} |d_i + \mln{3} - 2\rr - 2| + \grn{\rr + 1}.
            \end{align*}
            Hence,
            \begin{align*}
                & \sfdsb{\bxrs{\rr+1}}{\byrs{\rr+1}}  - 	\sfdsb{\bxrs{\rr}}{\byrs{\rr}} \\
                & = \sum_{\iconstant=1}^{\n - \mln{3}} |d_i + \mln{3} - 2\rr - 2| - \sum_{\iconstant=1}^{\n - \mln{3}} |d_i + \mln{3} - 2\rr| +  \grn{\rr+1} -  \grn{\rr}.
            \end{align*}
            Therefore, in order to prove our result,
            it is sufficient to show that
            \begin{align*}
                \grn{\rr+1} -  \grn{\rr} = |2\rr + 1 - \n| - |\n + 1 - 2(\mln{3} - \rr)|,
            \end{align*}
            and 
            \begin{align*}
                \sum_{\iconstant=1}^{\n - \mln{3}} |d_i + \mln{3} - 2\rr - 2| - \sum_{\iconstant=1}^{\n - \mln{3}} |d_i + \mln{3} - 2\rr|  = 2(\ssln{2\rr - \mln{3}} + \ssln{2\rr -\mln{3} + 1} - \n -\mln{3}).
            \end{align*}
            Notice that according to Lemma~\ref{supp:proposition:alg:3:lemma:1.1},
            \begin{align*}
                \grn{\rr+1} -  \grn{\rr} = |2\rr + 1 - \n| - |\n + 1 - 2(\mln{3} - \rr)|
            \end{align*}
            is true.
            Hence we only need to show
            \begin{align*}
                \sum_{\iconstant=1}^{\n - \mln{3}} |d_i + \mln{3} - 2\rr - 2| - \sum_{\iconstant=1}^{\n - \mln{3}} |d_i + \mln{3} - 2\rr|  = 2 (\ssln{ 2\rr + 1 - \mln{3}} + \ssln{2\rr - \mln{3}} - \n + \mln{3})
            \end{align*}
            is true.

            According to Lemma~\ref{supp:proposition:alg:3:lemma:2},
            we have
            \begin{align*}
                &\sum_{\iconstant=1}^{\n - \mln{3}} |d_i + \mln{3} - 2\rr - 2| - \sum_{\iconstant=1}^{\n - \mln{3}} |d_i + \mln{3} - 2\rr |\\
                &= 2(\n - \mln{3}) - 2 \sum_{\iconstant=1}^{\n -\mln{3}} \indicator{d_i + \mln{3} - 2\rr - 2 \ge 0} -2 \sum_{\iconstant=1}^{\n -\mln{3}} \indicator{d_i + \mln{3} - 2\rr - 2 + 1\ge 0} \\
                & = 2(\n - \mln{3}) - 2\left( \n - \mln{3} - \sum_{\iconstant=1}^{\n -\mln{3}} \indicator{d_i + \mln{3} - 2\rr - 2 < 0} \right)\\
                & -2 \left( \n - \mln{3} - \sum_{\iconstant=1}^{\n -\mln{3}} \indicator{d_i + \mln{3} - 2\rr - 1 < 0} \right)\\
                & = 2 \left(\sum_{\iconstant=1}^{\n -\mln{3}} \indicator{d_i < 2\rr + 2 - \mln{3}} + \sum_{\iconstant=1}^{\n -\mln{3}} \indicator{d_i < 2\rr + 1 - \mln{3}} - \n + \mln{3}\right).
            \end{align*}
            Since $\rr, \mln{3}$ are integers and for any $\iconstant \in \{1, \ldots, \n - \mln{3}\}$, $\dln{\iconstant}$ is also an integer, then we have 
            \begin{align*} 
                &\indicator{d_i < 2\rr + 2 - \mln{3}} = \indicator{d_i \le 2\rr + 1 - \mln{3}}\\
                \text{and }& \indicator{d_i < 2\rr + 1 - \mln{3}} = \indicator{d_i \le 2\rr - \mln{3}}.
            \end{align*}
            Hence, we have 
            \begin{align*}
                \sum_{\iconstant=1}^{\n - \mln{3}} |d_i + \mln{3} - 2\rr - 2| - \sum_{\iconstant=1}^{\n - \mln{3}} |d_i + \mln{3} - 2\rr | = 2 (\ssln{ 2\rr + 1 - \mln{3}} + \ssln{2\rr - \mln{3}} - \n + \mln{3}).
            \end{align*}
            This completes our proof.
        \end{proof}

        \begin{algorithm} 
            \caption{An efficient algorithm for computing exact upper bounds of Spearman's Footrule under Missing Case III} \label{supp:alg:3}
            \begin{algorithmic}[1]
                \Require{$\bx, \by \in \vndistinct$, where $\bx$ and $\by$ might be
                partially observed, and for any pair $(\bxln{\iconstant}, \byln{\iconstant})$,
                where $\iconstant \in \tonumber{\n}$, the two values
                are either both observed, or both missing.} 
                \Ensure{Maximum possible Spearman's footrule distance between $X$ and $Y$.}
                \State Denote $\mln{3}$ as the number of components such that
                both values $(\bxn{\iconstant}, \byn{\iconstant})$
                are missing. If $\mln{3} = \n$, then return $\sum_{\iconstant=1}^{\n} |\iconstant - (\n -\iconstant + 1)|$.
                \State Rank all observed data in $\bx$ and $\by$. Relabel $\bx$ and $\by$ such that
                $(\bxn{\n - \mln{3} + 1}, \byn{\n - \mln{3} + 1}), \ldots, (\bxn{\n}, \byn{\n})$ are missing. \label{supp:alg:3:line:1}
                \State Let $\bw = \{\n - \mln{3} + 1, \ldots, \n\}$, and for any $\iconstant \in \tonumber{\n} \setminus \bw$, let \label{supp:alg:3:line:2}
                \begin{align*}
                    \dln{\iconstant} = \rank{\byn{\iconstant}}{\svector{\by}{\lconstant}{\tonumber{\n} \setminus \bv}} - \rank{\bxn{\iconstant}}{\svector{\bx}{\lconstant}{\tonumber{\n} \setminus \bu}}.
                \end{align*}
                \State Run Algorithm~\ref{supp:alg:0} for computing $\ssln{\iconstant} = \sum_{\lconstant \in \tonumber{\n} \setminus \bw} \indicator{d_l \le \iconstant}$ for any $\iconstant \in \{-\mln{3}, \ldots, \mln{3} -1\}$. \label{supp:alg:3:line:3}
                \State Initialize $\bdln{0}$ = $\sum_{\iconstant = 1}^{\mln{3}} \left|  \n + 1 - 2\iconstant\right| + \sum_{i \in \tonumber{\n} \setminus \bw} |d_i|$. \label{supp:alg:3:line:4}
                \For{$\rr = 0,\ldots,\mln{3}-1$}
                \State Compute \label{supp:alg:3:line:5}
                \begin{align*}
                    \bdln{\rr + 1} &= \bdln{\rr} + 2(\ssln{ 2\rr + 1 - \mln{3}} + \ssln{2\rr - \mln{3}} - \n + \mln{3})\\
                    & + |2\rr + 1 - \n| - |\n + 1 - 2(\mln{3} - \rr)|.
                \end{align*}
                \EndFor
                \State Return $\max \{\bdln{0}, \ldots, \bdln{\mln{3}}\}$.
            \end{algorithmic}
        \end{algorithm}

        \begin{remark}
            A few comments of Algorithm~\ref{supp:alg:3} are made below:

            In Algorithm~\ref{supp:alg:3}, the initialization $\bdln{0}$ computed in
            line~\ref{supp:alg:3:line:4} equals to $\sfdsb{\bxrs{0}}{\byrs{0}}$ defined in 
            Proposition~\ref{supp:proposition:alg:3:1}.
            Spearman's footrule
            $\bdln{\rr + 1}$, and 
            $\bdln{\rr}$ equal to 
            $\sfdsb{\bxrs{\rr + 1}}{\byrs{\rr}}$, 
            $\sfdsb{\bxrs{\rr}}{\byrs{\rr}}$,
            respectively, and are updated according to
            Proposition~\ref{supp:proposition:alg:3}.

            By computing $\{\bdln{0}, \ldots, \bdln{\mln{3}}\}$, Algorithm~\ref{supp:alg:3} 
            finds all possible Spearman's footrule
            values $\sfd{\bxs}{\bys}$, where $(\bxs,\bys) \in \mythirdset{\bw}{\n}$ 
            are imputations of $\bx$ and $\by$ for indices $\bw$.
            See discussions following Theorem 2.20 in the main paper
            for explanations.
            Hence, according to Theorem~\ref{supp:theorem:2.20},
            the algorithm guarantees to find the maximum possible Spearman's footrule
            between $\bx$ and $\by$.

            The computational complexity of Algorithm~\ref{supp:alg:3} is analyzed as follows.				 
            Ranking and relabeling all observed components in $\bx$ and $\by$ in line~\ref{supp:alg:3:line:1} 
            requires $\bmo (\n \log \n)$ steps. 				
            Using these rankings, in line~\ref{supp:alg:3:line:2} computing each $\dln{\iconstant}$ is $\bmo(1)$, 
            and so overall line~\ref{supp:alg:3:line:2} is $\bmo(\n - \mln{3})$.
            According to Remark~\ref{supp:remark:alg:0}, the computational
            complexity of running Algorithm~\ref{supp:alg:0} in 
            line~\ref{supp:alg:3:line:3} is $\bmo(\n + \mln{3})$.
            Line~\ref{supp:alg:3:line:4} requires $\bmo (\n)$ steps.
            In line~\ref{supp:alg:3:line:5}, each iteration of the for loop is $\bmo(1)$, and since the loop runs $\mln{3}$ times, 
            the computational complexity for the loop is $\bmo(\mln{3})$. 
            Therefore, the overall computational complexity for Algorithm~\ref{supp:alg:3} 
            is $\bmo(\n\log \n)$.
        \end{remark}

        \subsection{General Missing Case}
        This subsection provides an efficient algorithm for computing
        upper bounds of Spearman's footrule under General Missing Case.

        We first prove the following result:

        \begin{lemma} \label{supp:proposition:alg:4:lemma:2}
            Suppose $\bx, \by \in \vndistinct$ and for $0 < \mln{1}, \mln{2}, \mln{3} < \n$ such that $\mln{1} + \mln{2} + \mln{3} < \n$, 
            let $\bv = \{1, \ldots, \mln{2}\}$, $\bu = \{\mln{2} + 1,\ldots, \mln{2} + \mln{1}\}$ and $\bw = \{\n - \mln{3} + 1, \ldots, \n\}$ be subsets of indices. Suppose
            $\bxn{1} < \ldots < \bxn{\mln{2}}$, and
            $\byn{\mln{2} + 1} < \ldots < \byn{\mln{2} + \mln{1}}$.
            Let $(\bxs^{(\rrln{1}, \rrln{3})}, \bys^{(\rrln{2}, \rrln{3})}) \in (\mysecondset{\by}{\bu}{\tonumber{\n} \setminus \bw}{\n}, \mysecondset{\bx}{\bv}{\tonumber{\n} \setminus \bw}{\n}) \cap \mythirdset{\bw}{\n}$ 
            be imputations of $\bx, \by$ for indices
            $\bu, \bv$ such that 
            \begin{align*}
                &\sum_{\iconstant \in \bu} \indicator{ \bxs^{(\rrln{1}, \rrln{3})}(\iconstant)   < \min \svector{\bxs^{(\rrln{1}, \rrln{3})}}{\lconstant}{ (\tonumber{\n} \setminus \bw) \setminus \bu} } = \rrln{1},\\
                &\sum_{\iconstant \in \bv} \indicator{ \bys^{(\rrln{2}, \rrln{3})}(\iconstant)   < \min \svector{\bys^{(\rrln{2}, \rrln{3})}}{\lconstant}{ (\tonumber{\n} \setminus \bw) \setminus \bv} } = \rrln{2},\\
                \text{and }&\sum_{\iconstant \in \bw} \indicator{ \bxs^{(\rrln{1}, \rrln{3})}(\iconstant)   < \min \svector{\bxs^{(\rrln{1}, \rrln{3})}}{\lconstant}{\tonumber{\n} \setminus \bw} } = \rrln{3}.
            \end{align*}
            Then, for any $\iconstant \in (\tonumber{\n} \setminus \bw) \setminus \bu$, we have
            \begin{align} \label{supp:proposition:alg:4:lemma:2:eqn:1}
                &\rank{\bxrsn{\rrln{1}, \rrln{3}}{{\iconstant}}}{\bxrs{\rrln{1}, \rrln{3}}}
                = \rrln{3} + \rrln{1} + \rank{\bxn{\iconstant}}{\svector{\bx}{\lconstant}{(\tonumber{\n} \setminus \bw) \setminus (\bu \cup \bv) } }.
            \end{align}
            For any $\iconstant \in (\tonumber{\n} \setminus \bw) \setminus \bv$, we have
            \begin{align} \label{supp:proposition:alg:4:lemma:2:eqn:2}
                &\rank{\byrsn{\rrln{2}, \rrln{3}}{{\iconstant}}}{\byrs{\rrln{2}, \rrln{3}}}
                = \mln{3} - \rrln{3} + \rrln{2} + \rank{\byn{\iconstant}}{\svector{\by}{\lconstant}{(\tonumber{\n} \setminus \bw) \setminus (\bu \cup \bv) } }.
            \end{align}
            For any $\rrln{1} \in \{0, \ldots, \mln{1}\}$, $\rrln{2} \in \{0, \ldots, \mln{2}\}$, $\rrln{3} \in \{0, \ldots, \mln{3}\}$, we have
            \begin{align}
                &\rank{\bxrsn{\rrln{1}, \rrln{3}}{{\iconstant + \mln{2}}}}{\bxrs{\rrln{1}, \rrln{3}}} = \pnn{\rrln{1}}{\iconstant} + \rrln{3}, \text{ for any } \iconstant \in \{1,\ldots,\mln{1}\}, \label{supp:proposition:alg:4:lemma:2:eqn:3}\\
                \text{ and }&\rank{\byrsn{\rrln{2}, \rrln{3}}{{\iconstant}}}{\byrs{\rrln{2}, \rrln{3}}} = \qnn{\rrln{2}}{\iconstant} + \mln{3} - \rrln{3}, \text{ for any } \iconstant \in \{1,\ldots, \mln{2}\}, \label{supp:proposition:alg:4:lemma:2:eqn:4}
            \end{align}
            where $\pnn{\rrln{1}}{\iconstant} = \indicator{\iconstant \le  \mln{1} - \rrln{1}} (\n - \mln{3} - \iconstant + 1) 
            + \indicator{\iconstant >  \mln{1} - \rrln{1}} (\mln{1} - \iconstant + 1) $, and $\qnn{\rrln{2}}{\iconstant}  = \indicator{\iconstant \le \mln{2} - \rrln{2}} (\n - \mln{3} - \iconstant + 1) 
            + \indicator{\iconstant > \mln{2} - \rrln{2}} (\mln{2} - \iconstant + 1)$.
        \end{lemma}

        \begin{proof}
            First, we show \eqref{supp:proposition:alg:4:lemma:2:eqn:1} is true,
            since $(\bxs^{(\rrln{1}, \rrln{3})}, \bys^{(\rrln{2}, \rrln{3})}) \in \mythirdset{\bw}{\n}$,
            then, according to the definition of $\mythirdset{\bw}{\n}$,
            we have $\bxs^{(\rrln{1}, \rrln{3})} \in \myfirstset{\bw}{\tonumber{\n}}{\n}$.
            Notice that, according to the definition of imputations,
            we have $\bxs^{(\rrln{1}, \rrln{3})}$ is an imputation of itself for indices $\bw$.
            Since $\sum_{\iconstant \in \bw} \indicator{ \bxs^{(\rrln{1}, \rrln{3})}(\iconstant)   < \min \svector{\bxs^{(\rrln{1}, \rrln{3})}}{\lconstant}{\tonumber{\n} \setminus \bw} } = \rrln{3}$,
            then according to Lemma~\ref{supp:proposition:alg:1:lemma:2},
            for any $\iconstant \in \tonumber{\n} \setminus \bw$, we have
            \begin{align} \label{supp:proposition:alg:4:lemma:2:eqn:5}
                \rank{\bxrsn{\rrln{1}, \rrln{3}}{{\iconstant}}}{\bxrs{\rrln{1}, \rrln{3}}} = \rank{\bxrsn{\rrln{1}, \rrln{3}}{{\iconstant}}}{\svector{\bxrs{\rrln{1}, \rrln{3}}}{\lconstant}{\tonumber{\n} \setminus\bw} } + \rrln{3}.
            \end{align}

            Further, since $\bxs^{(\rrln{1}, \rrln{3})} \in \mysecondset{\by}{\bu}{\tonumber{\n} \setminus \bw}{\n}$,
            according to the definition of $\mysecondset{\by}{\bu}{\tonumber{\n} \setminus \bw}{\n}$, we also have
            $\bxs^{(\rrln{1}, \rrln{3})} \in \myfirstset{\bu}{\tonumber{\n} \setminus \bw}{\n}$.
            According to the definition of $\myfirstset{\bu}{\tonumber{\n} \setminus \bw}{\n}$,
            for any $\iconstant \in \bu$, we have
            \begin{align*}
                \bxs^{(\rrln{1}, \rrln{3})}(\iconstant) > \max \svector{\bxs^{(\rrln{1}, \rrln{3})}}{\lconstant}{\tonumber{\n} \setminus\bw}, \text{ or } \bxs^{(\rrln{1}, \rrln{3})}(\iconstant) < \min \svector{\bxs^{(\rrln{1}, \rrln{3})}}{\lconstant}{\tonumber{\n} \setminus\bw}.
            \end{align*}
            Hence, we also have $\svector{\bxs^{(\rrln{1}, \rrln{3})}}{\lconstant}{\tonumber{\n} \setminus\bw} \in \myfirstset{\bu}{\tonumber{\n} \setminus \bw}{\n - \mln{3}}$.
            Since $\svector{\bxs^{(\rrln{1}, \rrln{3})}}{\lconstant}{\tonumber{\n} \setminus\bw}$
            is an imputation of itself for indices $\bu$, and we have
            \begin{align*}
                \sum_{\iconstant \in \bu} \indicator{ \bxs^{(\rrln{1}, \rrln{3})}(\iconstant)   < \min \svector{\bxs^{(\rrln{1}, \rrln{3})}}{\lconstant}{ (\tonumber{\n} \setminus \bw) \setminus \bu} } = \rrln{1},
            \end{align*}
            then according to Lemma~\ref{supp:proposition:alg:1:lemma:2},
            for any $\iconstant \in (\tonumber{\n} \setminus \bw) \setminus \bu$, we have
            \begin{align} \label{supp:proposition:alg:4:lemma:2:eqn:6}
                \begin{split}
                    &\rank{\bxrsn{\rrln{1}, \rrln{3}}{{\iconstant}}}{\svector{\bxrs{\rrln{1}, \rrln{3}}}{\lconstant}{\tonumber{\n} \setminus\bw} } \\
                    &= \rank{\bxrsn{\rrln{1}, \rrln{3}}{{\iconstant}}}{\svector{\bxrs{\rrln{1}, \rrln{3}}}{\lconstant}{(\tonumber{\n} \setminus\bw) \setminus \bu} } + \rrln{1}.
                \end{split}
            \end{align}
            Since $\bxrs{\rrln{1}, \rrln{3}}$ is an imputation of $\bx$ for indices $\bu \cup \bw$,
            for any $\iconstant \in (\tonumber{\n} \setminus \bw) \setminus \bu$, we have
            \begin{align} \label{supp:proposition:alg:4:lemma:2:eqn:7}
                \rank{\bxrsn{\rrln{1}, \rrln{3}}{{\iconstant}}}{\svector{\bxrs{\rrln{1}, \rrln{3}}}{\lconstant}{(\tonumber{\n} \setminus\bw) \setminus \bu} } = \rank{\bxn{\iconstant}}{\svector{\bx}{\lconstant}{(\tonumber{\n} \setminus\bw) \setminus \bu} }.
            \end{align}
            Combining \eqref{supp:proposition:alg:4:lemma:2:eqn:5},
            \eqref{supp:proposition:alg:4:lemma:2:eqn:6},
            and 
            \eqref{supp:proposition:alg:4:lemma:2:eqn:7}, for any $\iconstant \in (\tonumber{\n} \setminus \bw) \setminus \bu$, we have
            \begin{align*}
                \rank{\bxrsn{\rrln{1}, \rrln{3}}{{\iconstant}}}{\bxrs{\rrln{1}, \rrln{3}}}
                = \rrln{3} + \rrln{1} + \rank{\bxn{\iconstant}}{\svector{\bx}{\lconstant}{(\tonumber{\n} \setminus \bw) \setminus (\bu \cup \bv) } }.
            \end{align*}
            This proves \eqref{supp:proposition:alg:4:lemma:2:eqn:1}.

            Similarly, we can show \eqref{supp:proposition:alg:4:lemma:2:eqn:2} is true.
            Since $(\bxs^{(\rrln{1}, \rrln{3})}, \bys^{(\rrln{2}, \rrln{3})}) \in \mythirdset{\bw}{\n}$,
            then according to Lemma~\ref{supp:proposition:alg:3:lemma:1},
            we have $\bys^{(\rrln{2}, \rrln{3})} \in \myfirstset{\bw}{\tonumber{\n}}{\n}$,
            and
            \begin{align*}
                \sum_{\iconstant \in \bw} \indicator{ \bys^{(\rrln{2}, \rrln{3})}(\iconstant)   < \min \svector{\bys^{(\rrln{2}, \rrln{3})}}{\lconstant}{\tonumber{\n} \setminus \bw} } = \mln{3} - \rrln{3}.
            \end{align*}
            Notice that, according to the definition of imputations,
            $\bys^{(\rrln{2}, \rrln{3})} $ is an imputation of itself for indices $\bw$.
            Hence, according to Lemma~\ref{supp:proposition:alg:1:lemma:2},
            for any $\iconstant \in \tonumber{\n} \setminus \bw$, we have 
            \begin{align} \label{supp:proposition:alg:4:lemma:2:eqn:8}
                \rank{\byrsn{\rrln{2}, \rrln{3}}{{\iconstant}}}{\byrs{\rrln{2}, \rrln{3}}} = \rank{\byrsn{\rrln{2}, \rrln{3}}{{\iconstant}}}{\svector{\byrs{\rrln{2}, \rrln{3}}}{\lconstant}{\tonumber{\n} \setminus\bw} } + \mln{3} - \rrln{3}.
            \end{align}

            Further, since $\bys^{(\rrln{2}, \rrln{3})} \in \mysecondset{\bx}{\bv}{\tonumber{\n} \setminus \bw}{\n}$, we have
            $\bys^{(\rrln{2}, \rrln{3})} \in \myfirstset{\bv}{\tonumber{\n} \setminus \bw}{\n}$.
            Then according to the definition of $ \myfirstset{\bv}{\tonumber{\n} \setminus \bw}{\n}$,
            we have
            \begin{align*}
                \bys^{(\rrln{2}, \rrln{3})}(\iconstant) > \max \svector{\bys^{(\rrln{2}, \rrln{3})}}{\lconstant}{\tonumber{\n} \setminus\bw}, \text{ or } \bys^{(\rrln{2}, \rrln{3})}(\iconstant) < \min \svector{\bys^{(\rrln{2}, \rrln{3})}}{\lconstant}{\tonumber{\n} \setminus\bw}, \text{ for any } \iconstant \in \bv.
            \end{align*}
            Hence, we also have $\svector{\bys^{(\rrln{2}, \rrln{3})}}{\lconstant}{\tonumber{\n} \setminus\bw} \in \myfirstset{\bv}{\tonumber{\n} \setminus \bw}{\n - \mln{3}}$.
            Since $\svector{\bys^{(\rrln{2}, \rrln{3})}}{\lconstant}{\tonumber{\n} \setminus\bw}$
            is an imputation of itself for indices $\bv$, and we have
            \begin{align*}
                \sum_{\iconstant \in \bv} \indicator{ \bys^{(\rrln{2}, \rrln{3})}(\iconstant)   < \min \svector{\bys^{(\rrln{2}, \rrln{3})}}{\lconstant}{ (\tonumber{\n} \setminus \bw) \setminus \bv} } = \rrln{2}.
            \end{align*}
            then according to Lemma~\ref{supp:proposition:alg:1:lemma:2},
            for any $\iconstant \in (\tonumber{\n} \setminus \bw) \setminus \bv$, we have
            \begin{align} \label{supp:proposition:alg:4:lemma:2:eqn:9}
                \begin{split}
                    &\rank{\byrsn{\rrln{2}, \rrln{3}}{{\iconstant}}}{\svector{\byrs{\rrln{2}, \rrln{3}}}{\lconstant}{\tonumber{\n} \setminus\bw} } \\
                    &= \rank{\byrsn{\rrln{2}, \rrln{3}}{{\iconstant}}}{\svector{\byrs{\rrln{2}, \rrln{3}}}{\lconstant}{(\tonumber{\n} \setminus\bw) \setminus \bv} } + \rrln{2}.
                \end{split}
            \end{align}
            Since $\byrs{\rrln{2}, \rrln{3}}$ is an imputation of $\by$ for indices $\bv \cup \bw$,
            for any $\iconstant \in (\tonumber{\n} \setminus \bw) \setminus \bv$, we have
            \begin{align} \label{supp:proposition:alg:4:lemma:2:eqn:10}
                \rank{\byrsn{\rrln{2}, \rrln{3}}{{\iconstant}}}{\svector{\byrs{\rrln{2}, \rrln{3}}}{\lconstant}{(\tonumber{\n} \setminus\bw) \setminus \bv} }= \rank{\byn{\iconstant}}{\svector{\by}{\lconstant}{(\tonumber{\n} \setminus\bw) \setminus \bv} }.
            \end{align}
            Combining \eqref{supp:proposition:alg:4:lemma:2:eqn:8},
            \eqref{supp:proposition:alg:4:lemma:2:eqn:9},
            and 
            \eqref{supp:proposition:alg:4:lemma:2:eqn:10}, for any $\iconstant \in (\tonumber{\n} \setminus \bw) \setminus \bv$, we have
            \begin{align*}
                \rank{\byrsn{\rrln{2}, \rrln{3}}{{\iconstant}}}{\byrs{\rrln{2}, \rrln{3}}}
                = \mln{3} - \rrln{3} + \rrln{2} + \rank{\byn{\iconstant}}{\svector{\by}{\lconstant}{(\tonumber{\n} \setminus \bw) \setminus (\bu \cup \bv) } }.
            \end{align*}
            This proves \eqref{supp:proposition:alg:4:lemma:2:eqn:2}
            is true.

            Next, we show \eqref{supp:proposition:alg:4:lemma:2:eqn:3} is true,
            since $\bxs^{(\rrln{1}, \rrln{3})} \in \mysecondset{\by}{\bu}{\tonumber{\n} \setminus \bw}{\n}$,
            then we have
            \begin{align*}
                &\bxs^{(\rrln{1}, \rrln{3})}(\iconstant) > \max \svector{\bxs^{(\rrln{1}, \rrln{3})}}{\lconstant}{\tonumber{\n} \setminus\bw}, \text{ or } \bxs^{(\rrln{1}, \rrln{3})}(\iconstant) < \min \svector{\bxs^{(\rrln{1}, \rrln{3})}}{\lconstant}{\tonumber{\n} \setminus\bw}, \text{ for any } \iconstant \in \bu,\\
                &\text{and } \bxs^{(\rrln{1}, \rrln{3})}(\iconstant) > \bxs^{(\rrln{1}, \rrln{3})}(\jconstant), \text{ if } \byn{\iconstant} < \byn{\jconstant}, \text{ for any } \iconstant, \jconstant \in \bu.
            \end{align*}
            Hence, we have $\svector{\bxs^{(\rrln{1}, \rrln{3})}}{\lconstant}{\tonumber{\n} \setminus \bw} \in \mysecondset{\svector{\by}{\lconstant}{\tonumber{\n} \setminus \bw}}{\bu}{\tonumber{\n} \setminus \bw}{\n - \mln{3}}$. Since $$\sum_{\iconstant \in \bu} \indicator{ \bxs^{(\rrln{1}, \rrln{3})}(\iconstant)   < \min \svector{\bxs^{(\rrln{1}, \rrln{3})}}{\lconstant}{ (\tonumber{\n} \setminus \bw) \setminus \bu} } = \rrln{1},$$
            then according to Lemma~\ref{supp:proposition:alg:1:lemma:0}, for any $\iconstant \in \{1, \ldots, \mln{1}\}$, we have
            \begin{align*}
                \rank{\bxs^{(\rrln{1}, \rrln{3})}(\mln{2} + \iconstant)}{\svector{\bxs^{(\rrln{1}, \rrln{3})}}{\lconstant}{\tonumber{\n} \setminus \bw}} 
                = \left\{ \begin{array}{ll}
                    \n - \mln{3} - \iconstant + 1, & \text{ if } \rrln{1} = 0, \\
                    \pnn{\rrln{1}}{\iconstant}, & \text{ if } \mln{1} > \rr > 0, \\
                    \mln{1} - \iconstant + 1, & \text{ if } \rrln{1} = \mln{2},
                \end{array}\right.
            \end{align*}	
            where $\pnn{\rrln{1}}{\iconstant} = \indicator{\mln{2} + \iconstant \le \mln{2} + \mln{1} - \rrln{1}} (\n - \mln{3} - \iconstant + 1) 
            + \indicator{\mln{2} + \iconstant > \mln{2} + \mln{1} - \rrln{1}} (\mln{1} - \iconstant + 1) = \indicator{ \iconstant \le \mln{1} - \rrln{1}} (\n - \mln{3} - \iconstant + 1) 
            + \indicator{ \iconstant >  \mln{1} - \rrln{1}} (\mln{1} - \iconstant + 1)$.

            Notice that $	\rank{\bxs^{(\rrln{1}, \rrln{3})}(\iconstant)}{\svector{\bxs^{(\rrln{1}, \rrln{3})}}{\lconstant}{\tonumber{\n} \setminus \bw}}  = \pnn{\rrln{1}}{\iconstant}$
            is still true when $\rrln{1} = 0$, or $\rrln{1} = \mln{2}$.		
            Combining this result with \eqref{supp:proposition:alg:4:lemma:2:eqn:5},
            we have
            \begin{align*}
                \rank{\bxrsn{\rrln{1}, \rrln{3}}{{\iconstant}}}{\bxrs{\rrln{1}, \rrln{3}}} = \pnn{\rrln{1}}{\iconstant} + \rrln{3}, \text{ for any } \iconstant \in \bu.
            \end{align*}
            This completes our proof for \eqref{supp:proposition:alg:4:lemma:2:eqn:3}.

            Next, we show \eqref{supp:proposition:alg:4:lemma:2:eqn:3} is true,
            since $\bys^{(\rrln{2}, \rrln{3})} \in \mysecondset{\bx}{\bv}{\tonumber{\n} \setminus \bw}{\n}$,
            then we have
            \begin{align*}
                &\bys^{(\rrln{2}, \rrln{3})}(\iconstant) > \max \svector{\bys^{(\rrln{2}, \rrln{3})}}{\lconstant}{\tonumber{\n} \setminus\bw}, \text{ or } \bys^{(\rrln{2}, \rrln{3})}(\iconstant) < \min \svector{\bys^{(\rrln{2}, \rrln{3})}}{\lconstant}{\tonumber{\n} \setminus\bw}, \text{ for any } \iconstant \in \bv,\\
                &\text{and } \bys^{(\rrln{2}, \rrln{3})}(\iconstant) > \bys^{(\rrln{2}, \rrln{3})}(\jconstant), \text{ if } \bxn{\iconstant} < \bxn{\jconstant}, \text{ for any } \iconstant, \jconstant \in \bv.
            \end{align*}
            Hence, we have $\svector{\bys^{(\rrln{1}, \rrln{3})}}{\lconstant}{\tonumber{\n} \setminus \bw} \in \mysecondset{\svector{\bx}{\lconstant}{\tonumber{\n} \setminus \bw}}{\bv}{\tonumber{\n} \setminus \bw}{\n - \mln{3}}$.
            Since
            \begin{align*}
                \sum_{\iconstant \in \bv} \indicator{ \bys^{(\rrln{2}, \rrln{3})}(\iconstant)   < \min \svector{\bys^{(\rrln{2}, \rrln{3})}}{\lconstant}{ (\tonumber{\n} \setminus \bw) \setminus \bv} } = \rrln{2},
            \end{align*}
            then according to Lemma~\ref{supp:proposition:alg:1:lemma:0}, for any $\iconstant \in \{1, \ldots, \mln{2}\}$, we have
            \begin{align*}
                &\rank{\bys^{(\rrln{2}, \rrln{3})}(\iconstant)}{\svector{\bys^{(\rrln{2}, \rrln{3})}}{\lconstant}{\tonumber{\n} \setminus \bw}} \\
                &=  \left\{ \begin{array}{ll}
                    \n - \mln{3} - \iconstant + 1, & \text{ if } \rrln{2} = 0, \\
                    \indicator{\iconstant \le \mln{2} - \rrln{2}} (\n - \mln{3} - \iconstant + 1) 
                    + \indicator{\iconstant > \mln{2} - \rrln{2}} (\mln{2} - \iconstant + 1) , & \text{ if } \mln{2} > \rrln{2} > 0, \\
                    \mln{2} - \iconstant + 1, & \text{ if } \rrln{2} = \mln{2}
                \end{array}\right.\\
                & = \indicator{\iconstant \le \mln{2} - \rrln{2}} (\n - \mln{3} - \iconstant + 1) 
                + \indicator{\iconstant > \mln{2} - \rrln{2}} (\mln{2} - \iconstant + 1) \\
                & = \pnn{\rrln{2}}{\iconstant}.
            \end{align*}
            Combining this result with \eqref{supp:proposition:alg:4:lemma:2:eqn:8},
            we have
            \begin{align*}
                \rank{\byrsn{\rrln{2}, \rrln{3}}{{\iconstant}}}{\byrs{\rrln{2}, \rrln{3}}} = \pnn{\rrln{2}}{\iconstant} + \mln{3} -\rrln{3}.
            \end{align*}
            This proves \eqref{supp:proposition:alg:4:lemma:2:eqn:4} is true,
            and completes our proof.
        \end{proof}

        Using the results of Lemma~\ref{supp:proposition:alg:4:lemma:2},
        we can show the following result:
        \begin{proposition}  \label{supp:proposition:alg:4:1}
            Suppose $\bx, \by \in \vndistinct$ and for $0 < \mln{1}, \mln{2}, \mln{3} < \n$ such that $\mln{1} + \mln{2} + \mln{3} < \n$, 
            let $\bv = \{1, \ldots, \mln{2}\}$, $\bu = \{\mln{2} + 1,\ldots, \mln{2} + \mln{1}\}$ and $\bw = \{\n - \mln{3} + 1, \ldots, \n\}$ be subsets of indices. Suppose
            $\bxn{1} < \ldots < \bxn{\mln{2}}$, and
            $\byn{\mln{2} + 1} < \ldots < \byn{\mln{2} + \mln{1}}$.
            Let $(\bxs^{(\rrln{1}, \rrln{3})}, \bys^{(\rrln{2}, \rrln{3})}) \in (\mysecondset{\by}{\bu}{\tonumber{\n} \setminus \bw}{\n}, \mysecondset{\bx}{\bv}{\tonumber{\n} \setminus \bw}{\n}) \cap \mythirdset{\bw}{\n}$ 
            be imputations of $\bx, \by$ for indices
            $\bu \cup \bw, \bv \cup \bw$ such that 
            \begin{align*}
                &\sum_{\iconstant \in \bu} \indicator{ \bxs^{(\rrln{1}, \rrln{3})}(\iconstant)   < \min \svector{\bxs^{(\rrln{1}, \rrln{3})}}{\lconstant}{ (\tonumber{\n} \setminus \bw) \setminus \bu} } = \rrln{1},\\
                &\sum_{\iconstant \in \bv} \indicator{ \bys^{(\rrln{2}, \rrln{3})}(\iconstant)   < \min \svector{\bys^{(\rrln{2}, \rrln{3})}}{\lconstant}{ (\tonumber{\n} \setminus \bw) \setminus \bv} } = \rrln{2},\\
                \text{and }&\sum_{\iconstant \in \bw} \indicator{ \bxs^{(\rrln{1}, \rrln{3})}(\iconstant)   < \min \svector{\bxs^{(\rrln{1}, \rrln{3})}}{\lconstant}{\tonumber{\n} \setminus \bw} } = \rrln{3}.
            \end{align*}
            Denote $d_i = \rank{\byn{\iconstant}}{\svector{\by}{\lconstant}{ (\tonumber{\n} \setminus \bw) \setminus \bv }} - \rank{\bxn{\iconstant}}{\svector{\bx}{\lconstant}{ (\tonumber{\n} \setminus \bw) \setminus \bu }}$ for any $\iconstant \in (\tonumber{\n} \setminus \bw) \setminus (\bu \cup \bv)   $. Then for any
            $\rrln{1} \in \{0,\ldots, \mln{1}\}$,  $\rrln{2} \in \{0,\ldots, \mln{2}\}$, and $\rrln{2} \in \{0,\ldots, \mln{3}\}$, we have
            \begin{align*}
                \sfd{\bxrs{\rrln{1}, \rrln{3}}}{\byrs{\rrln{2}, \rrln{3}}} 
                & = \sum_{\iconstant \in \bv} \left|\rank{\bxn{\iconstant}}{\svector{\bx}{\lconstant}{ (\tonumber{\n} \setminus \bw) \setminus \bu }} + \rrln{1} -  \qnn{\rrln{2}}{\iconstant} + 2\rrln{3} - \mln{3} \right|\\
                & +  \sum_{\iconstant \in \bu} \left|\pnn{\rrln{1}}{\iconstant - \mln{2}} -  \rank{\byn{\iconstant}}{\svector{\by}{\lconstant}{ (\tonumber{\n} \setminus \bw) \setminus \bv }} - \rrln{2} + 2\rrln{3} - \mln{3} \right| \\
                & + \sum_{\iconstant \in (\tonumber{\n} \setminus \bw) \setminus (\bu \cup \bv)   } \left| \rrln{1} -  d_i - \rrln{2} + 2\rrln{3} - \mln{3} \right| + \grn{\rrln{3}},
            \end{align*}
            where $\pnn{\rrln{1}}{\iconstant} = \indicator{\iconstant \le  \mln{1} - \rrln{1}} (\n - \mln{3} - \iconstant + 1) 
            + \indicator{\iconstant >  \mln{1} - \rrln{1}} (\mln{1} - \iconstant + 1) $, $\qnn{\rrln{2}}{\iconstant}  = \indicator{\iconstant \le \mln{2} - \rrln{2}} (\n - \mln{3} - \iconstant + 1) 
            + \indicator{\iconstant > \mln{2} - \rrln{2}} (\mln{2} - \iconstant + 1)$,
            and 
            \begin{align*}
                \grn{\rrln{3}} = \sum_{\iconstant=1}^{\mln{3}} \left(\indicator{\iconstant \le \mln{3} - \rrln{3}} |\n + 1 - 2\iconstant| \right)
                + \sum_{\iconstant=1}^{\mln{3}}\left(\indicator{\iconstant > \mln{3} - \rrln{3}} |2\mln{3} - 2\iconstant + 1 -\n|\right).
            \end{align*}

        \end{proposition}

        \begin{proof}
            To start, according to Lemma~\ref{supp:proposition:alg:4:lemma:2}, we have
            \begin{align*}
                \rank{\bxrsn{\rrln{1}, \rrln{3}}{{\iconstant}}}{\bxrs{\rrln{1}, \rrln{3}}}
                = \rrln{3} + \rrln{1} + \rank{\bxn{\iconstant}}{\svector{\bx}{\lconstant}{(\tonumber{\n} \setminus \bw) \setminus (\bu \cup \bv) } }, \text{ for any } \iconstant \in \bv,
            \end{align*}
            and
            \begin{align*}
                \rank{\byrsn{\rrln{2}, \rrln{3}}{{\iconstant}}}{\byrs{\rrln{2}, \rrln{3}}} = \qnn{\rrln{2}}{\iconstant} + \mln{3} - \rrln{3}, \text{ for any } \iconstant \in \{1,\ldots, \mln{2}\}.
            \end{align*}
            Hence, we have
            \begin{align}
                \begin{split} \label{supp:proposition:alg:4:1:eqn:1}
                    &\sum_{\iconstant \in \bv } \left| \rank{\bxrsn{\rrln{1}, \rrln{3}}{\iconstant}}{\bxrs{\rrln{1}, \rrln{3}}} - \rank{\byrsn{\rrln{2}, \rrln{3}}{\iconstant}}{\byrs{\rrln{2}, \rrln{3}}}\right| \\
                    &=  \sum_{\iconstant \in \bv} \left|\rank{\bxn{\iconstant}}{\svector{\bx}{\lconstant}{ (\tonumber{\n} \setminus \bw) \setminus \bu }} + \rrln{1} -  \qnn{\rrln{2}}{\iconstant} + 2\rrln{3} - \mln{3} \right|.
                \end{split}
            \end{align}

            Next, according to Lemma~\ref{supp:proposition:alg:4:lemma:2}, we also have
            \begin{align*}
                \rank{\byrsn{\rrln{2}, \rrln{3}}{{\iconstant}}}{\byrs{\rrln{2}, \rrln{3}}}
                = \mln{3} - \rrln{3} + \rrln{2} + \rank{\byn{\iconstant}}{\svector{\by}{\lconstant}{(\tonumber{\n} \setminus \bw) \setminus (\bu \cup \bv) } }, \text{ for any } \iconstant \in \bu,
            \end{align*}
            and 
            \begin{align*}
                &\rank{\bxrsn{\rrln{1}, \rrln{3}}{{\iconstant + \mln{2}}}}{\bxrs{\rrln{1}, \rrln{3}}} = \pnn{\rrln{1}}{\iconstant} + \rrln{3}, \text{ for any } \iconstant \in \{1,\ldots,\mln{1}\}.\\
                \Rightarrow &\rank{\bxrsn{\rrln{1}, \rrln{3}}{{\iconstant }}}{\bxrs{\rrln{1}, \rrln{3}}} = \pnn{\rrln{1}}{\iconstant -\mln{2}} + \rrln{3}, \text{ for any } \iconstant \in \{\mln{2} + 1,\ldots,\mln{2} + \mln{1}\}.
            \end{align*}		
            Hence,
            \begin{align}
                \begin{split} \label{supp:proposition:alg:4:1:eqn:2}
                    &\sum_{\iconstant \in \bu } \left| \rank{\bxrsn{\rrln{1}, \rrln{3}}{\iconstant}}{\bxrs{\rrln{1}, \rrln{3}}} - \rank{\byrsn{\rrln{2}, \rrln{3}}{\iconstant}}{\byrs{\rrln{2}, \rrln{3}}}\right| \\
                    &=  \sum_{\iconstant \in \bu} \left|\pnn{\rrln{1}}{\iconstant - \mln{2}} + 2\rrln{3} - \mln{3} -\rrln{2} -   \rank{\byn{\iconstant}}{\svector{\by}{\lconstant}{(\tonumber{\n} \setminus \bw) \setminus (\bu \cup \bv)} } \right|.
                \end{split}
            \end{align}

            Next, according to Lemma~\ref{supp:proposition:alg:4:lemma:2}, for any $\iconstant \in (\tonumber{\n} \setminus \bw) \setminus (\bu \cup \bv)$, we have
            \begin{align*}
                &\rank{\bxrsn{\rrln{1}, \rrln{3}}{{\iconstant}}}{\bxrs{\rrln{1}, \rrln{3}}}
                = \rrln{3} + \rrln{1} + \rank{\bxn{\iconstant}}{\svector{\bx}{\lconstant}{(\tonumber{\n} \setminus \bw) \setminus (\bu \cup \bv) } },\\
                \text{ and }&\rank{\byrsn{\rrln{2}, \rrln{3}}{{\iconstant}}}{\byrs{\rrln{2}, \rrln{3}}}
                = \mln{3} - \rrln{3} + \rrln{2} + \rank{\byn{\iconstant}}{\svector{\by}{\lconstant}{(\tonumber{\n} \setminus \bw) \setminus (\bu \cup \bv) } }.
            \end{align*}
            Hence,
            \begin{align}
                \begin{split} \label{supp:proposition:alg:4:1:eqn:3}
                    &\sum_{\iconstant \in (\tonumber{\n} \setminus \bw) \setminus (\bu \cup \bv) } \left| \rank{\bxrsn{\rrln{1}, \rrln{3}}{\iconstant}}{\bxrs{\rrln{1}, \rrln{3}}} - \rank{\byrsn{\rrln{2}, \rrln{3}}{\iconstant}}{\byrs{\rrln{2}, \rrln{3}}}\right| \\
                    &=  \sum_{\iconstant \in (\tonumber{\n} \setminus \bw) \setminus (\bu \cup \bv)} \left|2\rrln{3} + \rrln{1} - \mln{3} - \rrln{2} - d_i \right|.
                \end{split}
            \end{align}

            Now, since
            $(\bxs^{(\rrln{1}, \rrln{3})}, \bys^{(\rrln{2}, \rrln{3})}) \in  \mythirdset{\bw}{\n}$
            and
            \begin{align*}
                \sum_{\iconstant \in \bw} \indicator{ \bxs^{(\rrln{1}, \rrln{3})}(\iconstant)   < \min \svector{\bxs^{(\rrln{1}, \rrln{3})}}{\lconstant}{\tonumber{\n} \setminus \bw} } = \rrln{3},
            \end{align*}
            then according to Lemma~\ref{supp:proposition:alg:3:lemma:1},
            we have 
            \begin{align}  \label{supp:proposition:alg:4:1:eqn:4}
                \sum_{\iconstant \in \bw } \left| \rank{\bxrsn{\rrln{1}, \rrln{3}}{\iconstant}}{\bxrs{\rrln{1}, \rrln{3}}} - \rank{\byrsn{\rrln{2}, \rrln{3}}{\iconstant}}{\byrs{\rrln{2}, \rrln{3}}}\right| = \grn{\rrln{3}}.
            \end{align}

            Finally, combining  \eqref{supp:proposition:alg:4:1:eqn:1},
            \eqref{supp:proposition:alg:4:1:eqn:2}, \eqref{supp:proposition:alg:4:1:eqn:3}
            and \eqref{supp:proposition:alg:4:1:eqn:4}, we have
            \begin{align*}
                \sfd{\bxrs{\rrln{1}, \rrln{3}}}{\byrs{\rrln{2}, \rrln{3}}} &= \sum_{\iconstant = 1}^{\n} \left| \rank{\bxrsn{\rrln{1}, \rrln{3}}{\iconstant}}{\bxrs{\rrln{1}, \rrln{3}}} - \rank{\byrsn{\rrln{2}, \rrln{3}}{\iconstant}}{\byrs{\rrln{2}, \rrln{3}}}\right| \\
                & = \sum_{\iconstant \in \bv } \left| \rank{\bxrsn{\rrln{1}, \rrln{3}}{\iconstant}}{\bxrs{\rrln{1}, \rrln{3}}} - \rank{\byrsn{\rrln{2}, \rrln{3}}{\iconstant}}{\byrs{\rrln{2}, \rrln{3}}}\right|  \\
                & + \sum_{\iconstant \in \bu } \left| \rank{\bxrsn{\rrln{1}, \rrln{3}}{\iconstant}}{\bxrs{\rrln{1}, \rrln{3}}} - \rank{\byrsn{\rrln{2}, \rrln{3}}{\iconstant}}{\byrs{\rrln{2}, \rrln{3}}}\right|  \\
                & + \sum_{\iconstant \in (\tonumber{\n} \setminus \bw) \setminus (\bu \cup \bv)} \left| \rank{\bxrsn{\rrln{1}, \rrln{3}}{\iconstant}}{\bxrs{\rrln{1}, \rrln{3}}} - \rank{\byrsn{\rrln{2}, \rrln{3}}{\iconstant}}{\byrs{\rrln{2}, \rrln{3}}}\right|  \\
                & + \sum_{\iconstant \in \bw} \left| \rank{\bxrsn{\rrln{1}, \rrln{3}}{\iconstant}}{\bxrs{\rrln{1}, \rrln{3}}} - \rank{\byrsn{\rrln{2}, \rrln{3}}{\iconstant}}{\byrs{\rrln{2}, \rrln{3}}}\right| \\
                & = \sum_{\iconstant \in \bv} \left|\rank{\bxn{\iconstant}}{\svector{\bx}{\lconstant}{ (\tonumber{\n} \setminus \bw) \setminus \bu }} + \rrln{1} -  \qnn{\rrln{2}}{\iconstant} + 2\rrln{3} - \mln{3} \right|\\
                & +  \sum_{\iconstant \in \bu} \left|\pnn{\rrln{1}}{\iconstant - \mln{2}} -  \rank{\byn{\iconstant}}{\svector{\by}{\lconstant}{ (\tonumber{\n} \setminus \bw) \setminus \bv }} - \rrln{2} + 2\rrln{3} - \mln{3} \right| \\
                & + \sum_{\iconstant \in (\tonumber{\n} \setminus \bw) \setminus (\bu \cup \bv)   } \left| \rrln{1} -  d_i - \rrln{2} + 2\rrln{3} - \mln{3} \right| + \grn{\rrln{3}}.
            \end{align*}
            This completes our proof.
        \end{proof}

        Finally, we show the following result:

        \begin{proposition} \label{supp:proposition:alg:4}
            Suppose $\bx, \by \in \vndistinct$ and for $0 < \mln{1}, \mln{2}, \mln{3} < \n$ such that $\mln{1} + \mln{2} + \mln{3} < \n$, 
            let $\bv = \{1, \ldots, \mln{2}\}$, $\bu = \{\mln{2} + 1,\ldots, \mln{2} + \mln{1}\}$ and $\bw = \{\n - \mln{3} + 1, \ldots, \n\}$ be subsets of indices. Suppose
            $\bxn{1} < \ldots < \bxn{\mln{2}}$, and
            $\byn{\mln{2} + 1} < \ldots < \byn{\mln{2} + \mln{1}}$.
            Let $(\bxs^{(\rrln{1}, \rrln{3})}, \bys^{(\rrln{2}, \rrln{3})}) \in (\mysecondset{\by}{\bu}{\tonumber{\n} \setminus \bw}{\n}, \mysecondset{\bx}{\bv}{\tonumber{\n} \setminus \bw}{\n}) \cap \mythirdset{\bw}{\n}$ 
            be imputations of $\bx, \by$ for indices
            $\bu \cup \bw$, and $\bv \cup \bw$, respectively, such that 
            \begin{align*}
                &\sum_{\iconstant \in \bu} \indicator{ \bxs^{(\rrln{1}, \rrln{3})}(\iconstant)   < \min \svector{\bxs^{(\rrln{1}, \rrln{3})}}{\lconstant}{ (\tonumber{\n} \setminus \bw) \setminus \bu} } = \rrln{1},\\
                &\sum_{\iconstant \in \bv} \indicator{ \bys^{(\rrln{2}, \rrln{3})}(\iconstant)   < \min \svector{\bys^{(\rrln{2}, \rrln{3})}}{\lconstant}{ (\tonumber{\n} \setminus \bw) \setminus \bv} } = \rrln{2},\\
                \text{and }&\sum_{\iconstant \in \bw} \indicator{ \bxs^{(\rrln{1}, \rrln{3})}(\iconstant)   < \min \svector{\bxs^{(\rrln{1}, \rrln{3})}}{\lconstant}{\tonumber{\n} \setminus \bw} } = \rrln{3}.
            \end{align*}
            For any
            $\rrln{1} \in \{0,\ldots, \mln{1}\}$, $\rrln{2} \in \{0, \ldots, \mln{2}\}$, and $\rrln{3} \in \{0, \ldots, \mln{3}\}$, 
            denote 
            \begin{align*}
                &\bs^{(\rrln{1}, \rrln{2}, \rrln{3})} = \sum_{\iconstant \in \bv} \indicator{\rank{\bxn{\iconstant}}{\svector{\bx}{\lconstant}{ (\tonumber{\n} \setminus \bw) \setminus \bu }} + \rrln{1} -  \qnn{\rrln{2}}{\iconstant} +2\rrln{3} - \mln{3} \ge 0},\\
                \text{and }&\br^{(\rrln{1}, \rrln{2}, \rrln{3})} = \sum_{\iconstant \in \bu} \indicator{\pnn{\rrln{1}}{\iconstant - \mln{2}} -  \rank{\byn{\iconstant}}{\svector{\by}{\lconstant}{ (\tonumber{\n} \setminus \bw) \setminus \bv }} - \rrln{2} - 1 + 2\rrln{3} - \mln{3} \ge 0},
            \end{align*}		
            where $\pnn{\rrln{1}}{\iconstant} = \indicator{\iconstant \le  \mln{1} - \rrln{1}} (\n - \mln{3} - \iconstant + 1) 
            + \indicator{\iconstant >  \mln{1} - \rrln{1}} (\mln{1} - \iconstant + 1) $, $\qnn{\rrln{2}}{\iconstant}  = \indicator{\iconstant \le \mln{2} - \rrln{2}} (\n - \mln{3} - \iconstant + 1) 
            + \indicator{\iconstant > \mln{2} - \rrln{2}} (\mln{2} - \iconstant + 1)$.
            For any $\iconstant \in \{-\mln{1}-1-2\mln{3}, \ldots, \mln{2}\}$,
            let $\ssln{\iconstant} = \sum_{\lconstant \in (\tonumber{\n} \setminus \bw) \setminus (\bu \cup \bv)} \indicator{\dln{\iconstant} \le \iconstant}$,
            and denote  $\np = \n - \mln{1} - \mln{2} - \mln{3}$.
            Then, for any $\rrln{1} \in \{0,\ldots, \mln{1} - 1\}$, $\rrln{2} \in \{0, \ldots, \mln{2}\}$, and $\rrln{3} \in \{0, \ldots, \mln{3}\}$, we have
            \begin{align} \label{supp:proposition:alg:4:eqn:1}
                \begin{split}
                    \sfdsb{\bxrs{\rrln{1}+1, \rrln{3}} }{\byrs{\rrln{2}, \rrln{3} }}  &= \sfdsb{\bxrs{\rrln{1}, \rrln{3}} }{\byrs{\rrln{2}, \rrln{3} }} \\
                    &+
                    2 S^{(\rrln{1}, \rrln{2}, \rrln{3})} - 2 \ssln{\mln{3} + \rrln{2} - \rrln{1} -2\rrln{3} - 1} + \bcnn{1}{\rrln{1}, \rrln{2}, \rrln{3}} ,
                \end{split}
            \end{align}
            for any $\rrln{1} \in \{0,\ldots, \mln{1}\}$, $\rrln{2} \in \{0, \ldots, \mln{2}-1\}$, and $\rrln{3} \in \{0, \ldots, \mln{3}\}$, we have
            \begin{align} \label{supp:proposition:alg:4:eqn:2}
                \begin{split}
                    \sfdsb{\bxrs{\rrln{1}, \rrln{3}} }{\byrs{\rrln{2}+1, \rrln{3} }}  &= \sfdsb{\bxrs{\rrln{1}, \rrln{3}} }{\byrs{\rrln{2}, \rrln{3} }} \\
                    & - 2 R^{(\rrln{1}, \rrln{2}, \rrln{3})} + 2 \ssln{\mln{3} + \rrln{2} - \rrln{1} -2\rrln{3}} + \bcnn{2}{\rrln{1}, \rrln{2}, \rrln{3}}.
                \end{split}
            \end{align}
            and for any $\rrln{1} \in \{0,\ldots, \mln{1}\}$, $\rrln{2} \in \{0, \ldots, \mln{2}\}$, and $\rrln{3} \in \{0, \ldots, \mln{3} - 1\}$, we have
            \begin{align} \label{supp:proposition:alg:4:eqn:3}
                \begin{split}
                    &\sfd{\bxrs{\rrln{1}, \rrln{3} + 1}}{\byrs{\rrln{2}, \rrln{3} + 1}}  = \sfd{\bxrs{\rrln{1}, \rrln{3}}}{\byrs{\rrln{2}, \rrln{3}}} + |2\rrln{3} + 1 - \n| \\
                    &- |\n + 1 - 2(\mln{3} - \rrln{3})| + 2S^{(\rrln{1}, \rrln{2}, \rrln{3})} + 2S^{(\rrln{1}+1, \rrln{2}, \rrln{3})} - 2\mln{2} \\
                    & + 2R^{(\rrln{1}, \rrln{2} - 1, \rrln{3})} + 2R^{(\rrln{1}, \rrln{2} - 2,  \rrln{3})} - 2\mln{1} \\
                    & + 2\ssln{2\rrln{3} + \rrln{1} - \rrln{2} - \mln{3}} + 2\ssln{2\rrln{3} + \rrln{1} - \rrln{2} -\mln{3} + 1} - 2\np.
                \end{split}
            \end{align}
            where for any $\rrln{1} \in \{0,\ldots, \mln{1}\}$, $\rrln{2} \in \{0, \ldots, \mln{2}\}$, and $\rrln{3} \in \{0, \ldots, \mln{3} \}$, 
            \begin{align*}
                &\bcnn{1}{\rrln{1}, \rrln{2}, \rrln{3}} = - \mln{2} + \np \\
                &+ \left|\rrln{1} + 1 - \rank{\byn{\mln{1} + \mln{2} - \rrln{1}}}{\svector{\by}{\lconstant}{ (\tonumber{\n} \setminus \bw) \setminus \bv }} - \rrln{2} +2\rrln{3} - \mln{3} \right|  \\
                & - \left|\n -\mln{3} - \mln{1} + \rrln{1} + 1 - \rank{\byn{\mln{1} + \mln{2} - \rrln{1}}}{\svector{\by}{\lconstant}{ (\tonumber{\n} \setminus \bw) \setminus \bv }} - \rrln{2} +2\rrln{3} - \mln{3} \right|,
            \end{align*}
            and
            \begin{align*}
                &\bcnn{2}{\rrln{1}, \rrln{2}, \rrln{3}} = \mln{1} - \np \\
                &+ \left|\rrln{2} + 1 -  \rank{\bxn{\mln{2} - \rr}}{\svector{\bx}{\lconstant}{ (\tonumber{\n} \setminus \bw) \setminus \bu }} - \rrln{1} -2\rrln{3} + \mln{3} \right|  \\
                & - \left|\n -\mln{3} - \mln{2} + \rrln{2} + 1 - \rank{\bxn{\mln{2} - \rr}}{\svector{\bx}{\lconstant}{ (\tonumber{\n} \setminus \bw) \setminus \bu }} - \rrln{1} - 2\rrln{3} + \mln{3} \right|.
            \end{align*}
        \end{proposition}

        \begin{proof}
            First, we prove \eqref{supp:proposition:alg:4:eqn:1} is true.
            According to Proposition~\ref{supp:proposition:alg:4:1},
            we have
            \begin{align*}
                \sfd{\bxrs{\rrln{1}, \rrln{3}}}{\byrs{\rrln{2}, \rrln{3}}} 
                & = \sum_{\iconstant \in \bv} \left|\rank{\bxn{\iconstant}}{\svector{\bx}{\lconstant}{ (\tonumber{\n} \setminus \bw) \setminus \bu }} + \rrln{1} -  \qnn{\rrln{2}}{\iconstant} + 2\rrln{3} - \mln{3} \right|\\
                & +  \sum_{\iconstant \in \bu} \left|\pnn{\rrln{1}}{\iconstant - \mln{2}} -  \rank{\byn{\iconstant}}{\svector{\by}{\lconstant}{ (\tonumber{\n} \setminus \bw) \setminus \bv }} - \rrln{2} + 2\rrln{3} - \mln{3} \right| \\
                & + \sum_{\iconstant \in (\tonumber{\n} \setminus \bw) \setminus (\bu \cup \bv)   } \left| \rrln{1} -  d_i - \rrln{2} + 2\rrln{3} - \mln{3} \right| + \grn{\rrln{3}},
            \end{align*}
            and
            \begin{align*}
                \sfd{\bxrs{\rrln{1} + 1, \rrln{3}}}{\byrs{\rrln{2}, \rrln{3}}} 
                & = \sum_{\iconstant \in \bv} \left|\rank{\bxn{\iconstant}}{\svector{\bx}{\lconstant}{ (\tonumber{\n} \setminus \bw) \setminus \bu }} + \rrln{1} + 1 -  \qnn{\rrln{2}}{\iconstant} + 2\rrln{3} - \mln{3} \right|\\
                & +  \sum_{\iconstant \in \bu} \left|\pnn{\rrln{1} + 1}{\iconstant - \mln{2}} -  \rank{\byn{\iconstant}}{\svector{\by}{\lconstant}{ (\tonumber{\n} \setminus \bw) \setminus \bv }} - \rrln{2} + 2\rrln{3} - \mln{3} \right| \\
                & + \sum_{\iconstant \in (\tonumber{\n} \setminus \bw) \setminus (\bu \cup \bv)   } \left| \rrln{1} + 1 -  d_i - \rrln{2} + 2\rrln{3} - \mln{3} \right| + \grn{\rrln{3}},
            \end{align*}
            Hence, we have
            \begin{align}
                \begin{split}  \label{supp:proposition:alg:4:eqn:4}
                    &\sfd{\bxrs{\rrln{1} + 1, \rrln{3}}}{\byrs{\rrln{2}, \rrln{3}}}  - \sfd{\bxrs{\rrln{1}, \rrln{3}}}{\byrs{\rrln{2}, \rrln{3}}} \\
                    & =  \sum_{\iconstant \in \bv} \left|\rank{\bxn{\iconstant}}{\svector{\bx}{\lconstant}{ (\tonumber{\n} \setminus \bw) \setminus \bu }} + \rrln{1} + 1 -  \qnn{\rrln{2}}{\iconstant} + 2\rrln{3} - \mln{3} \right| \\
                    & - \sum_{\iconstant \in \bv} \left|\rank{\bxn{\iconstant}}{\svector{\bx}{\lconstant}{ (\tonumber{\n} \setminus \bw) \setminus \bu }} + \rrln{1} -  \qnn{\rrln{2}}{\iconstant} + 2\rrln{3} - \mln{3} \right|\\
                    & + \sum_{\iconstant \in \bu} \left|\pnn{\rrln{1} + 1}{\iconstant - \mln{2}} -  \rank{\byn{\iconstant}}{\svector{\by}{\lconstant}{ (\tonumber{\n} \setminus \bw) \setminus \bv }} - \rrln{2} + 2\rrln{3} - \mln{3} \right| \\
                    & -  \sum_{\iconstant \in \bu} \left|\pnn{\rrln{1}}{\iconstant - \mln{2}} -  \rank{\byn{\iconstant}}{\svector{\by}{\lconstant}{ (\tonumber{\n} \setminus \bw) \setminus \bv }} - \rrln{2} + 2\rrln{3} - \mln{3} \right|\\
                    & +  \sum_{\iconstant \in (\tonumber{\n} \setminus \bw) \setminus (\bu \cup \bv)   } \left| \rrln{1} + 1 -  d_i - \rrln{2} + 2\rrln{3} - \mln{3} \right|  \\
                    & -  \sum_{\iconstant \in (\tonumber{\n} \setminus \bw) \setminus (\bu \cup \bv)   } \left| \rrln{1} -  d_i - \rrln{2} + 2\rrln{3} - \mln{3} \right|.
                \end{split}
            \end{align}

            Notice that, according to Lemma~\ref{supp:proposition:alg:1:lemma:0.0}, we have
            \begin{align}
                \begin{split} \label{supp:proposition:alg:4:eqn:5}
                    &\sum_{\iconstant \in \bv} \left|\rank{\bxn{\iconstant}}{\svector{\bx}{\lconstant}{ (\tonumber{\n} \setminus \bw) \setminus \bu }} + \rrln{1} + 1 -  \qnn{\rrln{2}}{\iconstant} +2\rrln{3} - \mln{3} \right| \\
                    & - \sum_{\iconstant \in \bv} \left|\rank{\bxn{\iconstant}}{\svector{\bx}{\lconstant}{ (\tonumber{\n} \setminus \bw) \setminus \bu }} + \rrln{1} -  \qnn{\rrln{2}}{\iconstant} + 2\rrln{3} - \mln{3} \right| \\ 
                    & =  2\sum_{\iconstant \in \bv} \indicator{\rank{\bxn{\iconstant}}{\svector{\bx}{\lconstant}{ (\tonumber{\n} \setminus \bw) \setminus \bu }} + \rrln{1} -  \qnn{\rrln{2}}{\iconstant} +2\rrln{3} - \mln{3} \ge 0} - \mln{2}, \\
                    & = 2 S^{(\rrln{1}, \rrln{2}, \rrln{3})} - \mln{2},
                \end{split}
            \end{align}	
            and
            \begin{align*} 
                \begin{split} 
                    &\sum_{\iconstant \in (\tonumber{\n} \setminus \bw) \setminus (\bu \cup \bv)   } \left| \dln{\iconstant} + \rrln{1} + 1 - \rrln{2} + 2\rrln{3} - \mln{3} \right| 
                    - \sum_{\iconstant \in (\tonumber{\n} \setminus \bw) \setminus (\bu \cup \bv)   } \left|\dln{\iconstant} +  \rrln{1} - \rrln{2}  + 2\rrln{3} - \mln{3} \right|\\
                    & = 2 \sum_{\iconstant \in (\tonumber{\n} \setminus \bw) \setminus (\bu \cup \bv)   } \indicator{\dln{\iconstant} + \rrln{1}  - \rrln{2} + 2\rrln{3} - \mln{3} \ge 0 } - (\n - \mln{1} - \mln{2} - \mln{3}) \\
                    & = 2 \sum_{\iconstant \in (\tonumber{\n} \setminus \bw) \setminus (\bu \cup \bv)} \indicator{\dln{\iconstant} \ge \mln{3} + \rrln{2} - \rrln{1} - 2\rrln{3}} - \np\\
                    & = 2\np - 2\sum_{\iconstant \in (\tonumber{\n} \setminus \bw) \setminus (\bu \cup \bv)} \indicator{\dln{\iconstant} < \mln{3} + \rrln{2} - \rrln{1} - 2\rrln{3}}  - \np.
                \end{split}
            \end{align*}
            Since $\mln{3}, \rrln{1}, \rrln{2}, \rrln{3}$ are integers, and for any $\iconstant \in (\tonumber{\n} \setminus \bw) \setminus (\bu \cup \bv)$,
            $d_i$ is an integer, we have
            \begin{align*}
                \indicator{\dln{\iconstant} < \mln{3} + \rrln{2} - \rrln{1} - 2\rrln{3} } = 
                \indicator{\dln{\iconstant} \le \mln{3} + \rrln{2} - \rrln{1} - 2\rrln{3} - 1}.
            \end{align*}
            Thus, we have
            \begin{align} 
                \begin{split} \label{supp:proposition:alg:4:eqn:6}
                    &\sum_{\iconstant \in (\tonumber{\n} \setminus \bw) \setminus (\bu \cup \bv)   } \left| \dln{\iconstant} + \rrln{1} + 1 - \rrln{2} + 2\rrln{3} - \mln{3} \right| \\ 
                    &- \sum_{\iconstant \in (\tonumber{\n} \setminus \bw) \setminus (\bu \cup \bv)   } \left|\dln{\iconstant} +  \rrln{1} - \rrln{2}  + 2\rrln{3} - \mln{3} \right|\\
                    & = \np - 2\sum_{\iconstant \in (\tonumber{\n} \setminus \bw) \setminus (\bu \cup \bv)} \indicator{\dln{\iconstant} \le \mln{3} + \rrln{2} - \rrln{1} - 2\rrln{3} - 1} \\
                    & = \np - 2 \ssln{\mln{3} + \rrln{2} - \rrln{1} -2\rrln{3} - 1}.
                \end{split}
            \end{align}

            Next, notice that
            \begin{align*}
                &\sum_{\iconstant \in \bu} \left|\pnn{\rrln{1} + 1}{\iconstant - \mln{2}} -  \rank{\byn{\iconstant}}{\svector{\by}{\lconstant}{ (\tonumber{\n} \setminus \bw) \setminus \bv }} - \rrln{2} +2\rrln{3} - \mln{3}\right| \\
                &- \sum_{\iconstant \in \bu} \left|\pnn{\rrln{1}}{\iconstant - \mln{2}} -  \rank{\byn{\iconstant}}{\svector{\by}{\lconstant}{ (\tonumber{\n} \setminus \bw) \setminus \bv }} - \rrln{2} +2\rrln{3} - \mln{3}\right| \\
                =	& \sum_{\iconstant = \mln{2} + 1}^{\mln{2} + \mln{1}} \left|\pnn{\rrln{1} + 1}{\iconstant - \mln{2}} -  \rank{\byn{\iconstant}}{\svector{\by}{\lconstant}{ (\tonumber{\n} \setminus \bw) \setminus \bv }} - \rrln{2} +2\rrln{3} - \mln{3}\right| \\
                - & \sum_{\iconstant = \mln{2} + 1}^{\mln{2} + \mln{1}} \left|\pnn{\rrln{1} }{\iconstant - \mln{2}} -  \rank{\byn{\iconstant}}{\svector{\by}{\lconstant}{ (\tonumber{\n} \setminus \bw) \setminus \bv }} - \rrln{2} +2\rrln{3} - \mln{3}\right| \\
                = & \sum_{\iconstant = 1}^{ \mln{1}} \left|\pnn{\rrln{1} + 1}{\iconstant} -  \rank{\byn{\mln{2} + \iconstant}}{\svector{\by}{\lconstant}{ (\tonumber{\n} \setminus \bw) \setminus \bv }} - \rrln{2} +2\rrln{3} - \mln{3}\right| \\
                - & \sum_{\iconstant = 1}^{ \mln{1}} \left|\pnn{\rrln{1} }{\iconstant} -  \rank{\byn{\mln{2} + \iconstant}}{\svector{\by}{\lconstant}{ (\tonumber{\n} \setminus \bw) \setminus \bv }} - \rrln{2} +2\rrln{3} - \mln{3}\right|,
            \end{align*}
            where $\pnn{\rrln{1}}{\iconstant} = \indicator{\iconstant \le  \mln{1} - \rrln{1}} (\n - \mln{3} - \iconstant + 1) + \indicator{\iconstant >  \mln{1} - \rrln{1}} (\mln{1} - \iconstant + 1)$
            for any $\iconstant \in \{1, \ldots, \mln{1}\}$ and $\rrln{1} \in \{0, \ldots, \mln{1}\}$.			
            Then, according to Lemma~\ref{supp:proposition:alg:2:lemma:3}, we have
            \begin{align*}
                & \sum_{\iconstant = 1}^{ \mln{1}} \left|\pnn{\rrln{1} + 1}{\iconstant} -  \rank{\byn{\mln{2} + \iconstant}}{\svector{\by}{\lconstant}{ (\tonumber{\n} \setminus \bw) \setminus \bv }} - \rrln{2} +2\rrln{3} - \mln{3}\right| \\
                - & \sum_{\iconstant = 1}^{ \mln{1}} \left|\pnn{\rrln{1} }{\iconstant} -  \rank{\byn{\mln{2} + \iconstant}}{\svector{\by}{\lconstant}{ (\tonumber{\n} \setminus \bw) \setminus \bv }} - \rrln{2} +2\rrln{3} - \mln{3}\right|\\
                & = \left|\rrln{1} + 1 - \rank{\byn{\mln{1} + \mln{2} - \rrln{1}}}{\svector{\by}{\lconstant}{ (\tonumber{\n} \setminus \bw) \setminus \bv }} - \rrln{2} +2\rrln{3} - \mln{3} \right|  \\
                & - \left|\n -\mln{3} - \mln{1} + \rrln{1} + 1 - \rank{\byn{\mln{1} + \mln{2} - \rrln{1}}}{\svector{\by}{\lconstant}{ (\tonumber{\n} \setminus \bw) \setminus \bv }} - \rrln{2} +2\rrln{3} - \mln{3} \right| \\
                & = \bcnn{1}{\rrln{1}, \rrln{2}, \rrln{3}} + \mln{2} - \np.
            \end{align*}
            Hence, we have
            \begin{align}
                \begin{split} \label{supp:proposition:alg:4:eqn:7}
                    & \sum_{\iconstant \in \bu} \left|\pnn{\rrln{1} + 1}{\iconstant - \mln{2}} -  \rank{\byn{\iconstant}}{\svector{\by}{\lconstant}{ (\tonumber{\n} \setminus \bw) \setminus \bv }} - \rrln{2} +2\rrln{3} - \mln{3}\right| \\
                    & - \sum_{\iconstant \in \bu} \left|\pnn{\rrln{1}}{\iconstant - \mln{2}} -  \rank{\byn{\iconstant}}{\svector{\by}{\lconstant}{ (\tonumber{\n} \setminus \bw) \setminus \bv }} - \rrln{2} +2\rrln{3} - \mln{3}\right| \\
                    & = \bcnn{1}{\rrln{1}, \rrln{2}, \rrln{3}} + \mln{2} - \np.
                \end{split}
            \end{align}

            Put \eqref{supp:proposition:alg:4:eqn:5}, \eqref{supp:proposition:alg:4:eqn:6},
            and  \eqref{supp:proposition:alg:4:eqn:7} back into  \eqref{supp:proposition:alg:4:eqn:4}, 
            we obtain 
            \begin{align*}
                &\sfdsb{\bxrs{\rrln{1}+1, \rrln{3}} }{\byrs{\rrln{2}, \rrln{3} }}  - \sfdsb{\bxrs{\rrln{1}, \rrln{3}} }{\byrs{\rrln{2}, \rrln{3} }}  \\
                &= 2 S^{(\rrln{1}, \rrln{2}, \rrln{3})} - \mln{2} +  \np - 2 \tln{\mln{3} + \rrln{2} - \rrln{1} -2\rrln{3} - 1} + \bcnn{1}{\rrln{1}, \rrln{2}, \rrln{3}} + \mln{2} - \np \\
                & = 2 S^{(\rrln{1}, \rrln{2}, \rrln{3})} - 2 \ssln{\mln{3} + \rrln{2} - \rrln{1} -2\rrln{3} - 1} + \bcnn{1}{\rrln{1}, \rrln{2}, \rrln{3}} .
            \end{align*}
            This proves \eqref{supp:proposition:alg:4:eqn:1} is true.

            Similarly, we can show \eqref{supp:proposition:alg:4:eqn:2} is true.
            According to Proposition~\ref{supp:proposition:alg:4:1},
            we have
            \begin{align*}
                \sfd{\bxrs{\rrln{1}, \rrln{3}}}{\byrs{\rrln{2}, \rrln{3}}} 
                & = \sum_{\iconstant \in \bv} \left|\rank{\bxn{\iconstant}}{\svector{\bx}{\lconstant}{ (\tonumber{\n} \setminus \bw) \setminus \bu }} + \rrln{1} -  \qnn{\rrln{2}}{\iconstant} + 2\rrln{3} - \mln{3} \right|\\
                & +  \sum_{\iconstant \in \bu} \left|\pnn{\rrln{1}}{\iconstant - \mln{2}} -  \rank{\byn{\iconstant}}{\svector{\by}{\lconstant}{ (\tonumber{\n} \setminus \bw) \setminus \bv }} - \rrln{2} + 2\rrln{3} - \mln{3} \right| \\
                & + \sum_{\iconstant \in (\tonumber{\n} \setminus \bw) \setminus (\bu \cup \bv)   } \left| \rrln{1} -  d_i - \rrln{2} + 2\rrln{3} - \mln{3} \right| + q^{(\rrln{3})},
            \end{align*}
            and
            \begin{align*}
                \sfd{\bxrs{\rrln{1}, \rrln{3}}}{\byrs{\rrln{2} + 1, \rrln{3}}} 
                & = \sum_{\iconstant \in \bv} \left|\rank{\bxn{\iconstant}}{\svector{\bx}{\lconstant}{ (\tonumber{\n} \setminus \bw) \setminus \bu }} + \rrln{1} -  \qnn{\rrln{2} + 1}{\iconstant} + 2\rrln{3} - \mln{3} \right|\\
                & +  \sum_{\iconstant \in \bu} \left|\pnn{\rrln{1}}{\iconstant - \mln{2}} -  \rank{\byn{\iconstant}}{\svector{\by}{\lconstant}{ (\tonumber{\n} \setminus \bw) \setminus \bv }} - \rrln{2} - 1  + 2\rrln{3} - \mln{3} \right| \\
                & + \sum_{\iconstant \in (\tonumber{\n} \setminus \bw) \setminus (\bu \cup \bv)   } \left| \rrln{1} -  d_i - \rrln{2} - 1 + 2\rrln{3} - \mln{3} \right| + q^{(\rrln{3})},
            \end{align*}
            Hence, we have
            \begin{align}
                \begin{split}  \label{supp:proposition:alg:4:eqn:8}
                    &\sfd{\bxrs{\rrln{1}, \rrln{3}}}{\byrs{\rrln{2} + 1, \rrln{3}}}  - \sfd{\bxrs{\rrln{1}, \rrln{3}}}{\byrs{\rrln{2}, \rrln{3}}} \\
                    & =  \sum_{\iconstant \in \bv} \left|\rank{\bxn{\iconstant}}{\svector{\bx}{\lconstant}{ (\tonumber{\n} \setminus \bw) \setminus \bu }} + \rrln{1} -  \qnn{\rrln{2} + 1}{\iconstant} + 2\rrln{3} - \mln{3} \right| \\
                    & - \sum_{\iconstant \in \bv} \left|\rank{\bxn{\iconstant}}{\svector{\bx}{\lconstant}{ (\tonumber{\n} \setminus \bw) \setminus \bu }} + \rrln{1} -  \qnn{\rrln{2}}{\iconstant} + 2\rrln{3} - \mln{3} \right|\\
                    & + \sum_{\iconstant \in \bu} \left|\pnn{\rrln{1}}{\iconstant - \mln{2}} -  \rank{\byn{\iconstant}}{\svector{\by}{\lconstant}{ (\tonumber{\n} \setminus \bw) \setminus \bv }} - \rrln{2} - 1+ 2\rrln{3} - \mln{3} \right| \\
                    & -  \sum_{\iconstant \in \bu} \left|\pnn{\rrln{1}}{\iconstant - \mln{2}} -  \rank{\byn{\iconstant}}{\svector{\by}{\lconstant}{ (\tonumber{\n} \setminus \bw) \setminus \bv }} - \rrln{2} + 2\rrln{3} - \mln{3} \right|\\
                    & +  \sum_{\iconstant \in (\tonumber{\n} \setminus \bw) \setminus (\bu \cup \bv)   } \left| \rrln{1} -  d_i - \rrln{2} - 1 + 2\rrln{3} - \mln{3} \right|  \\
                    & -  \sum_{\iconstant \in (\tonumber{\n} \setminus \bw) \setminus (\bu \cup \bv)   } \left| \rrln{1} -  d_i - \rrln{2} + 2\rrln{3} - \mln{3} \right|.
                \end{split}
            \end{align}

            Notice that, according to Lemma~\ref{supp:proposition:alg:1:lemma:0.0}, we have
            \begin{align} \label{supp:proposition:alg:4:eqn:9}
                \begin{split} 
                    &\sum_{\iconstant \in \bu} \left|\pnn{\rrln{1}}{\iconstant - \mln{2}} -  \rank{\byn{\iconstant}}{\svector{\by}{\lconstant}{ (\tonumber{\n} \setminus \bw) \setminus \bv }} - \rrln{2} - 1 + 2\rrln{3} - \mln{3} \right|  \\
                    & - \sum_{\iconstant \in \bu} \left|\pnn{\rrln{1}}{\iconstant - \mln{2}} -  \rank{\byn{\iconstant}}{\svector{\by}{\lconstant}{ (\tonumber{\n} \setminus \bw) \setminus \bv }} - \rrln{2} + 2\rrln{3} - \mln{3} \right|  \\ 
                    & = - 2\sum_{\iconstant \in \bu} \indicator{\pnn{\rrln{1}}{\iconstant - \mln{2}} -  \rank{\byn{\iconstant}}{\svector{\by}{\lconstant}{ (\tonumber{\n} \setminus \bw) \setminus \bv }} - \rrln{2} - 1 + 2\rrln{3} - \mln{3} \ge 0} + \mln{1}, \\
                    & = -2 R^{(\rrln{1}, \rrln{2}, \rrln{3})} + \mln{1},
                \end{split}
            \end{align}
            and
            \begin{align*}
                \begin{split}
                    &\sum_{\iconstant \in (\tonumber{\n} \setminus \bw) \setminus (\bu \cup \bv)   } \left|\dln{\iconstant} +  \rrln{1} - \rrln{2} - 1  + 2\rrln{3} - \mln{3} \right| - \sum_{\iconstant \in (\tonumber{\n} \setminus \bw) \setminus (\bu \cup \bv)   } \left|\dln{\iconstant} +  \rrln{1} - \rrln{2}  + 2\rrln{3} - \mln{3} \right|\\
                    & = -2 \sum_{\iconstant \in (\tonumber{\n} \setminus \bw) \setminus (\bu \cup \bv)   } \indicator{\dln{\iconstant} + \rrln{1}  - \rrln{2} -1 + 2\rrln{3} - \mln{3} \ge 0 } + (\n - \mln{1} - \mln{2} - \mln{3}) \\
                    & = -2 \sum_{\iconstant \in (\tonumber{\n} \setminus \bw) \setminus (\bu \cup \bv)} \indicator{\dln{\iconstant} \ge \mln{3} + \rrln{2} - \rrln{1} - 2\rrln{3} + 1} + \np \\
                    & = -2\left(\np - \sum_{\iconstant \in (\tonumber{\n} \setminus \bw) \setminus (\bu \cup \bv)} \indicator{\dln{\iconstant} < \mln{3} + \rrln{2} - \rrln{1} - 2\rrln{3} + 1}  \right) + \np.
                \end{split}
            \end{align*}
            Since $\mln{3}, \rrln{1}, \rrln{2}, \rrln{3}$ are integers, and for any $\iconstant \in (\tonumber{\n} \setminus \bw) \setminus (\bu \cup \bv)$,
            $d_i$ is an integer, we have
            \begin{align*}
                \indicator{\dln{\iconstant} < \mln{3} + \rrln{2} - \rrln{1} - 2\rrln{3} + 1} = 
                \indicator{\dln{\iconstant} \le \mln{3} + \rrln{2} - \rrln{1} - 2\rrln{3} }.
            \end{align*}
            Hence, we have
            \begin{align}
                \begin{split} \label{supp:proposition:alg:4:eqn:10}
                    &\sum_{\iconstant \in (\tonumber{\n} \setminus \bw) \setminus (\bu \cup \bv)   } \left|\dln{\iconstant} +  \rrln{1} - \rrln{2} - 1  + 2\rrln{3} - \mln{3} \right| \\
                    &- \sum_{\iconstant \in (\tonumber{\n} \setminus \bw) \setminus (\bu \cup \bv)   } \left|\dln{\iconstant} +  \rrln{1} - \rrln{2}  + 2\rrln{3} - \mln{3} \right|\\
                    & = 2 \ssln{\mln{3} + \rrln{2} - \rrln{1} -2\rrln{3}} - \np.
                \end{split}
            \end{align}

            Next, notice that
            \begin{align*}
                &\sum_{\iconstant \in \bv} \left|\rank{\bxn{\iconstant}}{\svector{\bx}{\lconstant}{ (\tonumber{\n} \setminus \bw) \setminus \bu }} + \rrln{1} -  \qnn{\rrln{2}+1}{\iconstant} +2\rrln{3} - \mln{3} \right| \\
                & - \sum_{\iconstant \in \bv} \left|\rank{\bxn{\iconstant}}{\svector{\bx}{\lconstant}{ (\tonumber{\n} \setminus \bw) \setminus \bu }} + \rrln{1} -  \qnn{\rrln{2}}{\iconstant} + 2\rrln{3} - \mln{3} \right| \\
                & = \sum_{\iconstant \in \bv} \left|\qnn{\rrln{2}+1}{\iconstant} - \rank{\bxn{\iconstant}}{\svector{\bx}{\lconstant}{ (\tonumber{\n} \setminus \bw) \setminus \bu }} - \rrln{1} -2\rrln{3} + \mln{3} \right| \\
                & -  \sum_{\iconstant \in \bv} \left| \qnn{\rrln{2}}{\iconstant} - \rank{\bxn{\iconstant}}{\svector{\bx}{\lconstant}{ (\tonumber{\n} \setminus \bw) \setminus \bu }} - \rrln{1} - 2\rrln{3} + \mln{3} \right|.
            \end{align*}
            Since $\qnn{\rrln{2}}{\iconstant}  = \indicator{\iconstant \le \mln{2} - \rrln{2}} (\n - \mln{3} - \iconstant + 1) + \indicator{\iconstant > \mln{2} - \rrln{2}} (\mln{2} - \iconstant + 1)$, then according to Lemma~\ref{supp:proposition:alg:2:lemma:3}, we have
            \begin{align*}
                &\sum_{\iconstant \in \bv} \left|\qnn{\rrln{2}+1}{\iconstant} - \rank{\bxn{\iconstant}}{\svector{\bx}{\lconstant}{ (\tonumber{\n} \setminus \bw) \setminus \bu }} - \rrln{1} -2\rrln{3} + \mln{3} \right| \\
                & -  \sum_{\iconstant \in \bv} \left| \qnn{\rrln{2}}{\iconstant} - \rank{\bxn{\iconstant}}{\svector{\bx}{\lconstant}{ (\tonumber{\n} \setminus \bw) \setminus \bu }} - \rrln{1} - 2\rrln{3} + \mln{3} \right| \\
                & =  \left|\rrln{2} + 1 -  \rank{\bxn{\mln{2} - \rr}}{\svector{\bx}{\lconstant}{ (\tonumber{\n} \setminus \bw) \setminus \bu }} - \rrln{1} -2\rrln{3} + \mln{3} \right|  \\
                & - \left|\n -\mln{3} - \mln{2} + \rrln{2} + 1 - \rank{\bxn{\mln{2} - \rr}}{\svector{\bx}{\lconstant}{ (\tonumber{\n} \setminus \bw) \setminus \bu }} - \rrln{1} - 2\rrln{3} + \mln{3} \right|\\
                & = \bcnn{2}{\rrln{1}, \rrln{2}, \rrln{3}}  - \mln{1} + \np.
            \end{align*}
            Hence, we have
            \begin{align}
                \begin{split} \label{supp:proposition:alg:4:eqn:11}
                    & \sum_{\iconstant \in \bv} \left|\rank{\bxn{\iconstant}}{\svector{\bx}{\lconstant}{ (\tonumber{\n} \setminus \bw) \setminus \bu }} + \rrln{1} -  \qnn{\rrln{2}+1}{\iconstant} +2\rrln{3} - \mln{3} \right| \\
                    & - \sum_{\iconstant \in \bv} \left|\rank{\bxn{\iconstant}}{\svector{\bx}{\lconstant}{ (\tonumber{\n} \setminus \bw) \setminus \bu }} + \rrln{1} -  \qnn{\rrln{2}}{\iconstant} + 2\rrln{3} - \mln{3} \right| \\
                    & = \bcnn{2}{\rrln{1}, \rrln{2}, \rrln{3}}  - \mln{1} + \np.
                \end{split}
            \end{align}

            Put \eqref{supp:proposition:alg:4:eqn:9}, \eqref{supp:proposition:alg:4:eqn:10},
            and  \eqref{supp:proposition:alg:4:eqn:11} back into  \eqref{supp:proposition:alg:4:eqn:8}, 
            we obtain 
            \begin{align*}
                &\sfdsb{\bxrs{\rrln{1}, \rrln{3}} }{\byrs{\rrln{2}+1, \rrln{3} }}  - \sfdsb{\bxrs{\rrln{1}, \rrln{3}} }{\byrs{\rrln{2}, \rrln{3} }}  \\
                &=- 2 R^{(\rrln{1}, \rrln{2}, \rrln{3})} + \mln{1} + 2 \tln{\mln{3} + \rrln{2} - \rrln{1} -2\rrln{3}} - \np + \bcnn{2}{\rrln{1}, \rrln{2}, \rrln{3}} - \mln{1} + \np\\
                & = - 2 R^{(\rrln{1}, \rrln{2}, \rrln{3})} + 2 \ssln{\mln{3} + \rrln{2} - \rrln{1} -2\rrln{3}} + \bcnn{2}{\rrln{1}, \rrln{2}, \rrln{3}}.
            \end{align*}
            This proves \eqref{supp:proposition:alg:4:eqn:2} is true.

            Finally, we show \eqref{supp:proposition:alg:4:eqn:3} is true.	
            According to Proposition~\ref{supp:proposition:alg:4:1},
            we have
            \begin{align*}
                \sfd{\bxrs{\rrln{1}, \rrln{3}}}{\byrs{\rrln{2}, \rrln{3}}} 
                & = \sum_{\iconstant \in \bv} \left|\rank{\bxn{\iconstant}}{\svector{\bx}{\lconstant}{ (\tonumber{\n} \setminus \bw) \setminus \bu }} + \rrln{1} -  \qnn{\rrln{2}}{\iconstant} + 2\rrln{3} - \mln{3} \right|\\
                & +  \sum_{\iconstant \in \bu} \left|\pnn{\rrln{1}}{\iconstant - \mln{2}} -  \rank{\byn{\iconstant}}{\svector{\by}{\lconstant}{ (\tonumber{\n} \setminus \bw) \setminus \bv }} - \rrln{2} + 2\rrln{3} - \mln{3} \right| \\
                & + \sum_{\iconstant \in (\tonumber{\n} \setminus \bw) \setminus (\bu \cup \bv)   } \left| \rrln{1} -  d_i - \rrln{2} + 2\rrln{3} - \mln{3} \right| + \grn{\rrln{3}},
            \end{align*}
            and
            \begin{align*}
                \sfd{\bxrs{\rrln{1}, \rrln{3} + 1}}{\byrs{\rrln{2}, \rrln{3} + 1}} 
                & = \sum_{\iconstant \in \bv} \left|\rank{\bxn{\iconstant}}{\svector{\bx}{\lconstant}{ (\tonumber{\n} \setminus \bw) \setminus \bu }} + \rrln{1} -  \qnn{\rrln{2} }{\iconstant} + 2\rrln{3} + 2 - \mln{3} \right|\\
                & +  \sum_{\iconstant \in \bu} \left|\pnn{\rrln{1}}{\iconstant - \mln{2}} -  \rank{\byn{\iconstant}}{\svector{\by}{\lconstant}{ (\tonumber{\n} \setminus \bw) \setminus \bv }} - \rrln{2}  + 2\rrln{3} + 2 - \mln{3} \right| \\
                & + \sum_{\iconstant \in (\tonumber{\n} \setminus \bw) \setminus (\bu \cup \bv)   } \left| \rrln{1} -  \dln{\iconstant} - \rrln{2}  + 2\rrln{3} + 2 - \mln{3} \right| + \grn{\rrln{3} + 1},
            \end{align*}
            Hence, we have
            \begin{align}
                \begin{split}  \label{supp:proposition:alg:4:eqn:12}
                    &\sfd{\bxrs{\rrln{1}, \rrln{3} + 1}}{\byrs{\rrln{2}, \rrln{3} + 1}}  - \sfd{\bxrs{\rrln{1}, \rrln{3}}}{\byrs{\rrln{2}, \rrln{3}}} \\
                    & =  \sum_{\iconstant \in \bv} \left|\rank{\bxn{\iconstant}}{\svector{\bx}{\lconstant}{ (\tonumber{\n} \setminus \bw) \setminus \bu }} + \rrln{1} -  \qnn{\rrln{2}}{\iconstant} + 2\rrln{3} + 2 - \mln{3} \right| \\
                    & - \sum_{\iconstant \in \bv} \left|\rank{\bxn{\iconstant}}{\svector{\bx}{\lconstant}{ (\tonumber{\n} \setminus \bw) \setminus \bu }} + \rrln{1} -  \qnn{\rrln{2}}{\iconstant} + 2\rrln{3} - \mln{3} \right|\\
                    & + \sum_{\iconstant \in \bu} \left|\pnn{\rrln{1}}{\iconstant - \mln{2}} -  \rank{\byn{\iconstant}}{\svector{\by}{\lconstant}{ (\tonumber{\n} \setminus \bw) \setminus \bv }} - \rrln{2} + 2\rrln{3} + 2 - \mln{3} \right| \\
                    & -  \sum_{\iconstant \in \bu} \left|\pnn{\rrln{1}}{\iconstant - \mln{2}} -  \rank{\byn{\iconstant}}{\svector{\by}{\lconstant}{ (\tonumber{\n} \setminus \bw) \setminus \bv }} - \rrln{2} + 2\rrln{3} - \mln{3} \right|\\
                    & +  \sum_{\iconstant \in (\tonumber{\n} \setminus \bw) \setminus (\bu \cup \bv)   } \left| \rrln{1} -  \dln{\iconstant} - \rrln{2} + 2\rrln{3} + 2 - \mln{3} \right|  \\
                    & -  \sum_{\iconstant \in (\tonumber{\n} \setminus \bw) \setminus (\bu \cup \bv)   } \left| \rrln{1} -  \dln{\iconstant} - \rrln{2} + 2\rrln{3} - \mln{3} \right| + \grn{\rrln{3} + 1} - \grn{\rrln{3}}.
                \end{split}
            \end{align}

            According to Lemma~\ref{supp:proposition:alg:3:lemma:1.1}, we have 
            \begin{align} \label{supp:proposition:alg:4:eqn:13}
                \begin{split}
                    \grn{\rrln{3} + 1} - \grn{\rrln{3}} = |2\rrln{3} + 1 - \n| - |\n + 1 - 2(\mln{3} - \rrln{3})|.
                \end{split}
            \end{align}

            According to Lemma~\ref{supp:proposition:alg:3:lemma:2}, we have
            \begin{align}  \label{supp:proposition:alg:4:eqn:14}
                \begin{split}
                    &  \sum_{\iconstant \in \bv} \left|\rank{\bxn{\iconstant}}{\svector{\bx}{\lconstant}{ (\tonumber{\n} \setminus \bw) \setminus \bu }} + \rrln{1} -  \qnn{\rrln{2}}{\iconstant} + 2\rrln{3} + 2 - \mln{3} \right| \\
                    & - \sum_{\iconstant \in \bv} \left|\rank{\bxn{\iconstant}}{\svector{\bx}{\lconstant}{ (\tonumber{\n} \setminus \bw) \setminus \bu }} + \rrln{1} -  \qnn{\rrln{2}}{\iconstant} + 2\rrln{3} - \mln{3} \right|\\
                    & =   2\sum_{\iconstant \in \bv} \indicator{\rrln{1} + 2\rrln{3} +  \rank{\bxn{\iconstant}}{\svector{\bx}{\lconstant}{(\tonumber{\n} \setminus \bw) \setminus (\bu \cup \bv) } } - \qnn{\rrln{2}}{\iconstant} - \mln{3} \ge 0}  \\
                    & + 2\sum_{\iconstant \in \bv} \indicator{\rrln{1} + 2\rrln{3} +  \rank{\bxn{\iconstant}}{\svector{\bx}{\lconstant}{(\tonumber{\n} \setminus \bw) \setminus (\bu \cup \bv) } } - \qnn{\rrln{2}}{\iconstant} - \mln{3} + 1 \ge 0} - 2\mln{2} \\
                    & = 2S^{(\rrln{1}, \rrln{2}, \rrln{3})} + 2S^{(\rrln{1}+1, \rrln{2}, \rrln{3})} - 2\mln{2}.
                \end{split}
            \end{align}			
            and
            \begin{align}  \label{supp:proposition:alg:4:eqn:15}
                \begin{split}
                    & \sum_{\iconstant \in \bu} \left|\pnn{\rrln{1}}{\iconstant - \mln{2}} -  \rank{\byn{\iconstant}}{\svector{\by}{\lconstant}{ (\tonumber{\n} \setminus \bw) \setminus \bv }} - \rrln{2} + 2\rrln{3} + 2 - \mln{3} \right| \\
                    & -  \sum_{\iconstant \in \bu} \left|\pnn{\rrln{1}}{\iconstant - \mln{2}} -  \rank{\byn{\iconstant}}{\svector{\by}{\lconstant}{ (\tonumber{\n} \setminus \bw) \setminus \bv }} - \rrln{2} + 2\rrln{3} - \mln{3} \right|\\
                    & =  2\sum_{\iconstant \in \bu} \indicator{\pnn{\rrln{1}}{\iconstant - \mln{2}} + 2\rrln{3}-\mln{3} - \rrln{2} - \rank{\byn{\iconstant}}{\svector{\by}{\lconstant}{(\tonumber{\n} \setminus \bw) \setminus (\bu \cup \bv) } }  \ge 0 } \\
                    & + 2\sum_{\iconstant \in \bu} \indicator{\pnn{\rrln{1}}{\iconstant - \mln{2}} + 2\rrln{3}-\mln{3} - \rrln{2} - \rank{\byn{\iconstant}}{\svector{\by}{\lconstant}{(\tonumber{\n} \setminus \bw) \setminus (\bu \cup \bv) } }  + 1\ge 0 } - 2\mln{1}\\
                    & = 2\br^{(\rrln{1}, \rrln{2} - 1, \rrln{3})} + 2\br^{(\rrln{1}, \rrln{2} - 2, \rrln{3})} - 2\mln{1},
                \end{split}
            \end{align}
            and
            \begin{align} \label{supp:proposition:alg:4:eqn:16}
                \begin{split}
                    &\sum_{\iconstant \in (\tonumber{\n} \setminus \bw) \setminus (\bu \cup \bv)   } \left| \rrln{1} -  \dln{\iconstant} - \rrln{2} + 2\rrln{3} + 2 - \mln{3} \right|  \\
                    & -  \sum_{\iconstant \in (\tonumber{\n} \setminus \bw) \setminus (\bu \cup \bv)   } \left| \rrln{1} -  \dln{\iconstant} - \rrln{2} + 2\rrln{3} - \mln{3} \right|   \\
                    & = 2\sum_{\iconstant \in (\tonumber{\n} \setminus \bw) \setminus (\bv \cup \bu) } \indicator{ \rrln{1} -  \dln{\iconstant} - \rrln{2} + 2\rrln{3} - \mln{3}   \ge 0} \\
                    & + 2\sum_{\iconstant \in (\tonumber{\n} \setminus \bw) \setminus (\bv \cup \bu) } \indicator{ \rrln{1} -  \dln{\iconstant} - \rrln{2} + 2\rrln{3} - \mln{3} +1 \ge 0} - 2(\n - \mln{1} -\mln{2} -\mln{3})\\
                    & = 2\sum_{\iconstant \in (\tonumber{\n} \setminus \bw) \setminus (\bv \cup \bu) } \indicator{\dln{\iconstant} \le 2\rrln{3} + \rrln{1} -\rrln{2} - \mln{3}}\\
                    & + 2\sum_{\iconstant \in (\tonumber{\n} \setminus \bw) \setminus (\bv \cup \bu) } \indicator{\dln{\iconstant} \le 2\rrln{3} + \rrln{1} -\rrln{2} - \mln{3} + 1} -2\np \\
                    & = 2\ssln{2\rrln{3} + \rrln{1} - \rrln{2} - \mln{3}} + 2\ssln{2\rrln{3} + \rrln{1} - \rrln{2} -\mln{3} + 1} - 2\np.
                \end{split}
            \end{align}

            Put \eqref{supp:proposition:alg:4:eqn:13},  \eqref{supp:proposition:alg:4:eqn:14},
            \eqref{supp:proposition:alg:4:eqn:15} and  \eqref{supp:proposition:alg:4:eqn:16}
            back into  \eqref{supp:proposition:alg:4:eqn:12}, we obtain
            \begin{align*}
                &\sfd{\bxrs{\rrln{1}, \rrln{3} + 1}}{\byrs{\rrln{2}, \rrln{3} + 1}}  - \sfd{\bxrs{\rrln{1}, \rrln{3}}}{\byrs{\rrln{2}, \rrln{3}}} \\
                & = |2\rrln{3} + 1 - \n| - |\n + 1 - 2(\mln{3} - \rrln{3})| + 2S^{(\rrln{1}, \rrln{2}, \rrln{3})} + 2S^{(\rrln{1}+1, \rrln{2}, \rrln{3})} - 2\mln{2} \\
                & +  2\br^{(\rrln{1}, \rrln{2} - 1, \rrln{3})} + 2\br^{(\rrln{1}, \rrln{2} - 2, \rrln{3})} - 2\mln{1} + 2\ssln{2\rrln{3} + \rrln{1} - \rrln{2} - \mln{3}} + 2\ssln{2\rrln{3} + \rrln{1} - \rrln{2} -\mln{3} + 1} - 2\np.
            \end{align*}
            This proves \eqref{supp:proposition:alg:4:eqn:3}, and completes our proof.
        \end{proof}

        \begin{algorithm} 
            \caption{An efficient algorithm for computing exact upper bounds of Spearman's Footrule under General Missing Case. } \label{supp:alg:4}
            \begin{algorithmic}[1]
                \Require{$\bx, \by \in \vndistinct$, where $\bx$ and $\by$ might be
                partially observed.} 
                \Ensure{Maximum possible Spearman's footrule distance between $X$ and $Y$.}
                \State Denote $\mln{1}$ as the number of pairs $(\bxn{\iconstant}, \byn{\iconstant})$ such that $\byn{\iconstant}$ is observed while
                $\bxn{\iconstant}$ is missing. Denote $\mln{2}$ as the number of pairs $(\bxn{\iconstant}, \byn{\iconstant})$ such that $\bxn{\iconstant}$ is observed while $\byn{\iconstant}$ is missing. 
                Denote $\mln{3}$ as the number of pairs such that
                both $(\bxn{\iconstant}, \byn{\iconstant})$
                are missing.
                \State If $\mln{1} + \mln{2} = 0$, then run Algorithm~\ref{supp:alg:3}. However, if $\mln{3} = 0$, then run Algorithm~\ref{supp:alg:2}. If $\mln{1} + \mln{2} + \mln{3} = \n$, then return $\sum_{\iconstant=1}^{\n} |\iconstant - (\n -\iconstant + 1)|$.			
                \State Rank all observed data in $\bx$ and $\by$.
                Relabel $\bx$ and $\by$ such that
                $\byn{1}, \ldots, \byn{\mln{2}}$ are missing, 
                while $\bxn{1}, \ldots, \bxn{\mln{2}}$ are observed,
                and $\bxn{1} < \ldots, \bxn{\mln{2}}$.
                Let $\bxn{\mln{2} + 1}, \ldots, \bxn{\mln{2} + \mln{2}}$
                be missing, while $\byn{\mln{2}+1}, \ldots, \byn{\mln{2}+\mln{1}}$
                are observed and  $\byn{\mln{2}+1}< \ldots < \byn{\mln{2}+\mln{1}}$.
                Let $(\bxn{\n - \mln{3} + 1}, \byn{\n - \mln{3} + 1}), \ldots, (\bxn{\n}, \byn{\n})$ be missing. \label{supp:alg:4:line:1}
                \State Denote $\bv = \{1,\ldots, \mln{2}\}$, $\bu = \{\mln{2} + 1, \ldots, \mln{2} + \mln{1}\}$, and $\bw = \{\n - \mln{3} + 1, \ldots, \n\}$. 
                \State For $\iconstant \in (\tonumber{\n} \setminus \bw) \setminus (\bu \cup \bv)$, let
                $\dln{\iconstant} = \rank{\byn{\iconstant}}{\svector{\by}{\lconstant}{(\tonumber{\n} \setminus \bw) \setminus \bv}} - \rank{\bxn{\iconstant}}{\svector{\bx}{\lconstant}{(\tonumber{\n} \setminus \bw) \setminus \bu}}$.\label{supp:alg:4:line:2}
                \State For any $\iconstant \in \{1, \ldots, \mln{2}\}$ and $\rrln{2} \in \{0, \ldots, \mln{2}\}$, let $\qnn{\rrln{2}}{\iconstant}  = \indicator{\iconstant \le \mln{2} - \rrln{2}} (\n - \mln{3} - \iconstant + 1) 
                + \indicator{\iconstant > \mln{2} - \rrln{2}} (\mln{2} - \iconstant + 1)$. \label{supp:alg:4:line:3}
                \State For any $\iconstant \in \{1, \ldots, \mln{1}\}$, let  $\pnn{0}{\iconstant} = \indicator{\iconstant \le  \mln{1}} (\n - \mln{3} - \iconstant + 1) 
                + \indicator{\iconstant >  \mln{1}} (\mln{1} - \iconstant + 1) $. \label{supp:alg:4:line:4}
                \State Run Algorithm~\ref{supp:alg:0} computing $\ssln{\iconstant} = \sum_{\lconstant \in (\tonumber{\n} \setminus \bw) \setminus (\bu \cup \bv)} \indicator{d_l \le \iconstant}$ for any $\iconstant \in \{-\mln{1} - \mln{2} - \mln{3} - 1, \ldots, \mln{1} + \mln{2} +  \mln{3} + 1\}$. \label{supp:alg:4:line:5}
                \State 	Run Algorithm~\ref{supp:alg:0} computing \label{supp:alg:4:line:6}
                \begin{align*}
                    \br^{(k)} = \sum_{\iconstant \in \bu} \indicator{\rank{\byn{\iconstant}}{\svector{\by}{\lconstant}{ (\tonumber{\n} \setminus \bw) \setminus \bv }} - \pnn{0}{\iconstant - \mln{2}}  \le k },
                \end{align*}		
                for any $k \in \{-\mln{2} - \mln{3} - 1, \ldots, \mln{3} - 1\}$.	
                \State For any $\rrln{2} \in \{0, \ldots, \mln{2}\}, \rrln{3} \in \{0, \ldots, \mln{3}\}$, let  \label{supp:alg:4:line:6.1}
                \begin{align*}
                    \br^{(0, \rrln{2}, \rrln{3})} = \br^{(k)}, \text{ if } - \rrln{2} - 1 + 2\rrln{3} - \mln{3} = k.
                \end{align*}
                \For{$\rrln{2} \in \{0, \ldots, \mln{2}\}$}
                \State  Run Algorithm~\ref{supp:alg:0} computing \label{supp:alg:4:line:7}
                \begin{align*}
                    \bs^{(k)} = \sum_{\iconstant \in \bv} \indicator{  \qnn{\rrln{2}}{\iconstant} - \rank{\bxn{\iconstant}}{\svector{\bx}{\lconstant}{ (\tonumber{\n} \setminus \bw) \setminus \bu }} \le  k },
                \end{align*}
                for any $k \in \{-\mln{3}, \ldots, \mln{1} + \mln{3}\}$.	
                \State For any $\rrln{1} \in \{0, \ldots, \mln{1}\}, \rrln{3} \in \{0, \ldots, \mln{3}\}$, let  \label{supp:alg:4:line:7.1}
                \begin{align*}
                    \bs^{(\rrln{1}, \rrln{2}, \rrln{3})} = \bs^{(k)}, \text{ if } \rrln{1} + 2\rrln{3} - \mln{3} = k.
                \end{align*}

                \EndFor		

                \algstore{alg:4:part:2}
            \end{algorithmic}

        \end{algorithm}

        \begin{algorithm} 
            \begin{algorithmic}[1]
                \algrestore{alg:4:part:2}
                \State Initialize \label{supp:alg:4:line:8}
                \begin{align*}
                    \bdln{0,0,0} &= \sum_{\iconstant = 1}^{\mln{2}} \left| \n + \iconstant - 1  - \rank{\bxn{\iconstant}}{\svector{\bx}{\lconstant}{(\tonumber{\n} \setminus \bw) \setminus \bu}}\right| \\
                    & + \sum_{\iconstant = 1}^{\mln{1}} \left| \n -\mln{3} + \iconstant - 1  - \rank{\byn{\mln{2} + \iconstant}}{\svector{\by}{\lconstant}{(\tonumber{\n} \setminus \bw) \setminus \bu}}\right| \\
                    & + \sum_{\iconstant = 1}^{\mln{3}} |\n + 1 - 2\iconstant| + \sum_{\iconstant \in (\tonumber{\n} \setminus \bw) \setminus (\bu \cup \bv)} |\dln{\iconstant} + \mln{3}|.
                \end{align*}
                \For{$\rrln{3} \in \{0,\ldots,\mln{3}\}$} \label{supp:alg:4:line:8.1}
                \For{$\rrln{2} \in \{0,\ldots,\mln{2}\}$} \label{supp:alg:4:line:8.2}
                \For{$\rrln{1} \in \{0, \ldots, \mln{1}\}$} \label{supp:alg:4:line:8.3}
                \State Denote $c_1 = \rrln{1} + 1  - \rrln{2} +2\rrln{3} - \mln{3}$, and compute \label{supp:alg:4:line:9}
                \begin{align*}
                    &\bcnn{1}{\rrln{1}, \rrln{2}, \rrln{3}} = - \mln{2} + \np + \left|c_1 - \rank{\byn{\mln{1} + \mln{2} - \rrln{1}}}{\svector{\by}{\lconstant}{ (\tonumber{\n} \setminus \bw) \setminus \bv }}  \right|  \\
                    & - \left|c_1 + \n -\mln{3} - \mln{1} - \rank{\byn{\mln{1} + \mln{2} - \rrln{1}}}{\svector{\by}{\lconstant}{ (\tonumber{\n} \setminus \bw) \setminus \bv }} \right|.
                \end{align*}
                \State Compute \label{supp:alg:4:line:10}
                \begin{align*}
                    \bdln{\rrln{1}+1, \rrln{2}, \rrln{3}} = \bdln{\rrln{1}, \rrln{2}, \rrln{3}} + 2 S^{(\rrln{1}, \rrln{2}, \rrln{3})} - 2 \ssln{\mln{3} + \rrln{2} - \rrln{1} -2\rrln{3} - 1} + \bcnn{1}{\rrln{1}, \rrln{2}, \rrln{3}}.
                \end{align*}
                \EndFor
                \State Let 
                \If{$\rrln{2} < \mln{2}$}
                \State Let $\rrln{1} = 0$. Denote $c_2 = \rrln{2} + 1  - \rrln{1} - 2\rrln{3} + \mln{3}$, and compute \label{supp:alg:4:line:11}
                \begin{align*}
                    &\bcnn{2}{\rrln{1}, \rrln{2}, \rrln{3}} = \mln{1} - \np + \left|c_2 -  \rank{\bxn{\mln{2} - \rr}}{\svector{\bx}{\lconstant}{ (\tonumber{\n} \setminus \bw) \setminus \bu }} \right|  \\
                    & - \left|c_2 + \n -\mln{3} - \mln{2} - \rank{\bxn{\mln{2} - \rr}}{\svector{\bx}{\lconstant}{ (\tonumber{\n} \setminus \bw) \setminus \bu }} \right|.
                \end{align*}
                \State Compute \label{supp:alg:4:line:12} 
                \begin{align*}
                    \bdln{\rrln{1}, \rrln{2}+1, \rrln{3}} = \bdln{\rrln{1}, \rrln{2}, \rrln{3}} - 2 R^{(\rrln{1}, \rrln{2}, \rrln{3})} + 2 \ssln{\mln{3} + \rrln{2} - \rrln{1} -2\rrln{3}} + \bcnn{2}{\rrln{1}, \rrln{2}, \rrln{3}}.
                \end{align*}
                \EndIf
                \EndFor
                \If{$\rrln{3} < \mln{3}$}
                \State Let $\rrln{1} = \rrln{2} = 0$ and compute \label{supp:alg:4:line:13}
                \begin{align*}
                    \bdln{\rrln{1}, \rrln{2}, \rrln{3} + 1} &= \bdln{\rrln{1}, \rrln{2}, \rrln{3}} 
                    + |2\rrln{3} + 1 - \n| - |\n + 1 - 2(\mln{3} - \rrln{3})| \\
                    &+ 2S^{(\rrln{1}, \rrln{2}, \rrln{3})} + 2S^{(\rrln{1}+1, \rrln{2}, \rrln{3})} - 2\mln{2}
                    + 2R^{(\rrln{1}, \rrln{2} - 1, \rrln{3})} + 2R^{(\rrln{1}, \rrln{2} - 2,  \rrln{3})} \\
                    &- 2\mln{1} + 2\ssln{2\rrln{3} + \rrln{1} - \rrln{2} - \mln{3}} + 2\ssln{2\rrln{3} + \rrln{1} - \rrln{2} -\mln{3} + 1} - 2\np.
                \end{align*}
                \EndIf
                \EndFor
                \State Return $\max \{\bdln{0,0, 0}, \ldots, \bdln{\mln{1},\mln{2}, \mln{3}}\}$.
            \end{algorithmic}
        \end{algorithm}	

        \begin{remark}
            A few comments of Algorithm~\ref{supp:alg:4} are made below:

            In Algorithm~\ref{supp:alg:4}, the initialization $\bdln{0, 0, 0}$ computed in
            line~\ref{supp:alg:4:line:8} equals to $\sfdsb{\bxrs{0,0}}{\byrs{0,0}}$ defined in 
            Proposition~\ref{supp:proposition:alg:4:1}.
            Spearman's footrule
            $\bdln{\rrln{1} + 1, \rrln{2}, \rrln{3}}$, $\bdln{\rrln{1}, \rrln{2} + 1, \rrln{3}}$,
            $\bdln{\rrln{1}, \rrln{2}, \rrln{3} + 1}$,
            and $\bdln{\rrln{1}, \rrln{2}, \rrln{3}}$ equal to 
            $\sfdsb{\bxrs{\rrln{1} + 1, \rrln{3}}}{\byrs{\rrln{2}, \rrln{3}}}$, 
            $\sfdsb{\bxrs{\rrln{1}, \rrln{3}}}{\byrs{\rrln{2} + 1, \rrln{3}}}$,
            $\sfdsb{\bxrs{\rrln{1}, \rrln{3} + 1}}{\byrs{\rrln{2}, \rrln{3} + 1}}$,
            and $\sfdsb{\bxrs{\rrln{1}, \rrln{3}}}{\byrs{\rrln{2}, \rrln{3}}}$,
            respectively, and are updated according to
            Proposition~\ref{supp:proposition:alg:4}.

            By computing $\{\bdln{0,0,0}, \ldots, \bdln{\mln{1}, \mln{2}, \mln{3}}\}$, Algorithm~\ref{supp:alg:4} 
            finds all possible Spearman's footrule
            values $\sfd{\bxs}{\bys}$, where $(\bxs,\bys) \in (\mysecondset{\by}{\bu}{\tonumber{\n} \setminus \bw}{\n},  
            \mysecondset{\bx}{\bv}{\tonumber{\n} \setminus \bw}{\n}) \cap \mythirdset{\bw}{\n}$ are imputations $\bx$ and $\by$ for indices $\bu\cup\bw$ and $\bv \cup \bw$, respectively.
            See discussions following Theorem 2.21 in the main paper
            for explanations.
            Hence, according to Theorem~\ref{supp:theorem:2.21},
            the algorithm guarantees to find the maximum possible Spearman's footrule
            between $\bx$ and $\by$.

            The computational complexity of Algorithm~\ref{supp:alg:4} is analyzed as follows. 
            Ranking and relabeling all observed components in $\bx$ and $\by$ in line~\ref{supp:alg:4:line:1} 
            requires $\bmo (\n \log \n)$ steps. 
            Using these ranks, in line~\ref{supp:alg:4:line:2} computing each $\dln{\iconstant}$ is $\bmo(1)$, 
            and so overall line~\ref{supp:alg:4:line:2} is $\bmo(\n - \mln{1} - \mln{2} - \mln{3})$.
            Line~\ref{supp:alg:4:line:3} and line~\ref{supp:alg:4:line:4} takes $\bmo (\mln{2}^2)$
            and $\bmo (\mln{1})$ steps, respectively.

            According to Remark~\ref{supp:remark:alg:0}, the computational
            complexity of running Algorithm~\ref{supp:alg:0} in 
            line~\ref{supp:alg:4:line:5} and line~\ref{supp:alg:4:line:6} is $\bmo(\mln{1} + \mln{2} + \mln{3})$,
            and $\bmo(\mln{2} + \mln{3})$, respectively. 		
            Line~\ref{supp:alg:4:line:6} takes $\bmo(\mln{2}\mln{3})$ steps. 
            In line~\ref{supp:alg:4:line:7}, each iteration of running Algorithm~\ref{supp:alg:0} 
            within the for loop is $\bmo(\mln{1} + \mln{3})$. 
            Line~\ref{supp:alg:4:line:7.1} takes $\bmo(\mln{1}\mln{3})$.
            Since the loop runs $\mln{2} + 1$ times, 
            the computational complexity for the loop is $\bmo(\mln{1}\mln{2}\mln{3})$. 

            Line~\ref{supp:alg:4:line:8} requires $\bmo (\n)$ steps using the ranks of observed components.
            Line~\ref{supp:alg:4:line:9}, line~\ref{supp:alg:4:line:10},
            line~\ref{supp:alg:4:line:11}, and line~\ref{supp:alg:4:line:12}
            all take $\bmo(1)$ steps. Since the loop in line~\ref{supp:alg:4:line:8.3}
            runs $\mln{1}$ times, 
            the computational complexity for the loop is $\bmo(\mln{1})$.
            Since line~\ref{supp:alg:4:line:13} takes $\bmo(1)$ steps,
            each iteration within the loop in line~\ref{supp:alg:4:line:8.2}
            takes $\bmo(\mln{1} + 1)$ steps. Since the loop in line~\ref{supp:alg:4:line:8.2}
            runs $\mln{2}$ times, 
            the computational complexity for this loop is $\bmo(\mln{1}\mln{2})$.
            Finally, the loop in line~\ref{supp:alg:4:line:8.1}
            runs $\mln{3}$ times. Hence, 
            the computational complexity for this 
            loop is $\bmo(\mln{1}\mln{2}\mln{3})$.

            Therefore, the overall computational complexity for Algorithm~\ref{supp:alg:2} is $\bmo(\n\log \n + \mln{2}^2 + \mln{1}\mln{2}\mln{3})$.
        \end{remark}

        \section{Complete table for Table 1 in the main paper} \label{appF}

        This section provides the following Table~\ref{supp:tab:1} which
        completes Table 1 in the main paper by showing the values of rank correlation 
        statistics of Spearman's footrule $D$, Spearman's rank correlation $\rho$
        and Kendall's $\tau$ when the rank of $d$ is imputed as $1, \ldots, 8$.

        \begin{table}[ht]
            \centering
            \caption{Compare the values of Spearman's footrule $D$, Spearman's rank correlation $\rho$
            and Kendall's $\tau$ coefficient between $\bxs$ and $\by$,
            where $\bxs$ is imputation of $\bx$ for object $d$.	
            The ranked $\bx$ when the value of object $d$ is missing is shown as the case when the value of $d$ is denoted as $*$. The values of ranked $\by$ for each object are included in bracelets. 	
            The minimum values of these statistics among all possible imputations of ranks are bolden.}
            \label{supp:tab:1}
            \begin{tabular}{cccccccccccccc}
                Rank of &  \multicolumn{8}{c}{Ranked }&  \multicolumn{3}{c}{Rank Correlation Statistics} \\
                Imputed value&  \multicolumn{8}{c}{imputation $\bxs$} &  \multicolumn{3}{c}{between $\bxs$ and $\by$}\\
                $d$ & $a$ & $b$ & $c$ & $d$ & $e$ & $f$ & $g$ & $h$ & $\sfd{\bxs}{\by}$ & $\rho(\bxs,\by)$ & $\tau(\bxs,\by)$ \\
                * & 7 & 3  & 6  & *  &2  &5 & 4 & 1 & -- & -- & --\\
                & (1) & (2) & (3) & (4) & (5)& (6)& (7) & (8) &-- & --& --	 \\
                1 & 8 & 4 & 7 & 1 & 3 & 6 & 5 & 2 & 26 & \textbf{122} & \textbf{19} \\ 
                2 & 8 & 4 & 7 & 2 & 3 & 6 & 5 & 1 & 26 & 130 & 20 \\ 
                3 & 8 & 4 & 7 & 3 & 2 & 6 & 5 & 1 & 26 & 132 & 21 \\ 
                4 & 8 & 3 & 7 & 4 & 2 & 6 & 5 & 1 & \textbf{24} & 128 & 20 \\ 
                5 & 8 & 3 & 7 & 5 & 2 & 6 & 4 & 1 & 26 & 134 & 21 \\ 
                6 & 8 & 3 & 7 & 6 & 2 & 5 & 4 & 1 & 28 & 138 & 22 \\ 
                7 & 8 & 3 & 6 & 7 & 2 & 5 & 4 & 1 & 28 & 136 & 21 \\ 
                8 & 7 & 3 & 6 & 8 & 2 & 5 & 4 & 1 & 28 & 130 & 20 \\ 
            \end{tabular}
        \end{table}

        \section{Additional simulation results} \label{appG}

        \subsection{Additional simulation results for the bounds with missing data}

\begin{figure}
            \centering
            \includegraphics[width=14cm]{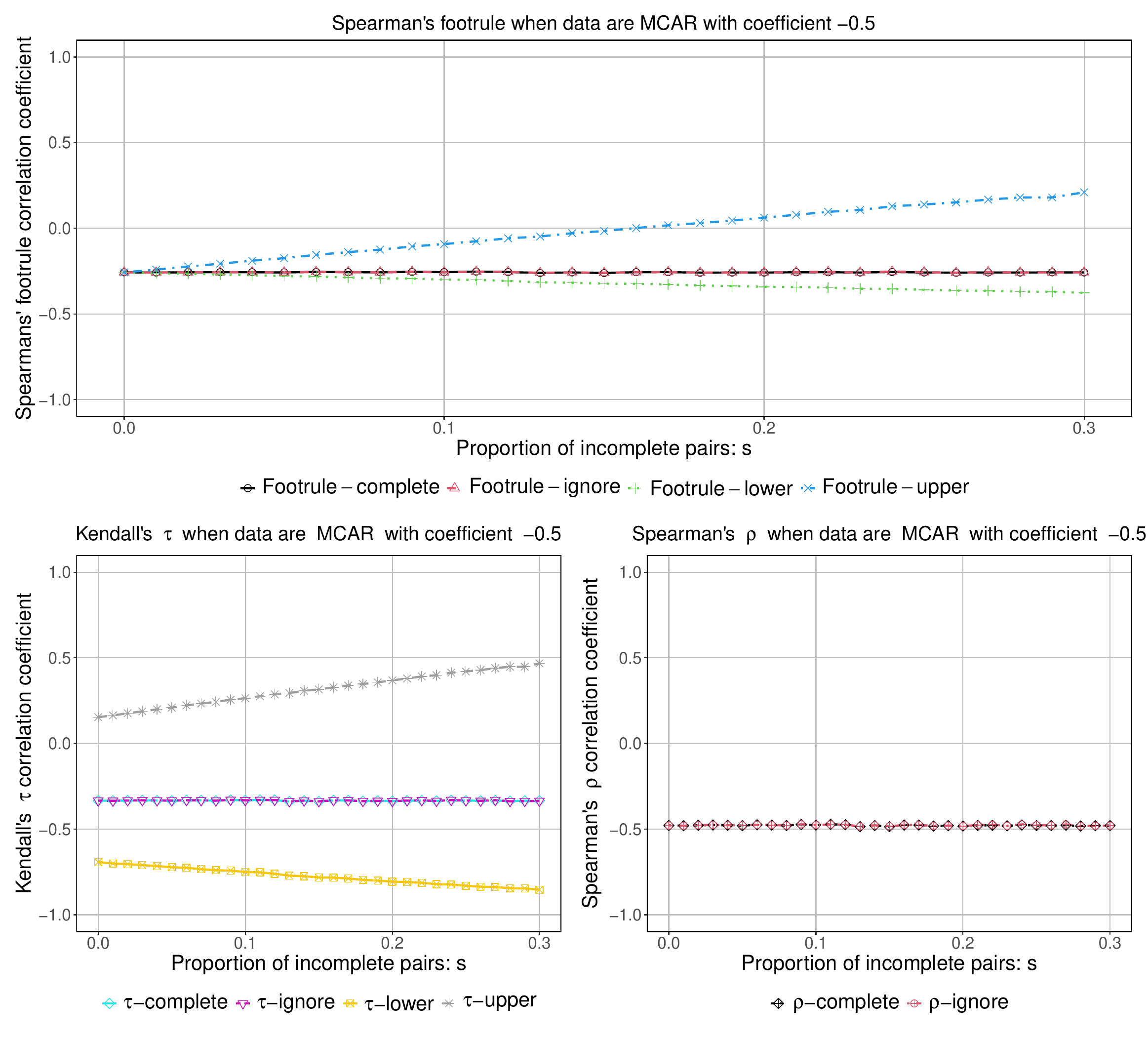}
            \caption{Rank correlation coefficients when data are missing completely at random (MCAR). 
            $\bx,\by$ are generated such that 
            $(X, Y) \overset{iid}{\sim} N(0, \Sigma)$, where
            $\Sigma = \begin{pmatrix} 1 &  \corrcoef  \\  \corrcoef & 1 \end{pmatrix}$,
                with covariance coefficient $\corrcoef=-0.5$. The methods are described in Table~\ref{supp:tab:2}.
                The sample size for $\bx$ and $\by$ is $\n = 100$.
                Results represent the average of 1000 Monte Carlo simulations.
                \textbf{Note}: this figure is generated following the same approach for generating 
                Figure~1 in the main paper, with different covariance coefficient $\corrcoef=-0.5$.}
                \label{supp:fig:1}
        \end{figure}

\begin{figure}
            \centering
            \includegraphics[width=14cm]{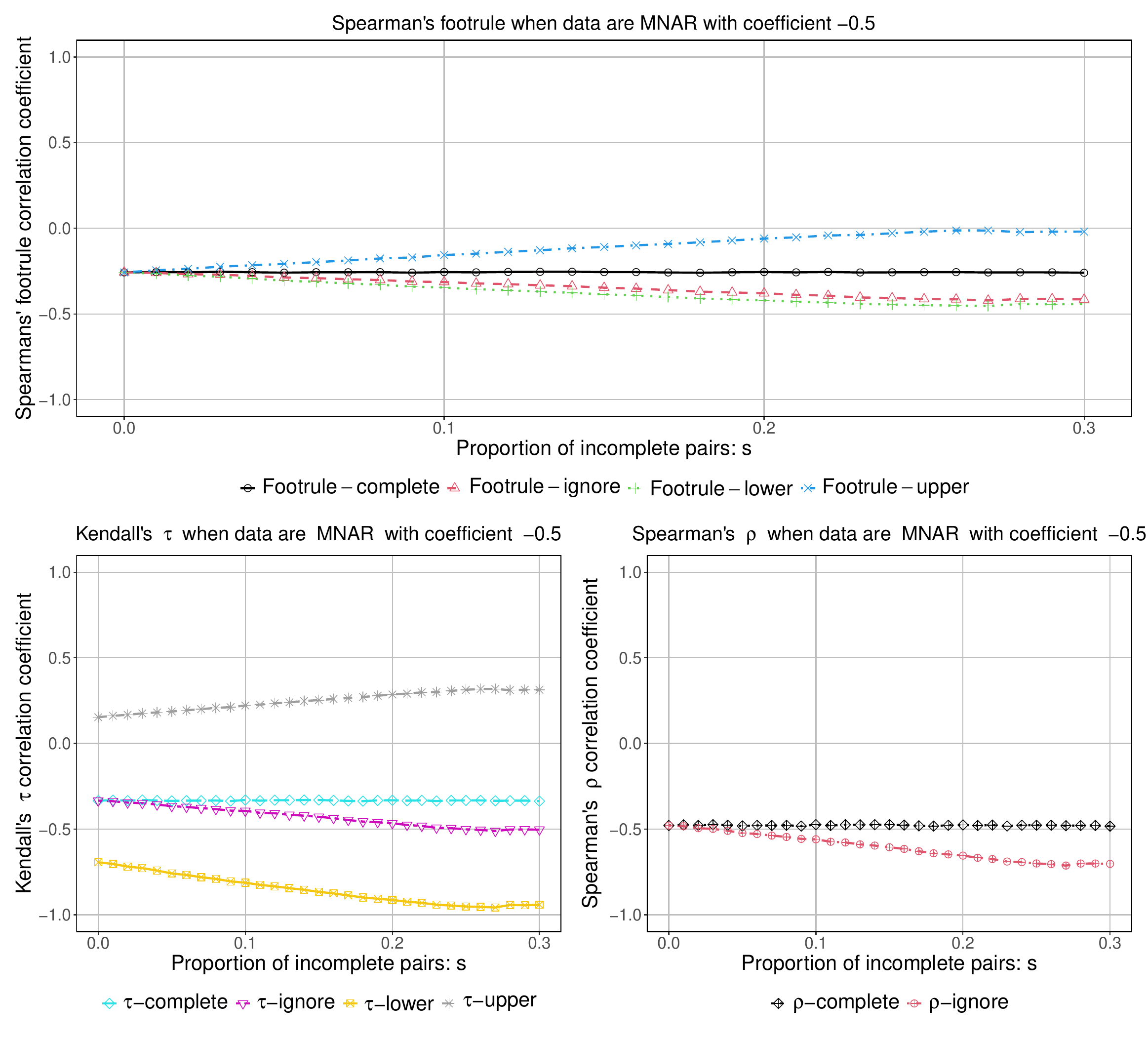}
            \caption{Rank correlation coefficients when data are missing not at random (MNAR). 
            $\bx,\by$ are generated such that 
            $(X, Y) \overset{iid}{\sim} N(0, \Sigma)$, where
            $\Sigma = \begin{pmatrix} 1 &  \corrcoef  \\  \corrcoef & 1 \end{pmatrix}$,
                with covariance coefficient $\corrcoef=-0.5$. The methods are described in Table~\ref{supp:tab:2}.
                The sample size for $\bx$ and $\by$ is $\n = 100$.
                Results represent the average of 1000 Monte Carlo simulations.
                \textbf{Note}: this figure is generated following the same approach for generating 
                Figure~2 in the main paper, with different covariance coefficient $\corrcoef=-0.5$.}
                \label{supp:fig:2}
        \end{figure}

\begin{figure}
            \centering
            \includegraphics[width=14cm]{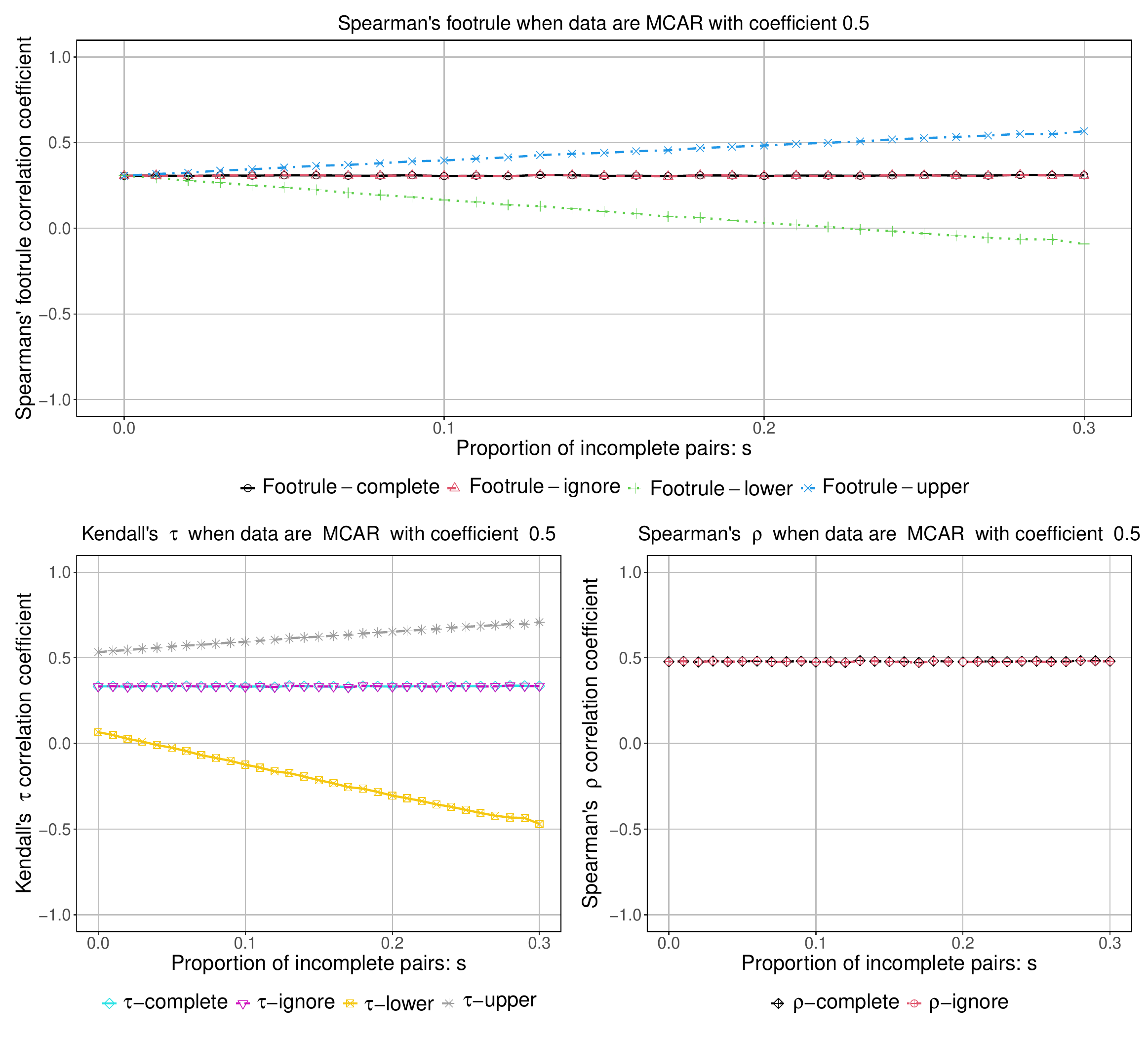}
            \caption{Rank correlation coefficients when data are missing completely at random (MCAR). 
            $\bx,\by$ are generated such that 
            $(X, Y) \overset{iid}{\sim} N(0, \Sigma)$, where
            $\Sigma = \begin{pmatrix} 1 &  \corrcoef  \\  \corrcoef & 1 \end{pmatrix}$,
                with covariance coefficient $\corrcoef=0.5$. The methods are described in Table~\ref{supp:tab:2}.
                The sample size for $\bx$ and $\by$ is $\n = 100$.
                Results represent the average of 1000 Monte Carlo simulations.
                \textbf{Note}: this figure is generated following the same approach for generating 
                Figure~1 in the main paper, with different covariance coefficient $\corrcoef=0.5$.}
                \label{supp:fig:3}
        \end{figure}

\begin{figure}
            \centering
            \includegraphics[width=14cm]{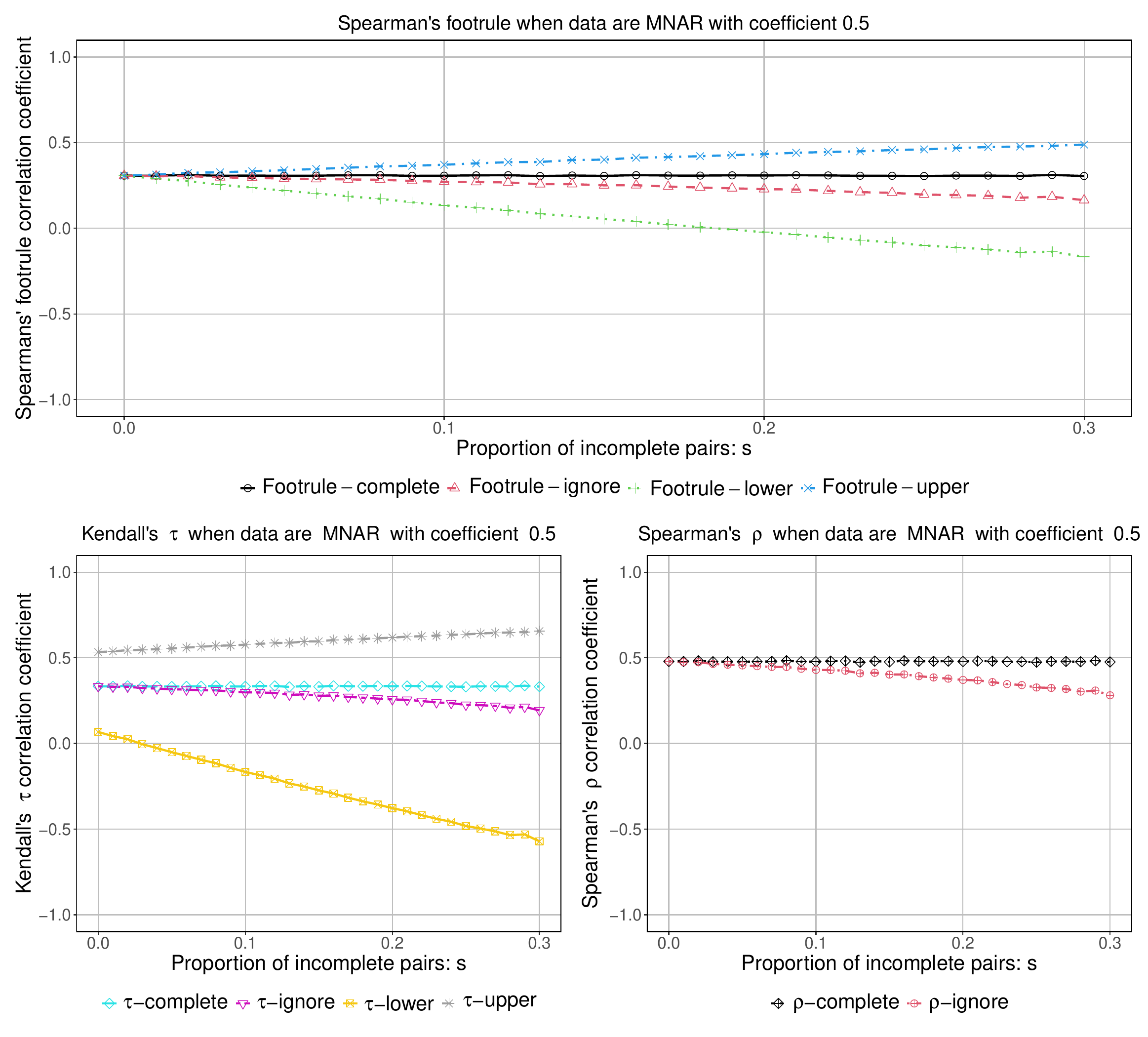}
            \caption{Rank correlation coefficients when data are missing not at random (MNAR). 
            $\bx,\by$ are generated such that 
            $(X, Y) \overset{iid}{\sim} N(0, \Sigma)$, where
            $\Sigma = \begin{pmatrix} 1 &  \corrcoef  \\  \corrcoef & 1 \end{pmatrix}$,
                with covariance coefficient $\corrcoef=0.5$. The methods are described in Table~\ref{supp:tab:2}.
                The sample size for $\bx$ and $\by$ is $\n = 100$.
                Results represent the average of 1000 Monte Carlo simulations.
                \textbf{Note}: this figure is generated following the same approach for generating 
                Figure~2 in the main paper, with different covariance coefficient $\corrcoef=0.5$.}
                \label{supp:fig:4}
        \end{figure}

        This subsection performs simulations following the same  way  
        as in Section 2.6 in the main paper, with different correlation 
        coefficients $\gamma \in \{-0.5, 0.5\}$.
        The results are shown in Figure~\ref{supp:fig:1}~--~\ref{supp:fig:4}.

        \begin{table*}
            \caption{Description of the rank correlation coefficients or their bounds, often in the presence of 
            missing data, shown in Figure~\ref{supp:fig:1}~--~\ref{supp:fig:4}.}
            \footnotesize
            \begin{tabular}{ll}
                \hline
                Coefficient/bound & Description \\
                \hline
                Footrule-upper & Upper bound for Spearman's footrule, when data is partially observed. \\
                Footrule-lower & Lower bound for Spearman's footrule, when data is partially observed. \\
                Footrule-ignore & Spearman's footrule, when any missing or partially observed data is ignored. \\
                Footrule-complete & Spearman's footrule, when data is fully observed. \\
                \hline
                $\tau$-upper & Upper bound for Kendall's $\tau$, when data is partially observed. \\
                $\tau$-lower & Lower bound for Kendall's $\tau$, when data is partially observed. \\
                $\tau$-ignore & Kendall's $\tau$, when any missing or partially observed data is ignored. \\
                $\tau$-complete & Kendall's $\tau$, when the data is fully observed. \\
                \hline
                $\rho$-ignore & Spearman's $\rho$, when any missing or partially observed data is ignored. \\
                $\rho$-complete & Spearman's $\rho$, when the data is fully observed. \\
                \hline
            \end{tabular}
            \label{supp:tab:2}
        \end{table*}

        \subsection{Additional independence testing results with missing data}	

        This subsection performs simulations following the same  way  
        as in Section 3.2.1, 3.2.2 and 3.2.3
        in the main paper, with potentially different correlation 
        coefficients $\gamma$, sample size $\n$, proportion of missing pairs $s$, 
        significance level $\alpha$, and missingness mechanisms.

        \begin{table}
            \caption{Description of the independence testing methods, often in the presence of 
            missing data, shown in Figure~\ref{supp:fig:5} -- \ref{supp:fig:23}.}
            \footnotesize
            \begin{tabular}{ll}
                \hline
                Testing method & Description \\
                \hline
                Proposed & Based on $p$-values computed from bounds Footrule-upper and Footrule-lower.  \\
                Footrule-ignore & Based on $p$-value of Spearman's footrule, when partially-observed data is ignored. \\
                Footrule-complete & Based on $p$-value of Spearman's footrule, when data is fully observed. \\
                \hline
                Footrule-mean &  Based on $p$-value of Spearman's footrule, using mean imputation for missing values. \\
                Footrule-median & Based on $p$-value of Spearman's footrule, using median imputation for missing values.\\
                Footrule-hot deck &Based on $p$-value of Spearman's footrule, using hot deck imputation for missing values.\\
                \hline
                $\tau$-ignore & Based on $p$-value of Kendall's $\tau$ coefficient, when partially observed data is ignored. \\
                $\tau$-complete & Based on $p$-value of Kendall's $\tau$ coefficient, when the data is fully observed. \\
                \hline
                $\tau$-mean &  Based on $p$-value of Kendall's $\tau$, using mean imputation for missing values.\\
                $\tau$-median & Based on $p$-value of Kendall's $\tau$, using median imputation for missing values.\\
                $\tau$-hot deck & Based on $p$-value of Kendall's $\tau$, using hot deck imputation for missing values.\\
                \hline
                $\rho$-ignore & Based on $p$-value of Spearman's $\rho$, when partially observed data is ignored. \\
                $\rho$-complete & Based on $p$-value of Spearman's $\rho$, when the data is fully observed. \\
                \hline
                $\rho$-mean &  Based on $p$-value of Spearman's $\rho$, using mean imputation for missing values.\\
                $\rho$-median & Based on $p$-value of Spearman's $\rho$, using median imputation for missing values.\\
                $\rho$-hot deck & Based on $p$-value of Spearman's $\rho$, using hot deck imputation for missing values.\\
                \hline
                Alvo and Cabilio's $\rho$ & Based on $p$-value of estimate of Spearman's rank correlation $\rho$, from \cite{Alvo1995RankCM}.\\
                Alvo and Cabilio's $\tau$ & Based on $p$-value of estimate of Kendall's $\tau$ coefficent, from \cite{Alvo1995RankCM}.\\
                \hline
            \end{tabular}
            \label{supp:tab:3}
        \end{table}

        \subsubsection{As the proportion of missing data increases}
\begin{figure}
            \centering
            \includegraphics[width=13.5cm]{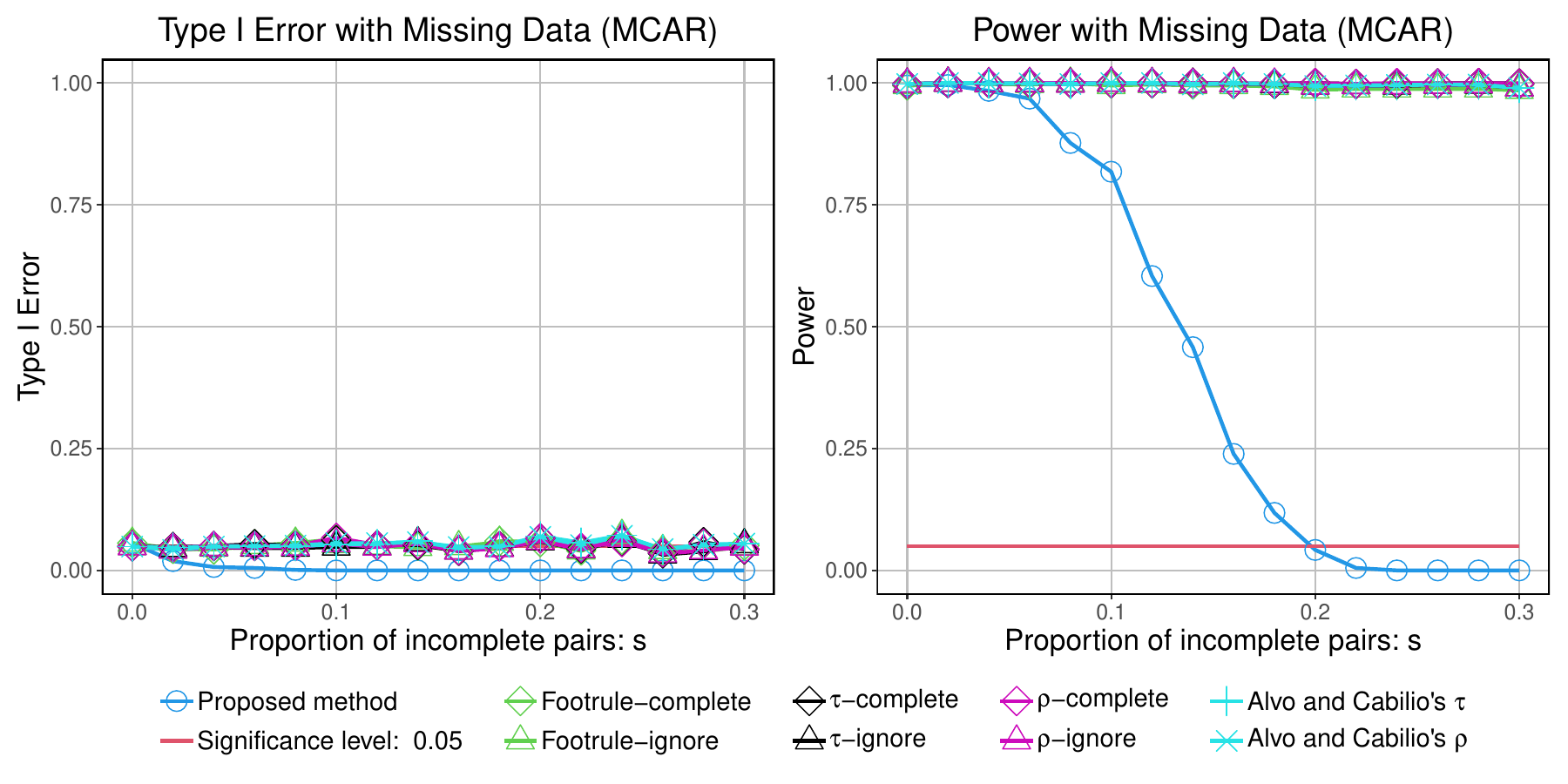}
            \vspace{-0.3cm}
            \caption{Statistical Type I error and power of the proposed method,
            methods that ignore missing data or use complete data,
            and Alvo and Cabilio's $\rho$ and Alvo and Cabilio's $\tau$ methods,
            as the proportion of missing data $s$ increases,
            where $s \in \{0.00, 0.02, \ldots, 0.30\}$.
            These methods are described in Table~\ref{supp:tab:3}.
            The data is missing completely at random (MCAR).
            (Left) Type I error: $(\bx,\by) \overset{iid}{\sim} N(0, I_2)$;
            (Right) Power: 	$(X, Y) \overset{iid}{\sim} N(0, \Sigma)$, where
            $\Sigma = \begin{pmatrix} 1 &  \corrcoef  \\  \corrcoef & 1 \end{pmatrix}$,
                with covariance coefficient $\corrcoef=0.5$.
                For both figures, a significance level $\alpha = 0.05$ is used and the sample
                size $\n = 100$. The results in the figures are average of 1000 trials. 
                \textbf{Note}: this figure is generated following the same approach for generating 
                Figure~3 in the main paper, with different sample size $\n = 100$.}
                \label{supp:fig:5}
        \end{figure}

\begin{figure}
            \centering
            \includegraphics[width=13.5cm]{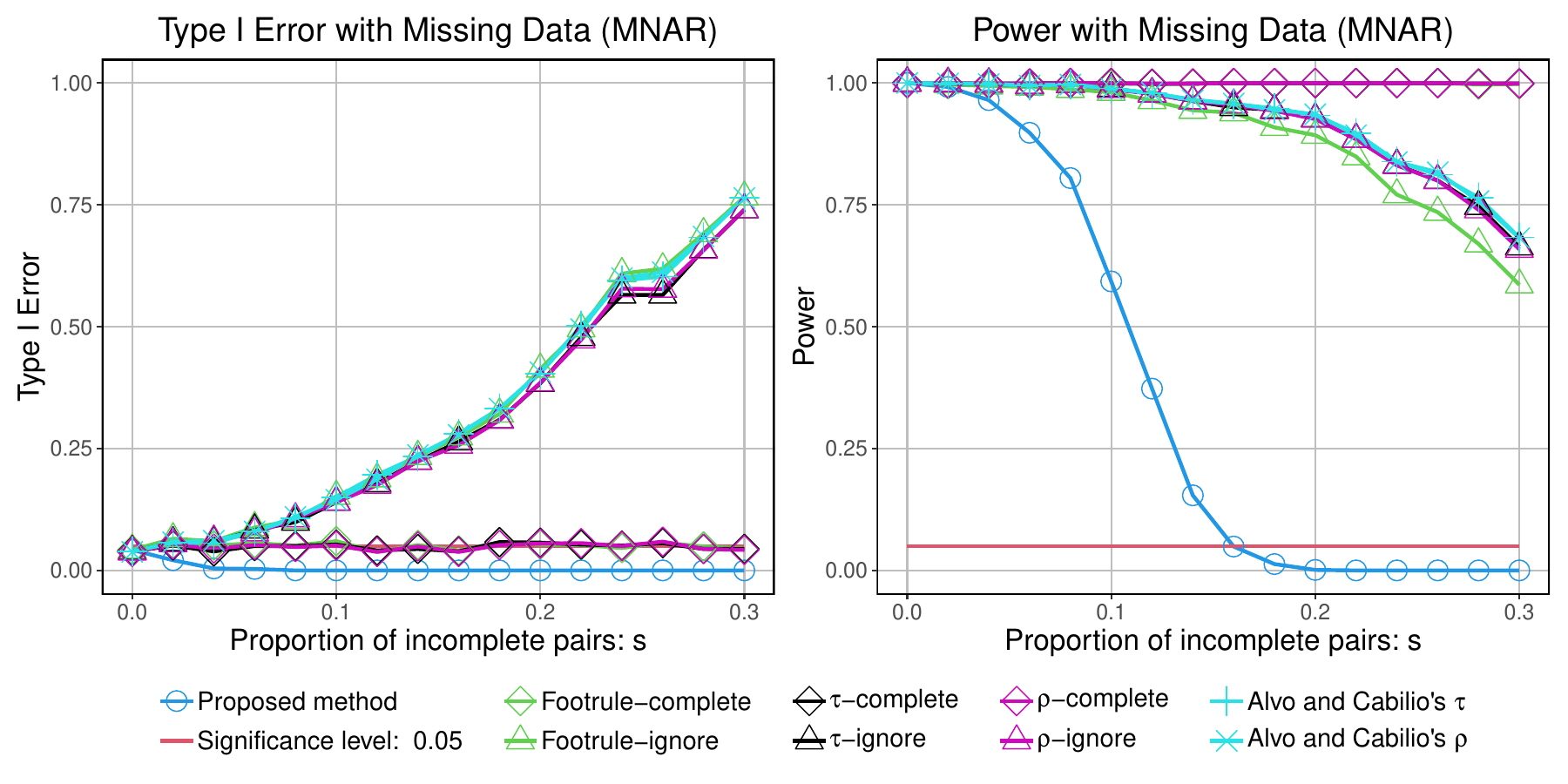}
            \vspace{-0.3cm}
            \caption{Statistical Type I error and power of the proposed method,
            methods that ignore missing data or use complete data,
            and Alvo and Cabilio's $\rho$ and Alvo and Cabilio's $\tau$ methods,
            as the proportion of missing data $s$ increases,
            where $s \in \{0.00, 0.02, \ldots, 0.30\}$.
            These methods are described in Table~\ref{supp:tab:3}.
            The data is missing not at random (MNAR).
            (Left) Type I error: $(\bx,\by) \overset{iid}{\sim} N(0, I_2)$;
            (Right) Power: 	$(X, Y) \overset{iid}{\sim} N(0, \Sigma)$, where
            $\Sigma = \begin{pmatrix} 1 &  \corrcoef  \\  \corrcoef & 1 \end{pmatrix}$,
                with covariance coefficient $\corrcoef=0.5$.
                For both figures, a significance level $\alpha = 0.05$ is used and the sample
                size $\n = 100$. The results in the figures are average of 1000 trials. \textbf{Note}: this figure is generated following the same approach for generating 
                Figure~4 in the main paper, with different sample size $\n = 100$.}
                \label{supp:fig:6}
        \end{figure}

\begin{figure}
            \centering
            \includegraphics[width=13.5cm]{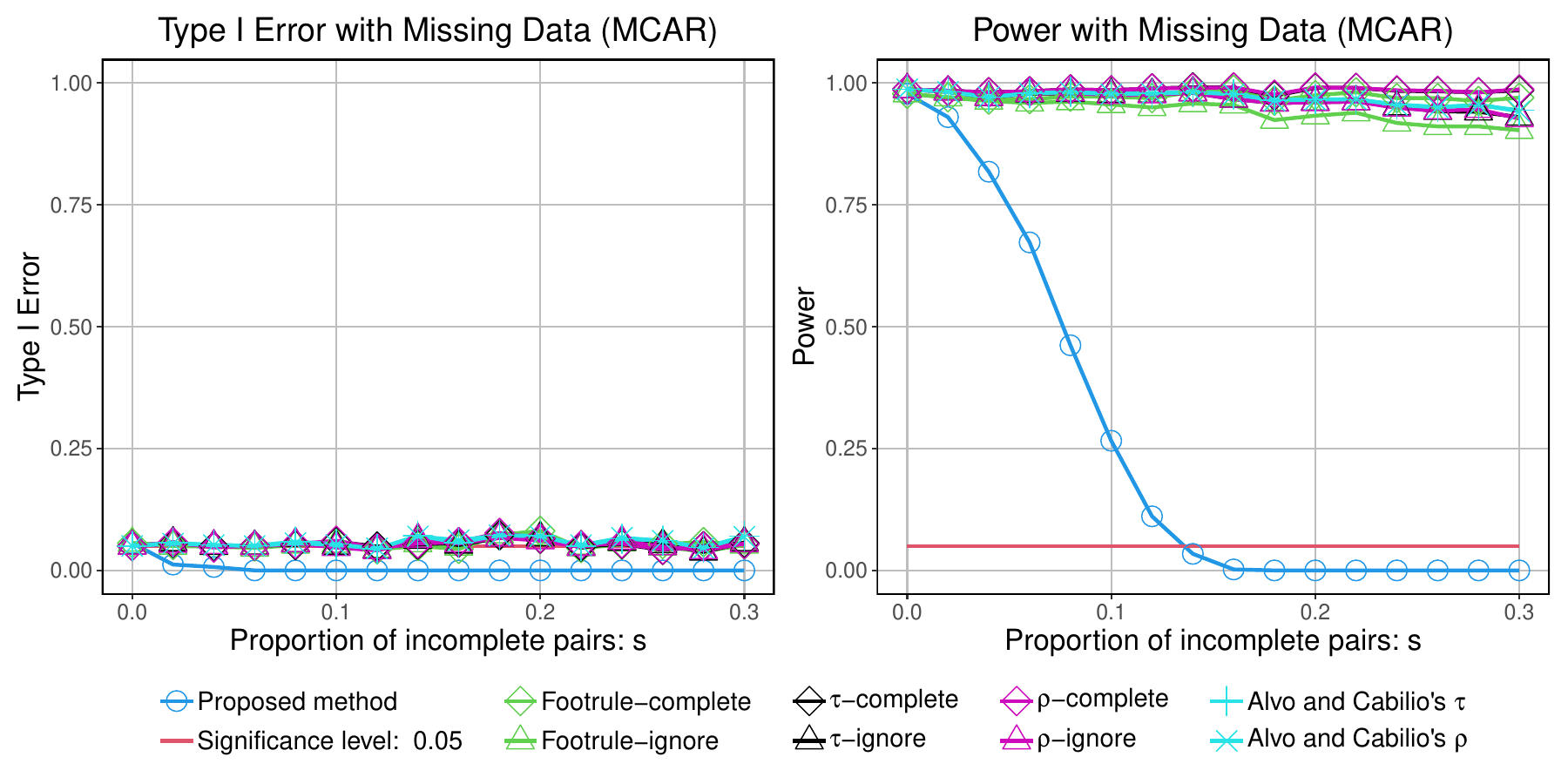}
            \vspace{-0.3cm}
            \caption{Statistical Type I error and power of the proposed method,
            methods that ignore missing data or use complete data,
            and Alvo and Cabilio's $\rho$ and Alvo and Cabilio's $\tau$ methods,
            as the proportion of missing data $s$ increases,
            where $s \in \{0.00, 0.02, \ldots, 0.30\}$.
            These methods are described in Table~\ref{supp:tab:3}.
            The data is missing completely at random (MCAR).
            (Left) Type I error: $(\bx,\by) \overset{iid}{\sim} N(0, I_2)$;
            (Right) Power: 	$(X, Y) \overset{iid}{\sim} N(0, \Sigma)$, where
            $\Sigma = \begin{pmatrix} 1 &  \corrcoef  \\  \corrcoef & 1 \end{pmatrix}$,
                with covariance coefficient $\corrcoef=0.3$.
                For both figures, a significance level $\alpha = 0.05$ is used and the sample
                size $\n = 200$. The results in the figures are average of 1000 trials. 
                \textbf{Note}: this figure is generated following the same approach for generating 
                Figure~3 in the main paper, with different correlation coefficient $\gamma = 0.3$.}
                \label{supp:fig:7}
        \end{figure}

\begin{figure}
            \centering
            \includegraphics[width=13.5cm]{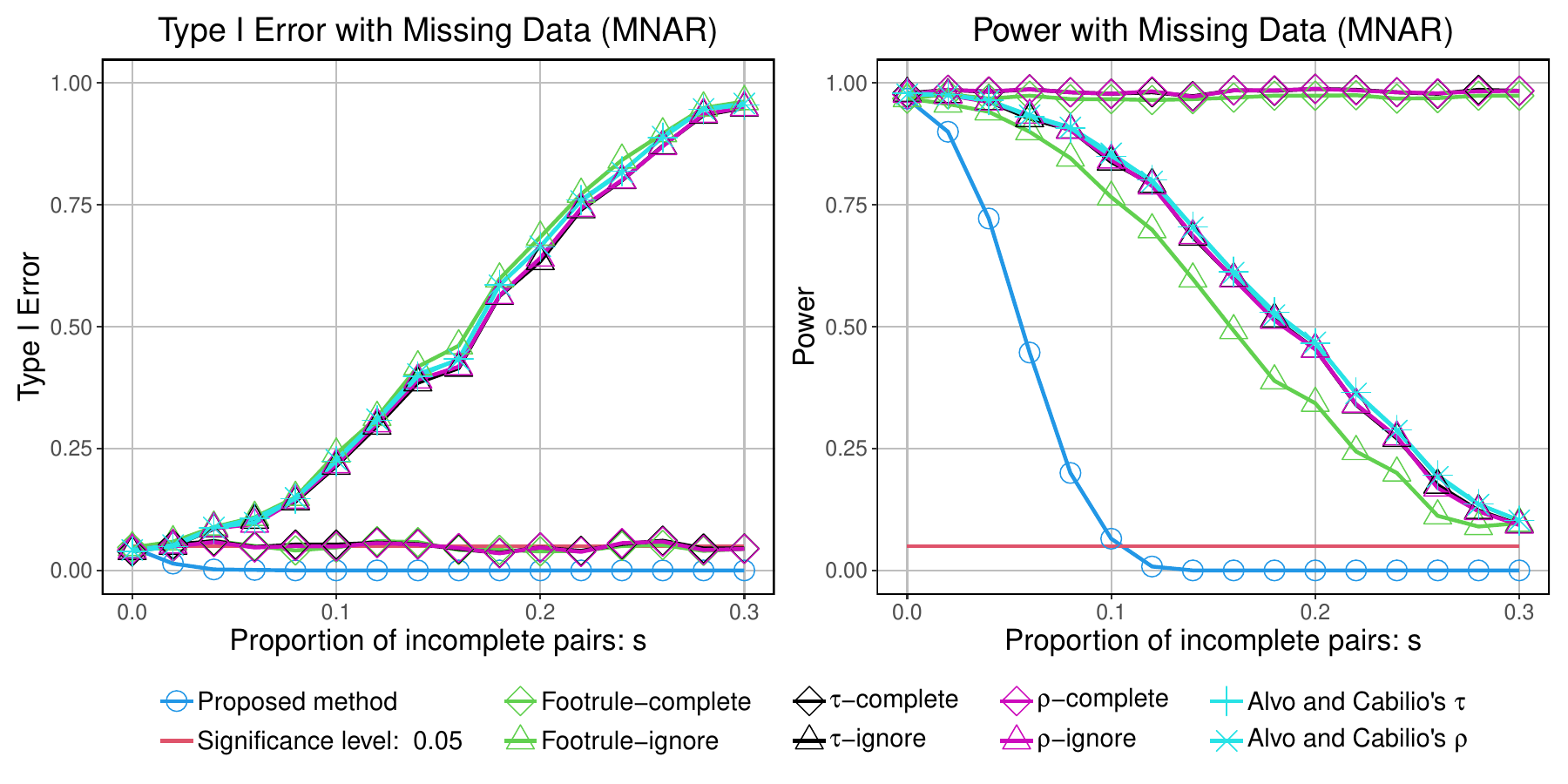}
            \vspace{-0.3cm}
            \caption{Statistical Type I error and power of the proposed method,
            methods that ignore missing data or use complete data,
            and Alvo and Cabilio's $\rho$ and Alvo and Cabilio's $\tau$ methods,
            as the proportion of missing data $s$ increases,
            where $s \in \{0.00, 0.02, \ldots, 0.30\}$.
            These methods are described in Table~\ref{supp:tab:3}.
            The data is missing not at random (MNAR).
            (Left) Type I error: $(\bx,\by) \overset{iid}{\sim} N(0, I_2)$;
            (Right) Power: 	$(X, Y) \overset{iid}{\sim} N(0, \Sigma)$, where
            $\Sigma = \begin{pmatrix} 1 &  \corrcoef  \\  \corrcoef & 1 \end{pmatrix}$,
                with covariance coefficient $\corrcoef=0.3$.
                For both figures, a significance level $\alpha = 0.05$ is used and the sample
                size $\n = 200$. The results in the figures are average of 1000 trials. \textbf{Note}: this figure is generated following the same approach for generating 
                Figure~4 in the main paper, with different correlation coefficient $\gamma = 0.3$.}
                \label{supp:fig:8}
        \end{figure}

\begin{figure}
            \centering
            \includegraphics[width=13.5cm]{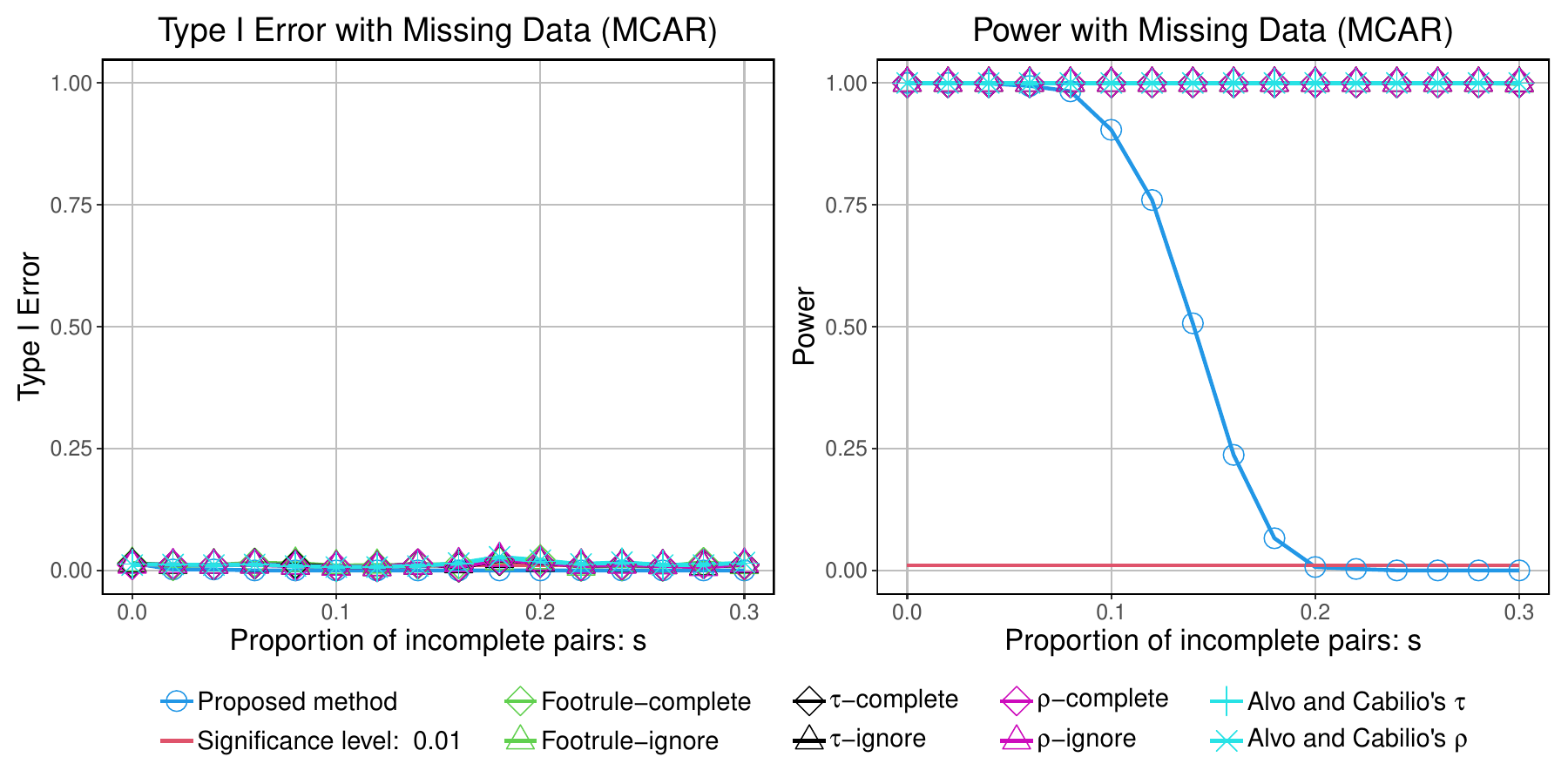}
            \vspace{-0.3cm}
            \caption{Statistical Type I error and power of the proposed method,
            methods that ignore missing data or use complete data,
            and Alvo and Cabilio's $\rho$ and Alvo and Cabilio's $\tau$ methods,
            as the proportion of missing data $s$ increases,
            where $s \in \{0.00, 0.02, \ldots, 0.30\}$.
            These methods are described in Table~\ref{supp:tab:3}.
            The data is missing completely at random (MCAR).
            (Left) Type I error: $(\bx,\by) \overset{iid}{\sim} N(0, I_2)$;
            (Right) Power: 	$(X, Y) \overset{iid}{\sim} N(0, \Sigma)$, where
            $\Sigma = \begin{pmatrix} 1 &  \corrcoef  \\  \corrcoef & 1 \end{pmatrix}$,
                with covariance coefficient $\corrcoef=0.5$.
                For both figures, a significance level $\alpha = 0.05$ is used and the sample
                size $\n = 200$. The results in the figures are average of 1000 trials. 
                \textbf{Note}: this figure is generated following the same approach for generating 
                Figure~3 in the main paper, with different significance level $\alpha = 0.01$.}
                \label{supp:fig:9}
        \end{figure}

\begin{figure}
            \centering
            \includegraphics[width=13.5cm]{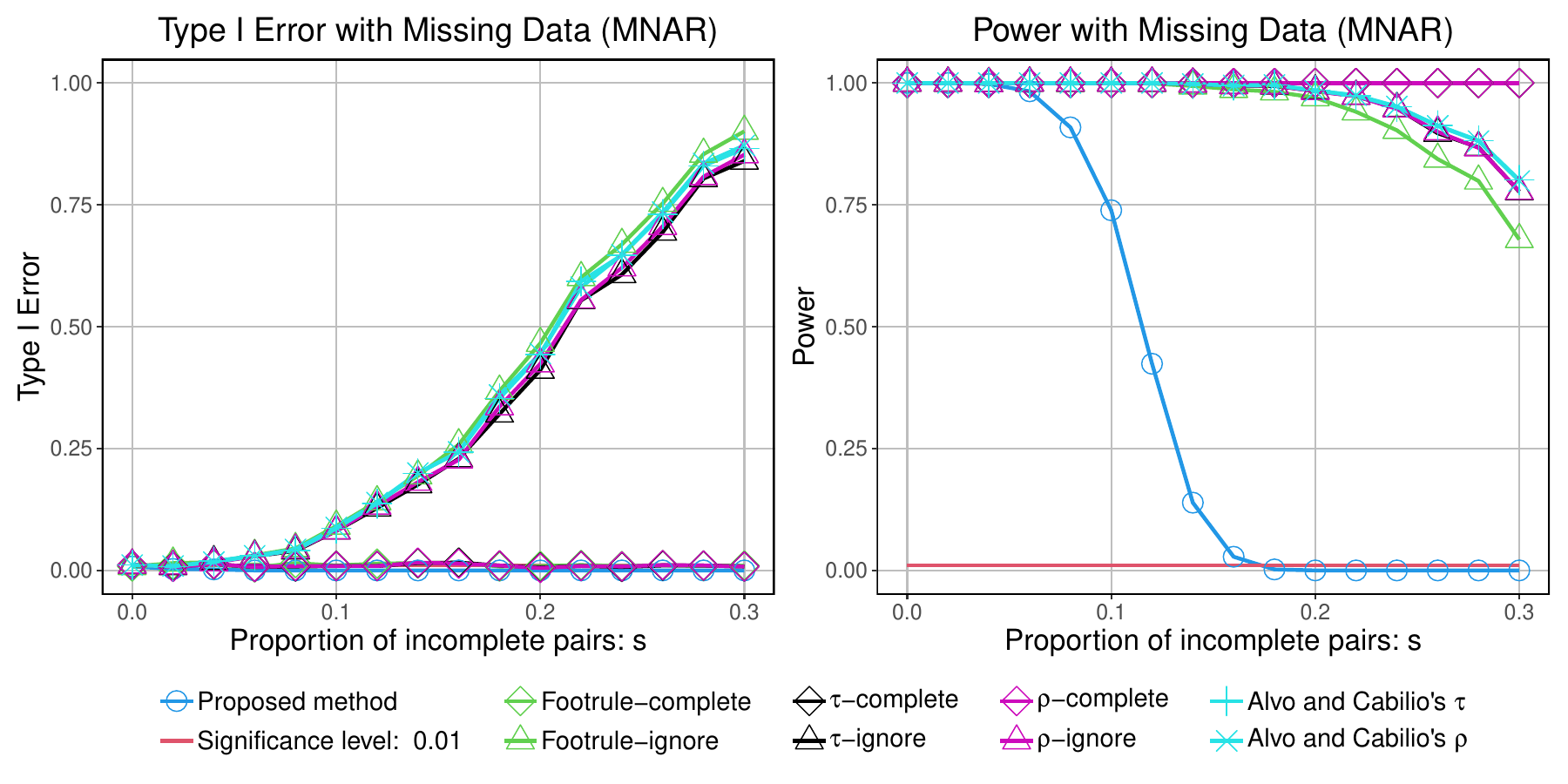}
            \vspace{-0.3cm}
            \caption{Statistical Type I error and power of the proposed method,
            methods that ignore missing data or use complete data,
            and Alvo and Cabilio's $\rho$ and Alvo and Cabilio's $\tau$ methods,
            as the proportion of missing data $s$ increases,
            where $s \in \{0.00, 0.02, \ldots, 0.30\}$.
            These methods are described in Table~\ref{supp:tab:3}.
            The data is missing not at random (MNAR).
            (Left) Type I error: $(\bx,\by) \overset{iid}{\sim} N(0, I_2)$;
            (Right) Power: 	$(X, Y) \overset{iid}{\sim} N(0, \Sigma)$, where
            $\Sigma = \begin{pmatrix} 1 &  \corrcoef  \\  \corrcoef & 1 \end{pmatrix}$,
                with covariance coefficient $\corrcoef=0.5$.
                For both figures, a significance level $\alpha = 0.05$ is used and the sample
                size $\n = 200$. The results in the figures are average of 1000 trials. \textbf{Note}: this figure is generated following the same approach for generating 
                Figure~4 in the main paper, with different significance level $\alpha = 0.01$.}
                \label{supp:fig:10}
        \end{figure}

\begin{figure}
            \centering
            \includegraphics[width=13.5cm]{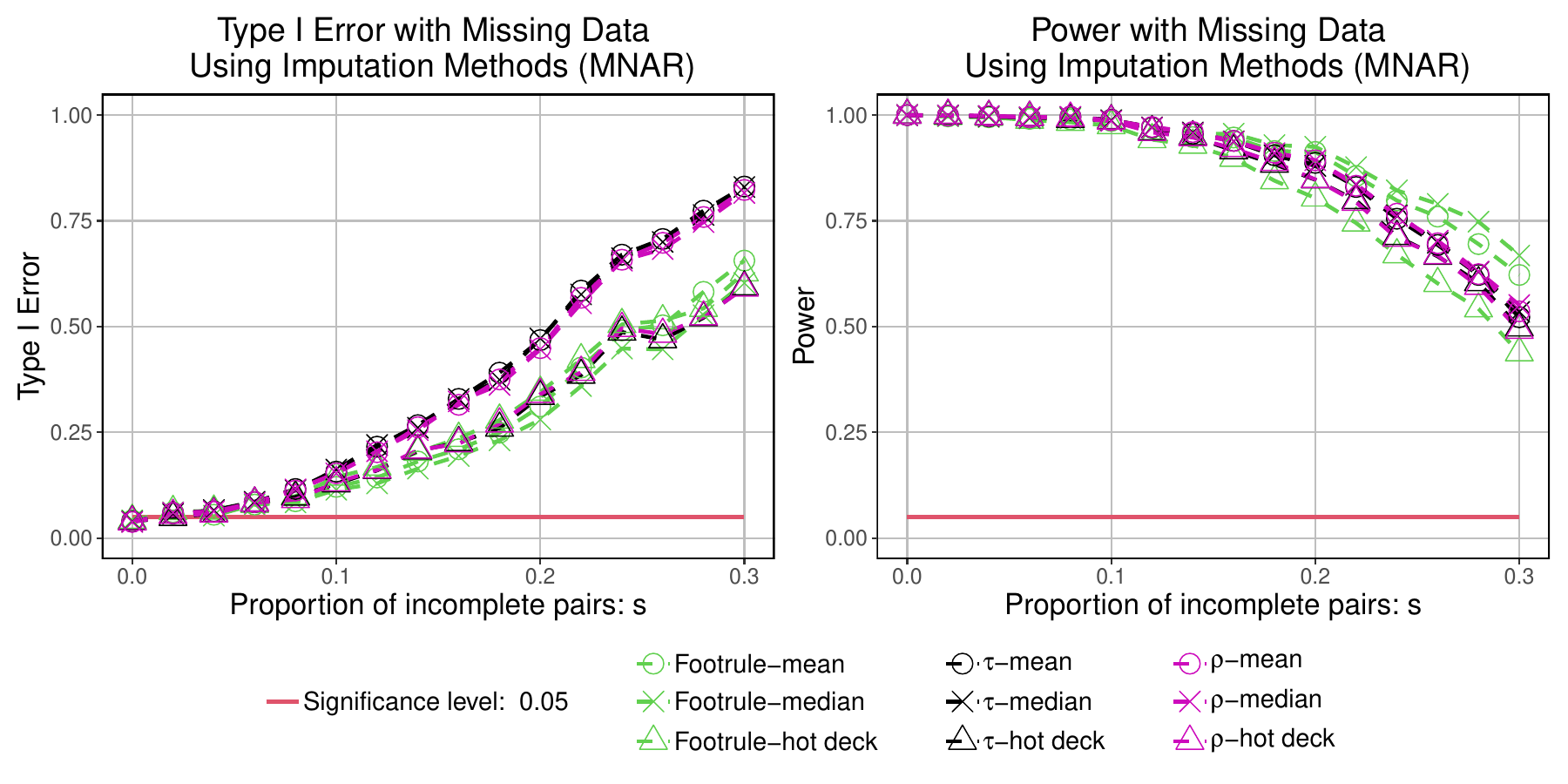}
            \vspace{-0.3cm}
            \caption{Statistical Type I error and power of imputation methods
            as the proportion of missing data $s$ increases,
            where $s \in \{0.00, 0.02, \ldots, 0.30\}$.
            The imputation methods impute missing data using either mean, median or 
            randomly selected (hot deck) values of observed data.
            These methods are described in Table~\ref{supp:tab:3}.
            The data is missing completely at random (MNAR).
            (Left) Type I error: $(\bx,\by) \overset{iid}{\sim} N(0, I_2)$;
            (Right) Power: 	$(X, Y) \overset{iid}{\sim} N(0, \Sigma)$, where
            $\Sigma = \begin{pmatrix} 1 &  \corrcoef  \\  \corrcoef & 1 \end{pmatrix}$,
                with covariance coefficient $\corrcoef=0.5$.
                For both figures, a significance level $\alpha = 0.05$ is used and the sample
                size $\n = 100$. The results in the figures are average of 1000 trials.
                \textbf{Note}: this figure is generated following the same approach for generating 
                Figure~5 in the main paper, with different sample size $\n = 100$.}
                \label{supp:fig:11}
        \end{figure}

\begin{figure}
            \centering
            \includegraphics[width=13.5cm]{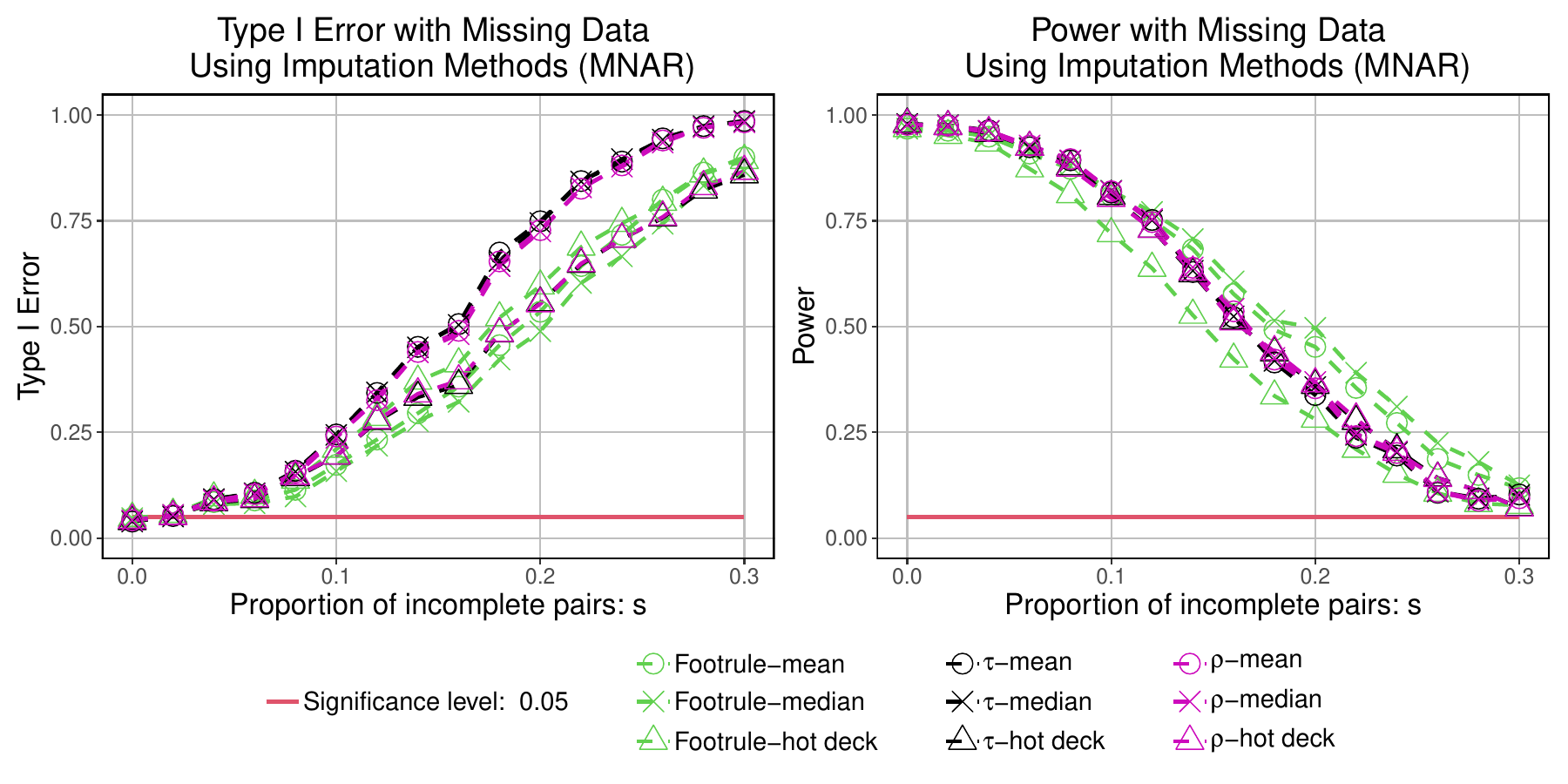}
            \vspace{-0.3cm}
            \caption{Statistical Type I error and power of imputation methods
            as the proportion of missing data $s$ increases,
            where $s \in \{0.00, 0.02, \ldots, 0.30\}$.
            The imputation methods impute missing data using either mean, median or 
            randomly selected (hot deck) values of observed data.
            These methods are described in Table~\ref{supp:tab:3}.
            The data is missing completely at random (MNAR).
            (Left) Type I error: $(\bx,\by) \overset{iid}{\sim} N(0, I_2)$;
            (Right) Power: 	$(X, Y) \overset{iid}{\sim} N(0, \Sigma)$, where
            $\Sigma = \begin{pmatrix} 1 &  \corrcoef  \\  \corrcoef & 1 \end{pmatrix}$,
                with covariance coefficient $\corrcoef=0.5$.
                For both figures, a significance level $\alpha = 0.05$ is used and the sample
                size $\n = 200$. The results in the figures are average of 1000 trials.
                \textbf{Note}: this figure is generated following the same approach for generating 
                Figure~5 in the main paper, with different correlation coefficient $\gamma = 0.3$.}
                \label{supp:fig:12}
        \end{figure}

\begin{figure}
            \centering
            \includegraphics[width=13.5cm]{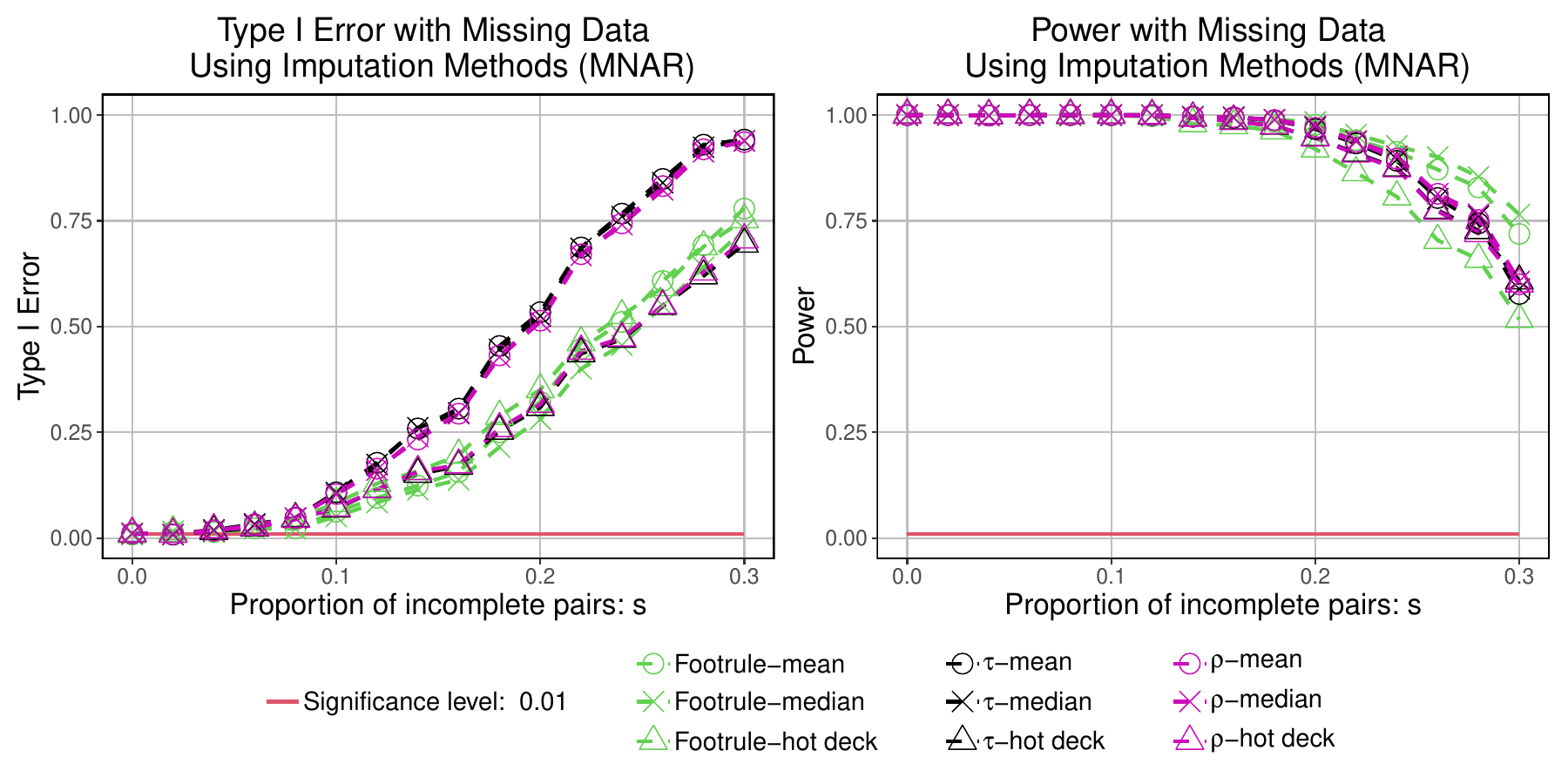}
            \vspace{-0.3cm}
            \caption{Statistical Type I error and power of imputation methods
            as the proportion of missing data $s$ increases,
            where $s \in \{0.00, 0.02, \ldots, 0.30\}$.
            The imputation methods impute missing data using either mean, median or 
            randomly selected (hot deck) values of observed data.
            These methods are described in Table~\ref{supp:tab:3}.
            The data is missing completely at random (MNAR).
            (Left) Type I error: $(\bx,\by) \overset{iid}{\sim} N(0, I_2)$;
            (Right) Power: 	$(X, Y) \overset{iid}{\sim} N(0, \Sigma)$, where
            $\Sigma = \begin{pmatrix} 1 &  \corrcoef  \\  \corrcoef & 1 \end{pmatrix}$,
                with covariance coefficient $\corrcoef=0.5$.
                For both figures, a significance level $\alpha = 0.05$ is used and the sample
                size $\n = 100$. The results in the figures are average of 1000 trials.
                \textbf{Note}: this figure is generated following the same approach for generating 
                Figure~5 in the main paper, with different significance level $\alpha = 0.01$.}
                \label{supp:fig:13}
        \end{figure}

        We perform simulations following the same way as in Section 3.2.1
        in the main paper when the proportion of missing data $s$ increases, with potentially different correlation 
        coefficients $\gamma$, sample size $\n$, and significance level $\alpha$.
        The results are shown in Figure~\ref{supp:fig:5}--\ref{supp:fig:13}.

        \subsubsection{As the sample size increases}
\begin{figure}
            \centering
            \includegraphics[width=13.5cm]{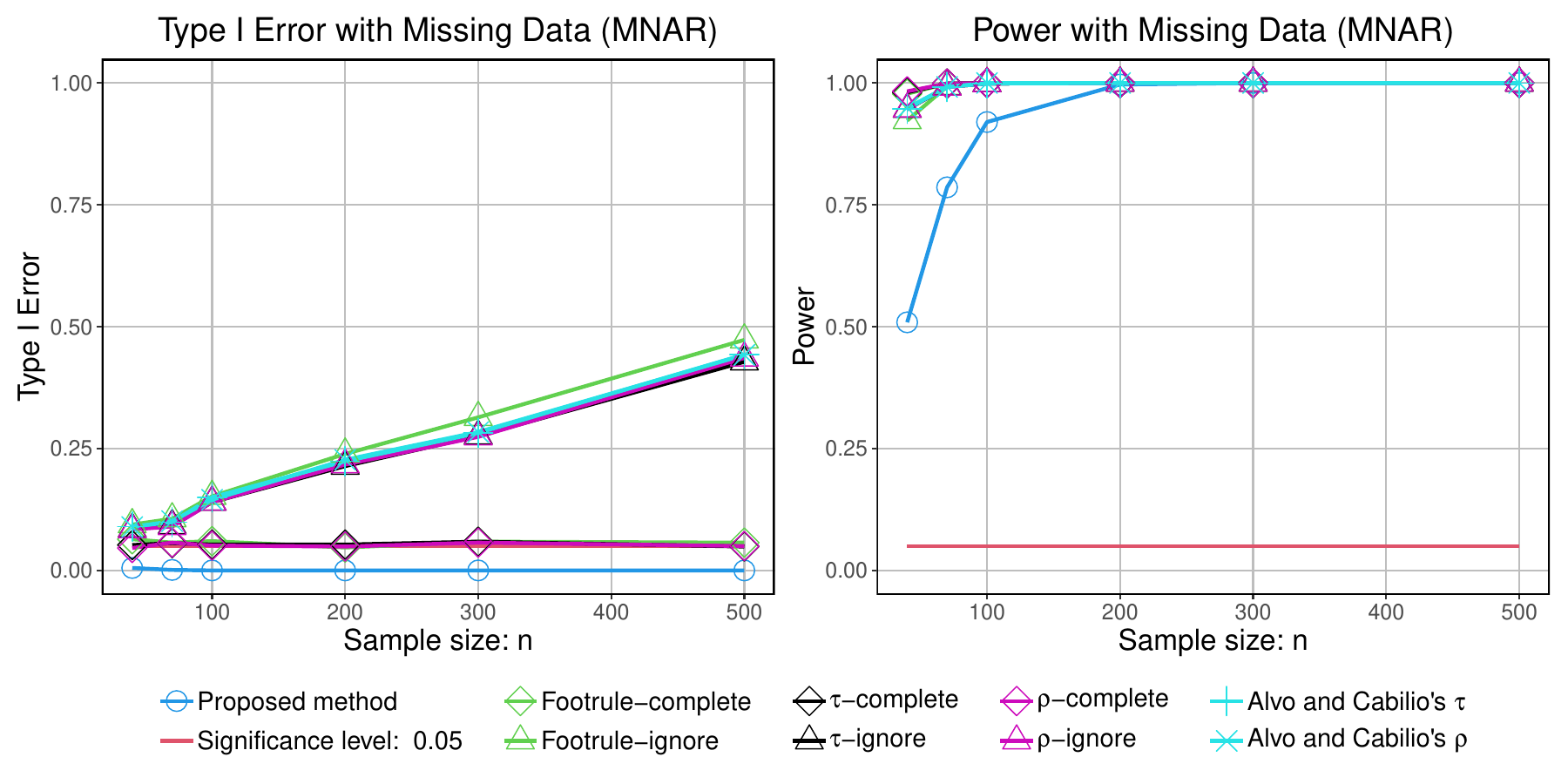}
            \vspace{-0.3cm}
            \caption{
                Statistical Type I error and power of the proposed method,
                methods that ignore missing data or use complete data,
                and Alvo and Cabilio's $\rho$ and Alvo and Cabilio's $\tau$ methods,
                as the sample size $\n$ increases,
                where $\n \in \{40, 70, 100, 200, 300, 500\}$.
                These methods are described in Table~\ref{supp:tab:3}.
                The data is missing not at random (MNAR).
                (Left) Type I error: $(\bx,\by) \overset{iid}{\sim} N(0, I_2)$;
                (Right) Power: 	$(X, Y) \overset{iid}{\sim} N(0, \Sigma)$, where
                $\Sigma = \begin{pmatrix} 1 &  \corrcoef  \\  \corrcoef & 1 \end{pmatrix}$,
                    with covariance coefficient $\corrcoef=0.6$.
                    For both figures, a significance level $\alpha = 0.05$ is used and the proportion
                    of missing pairs $s = 0.1$. The results in the figures are average of 1000 trials.
                    \textbf{Note}: this figure is generated following the same approach for generating 
                    Figure~6 in the main paper, with different correlation coefficient $\gamma = 0.6$.}
                \label{supp:fig:14}
        \end{figure}

\begin{figure}
            \centering
            \includegraphics[width=13.5cm]{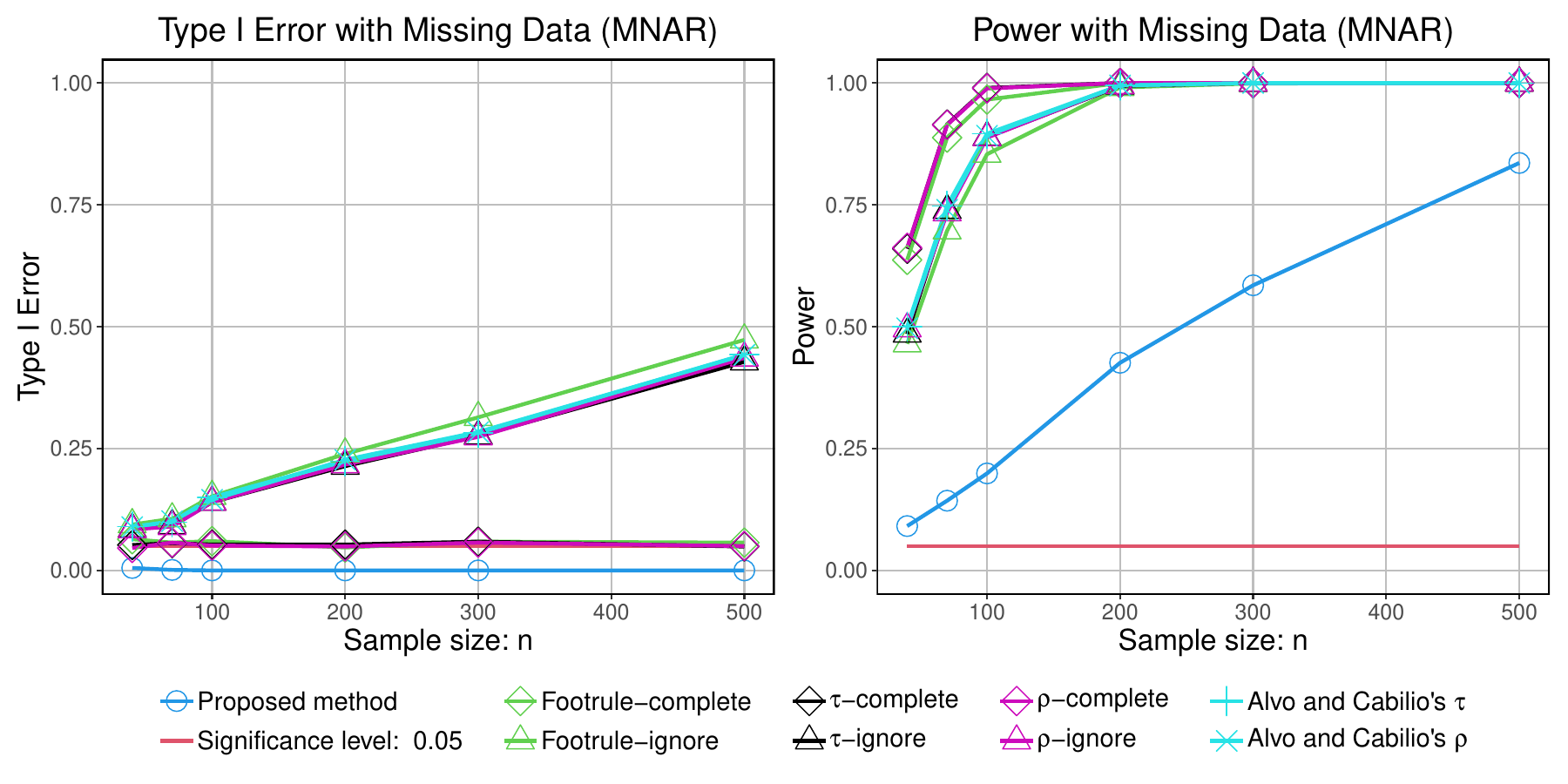}
            \vspace{-0.3cm}
            \caption{
                Statistical Type I error and power of the proposed method,
                methods that ignore missing data or use complete data,
                and Alvo and Cabilio's $\rho$ and Alvo and Cabilio's $\tau$ methods,
                as the sample size $\n$ increases,
                where $\n \in \{40, 70, 100, 200, 300, 500\}$.
                These methods are described in Table~\ref{supp:tab:3}.
                The data is missing not at random (MNAR).
                (Left) Type I error: $(\bx,\by) \overset{iid}{\sim} N(0, I_2)$;
                (Right) Power: 	$(X, Y) \overset{iid}{\sim} N(0, \Sigma)$, where
                $\Sigma = \begin{pmatrix} 1 &  \corrcoef  \\  \corrcoef & 1 \end{pmatrix}$,
                    with covariance coefficient $\corrcoef=0.4$.
                    For both figures, a significance level $\alpha = 0.05$ is used and the proportion
                    of missing pairs $s = 0.1$. The results in the figures are average of 1000 trials.
                    \textbf{Note}: this figure is generated following the same approach for generating 
                    Figure~6 in the main paper, with different correlation coefficient $\gamma = 0.4$.}
                \label{supp:fig:15}
        \end{figure}

\begin{figure}
            \centering
            \includegraphics[width=13.5cm]{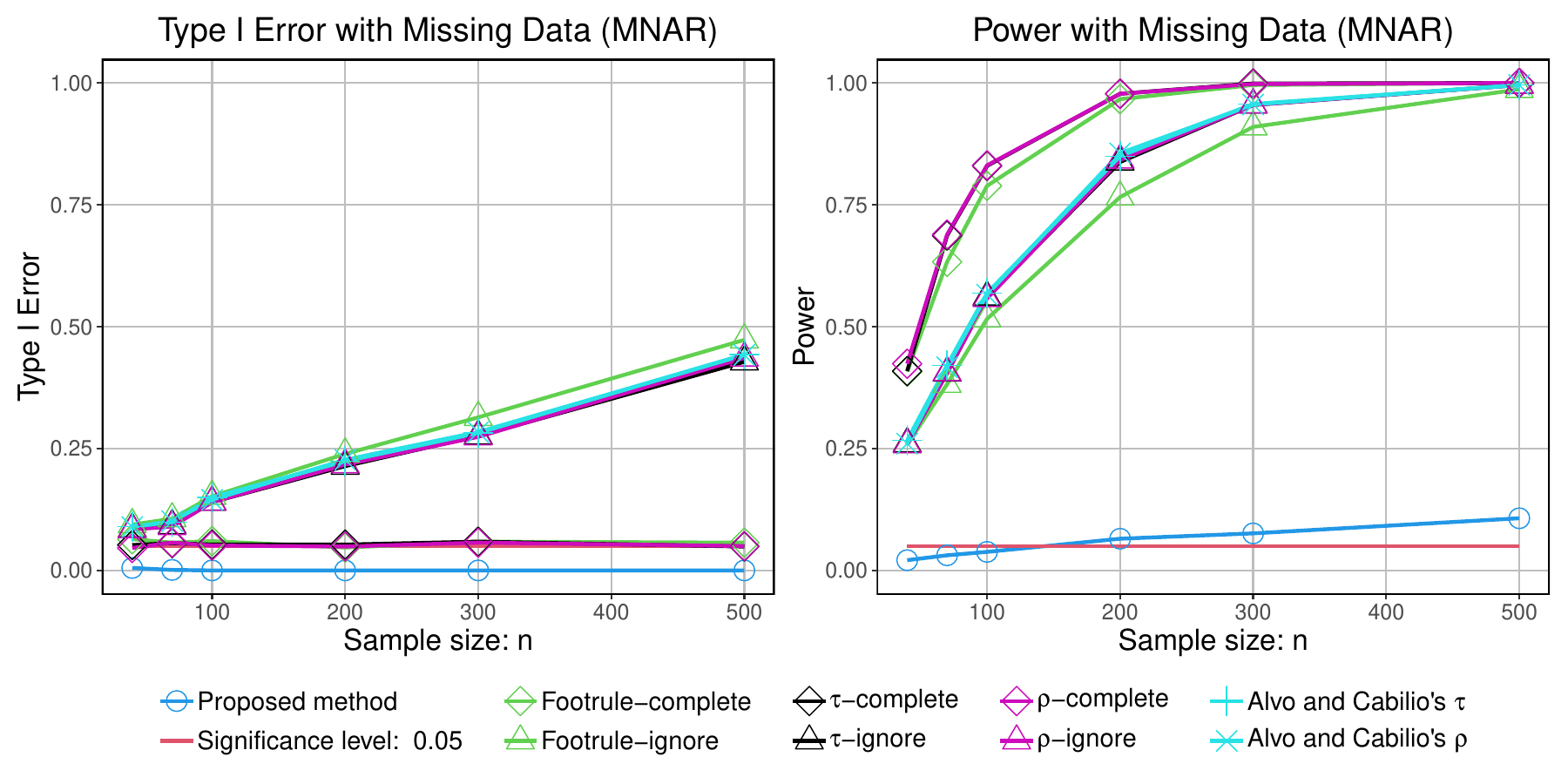}
            \vspace{-0.3cm}
            \caption{
                Statistical Type I error and power of the proposed method,
                methods that ignore missing data or use complete data,
                and Alvo and Cabilio's $\rho$ and Alvo and Cabilio's $\tau$ methods,
                as the sample size $\n$ increases,
                where $\n \in \{40, 70, 100, 200, 300, 500\}$.
                These methods are described in Table~\ref{supp:tab:3}.
                The data is missing not at random (MNAR).
                (Left) Type I error: $(\bx,\by) \overset{iid}{\sim} N(0, I_2)$;
                (Right) Power: 	$(X, Y) \overset{iid}{\sim} N(0, \Sigma)$, where
                $\Sigma = \begin{pmatrix} 1 &  \corrcoef  \\  \corrcoef & 1 \end{pmatrix}$,
                    with covariance coefficient $\corrcoef=0.3$.
                    For both figures, a significance level $\alpha = 0.05$ is used and the proportion
                    of missing pairs $s = 0.1$. The results in the figures are average of 1000 trials.
                    \textbf{Note}: this figure is generated following the same approach for generating 
                    Figure~6 in the main paper, with different correlation coefficient $\gamma = 0.3$.}
                \label{supp:fig:16}
        \end{figure}

\begin{figure}
            \centering
            \includegraphics[width=13.5cm]{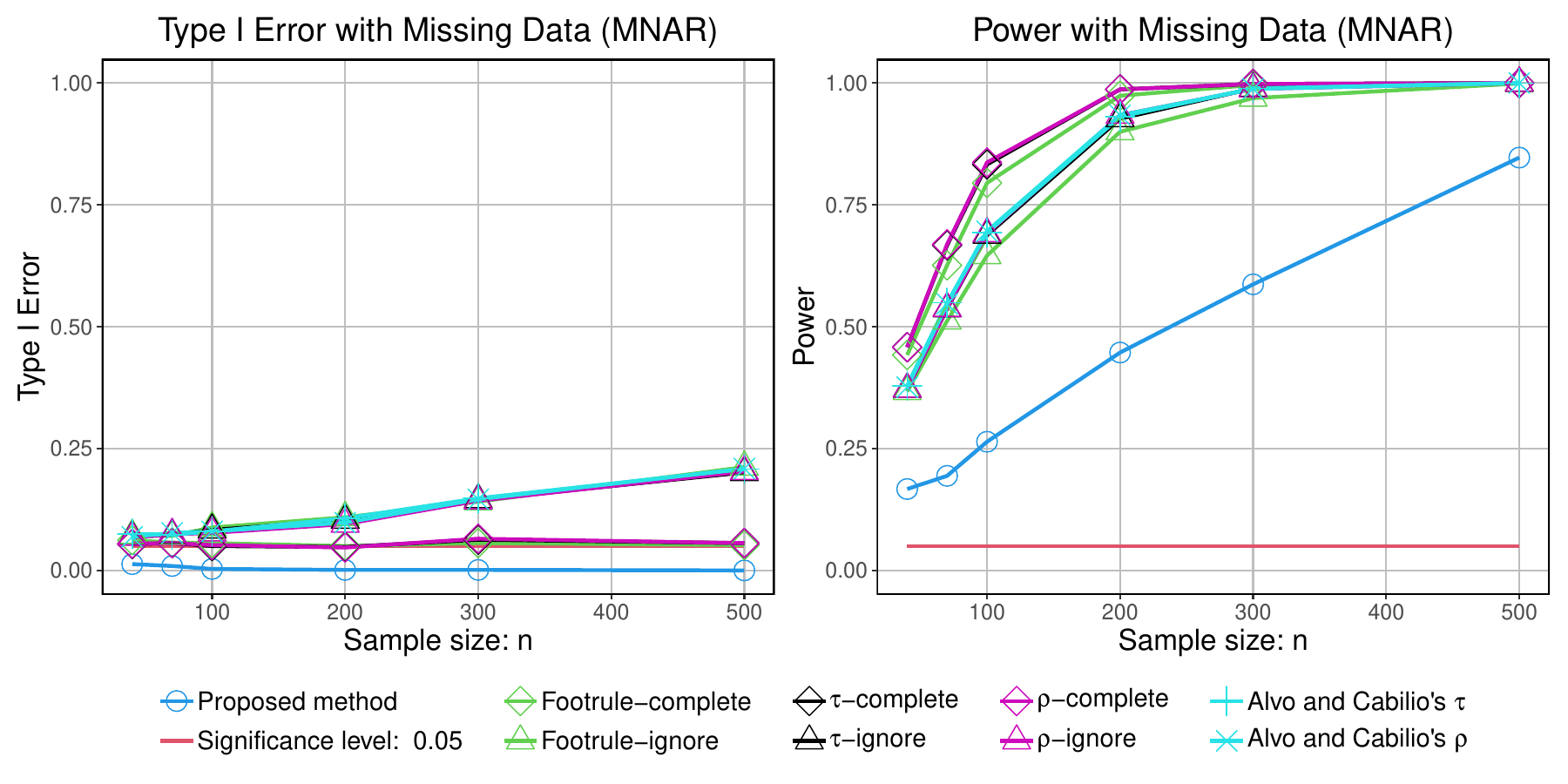}
            \vspace{-0.3cm}
            \caption{
                Statistical Type I error and power of the proposed method,
                methods that ignore missing data or use complete data,
                and Alvo and Cabilio's $\rho$ and Alvo and Cabilio's $\tau$ methods,
                as the sample size $\n$ increases,
                where $\n \in \{40, 70, 100, 200, 300, 500\}$.
                These methods are described in Table~\ref{supp:tab:3}.
                The data is missing not at random (MNAR).
                (Left) Type I error: $(\bx,\by) \overset{iid}{\sim} N(0, I_2)$;
                (Right) Power: 	$(X, Y) \overset{iid}{\sim} N(0, \Sigma)$, where
                $\Sigma = \begin{pmatrix} 1 &  \corrcoef  \\  \corrcoef & 1 \end{pmatrix}$,
                    with covariance coefficient $\corrcoef=0.3$.
                    For both figures, a significance level $\alpha = 0.05$ is used and the proportion
                    of missing pairs $s = 0.06$. The results in the figures are average of 1000 trials.
                    \textbf{Note}: this figure is generated following the same approach for generating 
                    Figure~6 in the main paper, with different correlation coefficient $\gamma = 0.3$, and
                    different proportion of missing pairs $s = 0.06$.}
                \label{supp:fig:17}
        \end{figure}

\begin{figure}
            \centering
            \includegraphics[width=13.5cm]{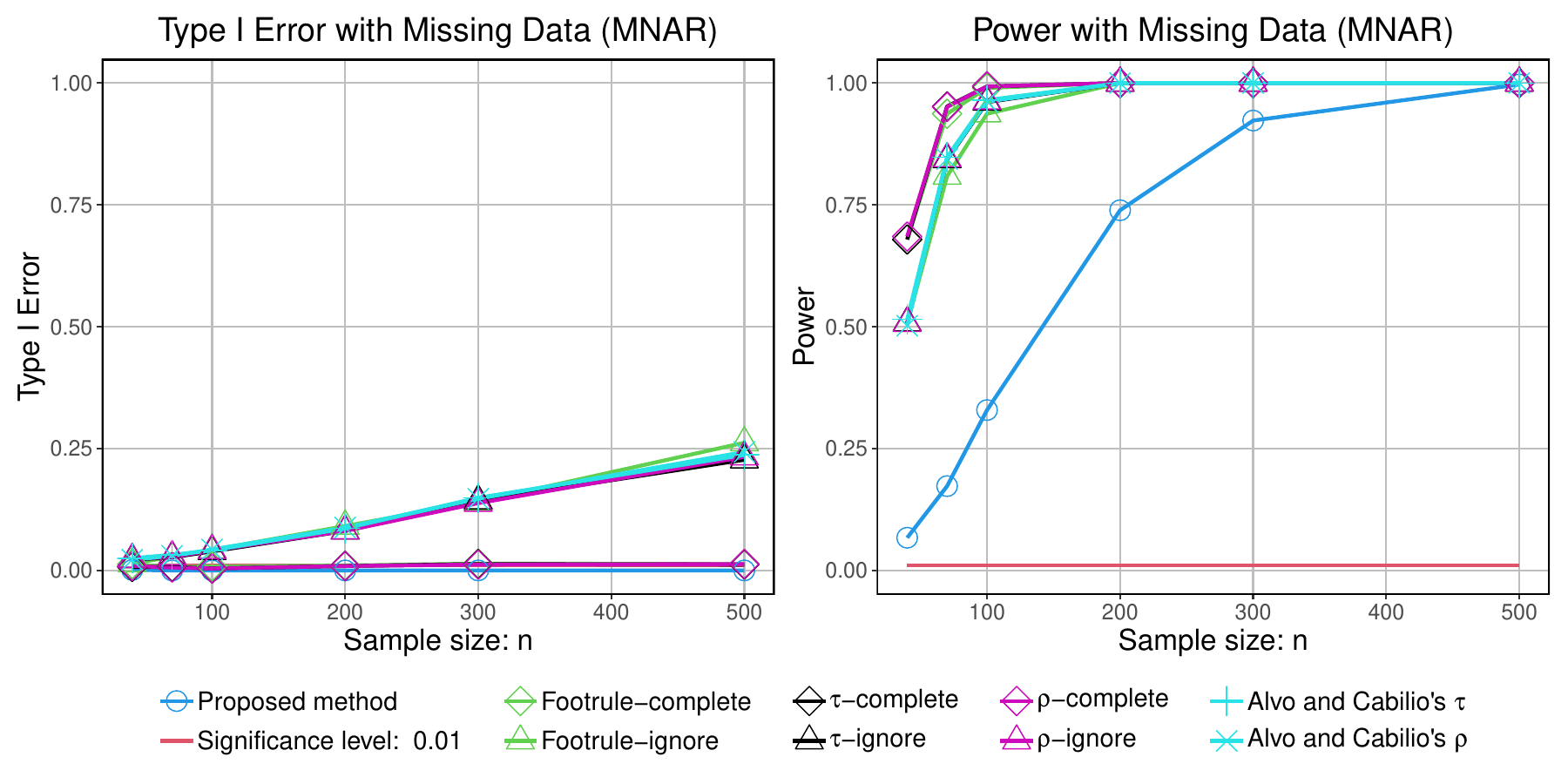}
            \vspace{-0.3cm}
            \caption{
                Statistical Type I error and power of the proposed method,
                methods that ignore missing data or use complete data,
                and Alvo and Cabilio's $\rho$ and Alvo and Cabilio's $\tau$ methods,
                as the sample size $\n$ increases,
                where $\n \in \{40, 70, 100, 200, 300, 500\}$.
                These methods are described in Table~\ref{supp:tab:3}.
                The data is missing not at random (MNAR).
                (Left) Type I error: $(\bx,\by) \overset{iid}{\sim} N(0, I_2)$;
                (Right) Power: 	$(X, Y) \overset{iid}{\sim} N(0, \Sigma)$, where
                $\Sigma = \begin{pmatrix} 1 &  \corrcoef  \\  \corrcoef & 1 \end{pmatrix}$,
                    with covariance coefficient $\corrcoef=0.5$.
                    For both figures, a significant level $\alpha = 0.01$ is used and the proportion
                    of missing pairs $s = 0.1$. The results in the figures are average of 1000 trials.
                    \textbf{Note}: this figure is generated following the same approach for generating 
                    Figure~6 in the main paper, with different significance level $\alpha = 0.01$.}
                \label{supp:fig:18}
        \end{figure}

        We perform simulations following the same approach as in Section 3.2.2
        in the main paper when the sample size $\n$ increases, with potentially different correlation 
        coefficients $\gamma$, proportion of missing data $s$, and significance level $\alpha$.
        The results are shown in Figure~\ref{supp:fig:14}~--~\ref{supp:fig:18}.

        \subsubsection{As the correlation coefficient increases}

\begin{figure}
            \centering
            \includegraphics[width=13cm]{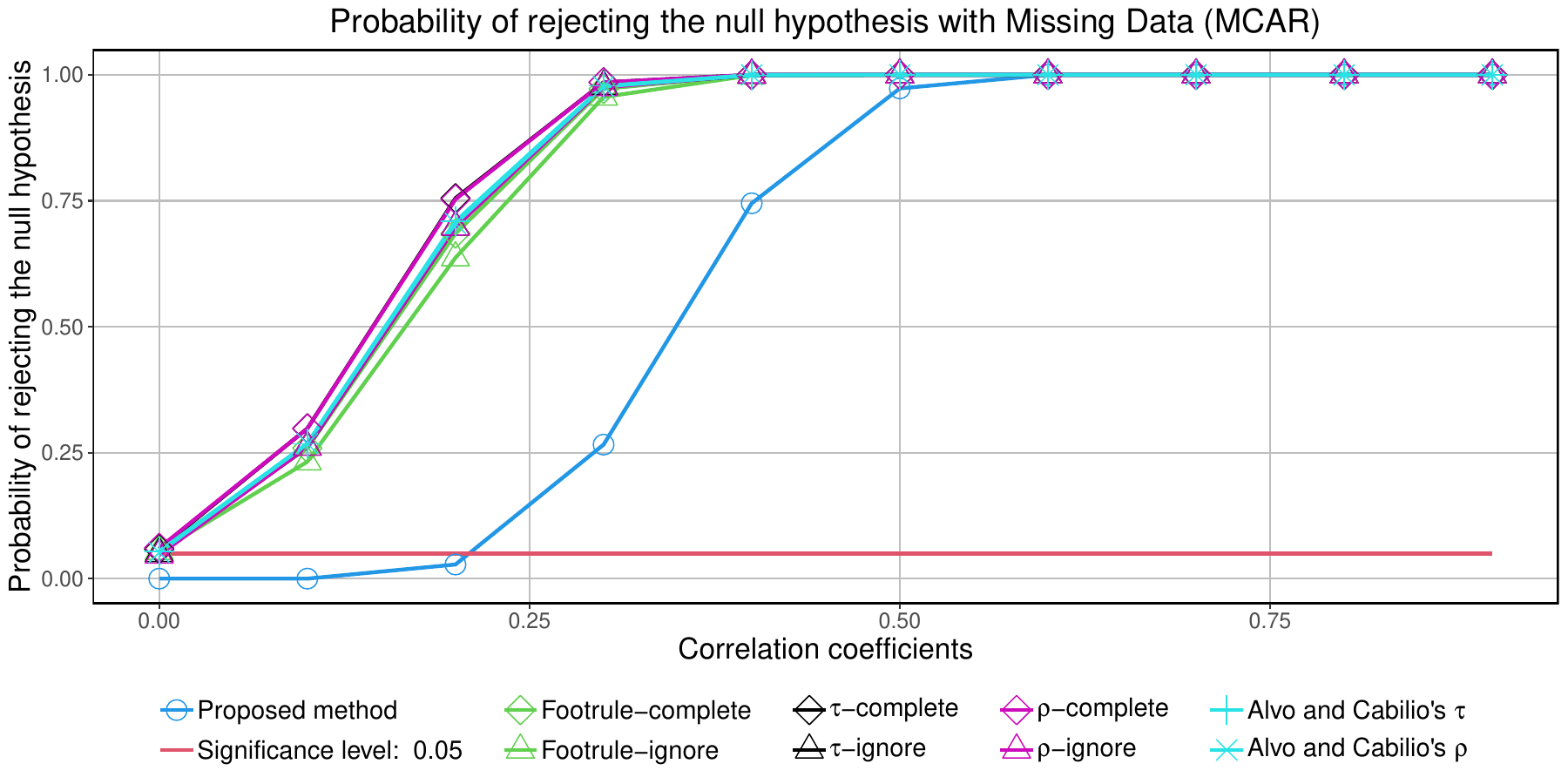}
            \caption{Statistical Type I error and power of the proposed method,
            methods that ignore missing data or use complete data,
            and Alvo and Cabilio's $\rho$ and Alvo and Cabilio's $\tau$ methods,
            as the correlation coefficient $\gamma$ increases,
            where $\gamma \in \{0, 0.1, \ldots, 0.9\}$.
            These methods are described in Table~\ref{supp:tab:3}.
            The data is missing completely at random (MCAR).
            $(X, Y) \overset{iid}{\sim} N(0, \Sigma)$, where
            $\Sigma = \begin{pmatrix} 1 &  \corrcoef  \\  \corrcoef & 1 \end{pmatrix}$,
                with $\corrcoef$ denoting covariance coefficient.
                For both figures, a significance level $\alpha = 0.05$ is used and the proportion
                of missing pairs $s = 0.1$. The results in the figures are average of 1000 trials.}
                \label{supp:fig:19}
        \end{figure}

\begin{figure}
            \centering
            \includegraphics[width=13cm]{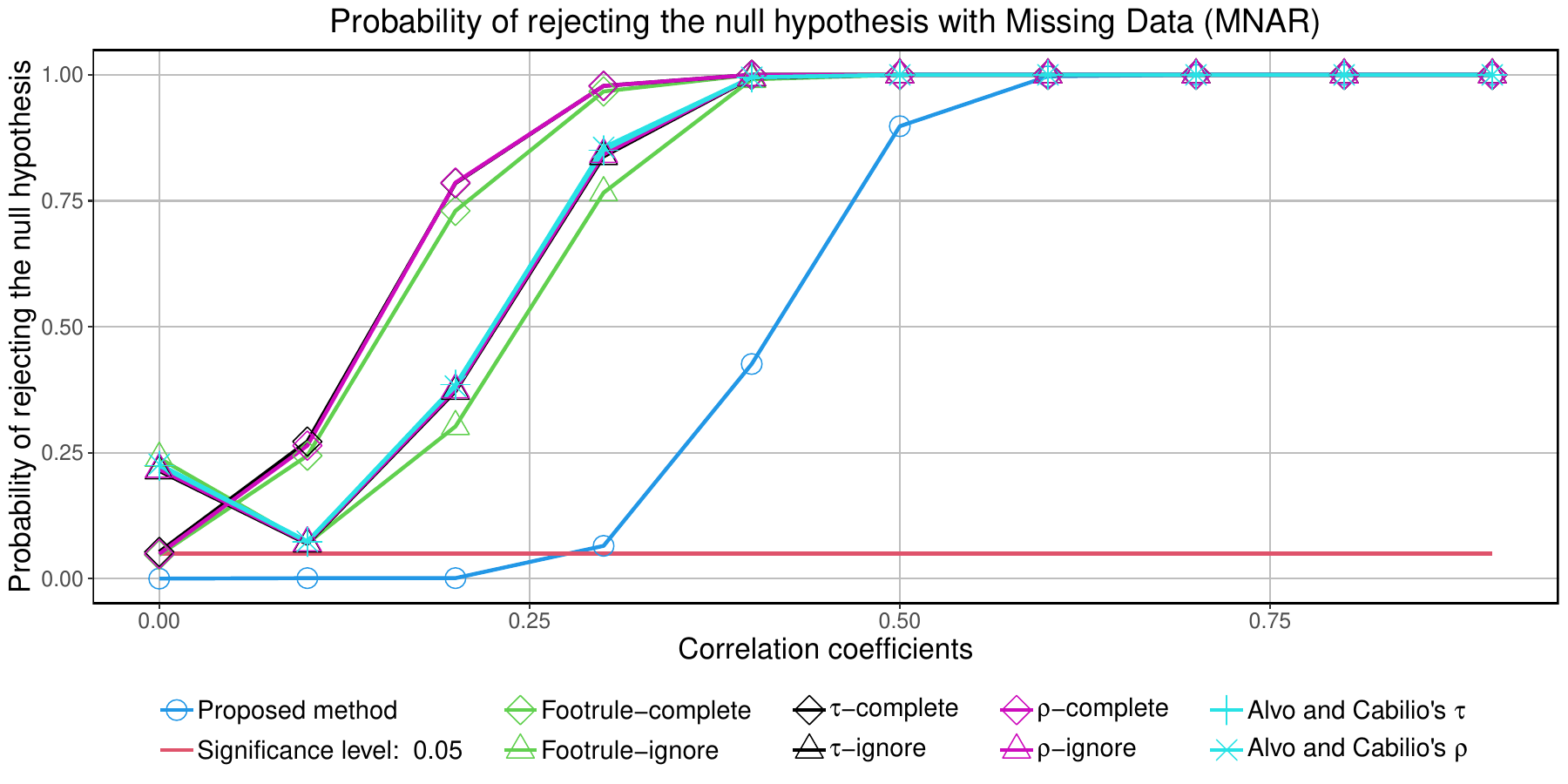}
            \caption{Statistical Type I error and power of the proposed method,
            methods that ignore missing data or use complete data,
            and Alvo and Cabilio's $\rho$ and Alvo and Cabilio's $\tau$ methods,
            as the correlation coefficient $\gamma$ increases,
            where $\gamma \in \{0, 0.1, \ldots, 0.9\}$.
            These methods are described in Table~\ref{supp:tab:3}.
            The data is missing not at random (MNAR).
            $(X, Y) \overset{iid}{\sim} N(0, \Sigma)$, where
            $\Sigma = \begin{pmatrix} 1 &  \corrcoef  \\  \corrcoef & 1 \end{pmatrix}$,
                with $\corrcoef$ denoting covariance coefficient.
                For both figures, a significance level $\alpha = 0.05$ is used and the proportion
                of missing pairs $s = 0.1$. The results in the figures are average of 1000 trials.}
                \label{supp:fig:20}
        \end{figure}

        Here we consider the case as in Section 3.2.1 when data are 
        either missing completely at random (MCAR), or missing not at random (MNAR), 
        but now as the correlation coefficient $\gamma$ increases, with fixed proportion of 
        missing pairs $s=0.1$, sample size $\n = 200$ and significance level $\alpha = 0.05$.
        The results are shown in Figure~\ref{supp:fig:19}--\ref{supp:fig:20}.

        \subsubsection{A different missing not at random missingness mechanism}

        Now we perform simulations following the same was as in Section 3.2
        in the main paper when data are missing not at random (MNAR), with
        a different MNAR missingness mechanism described below.

\begin{figure}
            \centering
            \includegraphics[width=13.5cm]{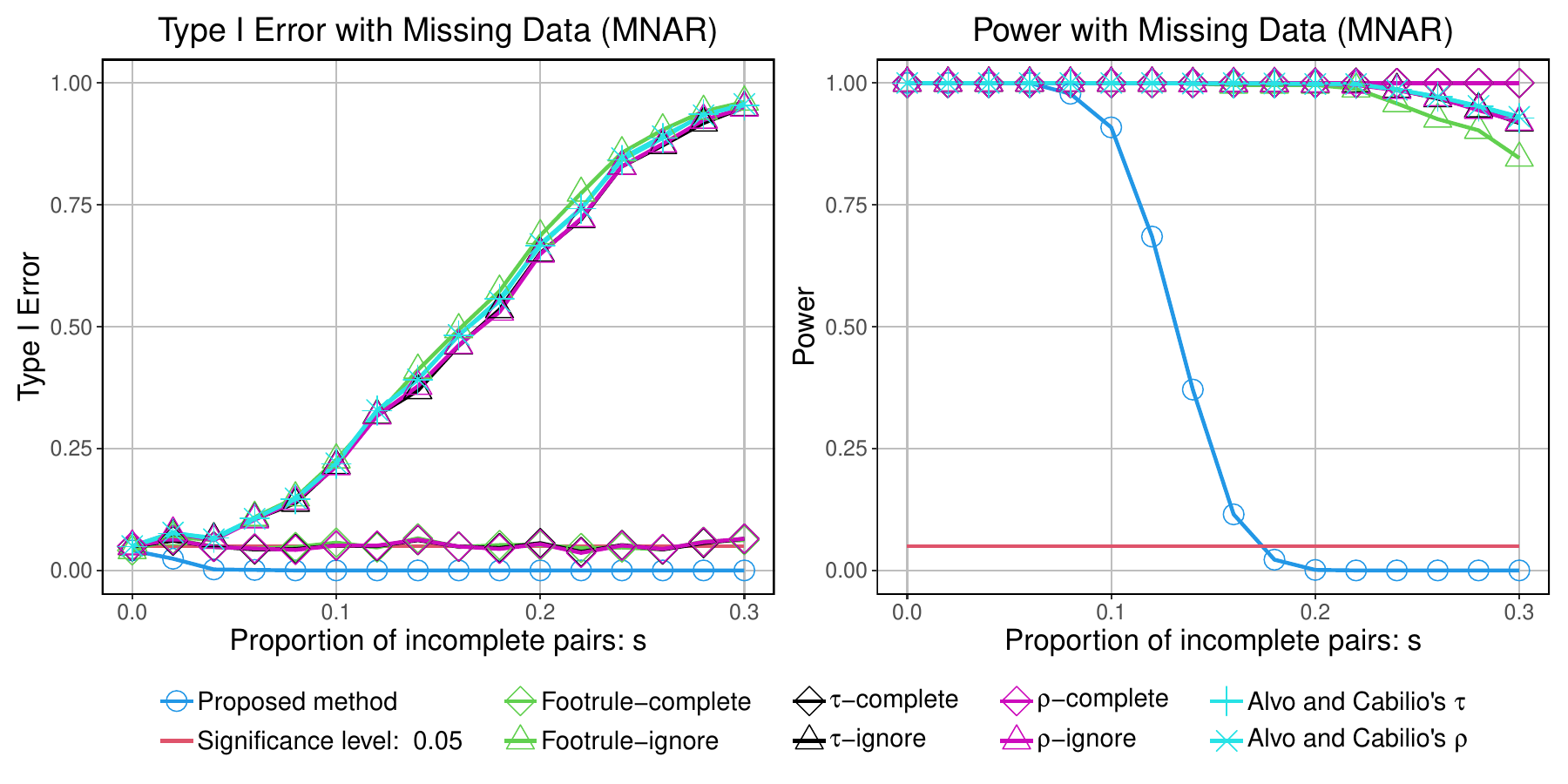}
            \vspace{-0.3cm}
            \caption{Statistical Type I error and power of the proposed method,
            methods that ignore missing data or use complete data,
            and Alvo and Cabilio's $\rho$ and Alvo and Cabilio's $\tau$ methods,
            as the proportion of missing data $s$ increases,
            where $s \in \{0.00, 0.02, \ldots, 0.30\}$.
            These methods are described in Table~\ref{supp:tab:3}.
            The data is missing completely at random (MCAR).
            (Left) Type I error: $(\bx,\by) \overset{iid}{\sim} N(0, I_2)$;
            (Right) Power: 	$(X, Y) \overset{iid}{\sim} N(0, \Sigma)$, where
            $\Sigma = \begin{pmatrix} 1 &  \corrcoef  \\  \corrcoef & 1 \end{pmatrix}$,
                with covariance coefficient $\corrcoef=0.5$.
                For both figures, a significance level $\alpha = 0.05$ is used and the sample
                size $\n = 200$. The results in the figures are average of 1000 trials. 
                \textbf{Note}: this figure is generated following the same approach for generating 
                Figure~4 in the main paper, with a different missingness mechanism defined in \eqref{supp:missingnessmechanism:1}.}
                \label{supp:fig:21}
        \end{figure}

\begin{figure}
            \centering
            \includegraphics[width=13.5cm]{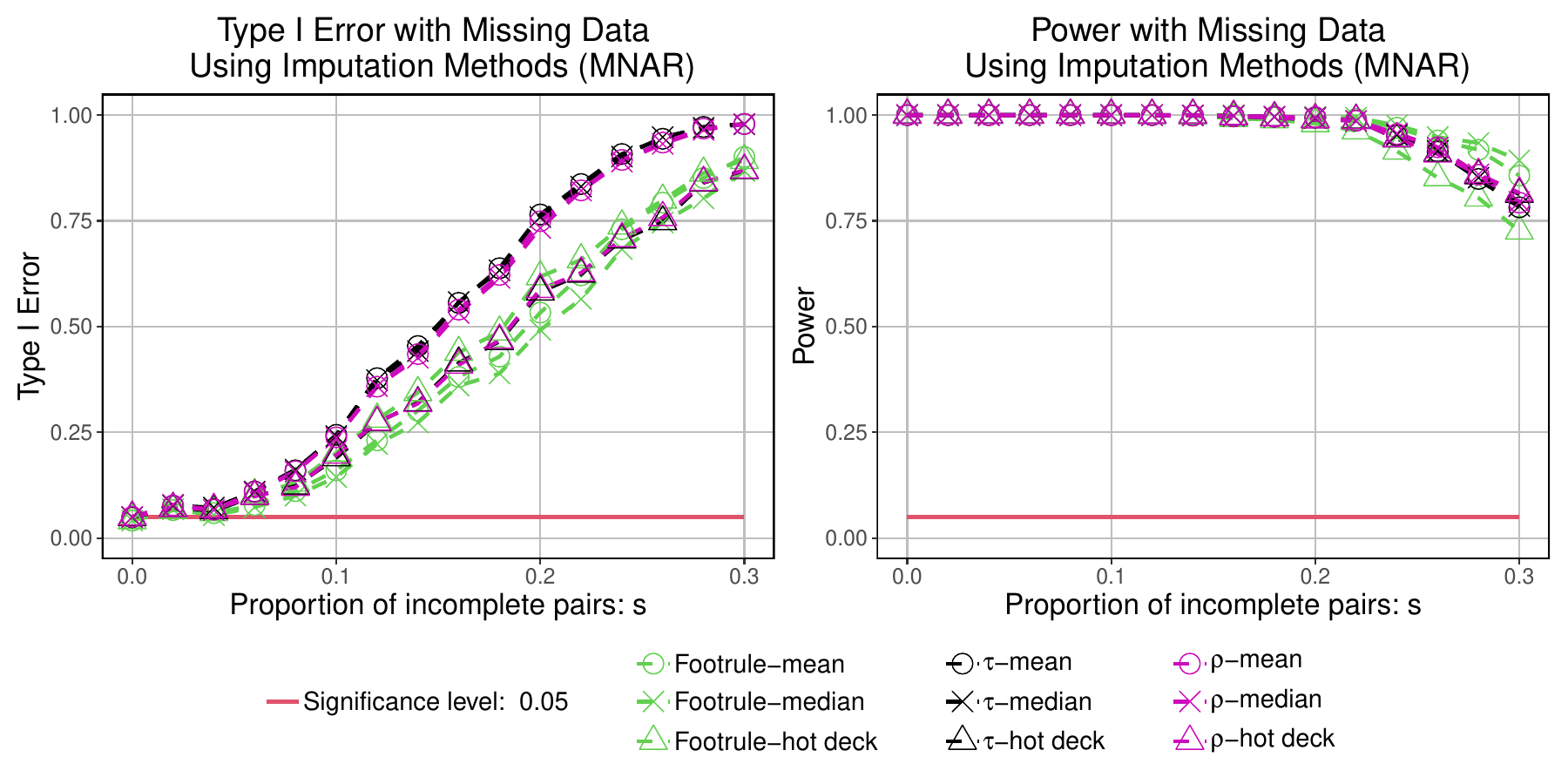}
            \vspace{-0.3cm}
            \caption{Statistical Type I error and power of imputation methods
            as the proportion of missing data $s$ increases,
            where $s \in \{0.00, 0.02, \ldots, 0.30\}$.
            The imputation methods impute missing data using either mean, median or 
            randomly selected (hot deck) values of observed data.
            These methods are described in Table~\ref{supp:tab:3}.
            The data is missing completely at random (MNAR).
            (Left) Type I error: $(\bx,\by) \overset{iid}{\sim} N(0, I_2)$;
            (Right) Power: 	$(X, Y) \overset{iid}{\sim} N(0, \Sigma)$, where
            $\Sigma = \begin{pmatrix} 1 &  \corrcoef  \\  \corrcoef & 1 \end{pmatrix}$,
                with covariance coefficient $\corrcoef=0.5$.
                For both figures, a significance level $\alpha = 0.05$ is used and the sample
                size $\n = 200$. The results in the figures are average of 1000 trials.
                \textbf{Note}: this figure is generated following the same approach for generating 
                Figure~5 in the main paper, with a different missingness mechanism defined in \eqref{supp:missingnessmechanism:1}.}
                \label{supp:fig:22}
        \end{figure}

\begin{figure}
            \centering
            \includegraphics[width=13.5cm]{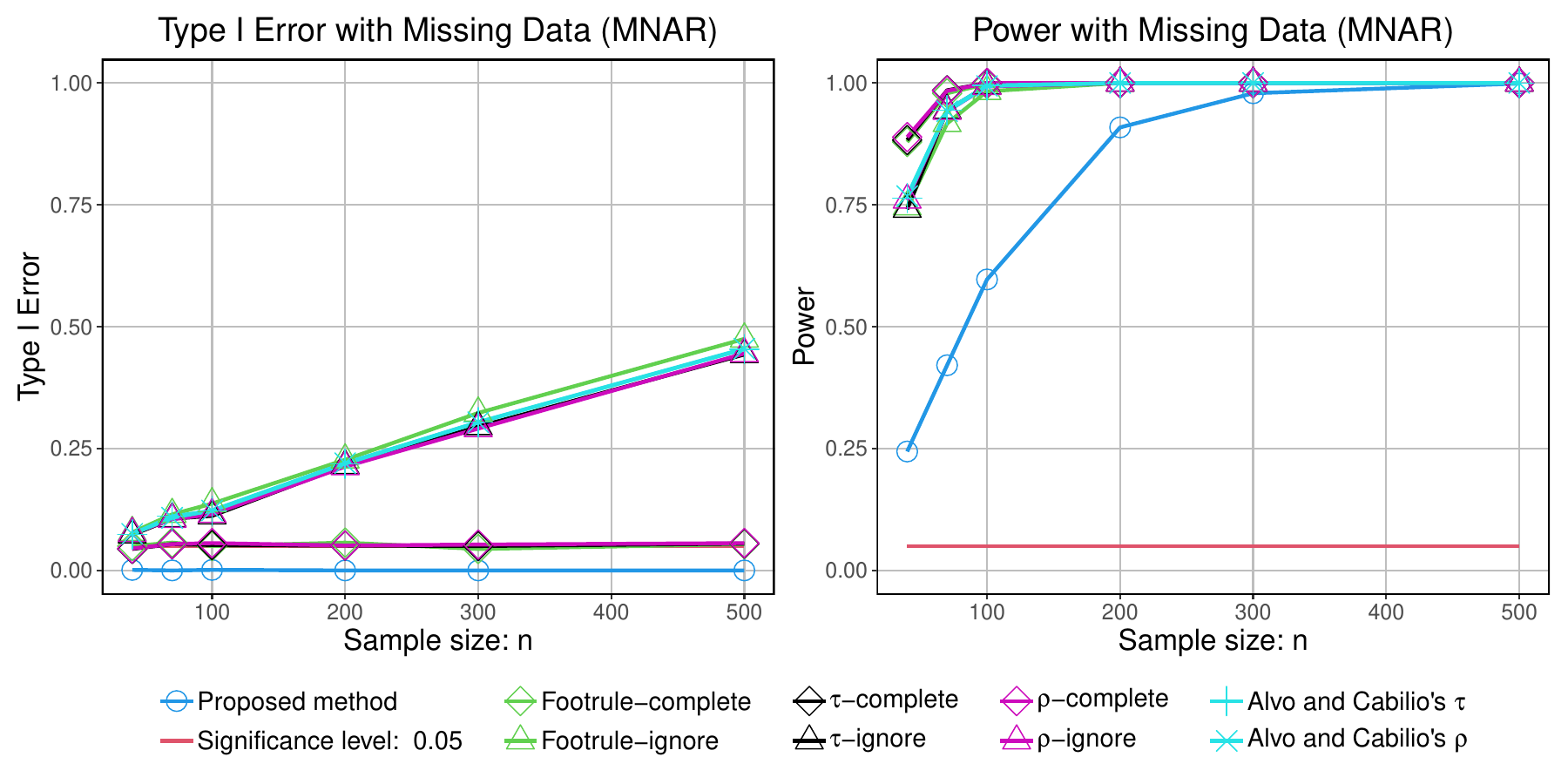}
            \vspace{-0.3cm}
            \caption{
                Statistical Type I error and power of the proposed method,
                methods that ignore missing data or use complete data,
                and Alvo and Cabilio's $\rho$ and Alvo and Cabilio's $\tau$ methods,
                as the sample size $\n$ increases,
                where $\n \in \{40, 70, 100, 200, 300, 500\}$.
                These methods are described in Table~\ref{supp:tab:3}.
                The data is missing not at random (MNAR).
                (Left) Type I error: $(\bx,\by) \overset{iid}{\sim} N(0, I_2)$;
                (Right) Power: 	$(X, Y) \overset{iid}{\sim} N(0, \Sigma)$, where
                $\Sigma = \begin{pmatrix} 1 &  \corrcoef  \\  \corrcoef & 1 \end{pmatrix}$,
                    with covariance coefficient $\corrcoef=0.5$.
                    For both figures, a significance level $\alpha = 0.05$ is used and the proportion
                    of missing pairs $s = 0.1$. The results in the figures are average of 1000 trials.
                    \textbf{Note}: this figure is generated following the same approach for generating 
                    Figure~6 in the main paper, with a different missingness mechanism defined in \eqref{supp:missingnessmechanism:1}.}
                \label{supp:fig:23}

        \end{figure}

        The way that the index set $\bt \subset \{1, \dots, \n\}$ of the missing components 
        is chosen depends on the values of $\bx$ and $\by$.
        Let $q = \sum_{\iconstant=1}^{\n} \indicator{ \left|\rank{\bxn{\iconstant}}{\bx} - \rank{\byn{\iconstant}}{\by}\right| < \n/2 }$
        be the number of pairs of components in $\bx$ and $\by$ such that $\left|\rank{\bxn{\iconstant}}{\bx} - \rank{\byn{\iconstant}}{\by}\right| < \n/2$.
        Each index $\iconstant \in \tonumber{\n}$ 
        is selected to be in the set $\bt$ of indices of missing components according to the 
        following probability conditional on $C = |T|$, the size of the set $T$:
        \begin{align} \label{supp:missingnessmechanism:1}
            p\left( i \in T \, | \, C = \lfloor s \cdot \n  \rfloor  \right) = \left\{ \begin{array}{ll}
                \min\left\{1, sn/q   \right\},~~& \mbox{if}  \left|\rank{\bxn{\iconstant}}{\bx} - \rank{\byn{\iconstant}}{\by}\right| < \n/2,  \\  \max\{0, (sn - q) / (n - q ) \},~~& \mbox{otherwise,} 
            \end{array}\right.
        \end{align}
        for any given $s$. The results are shown in Figure~\ref{supp:fig:21}--\ref{supp:fig:23}.

    \end{appendix}

\clearpage

    \bibliographystyle{plainnat}
    \bibliography{bibliography}       

\begin{thebibliography}{44}
\providecommand{\natexlab}[1]{#1}
\providecommand{\url}[1]{\texttt{#1}}
\expandafter\ifx\csname urlstyle\endcsname\relax
  \providecommand{\doi}[1]{doi: #1}\else
  \providecommand{\doi}{doi: \begingroup \urlstyle{rm}\Url}\fi

\bibitem[Alvo and Cabilio(1995)]{Alvo1995RankCM}
M.~Alvo and P.~Cabilio.
\newblock Rank correlation methods for missing data.
\newblock \emph{The Canadian Journal of Statistics}, 23:\penalty0 345--358,
  1995.

\bibitem[Alvo and Charbonneau(1997)]{alvo1997use}
M.~Alvo and M.~Charbonneau.
\newblock The use of spearman's footrule in testing for trend when the data are
  incomplete.
\newblock \emph{Communications in Statistics-Simulation and Computation},
  26:\penalty0 193--213, 1997.

\bibitem[Bar-Ilan(2005)]{bar2005comparing}
J.~Bar-Ilan.
\newblock Comparing rankings of search results on the web.
\newblock \emph{Information Processing \& Management}, 41\penalty0
  (6):\penalty0 1511--1519, 2005.

\bibitem[Bar-Ilan et~al.(2006)Bar-Ilan, Levene, and Mat-Hassan]{bar2006methods}
J.~Bar-Ilan, M.~Levene, and M.~Mat-Hassan.
\newblock Methods for evaluating dynamic changes in search engine rankings: a
  case study.
\newblock \emph{Journal of Documentation}, 62\penalty0 (6):\penalty0 708--729,
  2006.

\bibitem[Baraldi and Enders(2010)]{baraldi2010introduction}
A.~N. Baraldi and C.~K. Enders.
\newblock An introduction to modern missing data analyses.
\newblock \emph{Journal of School Psychology}, 48:\penalty0 5--37, 2010.

\bibitem[Bennett(2001)]{Derrick2009How}
D.~A. Bennett.
\newblock How can i deal with missing data in my study?
\newblock \emph{Australian and New Zealand Journal of Public Health},
  25:\penalty0 464--469, 2001.

\bibitem[Bennett(2020)]{bennett2020changes}
V.~M. Bennett.
\newblock Changes in persistence of performance over time.
\newblock \emph{Strategic Management Journal}, 41:\penalty0 1745--1769, 2020.

\bibitem[Brandenburg et~al.(2013)Brandenburg, Glei{\ss}ner, and
  Hofmeier]{Brandenburg2013The}
F.~J. Brandenburg, A.~Glei{\ss}ner, and A.~Hofmeier.
\newblock The nearest neighbor spearman footrule distance for bucket, interval,
  and partial orders.
\newblock \emph{Journal of Combinatorial Optimization}, 26:\penalty0 310--332,
  2013.

\bibitem[Cabilio and Tilley(1999)]{cabilio1999power}
P.~Cabilio and J.~Tilley.
\newblock Power calculations for tests of trend with missing observations.
\newblock \emph{Environmetrics}, 10:\penalty0 803--816, 1999.

\bibitem[Chen et~al.(2023)Chen, Xu, Zhang, Zhu, and Dai]{chen2023asymptotic}
C.~Chen, W.~Xu, W.~Zhang, H.~Zhu, and J.~Dai.
\newblock Asymptotic properties of spearman’s footrule and gini’s gamma in
  bivariate normal model.
\newblock \emph{Journal of the Franklin Institute}, 360:\penalty0 9812--9843,
  2023.

\bibitem[Cook and Zea(2020)]{cookzea2020}
T.~Cook and R.~Zea.
\newblock Missing data and sensitivity analysis for binary data with
  implications for sample size and power of randomized clinical trials.
\newblock \emph{Statistics in Medicine}, 39\penalty0 (2):\penalty0 192--204,
  2020.

\bibitem[Dempster et~al.(1977)Dempster, Laird, and Rubin]{1977Maximum}
A.~P. Dempster, N.~M. Laird, and D.~B. Rubin.
\newblock Maximum likelihood from incomplete data via the \emph{EM} algorithm.
\newblock \emph{Journal of the Royal Statistical Society: Series B
  (Methodological)}, 39:\penalty0 1--22, 1977.

\bibitem[Diaconis and Graham(1977)]{diaconis1977spearman}
P.~Diaconis and R.~L. Graham.
\newblock Spearman's footrule as a measure of disarray.
\newblock \emph{Journal of the Royal Statistical Society: Series B
  (Methodological)}, 39:\penalty0 262--268, 1977.

\bibitem[Dong and Peng(2013)]{Dong2013PrincipledMD}
Y.~Dong and C.-Y.~J. Peng.
\newblock Principled missing data methods for researchers.
\newblock \emph{SpringerPlus}, 2:\penalty0 1--7, 2013.

\bibitem[Dwork et~al.(2001)Dwork, Kumar, Naor, and Sivakumar]{dwork2001rank}
C.~Dwork, R.~Kumar, M.~Naor, and D.~Sivakumar.
\newblock Rank aggregation methods for the web.
\newblock In \emph{Proceedings of the 10th International Conference on World
  Wide Web}, pages 613--622, 2001.

\bibitem[Fagin et~al.(2003)Fagin, Kumar, and Sivakumar]{fagin2003comparing}
R.~Fagin, R.~Kumar, and D.~Sivakumar.
\newblock Comparing top k lists.
\newblock \emph{SIAM Journal on Discrete Mathematics}, 17:\penalty0 134--160,
  2003.

\bibitem[Genest et~al.(2010)Genest, Ne{\v{s}}lehov{\'a}, and
  Ben~Ghorbal]{genest2010spearman}
C.~Genest, J.~Ne{\v{s}}lehov{\'a}, and N.~Ben~Ghorbal.
\newblock Spearman's footrule and gini's gamma: a review with complements.
\newblock \emph{Journal of Nonparametric Statistics}, 22:\penalty0 937--954,
  2010.

\bibitem[Goldberg et~al.(2021)Goldberg, Bolt, and Davidson]{goldberg2021data}
S.~B. Goldberg, D.~M. Bolt, and R.~J. Davidson.
\newblock Data missing not at random in mobile health research: Assessment of
  the problem and a case for sensitivity analyses.
\newblock \emph{Journal of Medical Internet Research}, 23\penalty0
  (6):\penalty0 e26749, 2021.

\bibitem[Heymans and Twisk(2022)]{heymans2022handling}
M.~W. Heymans and J.~W.~R. Twisk.
\newblock Handling missing data in clinical research.
\newblock \emph{Journal of Clinical Epidemiology}, 151:\penalty0 185--188,
  2022.

\bibitem[Horowitz and Manski(2000)]{horowitzmanski2000}
J.~L. Horowitz and C.~F. Manski.
\newblock Nonparametric analysis of randomized experiments with missing
  covariate and outcome data.
\newblock \emph{Journal of the American Statistical Association}, 95\penalty0
  (449):\penalty0 77--84, 2000.

\bibitem[Kendall(1948)]{kendall1948rank}
M.~G. Kendall.
\newblock \emph{Rank correlation methods.}
\newblock Griffn, London, 4 edition, 1948.

\bibitem[Kim et~al.(2004)Kim, Rha, Cho, and Chung]{kim2004spearman}
B.~S. Kim, S.~Y. Rha, G.~B. Cho, and H.~C. Chung.
\newblock Spearman's footrule as a measure of cdna microarray reproducibility.
\newblock \emph{Genomics}, 84:\penalty0 441--448, 2004.

\bibitem[Kumar and Vassilvitskii(2010)]{2010Generalized}
R.~Kumar and S.~Vassilvitskii.
\newblock Generalized distances between rankings.
\newblock In \emph{Proceedings of the 19th International Conference on World
  Wide Web}, pages 571--580, 2010.

\bibitem[Lin(2010)]{lin2010rank}
S.~Lin.
\newblock Rank aggregation methods.
\newblock \emph{Wiley Interdisciplinary Reviews: Computational Statistics},
  2:\penalty0 555--570, 2010.

\bibitem[Little and Rubin(2019)]{little2019statistical}
R.~J.~A. Little and D.~B. Rubin.
\newblock \emph{Statistical Analysis with Missing Data}, volume 793.
\newblock John Wiley \& Sons, Hoboken, 3 edition, 2019.

\bibitem[Loukas and Papaioannou(1991)]{loukas1991rank}
S.~Loukas and T.~Papaioannou.
\newblock Rank correlation inequalities with ties and missing data.
\newblock \emph{Statistics \& Probability Letters}, 11:\penalty0 53--56, 1991.

\bibitem[Luigi~Conti and Nikitin(1999)]{luigi1999asymptotic}
P.~Luigi~Conti and Y.~Nikitin.
\newblock Asymptotic efficiency of independence tests based on gini's rank
  association coefficient, spearman's footrule and their generalizations.
\newblock \emph{Communications in Statistics-Theory and Methods}, 28:\penalty0
  453--465, 1999.

\bibitem[Madley-Dowd et~al.(2019)Madley-Dowd, Hughes, Tilling, and
  Heron]{Madley2019The}
P.~Madley-Dowd, R.~Hughes, K.~Tilling, and J.~Heron.
\newblock The proportion of missing data should not be used to guide decisions
  on multiple imputation.
\newblock \emph{Journal of Clinical Epidemiology}, 110:\penalty0 63--73, 2019.

\bibitem[Papaioannou and Loukas(1984)]{papaioannou1984inequalities}
T.~Papaioannou and S.~Loukas.
\newblock Inequalities on rank correlation with missing data.
\newblock \emph{Journal of the Royal Statistical Society: Series B
  (Methodological)}, 46:\penalty0 68--71, 1984.

\bibitem[Parzen et~al.(2010)Parzen, Lipsitz, Metters, and
  Fitzmaurice]{ParzenCorrelation}
M.~Parzen, S.~Lipsitz, R.~Metters, and G.~Fitzmaurice.
\newblock Correlation when data are missing.
\newblock \emph{Journal of the Operational Research Society}, 61:\penalty0
  1049--1056, 2010.

\bibitem[Powell and Reinhardt(2010)]{powell2010rank}
T.~C. Powell and I.~Reinhardt.
\newblock Rank friction: an ordinal approach to persistent profitability.
\newblock \emph{Strategic Management Journal}, 31:\penalty0 1244--1255, 2010.

\bibitem[Pérez and Prieto-Alaiz(2016)]{P2016Measuring}
A.~Pérez and M.~Prieto-Alaiz.
\newblock Measuring the dependence among dimensions of welfare: A study based
  on spearman's footrule and gini's gamma.
\newblock \emph{International Journal of Uncertainty, Fuzziness and
  Knowledge-Based Systems}, 24:\penalty0 87--105, 2016.

\bibitem[Quade and Salama(2006)]{quade2006concordance}
D.~Quade and I.~A. Salama.
\newblock Concordance of complete or right-censored rankings based on
  spearman's footrule.
\newblock \emph{Communications in Statistics-Theory and Methods}, 35:\penalty0
  1059--1069, 2006.

\bibitem[Raykov et~al.(2014)Raykov, Schneider, Marcoulides, and
  Lichtenberg]{2014Examining}
T.~Raykov, B.~C. Schneider, G.~A. Marcoulides, and P.~A. Lichtenberg.
\newblock Examining measure correlations with incomplete data sets.
\newblock \emph{Structural Equation Modeling: A Multidisciplinary Journal},
  21:\penalty0 318--324, 2014.

\bibitem[Salama and Quade(1990)]{salama1990note}
I.~A. Salama and D.~Quade.
\newblock A note on spearman's footrule.
\newblock \emph{Communications in Statistics-Simulation and Computation},
  19\penalty0 (2):\penalty0 591--601, 1990.

\bibitem[Salama and Quade(2004)]{salama2004agreement}
I.~A. Salama and D.~Quade.
\newblock Agreement among censored rankings using spearman's footrule.
\newblock \emph{Communications in Statistics-Theory and Methods}, 33:\penalty0
  1837--1850, 2004.

\bibitem[Schafer(1999)]{Schafer1999Multiple}
J.~L. Schafer.
\newblock Multiple imputation: a primer.
\newblock \emph{Statistical Methods in Medical Research}, 8:\penalty0 3--15,
  1999.

\bibitem[Schafer and Graham(2002)]{2002Missing}
J.~L. Schafer and J.~W. Graham.
\newblock Missing data: Our view of the state of the art.
\newblock \emph{Psychological Methods}, 7:\penalty0 147--177, 2002.

\bibitem[Scheffer(2002)]{scheffer2002dealing}
J.~Scheffer.
\newblock Dealing with missing data.
\newblock \emph{Research Letters in the Information and Mathematical Sciences},
  3\penalty0 (1):\penalty0 153--160, 2002.

\bibitem[Sen et~al.(2003)Sen, Salama, and Quade]{sen2003spearman}
P.~Sen, I.~Salama, and D.~Quade.
\newblock Spearman's footrule under progressive censoring.
\newblock \emph{Journal of Nonparametric Statistics}, 15:\penalty0 53--60,
  2003.

\bibitem[Smuk(2015)]{smuk2015}
M.~Smuk.
\newblock \emph{Missing data methodology: sensitivity analysis after multiple
  imputation}.
\newblock PhD thesis, London School of Hygiene \& Tropical Medicine, 2015.

\bibitem[Spearman(1906)]{spearman1906footrule}
C.~Spearman.
\newblock Footrule for measuring correlation.
\newblock \emph{British Journal of Psychology}, 2:\penalty0 89, 1906.

\bibitem[Speevak(2017)]{speevak2017inequalities}
T.~Speevak.
\newblock Inequalities for sums of squares of reranked differences involving
  ties and missing data.
\newblock \emph{Communications in Statistics-Theory and Methods}, 46:\penalty0
  8419--8429, 2017.

\bibitem[Thabane et~al.(2013)Thabane, Mbuagbaw, Zhang, Samaan, Marcucci, Ye,
  Thabane, Giangregorio, Dennis, and Kosa]{thabaneetal2013}
L.~Thabane, L.~Mbuagbaw, S.~Zhang, Z.~Samaan, M.~Marcucci, C.~Ye, M.~Thabane,
  L.~Giangregorio, B.~Dennis, and D.~et~al. Kosa.
\newblock A tutorial on sensitivity analyses in clinical trials: the what, why,
  when and how.
\newblock \emph{BMC medical research methodology}, 13:\penalty0 1--12, 2013.

\end{thebibliography}

\end{document}